\documentclass[aps,rmp,reprint,amsmath,amssymb,graphicx,floatfix]{revtex4-2}

\usepackage{graphicx}	
\usepackage{mathtools}
\usepackage{braket}
\usepackage{physics}
\usepackage[table,xcdraw]{xcolor}
\usepackage{amssymb}
\usepackage[normalem]{ulem}
\usepackage{makecell}

\newcommand{\BKFA}{Ba$_{1-x}$K$_{x}$Fe$_2$As$_2$}
\newcommand{\KFA}{KFe$_2$As$_2$}
\newcommand{\BRFA}{Ba$_{1-x}$Rb$_{x}$Fe$_2$As$_2$}

\newcommand{\BFCA}{Ba(Fe$_{1-x}$Co$_{x})_2$As$_2$}
\newcommand{\BFAP}{BaFe$_2$(As$_{1-x}$P$_x)_2$}
\newcommand{\BFA}{BaFe$_2$As$_2$}

\newcommand{\FSS}{FeSe$_{1-x}$S$_{x}$}
\newcommand{\FST}{FeSe$_{1-x}$Te$_{x}$}

\begin{document}

\title{Iron-Based Superconductors: A Decade of Materials, Magnetism, and Mechanisms}

\begin{abstract}
Since its discovery in 2008, iron-based superconductors (FeSCs) 
have become a central platform for exploring high-temperature superconductivity in multiband, electron-correlated materials. This review focuses on major developments over the past decade or so, emphasizing experimental advances, pairing mechanisms, and emerging applications. Structural tuning through chemical substitution, pressure, and epitaxial growth enables precise control of the electronic, magnetic, and superconducting ground states, thereby revealing their interplay. In particular, the electronic nematic phase and stripe-type antiferromagnetic order-often coexisting or competing-are central to understanding the phase diagrams. Spin waves in magnetically ordered parent compounds and spin excitations (fluctuations) in doped superconductors are extensively characterized by inelastic neutron scattering. While high-energy spin excitations in doped superconductors retain substantial spectral weight across a wide energy range reminiscent of spin waves in their undoped parents, the low-energy response reveals a collective spin excitation termed ``resonance'' coupled to superconductivity. 
The momentum structure of superconductivity-induced resonance provides strong evidence for sign-changing pairing in many FeSCs, while disorder effects, orbital-fluctuation scenarios, quasiparticle damping, and compound-dependent gap structures indicate that $s_{\pm}$, $s_{++}$, nodal $s$, $d$-wave, and multicomponent states must be discussed in a material-specific framework.
Advances in thin-film growth, intercalation chemistry, and interface engineering-particularly in FeSe-based systems-have enabled enhanced $T_{c}$ and novel device geometries. With high upper critical fields, moderate anisotropy, and improving current densities, FeSCs continue to drive both fundamental insight and technological applications in superconductivity.
\end{abstract}

\author{Xingye Lu\textsuperscript{\S}}
\email{luxy@bnu.edu.cn}

\affiliation{Center for Advanced Quantum Studies, School of Physics and Astronomy, Beijing Normal University, and Key Laboratory of Multiscale Spin Physics (Beijing Normal University), Ministry of Education, Beijing 100875, China}

\author{Hechang Lei\textsuperscript{\S}}
\email{hlei@ruc.edu.cn}
\affiliation{School of Physics and Key Laboratory of Quantum State Construction and Manipulation (Ministry of Education), Renmin University of China, Beijing 100872, China}

\author{Jun Zhao}
\email{zhaoj@fudan.edu.cn}
\affiliation{State Key Laboratory of Surface Physics and Department of Physics, Fudan University, Shanghai 200433, China}

\author{Hideo Hosono}
\email{hosono@mces.titech.ac.jp}
\affiliation{MDX Research Center for Element Strategy, Institute of Science Tokyo, Nagatsuta, Midori, Yokohama 226-8501, Japan \& Department of Physics, Korea University, Anam-dong 5, Seongbuk-gu, Seoul 02841, Republic of Korea}

\author{Pengcheng Dai}
\email{pdai@rice.edu}
\affiliation{Department of Physics and Astronomy, Rice Laboratory for Emergent Magnetic Materials and Smalley-Curl Institute, Rice University, Houston, TX 77005, USA}

\maketitle

\tableofcontents

\section{Introduction}

The discovery in 2008 of superconductivity (SC) at $T_{c}=26$~K in fluorine-doped LaFeAsO \cite{Kamihara2008,Kamihara2006iron} and the rapid rise of $T_{c}$ above $50$~K in related rare-earth oxypnictides quickly established the iron-based superconductors (FeSCs) as a second high-temperature superconducting family alongside the cuprates \cite{WOS:000349190300029}. Within months, superconductivity was realized across several crystal families by charge doping, isovalent substitution, and pressure, including the iron-pnictide FePn-122 (AEFe$_2$As$_2$, where AE = Ba, Sr, Ca), FePn-111 ({\it A}FeAs, where {\it A} = Li, Na), and iron-chalcogenide FeCh-11 (FeSe/FeTe) systems \cite{Ren2008,Chen2008a,Chen2008b,Hsu2008} [Fig.~\ref{Fig_structure}]. Subsequent breakthroughs-alkali-intercalated selenides and ammonia/organic-intercalated FeSe derivatives \cite{Guo2010,Burrard-Lucas2013}, and monolayer FeSe on SrTiO$_3$ with dramatically enhanced $T_{c}$ \cite{Liu2012monoFeSe}-demonstrated the unusual tunability of superconductivity in this materials platform.

The structural nomenclature used throughout this Review is summarized in Fig.~\ref{Fig_structure}. The simplest Fe-pnictide families are FePn-111 ($A$FeAs), FePn-122 (AEFe$_2$As$_2$), and FePn-1111 (LnOFeAs), followed by structurally derived or cation-ordered variants including FePn-112, FePn-1144, FePn-12244, FePn-32522, FePn-42622, FePn-22241, FePn-1048, and FePn-1038 [Fig.~\ref{Fig_structure}(a)--(k); Sec.~II.2]. The corresponding Fe-chalcogenide families comprise FeCh-11 (FeSe/FeTe), alkali- or Tl-intercalated $A$FeCh-122 compounds, hydroxide-intercalated FeCh-11111 materials, ammonia-containing $M$NHFeCh-122 compounds, and organic-intercalated FeCh-Org systems [Fig.~\ref{Fig_structure}(l)--(p); Secs.~II.2 and III.A.1]. Here, the numerical labels identify the stoichiometric ratios of the principal structural blocks, whereas FePn and FeCh denote Fe-pnictide and Fe-chalcogenide compounds, respectively.

Despite their chemical variety, FeSCs share a universal structural unit: nearly square nets of Fe coordinated tetrahedrally by pnictogen (Pn) or chalcogen (Ch) anions. This layered motif yields quasi-two-dimensional (quasi-2D), multiband electronic structures with multiple hole and electron Fermi-surface sheets derived primarily from the Fe $3d$ orbitals. The low-energy band topology and electron correlation strength are exquisitely sensitive to structural parameters such as the anion height above the Fe plane and the Fe--X--Fe (X = Pn/Ch) bond angle, which can be tuned by chemical substitution, pressure, strain, and intercalation. Empirical trends linking these geometric parameters to superconducting $T_{c}$ provided early materials-design heuristics and continue to guide the search for new compounds and optimized compositions \cite{Hosono2015,Hosono2016}.

A central theme of FeSC physics is the intimate interplay among magnetism, electronic nematicity, and superconductivity \cite{fernandes2014what,bohmer2022nematicity,si2023iron}. The undoped FePn-1111 and FePn-122 parent compounds are metallic antiferromagnets with a stripe-type magnetic [or spin-density-wave (SDW)] order that is often preceded or accompanied by a tetragonal-to-orthorhombic lattice distortion associated with an electronic nematic phase. Suppression of these orders by doping or pressure typically produces a superconducting dome and, in many families, a region where nematic fluctuations, magnetic fluctuations, and superconductivity coexist and compete \cite{dai2012magnetism,dai2015antiferromagnetic,yi2017role}. In FeSe-based materials, nematic order emerges in the absence of long-range magnetism at ambient pressure, offering a particularly clean venue to disentangle orbital, spin, and lattice degrees of freedom and to benchmark theory against spectroscopic measurements \cite{coldea2018thekey,coldea2021electronic}.

In terms of the pairing problem, the multiband nature of FeSCs allows several competing superconducting channels. A sign-changing $s$-wave state ($s_{\pm}$), in which the order parameter reverses the sign between hole and electron pockets near the $\Gamma$ and $M$ points, respectively \cite{dai2012magnetism,dai2015antiferromagnetic,yi2017role}, naturally arises from repulsive inter-band interactions mediated by spin fluctuations and has strong support across many materials \cite{Hirschfeld2011,si2016high}. At the same time, material-specific details--such as Fermi-surface reconstructions that remove hole pockets, the strength of orbital fluctuations, and the degree of electronic correlation--can tip the balance toward alternative gap structures, including highly anisotropic or nodal $s$-wave states and, in special cases, $d$-wave-like states. This diversity, now well documented by 
angle-resolved photoemission spectroscopy (ARPES), quasiparticle-interference (QPI) imaging, thermodynamic/transport probes, and inelastic neutron scattering (INS), is both a challenge and an opportunity: it ties the superconducting gap symmetry directly to the underlying normal-state electronic structure and fluctuations \cite{dai2015antiferromagnetic,yi2017role,Hirschfeld2011,si2016high}.

The flexibility of the FeSC platform has enabled several frontiers. First, interface and low-dimensional engineering (e.g., FeSe/SrTiO$_3$) has revealed avenues to enhance pairing via interfacial modes and forward-focused interactions \cite{Liu2012monoFeSe}. Second, Fe(Te,Se) has emerged as a candidate host of topological superconductivity and Majorana bound states due to nontrivial band inversions and strong spin--orbit coupling, stimulating intense activity at the nexus of topology and unconventional pairing \cite{wang2018evidence}. Third, progress in synthesis-ranging from high-pressure routes and flux growth to epitaxy and soft-chemistry intercalation-has expanded the accessible phase space and delivered crystals and films of the quality needed for decisive spectroscopic and bulk measurements \cite{Hosono2015,Hosono2016}.

Beyond fundamental interest, FeSCs exhibit several attributes favorable for applications: high upper critical fields ($H_{c2}$), relatively modest in- and out-of-plane
anisotropy compared with cuprates, and, in many compositions, robustness against disorder. Steady advances in wires, tapes, and coated conductors-especially within the 122 and FeSe-derived families-have improved grain connectivity, chemical stability, and critical current density, while thin-film heterostructures have opened routes toward devices that leverage high upper critical field $\mu_{0}H_{c2}$ and interfacial effects.

\textit{Scope and organization.} This Review emphasizes a materials-to-mechanism perspective on FeSCs. Section~II discusses the universal structural unit and the structural diversity across the major families, highlighting how geometric parameters correlate with the electronic structure and $T_{c}$. Section~III surveys materials design and synthesis, including chemical substitution, pressure, intercalation, and epitaxial growth strategies. Section~IV highlights application-relevant properties and the recent progress in coated conductors, superconducting wires, and bulk magnets. Section~V focuses on intertwined orders and exotic phases--nematicity, magnetism, and topology--and their evolution with tuning parameters. Section~VI reviews the superconducting gap symmetry and pairing mechanisms, synthesizing results from bulk, spectroscopic, and theoretical studies. Section~VII summarizes antiferromagnetic (AFM) spin excitations and their connection to superconductivity. 
Section~VIII concludes with an outlook on open questions and promising directions.

\section{Lattice Structure}

\subsubsection{Universal structural unit of FeSCs and influence on superconductivity}\label{sec_structure_tc}

\begin{figure}
\centering
\includegraphics[width=8 cm]{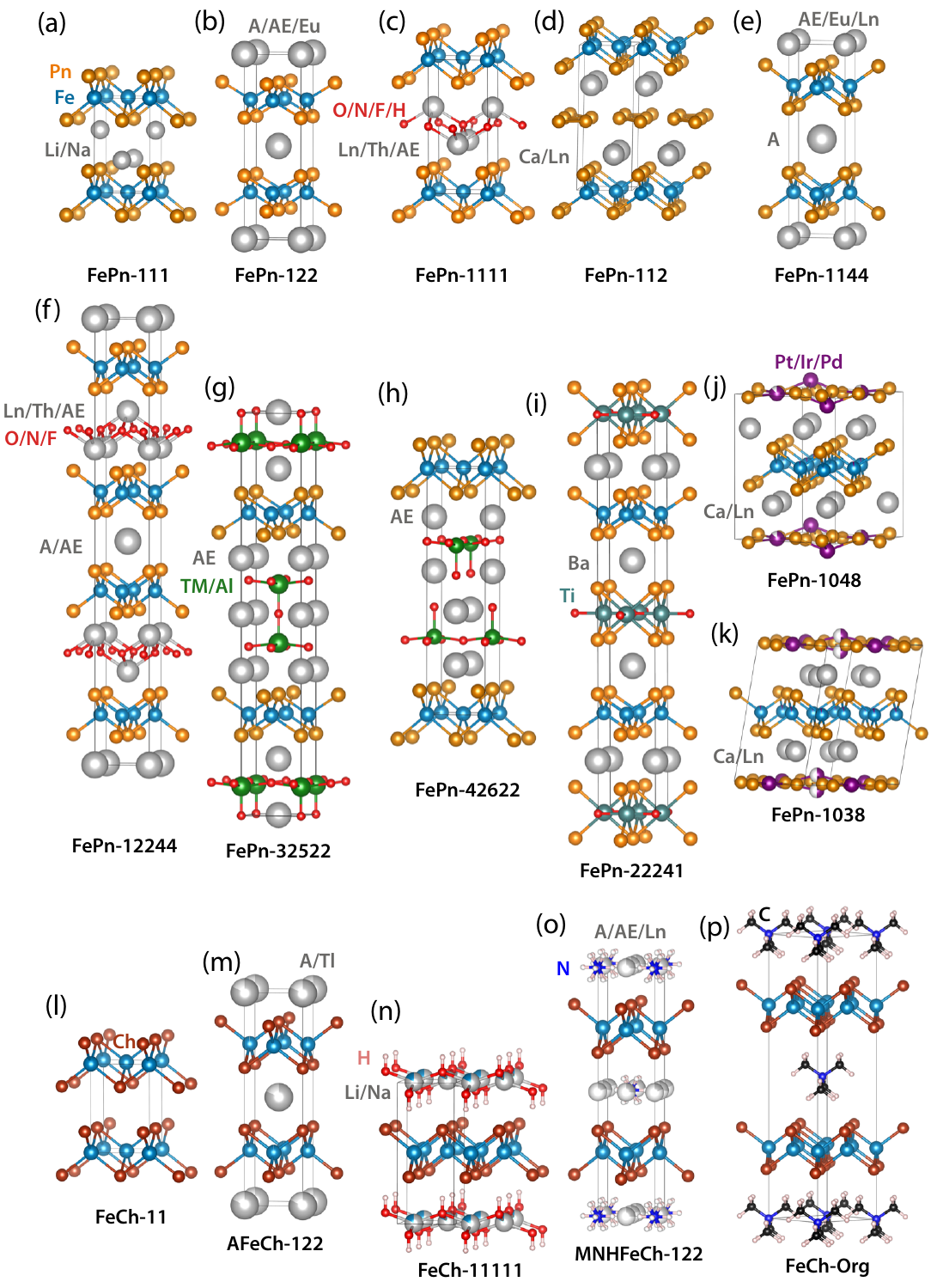}
\caption[]{Crystal structures and structural-family notation of
representative iron-based superconductors (FeSCs). The numerical labels in the family names denote the stoichiometric ratios of the principal structural blocks in the corresponding compounds. Here, $A$ denotes an alkali metal (Li, Na, K,
Rb, or Cs); AE an alkaline-earth metal (Ca, Sr, or Ba); Ln a lanthanide
element; TM a transition-metal element; $M$ metallic ion in $M$NHFeCh-122; Pn a pnictogen (P or As); Ch a
chalcogen (S, Se, or Te); and Org an organic molecular spacer.
Panels (a)--(p) show representative members of the major FeSC
structural families.
\label{Fig_structure}}
\end{figure}

\begin{figure}
\centering
\includegraphics[width=8.5cm]{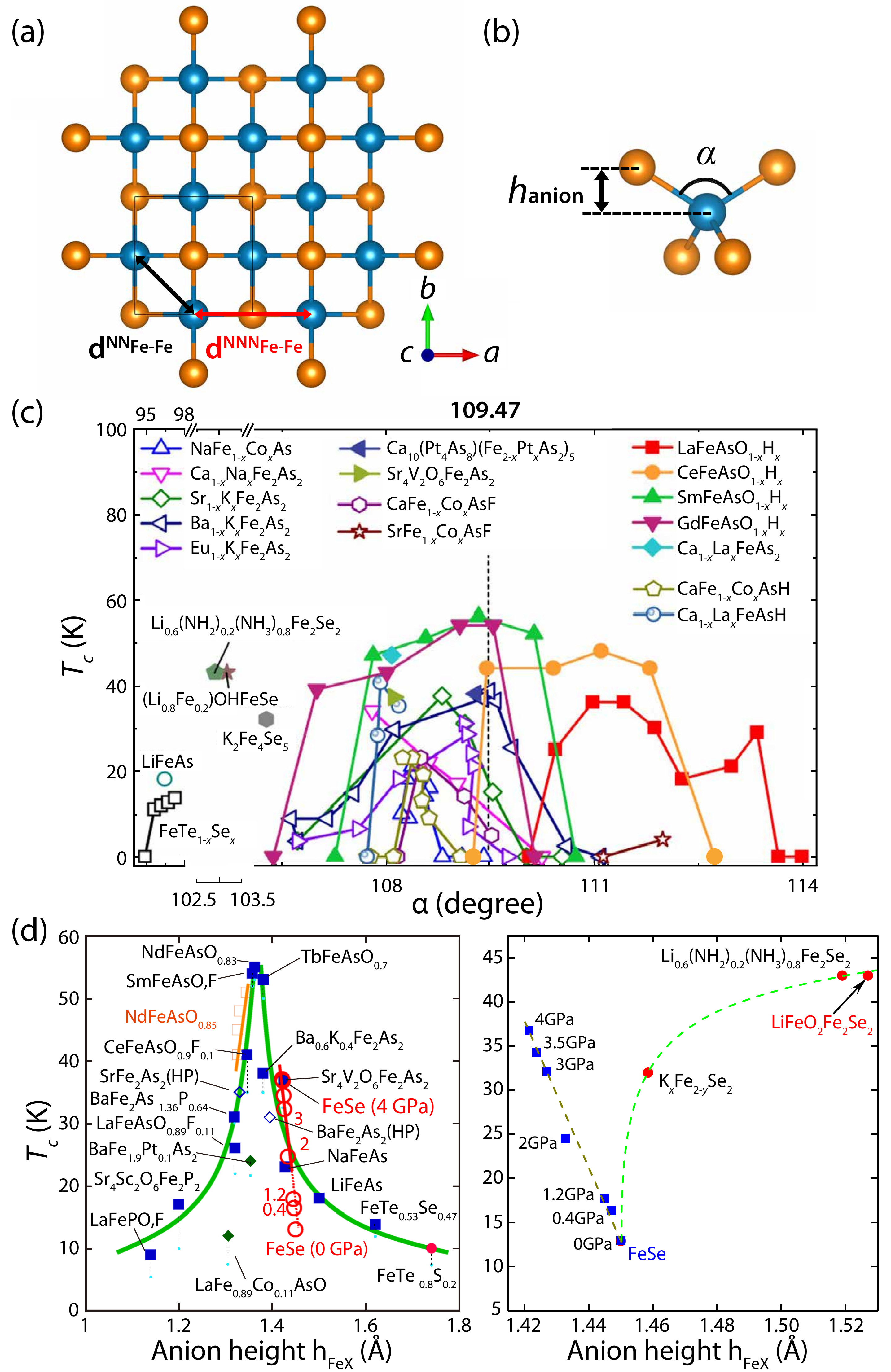}
\caption[]{(a) Local structure of 2D Fe layer with highlighted nearest and next-nearest neighbor Fe--Fe atomic distance	$d^{\rm {NN}}_{\rm {Fe--Fe}}$ and $d^{\rm {NNN}}_{\rm {Fe--Fe}}$. (b) Definitions of X anion height $h_{\rm FeX}$ and X--Fe--X bond angle $\alpha$. (c) Correlation between $T_{c}$ and $\alpha$ in various FeSCs. The dotted line denotes the bond angle for a regular tetrahedron.  From \cite{Hosono2015}. (d) $T_{c}$ as a function of $h_{\rm FeX}$ for representative FeSCs. From \cite{Mizuguchi2010,Lu2013}.
\label{Fig_Structure_sc_relation}}
\end{figure}

Like the cuprates \cite{WOS:000349190300029}, FeSCs are layered materials composed of FePn/FeCh conducting layers and spacer layers. So far, hundreds of FeSCs have been discovered, and they can be classified into more than a dozen types in general, as shown in Fig.~\ref{Fig_structure}. The key superconducting structural unit of FeSCs is composed of FePn or FeCh layers, where Fe atoms form a 2D square lattice, and each Fe atom is coordinated by four X atoms in a tetrahedral arrangement and FeX$_{4}$ tetrahedra are connected to each other by edge sharing [Fig.~\ref{Fig_Structure_sc_relation}(a)]. This structural configuration resembles that of Li$_{2}$O with anti-CaF$_{2}$-type structure. From the chemical point of view, the bonding between Fe and X atoms is not purely ionic. In most FeSCs, the $3d$ orbitals of Fe$^{2+}$ hybridize with the $p$ orbitals of Pn$^{3-}$ or Ch$^{2-}$ through polar covalent bonds. This leads to much weaker crystal-field splitting in FeSCs than in the cuprates. The weak crystal-field splitting places multiple Fe $3d$ orbitals near the Fermi level $E_{\rm F}$. Consequently, the multiorbital character strongly affects the physical properties of FeSCs, especially their superconductivity \cite{Ma2008,Kuroki2009a,Graser2009,Zhang2010,Kontani2010,Hosono2015,yi2013observation,yi2015observation,yi2017role,si2008strong}.

In the nearly square two-dimensional Fe lattice, the nearest-neighbor (NN) and next-nearest neighbor (NNN) Fe--Fe atomic distances $d^{\rm {NN}}_{\rm {Fe--Fe}}$ and $d^{\rm {NNN}}_{\rm {Fe--Fe}}$ are approximately $2.7 -- 2.8$ \AA\ and $3.8 -- 4.0$ \AA, respectively. The X atoms are located above and below the Fe plane with X height $h_{\rm FeX}$ of about $1.2 -- 1.8$ \AA\ and the X--Fe--X bond angle $\alpha$ distributed over a wide range of $95^{\circ} -- 125^{\circ}$ [Fig.~\ref{Fig_Structure_sc_relation}(b)]. 
The polar covalency of the Fe-X bond makes the geometry of the FeX$_4$ tetrahedron a key control parameter for superconductivity, magnetism, electronic correlation strength, and, in some cases, the topology of the bands.
Microscopically, variations in $h_{\rm FeX}$, $\alpha$, and Fe--Fe distances modify the hybridization between Fe $3d$ and Pn/Ch $p$ orbitals, thereby changing the relative energies and orbital characters of the low-energy Fe-derived bands \cite{Kuroki2009a,Hosono2015,yin2011kinetic}. This orbital sensitivity provides the basis for understanding why structural tuning affects
superconductivity, magnetism, electronic correlations, and nematicity.

For superconducting properties, there are some empirical correlations between the structural parameters and the superconducting transition temperature $T_{c}$ in FeSCs. First, the maximum $T_{c}$ is achieved when the FeX$_{4}$ tetrahedra are perfectly regular ($\alpha \approx 109.47^{\circ}$) [Fig.~\ref{Fig_Structure_sc_relation}(c)] \cite{WOS:000261127100016,Lee2008,Lee2012,Hosono2015}. 
Second, it is found that the $T_{c}$ depends on the anion height $h_{\rm FeX}$ non-monotonically and the highest $T_{c}$ is obtained when $h_{\rm FeX}$ $\sim$ 1.38 \AA\ [Fig.~\ref{Fig_Structure_sc_relation}(d)] \cite{Mizuguchi2010}. Kuroki \textit{et al.} proposed that because the position of the $d_{xy}$ band is closely linked to the $h_{\rm FeX}$ in FePn-1111 system, the change of $h_{\rm FeX}$ can effectively tune the strength of spin fluctuations and thus the $T_{c}$ can be enhanced when increasing $h_{\rm FeX}$ \cite{Kuroki2009a,Hosono2015}. It should be noted that there are some exceptions to these empirical rules. For example, the $T_{c}$ decreases as $h_{\rm FeX}$ approaches 1.38 \AA\ for the electron-doped FeCh-based superconductors [Fig.~\ref{Fig_Structure_sc_relation}(d)] \cite{Lu2013}. Moreover, for (La/Ce)(O,H)FeAs and Fe(Se,Te), the bond angles at optimal $T_c$ deviate substantially from 109.47 $^{\circ}$ [Fig.~\ref{Fig_Structure_sc_relation}(c)] \cite{Lee2012,Iimura2012,Hosono2015}.
These discrepancies suggest that the $T_{c}$ is not determined solely by the local structure of FeX$_{4}$ and/or the underlying physics is somewhat different in FeCh-based superconductors compared to FePn-based ones.

The diversity of FeSC materials provides a broad parameter space for tuning the local geometry of the FeX layer, including the Fe--Fe distance, the anion height $h_{\rm FeX}$, and the X--Fe--X bond angle $\alpha$. These structural parameters modify the hybridization between Fe 3$d$ and pnictogen/chalcogen $p$ orbitals, thereby changing the orbital characters and relative energy levels of the low-energy Fe-derived bands. Together with spin--orbit coupling, such structural fine tuning can invert bands with opposite parities and drive a topological phase transition \cite{hao2014topological,Hao2015,wang2015nematicity,Hao2019}. Coexistence of superconductivity and non-trivial band topology can lead to intrinsic or self-proximity-effect-induced topological superconductivity in FeSCs. In the last several years, the theoretically predicted band inversions with topologically protected surface states, as well as Majorana zero modes (MZMs) in vortex cores have been experimentally observed \cite{zhang2018observation,zhang2019multiple,wang2018evidence,liu2018robust} (Sec. \ref{sec_topological}). Thus, FeSCs have become an important platform to study topological superconductivity.

\subsubsection{Structural diversity of FeSCs}\label{sec_str_diversity}

Compared with the universal FeX superconducting layer, the spacer layer of FeSCs exhibits remarkable diversity and structural flexibility. For FePn-based superconductors, the valence states of Fe and Pn are usually close to $+2$ and $-3$. The Fe$_{2}$Pn$_{2}$] layer therefore carries a net charge of $2-$, and the spacer layer must be positively charged ($2+$ in general) in order to maintain the charge neutrality. Different from the dominant covalent bond between Fe and X atoms, the bonding between superconducting and spacer layers is usually ionic in nature. Due to the requirement of charge neutrality, a bilayer alkali [A$_{2}$]$^{2+}$ and monolayer alkaline earth AE$^{2+}$ or Eu$^{2+}$ can be inserted between [Fe$_2$Pn$_2$]$^{2-}$ layers, yielding the FePn-111 and FePn-122 architectures
[Figs.~\ref{Fig_structure}(a) and \ref{Fig_structure}(b)]. 
The representative materials are LiFeAs, BaFe$_{2}$As$_{2}$ and EuFe$_{2}$As$_{2}$ \cite{Tapp2008,Wang2008LiFeAs,Rotter2008a,Miclea2009}. Note that even monolayer {\it A}$^{+}$ such as K$^{+}$/Rb$^{+}$/Cs$^{+}$ ions can form the FePn-122 structure \cite{Rotter2008b,Sasmal2008,Bukowski2010}, but the $T_{c}$ values are rather low ($\sim$ 2.6~K -- 3.8~K) due to heavily hole doping.

The principal FePn and FeCh structural families, their representative formulas, and approximate ambient-pressure maximum transition temperatures are summarized in Table~\ref{tab:FeSC_families}. Together with Fig.~\ref{Fig_structure} and Fig.~\ref{Fig_Structure_sc_relation}, this table provides a structural and materials-based guide to the detailed discussion below.

\begin{table}[t]
\caption{Major structural families of iron-based
superconductors. The formulas indicate representative structural
compositions. The listed $T_c^{\max}$ values are approximate
ambient-pressure maxima for each family and are intended as a compact
guide rather than an exhaustive record of all compositions.}
\bigskip
\label{tab:FeSC_families}
\centering
\small
\renewcommand{\arraystretch}{1.10}
\begin{tabular}{lll}
\hline\hline
\textbf{Family} & \textbf{Formula} & \textbf{$T_c^{\max}$ (K)} \\
\hline
\multicolumn{3}{l}{\textit{Iron-pnictide families}} \\
\hline
FePn-111    & $A$FeAs                                  & $\sim25$ \\
FePn-122    & AEFe$_2$As$_2$                           & $\sim39$ \\
FePn-1111   & LnOFeAs                                  & $\sim56$ \\
FePn-112    & Ca$_{1-x}$Ln$_x$FeAs$_2$                & $\sim47$ \\
FePn-1144   & $A$AEFe$_4$As$_4$\textsuperscript{a}     & $\sim35$ \\
FePn-12244  & $A$Ca$_2$F$_2$Fe$_4$As$_4$               & $28$--$37$ \\
FePn-32522  & AE$_3$TM$_2$O$_5$Fe$_2$Pn$_2$            & $\sim30$ \\
FePn-42622  & AE$_4$TM$_2$O$_6$Fe$_2$Pn$_2$            & $\sim45$ \\
FePn-22241  & Ba$_2$Ti$_2$Fe$_2$As$_4$O                & $\sim21$ \\
FePn-1038   & Ca$_{10}$Pt$_3$As$_8$(Fe$_2$As$_2$)$_5$  & $\sim13$ \\
FePn-1048   & Ca$_{10}$Pt$_4$As$_8$(Fe$_2$As$_2$)$_5$  & $\sim38$ \\
\hline
\multicolumn{3}{l}{\textit{Iron-chalcogenide families}} \\
\hline
FeCh-11       & FeSe, Fe(Se,Te)                          & $\sim15$ \\
$A$FeCh-122   & $A_x$Fe$_{2-y}$Se$_2$\textsuperscript{b} & $\sim32$ \\
FeCh-11111    & (Li,Fe)OHFeSe                            & $\sim42$ \\
$M$NHFeCh-122   & $M_x$(NH$_2$)$_y$(NH$_3$)$_z$Fe$_2$Se$_2$ & $\sim44$ \\
FeCh-Org      & organic-intercalated FeSe                & $\sim50$ \\
\hline\hline
\end{tabular}
\par\bigskip
\begin{minipage}{0.94\columnwidth}
\footnotesize
\raggedright
\textsuperscript{a}\,$A$ denotes K, Rb, or Cs. In the
FePn-1144 family, AE denotes
the divalent spacer-site cation Ca, Sr, or Eu; here AE is used as
a structural and valence designation, and therefore includes Eu
despite its lanthanide classification.

\textsuperscript{b}\, For the FeCh-122 family, $A$ usually denotes K, Rb, or Cs, but is conventionally also used to include Tl-containing members, such as (Tl,Rb)$_x$Fe$_{2-y}$Se$_2$.
\end{minipage}
\end{table}

Besides simple $A^{+}$/AE$^{2+}$/Eu$^{2+}$ cations, more complex structural units can be used as spacer layers to form various FePn-based superconductors. Because [Fe$_{2}$Pn$_{2}$]$^{2-}$ layer has the anti-CaF$_{2}$-type structure, naturally the CaF$_{2}$-type spacer layers [Ln$_{2}$O$_{2}$]$^{2+}$, [Th$_{2}$N$_{2}$]$^{2+}$, [AE$_{2}$F$_{2}$]$^{2+}$ and [Ca$_{2}$H$_{2}$]$^{2+}$ could intergrow with [Fe$_{2}$Pn$_{2}$]$^{2-}$ layers and LnOFePn, ThNFePn, AEFFePn and CaHFePn with ZrCuSiAs-type structure FePn-1111 can be formed [Fig.~\ref{Fig_structure}(c)] \cite{Kamihara2008,Chen2008a,Chen2008b,Ren2008,Wang2008b,Wen2008,Fujioka2013,Iimura2012,Matsuishi2008,Han2008,Wu2009,Wang2013,Hanna2011,Muraba2014}. 
Interestingly, a special fluorite-type spacer layer, [(Ca/Eu)$_2$As$_2$]$^{2+}$, can also be inserted between two [Fe$_2$Pn$_2$]$^{2-}$ layers, giving rise to the FePn-112 structure (Ca/Eu)FeAs$_2$ [Fig.~\ref{Fig_structure}(d)]. In this spacer layer, the formal valence of As is close to $-1$, rather than the usual $-3$ state in the FePn layers. As a result, As--As covalent bonding is formed within the spacer layer, which distorts the As square net and produces zigzag As chains. Correspondingly, the crystal symmetry is lowered from tetragonal to monoclinic, with space group
$P$12$_1$1 or $P$12$_1$/$m$.
Experimentally, CaFeAs$_2$ is not stable in this structure in its undoped form and can only be stabilized by partial substitution of Ln ions for Ca. Superconductivity with $T_c$ up to about 47~K has been observed in Ca$_{1-x}$La$_x$Fe(As$_{1-y}$Sb$_y$)$_2$ \cite{katayama2013superconductivity,Yakita2014,Sala2014, Sala2014,Kudo2014a,Kudo2014b}. By contrast, EuFeAs$_2$ can be synthesized without La substitution and exhibits an AFM transition at 41~K. La substitution in EuFeAs$_2$ induces superconductivity, which appears to coexist with the AFM order at low temperature \cite{Yu2017}.

Later on, it is realized that the FePn-based superconductors with
ordered spacer layers can also be stabilized. For example, $A^{+}$ and AE$^{2+}$/Eu$^{2+}$ ions occupy the
inequivalent sites and the so-called FePn-1144 materials can be synthesized [Fig.~\ref{Fig_structure}(e)] \cite{Iyo2016,Kawashima2016,Liu2016a,Liu2016b}. Following this strategy, another kind of FePn-based superconductors with ordered spacer layers ACa$_{2}$F$_{2}$Fe$_{4}$As$_{4}$, ALn$_{2}$O$_{2}$Fe$_{4}$As$_{4}$ and BaTh$_{2}$(N,O)$_{2}$Fe$_{4}$As$_{4}$ with $T_{c}$ of 28--37~K were also synthesized, forming ordered-spacer FePn-12244 phases [Fig.~\ref{Fig_structure}(f)] and these materials are formed with the intergrowth of one FePn-122 and double FePn-1111  \cite{Wang2016,Wang2017,Wu2017a,Wu2017b,Wu2017c,Shao2019}. Moreover, NiAs-based superconductor La$_{3}$O$_{2}$Ni$_{4}$As$_{4}$ with $T_{c}$ $\sim$ 2.2~K also has the FePn-1144 structure, where one of the three La$^{3+}$ cations occupies the $A^{+}$/Ba$^{2+}$ position whereas the other two reside in the [La$_{2}$O$_{2}$]$^{2+}$ layer \cite{Klimczuk2009}.

Usually, the $a$-axis lattice parameter of the FePn-based superconductors ranges from about 3.8 \AA\ to 4.1 \AA\, and it makes the ABO$_{3}$ perovskite-like spacer layers compatible with the FePn layer and thus the intergrowth between these two structural units is possible. The homologous series which can be written as AE$_{n+1}$B$_{n}$O$_{3n-1}$Fe$_{2}$Pn$_{2}$ [denoted as FePn-($n$+1)$n$(3$n$-1)22] has been synthesized. For $n$ = 2 [FePn-32522, Fig.~\ref{Fig_structure}(g)], the compounds AE$_{3}$(Sc,Ti)$_{2}$O$_{5}$Fe$_{2}$As$_{2}$ and Ca$_{3}$Al$_{2}$O$_{5-\delta}$Fe$_{2}$Pn$_{2}$ with $T_{c}$ $\sim$ 17~K -- 30~K were reported \cite{Zhu2009a,Chen2009,Shirage2011}. For the member of $n$ = 1, Sr$_{2}$CrO$_{2}$Fe$_{2}$As$_{2}$ was successfully synthesized but it does not exhibit superconductivity above 3~K \cite{Eguchi2013}. 
In contrast, other intergrowth structures with thicker perovskite-like spacer layers ($n$ = 3 -- 5) were also synthesized and all of them exhibit superconductivity with $T_{c}$ $\sim$ 28~K -- 42~K \cite{Ogino2010a,Ogino2010b,Ogino2011,Kawaguchi2010}. 

In addition, perovskite structure belongs to a large family of oxides called Ruddlesden-Popper (RP) phase with the formula A$_{n+1}$B$_{n}$O$_{3n+1}$, which contains a rocksalt layer in addition to the perovskite layer. Therefore, the RP phase spacer layers can also intergrow with FePn layer
to form another homologous series of FePn-based superconductors with the formula A$_{n+2}$B$_{n}$O$_{3n}$Fe$_{2}$Pn$_{2}$. Sr$_{4}$Sc$_{2}$O$_{6}$Fe$_{2}$P$_{2}$
and Ca$_{4}$Al$_{2}$O$_{6}$Fe$_{2}$P$_{2}$ with $n$ = 2 [FePn-42622, Fig.~\ref{Fig_structure}(h)] show superconductivity at about 17~K, which has the highest $T_{c}$ among phosphides so far \cite{Ogino2009,Shirage2010}.
Moreover, AE$_{4}$Tm$_{2}$O$_{6}$Fe$_{2}$As$_{2}$ (Tm = Sc, Ti, V, Cr) and Ca$_{4}$(Al,Ti)$_{2}$O$_{6}$Fe$_{2}$As$_{2}$ were also discovered and some of these materials show superconductivity with the $T_{c}$ $\sim$ 28--45~K \cite{Zhu2009b,Shirage2010,Sato2010,Chen2009,Ogino2009}.
In this homologous series, $n= 3$ and 4 members were also successfully prepared with Ca$_{n+2}$(Sc, Ti, Al)$_{n}$O$_{3n}$Fe$_{2}$As$_{2}$ and bulk superconductivity is observed with $T_{c}$ up to 39~K \cite{Ogino2010c}.

Oxypnictides BaTi$_{2}$As$_{2}$O containing Ti$_{2}$O square lattice has tetragonal symmetry and its $a$-axial lattice parameter is about 4 \AA, which is comparable with those values of FeSCs. Thus, Ti$_{2}$As$_{2}$O layer can be used as the spacer layer for FeSCs. In 2012, the intergrowth structure of BaTi$_{2}$As$_{2}$O and BaFe$_{2}$As$_{2}$ was synthesized successfully \cite{Sun2012} and the chemical formula is Ba$_{2}$Ti$_{2}$Fe$_{2}$As$_{4}$O
[FePn-22241, Fig.~\ref{Fig_structure}(i)]. Moreover, BaTi$_{2}$As$_{2}$O exhibits a density-wave (DW) transition at 200~K, which is related to the change of Ti$_{2}$O sheets \cite{Wang2010}. 
Interestingly, such DW anomaly also appears in Ba$_{2}$Ti$_{2}$Fe$_{2}$As$_{4}$O at lower temperature ($T_{\rm {DW}}$ = 125~K) and coexists with bulk superconductivity with $T_{c}$ = 21.5 K. These phenomena are attributed to a self-doping effect due to the charge transfer between Ti$_{2}$As$_{2}$O and FeAs layers \cite{Sun2012}.

Besides the spacer layers mentioned above, other block layers are also able to be inserted into the FePn layers. As shown in Figs.~\ref{Fig_structure}(j) and \ref{Fig_structure}(k), two kinds of FeSCs with the chemical formulas Ca$_{10}$Pt$_{3}$As$_{8}$(Fe$_{2}$As$_{2}$)$_{5}$ (FePn-1038) and Ca$_{10}$Pt$_{4}$As$_{8}$(Fe$_{2}$As$_{2}$)$_{5}$ (FePn-1048), have been reported \cite{Ni2011,Lohnert2011,Kakiya2011}. 
For FePn-1048, the PtAs layer consists of a square lattice of corner-sharing PtAs$_{4}$ squares with a rotation of approximately 25$^{\circ}$ about an axis perpendicular to the plane, governed by the formation of intraplanar As-As dimers, and Ca atoms are located below and above PtAs layer. Because the $2\times2$ superlattice of Pt (i.e., Pt$_{4}$As$_{8}$) matches the \(\sqrt{5} \times \sqrt{5}\) superlattice of Fe$_{2}$As$_{2}$, the Fe$_{2}$As$_{2}$ and Pt$_{4}$As$_{8}$ layers become commensurate in the (210) Fe$_{2}$As$_{2}$ direction, forming FePn-1048 structure \cite{Ni2011}. 
For FePn-1038 phase, one of the four Pt sites is absent, leaving the Pt$_{3}$As$_{8}$ planes between Fe$_{2}$As$_{2}$ layers. Superconductivity with $T_{c} = 11$~K -- 13~K and 26~K -- 38~K is observed in the FePn-1038 and FePn-1048 phases, respectively \cite{Ni2011,Lohnert2011,Kakiya2011}.
Moreover, the Pt can be replaced by Pd and Ir. Superconductivity have also been reported in Ca$_{10}$Pd$_{3}$As$_{8}$(Fe$_{2}$As$_{2}$)$_{5}$ and Ca$_{10}$Ir$_{4}$As$_{8}$(Fe$_{2}$As$_{2}$)$_{5}$ with $T_{c}$ $\sim$ 16~K and 17~K \cite{Hieke2013,Kudo2013}.

Superconductivity in the iron chalcogenide $\beta$-FeSe
($T_c\sim8$~K) was reported soon after the discovery of the
FePn-based superconductors \cite{Hsu2008}. The key structural unit of FeSe is similar to that of FePn-based superconductors, i.e., the square-planar lattice of Fe with tetrahedral coordination which is regarded as the superconducting layer. But because the valence state of Se is usually $-2$, the FeSe layer is supposed to be charge neutral and the spacer layer is not needed, as shown in Fig.~\ref{Fig_structure}(l). Thus FeCh-11 is the simplest FeSC family. On the one hand, this seems to lead to the invalidity of carrier doping from charge spacer layer usually used in FePn-based superconductors, taking K doped BaFe$_{2}$As$_{2}$ and F doped LaOFeAs as examples. Therefore, early studies mainly focused on the aliovalent doping at Fe site \cite{Williams2009,Mizuguchi2009a} or isovalent substitution at Se site using S or Te \cite{Fang2008,Mizuguchi2009b,Liu2010,Lai2015} because there is no spacer layer in FeSe. The most prominent result is that the $T_{c}$ increases to about 15~K upon Te substitution in FeSe \cite{Liu2010}. 
On the other hand, FeSe can be regarded as a van der Waals (vdW) material due to the rather weak interlayer interaction. 
This feature enables the insertion of various ionic, molecular, or
oxide spacer layers between FeCh sheets.
For example, $A^{+}$/Tl$^{+}$ cations [$A$FeCh-122, Fig.~\ref{Fig_structure}(m)] \cite{Guo2010}, [(Li,Fe)OH]$^{\delta+}$ [FeCh-11111, Fig.~\ref{Fig_structure}(n)] \cite{Lu2013}, $A^{+}$/AE$^{2+}$/Ln$^{3+}$/Ch$^{2-}$ ions combined with charge-neutral organic molecules [{\it M}NHFeCh-122, Fig.~\ref{Fig_structure}(o)]  \cite{Ying2012}, or pure organic cations [FeCh-Org, Fig.~\ref{Fig_structure}(p)] \cite{shi2018organicionintercalated,Shi2018b,Rendenbach2021} can be intercalated into FeCh layers.
The weak vdW interlayer interaction also enables FeCh-11 materials to be exfoliated down to a few layers, thus some novel techniques can be used to dope carriers into these materials, including ionic liquid gating and solid-state lithium-ion gating/intercalation \cite{lei2016evolution,shiogai2016electricfieldinduced,Hanzawa2016}. Moreover, due to the absence of spacer layer, the monolayer FeSe can be grown directly on a substrate and it provides an ideal platform for studying the effects of interface interaction and dimensionality on superconductivity \cite{Wang2012}.

For exploring novel spacer layers in FeSCs, several empirical rules or models have been proposed \cite{Jiang2013,wang2018evidence,Wang2021}. Among those rules, ``lattice match'' and ``charge transfer'' are two of the most important design criteria. The former brings about an energy rise (positive elastic energy), and thus good lattice match between the spacer and FePn/Ch layers should be satisfied and the degree of lattice mismatch is normally less than 2\% \cite{wang2018evidence,Wu2017b}. The latter lowers the internal energy and a substantial charge transfer between spacer and FePn/FeCh layers is necessary, which acts as the glue that combines them together \cite{wang2018evidence}. 
Once a balance between these two effects is achieved, a thermodynamically stable intergrowth phase of FeSC can form. Moreover, the spacer layers can tune the physical properties of the superconducting layer not only via transferring  charge carriers (electrons or holes) into FePn/FeCh layers but also through lattice mismatch effect. For
example, in FePn-1111 system, replacing La$^{3+}$ ions by Sm$^{3+}$ in [Ln$_{2}$O$_{2}$]$^{2+}$ layer can compress the FePn layer, shortening the Fe--Fe bond length and enhancing $T_{c}$ dramatically \cite{Kamihara2008,Chen2008a}. Consequently, precise engineering of the local coordination geometry of FePn/Ch layer, interlayer coupling strength, and carrier doping level through tailored spacer layers may enable the realization of FeSCs with enhanced $T_{c}$.

\section{Material Design and Synthesis}

Materials design in FeSCs changes not only the carrier
density, crystal structure, and superconducting transition temperature,
but also the magnetic state of the Fe layers. In the parent iron
pnictides, stripe-type Fe SDW order is commonly
suppressed or reconstructed by chemical substitution, pressure, or
structural modification; in several families, superconductivity can
coexist or compete microscopically with the Fe-derived magnetic order.
The following sections emphasize the materials routes that control these
ground states. The broader phenomenology of Fe-layer magnetism,
nematicity, and their interplay with superconductivity is discussed in
Secs.~V and VII \cite{dai2012magnetism,dai2015antiferromagnetic,yi2017role}.

\subsection{Electron-doping routes and heavily electron-doped FeSCs}

Section~II.2 classified FeSCs according to their spacer-layer and structural chemistry. Here, we turn from this structural classification to the materials-design routes by which carrier density, interlayer coupling, and dimensionality can be tuned. The discussion focuses primarily on FeCh-derived systems, where chemical intercalation, ionic gating, epitaxial growth, and interface engineering provide particularly effective routes to electron doping, but also includes H$^-$ substitution in LnOFeAs as a complementary high-electron-doping route in FePn systems.

A particularly important limit is heavy electron doping, in which the hole pockets near $\Gamma$ are strongly reduced or removed and the low-energy electronic structure becomes dominated by electron pockets. This regime provides a stringent test of pairing descriptions based primarily on hole--electron scattering and has motivated extensive studies of inter-pocket pairing, spin excitations, and superconducting gap structure. Sections~III.A.1 and III.A.2 first describe chemical intercalation and ionic-gating approaches to electron doping. Sections~III.A.3 and III.A.4 then discuss FeSe and FeTe
thin-film platforms, respectively, including the roles of epitaxial strain and interface effects. Sec. III.A.5 compares the common electronic and superconducting properties of heavily electron-doped FeSe-derived systems; Sec. III.A.6 discusses charge-neutral and hole-doped FeSe intercalates as control systems for disentangling the effects of charge transfer and interlayer
spacing; and Sec. III.A.7 considers H$^-$ substitution in LnOFeAs as a complementary high-electron-doping route in iron pnictides.

\subsubsection{Intercalated FeCh-based superconductors}

\begin{figure}
\centering
\includegraphics[width=8.5cm]{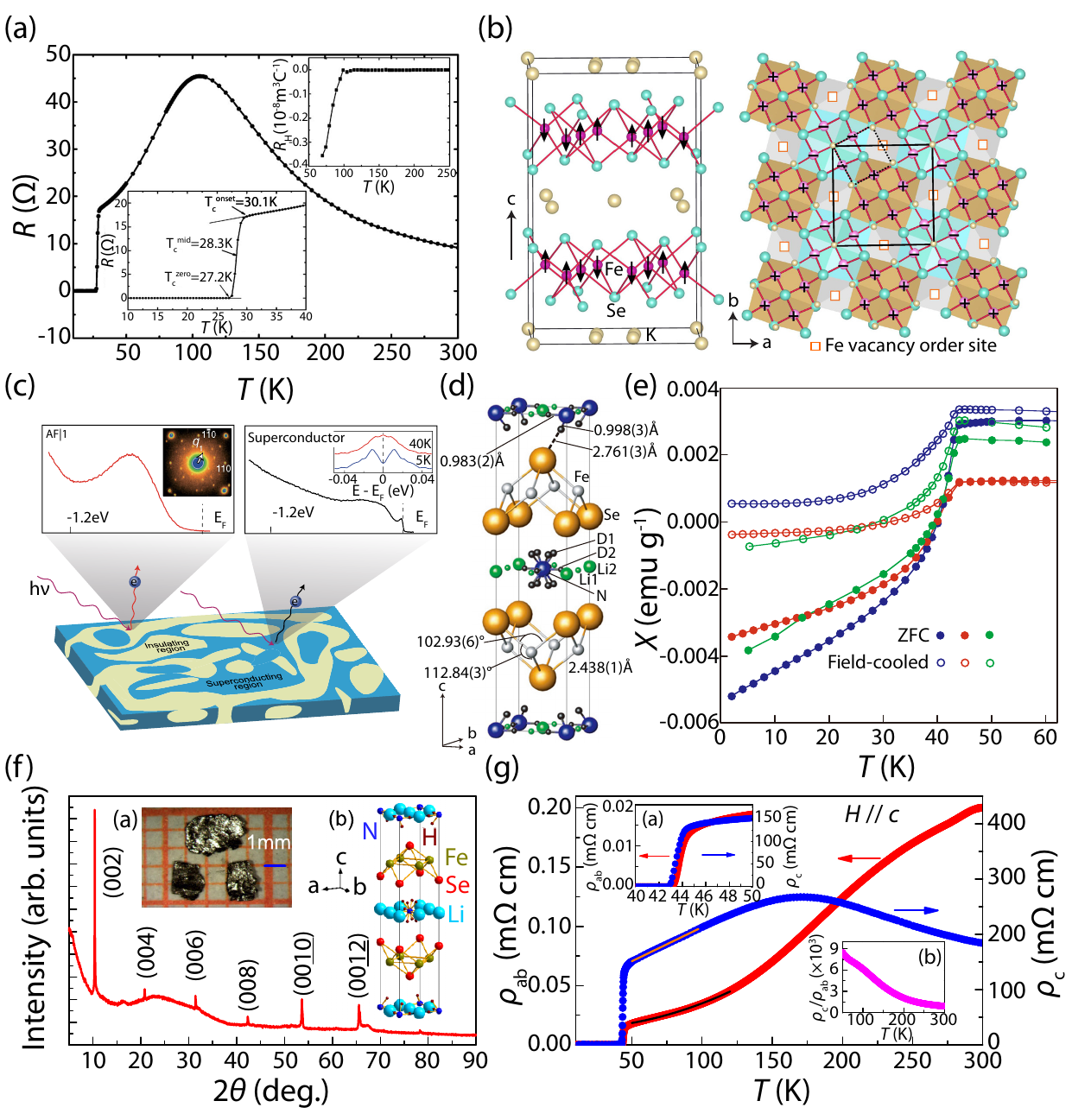}
\caption[]{(a) Temperature dependence of $\rho$($T$) for KFeSe-122. From \cite{Guo2010}. (b) Crystal structure of KFeSe-122. K ions are intercalated between FeSe layers with the existence of ordered Fe vacancies. From \cite{Bao2011}. (c)	Cartoon for mesoscopic phase separation in KFeSe-122. Different regions exhibit different photoemission-spectroscopic signatures. From \cite{chen2011electronic}. (d) Crystal structure of Li$_{0.6(1)}$(ND$_{2}$)$_{0.2(1)}$(ND$_{3}$)$_{0.8(1)}$Fe$_{2}$Se$_{2}$. (e) Magnetic susceptibility of three 		Li$_{0.6(1)}$(ND$_{2}$)$_{0.2(1)}$(ND$_{3}$)$_{0.8(1)}$Fe$_{2}$Se$_{2}$ polycrystals. From \cite{Burrard-Lucas2013}. (f) XRD pattern of a LiNHFeSe-122 single crystal. Insets (a) and (b) show a photograph and crystal structure of LiNHFeSe-122 single crystals, respectively. (g) Temperature dependence of in-plane resistivity $\rho_{ab}(T)$ and $c$-axial resistivity $\rho_{c}(T)$ at zero field. Insets (a) and (b) show enlarged resistivity curves near $T_{c}$ and the	ratio of $\rho_{c}/\rho_{ab}$ as a function of temperature, respectively. From \cite{sun2017extreme}.
\label{Fig_FeSe122}}
\end{figure}

The discovery of K$_{x}$Fe$_{2-y}$Se$_{2}$ (KFeSe-122) with remarkably enhanced $T_{c}$ ($\sim$ 30~K) marked an important breakthrough in FeSC research [Fig.~\ref{Fig_FeSe122}(a)] \cite{Guo2010}, which shed light on the exploration of novel FeCh-based materials using intercalation method and the
realization of high $T_{c}$ in the heavily-electron-doped (HED) FeCh-based superconductors. After that, Rb-, Cs- and Tl- intercalated FeSe with $T_{c}$ $\sim$ 27 K--32~K, and the sister compounds K$_{x}$Fe$_{2-y}$S$_{2}$ with spin glass state and AFM K$_{x}$(Fe, Ag)$_{2}$Te$_{2}$ have also been discovered \cite{Krzton-Maziopa2011,Li2011,Liu2011,Wang2011, Lei2011a,Lei2011b,song2019evidence,song2019intertwined}.
However, it was soon recognized that $A$FeCh-122 systems prepared using high-temperature reaction method exhibit 
Fe-vacancy order coupled to long-range magnetic order and mesoscopic phase separation of superconducting and insulating phases, as shown in Figs.~\ref{Fig_FeSe122}(b) and \ref{Fig_FeSe122}(c) \cite{Bao2011,chen2011electronic}. Such material complexity makes it difficult to establish intrinsic properties of $A$FeSe-122 systems. The detailed results of these complexities are reviewed by Krzton-Maziopa \textit{et al.} \cite{krzton-maziopa2016superconductivity}.

Later on, a series of {\it M}$_{x}$(NH$_{3}$)$_{y}$Fe$_{2}$Se$_{2}$ ({\it M} = alkali-metal, alkaline-earth-metal, and lanthanide ions, {\it M}NHFeSe-122) superconductors were synthesized using the low-temperature ammonothermal method \cite{Ying2012}. As shown in Fig.~\ref{Fig_FeSe122}(d), in
{\it M}NHFeSe-122, the $M$ ions and NH$_{3}$
molecules/NH$^{2-}$ ions are co-intercalated between FeSe layers and the Fe plane is reported to be nearly defect-free \cite{Burrard-Lucas2013}.
This low-temperature synthesis method 
introduces electrons into the FeSe layers while preserving their structural integrity, thereby avoiding the aforementioned problems and raising $T_c$ to about 44~K [Fig.~\ref{Fig_FeSe122}(e)].
Moreover, upon removing NH$_3$ molecules from Na-NH$_{3}$ layer, Na$_{x}$Fe$_{2-y}$Se$_{2}$ can be obtained, which cannot be prepared using conventional high-temperature method 
\cite{guo2014superconductivity}. Its $T_{c}$ is about 37~K with much less Fe vacancies than that in KFeSe-122. The lower $T_{c}$ and smaller interlayer distance $d$ of Na$_{x}$Fe$_{2-y}$Se$_{2}$ than those of 
{\it M}NHFeSe-122 imply that the change of interlayer interaction would have some influence on superconducting properties. 
On the other hand, the preparation of single crystals is indispensable in order to understand the intrinsic properties. A two-step method is developed to overcome the difficulty in growth of 
{\it M}NHFeSe-122 single crystals \cite{sun2017extreme}. The obtained LiNHFeSe-122 single crystals exhibit the same $(0,0,l)$ orientation as $\beta$-FeSe [Fig.~\ref{Fig_FeSe122}(f)]. In addition, transport measurements indicate high sample quality
, a $T_{c}$ of $\sim$ 44~K, and large resistivity anisotropy, as shown in Fig.~\ref{Fig_FeSe122}(g).

\begin{figure}
\centering
\includegraphics[width=8.5cm]{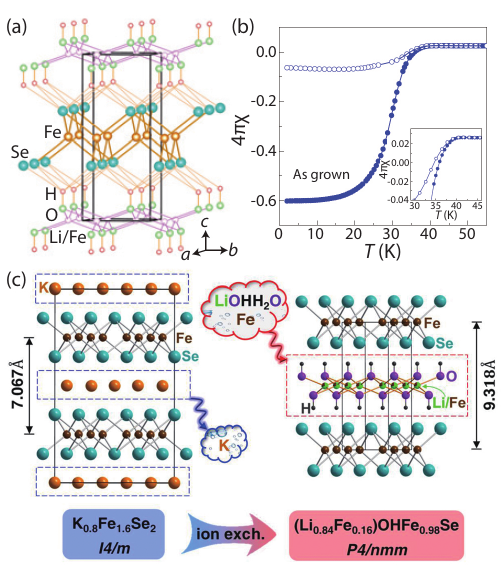}
\caption[]{(a) Crystal structure of FeSe-11111. (b) Temperature dependence of magnetic susceptibility 4$\pi$$\chi$ of FeSe-11111 polycrystal. From \cite{lu2015coexistence}. (c) A schematic illustration of the hydrothermal ionic exchange reaction. From \cite{dong2015li}.
\label{Fig_FeSe11111}}
\end{figure}

Besides the low-temperature ammonothermal method to synthesize the 
{\it M}-NH$_{3}$ co-intercalated FeCh, another intercalated FeCh-based superconductor (Li,Fe)OHFeSe (FeSe-11111) is discovered by using the low-temperature hydrothermal method \cite{Lu2013,Pachmayr2015}. For FeSe-11111, the (Li,Fe)OH layer is intercalated between FeSe layers [Fig.~\ref{Fig_FeSe11111}(a)]. Similar to 
{\it M}NHFeSe-122, there are almost no defects in the Fe plane and the $T_{c}$ is as high as $\sim$40~K [Fig.~\ref{Fig_FeSe11111}(b)]. Moreover, some Fe ions occupy the Li site in the (Li,~Fe)OH layer, which contributes to the electron doping into the FeSe layers because of the different valences of Fe$^{2+}$ and Li$^{+}$ ions. These doped Fe$^{2+}$ ions at Li site also result in the canted AFM order, which could coexist with superconductivity 
\cite{lu2015coexistence}. Later on, (Li,Fe)OHFeS with $T_{c}$ up to 8~K has also been reported \cite{Zhou2017}. 
For the growth of large FeSe-11111 single crystals, a novel hydrothermal ion-exchange reaction method is developed [Fig.~\ref{Fig_FeSe11111}(c)] \cite{dong2015phase}. First, large, high-quality crystals of insulating K$_{0.8}$Fe$_{1.6}$Se$_{2}$ (KFeSe-245) were grown by a high-temperature method as the mother compound. Then, under hydrothermal reaction conditions, the K ions in KFeSe-245 are completely released into solution. Simultaneously, (Li,Fe)OH layers constructed by ions from the solution are intercalated in between FeSe layers, linking the adjacent edge-sharing FeSe tetrahedra via weak hydrogen bonds. The obtained high-quality FeSe-11111 single crystals with $T_{c}$ $\sim$ 42~K exceed 10 mm in length and are about 0.4 mm thick \cite{dong2015li}.

\begin{figure}
\centering
\includegraphics[width=8.5cm]{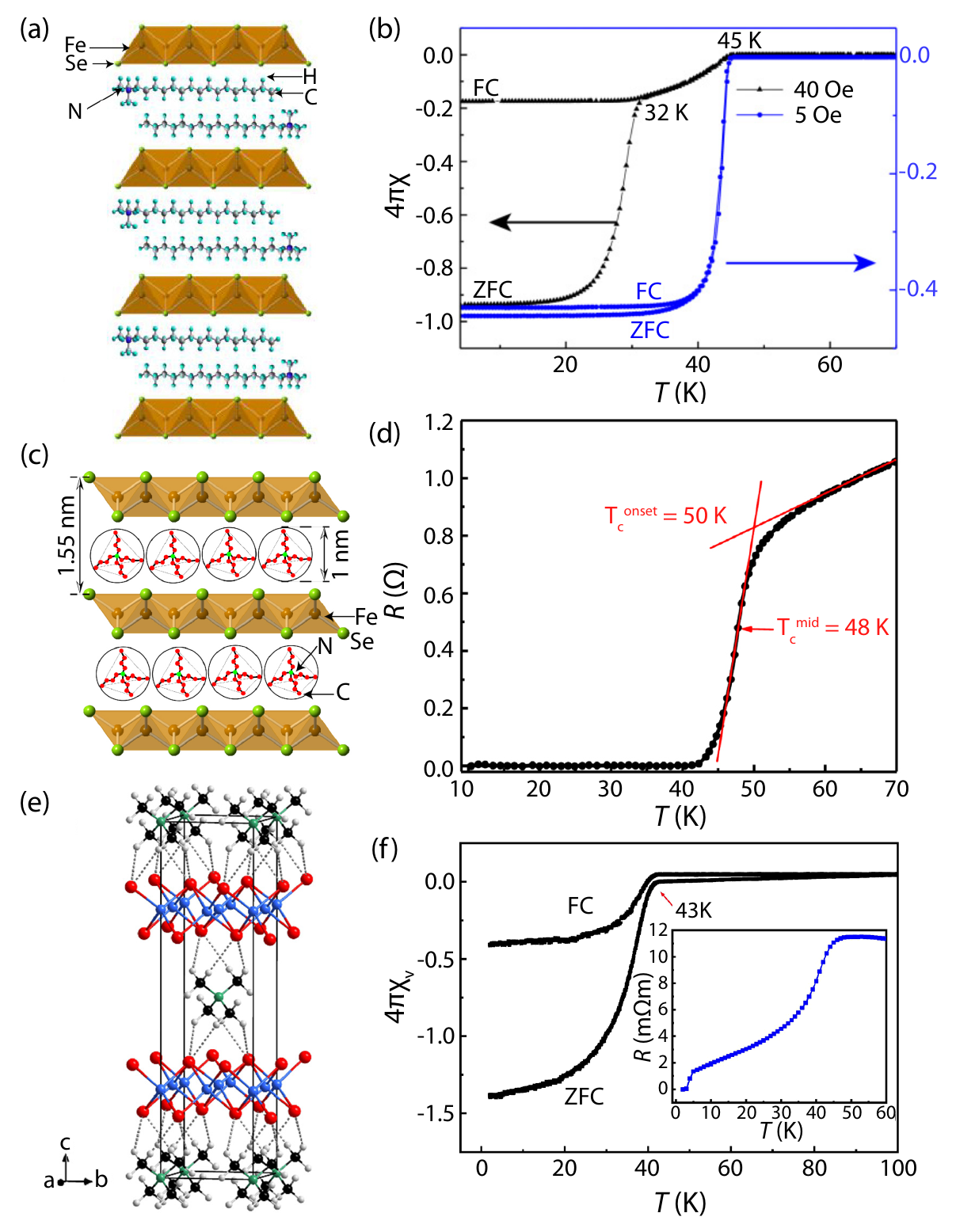}
\caption[]{(a) Crystal structure of (CTA)$_{0.3}$FeSe. From \cite{shi2018organicionintercalated}. (b) Temperature dependence of 4$\pi\chi$ of (CTA)$_{0.3}$FeSe. From \cite{shi2018organicionintercalated}. (c) Crystal structure of (TBA)$_{0.3}$FeSe with hydrogen atom neglected. (d) Resistance as a function of temperature for (TBA)$_{0.3}$FeSe. From \cite{Shi2018b}. (e) Crystal structure of (TMA)$_{0.5}$Fe$_2$Se$_2$. (f) 4$\pi\chi$ of (TMA)$_{0.5}$Fe$_2$Se$_2$. Inset: resistivity of a cold-pressed pellet. From \cite{Rendenbach2021}.
\label{Fig_FeSe_Org}}
\end{figure}

Above systems strongly indicate that once electrons are doped into FeSe layers with proper concentration, $T_{c}$ is usually enhanced irrespective of the type of charged spacer layer. Such universal relation is further confirmed in the organic-inorganic hybridized FeSe-based superconductors (FeCh-Org) in which pure organic ions are used as the charge spacer layers \cite{shi2018organicionintercalated,Shi2018b,Rendenbach2021}. 
Such hybridized FeSe-based superconductors can be prepared by intercalating FeSe single crystal with cetyltrimethyl ammonium (CTA$^{+}$), tetrabutyl ammonium (TBA$^{+}$) or tetramethyl ammonium cations (TMA$^{+}$) via an electrochemical intercalation method. The electrochemical reaction can be described as the following equations: FeSe + $x$CTA$^{+}$/TBA$^{+}$/TMA$^{+}$ + $x$e$^{-}$ $\rightarrow$ (CTA/TBA/TMA)$_{x}$FeSe (cathode); Br$^{-}$ - e$^{-}$ $\rightarrow$ 1/2Br$_{2}$, Br$^{-}$ - e$^{-}$ + Ag $\rightarrow$ AgBr, or 3I$^{-}$ - 2e$^{-}$ $\rightarrow$ I$^{-}_{3}$ (anode) \cite{shi2018organicionintercalated,Shi2018b,Rendenbach2021}. 
The organic-ion content $x$ ranges from $0.3$ to $0.5$. Similar to other FeSCs, the structures of FeCh-Org superconductors consist of the alternate stacking of monolayer FeSe and double layers of CTA$^{+}$ or one layer of TBA$^{+}$/TMA$^{+}$ cations [Figs.~\ref{Fig_FeSe_Org}(a), \ref{Fig_FeSe_Org}(c), and \ref{Fig_FeSe_Org}(e)]. The $T_{c}$ values of (TMA)$_{0.5}$Fe$_{2}$Se$_{2}$, (CTA)$_{0.3}$FeSe~and (TBA)$_{0.3}$FeSe are about 43~K, 45~K and 50~K, respectively [Figs.~\ref{Fig_FeSe_Org}(b), \ref{Fig_FeSe_Org}(d), and \ref{Fig_FeSe_Org}(f)]. The (TBA)$_{0.3}$FeSe has the highest $T_{c}$ among FeCh-based bulk superconductors \cite{Shi2018b}. The intercalation of chemically inert organic ions into FeSe paves a novel way to discover high-$T_{c}$ FeCh-based superconductors with good crystallinity.

\subsubsection{Gating-tuned FeCh-based superconductors}

\begin{figure}
\centering
\includegraphics[width=8.5cm]{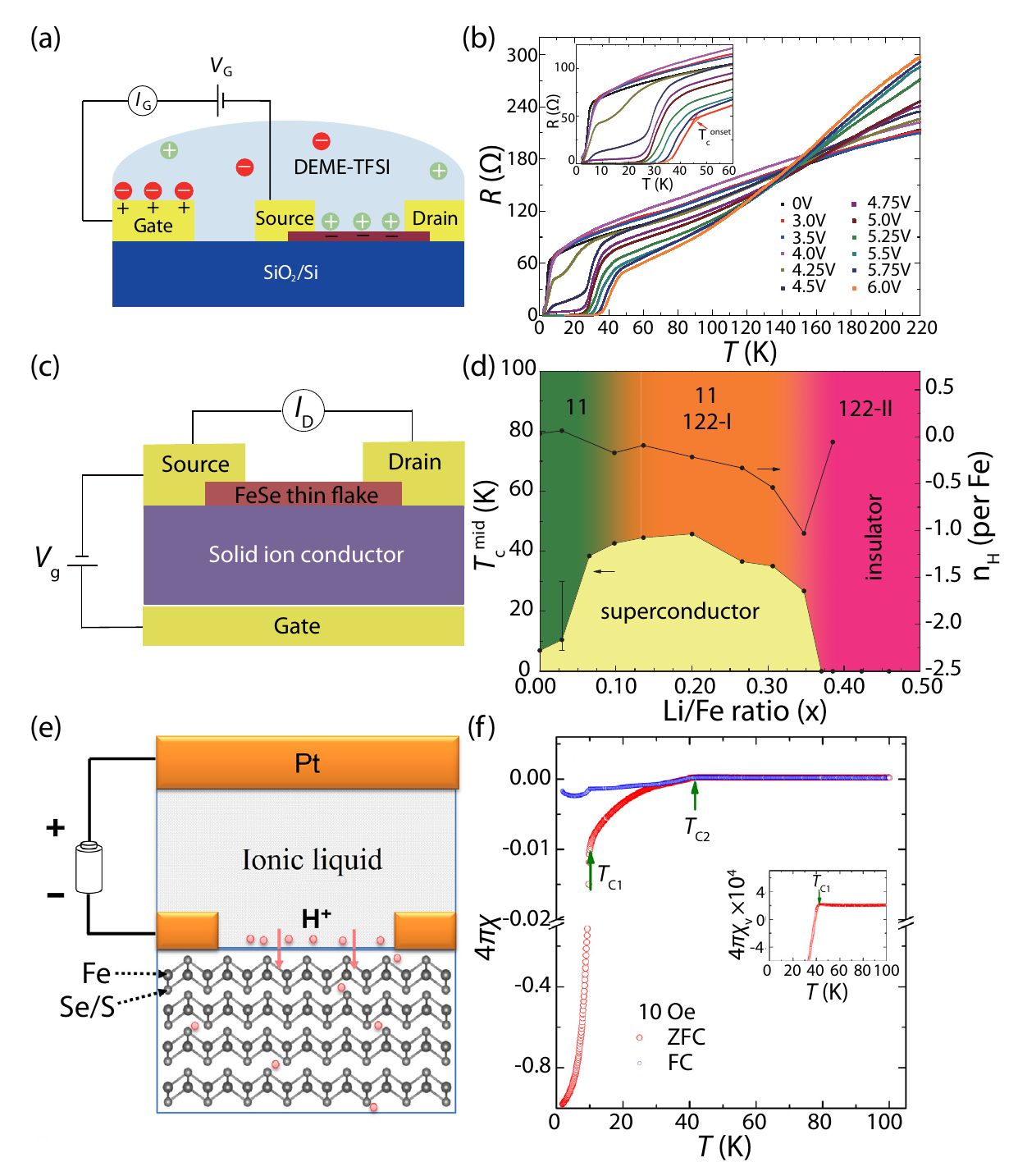}
\caption[]{(a) A schematic illustration of the FeSe thin flake	EDL-FET device. The ionic liquid DEME-TFSI serves as the dielectric, covering the sample and gate electrodes. (b) Temperature dependence of the resistance $R(T)$ for FeSe thin flake at different $V_{g}$. Inset: enlarged view of $R(T)$ near $T_{c}$. From \cite{lei2016evolution}. (c) A schematic diagram of a SIC-FET device with a solid ion conductor as the gate dielectric. (d) Phase diagram of the Li-intercalated FeSe thin flake as a function of the Li/Fe ratio. A series of structural phase transitions from the FeSe-11 phase to the LiFeSe-122 I phase, then to the LiFeSe-122 II phase, takes place. The Hall number $n_{\rm H}$ is plotted to show the variation in the effective carrier density. From \cite{Lei2017}. (e) A schematic illustration for protonation process using ion liquid. (f) Temperature dependence of 4$\pi\chi$ of H$^{+}$ intercalated FeSe$_{0.93}$S$_{0.07}$. From \cite{Cui2018}.
\label{Fig_Gating_FeSe}}					
\end{figure}

For the electron-doping methods described above, the intercalation process is usually irreversible, i.e., once the charge spacer layers are formed, they cannot be deintercalated or tuned freely. In contrast, because FeCh--11 can be readily exfoliated owing to its weak vdW interlayer coupling, carrier density can be controlled by electrostatic gating \cite{lei2016evolution,jiang2023interplay,shiogai2016electricfieldinduced,Hanzawa2016}. Because this method can tune the carrier concentration continuously and reversibly, it provides a powerful way to study the detailed electronic phase diagram of FeCh-based superconductors and the dynamic process of doping. The superconducting
$T_{c}$ higher than 35~K is successfully realized in FeSe thin flakes using the electrostatic gating and ionic liquid as the gate dielectric \cite{lei2016evolution,shiogai2016electricfieldinduced,Hanzawa2016}. 
This device is called an electric-double-layer field-effect transistor (EDL-FET) [Fig.~\ref{Fig_Gating_FeSe}(a)], which is widely used for tuning physical properties of 2D vdW materials, especially superconductivity \cite{Saito2016}. A jump of $T_{c}$ from 8~K to 30~K has been found at a certain gate voltage, ascribed to a Lifshitz transition, followed by a gradual increase of $T_{c}$ from $\sim$ 30~K to 48~K [Fig.~\ref{Fig_Gating_FeSe}(b)] \cite{lei2016evolution}. Importantly, when compared to the pristine bulk FeSe crystal and FeSe thin flake at gating voltage $V_{g}$ = 0 V, the Hall coefficient of FeSe flake at $V_{g} = 6$ V is negative in the whole
measuring temperature range, indicating this system is HED at high-$V_{g}$ region \cite{lei2016evolution}.

Although the EDL-FET is an efficient configuration to tune the physical properties of FeSe, there are still some limitations of this method. First, the charge accumulation in EDL-FET is only confined close to the surface of FeSe thin flake due to the Thomas-Fermi screening effect. Second, the doping level is limited due to sample damage at high gate voltages. Third, the existence of a liquid electrolyte on the sample hinders detailed characterization of the device using many experimental probes. Later on, a novel FET device using a solid ion conductor (SIC) as the gate dielectric (SIC-FET) was developed [Fig.~\ref{Fig_Gating_FeSe}(c)] \cite{Lei2017}. 
For the SIC-FET, the sample surface is exposed, favoring the surface-sensitive experiments. More importantly, the doping process produces bulk, reversible carrier doping, and the effective carrier concentration introduced by ion intercalation can go beyond the upper limit of the EDL-FET device. Using the Li-SIC-FET, a dome-shaped superconducting phase diagram is mapped out. The maximum $T_{c}$ is about 46.6~K and more interestingly, an insulating phase accompanied by structural transitions (ordering of Li$^{+}$ ions) is found in the extremely overdoped regime [Fig.~\ref{Fig_Gating_FeSe}(d)], attributed to charge localization effect \cite{Lei2017}. 
Another detailed experimental work indicates that Li or Na-SIC-FET can effectively enhance $T_{c}$ of FeSe with several discrete superconducting phases. Such feature should be intrinsic and universal in metal-intercalated FeSe, distinct from other unconventional superconductors \cite{ying2018discrete}.

Besides the electrostatic gating effect of an ionic liquid near the sample surface, the residual water in the ionic liquid can also lead to the penetration of oxygen- and hydrogen-containing ions into materials up to tens of nanometers via an electrochemical process \cite{Lu2017}. More importantly, the dopants reside in the sample permanently when the electrode voltage is switched off \cite{Lu2017}. Using this electrochemical process, the intercalation of protons (H$^{+}$ cations)
into FeSCs becomes feasible. The schematic of the protonation process for the FeSC is shown in Fig.~\ref{Fig_Gating_FeSe}(e) \cite{Cui2018}.
Two platinum plates are used as cathode and anode, and FeSe single crystals are stuck to the cathode using silver paint. The voltage between two electrodes is set to $\sim3$ V. A typical protonation treatment lasts several days. Two types of ionic liquids DEME-TFSI and FMIM-BF4 can be used as the electrolytes. It is found that protons can be successfully doped into FeSe$_{0.93}$S$_{0.07}$ and FeS single crystals over a macroscopic length scale and the $T_{c}$ is enhanced significantly from 9~K to 42.5~K [Fig.~\ref{Fig_Gating_FeSe}(f)] and from 5~K to 18~K, respectively \cite{Cui2018}.
It is proposed that the implanted protons serve as electron dopants. Further studies indicate that the proton-intercalated FeSe also shows discrete superconducting phases, like those observed in Li or Na intercalated FeSe by using the SIC gating technique \cite{Meng2022,Wang2022}. 
The generality of proton intercalation method to induce or enhance superconductivity has been proved in various systems, such as BaFe$_{2}$As$_{2}$, FeSe, ZrNCl, and Bi$_{2}$Se$_{3}$ \cite{Cui2019}.

\subsubsection{Thin FeSe films}

\begin{figure}
\centering
\includegraphics[width=8.5cm]{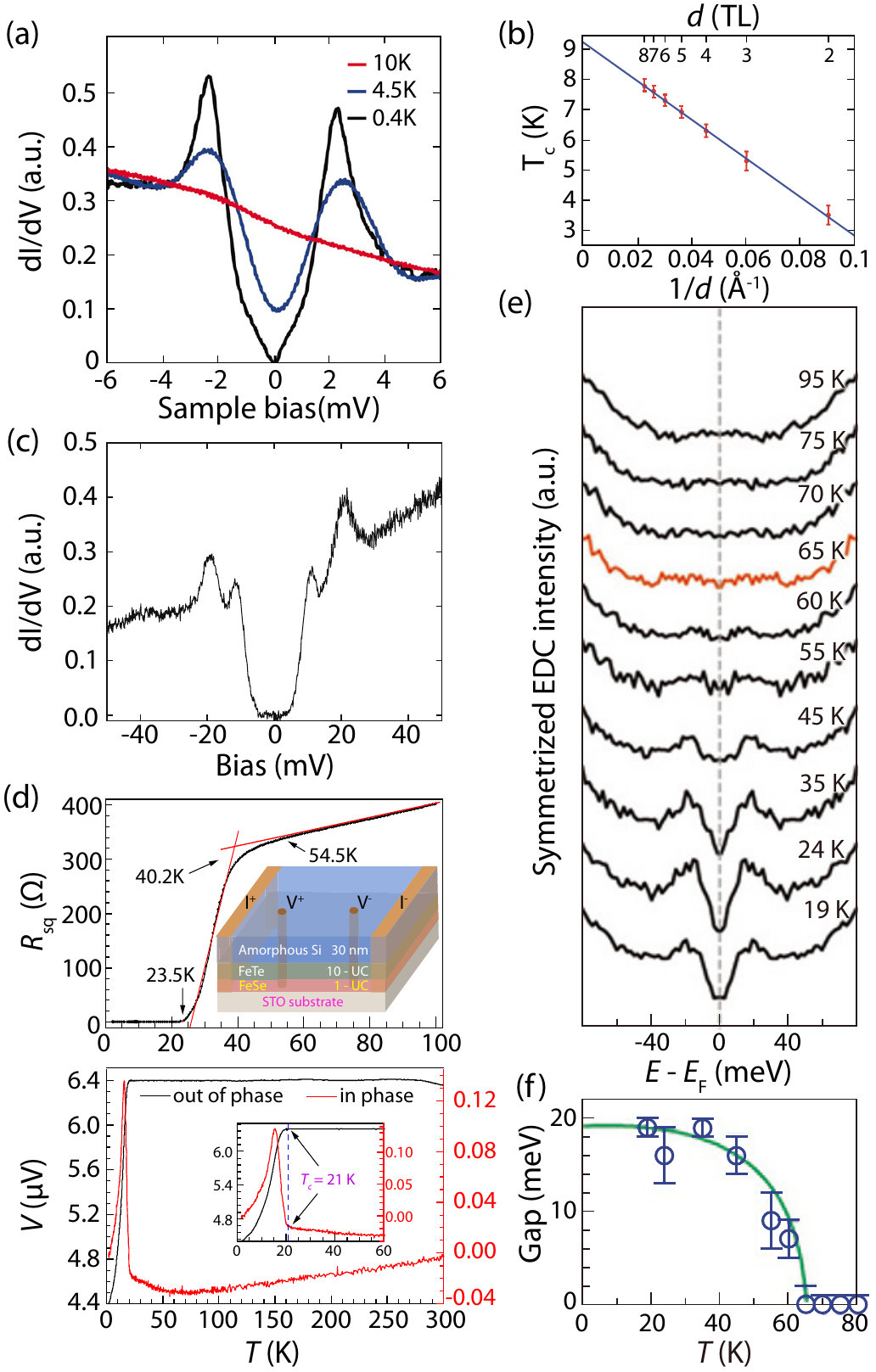}
\caption[]{(a) Temperature dependence of differential conductance spectra. From \cite{Song2011a}. (b) Superconducting transition temperature $T_{c}$ vs.	the inverse of the film thickness $d$. From \cite{Song2011b}. (c) Tunneling spectrum taken on the 1-UC-FeSe film on SrTiO$_{3}$(001) at 4.2~K. From \cite{Wang2012}. (d) Transport and diamagnetic measurements of 1-UC FeSe films grown on insulating SrTiO$_{3}$(001). From \cite{zhang2014direct}. (e) Symmetrized photoemission spectra of electron-like Fermi surface near the M point for the annealed 1-UC-FeSe/SrTiO$_{3}$. (f) The temperature dependence of the	superconducting gap. The green line represents the fit using BCS gap form. From \cite{He2013}.
\label{Fig_mono_FeSe}}
\end{figure}

Because of the weak vdW interlayer interaction of FeSe, the growth of few-layer or even monolayer FeSe on various substrates is feasible. FeSe thin films can be grown on bilayer graphene substrates through layer-by-layer epitaxy with preferential orientation along $c$ axis \cite{Song2011a,Song2011b}. Due to the weak interaction between FeSe and graphene, the in-plane lattice parameter of FeSe film is nearly the same as that of the bulk material (3.8 \AA) and for 8-unit cell (UC) FeSe films, the V-shape superconducting gap of 2.1 meV is observed from differential tunneling conductance (d$I$/d$V$) spectra [Fig.~\ref{Fig_mono_FeSe}(a)], in agreement with those in bulk FeSe \cite{Song2011a,Song2011b,Hsu2008,kasahara2014field}. Moreover, with decreasing the thickness of FeSe film on graphene, $T_{c}$ decreases gradually. The $T_{c}$ values for 8-unit cell (UC) and 2-UC FeSe films are 7.8~K and 3.7~K, respectively [Fig.~\ref{Fig_mono_FeSe}(b)], and superconductivity cannot be observed in 1-UC FeSe film down to 2.2~K \cite{Song2011b}.
As shown in Fig.~\ref{Fig_mono_FeSe}(b), the relationship between the $T_{c}$ and film thickness $d$ can be fitted well by using the empirical formula $T_{c}(d) = T_{c0}(1-d_{c}/d)$, where $T_{c0}$ is the critical temperature of the bulk material and $d_{c}$ is the threshold for the onset of superconductivity \cite{Song2011b}. Similar behavior has been observed in Pb and YBa$_{2}$Cu$_{3}$O$_{x}$ films \cite{Ozer2006,Tang2000}, and theoretically, it has generally been interpreted by adding a surface-energy term in the Ginzburg-Landau free energy of a superconductor \cite{simonin1986surface}. The fitted $T_{c0}$ and $d_{c}$ are about 9.3~K and 7 \AA, consistent with the $T_{c}$ of bulk FeSe \cite{Song2011b,Hsu2008}. These results imply that the graphene substrate has only a minor effect on the superconducting properties of FeSe films.

Distinct from 1-UC FeSe film on graphene, the 1-UC FeSe film grown on Nb-doped SrTiO$_{3}$(001) substrates by molecular beam epitaxy (MBE) with a proper annealing process exhibits significantly enhanced superconductivity \cite{Wang2012}. An overall U-shaped gap of 20.1 meV is observed [Fig.~\ref{Fig_mono_FeSe}(c)], which is nearly one order of magnitude higher than that of bulk FeSe and the superconducting gap remains visible at $\sim$43~K \cite{Wang2012}. Subsequently, STM/STS \cite{Wang2012,Zhang2014a}, transport and magnetization measurements \cite{Zhang2014a,Deng2014,Sun2014HighTemperature,ge2015superconductivity,Zhang2015,Li2015,Zhou2016}, and ARPES \cite{Liu2012monoFeSe,He2013,Tan2013,Peng2014} have all reported high-temperature superconducting signatures in 1-UC FeSe/SrTiO$_3$. However, the characteristic temperatures inferred from different probes vary substantially with sample preparation, capping layers, and the criterion used to identify superconductivity. ARPES typically finds gap-closing temperatures of order $60$--$70$~K, whereas transport and diamagnetic measurements in many studies yield lower transition temperatures. An \textit{in-situ} four-point-probe transport study also reported superconductivity above 100~K \cite{ge2015superconductivity}.

As shown in Fig.~\ref{Fig_mono_FeSe}(d), the \textit{ex-situ} resistance measurement on 10-UC-FeTe/1-UC-FeSe/SrTiO$_{3}$ shows $T_{c,\rm {onset}}$ at 54.5~K and $T_{c,\rm {zero}}$ at 23.5 K. Moreover, the two-coil mutual inductance measurement confirms the superconductivity with $T_{c}$ = 21~K, in agreement with $T_{c,\rm {zero}}$ measured from resistance \cite{zhang2014direct}. From the temperature dependence of symmetrized photoemission spectra of the annealed
1-UC-FeSe/SrTiO$_{3}$ [Fig.~\ref{Fig_mono_FeSe}(e)], the fitted gap size and $T_{c}$ are $\sim$ 19 meV and 65~K [Fig.~\ref{Fig_mono_FeSe}(f)], respectively, which is consistent with the STS results \cite{Wang2012}. 
Further ARPES study suggests that the pairing correlations may persist up to 83~K and the pairing state can be further divided into two temperature regions: the 64--83~K region may correspond to superconducting fluctuations while the region below 64~K may correspond to the phase coherence of Cooper pairs \cite{xu2021spectroscopic}. In contrast to 1-UC-FeSe/SrTiO$_{3}$, the 2-UC or thicker FeSe films on SrTiO$_{3}$ show no evidence of superconductivity, indicating that the interface between FeSe and SrTiO$_{3}$ substrate plays a key role in the enhancement of superconductivity in 1-UC FeSe/SrTiO$_{3}$ \cite{Wang2012}.

\subsubsection{Thin FeTe films}

\begin{figure}
\centering
\includegraphics[width=8cm]{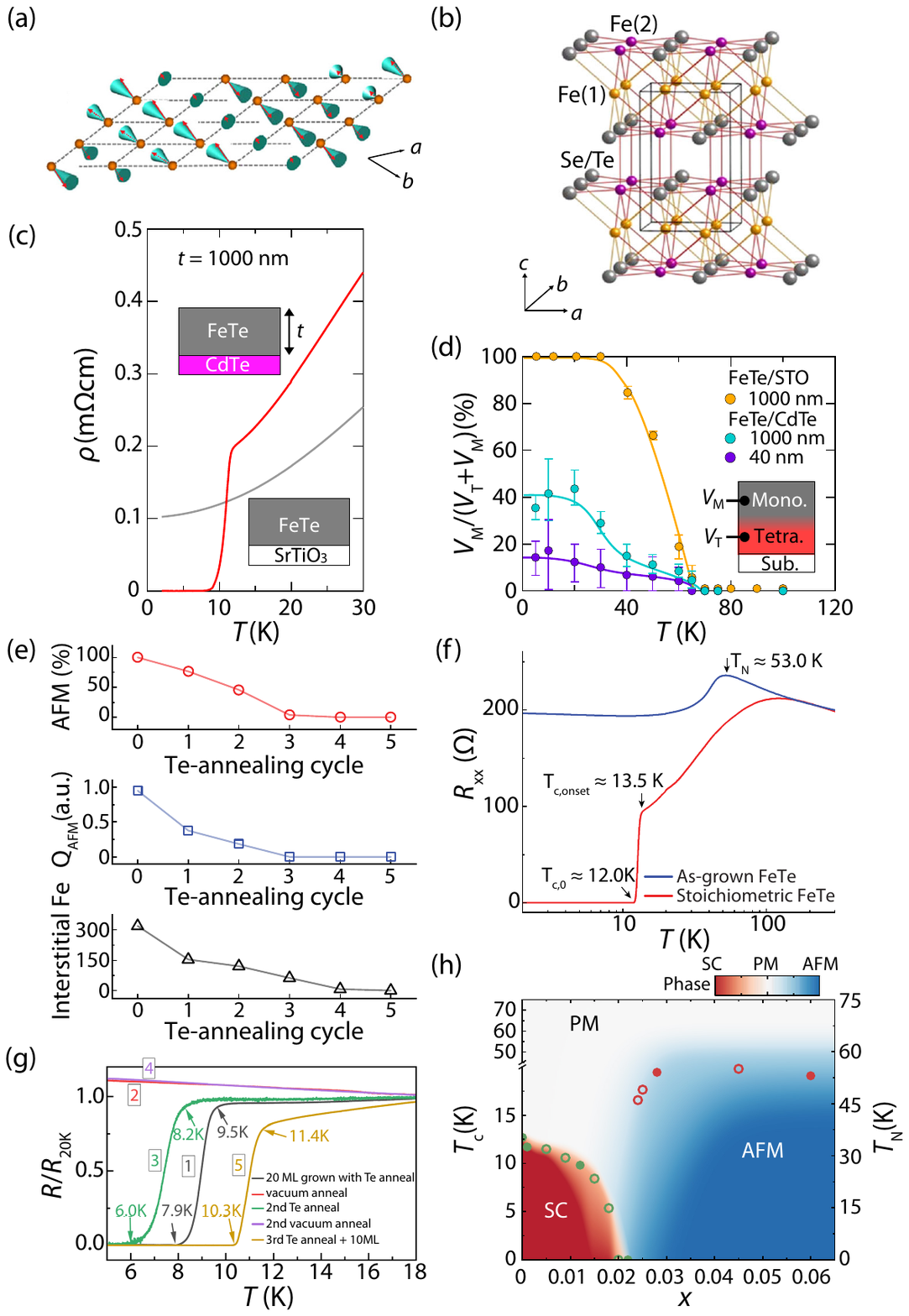}
\caption[]{(a) Crystal structure of $\alpha$-Fe(Te,Se), in which interstitial Fe atoms occupy the Fe(2) site. (b) Bicollinear magnetic structure of $\alpha$-FeTe. From \cite{Bao2009}. (c) Temperature dependence of the resistivity of 1000-nm-thick FeTe films grown on CdTe (red) and STO (gray). (d) Temperature dependence of the monoclinic-phase volume fraction, $V_{\rm M}/(V_{\rm T}+V_{\rm M})$. The inset schematically depicts phase separation into monoclinic and tetragonal regions. From \cite{Sato2025}. (e) AFM volume fraction, normalized intensity of the $Q_{\rm AFM}$ peak, and number of interstitial Fe atoms as functions of the Te-annealing cycle number. (f) Temperature-dependent sheet resistance $R_{xx}$ of as-grown (blue) and stoichiometric (red) 40-UC FeTe films. From \cite{Yan2026}. (g) Reversible tuning between the AFM and superconducting states by vacuum and Te-vapor annealing. From \cite{Xu2026}. (h) Phase diagram of Fe$_{1+x}$Te. Dark red, dark blue, and white denote the superconducting, AFM, and paramagnetic regions, respectively; light red and light blue indicate the corresponding transition regimes. From \cite{Yan2026}.
\label{Fig_FeTe}}
\end{figure}

Compared with FeSe, isostructural metallic FeTe has conventionally been regarded as nonsuperconducting in its bulk form \cite{Liu2010}. Instead, it undergoes a structural transition from the tetragonal phase to an orthorhombic or monoclinic phase near 70~K, accompanied by long-range bicollinear antiferromagnetic (AFM) order [Fig.~\ref{Fig_FeTe}(a)] \cite{PhysRevB.79.054503,Ma2009,Bao2009,Rodriguez2011}. Bulk Fe$_{1+x}$Te crystals typically contain excess Fe at the interstitial Fe(2) site [Fig.~\ref{Fig_FeTe}(b)], with $0.04<x<0.17$ \cite{Liu2010,Bao2009,Rodriguez2011,Koz2013,Joon2009}. These interstitial Fe atoms stabilize the bicollinear AFM order and strongly affect the crystal structure, magnetic order, and transport properties.

Nevertheless, superconductivity appears to be in close proximity to the bicollinear AFM phase. Chemical substitution and epitaxial strain can induce superconductivity when the monoclinic phase and bicollinear AFM order are suppressed [Figs.~\ref{Fig_FeTe}(c)--\ref{Fig_FeTe}(d)] \cite{Liu2010,Han2010,Sato2025}. Superconductivity with transition temperatures near 10~K has also been reported in FeTe films grown in an oxygen-containing atmosphere or exposed subsequently to air or oxygen
\cite{Si2010,Nie2010}. In addition, superconductivity with $T_c\sim 9$--13~K has been observed in FeTe/telluride heterostructures involving a range of materials, including Bi$_2$Te$_3$ \cite{He2014,manna2017interfacial}, MnTe \cite{Yao2022},
Fe$_3$GeTe$_2$ \cite{Li2026}, and CdTe \cite{Sato2025}. Although the microscopic origin of superconductivity in these systems remains under discussion, these observations suggest that the local chemical environment, substrate strain, and the concentration of interstitial Fe are all important control parameters.

Recent post-growth annealing studies have provided more direct evidence for the role of interstitial Fe. Te-flux annealing of FeTe films was reported to reduce the interstitial Fe concentration, suppress the AFM order [Fig.~\ref{Fig_FeTe}(e)], and induce superconductivity with $T_c\sim 11.4$--13.5~K [Fig.~\ref{Fig_FeTe}(f)] \cite{Yan2026,Xu2026}. The superconducting and AFM states were further reported to be reversibly tuned by alternating Te-vapor and vacuum annealing [Fig.~\ref{Fig_FeTe}(g)] \cite{Xu2026}. These results strongly support the view that excess Fe plays a central role in stabilizing the bicollinear AFM state and suppressing superconductivity. The phase diagram of Fe$_{1+x}$Te in Fig.~\ref{Fig_FeTe}(h) was interpreted as
indicating that superconductivity may emerge in nearly stoichiometric FeTe films, whereas interstitial Fe concentrations above
$x\sim0.02$ favor AFM order \cite{Yan2026}. Whether superconductivity can be realized in strain-free bulk stoichiometric FeTe remains an open question, because substrate strain, interfacial effects, and oxygen or Te chemical potentials may all contribute in thin films.

\subsubsection{Common properties of heavily electron-doped FeSe-derived superconductors}

\begin{figure}
\centering
\includegraphics[width=8.5cm]{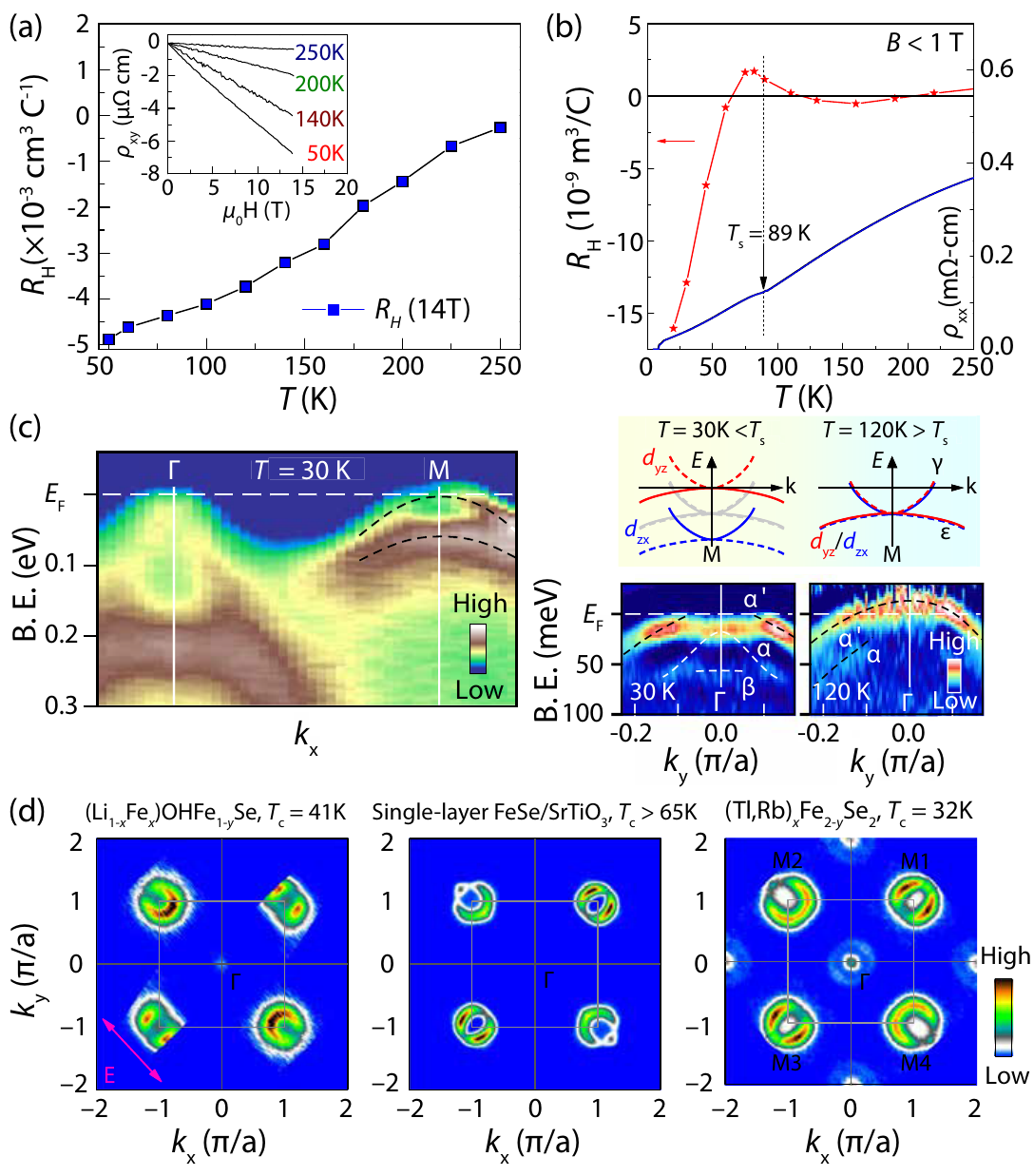}
\caption[]{(a) Temperature dependence of the $R_{\rm H}(T)$ at the field of 14 T for LiNHFeSe-122 single crystal. Inset: field dependence of Hall resistivity		$\rho_{xy}(\mu_{0}H)$ at various temperatures. From \cite{ sun2017extreme}. (b) Temperature dependence of low-field Hall coefficient	$R_{\rm H}(T)$ and in-plane resistivity $\rho_{xx}(T)$ for $\beta$-FeSe single crystal. From \cite{watson2015}. (c) ARPES results for $\beta$-FeSe below and above $T_{s}$ ($T$ = 30~K and 120~K). From \cite{Nakayama2014}. (d) Fermi surface mappings of	(Li$_{0.84}$Fe$_{0.16}$)OHFe$_{0.98}$Se, superconducting monolayer FeSe film and (Tl, Rb)$_{x}$Fe$_{2-y}$Se$_{2}$. From \cite{zhao2016common}.
\label{Fig_FeSe}}
\end{figure}

\begin{figure}
\centering
\includegraphics[width=8.5cm]{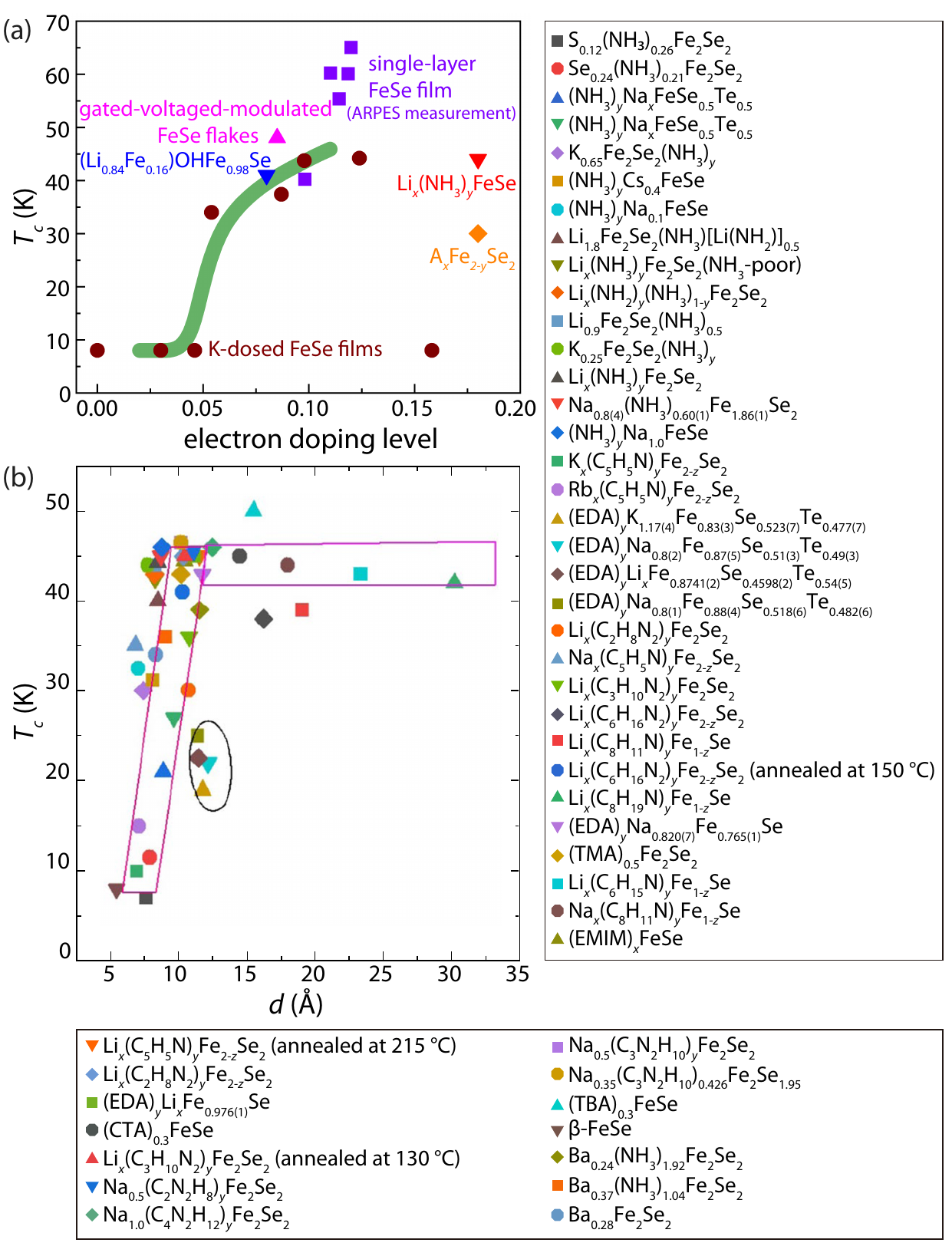}
\caption[]{(a) $T_{c}$ as a function of electron doping level in FeCh-based superconductors. The green line serves as a guide to the eye. (b) Relationship between $T_{c}$ and $d$ in FeCh-based superconductors. From \cite{xu2022research}.
\label{Fig_FeCh_PhaseDiagram}}
\end{figure}

Most HED FeSe-based superconductors exhibit some common features. First, the Hall coefficient $R_{\rm H}$ is negative in the low-temperature region \cite{sun2017extreme,dong2015li,lei2016evolution,Ding2014,Sun2014HighTemperature}. Figure  \ref{Fig_FeSe}(a) shows the temperature dependence of $R_{\rm H}(T)$ of LiNHFeSe-122 at $\mu_0H = 14$ T as an example. The dominant carriers are electrons, and this is distinctly different from FeSe, which is a compensated semimetal with nearly equal electron and hole concentrations at low temperature and shows a sign change of $R_{\rm H}$ near the structural transition ($T_{s}$ $\sim$ 89~K) [Fig.~\ref{Fig_FeSe}(b)] \cite{watson2015}.
The general behavior suggests that the high-$T_{c}$ superconductivity can be achieved in FeCh-based superconductors by heavy electron-doping \cite{Tan2013}. Various experimental results indicate that the electron-doping level is a key parameter governing $T_{c}$. After summarizing the relation of $T_{c}$ and carrier concentration for most HED FeSe-based superconductors, Sun \textit{et al.} proposed that the high-$T_{c}$ superconductivity above 40~K emerges and is robust against variations in electron-pocket size and interlayer spacing $d$ once the electron doping level is larger than 0.05 e$^{-}$/Fe [Fig.~\ref{Fig_FeCh_PhaseDiagram}(a)] \cite{sun2017extreme}. 
For 1-UC-FeSe/SrTiO$_{3}$ film, the optimal carrier doping level is $\sim$ 0.12 e$^{-}$/Fe \cite{He2013,wen2016anomalous}, which is between those of FeSe-11111 and LiNHFeSe-122  \cite{sun2017extreme} [Fig.~\ref{Fig_FeCh_PhaseDiagram}(a)], but it exhibits a much higher $T_{c}$ ($\sim$ 65~K) \cite{Wang2012,He2013}. 
Such enhancement of superconductivity should originate from other factors. The interfacial strong electron-phonon coupling between monolayer FeSe film and SrTiO$_{3}$ substrate is considered to be one of the key factors to boost $T_{c}$ \cite{Lee2014,Zhang2017a,song2019evidence,rebec2017coexistence,Tang2016,Tian2016,zhang2016superconducting,zhang2019enhanced,liu2021highorder,faeth2021interfacial,Zhang2017b,Lee2015,Coh2015,Rosenstein2016,Xie2015,Li2016}, i.e., the high-$T_{c}$ in monolayer FeSe film is a synergistic effect of the intrinsic pairing mechanism in the HED FeSe and interactions between the FeSe electrons and SrTiO$_{3}$ phonons.

Second, electron doping changes the electronic structures of the HED FeSe-based superconductors dramatically. ARPES measurements show that, in $\beta$-FeSe below $T_{s}$, two hole-like bands occur around $\Gamma$ (Brillouin zone (BZ) center), whereas near $M$ point (BZ
corner), an electron-like band crossing $E_{\rm F}$ and a hole-like band at higher binding energy are observed [Fig.~\ref{Fig_FeSe}(c)] \cite{Nakayama2014}. An energy splitting of bands up to 50 meV around $M$ point is observed below $T_{s}$. This originates from the splitting of the $d_{yz}$ and $d_{xz}$ orbital bands [Fig.~\ref{Fig_FeSe}(c)], indicating that an orbital ordering (nematic phase) appears below structural transition \cite{Nakayama2014}.

However, ARPES results reveal that all of HED FeSe-based superconductors exhibit a common electron-only Fermi-surface topology, remarkably different from that of $\beta$-FeSe. Electron-like pockets still exist around the $M$ point but hole-like pockets near the $\Gamma$ point disappear [Fig.~\ref{Fig_FeSe}(d)] \cite{Mou2011,Qian2011,chen2011electronic,Zhang2011,Niu2015,zhao2016common,wen2016anomalous,Liu2012monoFeSe,He2013,Tan2013,Peng2014,miyata2015hightemperature,zhang2016superconducting}. The superconducting gap structure and its implications for pairing in these electron-only systems are discussed collectively in Sec.~VI.C.3.

Third, besides the electron doping effect of intercalants, in bulk HED FeSe-based superconductors, intercalation also substantially increases the separation of adjacent FeSe layers \cite{Guo2010,Ying2012,lu2015coexistence,sun2017extreme}, which strongly influences the physical properties and $T_{c}$. The increased interlayer distance $d$ generally increases the anisotropy of physical properties for the HED FeCh-based superconductors with intercalants. For example, the resistivity anisotropies $\gamma_{\rho} =
\rho_{c}/\rho_{ab}$ for FeSe-11111 and LiNHFeSe-122 are larger than 1000 [Fig.~\ref{Fig_FeSe122}(g)] \cite{dong2015li,sun2017extreme}, much larger than that of $\beta$-FeSe ($\gamma_{\rho}\sim$ 25) \cite{Sadakov2015,tanatar2016origin}, and the anisotropy of upper critical field $\gamma_{Hc2} = H_{c2}^{ab}/H_{c2}^{c}$ near $T_{c0}$ is about 16 for LiNHFeSe-122, which is also remarkably higher than that of $\beta$-FeSe ($\gamma_{Hc2}\sim$ 5) \cite{Her2015}. 
On the other hand, the relationship between $T_{c}$ and $d$ for various FeCh-based SC is summarized in Fig.~\ref{Fig_FeCh_PhaseDiagram}(b) \cite{xu2022research}. 
It can be seen that $T_{c}$ increases quickly with increasing $d$ until $d$ $\sim$ 8.5 -- 9 \AA, above which $T_{c}$ saturates with the value between 40~K -- 45~K \cite{Noji2014,Hosono2016,sun2017extreme,xu2022research}. 
Density functional theory (DFT) suggests that the Fermi surfaces of bulk HED FeSe-based superconductors become more two-dimensional with the increase of $d$ and turn completely two-dimensional at $d$ = 8.1 -- 10.3 \AA\ \cite{Guterding2015}, in line with the threshold value of $d$ where $T_{c}$ starts to saturate. However,
because all high-$T_{c}$ FeSe-based superconductors at ambient pressure with increased $d$ values are HED and the high $T_{c}$ can also be realized in gated-voltage-modulated FeSe flakes, the increased $d$ is an important but not a decisive factor in the enhancement of superconductivity in the HED FeSe-based superconductors in contrast to the change of electron concentration.

\subsubsection{Charge-neutral and hole-doped FeSe intercalates}

The high-$T_c$ FeSe-derived superconductors discussed above
generally involve electron transfer into the FeSe layers. Complementary
intercalates with charge-neutral spacer layers, or spacer layers that
induce hole doping, provide useful control systems for separating the
effects of carrier concentration, interlayer separation, and
dimensionality.

\begin{figure}
\centering
\includegraphics[width=8.5cm]{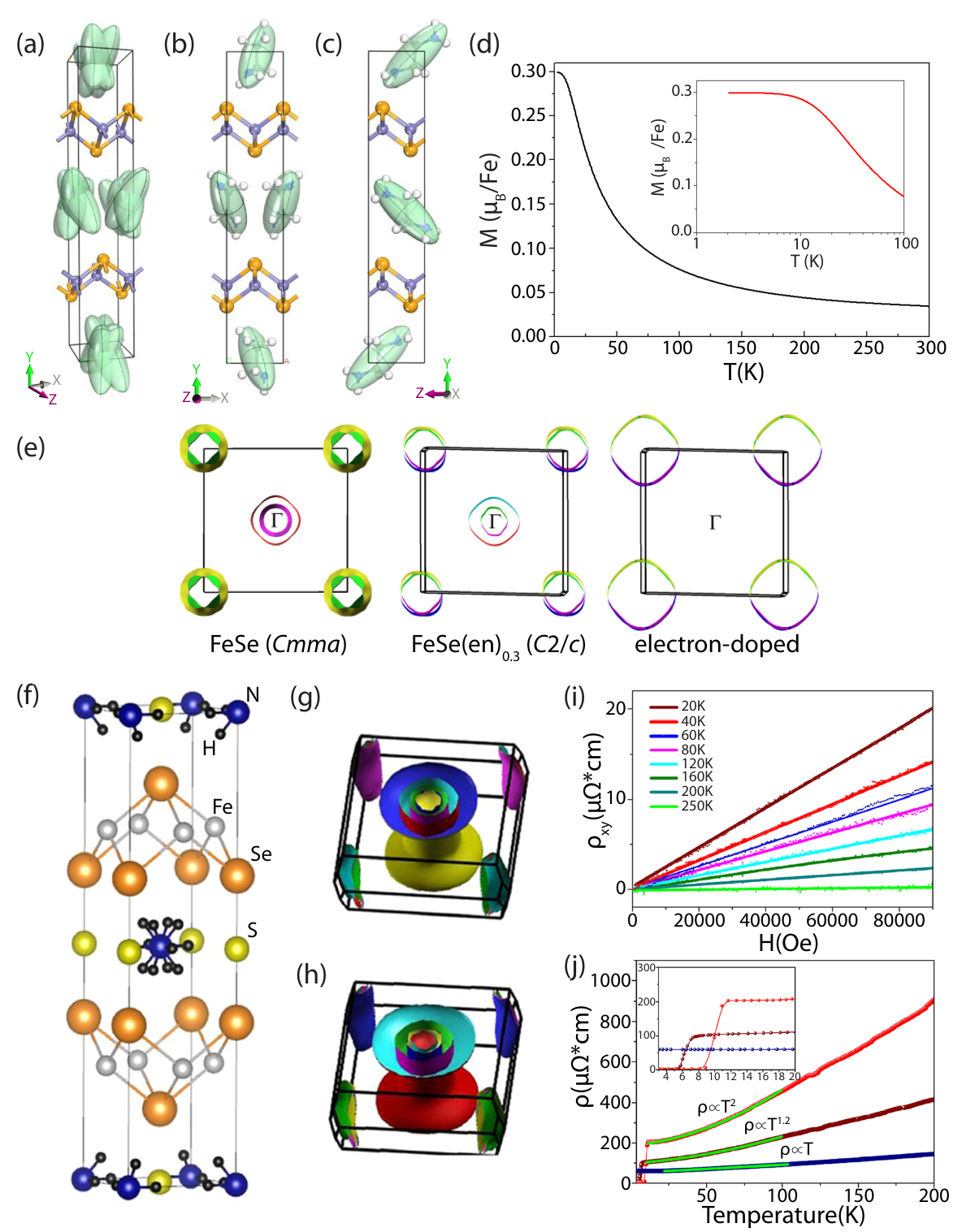}
\caption[]{(a) Crystal structure of (En)$_{0.3}$Fe$_{2}$Se$_{2}$ with En molecule represented as ellipsoids. Disorder of En molecules is superimposed within a unit cell. (b) and (c) Structure projected along the [001] direction and [100] direction, respectively. The orientational disorder is omitted for clarity. (d) ZFC magnetic susceptibilities of monoclinic (En)$_{0.3}$Fe$_{2}$Se$_{2}$. Inset shows the enlarged view at low temperature region. From \cite{Jin2016} (e) Fermi surfaces of	$\beta$-FeSe, (En)$_{0.3}$Fe$_{2}$Se$_{2}$ (En molecules omitted) and hypothetically electron-doped (En)$_{0.3}$Fe$_{2}$Se$_{2}$. From \cite{Stahl2017}. (f) Structure model for Se/S$_{x}$(NH$_{3}$)$_{y}$Fe$_{2}$Se$_{2}$. (g) and (h) Fermi surfaces of S$_{0.24}$(NH$_{3}$)$_{0.26}$Fe$_{2}$Se$_{2}$ (g) and Se$_{0.24}$(NH$_{3}$)$_{0.21}$Fe$_{2}$Se$_{2}$ (h). (i) Field dependence of Hall resistivity	$\rho_{xy}$ for	S$_{0.24}$(NH$_{3}$)$_{0.26}$Fe$_{2}$Se$_{2}$.	(j) Temperature dependence of the electric resistivity $\rho$ of S$_{0.12}$(NH$_{3}$)$_{0.26}$Fe$_{2}$Se$_{2}$	(wine),	S$_{0.24}$(NH$_{3}$)$_{0.26}$Fe$_{2}$Se$_{2}$ (red), and	Se$_{0.24}$(NH$_{3}$)$_{0.21}$Fe$_{2}$Se$_{2}$	(blue). From \cite{Sun2019}.
\label{Fig_FeSe_Neutral}}
\end{figure}

Usually, the intercalated charge spacer layers transfer electrons into the FeSe layers, accompanied by the expansion of $d$. This raises the question of how the physical properties evolve when $d$ is increased without changing the carrier density. Intercalation of charge-neutral spacer layers would shed light on the role of $d$ on physical properties. Using hydrothermal method, the charge-neutral ethylenediamine-intercalated FeSe material (En)$_{x}$Fe$_{2}$Se$_{2}$ (En = ethylenediamine, C$_{2}$N$_{2}$H$_{8}$) was successfully synthesized \cite{Jin2016,Stahl2017}. 
The air-stable (En)$_{x}$Fe$_{2}$Se$_{2}$ has an orthorhombic lattice cell. The FeSe and disordered En molecular layers stack alternately along the $c$-axis with expanded $c$-axis lattice parameters ($c$ $\sim$ 21.6 \AA) [Figs.~\ref{Fig_FeSe_Neutral}(a) - \ref{Fig_FeSe_Neutral}(c)]. No superconductivity is observed down to 2 K. Instead, the temperature-dependent magnetic susceptibility exhibits a Curie-Weiss (CW) behavior above 40~K with an effective moment of 1.92(1) $\mu_{\rm B}$ and a negative Weiss temperature of -5.7~K, indicating a weak AFM interaction [Fig.~\ref{Fig_FeSe_Neutral}(d)] \cite{Jin2016}. 
At low temperature, there may be a spin-disordered state with highly degenerate ground states \cite{Jin2016}, supporting a frustration-induced quantum paramagnetic ground state proposed in theory \cite{glasbrenner2015effect,wang2015nematicity}. As shown in Fig.~\ref{Fig_FeSe_Neutral}(e), theoretical calculations suggest that the Fermi surfaces of (En)$_{0.3}$Fe$_{2}$Se$_{2}$ largely retain the typical Fermi surface topology of $\beta$-FeSe with the more pronounced two-dimensional character due to the much bigger layer separation \cite{Stahl2017}. Adding about 0.2 electrons per formula unit increases the Fermi energy and the hole-like Fermi surfaces around the $\Gamma$-point vanish [Fig.~\ref{Fig_FeSe_Neutral}(e)], similar to those HED FeSe-based superconductors. Thus, the (En)$_{0.3}$Fe$_{2}$Se$_{2}$ can be regarded as a bulk analogue to the 1-UC-FeSe film \cite{Wang2012} without electron doping and interfacial electron-phonon coupling, both of which have been proposed to be important for high-$T_{c}$ superconductivity 
\cite{He2013,Tan2013,Lee2014}. 

On the other hand, because the Fe layers are encapsulated by negatively charged Se layers, the intercalation of anions or negatively charged layer seems to be infeasible because of the requirements for the stabilization of the rare anion-anion bonding. However, chalcogenide anions S or Se along with NH$_{3}$ molecules were successfully intercalated into FeSe using a hydrothermal ion-exchange redox reaction \cite{Sun2019}. The structure of (Se/S)$_{x}$(NH$_{3}$)$_{y}$Fe$_{2}$Se$_{2}$ is shown in Fig.~ \ref{Fig_FeSe_Neutral}(f). Theoretical calculations show that there are strong ionic characteristics in the unusual S-Se and Se-Se bonds, accompanied by the significantly enlarged hole pockets after
intercalation [Figs.~\ref{Fig_FeSe_Neutral}(g) and \ref{Fig_FeSe_Neutral}(h)], suggesting that charge transfer occurs between the FeSe layer and S (or Se) intercalants. These results are supported by the enhanced oxidation state of Fe ions probed by both x-ray photoelectron and $^{57}$Fe M\"{o}ssbauer spectra as well as the dominant hole-type carriers revealed by Hall resistivity measurements [Fig.~\ref{Fig_FeSe_Neutral}(i)]. More importantly, the bulk superconductivity appears in hole-doped S$_{x}$(NH$_{3}$)$_{y}$Fe$_{2}$Se$_{2}$ single crystals with the $T_{c}$ = 7~K and 11.5~K at $x$ = 0.12 and 0.24, respectively [Fig.~\ref{Fig_FeSe_Neutral}(j)].

\subsubsection{H$^{-}$-doped LnOFeAs systems}

Electron doping of LnOFeAs was initially achieved by replacing O$^{2-}$ with F$^{-}$. However, the solubility of F$^{-}$ is limited to approximately $x\sim0.15$, in part because of the preferential formation of LnOF, which long prevented access to the highly electron-doped regime of FeAs-1111 compounds. Hydride substitution provides an effective alternative: H$^{-}$ can occupy the O site in the LnO spacer layer and, relative to O$^{2-}$, donates approximately one electron to the FeAs layer. Because H$^{-}$ has an ionic size comparable to O$^{2-}$ and is stabilized in the electropositive Ln--O environment, high-pressure synthesis of Ln(O$_{1-x}$H$_x$)FeAs extends the accessible electron-doping range to $x\sim0.7$ \cite{Hanna2011,Iimura2016}.

\begin{figure}
\centering
\includegraphics[width=8.5cm]{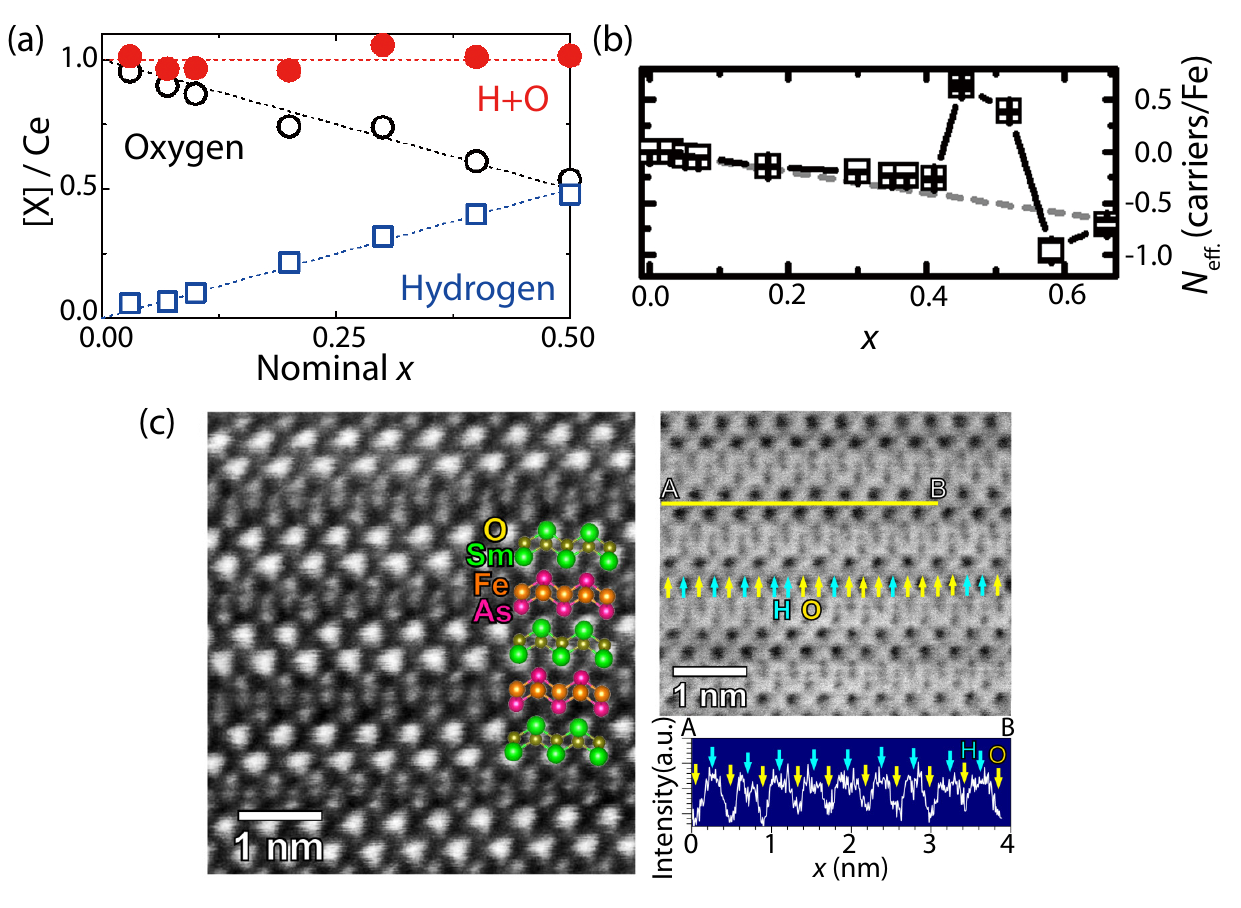}
\caption[]{Evidence for O$^{2-}$ substitution by H$^{-}$ and the accompanying electron doping in	Ln(O$_{1-x}$H$_{x}$)FeAs. (a) Chemical composition. (H + O) remains unity. (b) Increased carrier concentration with H$^{-}$ substitution. Data are evaluated by Hall measurement. Negative $n_{eff}$ means the	increased carriers are electrons. The deviation of linear behavior for $x$ = 0.45 and 0.53 is due to the local magnetism. Dashed line	denotes calculated $n_{eff}$ following the case that	substitution to an O$^{2-}$ with a H$^{-}$	generates a carrier electron. (c) Atomic images by STEM. Right figure shows ABF image sensitive to light atom. From \cite{Matsumoto2019}.
\label{Fig_LaOHFeAs_comp}}
\end{figure}

Ln(O$_{1-x}$H$_x$)FeAs samples are synthesized under high pressure using a hydrogen source such as LiAlH$_4$. The substitutional incorporation of H at the O site and the associated electron doping have been established by composition analysis, neutron diffraction, STEM, and Hall measurements
\cite{Hanna2011,Matsuishi2012,Matsumoto2019,Iimura2016} [Fig.~\ref{Fig_LaOHFeAs_comp}]. This method enabled systematic studies over a broad composition range, $0\lesssim x\lesssim0.7$.

\begin{figure}
\centering
\includegraphics[width=8.5cm]{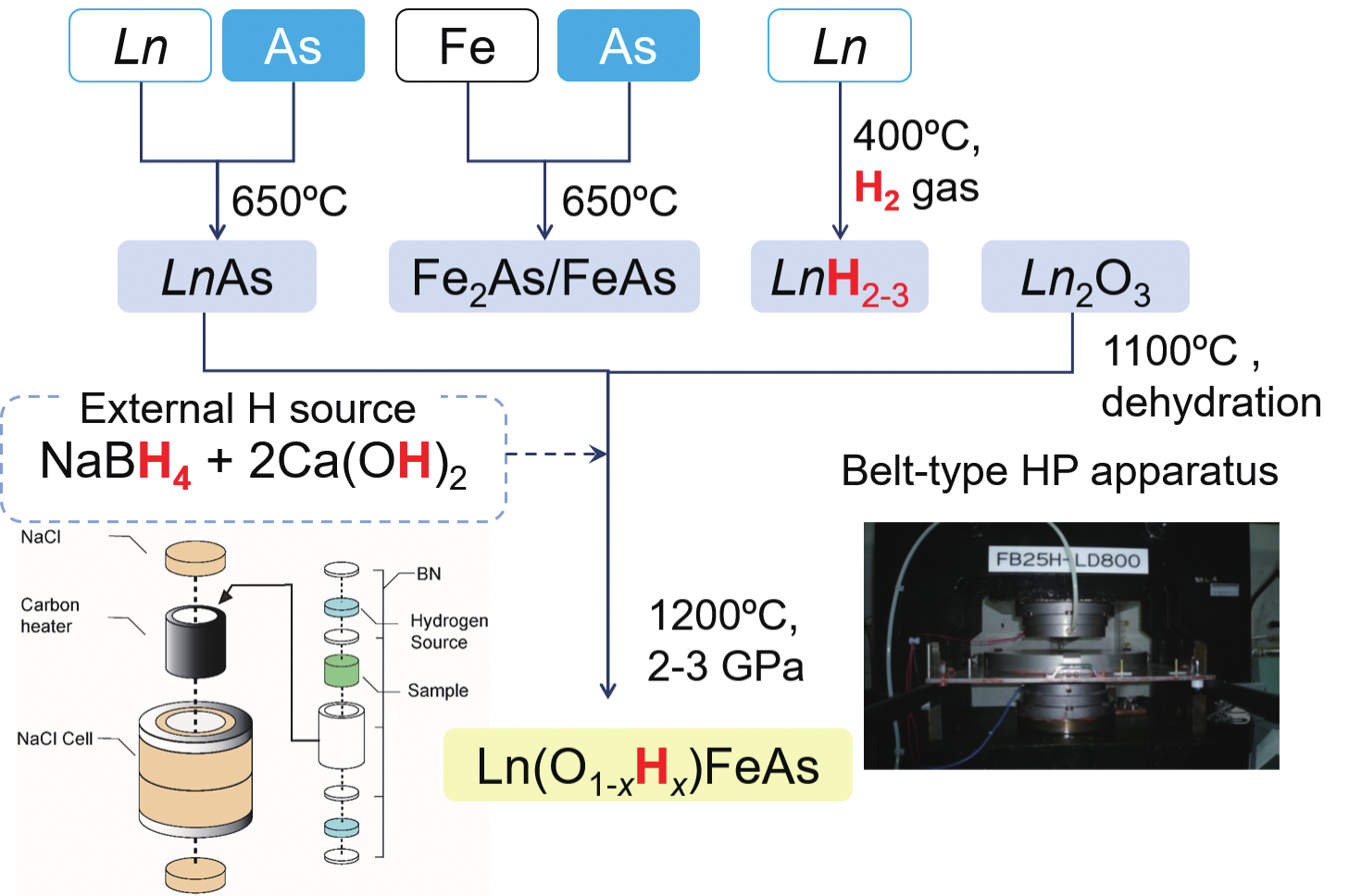}
\caption[]{Synthetic procedures of	Ln(O$_{1-x}$H$_{x}$)FeAs.
\label{Fig_LaOHFeAs_syn}}
\end{figure}

\begin{figure}
\centering
\includegraphics[width=8.5cm]{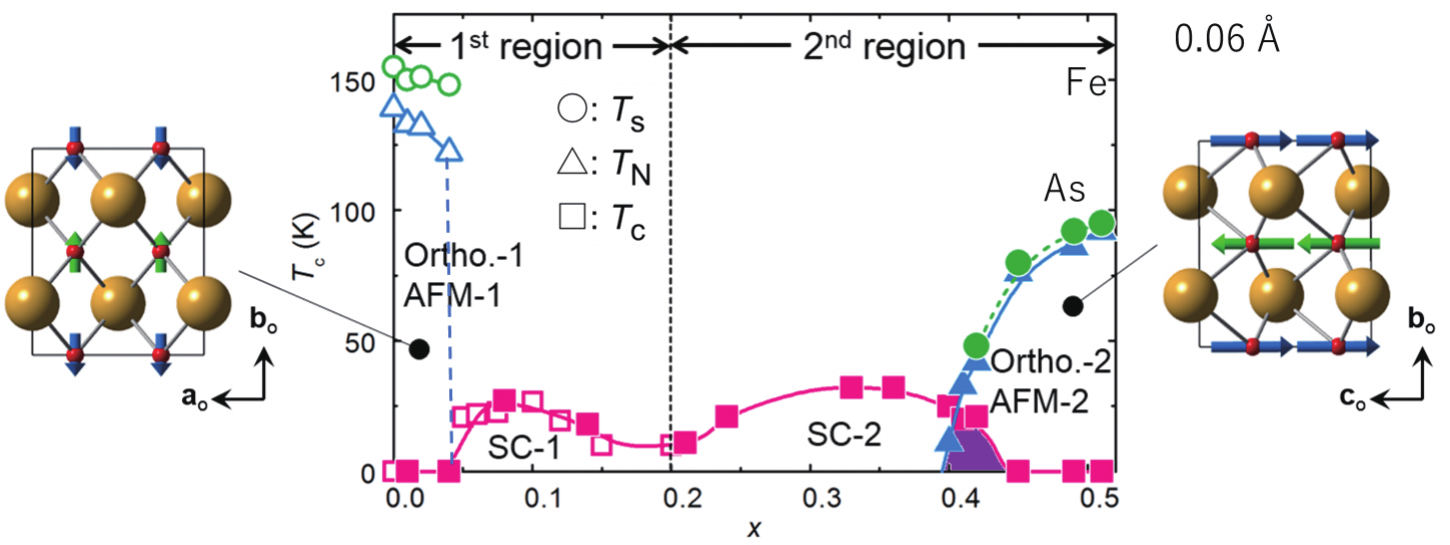}
\caption[]{Phase diagram of	La(O$_{1-x}$H$_{x}$)FeAs and spin arrangement of two AFM phases.
\label{Fig_LaOHFeAs_phase_diagram}}
\end{figure}

Figure \ref{Fig_LaOHFeAs_phase_diagram} shows the phase diagram of electron-doped La(O$_{1-x}$H$_{x}$)FeAs \cite{Iimura2012,hiraishi2014bipartite}. Two $T_{c}$-dome structures are seen for the H-doping, SC-1 ($x$ = 0.04 - 0.20) and SC-2 ($x$ = 0.20 - 0.45) phases. The $T_{c}$-dome in SC-1 phase almost agrees with that in the F-doped system but further electron doping over $x$ = 0.2 raises $T_{c}$ again to 36~K at $x$ = 0.36. The temperature dependences of resistivity above $T_{c}$ of SC-1 and SC-2 phases are entirely different. The former behaves as a Fermi liquid, whereas the latter behaves as a non-Fermi liquid in which the resistivity increases linearly with temperature. 
There are two parent phases, AFM-1 and AFM-2 adjacent to SC-1 and SC-2 phases at $x$ = 0 and $\sim$ 0.5, and both phases exhibit AFM order but the spin texture and spin moment are different, as shown in Fig.~\ref{Fig_LaOHFeAs_phase_diagram} and Table \ref{tab:LaOHFeAs} \cite{hiraishi2014bipartite}. 
These results indicate that the two superconducting regions are adjacent to distinct AFM phases. SC-1 evolves on electron doping away from AFM-1, whereas SC-2 is approached on reducing the electron concentration from the highly electron-doped AFM-2 regime.

\begin{figure}
\centering
\includegraphics[width=8.5cm]{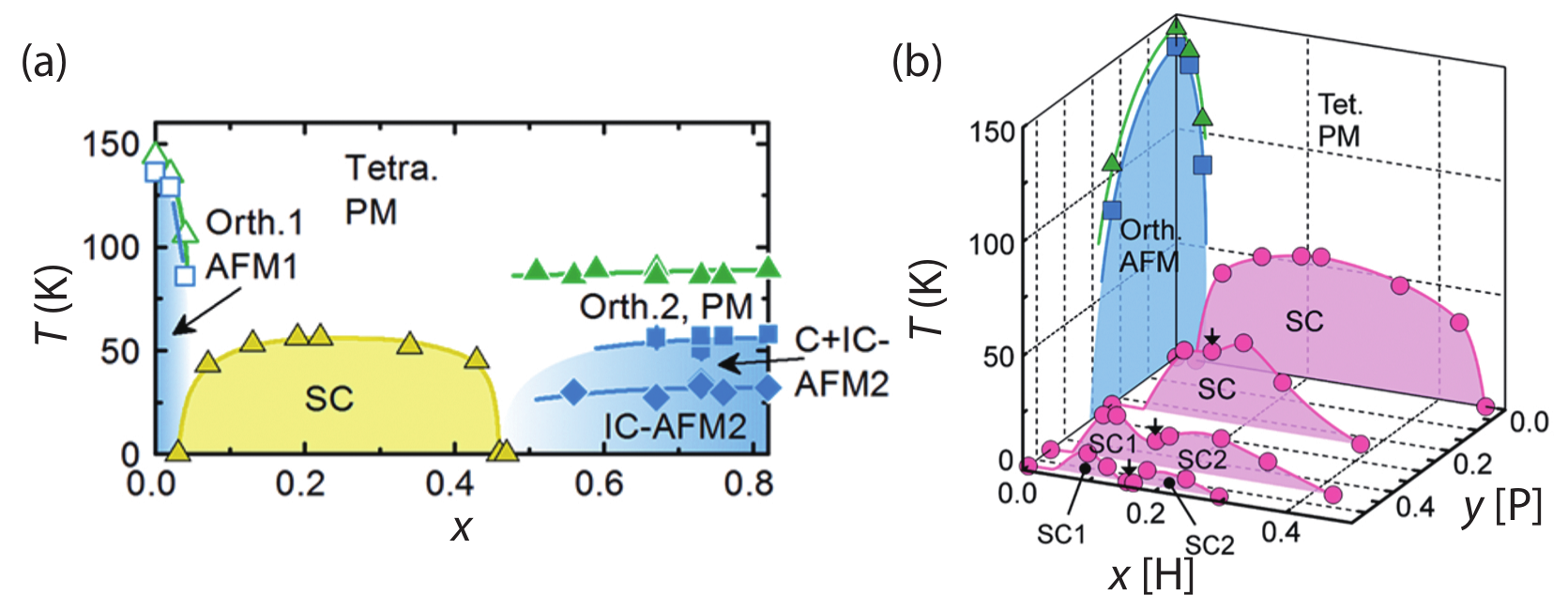}
\caption[]{Phase diagram of H-doped SmOFe(As, P) system. (a) Sm(O$_{1-x}$H$_{x}$)FeAs, (b)	Sm(O$_{1-x}$H$_{x}$)Fe(As$_{1-y}$P$_{y}$).
\label{Fig_LnOHFePn_phase_diagram}}
\end{figure}

\begin{figure}
\centering
\includegraphics[width=8.5cm]{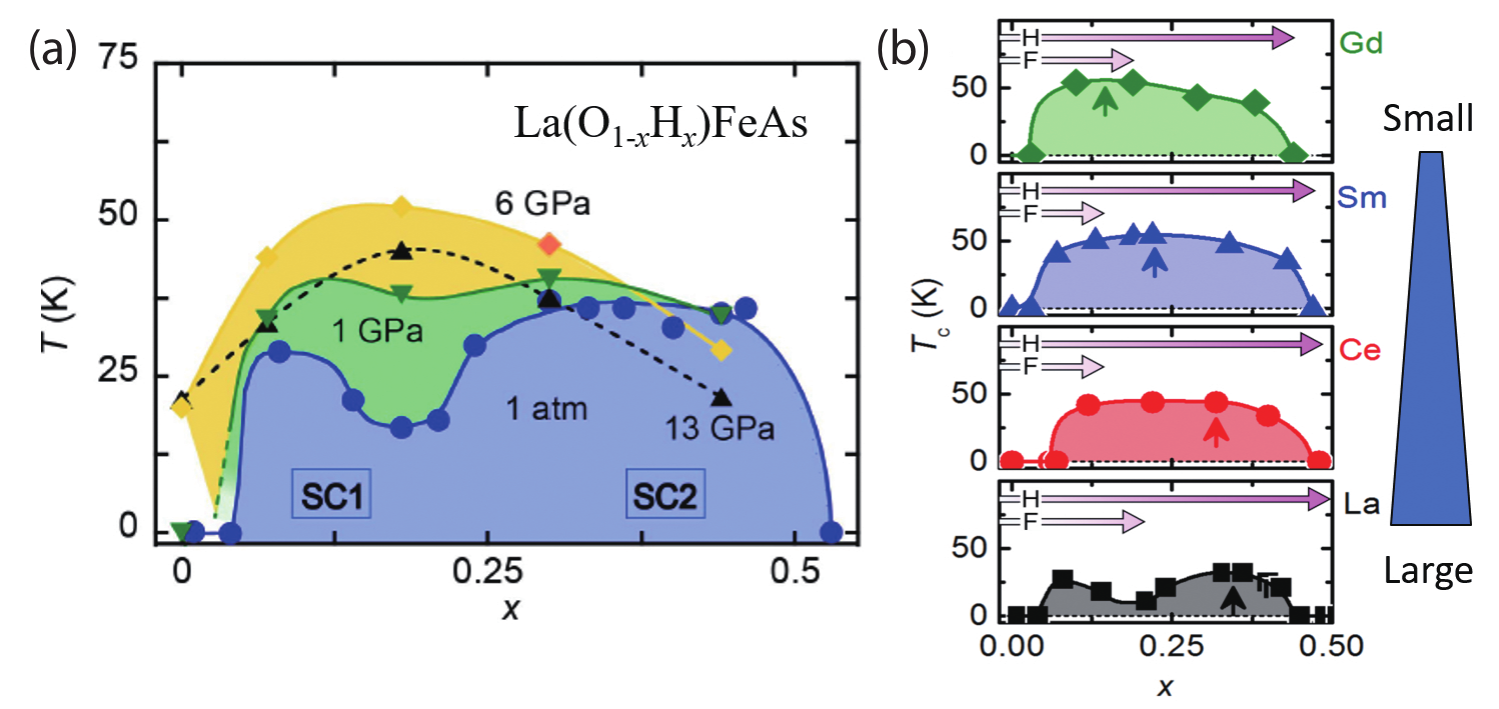}
\caption[]{The $T_{c}$-dome changes in LnOFeAs system. (a) Change in La(O$_{1-x}$H$_{x}$)FeAs with	pressure. From \cite{Takahashi2015}. (b) The evolution of $T_{c}$-dome in FeAs-1111 systems with electron doping	with F$^{-}$ or H$^{-}$.
\label{Fig_LnOHFeAs_size}}
\end{figure}

Figure \ref{Fig_LnOHFePn_phase_diagram}(a) shows a complete phase diagram of SmOFeAs with the highest $T_{c}$ obtained by H substitution. Although the apparent shape of $T_{c}$-dome is single, there exist two AFM phases located around $x$ = 0 and 0.5 like the LaOFeAs system \cite{Iimura2017}. Such a single-dome structure is commonly seen in Ln(O$_{1-x}$H$_{x}$)FeAs except La. An instructive $T_{c}$-dome structure is seen for Sm(O$_{1-x}$H$_{x}$)Fe(As$_{1-y}$P$_{y}$) system in which carrier concentration and chemical pressure can be tuned separately by H and P substitutions \cite{Matsuishi2014}.
When $T_{c}$ is lower than $\sim$ 30~K, the double-dome structure appears even for the Sm(O$_{1-x}$H$_{x}$)Fe(As$_{1-y}$P$_{y}$) system, as shown in Fig.~\ref{Fig_LnOHFePn_phase_diagram}(b). It is of interest to note that when high pressure is applied to La(O$_{1-x}$H$_{x}$)FeAs, the double dome transforms into a single dome and the maximal $T_{c}$ exceeds 50~K at $x$ $\sim$ 0.14 [Fig.~\ref{Fig_LnOHFeAs_size}(a)] \cite{Takahashi2015}, the composition corresponding to the valley at ambient pressure. These experimental findings suggest that the $T_{c}$ higher than 50~K in the FeAs-1111 system is realized when two different mechanisms controlling the SC-1 (low electron doping region) and the SC-2 (high electron-doped region) phases are superimposed [Fig.~\ref{Fig_LnOHFeAs_size}(b)]. This interpretation is consistent with the observation that the existence of two AFM phases adjacent to the end of $T_{c}$-domes is restricted to the FeAs-1111 system.

\begin{table}
\begin{center}
\caption{Characteristic properties of two regions in La(O$_{1-x}$H$_{x}$)FeAs.}
\label{tab:LaOHFeAs}
\begin{tabular}{c|c|c} 
\hline
\textbf{ } & \textbf{1$^{\rm st}$ region} & \textbf{2$^{\rm nd}$ region}\\
\hline
$x$ & 0.0 - 0.2 & 0.2 - 0.5 \\
\hline
$T_{c}^{\rm {max}}$ (K) & 29 & 36 \\
\hline
$T_{s}^{\rm {max}}$ (K) & 155 & 95 \\
\hline
$T_{\rm N}^{\rm {max}}$ (K) & 140 & 89 \\
\hline
Fe-Moment$^{\rm {max}}$ ($\mu_{\rm B}$) & 0.63 (2~K) & 1.21 (10~K) \\
\hline
$\rho(T)$ & $\sim$ $T^{1.9}$ & $\sim$ \ $T^{1.1}$ \\
\hline
SC/Magnetism & Separated & Overlapping \\
\hline
\end{tabular}
\end{center}

\end{table}

Other examples of two-dome superconductivity in heavily electron-doped FeSe-derived systems are discussed in Sec.~III.D.2. More generally, a nonmonotonic evolution of $T_c$ can arise when superconductivity competes or intertwines with another ordered state, or when distinct pairing interactions change in relative importance with doping or pressure. The cuprate La$_{2-x}$Ba$_x$CuO$_4$ provides a classic example: near $x\sim0.125$, superconductivity is strongly suppressed in association with charge and spin order, producing the well-known 1/8 anomaly \cite{Moodenbaugh1988,Tranquada01102020,fradkin2015colloquium,Hayden2024}. Related phenomena involving competing and intertwined orders in FeSCs are discussed in Sec.~V. A different possibility has been discussed for CeCu$_2$(Si$_{1-x}$Ge$_x$)$_2$, where a two-dome phase diagram was attributed to pairing channels associated with antiferromagnetic and valence/charge fluctuations \cite{Yuan2003}.

Several theoretical studies have examined the electronic, magnetic, and correlation effects that may underlie the two-dome phase diagram of H-doped FeAs-1111.
Yamakawa \textit{et al.} identified possible spin and orbital instabilities associated with incommensurate Fermi-surface nesting in La(O$_{1-x}$H$_x$)FeAs near $x=0.4$ \cite{Yamakawa2013}. Suzuki \textit{et al.} emphasized the increasing importance of next-nearest-neighbor hopping between Fe $3d_{xy}$ orbitals, for which $t_2>t_1$ can favor stripe-type spin fluctuations through $J_2>J_1$ \cite{Suzuki2014}. Using DFT+DMFT calculations, Moon \textit{et al.} showed that electron filling and structural modification induced by H substitution can have opposing effects on electronic correlations and magnetism; at high doping, the $3d_{xy}$ orbital acquires an enhanced role in the magnetic response \cite{Moon2016}. In a distinct pairing scenario, Misawa and Imada obtained superconductivity in an \textit{ab initio} multiorbital model and interpreted the two-dome structure in terms of enhanced uniform density fluctuations associated with instabilities toward first-order AFM and nematic transitions \cite{misawa2014superconductivity}. These theoretical proposals illustrate that the microscopic origin of the two superconducting regions remains unsettled.

\subsection{Cation-ordered FeSCs}

\begin{figure}
\centering
\includegraphics[width=8.5cm]{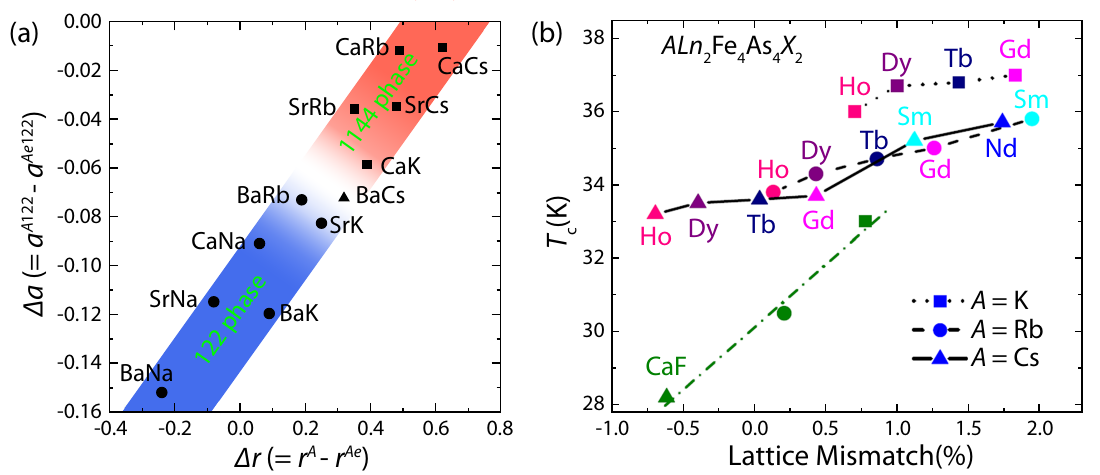}
\caption[]{(a) Phase boundary of ordered FePn-1144 and disordered FePn-122. From \cite{Iyo2016}. (b) Lattice	mismatch dependent $T_{c}$ in KCa$_{2}$F$_{2}$Fe$_{4}$As$_{4}$ and ALn$_{2}$O$_{2}$Fe$_{4}$As$_{4}$.	From \cite{Wu2017c}.
\label{Fig_FePn-1144}}
\end{figure}

Unlike conventional chemical doping, which randomly substitutes ions while preserving the parent space group (e.g., tetragonal $I4/mmm$) to continuously tune the superconducting phase diagram \cite{Rotter2008a,Cortes-Gil2010,Bukowski2009,Sefat2008,WOS:000349190300029}, cation-ordered FeSCs present a distinctly different structural paradigm. In 1144-type FeSCs ($A$AEFe$_{4}$As$_{4}$, {\it A} = K, Rb, Cs; AE = Ca, Sr, Eu) with $T_c \approx 30$--$36$~K [Fig.~\ref{Fig_structure}(e)], $A^{+}$ and AE$^{2+}$ layers strictly alternate along the $c$ axis, reducing the symmetry to $P4/mmm$ \cite{Iyo2016,Kawashima2016,Liu2016a,Liu2016b}. Forming this ordered phase requires a large ionic radius difference ($\Delta r > 0.4$ \AA) and a small in-plane lattice mismatch ($\Delta a/a \lesssim 1.8\%$) [Fig.~\ref{Fig_FePn-1144}(a)] to overcome the mixing entropy \cite{Iyo2016,Song2018a}. Attempts to incorporate trivalent Ln$^{3+}$ with $A^{+}$ generally revert to thermodynamically stable disordered $I4/mmm$ solid solutions (e.g., La$_{0.5}$Na$_{0.5}$Fe$_{2}$As$_{2}$) dictated by charge balance \cite{Song2018b,Yan2015,Iyo2018}.

The reduced symmetry of the 1144 phase yields unique physical properties. Transition-metal doping induces a novel hedgehog spin-vortex crystal magnetic order coexisting with superconductivity \cite{meier2018hedgehog}. Furthermore, the asymmetric A and AE layers break glide-mirror symmetry, inducing a topological band inversion that hosts Dirac surface states and MZMs \cite{Liu2020}. Practically, 1144 materials exhibit strong high-field performance, featuring low upper-critical-field anisotropy ($\gamma_{Hc2} < 2$) \cite{Meier2016}, a large zero-temperature depairing current densities ($J_0 = 2.65\times 10^8$ A cm$^{-2}$) surpassing 122-type FeSCs \cite{Cheng2019,Mishev2016}, and robust critical current densities ($J_c \sim 1$ MA cm$^{-2}$ at 15 T and 5~K) \cite{Singh2018}.

This structural ordering strategy was successfully extended to ``12244-type'' FeSCs [Fig.~\ref{Fig_structure}(f)]---e.g., ACa$_{2}$F$_{2}$Fe$_{4}$As$_{4}$ \cite{Wang2016,Wang2017} and ALn$_{2}$O$_{2}$Fe$_{4}$As$_{4}$ \cite{Wu2017a,Wu2017b,Wu2017c} ($T_c \sim 28$--$37$~K)---which intergrow ThCr$_{2}$Si$_{2}$-type (122) and ZrCuSiAs-type (1111) blocks [prototype La$_{3}$Ni$_{4}$As$_{4}$O$_{2}$ \cite{Klimczuk2009}]. Higher $T_c$ in these compounds positively correlates with a larger inter-block $a$-axis mismatch [Fig.~\ref{Fig_FePn-1144}(b)] and a smaller $c/a$ ratio, reflecting the critical role of intra-bilayer coupling and lattice instability \cite{Wu2017c}. Since their parent blocks possess very low or zero $T_c$, superconductivity inherently stems from structural modulations and self-hole-doping (average Fe valence of +2.25) \cite{Wu2017c}. Conversely, electron-doped 12244 phases are realized in BaTh$_{2}$(N$_{0.7}$O$_{0.3}$)$_{2}$Fe$_{4}$As$_{4}$ ($T_{c, \rm onset} \approx 30$~K), where oxygen substitution enables inter-block charge transfer \cite{Shao2019}.

\subsection{Europium-based FeSCs}

Europium-based (Eu-based) FeSCs provide a distinctive setting
in which itinerant magnetism of the FeAs layers coexists with localized
Eu $4f$ moments \cite{Zapf2017,liu2022ironbased}. In the parent
compound EuFe$_2$As$_2$ (Eu-122), the Fe sublattice undergoes a
stripe-type SDW transition near 190~K, together with the
tetragonal-to-orthorhombic structural transition
\cite{Tegel2008}, whereas the localized Eu$^{2+}$ moments order
separately into an A-type antiferromagnetic state near 18 K
\cite{Raffius1993,Ren2008,Xiao2009}. Pressure or chemical substitution
can suppress the Fe-derived SDW state and induce superconductivity with
$T_c\sim20$--$30$~K in Eu-122 derivatives
\cite{Miclea2009,Kurita2011,Matsubayashi2011,Terashima2009,Uhoya2010,
Ren2009,Cao2011,Jeevan2011,Tokiwa2012,Zapf2013,Zapf2011,Jiao2012,
Paramanik2013,Jiao2013,Blachowski2011,Matusiak2011,Jeevan2008,
Qi2012,Zhang2012}.

The Eu magnetic order should be distinguished from the
itinerant Fe-layer SDW. The Eu$^{2+}$ moments couple to the conduction
electrons through an RKKY interaction
\cite{Akbari2011,Akbari2013}, as supported by NMR studies
\cite{Guguchia2011,Dey2013}. Depending on the doping or pressure route,
this coupling can modify the Eu-spin orientation, produce canted or
ferromagnetic Eu configurations, and affect the low-temperature
superconducting response, including the upper critical field and gap
properties \cite{Nowik2011EuFeAsP,Nowik2011EuFeCoAs,Jin2013,Jin2015,
Kurita2011b,Yuan2009,Wu2009,Xia2014}. Eu-122 derivatives therefore
provide an instructive example in which local-moment magnetism is added
to, rather than identified with, the intrinsic Fe-layer magnetic
degrees of freedom.

\begin{figure}
\centering
\includegraphics[width=8 cm]{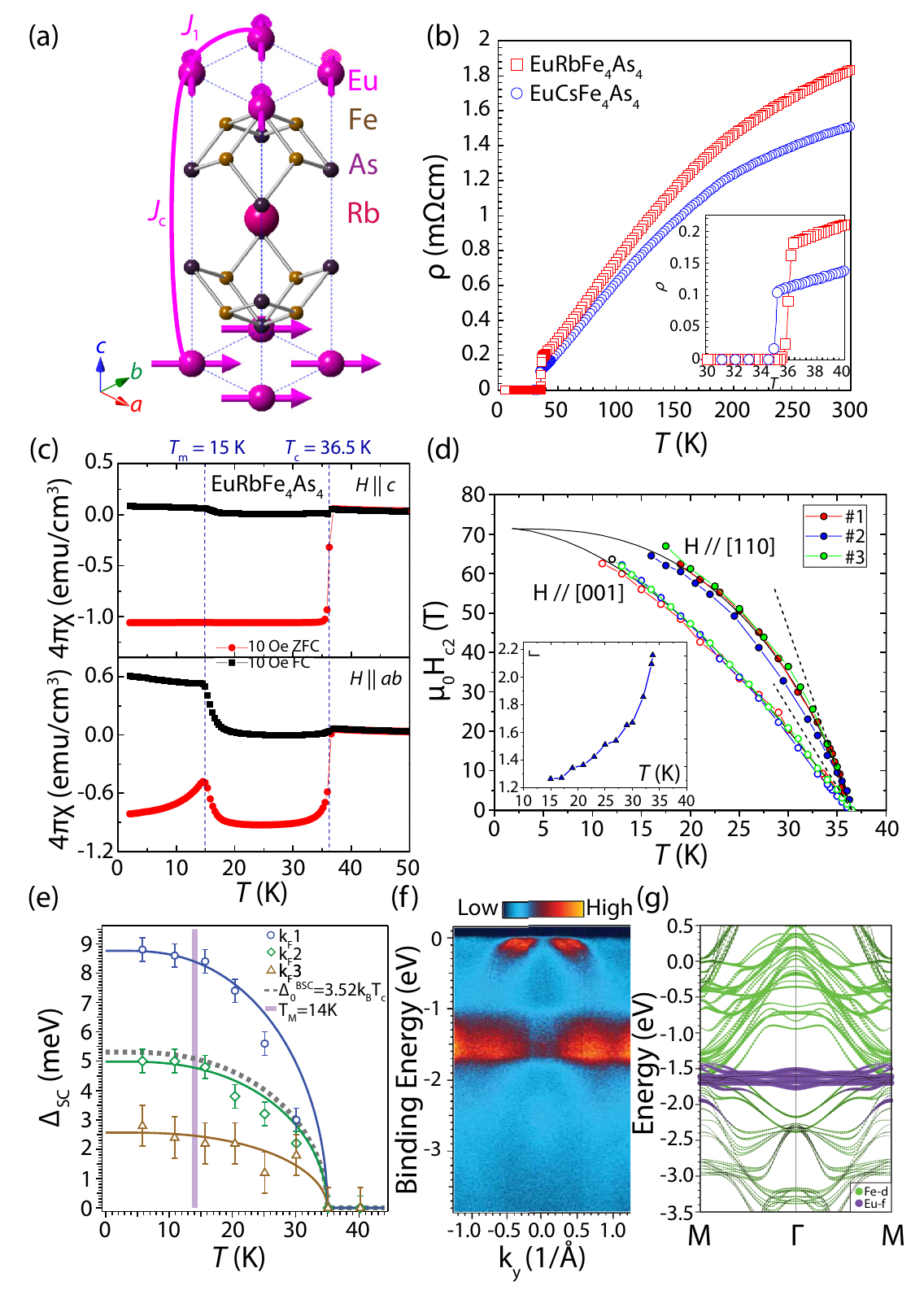}
\caption[]{(a) Magnetic structure of the Eu sublattice in RbEuFe$_{4}$As$_{4}$ with the magnetic propagation vector $k$ = (0, 0, 0.25). Dashed lines represent the chemical unit cell. The exchange couplings $J_{1}$ and $J_{c}$ are also described. From \cite{iida2019coexisting}. (b) Temperature dependence of electrical resistivity $\rho(T)$ for AEuFe$_{4}$As$_{4}$ ($A$ = Rb, Cs). Inset: enlarged view of $\rho(T)$ near $T_{c}$. From \cite{Kawashima2016}. (c) Temperature dependence of magnetic susceptibility $\chi$($T$) of RbEuFe$_{4}$As$_{4}$ in a field of 10 Oe applied along the $c$ axis and the $ab$ plane. From \cite{Hemmida2021}. (d) $\mu_{0}H_{c2}(T)$ of RbEuFe$_{4}$As$_{4}$ along [110] and [001] axes. The inset shows temperature dependence of the anisotropy ratio $\Gamma$ of $\mu_{0}H_{c2}$. From \cite{Smylie2019}. (e) Temperature dependence of the superconducting gap obtained for three different FS sheets of the hole pocket at $Z$-point. (f) Band dispersions in $M - \Gamma - M$ direction measured at 40 K. (g) Calculated band dispersions in $M - \Gamma - M$ direction showing Fe 3$d$ (green) and Eu 4$f$ (blue) bands. From \cite{Kim2021}.
\label{Fig_Eu1144Proper}}
\end{figure}

A particularly clear contrast is provided by the self-hole-doped
1144 compounds $A$EuFe$_4$As$_4$ ($A=$ Rb or Cs), in which robust bulk
superconductivity and Eu magnetic order coexist without extrinsic
chemical doping \cite{Iyo2016,Kawashima2016,Liu2016a,Liu2016b}. Featuring an optimal As--Fe--As geometry \cite{WOS:000261127100016,Lee2008,Mizuguchi2010} and intrinsic self-hole doping associated with charge transfer between the alternating
$A$ and Eu spacer layers, these intergrowth materials exhibit bulk
superconductivity at $T_c\sim35$--36.5~K without extrinsic doping
[Fig.~\ref{Fig_Eu1144Proper}(b)]. Below $T_c$, the Eu$^{2+}$ moments order at $T^{\rm Eu} \sim 15$~K \cite{Hemmida2021,Albedah2018a,Albedah2018b,Liu2016b} [Fig.~\ref{Fig_Eu1144Proper}(c)]. Unlike the A-type AFM in Eu-122, neutron diffraction and first-principles calculations reveal a non-trivial three-dimensional helical AFM structure: Eu$^{2+}$ spins align ferromagnetically within the $ab$ plane but rotate by $90^\circ$ across adjacent layers [Fig.~\ref{Fig_Eu1144Proper}(a)]. This helical configuration reflects a delicate interplay between magnetism and superconductivity, and it can be easily polarized into a ferromagnetic state under weak external fields \cite{iida2019coexisting,Kim2021,Devizorova2019,Koshelev2019,Smylie2018}.

\begin{figure}
\centering
\includegraphics[width=6cm]{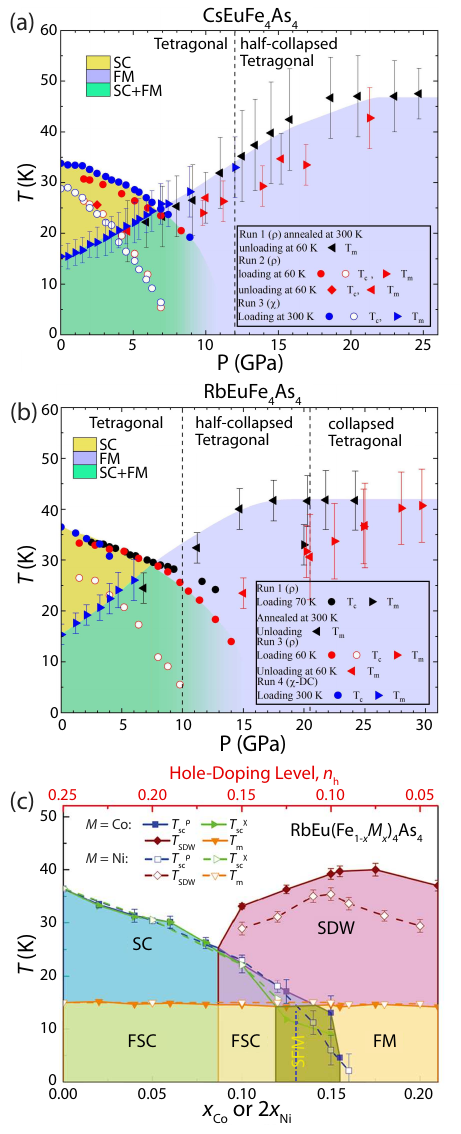}
\caption[]{(a) and (b) Phase diagrams of CsEuFe$_{4}$As$_{4}$ and RbEuFe$_{4}$As$_{4}$. The open and solid circles correspond to the midpoints and onset of superconducting transition. Solid triangles are magnetic transition temperatures. The dashed, vertical lines indicate onset pressures for the structural transitions. From \cite{Jackson2018}. (c) The phase diagrams of RbEu(Fe$_{1-x}$Co$_{x}$)$_{4}$As$_{4}$ (solid lines) and RbEu(Fe$_{1-x}$Ni$_{x}$)$_{4}$As$_{4}$ (dashed lines). FSC and SFM are the abbreviations of ferromagnetic superconductor and superconducting ferromagnet. From \cite{Liu2020}.
\label{Fig_Eu1144_PhaseDiagram}}
\end{figure}

In stark contrast to Eu-122, the coexistence of superconductivity and Eu-magnetism in AEuFe$_{4}$As$_{4}$ is exceptionally robust. Transport measurements reveal hole-dominant carriers without re-entrant resistivity. The $H_{c2}(T)$ is strongly Pauli-limited ($\mu_{0}H_{\rm P}(0) \sim 66$ T) and shows no anomalous suppression at $T^{\rm Eu}$ \cite{Smylie2019,Bristow2020} [Fig.~\ref{Fig_Eu1144Proper}(d)]. Furthermore, spectroscopic probes (ARPES, STS) confirm that the multiple superconducting gaps remain largely unperturbed by the magnetic transition [Fig.~\ref{Fig_Eu1144Proper}(e)] \cite{Kim2021,Stolyarov2020}. This striking decoupling has been attributed to the deep Eu 4$f$ level [Figs.~\ref{Fig_Eu1144Proper}(f) and \ref{Fig_Eu1144Proper}(g)] and the spatial separation between the Eu moments and the
FeAs superconducting layers \cite{Liu2016a,Liu2016b}. Nonetheless, a subtle dip in superfluid density near $T^{\rm Eu}$ remains observable, indicating that weak exchange interactions still persist \cite{Stolyarov2018,collomb2021observing}.

The independence of these two orderings is further corroborated by pressure and doping studies. Under hydrostatic pressure, $T_c$ monotonically decreases while $T^{\rm Eu}$ increases, crossing near 7 GPa [Figs.~\ref{Fig_Eu1144_PhaseDiagram}(a) and \ref{Fig_Eu1144_PhaseDiagram}(b)] \cite{Jackson2018}. Remarkably, pressure-induced structural transitions to half-collapsed or collapsed tetragonal phases minimally impact this magnetic phase diagram \cite{Holenstein2019,Borisov2018}. Chemically, non-magnetic Ca doping at the Eu site dilutes magnetism without altering $T_c$ \cite{Kawashima2018}. Conversely, in-plane doping with Ni or Co compensates holes and induces disorder, which suppresses $T_c$ and revives the Fe-SDW order \cite{Liu2017,Liu2020} [Fig.~\ref{Fig_Eu1144_PhaseDiagram}(c)]. Strikingly, $T^{\rm Eu}$ remains largely unchanged during in-plane transition-metal doping, as theoretical calculations reveal that the RKKY interaction is primarily mediated by Fe 3$d_{z^2}$ bands, which are largely insensitive to these dopants \cite{xu2019unique}.

Beyond 122 and 1144 phases, robust SC coexisting with Eu magnetism has also been realized in 1111-type EuFFeAs \cite{Zhu2009} and the recently discovered 112-type EuFeAs$_{2}$ ($T^{\rm Eu} \sim 40$~K, with $T_{c}$ up to 28~K via La, Ni, or Co doping) \cite{Yu2017,liu2021highorder,Yu2021}. These emerging systems continuously expand the fertile ground for exploring the complex phenomenology of localized spins embedded within strongly correlated superconductors.

\subsection{Hydrostatic-pressure tuning of FeCh-based superconductors}\label{sec_hp}

As shown in Figs.~\ref{Fig_Eu1144_PhaseDiagram} and ~\ref{Fig_HP_FeSe}, hydrostatic pressure
provides a disorder-free route to tune the lattice parameters,
electronic bandwidths, orbital overlaps, and magnetic exchange
couplings of FeSCs. It can therefore modify structural, nematic,
magnetic, and superconducting phase boundaries without changing the
chemical composition \cite{sun2017recent}. In many systems, modest
pressures suppress or induce structural and magnetic order, generate
nonmonotonic superconducting phase diagrams, and access nearby critical
regimes. At higher pressures, structural reconstructions can instead
stabilize new phases or suppress superconductivity
\cite{cheng2015pressure}.

Pressure experiments are particularly well suited to transport,
thermodynamic, diffraction, and several bulk spectroscopic probes.
However, pressure-cell geometries substantially restrict
surface-sensitive and momentum-resolved techniques, notably ARPES and
STM/STS. Hydrostatic-pressure phase diagrams should therefore be viewed
as complementary to, rather than interchangeable with, results from
chemical substitution and ambient-pressure spectroscopies.

\subsubsection{FeSe under hydrostatic pressure}

\begin{figure}[htbp!]
\centering
\includegraphics[width=8.5cm]{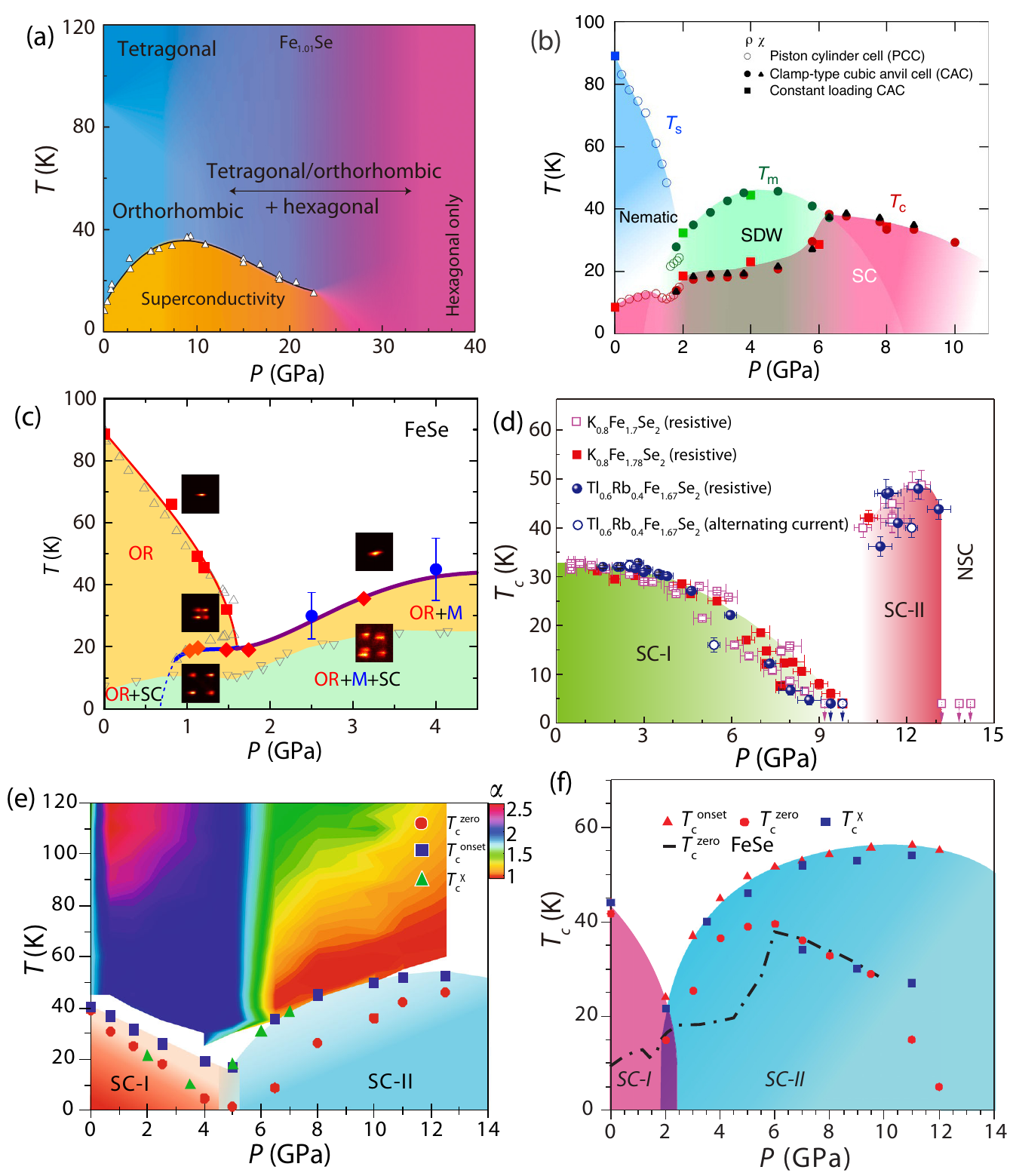}
\caption[]{Pressure--temperature phase diagrams of FeSe and heavily electron-doped FeSe-derived superconductors. (a) Broad-pressure-range phase diagram
of $\beta$-Fe$_{1.01}$Se. From \cite{Medvedev2009}. (b) Transport-derived phase diagram of FeSe. From \cite{sun2016dome}. (c) Structural, magnetic, and superconducting phase diagram of FeSe determined by x-ray diffraction and M\"ossbauer spectroscopy. From \cite{kothapalli2016strong}. (d) Pressure dependence of superconductivity in $A$FeSe-122. From \cite{Sun2012}. (e) Pressure-induced SC-I and SC-II phases in FeSe-11111. From \cite{sun2018reemergence}. (f) Pressure--temperature phase diagram of LiNHFeSe-122. From \cite{shahi2018hightc}.
\label{Fig_HP_FeSe}}
\end{figure}

Bulk FeSe provides a particularly important reference system
for understanding pressure tuning in iron-based superconductors. At
ambient pressure, it undergoes a tetragonal-to-orthorhombic transition
at $T_s\sim90$~K and becomes superconducting below
$T_c\sim8$--9~K without static long-range magnetic order. Hydrostatic
pressure strongly reconstructs this phase diagram, exposing the
interplay among nematicity, stripe antiferromagnetism, and
superconductivity
\cite{terashima2015pressure,sun2016dome,kothapalli2016strong,bohmer2019distinct}.

Early high-pressure measurements on
$\beta$-Fe$_{1.01}$Se established that superconductivity is strongly
enhanced under pressure, with $T_c$ increasing to approximately
37~K at intermediate pressures before decreasing at higher pressures
[Fig.~\ref{Fig_HP_FeSe}(a)] \cite{Medvedev2009}. This broad-pressure
study also revealed a structural evolution toward the hexagonal phase,
which ultimately suppresses superconductivity. Although static
pressure-induced magnetic order was not resolved in this early work, it
established FeSe as one of the most pressure-sensitive FeSCs.

Subsequent transport measurements revealed that the pressure
evolution of FeSe is more complex than a simple enhancement of $T_c$.
They identified the suppression of the nematic transition $T_s$, the
emergence of a dome-shaped SDW transition $T_m$, and a
strongly nonmonotonic superconducting transition temperature $T_c$
[Fig.~\ref{Fig_HP_FeSe}(b)] \cite{sun2016dome}. The pressure-induced
magnetic phase overlaps with superconductivity over part of the phase
diagram and provides a natural framework for understanding the multiple
maxima of $T_c$ under pressure.

High-energy x-ray diffraction and time-domain M\"ossbauer
spectroscopy further resolved the relation between the structural and magnetic transitions [Fig.~\ref{Fig_HP_FeSe}(c)]. At low pressures, separate orthorhombic and magnetic transitions are observed; with increasing pressure, they approach each other and merge into a coupled first-order transition above approximately 1.7 GPa \cite{kothapalli2016strong,gati2019bulk}. These results demonstrate a strong coupling between nematic/orthorhombic order and stripe-type magnetism. Although the detailed phase boundaries vary quantitatively among studies because of differences in sample composition, pressure conditions, experimental probes, and transition criteria, the common picture is that pressure tunes FeSe through closely intertwined nematic, magnetic, and superconducting regimes. The observed evolution is consistent with a pressure-driven quantum transition connecting a nematic antiferroquadrupolar state to stripe antiferromagnetism in a spin-1 bilinear--biquadratic model \cite{hu2020quantum}.

\subsubsection{Heavily electron-doped FeSe-based superconductors}

The pressure dependence of HED FeSe-derived
superconductors exhibits several distinctive features. In
$A$FeSe-122, the superconducting transition temperature of the
low-pressure SC-I phase ($T_c\sim32$~K) decreases with increasing
pressure and superconductivity is suppressed near 9.2 GPa. At higher
pressures, a second superconducting phase (SC-II) emerges above
approximately 11.5 GPa, reaches a maximum $T_c$ of 48.7~K near
12.5 GPa, and disappears above 13.2 GPa
[Fig.~\ref{Fig_HP_FeSe}(d)] \cite{Sun2012}. Thus, SC-II occurs only
within a narrow pressure range, in contrast to the more usual smooth,
dome-like pressure dependence of $T_c$.

Related pressure-induced SC-I and SC-II phases have also been observed
in FeSe-11111 and LiNHFeSe-122, but at substantially lower critical
pressures [Figs.~\ref{Fig_HP_FeSe}(e) and
\ref{Fig_HP_FeSe}(f)] \cite{sun2018reemergence,shahi2018hightc}.
The SC-II phases in these materials reach $T_c$ values of approximately
52~K and 55~K, respectively, representing the highest reported
transition temperatures for bulk FeCh-based superconductors. Structural
and transport measurements indicate no obvious structural transition
across the onset of SC-II, whereas the electron-carrier concentration
in SC-II is substantially higher than that in SC-I
\cite{Sun2012,sun2018reemergence,shahi2018hightc}. These observations
suggest that the emergence of SC-II is more closely associated with an
electronic reconstruction, possibly a Lifshitz transition driven by
changes in the local FeSe-tetrahedron environment, than with a
crystallographic phase transition \cite{shahi2018hightc}.

Theoretical calculations for $A$FeCh-122 further suggest that the SC-I
and SC-II phases may involve different pairing symmetries associated
with their distinct Fermi-surface topologies \cite{Das2013}. Together,
these results show that the combination of electron doping and
hydrostatic pressure provides an effective route for accessing
higher-$T_c$ phases in FeCh-based superconductors.

\subsubsection{FeSe$_{1-x}$S$_{x}$ and FeSe$_{1-x}$Te$_x$}
\begin{figure}[htbp!]
\centering
\includegraphics[width=8.5cm]{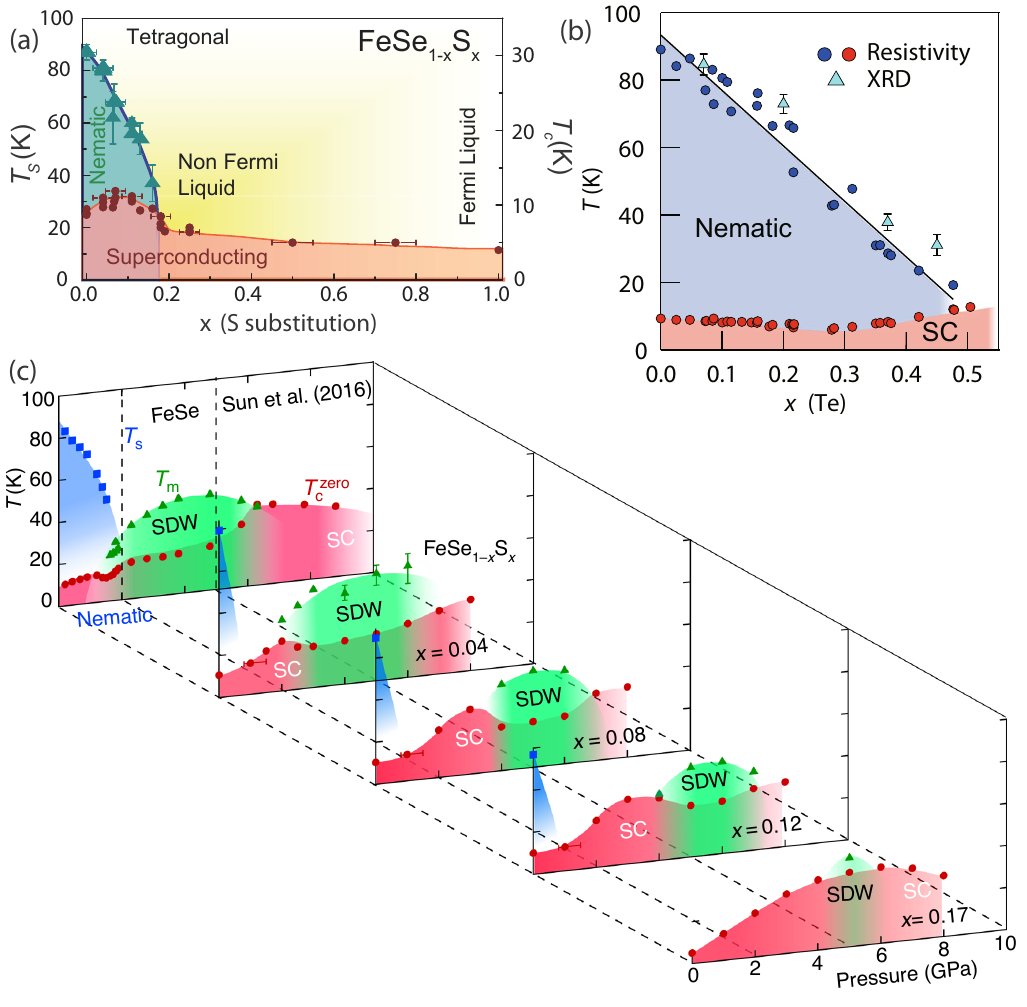}
\caption[]{
Electronic and pressure--composition phase diagrams of
FeSe$_{1-x}$S$_x$ and FeSe$_{1-x}$Te$_x$. (a) Ambient-pressure phase
diagram of FeSe$_{1-x}$S$_x$. From \cite{reiss2017suppression}. (b)
Ambient-pressure phase diagram of FeSe$_{1-x}$Te$_x$. From
\cite{mukasa2021highpressure}. (c) Evolution of the nematic,
spin-density-wave, and superconducting phases in
FeSe$_{1-x}$S$_x$ under hydrostatic pressure. From
\cite{matsuura2017maximizing,sun2016dome}.
\label{Fig_HP_FeSe_STe}}
\end{figure}

Sulfur- and tellurium-substituted FeSe provide chemically tuned
extensions of the pressure-tuned FeSe reference system. These materials
were investigated early in the development of the FeSC field
\cite{paglione2010hightemperature,stewart2011superconductivity}.
The subsequent availability of high-quality chemical-vapor-transport
single crystals enabled the establishment of well-defined phase diagrams
over broad composition ranges
[Figs.~\ref{Fig_HP_FeSe_STe}(a) and (b)]
\cite{reiss2017suppression,coldea2021electronic,mukasa2021highpressure,
shibauchi2020exotic,ishida2022pure}.

Isoelectronic substitution of S or Te for Se changes the
lattice parameters, anion height, and local Fe--Ch geometry without
directly changing the nominal carrier count. It therefore provides a
useful route for tuning the electronic structure and for comparison
with hydrostatic pressure. However, it is not a disorder-free tuning
parameter: random chalcogen-site substitution introduces local
structural and potential disorder. In FeSe$_{1-x}$Te$_x$, transmission
electron microscopy, x-ray scattering, and scanning tunneling
microscopy have further reported nanoscale Se-rich and Te-rich
compositional inhomogeneity
\cite{hu2011phase,louca2011suppression,he2011nanoscale}. The effects
of S/Te substitution should therefore not be identified solely with
chemical pressure.

Electronic nematicity denotes the spontaneous breaking of the
tetragonal $C_4$ rotational symmetry to $C_2$ without breaking
translational symmetry; its phenomenology and microscopic origins are
discussed in Sec.~\ref{sec_nematicity}. In FeSe$_{1-x}$S$_x$, nematic
order is strongly coupled to the tetragonal-to-orthorhombic lattice
distortion. At ambient pressure, FeSe$_{1-x}$S$_x$ does not develop
long-range magnetic order, thereby providing an important setting for
examining the relation between nematicity and superconductivity.
The suppression of the nematic transition with increasing $x$ drives
the system toward a putative nematic quantum critical point near
$x\approx0.17$, enabling studies of critical nematic
fluctuations and their consequences for quasiparticle dynamics and
pairing (Sec.~\ref{sec_strain_tc}) \cite{coldea2021electronic}.

By combining chemical and hydrostatic pressure,
FeSe$_{1-x}$S$_x$ disentangles nematic and magnetic instabilities
[Fig.~\ref{Fig_HP_FeSe_STe}(c)]
\cite{matsuura2017maximizing,reiss2020quenched}. S substitution
suppresses the nematic phase without stabilizing magnetic order,
whereas hydrostatic pressure induces a stripe-type AFM dome. The
resulting three-dimensional $(P,T,x)$ phase diagram reveals two distinct
superconducting regions, with $T_c$ reaching $\sim30$~K on either side
of the magnetic dome, thereby underscoring the
non-equivalence of chemical and physical pressure and supporting a role
for spin fluctuations in the enhanced pairing
\cite{matsuura2017maximizing}.

In FeSe$_{1-x}$Te$_x$, Te substitution expands the average
lattice parameters and extends nematic order to higher compositions
($x\approx0.5$) while avoiding long-range magnetic order at ambient
pressure [Fig.~\ref{Fig_HP_FeSe_STe}(b)]. This provides a useful
contrast to the S-substituted series, where magnetism emerges under
pressure, although the Te-substituted materials also involve substantial
local structural disorder and cannot be regarded as a simple
negative-chemical-pressure analogue of FeSe. Systematic high-pressure
studies reveal a single superconducting dome with $T_c$ up to
$\sim20$--$30$~K, whose maximum coincides with the suppression of
nematicity rather than proximity to magnetism
\cite{mukasa2021highpressure}. High-field measurements near the
nematic end point show enhanced pairing interactions and anomalous
scattering consistent with critical nematic fluctuations
\cite{mukasa2023enhanced}. Taken together, these results are
consistent with an important role for nematic quantum criticality in
the superconductivity of FeSe$_{1-x}$Te$_x$.

\section{Applications}

The preceding sections have systematically detailed the rich structural chemistry, diverse material families, and fascinating physical properties of FeSCs. Fundamental studies---ranging from the precise structural control in 1144-type compounds to the robust coexistence of superconductivity and local magnetism in Eu-based systems---have uncovered a wealth of exceptional intrinsic properties. Ultimately, the discovery and optimization of these novel materials inevitably drive the quest to harness their microscopic quantum properties---such as high-$T_c$, massive $H_{c2}$, and relatively low electromagnetic anisotropy---for macroscopic technological applications. Bridging the gap between fundamental materials discovery and real-world utility represents a natural and critical progression for the field.

In applications, superconductors in most cases are used in the form of polycrystalline aggregates. The major application of superconductors is for high magnetic field generation, which requires a high critical current density ($J_{c}$). Recently, strong demand for high-$T_{c}$ superconductors is emerging due to the rapidly increasing cost of helium. Advances in cryocooler technology make it practical to operate superconductors with $T_{c}>\sim 20$~K instead of traditional metal alloy superconductors.

Table~\ref{tab:comparison_hightc} compares the properties associated with the application of three representative high-$T_{c}$ superconductors, FeSCs, MgB$_{2}$, and cuprates (Y and Bi-based systems). FeSCs have several advantages over conventional metal alloy-based materials and high-$T_{c}$ cuprates: (1) Higher crystal symmetry (tetragonal) of FeSCs than cuprates (orthorhombic). This means that 3-axis alignment of each crystal is not needed, unlike cuprates, to achieve high $J_{c}$. The higher gap symmetry ($s_{\pm}$) of FeSCs works favorably to realize the small anisotropy. (2) Advantageous grain boundary nature with respect to the angle of misorientation and higher electrical conductivity even in the normal state. (3) Higher $\mu_{0}H_{c2}$ than that of MgB$_{2}$. These features are favorable for high-field applications and cost reduction for industrial applications. The following describes the current status of FeSCs towards application.

\begin{table*}
\begin{center}
\caption{Comparison among three representative high-$T_{c}$ superconductors.}
\label{tab:comparison_hightc}
\begin{tabular}{c|c|c|c} 
\hline
\textbf{ }                                                & FeSCs                                                                                    & MgB$_{2}$                                & Cuprates\\
\hline
Parent Materials                                          & \makecell{AFM-semimetal \\($T_{\rm N} \sim$ 150~K)}                                      & Pauli paramagnetic metal                 & \makecell{AFM-Mott insulator \\($T_{\rm N} \sim$ 400~K)}\\
\hline
Fermi Levels                                              & Fe 3$d$ 5-orbitals                                                                       & B 2$p$ 2-orbitals                        & Cu 3$d$ single orbital\\
\hline 
\makecell{Max $T_{c}$ (K) in bulk \\at ambient pressure}  & \makecell{$\sim$ 56 (FePn-1111), \\$\sim$ 39 (FePn-122)}                                 & $\sim$ 39                                & \makecell{$\sim$ 93 (YBCO), \\$\sim$ 110 (Bi2223)}\\
\hline
Impurity sensitivity                                      & Robust                                                                                   & Sensitive                                & Sensitive\\
\hline
SC gap symmetry                                           & $s_{\pm}$--wave                                                                          & $s$--wave                                & $d$--wave\\
\hline  
$\mu_{0}H_{c2}(0)$ (T)                                    & \makecell{100 -- 200 (FePn-1111), \\50 -- 100 (FePn-122), \\$\sim$ 50 (FeCh-11)}         & $\sim$ 40                                & $>$ 100\\
\hline
Irreversible Field (T)                                    & \makecell{$>$ 50 (4~K), \\$>$ 15 (20~K)}                                                 & \makecell{$>$ 25 (4~K), \\$>$ 10 (20~K)} & $>$ 10 (77~K, YBCO)\\
\hline
Anisotropy of $\mu_0H_{c2}(0)$, $\gamma_{H_{c2}}$                                     & \makecell{4 -- 5 (FePn-1111), \\1 -- 2 (FeCh-11, FePn-122)}                              & $\sim$ 2                                 & \makecell{5 -- 7 (YBCO), \\50 -- 90 (Bi-system)}\\
\hline
\makecell{Crystallographic symmetry \\in SC phase}        & Tetragonal                                                                               & Hexagonal                                & Orthorhombic\\
\hline
\makecell{Critical GB \\misorientation angle($^{\circ}$)} & 8 -- 9                                                                                   & No data                                  & 3 -- 7 (YBCO)\\
\hline
Advantage                                                 & \makecell{High $\mu_{0}H_{c2}(0)$, \\Easy fabrication}                                   & Easy fabrication                         & High $T_{c}$ and $\mu_{0}H_{c2}(0)$\\
\hline
Major disadvantage                                        & Toxicity                                                                                 & Low $\mu_{0}H_{c2}(0)$                   & \makecell{High cost due to \\3D alignment of crystallites}\\
\hline
\end{tabular}
\end{center}
\end{table*}

\subsection{Coated conductors}

\begin{figure}
\centering
\includegraphics[width=8.5cm]{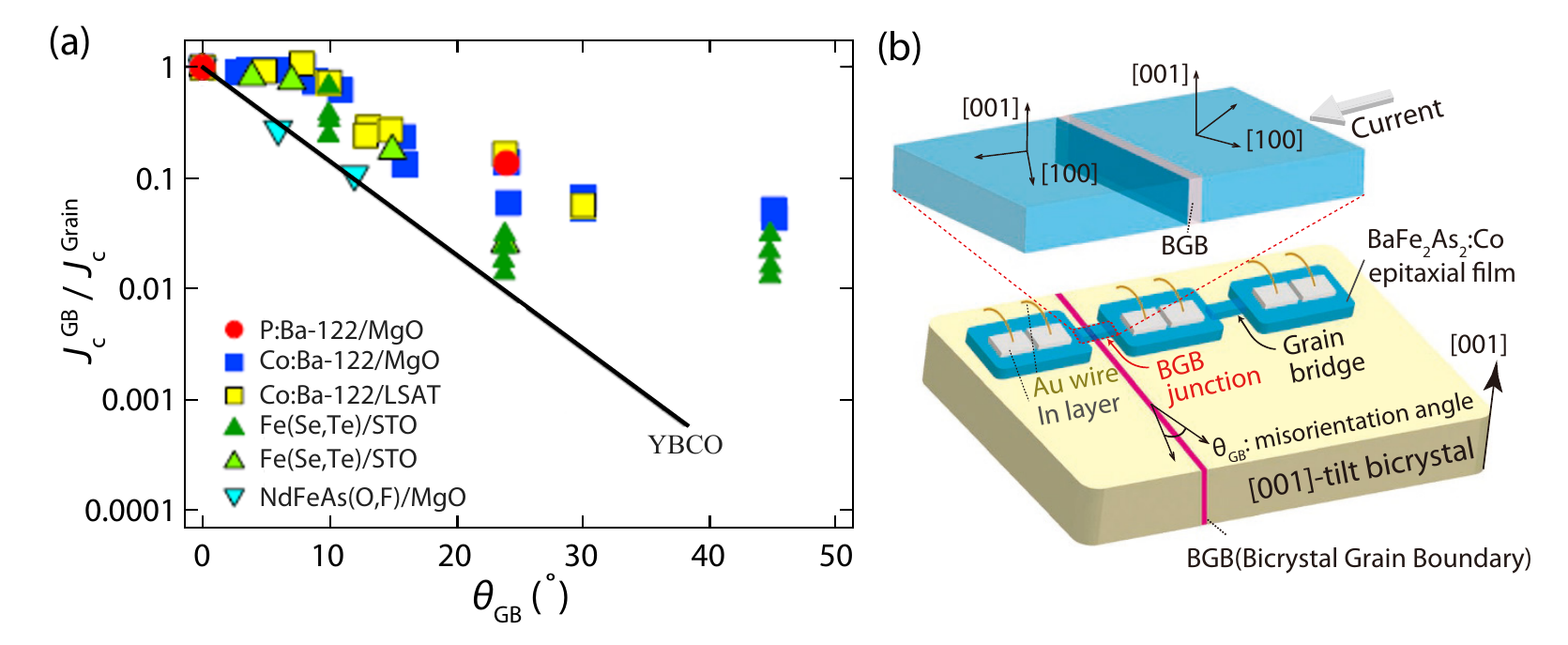}
\caption[]{(a) GB transparency $\epsilon$ of epitaxial thin films of FeSCs deposited on bicrystal substrates with different misorientation angle ($\theta_{\rm {GB}}$). (b) Schematic of the sample and $J_c$ measurement geometry \cite{Katase2011}.
\label{Fig_Jc_GB}}
\end{figure}

The advantageous grain boundary (GB) nature of FeSCs was experimentally revealed in 2011 utilizing the epitaxial thin films of BaFe$_{2}$As$_{2}$ deposited on the twinned bicrystal substrates with different misorientation angle ($\theta_{\rm {GB}}$) \cite{Katase2011}. The critical current density through bicrystal GB ($J_{c}^{\rm {BGB}}$) remained high ($>$ 1 MA cm$^{-2}$) and nearly constant up to a critical angle $\theta_{c}$ of $\sim$ 9$^{\circ}$, which is substantially larger than the $\theta_{c}$ of $\sim$ 5$^{\circ}$ for YBa$_{2}$Cu$_{3}$O$_{7-\delta}$ (YBCO). Even at $\theta_{\rm {GB}}>\theta_{c}$ , the decay of $J_{c}^{\rm {BGB}}$ was much slower than that of YBCO. Almost the same $\theta_{c}$ ($\sim$ 9$^{\circ}$) was found for other FeSCs (FeAs-1111 and FeCh-11) by the same approach. Figure \ref{Fig_Jc_GB} summarizes the ratio of inter- to intra-grain $J_{c}$ ($J_{c}^{\rm {inter}}$ and $J_{c}^{\rm {intra}}$), GB transparency ($\epsilon=J_{c}^{\rm {inter}}/J_{c}^{\rm {intra}}$), for the [001]-tilt GB as a function of $\theta_{\rm {GB}}$ for various FeSC thin films \cite{Iida2025}.

\begin{figure}
\centering
\includegraphics[width=8.5cm]{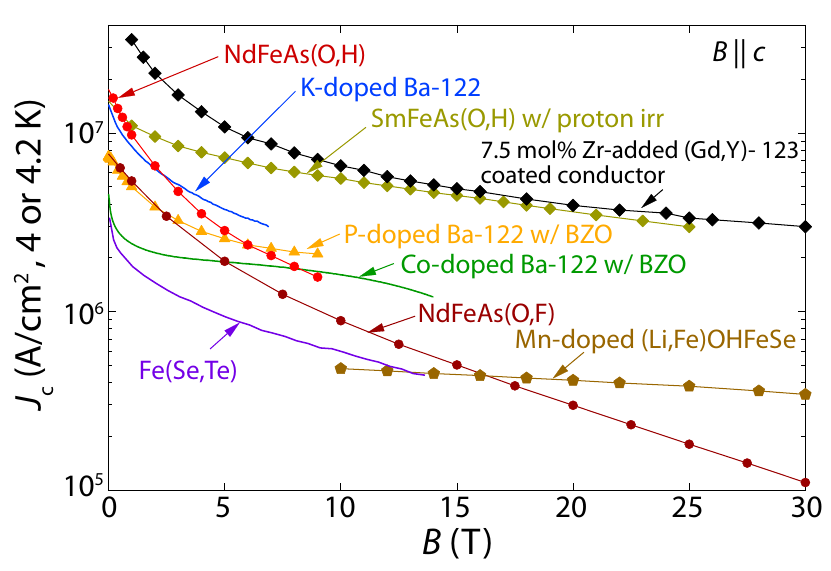}
\caption[]{Field dependence of $J_c$ for FeSC coated conductors. For comparison, YBCO coated conductor is included. Adapted from \cite{Iida2025}.
\label{Fig_CC_Jc_B}}
\end{figure}

Figure \ref{Fig_CC_Jc_B} summarizes the $J_{c}$ of various high-$T_{c}$ superconductors including coated conductors and wires. The highest self-field $J_{c}$ exceeds 17 MA cm$^{-2}$, achieved with Nd(O,H)FeAs \cite{Iida2025}. Although the $T_{c}$s of both H-doped and F-doped NdOFeAs are around 45~K, the $J_{c}$ of the H-doped thin film is higher by an order of magnitude than that of the F-doped sample \cite{Kondo2020}. This enhancement is attributed to the improvement in the depairing current density ($J_{d}$), achieved through HED, as the self-field $J_{c}$ is almost proportional to $J_{d}$, which in turn is proportional to the condensation energy. This thermodynamic approach, combined with the introduction of artificial pinning centers by low-energy proton irradiation, significantly improves $J_{c}$ -- $H$ behavior, as evidenced in Sm(O,H)FeAs \cite{Miura2024}. It is worth noting that the $J_{c}$ of proton-irradiated Sm(O,H)FeAs is comparable to that of YBCO coated conductors at high fields. Although H$^{-}$ and F$^{-}$ have a similar ion size and the same valence, and both work as a dopant for aliovalent substitution of O$^{2-}$ to generate electron carriers in the FeAs-1111 system, magnetic interaction between the FeAs layers is distinctly stronger in the H-doped system than that in the F-system \cite{Muraba2014}. These features originate from the covalent interaction between H-1$s$ and As-4$p$, which is negligibly small for F with deeper F-2$p$.

\subsection{Superconducting wires and tapes}

\begin{figure}
\centering
\includegraphics[width=8.5cm]{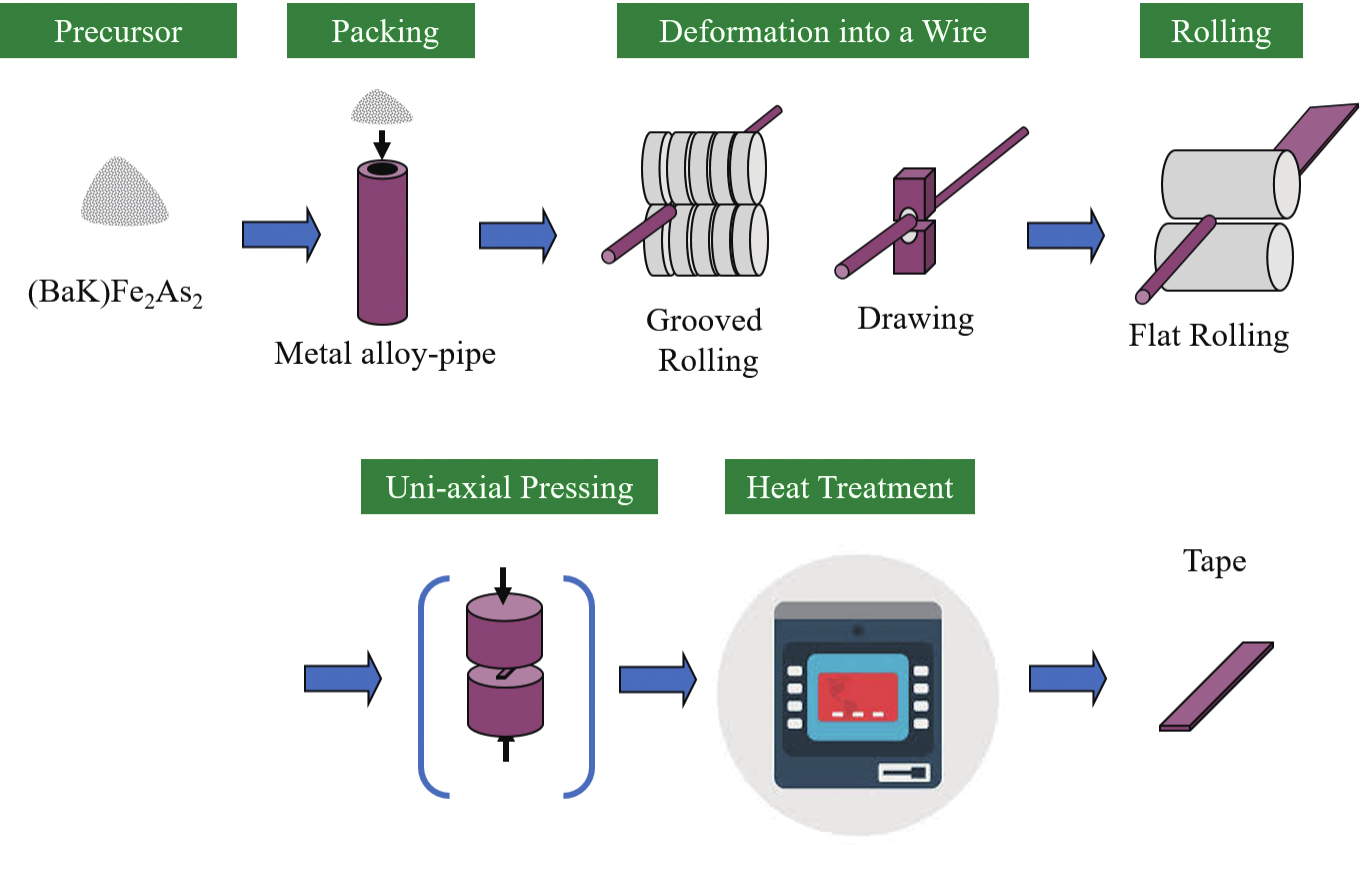}
\caption[]{Powder-in-tube methods for FeSC tape fabrication.
\label{Fig_PIT}}
\end{figure}

The powder-in-tube (PIT) method has been widely used for the fabrication of Bi2223 and MgB$_{2}$ wires with kilometer length. The PIT process schematically shown in Fig.~\ref{Fig_PIT} takes advantage of the low costs and relatively simple deformation techniques. There are two PIT methods: an \textit{in-situ} method, in which a powder mixture of raw precursor is used as a starting material, and an \textit{ex-situ} method, in which powder of the reacted superconducting phase is used. In the early years of FeSC wire development, the \textit{in-situ} PIT method was used; however, the transport $J_{c}$ is very low because of a large amount of impurities, cracks, and voids within the superconducting core. To overcome these issues, Qi \textit{et al.} fabricated the first \textit{ex-situ} PIT (Sr, K)Fe$_{2}$As$_{2}$ wires by filling Fe/Ag tubes with Sr$_{0.6}$K$_{0.4}$Fe$_{2}$As$_{2}$ powder and deforming them into fine wires \cite{Qi2010}.
Since then, the \textit{ex-situ} PIT approach has become the most successful way for fabricating high-performance FeSC wires. Nowadays, remarkably high $J_{c}$ values of 10$^{4}$ -- 10$^{5}$ A cm$^{-2}$ at 4~K have been obtained in \textit{ex-situ} PIT FeSC wires and tapes.

The improvement of $J_{c}$ for FeSC wires and tapes fabricated using the PIT method has been a focus of research and development over the past years, with the value increasing rapidly over time \cite{Pyon2021,Pyon2023,Weiss2012,Dong2024,Gao2015, Hosono2018}. To date, the $J_{c}$ of FeAs-122-based superconducting tapes has exceeded 10$^{5}$ A cm$^{-2}$ at 4.2~K under 14 T through sustained efforts by several groups (Institute of Electrical Engineering, Chinese Academy of Sciences (IEE-CAS, China), National Institute for Materials Science (Japan), the University of Tokyo (Japan), Florida State University (USA))
\cite{Hosono2018}, demonstrating their excellent potential for high-field applications.

\begin{figure}
\centering
\includegraphics[width=8cm]{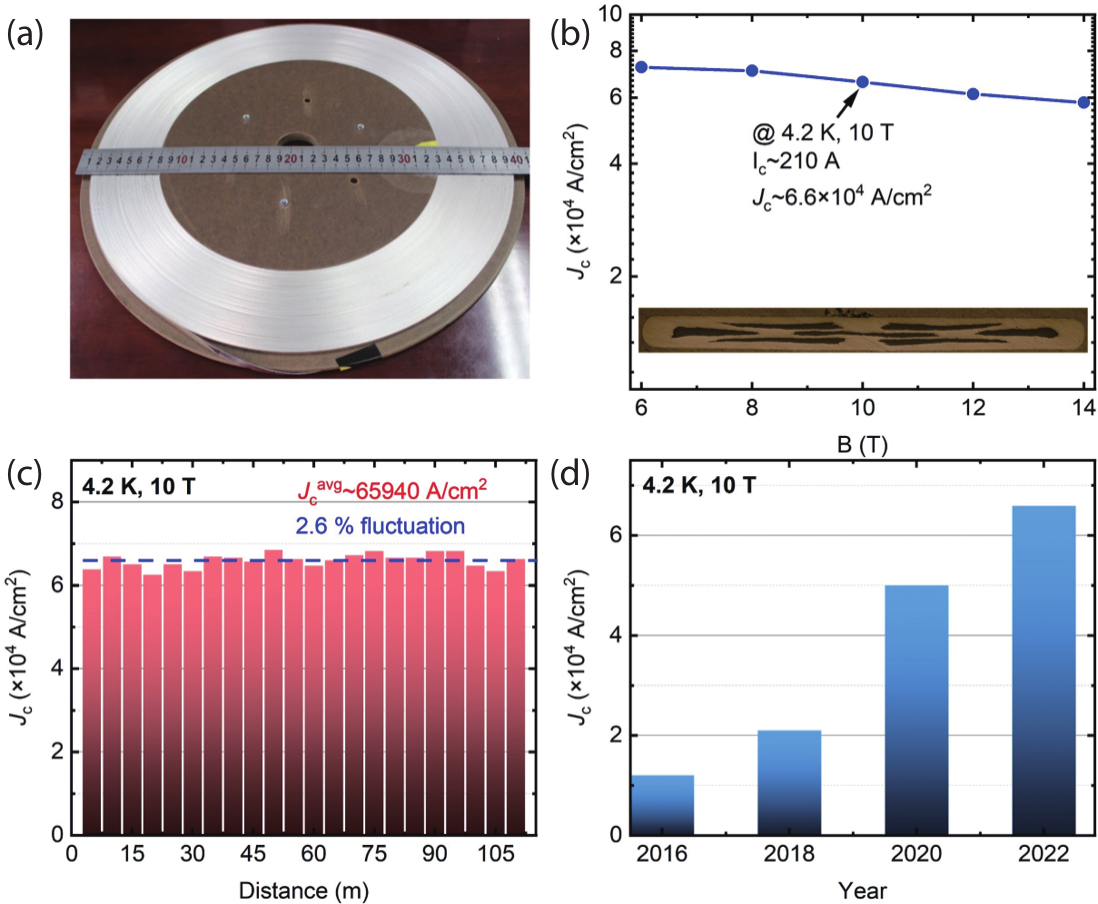}
\caption[]{Progress in 100-m-long (Ba, K)Fe$_{2}$As$_{2}$ tape. (a) Photo of tape. (b) Field dependence of $J_{c}$. The inset shows the transverse cross-section of the tape. (c) The homogeneity of $J_{c}$ for the long tape. (d) The progress of	$J_{c}$ at 4.2~K and 10 T of the 100-m-long tapes since 2016. From \cite{Dong2024}.
\label{Fig_Long_tape}}
\end{figure}

In 2016, Ma's group of IEE-CAS has successfully fabricated a 100-m-long, 7-core (Sr, K)Fe$_{2}$As$_{2}$ superconducting tape with a uniform $J_{c}$ along its entire length, reaching a value of 1.3$\times$10$^{4}$ A cm$^{-2}$ at 4.2~K under 10 T \cite{Zhang2017c}. Since then, the performance has been enhanced through improvement of the precursors. Figure \ref{Fig_Long_tape} shows the tape fabricated using (Ba, K)Fe$_{2}$As$_{2}$ powders and AgSb/Ag tube. The $J_{c}$ of the 100-m-long tape reaches 6.6$\times$10$^{4}$ A cm$^{-2}$ ($I_{c}$ = 210 A) at 4.2~K and 10 T, and the $J_{c}$ fluctuation is $\sim$ 2.6\% \cite{Dong2024}. This performance is a milestone in the fabrication of FeSC tapes. Subsequently, they manufactured double pancake coils (DPCs) using these tapes, resulting in a total of nine DPCs, from which seven coils with superior critical current $I_{c}$ characteristics are selected. These coils are subsequently assembled into high-field insert coils, with each coil connected using YBCO coated conductor joined by a low melting alloy, and achieved a remarkable magnetic field of 21 T at 4.2~K \cite{Ding2023}. This is the first realization of a Tesla-class coil using FeSCs in the world.

A distinct progress is reported for the FeCh-11 system in 2023 by a group at Shanghai Jiaotong University \cite{Liu2023}. They deposited a 1-meter-long, 320-nm-thick Fe(Se,Te) film with $T_c=17.5$~K on CeO$_2$-buffered metal tapes using a reel-to-reel pulsed laser deposition operating at a frequency of 20 Hz to 60 Hz. The end-to-end $I_{c}$ is 108 A, corresponding $J_{c}$ of 1.54 MA cm$^{-2}$, which is above the threshold for industrial applications. The FeCh-11 system with the simplest composition has an advantage in synthesis. If the $T_{c}$ is further increased as reported for the very thin films, the tape application would be competitive to the FeAs-122 tapes.

\subsection{Bulk superconducting magnet}

Superconducting bulk magnets have great potential to largely improve the performance of motors, generators, and analytical instruments by replacing conventional permanent magnets. The magnetization of superconductors with a high $J_{c}$ results in a strong bulk magnet at temperature below $T_{c}$ with refrigeration. The superconducting bulk magnet can generate a higher magnetic field with increasing $J_{c}$ and volume. Therefore, a superconducting bulk magnet is promising for applications such as table-top, liquid He-free NMR. For this application, FeSCs are superior to conventional Nb-based superconductors because they operate at higher temperatures and in higher magnetic fields. FeSCs exhibit minimal electromagnetic anisotropy and have an exceptionally elevated
$\mu_{0}H_{c2}$ exceeding 50 T, which is more than double that of Nb-based superconductors. In comparison with high-$T_{c}$ cuprate magnets, FeSC magnets can be synthesized by sintering polycrystalline materials that can be fabricated through a more scalable process than the cuprate magnets, i.e., the latter is grown from the melt as the nucleation of the superconducting crystal from the melt. Consequently, FeSCs can significantly enhance the performance of particle accelerators, medical magnetic resonance imaging (MRI) scanners, MAGLEV trains, and other devices that rely on conventional Nb-based superconducting magnets.

\begin{figure}
\centering
\includegraphics[width=8.5cm]{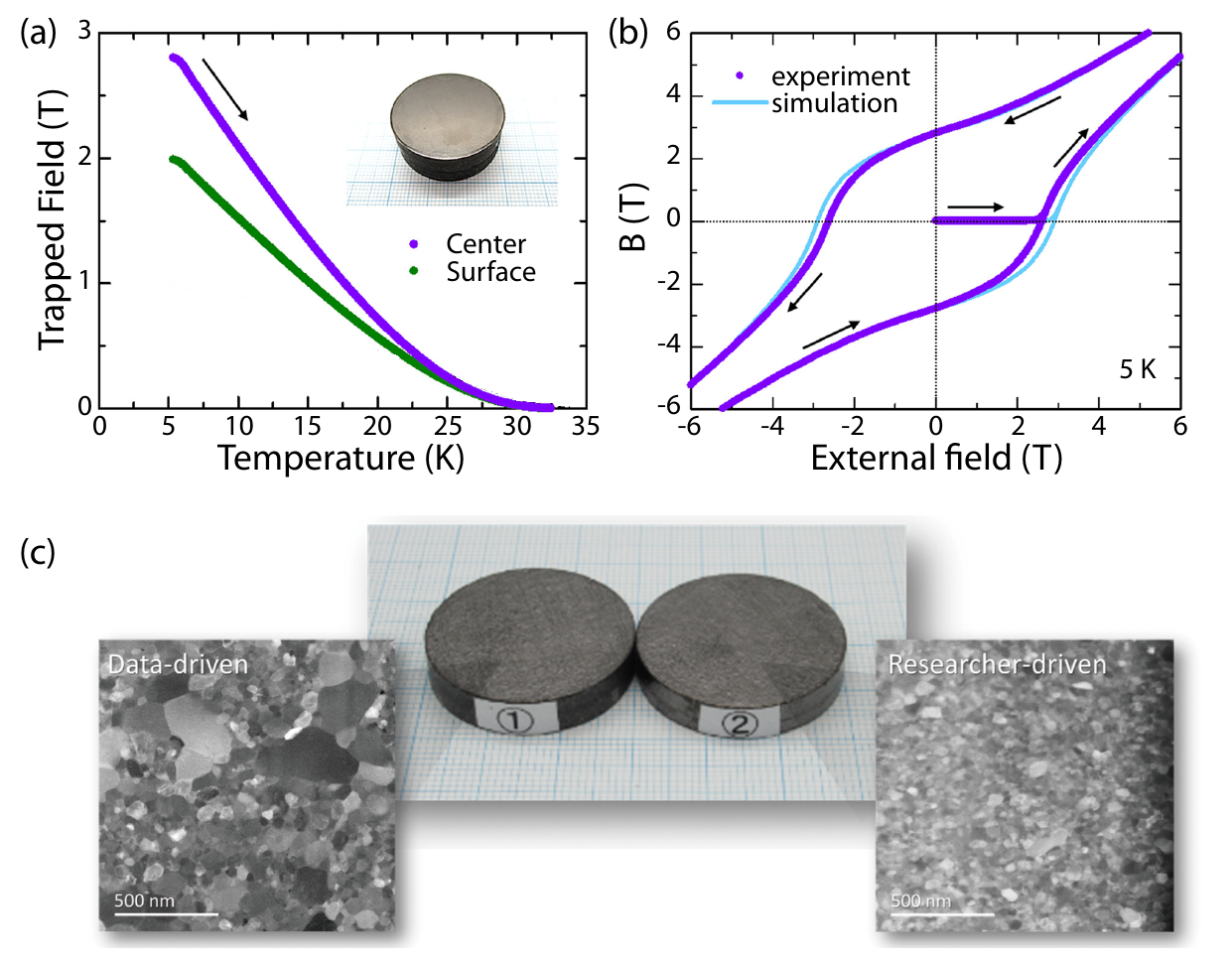}
\caption[]{Permanent bulk magnet of (Ba, K)Fe$_{2}$As$_{2}$. (a) Variations in the trapped magnetic field with temperature, as measured experimentally at both the center and the surface of the bulk pair. (b) Experimental and simulated magnetic hysteresis loops at a temperature of 5~K obtained through measurements at the center of the bulk pair. From \cite{Yamamoto2024}. (c) Photo and micro-structures of the (Ba, K)Fe$_{2}$As$_{2}$ bulk samples. The process design for the Bulk1 (left) and Bulk2 (right) are data- and researcher-driven, respectively.\label{Fig_BKFA_bulk}}
\end{figure}

In recent years, machine learning, which is a subset of artificial intelligence (AI), has been integrated with materials research to solve a variety of problems, including the modeling of $T_{c}$ of superconductors via database analysis \cite{Konno2021} and the optimization of material properties \cite{Mueller2016}. Nonetheless, the effective application to materials synthesis is somewhat restricted compared with its application in the construction of comprehensive materials databases. Synthesis processes involve an extensive array of parameters, ranging from factors related to equipment configurations to empirical variables manipulated by researchers. Yamamoto \textit{et al.} applied machine learning to manipulate polycrystalline microstructures of Ba$_{0.6}$K$_{0.4}$Fe$_{2}$As$_{2}$ through a process design that integrates researcher- and data-driven approaches using tailored software \cite{Yamamoto2024,Yamamoto2025}. The resulting bulk permanent magnet exhibits a trapped field 2.7 times larger than the previously reported value and a field stability better than 0.1 ppm h$^{-1}$ as shown in Figs.~\ref{Fig_BKFA_bulk}(a) and \ref{Fig_BKFA_bulk}(b). These performances are beyond a practical 1.5 T permanent magnet for a medical MRI. Micro-structural analysis revealed that the researcher-driven sample contained fine grains of approximately
20--30 nm, whereas the data-driven sample exhibited a bimodal distribution of grain sizes [Fig.~\ref{Fig_BKFA_bulk}(c)]. Such contrasting outcomes from data- and researcher-driven processes show that high-density defects and bimodal GB spacing distributions are primary contributors to the excellent strength and stability of bulk magnet.

\subsection{Perspective}

FeSCs are capable of operating effectively under magnetic fields below 20 T at 4.2~K, or below 10 T at temperatures ranging from 20~K to 30 K. They have the potential to be competitive with Nb$_{3}$Sn in terms of both cost and performance. However, many challenges remain, reflecting the comparatively short
history of this research. These include (1) mass production of
high-quality precursor powders; (2) texture development and current-
transport models; (3) flux pinning; (4) large-scale production of
high-strength composite tapes; and (5) technologies for joints, coils, and cables \cite{Dong2024}. Once these challenges are addressed, high-performance and low-cost FeSC wires and tapes could be widely used in fusion reactors, next-generation accelerators, and advanced MRI systems. 
Very recently, an asymmetric stress field strategy was proposed using extrusion to directly nucleate a high density of dislocations. These dislocations serve as strong pinning centers that lead to a fivefold enhancement in the current-carrying capacity of FeSC (Ba$_{0.6}$K$_{0.5}$Fe$_{2}$As$_{2}$, nominal composition) at 33 T, along with low anisotropy and a large irreversibility field \cite{Han2025}.

\section{Intertwined orders and exotic phases}

While fabricating high-performance FeSC wires, tapes, and magnets marks a practical milestone, further enhancing $T_c$ and optimizing critical currents relies on unraveling their microscopic mechanisms. These macroscopic properties stem from a complex interplay among lattice, spin, charge, and orbital degrees of freedom. This section explores these intertwined orders and exotic phases, which are essential for deciphering the pairing mechanism and guiding the design of next-generation superconductors. 

Before proceeding, we clarify the momentum-space nomenclature. For scattering techniques, the momentum transfer $\mathbf{Q}$, with Cartesian components ($q_x$, $q_y$, $q_z$), is expressed as $(H, K, L) = (q_xa/2\pi, q_yb/2\pi, q_zc/2\pi)$ in reciprocal lattice units (r.l.u.) of the orthorhombic (4-Fe) unit cell. Conversely, spectroscopies like ARPES typically adopt the unfolded 1-Fe or folded 2-Fe (tetragonal) BZs, expressing wave vectors as $(\pi, 0)$ and $(\pi, \pi)$ within the 1-Fe BZ. Geometrically, ${\bf Q} = (1, 0)$ and  ${\bf Q} = (1, 1)$ in orthorhombic r.l.u. map directly to the $(\pi, 0)$ and $(\pi, \pi)$ points in the unfolded 1-Fe BZ, respectively. Throughout this section, we use these notations where appropriate to align with the specific experimental frameworks under discussion.

\subsection{Electronic nematicity and strain tuning}\label{sec_nematicity}

Electronic nematicity, a correlated state spontaneously breaking the discrete rotational lattice symmetry (such as tetragonal $C_4$ to orthorhombic $C_2$) without altering translational symmetry, is a prominent intertwined phase in FeSCs \cite{fernandes2014what,bohmer2022nematicity,si2016high}. Experimentally, it manifests as pronounced in-plane electronic anisotropies far exceeding the subtle lattice distortion. This symmetry breaking has been extensively visualized across multiple probes, including anisotropic transport \cite{chu2010,chu2012,kuo2016ubiquitous}, orbital-dependent band shifts in ARPES \cite{yi2011symmetry,pfau2019momentum,yi2019nematic}, twofold magnetic excitations in INS and resonant inelastic x-ray scattering (RIXS) \cite{lu2014nematic,lu2018spin,chen2019anisotropic,tam2020orbital,lu2022spinexcitation,liu2024nematic,liu2025spin}, and $C_2$-symmetric electronic states observed by STM, together with
nematic fluctuations and anisotropic electronic responses detected by NMR and Raman scattering
\cite{gallais2016nematic,thorsmolle2016critical,kissikov2016nmr,li2017stripes}.

In the tetragonal phase, the $d_{xz}$ and $d_{yz}$ orbitals are related by $C_4$ symmetry and are therefore nearly degenerate in the absence of spin--orbit coupling and other symmetry-lowering effects. Upon entering the nematic phase, the breaking of $C_4$ rotational symmetry lifts or reconstructs this near degeneracy, producing momentum- and orbital-dependent shifts of the $d_{xz}$- and
$d_{yz}$-derived bands \cite{fernandes2014what,bohmer2022nematicity,yi2011symmetry,
pfau2019momentum,yi2019nematic}. This provides one of the most direct
spectroscopic signatures of electronic nematicity.

Despite this well-established phenomenology, its microscopic origin remains a subject of intense debate. The core controversy centers on whether the nematicity is primarily driven by spin fluctuations (acting as a vestigial Ising-nematic phase of the stripe magnetic order) \cite{fernandes2012preemptive, fernandes2014what,si2016high} or by spontaneous orbital polarization (a ferro-orbital instability). Because spin, orbital, and lattice degrees of freedom couple bilinearly, ordering in one channel inevitably induces the others, making it highly challenging to unambiguously isolate the primary driving force \cite{chubukov2016magnetism,yamakawa2016nematicity}.

Recent theoretical efforts have attempted to reconcile these contrasting views by treating both spin and orbital channels on an equal footing. Advanced theoretical models reveal a profound microscopic interplay: strong spin fluctuations could mediate an effective attraction in the orbital channel, potentially driving spontaneous orbital order even if the bare orbital interaction is repulsive \cite{chubukov2016magnetism, yamakawa2016nematicity}. Conversely, orbital fluctuations could strongly feed back into the spin channel. In this unified framework, the dominant character of the nematic phase is not universally fixed but is highly sensitive to the delicate details of the underlying electronic structure, such as Fermi surface topology, band depth, and magnetic frustration. Consequently, nematicity can manifest across a broad spectrum---from acting primarily as a precursor to static magnetic order in some families, to emerging alongside intense, dynamically frustrated spin fluctuations without long-range magnetism in others. This perspective conceptually captures the diverse nematic phenomena observed across different FeSCs, though the precise balance of microscopic driving forces continues to be actively investigated.

To experimentally disentangle these intricate couplings, uniaxial strain serves as a crucial tuning knob. Because the lattice and electronic sectors are linearly coupled, strain acts as a conjugate field to the nematic order parameter. In the following sections, we will first introduce the essential aspects of strain tuning, and discuss how elastoresistivity quantitatively probes nematic susceptibility and how \textit{in-situ} strain manipulates superconductivity and intertwined orders. Subsequently, we will also highlight the diverse manifestations of electronic nematicity across various physical phenomena and length scales.

\subsubsection{Introduction to strain tuning}

\begin{figure}
\centering
\includegraphics[width=8cm]{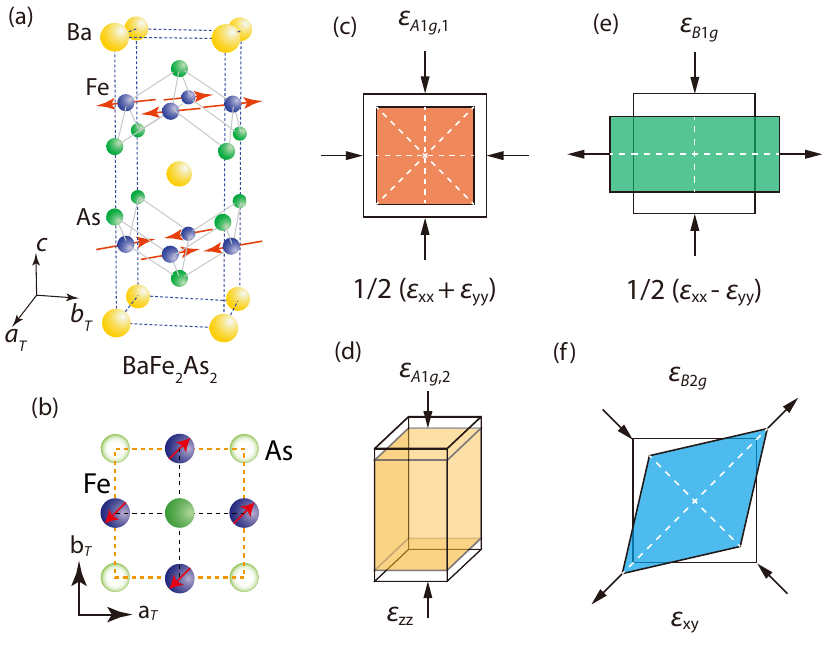}
\caption[]{Schematic depiction of the irreducible strains of the $D_{4h}$ point group. (a), (b) Tetragonal crystal structure (at $T>T_s$) and stripe-type AFM order for {\BFA}. The AFM order sets in below $T_{N}(<T_{s})$. (c)--(f) Irreducible strains in the $A_{1g,1}$, $A_{1g,2}$, $B_{1g}$, and $B_{2g}$ symmetry channels. The $A_{1g}$ strains preserve the $C_4$ rotational symmetry, while the $B_{1g}$ and $B_{2g}$ strains lower the primary symmetry to $C_2$.
\label{Fig_strain_D4h}}
\end{figure}

While hydrostatic pressure and epitaxial strain (from lattice-mismatched substrates) are widely used to tune quantum materials \cite{sun2016dome,matsuura2017maximizing,kothapalli2016strong,mukasa2021highpressure,sun2023signatures}, they predominantly generate symmetric ($A_{1g}$) deformations that preserve the crystal's fourfold ($D_{4h}$) lattice symmetry [Figs.~\ref{Fig_strain_D4h}(c) and (d)]. Because FeSCs retain a tetragonal structure above the structural transition temperature $T_s$, such symmetry-preserving strains do not couple linearly to the electronic nematic order parameter, rendering them incapable of directly driving or probing the nematic order. By contrast, applying in-plane uniaxial stress produces anisotropic strain components---specifically $B_{1g}=\frac{1}{2}(\varepsilon_{xx}-\varepsilon_{yy})$ or shear $B_{2g}=\varepsilon_{xy}$ [Figs.~\ref{Fig_strain_D4h}(e) and (f)]---that lower the lattice symmetry to $C_2$. These symmetry-breaking strains couple bilinearly to the nematicity \cite{fernandes2014what,bohmer2022nematicity,si2016high}.

Over the past decade, uniaxial strain techniques have evolved into an indispensable tool for exploring emergent phases in FeSCs and other correlated systems (e.g., Sr$_2$RuO$_4$) \cite{hicks2014strong,pustogow2019constraints}. Initially utilized as a static field to detwin crystals and expose intrinsic in-plane electronic anisotropies \cite{chu2010,yi2011symmetry,lu2014nematic}, precisely controlled uniaxial strain is now applied as a continuously tunable thermodynamic parameter. It enables the direct quantification of nematic susceptibility via elastoresistivity \cite{chu2012,kuo2016ubiquitous} and the systematic manipulation of the $T_s$, $T_N$, and $T_c$ phase boundaries \cite{worasaran2021nematic,bartlett2021relationship}.

In tetragonal FeSCs, uniaxial stress generally produces a combination of symmetry-allowed $A_{1g}$ and nematic $B_{1g}/B_{2g}$ strain components, whose partitioning is determined by the elastic tensor \cite{chu2012,kuo2016ubiquitous,li2024uniaxial}. In particular, stress along the tetragonal $[110]$ direction contains a symmetry-breaking shear component that couples strongly to the electronic nematic order parameter. Elastic-tensor analysis shows that the corresponding Young's modulus $E_{[110]}$ contains a dominant contribution from the shear modulus $C_{66}$, in addition to a comparatively nonsingular longitudinal background \cite{kuo2013measurement,li2024uniaxial}. Because $C_{66}$ softens markedly on approaching a nematic or structural transition, $E_{[110]}$ provides a useful thermodynamic proxy for the nematic susceptibility \cite{bohmer2014nematic,bohmer2016electronic, fujii2018anisotropic}. This shear softening has been characterized by ultrasonic sound-velocity measurements \cite{simayi2013strange}, high-resolution capacitive dilatometry \cite{bohmer2014nematic}, and INS measurements of in-plane transverse acoustic phonons. In BaFe$_2$As$_2$ and NaFeAs, the latter exhibit pronounced softening on cooling toward $T_s$, demonstrating a strong coupling between lattice
shear and electronic nematicity \cite{PhysRevB.91.134426,PhysRevX.8.021056}.

\subsubsection{Elastoresistivity as a probe of nematicity}\label{sec_strain_tc}

\begin{figure*}
\centering
\includegraphics[width=16cm]{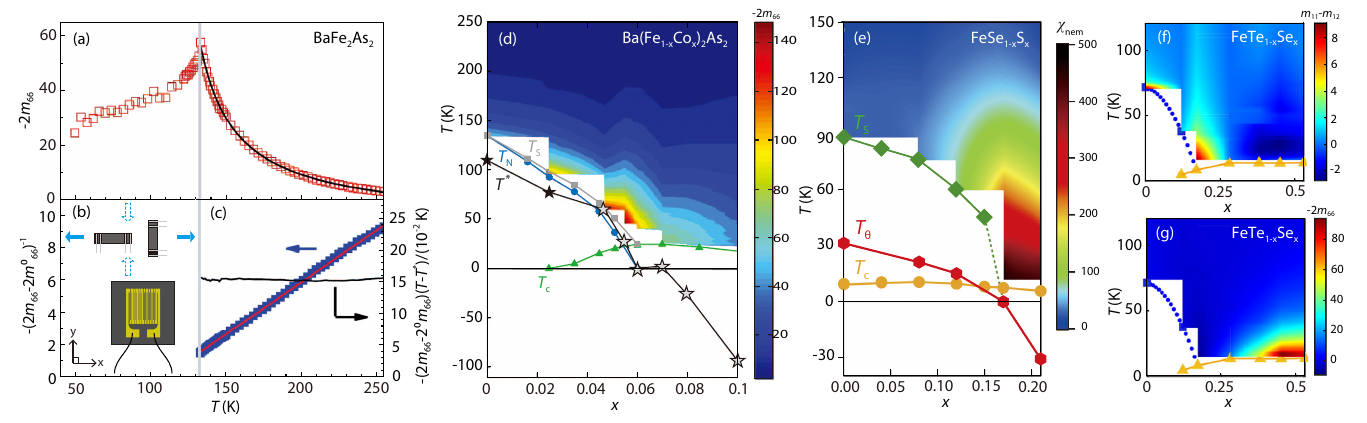}
\caption[]{Elastoresistivity coefficient as a measure of nematic susceptibility in FeSCs. (a) Temperature dependence of the $B_{2g}$ elastoresistance $-2m_{66}$, which is proportional to the nematic susceptibility $\chi_{N\left(B_{2 g}\right)}$ of {\BFA}. The black curve shows the CW fit, $2 m_{66}=\frac{\lambda}{a_0\left(T-T^*\right)}+2 m_{66}^0$. (b) Schematic diagram of long, thin crystals affixed on the side surface of a piezoelectric stack, with the strain characterized by a resistive strain gauge. (c) The quality of fit can be better appreciated by considering the inverse susceptibility $-\left(2 m_{66}-2 m_{66}^0\right)^{-1}$ (left axis) and the Curie constant $-\left(2 m_{66}-2 m_{66}^0\right)^{-1}\left(T-T^*\right)$ (right axis) \cite{kuo2016ubiquitous}. (d) Phase diagram of {\BFCA}, showing the variation of $-2m_{66}$ in the $x-T$ plane (color scale). Stars indicate the bare mean-field nematic critical temperatures ($T^*$) extracted from the CW fits of $2m_{66}$. (e) Phase diagram and quantum criticality in {\FSS}. The Weiss temperature ($T_{\theta}$, red hexagons) obtained by the CW analysis of the nematic susceptibility data. The magnitude of $\chi_{\rm nem}$ in the tetragonal phase is superimposed in the phase diagram by a color contour. (f), (g) Colormap of $m_{11}-m_{12}$ and of $2m_{66}$ as a function of temperature and doping. The double-spin stripe and the superconducting transition temperatures are denoted as blue squares and yellow triangles, respectively. Adapted from \cite{kuo2016ubiquitous,hosoi2016nematic,jiang2023nematic}.
\label{Fig_2m66}}
\end{figure*}

Electronic nematicity in FeSCs lowers the tetragonal $C_4$ symmetry to $C_2$ and can be described by an Ising-like order parameter $\psi$. In Landau theory, a symmetry-allowed bilinear coupling $-\lambda\psi\varepsilon_\Gamma$ between $\psi$ and the strain $\varepsilon_\Gamma$ in the irreducible representation $\Gamma$ ($B_{1g}$ or $B_{2g}$) makes strain a conjugate field to the nematic order. Consequently, the nematic susceptibility $\chi_{\rm nem} \equiv \partial\psi/\partial\varepsilon_\Gamma$ follows a CW temperature dependence, $\chi_{\rm nem} \propto \lambda/[a_0(T-T^*)]$ [Fig.~\ref{Fig_2m66}(a)], governed by electronic degrees of freedom \cite{chu2012}. Chu \textit{et al.} \cite{chu2010,chu2012} and Kuo \textit{et al.} \cite{kuo2013measurement,kuo2016ubiquitous} pioneered the use of elastoresistivity to directly measure this conjugate-field response. By bonding thin single crystals to a piezoelectric stack [Fig.~\ref{Fig_2m66}(b)] and tracking the fractional resistivity change $(\Delta\rho/\rho)_i = \sum_{k=1}^6 m_{ik} \varepsilon_k$ induced by \textit{in-situ} strains, one can extract the elastoresistivity tensor $m_{ik}$. Because the resistivity anisotropy $N \equiv (\rho_{xx}-\rho_{yy})/(\rho_{xx}+\rho_{yy}) \propto \psi$, the symmetry-resolved nematic susceptibilities are directly proportional to specific elastoresistivity coefficients: $\chi_{B_{1g}} \propto (m_{11}-m_{12})$ and $\chi_{B_{2g}} \propto 2m_{66}$ \cite{kuo2013measurement,kuo2016ubiquitous}. This approach avoids artifacts from static detwinning and provides a quantitative thermodynamic probe of nematic fluctuations.

Systematic elastoresistivity studies have revealed a ubiquitous CW-like divergence of the $B_{2g}$ nematic susceptibility upon cooling across a vast number of FeSC families, described by $|2m_{66}|=2m^{0}_{66}+\frac{\lambda}{a_0\,(T-T^*)}$ \cite{kuo2016ubiquitous}. A profound outcome is that the bare nematic transition temperature $T^*$ typically extrapolates to zero near optimal doping across diverse systems. These include {\BFCA} [Figs.~\ref{Fig_2m66}(a) and (d)] \cite{chu2012,kuo2016ubiquitous}, Ba(Fe$_{1-x}$Ni$_x$)$_2$As$_2$ \cite{liu2016nematic}, {\BFAP} \cite{kuo2016ubiquitous}, {\FSS} [Fig.~\ref{Fig_2m66}(e)] \cite{hosoi2016nematic,tanatar2016origin}, {\FST} [Figs.~\ref{Fig_2m66}(f) and (g)] \cite{jiang2023nematic,ishida2022pure}, and LaFe$_{1-x}$Co$_x$AsO \cite{hong2020evolution}. These results are consistent with a nematic quantum critical regime concealed beneath the superconducting dome in several FeSCs and provide macroscopic evidence that critical nematic fluctuations may contribute
to enhanced pairing \cite{kuo2016ubiquitous}. Furthermore, a cross-family analysis revealed an inverse linear scaling between the nematic Curie constant and the ordered magnetic moment, unifying the diverse parent compounds within a general phase diagram governed by nematic fluctuations \cite{gu2017unified}.

While the $B_{2g}$ nematicity is well established near optimal doping, its evolution in the heavily hole-overdoped regime has recently sparked intense debate. On one hand, measurements on heavily hole-doped {\BRFA} and $A$Fe$_2$As$_2$ ($A$ = Cs, Rb) reported the emergence of a novel electronic nematicity \cite{liu2019evidence,ishida2020novel}. Here, the nematic director reportedly rotates by $45^\circ$, manifesting as a $B_{1g}$ Ising-nematic state in the fully substituted limit \cite{liu2019evidence,ishida2020novel,mizukami2025thermodynamic}. However, this scenario was challenged by subsequent studies on overdoped {\BKFA} and $A$Fe$_2$As$_2$ ($A$ = K, Rb, Cs) \cite{wiecki2020dominant, wiecki2021emerging}. By decomposing the strain response into irreducible symmetry channels, these measurements argued that the elastoresistance in this regime is dominated by an in-plane symmetric ($A_{1g}$) response tied to a coherence-incoherence crossover of strong electronic correlations, finding no evidence for a divergent $B_{1g}$ nematic susceptibility \cite{wiecki2020dominant, wiecki2021emerging}. These contradictory observations indicate that the true symmetry and evolution of electronic nematicity in the extreme hole-overdoped limit remain highly controversial and require further clarification.

In contrast to the magnetically intertwined iron pnictides, the isovalently substituted {\FSS} series provides a pristine platform to isolate the role of pure nematic quantum criticality. In this system, elastoresistivity measurements pinpoint an NQCP at $x_c \approx 0.17$ that occurs in the absence of AFM order \cite{hosoi2016nematic,coldea2021electronic}. Furthermore, this NQCP is distinctly separated from the maximum $T_c$, which peaks at a lower substitution level of $x \approx 0.08$ [Fig.~\ref{Fig_2m66}(e)]. Even though the critical nematic fluctuations do not maximize $T_c$ in this system, they still exert a dramatic influence on the low-energy physics. In the normal state, high-field transport measurements reveal an anomalous non-Fermi liquid scaling localized exactly at $x_c$, underscoring the intense quasiparticle scattering driven by nematic fluctuations \cite{licciardello2019electrical}. Deep within the superconducting state, a suite of thermodynamic and high-resolution spectroscopic probes demonstrates that crossing this nematic endpoint triggers an abrupt and profound reconstruction of the superconducting gap structure \cite{hanaguri2018two,sato2018abrupt,mizukami2023unusual,matsuura2023two}. 
For comparison, in Ba$_{1 - x}$Sr$_x$Ni$_2$As$_2$ system without iron and magnetism, a NQCP is found to dramatically enhance superconductivity \cite{eckberg2020}. Together, these findings highlight that electronic nematicity can be tuned to a quantum critical regime entirely decoupled from a proximate magnetic QCP, exerting a distinct and powerful influence on both normal-state quasiparticle dynamics and superconducting pairing symmetry \cite{kreisel2020on}.

Finally, elastoresistivity has proved valuable beyond FeSCs. Quadrupole-strain susceptibility measurements in YbRu$_2$Ge$_2$ traced the interactions behind quadrupolar order \cite{rosenberg2019divergence}; quadratic elastoresistivity in ZrTe$_5$ revealed signatures of a strain-tuned topological transition \cite{mutch2019evidence}; and $D_{6h}$-adapted measurements have probed putative $E_{2g}$ nematicity in kagome superconductors (e.g., CsV$_3$Sb$_5$) \cite{Wilson2024AV3Sb5,1g9n-wm38}, yielding both positive and null results \cite{nie2022charge,asaba2024evidence,frachet2024colossal,liu2024absence}. These developments highlight elastoresistivity as a broadly applicable, symmetry-resolved probe of electronic order and fluctuations in correlated quantum materials.

\subsubsection{Strain tuning of critical temperatures}

\begin{figure}
\centering
\includegraphics[width=8.5cm]{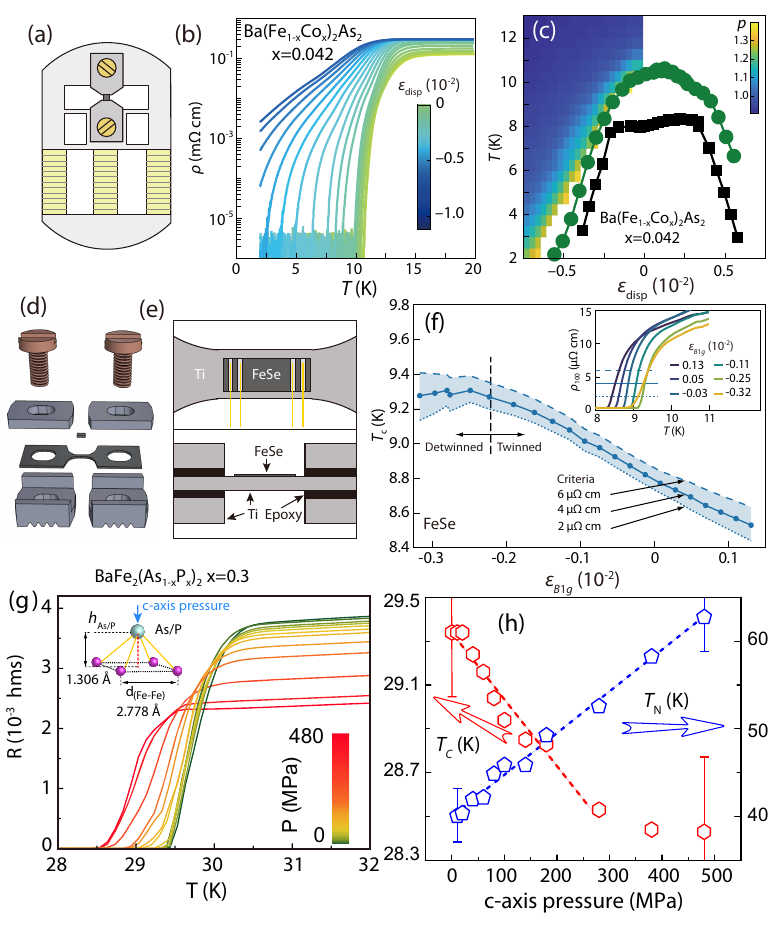}
\caption[]{Experimental configurations and uniaxial-strain tuning of superconductivity in FeSCs. 
(a) Schematic of a piezoelectric strain cell. Bar-shaped crystals (e.g., 122-type pnictides) are clamped across the gap between titanium plates. 
(b) Resistivity versus temperature under uniaxial stress for underdoped {\BFCA} ($x=0.042$). 
(c) Strain-temperature phase diagram for $x = 0.042$. Black squares and green circles denote the Meissner and resistive transitions, respectively. 
(d) Schematic of a strain-transmission platform for fragile crystals (e.g., FeSe), utilizing a titanium bow-tie bridge. 
(e) Top and side views of an FeSe crystal bonded to the titanium bridge. 
(f) $T_c$ of FeSe as a function of $B_{1g}$ strain, evaluated using varying resistivity threshold criteria (inset). 
(g) In-plane resistivity versus temperature under $c$-axis pressures up to 480~MPa for BaFe$_2$(As$_{0.70}$P$_{0.30}$)$_2$. The inset illustrates the FeAs tetrahedron under $c$-axis pressure at 300~K. $h_{\mathrm{As/P}}\approx1.304$~\AA{} and the in-plane $d_{\mathrm{(Fe--Fe)}}\approx 2.774$~\AA{} change to $1.306$~\AA{} and 2.778~\AA{}, respectively. (h) $c$-axis pressure dependence of $T_c$ and $T_N$ determined from transport.
Adapted from \cite{malinowski2020, park2020rigid, bartlett2021relationship,hu2018caxis}.
\label{Fig_strain_tc}}
\end{figure}

Early investigations into the intertwined phases of FeSCs relied heavily on static mechanical clamps and pressure devices \cite{chu2010, lu2014nematic, man2015electronic, tam2017uniaxial, man2018direct}, which primarily served to detwin macroscopic crystals below the structural transition. Subsequently, piezoelectrically driven strain cells substantially expanded the available tuning strategies \cite{hicks2014piezoelectric, liu2019nonlinear}, which enable \emph{in-situ}, continuously tunable, and bidirectional (tensile and compressive) strain control at cryogenic temperatures \cite{hicks2014strong, barber2019piezoelectric, ghosh2020piezoelectric}. Depending on the inherent mechanical properties of the target materials, two primary experimental configurations are typically employed. For mechanically robust bulk crystals, such as the 122-type iron pnictides, the sample is generally suspended across a gap and clamped directly between titanium plates [Fig.~\ref{Fig_strain_tc}(a)]. Conversely, to extend this powerful tuning capability to mechanically fragile or micaceous compounds like FeSe, advanced strain-transmission platforms utilizing flexible titanium bow-tie substrates [Figs.~\ref{Fig_strain_tc}(d) and (e)] were subsequently engineered to ensure uniform deformation without catastrophic exfoliation \cite{park2020rigid, bartlett2021relationship}. Equipped with these versatile mechanical architectures, uniaxial strain has evolved far beyond a mere detwinning tool. It now serves as a pristine, symmetry-selective thermodynamic tuning axis, allowing researchers to deterministically shift the critical temperatures ($T_s$, $T_N$, and $T_c$) and disentangle intertwined phases without introducing the chemical disorder inherently associated with elemental doping.

\paragraph*{Symmetry analysis}
In FeSCs, the primary phase transitions of interest are the structural (nematic) transition at $T_s$, the AFM transition at $T_N$, and the superconducting transition at $T_c$. To leading order in elastic strain, the shift of any of these critical temperatures, denoted generically as $T_{\rm crit}\in\{T_s,T_N,T_c\}$, obeys a symmetry-constrained expansion based on the irreducible representations of the $D_{4h}$ point group ($A_{1g}$, $B_{1g}$, and $B_{2g}$) \cite{ikeda2018symmetric,worasaran2021nematic}:
\begin{equation}
\Delta T_{\rm crit}(\varepsilon)=\sum_{i=1}^2 \lambda_{A_{1g,i}}\varepsilon_{A_{1g,i}}+\lambda_{B_{1g}}\varepsilon_{B_{1g}}^2+\lambda_{B_{2g}}\varepsilon_{B_{2g}}^2+\mathcal{O}(\varepsilon^3) .
\label{eq:general_symmetry_shift}
\end{equation}
Crucially, symmetry-preserving $A_{1g}$ strains shift $T_{\rm crit}$ linearly, whereas symmetry-breaking $B_{1g}$ and $B_{2g}$ strains enter strictly as even-power (predominantly quadratic) terms.

In experiments where uniaxial stress is applied along the tetragonal $[100]$ axis, transverse deformations emerge due to the Poisson effect ($\varepsilon_{yy}=-\nu_{xy}\varepsilon_{xx}$ and $\varepsilon_{zz}=-\nu_{xz}\varepsilon_{xx}$). This strain state cleanly decomposes into two symmetric channels, $\varepsilon_{A_{1g,1}}=(\varepsilon_{xx}+\varepsilon_{yy})/2$ and $\varepsilon_{A_{1g,2}}=\varepsilon_{zz}$, alongside the orthogonal symmetry-breaking channel $\varepsilon_{B_{1g}}=(\varepsilon_{xx}-\varepsilon_{yy})/2$. Conversely, the $B_{2g}$ pure shear channel ($\varepsilon_{xy}$) is most effectively isolated by cutting and straining the crystal parallel to the $[110]$ direction ($\varepsilon_{[110]}$). This configuration yields the decomposition $\varepsilon_{A_{1g,1}}=\tfrac{1-\nu'}{2}\varepsilon_{[110]}$, $\varepsilon_{A_{1g,2}}=\varepsilon_{zz}=-\nu''\varepsilon_{[110]}$, and $\varepsilon_{B_{2g}}=\tfrac{1+\nu'}{2}\varepsilon_{[110]}$.

For practical analysis, the net shift of $T_{\rm crit}$ is routinely fitted to a phenomenological parabola in the applied in-plane strain $\varepsilon$ (e.g., along $[100]$ or $[110]$):
\begin{equation}
T_{\rm crit}(\varepsilon)=T_0+\alpha\,\varepsilon+\beta\,\varepsilon^2,
\label{eqn-tc_strain}
\end{equation}
where the linear coefficient $\alpha$ captures the symmetry-preserving $A_{1g}$ response ($\alpha\propto \partial T_{\rm crit}/\partial \varepsilon_{A_{1g}}$), and the quadratic coefficient $\beta$ reflects the dominant symmetry-breaking response ($\beta\propto \partial^2 T_{\rm crit}/\partial \varepsilon_{B_{ig}}^2$, with $i=1, 2$) \cite{ikeda2018symmetric,worasaran2021nematic}. It is important to emphasize that $\alpha$ and $\beta$ are effective tuning parameters that inherently combine the bare thermodynamic couplings from Eq.~\eqref{eq:general_symmetry_shift} with material-specific, temperature-dependent Poisson ratios, and thus must be interpreted carefully.

\begin{figure}
\centering
\includegraphics[width=8.5cm]{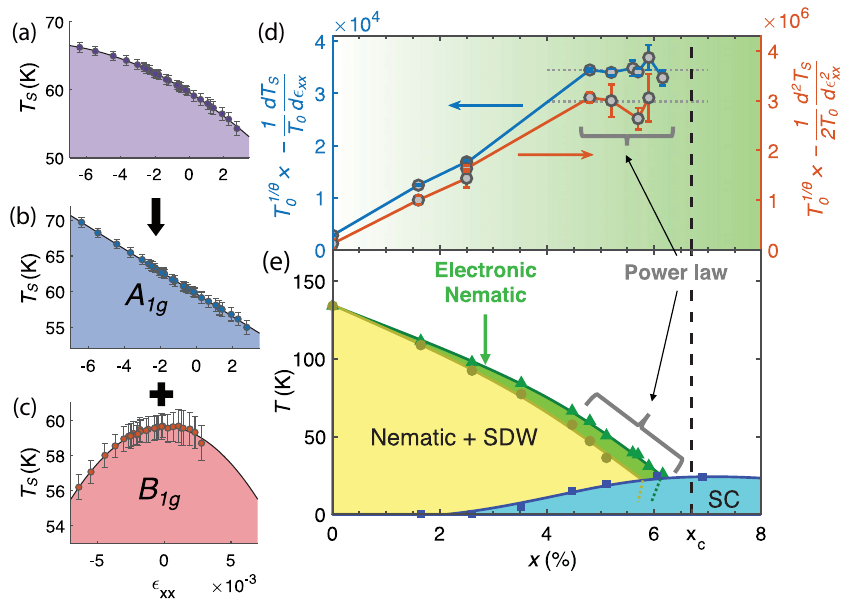}
\caption[]{Universal scaling and nematic quantum criticality in {\BFCA}. 
(a)--(c) $T_s$ ($=T_{\rm nem}$) as a function of the measured $[100]$ strain ($\varepsilon_{xx}$) for a {\BFCA} ($x = 4.8 \pm 0.2\%$) sample. The net variation in (a) is decomposed into a linear contribution from the $A_{1g}$ strain (b) and a quadratic contribution from the $B_{1g}$ strain (c). 
(d) Scaling collapse of the normalized linear and quadratic strain tuning coefficients. The data collapse onto a single curve when scaled by $T_0^{-1/\theta}$, yielding a critical exponent $\theta \approx 0.52$. 
(e) Temperature-doping phase diagram of {\BFCA}. The green shaded region highlights the parameter space governed by the power-law divergence of nematic critical fluctuations. 
Adapted from \cite{worasaran2021nematic}.
\label{Fig_strain_nqcp}}
\end{figure}

\paragraph*{Strain tuning of $T_{\rm crit}$}
Following the establishment of a rigorous symmetry framework that deconvolved the $A_{1g}$ and $B_{1g}$ couplings in Ba(Fe$_{1-x}$Co$_x$)$_2$As$_2$ \cite{ikeda2018symmetric}, continuous uniaxial strain along the $[110]$ direction (accessing the strongly coupled $B_{2g}$ channel) has been extensively utilized to directly tune the superconducting phase boundaries. In underdoped and optimally doped {\BFCA}, both tensile and compressive strains robustly suppress $T_c$; for example, $\sim 1\%$ nominal strain reduces $T_c$ by a factor of $\sim 5$ at $x=0.071$, and $\sim 0.5\%$ is sufficient to drive a complete superconductor-to-metal transition at $x=0.042$ [Figs.~\ref{Fig_strain_tc}(b) and \ref{Fig_strain_tc}(c)]. Fits of $\Delta T_c(\varepsilon_{[110]})$ to Eq.~\eqref{eqn-tc_strain} reveal a dominant negative quadratic contribution ($\beta < 0$), consistent with competition between nematicity and superconductivity \cite{malinowski2020}. Similarly, $\varepsilon_{[110]}$ has been shown to continuously tune $T_s$, $T_N$, and $T_c$ in {\BFAP}, acting in a manner broadly analogous to isovalent P doping and establishing uniaxial strain as a pristine, chemical-disorder-free control parameter for the entire phase diagram \cite{zhao2023uniaxial}.

This robust tuning capability extends to mechanically fragile 11-type chalcogenides. Early direct piezo-cell studies initially reached $|\varepsilon|\sim 0.1\%$ \cite{ghini2021strain}, below the detwinning threshold ($\sim 0.2\%$); in this regime, both $T_s$ and $T_c$ vary approximately linearly with $\varepsilon_{[110]}$, consistent with prior differential measurements \cite{tanatar2016origin}. By utilizing the aforementioned titanium bow-tie platforms [Figs.~\ref{Fig_strain_tc}(d) and \ref{Fig_strain_tc}(e)], researchers can overcome sample cleaving and extract the intrinsic $T_c(\varepsilon)$ behavior deep within the fully detwinned state ($|\varepsilon_{[110]}|\gtrsim 0.2\%$) [Fig.~\ref{Fig_strain_tc}(f)] \cite{park2020rigid, bartlett2021relationship, occhialini2023spontaneous, liu2024nematic}. Note that twin boundaries contribute $\sim 10\%$ to the residual resistivity, which must be explicitly accounted for when analyzing elastoresistivity data \cite{bartlett2021relationship, wiecki2021emerging}. Strikingly, systematic measurements across the isovalently substituted FeSe$_{1-x}$S$_x$ series reveal a profound evolutionary shift: the quadratic coefficient $\beta$ undergoes a strict sign reversal from negative in pure FeSe to positive with increasing S content $x$ \cite{liu2024evolution}. In contrast to the uniform $T_c$ suppression observed in iron pnictides \cite{malinowski2020,zhao2023uniaxial,strain1144}, this anomalous positive nonlinearity provides crucial macroscopic evidence for a growing $d$-wave admixture, consistent with theoretical proposals of nearly degenerate $s$- and $d$-wave pairing instabilities in heavily chalcogen-substituted systems \cite{fernandes2013nematicity, kang2018time, kang2018superconductivity, hicks2025probing}.

Complementing these in-plane strain effects, $c$-axis uniaxial pressure provides another powerful knob to tune the intertwined phase boundaries \cite{hu2018caxis,hu2020uniaxial}. In optimally doped BaFe$_2$(As$_{0.70}$P$_{0.30}$)$_2$,  $c$-axis uniaxial pressure induces a spontaneous reemergence of three-dimensional collinear AFM order \cite{hu2018caxis}. This transition occurs with both $T_N$ and $T_s$ exceeding 30~K [Figs.~\ref{Fig_strain_tc}(g) and \ref{Fig_strain_tc}(h)], while superconductivity is slightly suppressed. Notably, a $\sim400$ MPa pressure along the $c$ axis compresses the $c$ lattice parameter and expands the in-plane lattice, while leaving the average pnictogen height essentially unchanged within experimental uncertainty [inset of Fig.~\ref{Fig_strain_tc}(g)]. Thus, the pressure-induced AFM order is not attributable to a reduced pnictogen height, but instead points to a strong magnetoelastic sensitivity to the in-plane Fe--Fe distance near optimal doping, highlighting the delicate competition between superconductivity, nematicity, and stripe AFM order.

Beyond simply shifting macroscopic phase boundaries, uniaxial strain has provided thermodynamic evidence consistent with the extensive reach of nematic quantum criticality in FeSCs. In a detailed $[100]$-strain ($\varepsilon_{xx}$) study of {\BFCA} \cite{worasaran2021nematic}, the strain-induced shift of the nematic transition, $T_s(\varepsilon_{xx})$, was cleanly decomposed into a linear contribution originating from the symmetry-preserving $A_{1g}$ strain and a quadratic contribution from the orthogonal symmetry-breaking $B_{1g}$ strain [Figs.~\ref{Fig_strain_nqcp}(a)--(c)]. As the Co concentration approaches the optimal doping ($x_c \simeq 6.7\%$), the normalized coefficients from both orthogonal symmetry channels grow strongly and diverge. Interpreted through a generic critical tuning law $T_s\propto (g_c-g)^\theta$, the responses from both $A_{1g}$ and $B_{1g}$ channels remarkably collapse onto an identical universal curve with a single critical exponent $\theta \approx 0.52$ [Fig.~\ref{Fig_strain_nqcp}(d)]. When mapped onto the macroscopic phase diagram [Fig.~\ref{Fig_strain_nqcp}(e)], this robust scaling collapse reveals a physical picture: quantum critical nematic fluctuations are not tightly confined to a narrow singularity near the QCP, but rather extend pervasively over a remarkably broad parameter space enveloping much of the superconducting dome \cite{worasaran2021nematic}.

\subsubsection{Manifestations of the electronic nematicity}

To understand the impact of electronic nematicity on the physical properties of FeSCs, it is essential to characterize its manifestations across multiple degrees of freedom \cite{fernandes2012preemptive,fernandes2012manifestations,bohmer2022nematicity,mirri2015origin}. The integration of mechanical uniaxial pressure devices \cite{man2015electronic,tam2017uniaxial,man2018direct,chu2010,lu2014nematic} and piezoelectrically driven strain cells \cite{hicks2014piezoelectric,liu2019nonlinear,park2020rigid,barber2019piezoelectric} with a sophisticated array of probes has revolutionized the experimental characterization of these symmetry-breaking signatures \cite{hicks2025probing,bohmer2022nematicity}.

High-resolution ARPES has provided a direct, momentum-resolved view of the electronic reconstruction associated with nematicity \cite{sobota2021angle}. Measurements on detwinned iron pnictides, including BaFe$_2$As$_2$ and NaFeAs, revealed pronounced inequivalence between the two orthogonal in-plane directions below $T_s$. In particular, the $d_{xz}$- and
$d_{yz}$-derived bands near the zone boundary undergo opposite energy
shifts, producing strongly anisotropic and orbital-polarized electron
pockets \cite{yi2011symmetry,shimojima2014lifting,Zhang2012,
pfau2019momentum}. The associated nematic energy scale can reach
approximately 40--50 meV, corresponding to 500--600~K in
representative compounds and thus several times the nematic transition
temperature. ARPES therefore establishes that a sizable orbital
reconstruction is an essential component of the nematic state and
places a stringent constraint on any microscopic description of
nematicity. Beyond this uniform nematic band reconstruction,
high-resolution ARPES on Ba$_{1-x}$K$_x$Fe$_2$As$_2$ reported a
folded electronic structure suggestive of an antiferroic instability
in part of the nonmagnetic superconducting region
\cite{shimojima2017}. Its microscopic origin and relation to SDW and
nematic fluctuations remain unsettled.

Importantly, the ARPES band shifts are strongly momentum and
orbital dependent. The nematic state cannot therefore be described
simply as a momentum-independent on-site splitting of the $d_{xz}$ and
$d_{yz}$ levels. The distinct reconstruction near $\Gamma$ and the zone
corners has motivated descriptions in terms of nonlocal orbital or
bond order, momentum-dependent self-energies, and intertwined
spin--orbital correlations \cite{shimojima2014lifting,suzuki2015momentum,
watson2015,pfau2019momentum}. Near $\Gamma$, the nematic contribution
must also be distinguished from the spin--orbit-induced splitting already
present in the tetragonal phase. While the large ARPES band shifts show
that orbital degrees of freedom are strongly involved, they do not by
themselves uniquely identify a universal primary order parameter:
spin-nematic and orbital scenarios can generate closely coupled
symmetry-breaking orbital responses. In FeSe, the resulting orbital
texture of the Fermi surface provides an important basis for the
orbital-selective pairing and spin-fluctuation phenomenology discussed
in Secs.~VI.B and VII.D. ARPES thus defines the single-particle,
momentum-resolved manifestation of nematicity, whose thermodynamic,
magnetic, and real-space consequences are characterized by the
complementary probes discussed below.

Complementary bulk thermodynamic and transport probes establish how nematic electronic reconstruction couples to lattice and charge degrees of freedom. Transport and elastoresistivity measurements have quantified the divergent nematic susceptibility and pronounced in-plane electronic anisotropy \cite{chu2010,chu2012,kuo2016ubiquitous}, while shear modulus measurements of $C_{66}$ provide a thermodynamic measure of the nematic susceptibility \cite{bohmer2014nematic,bohmer2015origin,he2017dichotomy}. AC elastocaloric measurements offer a complementary thermodynamic probe of nematic fluctuations and have revealed strong strain-dependent responses in Ba(Fe$_{1-x}$Co$_x$)$_2$As$_2$ \cite{ikeda2019ac,ikeda2021elastocaloric}; magnetic-torque measurements detected in-plane susceptibility anisotropy extending above the bulk structural and superconducting transitions in BaFe$_2$(As$_{1-x}$P$_x$)$_2$ \cite{Kasahara2012}.

Spectroscopic probes further reveal the electronic, orbital, and spin channels involved in nematicity. Infrared optical spectroscopy clarified the origin of transport anisotropy \cite{mirri2015origin,chinotti2017optical,pal2019optical}; electronic Raman scattering captured critical charge fluctuations and their evolution into a symmetry-resolved nematic resonance in the superconducting state \cite{gallais2013observation,ren2015nematic,massat2016charge, gallais2016nematic,massat2018collapse,philippe2022nematic, gallais2016charge}; and NMR under uniaxial strain probed anisotropic spin-lattice relaxation, revealing a complex internal spin-space structure of the nematic order \cite{kissikov2016nmr,kissikov2018uniaxial}. INS mapped spin-excitation anisotropy and the orbital-selective character of spin excitations in detwinned nematic states \cite{lu2014nematic,lu2018spin,liu2025spin,tam2020orbital}, while RIXS showed that pronounced spin-excitation anisotropy can persist to unusually high energies, supporting a strong connection between spin correlations and nematicity \cite{lu2022spinexcitation,liu2024nematic}.

Diffraction, imaging, and combined scattering/transport measurements provide real- and reciprocal-space views of nematic symmetry breaking. SI-STM visualized mesoscopic and atomic-scale nematic textures \cite{Cheng2025,li2017stripes}, while neutron diffraction measurements of anomalous $d$-spacing distributions revealed local orthorhombicity persisting into the nominally tetragonal phase \cite{lu2016impact,wang2018local,frandsen2017local, frandsen2019quantitative}. Polarized and unpolarized neutron diffraction under uniaxial strain demonstrated that symmetry-breaking fields can enhance in-plane ordered moments and induce out-of-plane magnetic components \cite{tam2017uniaxial,liu2020inplane}; inelastic x-ray and neutron scattering extracted the divergent nematic correlation length \cite{weber2018soft,merritt2020nematic}; and simultaneous x-ray and elastoresistivity measurements correlated transport anisotropy with lattice distortions and spontaneous orbital polarizations \cite{sanchez2021transport,occhialini2023spontaneous}.

While macroscopic and spectroscopic probes have established the main phenomenology of electronic nematicity, recent spatial-imaging experiments point to a rich mesoscopic structure of the nematic state. Linear dichroic photoemission electron microscopy (LD-PEEM) measurements reported that the electronic nematic contrast does not simply form rigid domains with sharp boundaries, but can be described by smooth, long-wavelength sinusoidal ``nematicity waves'' with characteristic wavelengths of 400--1300 nm \cite{shimojima2021discovery}. This interpretation suggests an unusually large spatial stiffness of the nematic order. Subsequent energy-resolved LD-PEEM revealed a sign reversal of the dichroic contrast at binding energies greater than approximately $0.4$~eV \cite{onishi2026energy}, pointing to orbital-selective spectral-weight redistribution within the mesoscopic textures. Complementary site-selective NMR studies were interpreted as evidence for a strongly spin--orbital-intertwined nematic state beyond a simple $d_{xz}/d_{yz}$ ferro-orbital description. In this picture, the correlated $3d_{xy}$ orbital plays an essential role through a nonlocal nematic component \cite{li2020prx, li2025npjqm}. Taken
together, these results support the view that mesoscopic nematic textures are closely tied to multiorbital correlations.

Dark-field x-ray microscopy (DFXM) reported coherent mesoscopic modulations of the local shear strain ($\Delta\varepsilon_{xy}$) within the nematic state \cite{yay2026discovery}. The lattice was interpreted as developing intra-domain strain waves rather than only flat ferroelastic domains, whereas more irregular strain textures appear above $T_s$. These observations may provide a real-space counterpart to the anomalous $d$-spacing broadening seen by neutron (Larmor) diffraction, previously discussed in terms of short-range local orthorhombicity and quenched disorder \cite{lu2016impact,frandsen2017local,wang2018local, frandsen2019quantitative}.

\subsection{Emergent exotic electronic states}

\subsubsection{Topological superconductivity and Majorana zero mode}\label{sec_topological}

Topological superconductivity (TSC) and MZMs have emerged as significant frontiers in condensed matter physics \cite{fu2008superconducting}, and FeSCs have become pivotal platforms for exploring these exotic states \cite{wang2015topological,wu2016topological,xu2016topological,zhang2018observation,zhang2019multiple,wang2018evidence,zhu2020nearly,wang2020evidence,li2022ordered,liu2022tunable,liu2020new,kong2019halfinteger,wang2022stm,liu2018robust,zhang2021observation,chen2020atomic}. The topological superconducting state in FeSCs arises due to spin-helical Dirac surface states formed by a topological band inversion between the
Fe-$d$ orbitals and anion-$p$ orbitals, combined with proximity-induced superconductivity from the bulk, typically characterized by a nodeless $s_\pm$ pairing symmetry \cite{fu2008superconducting,wang2015topological,wu2016topological,xu2016topological}. Within this scenario, localized quasiparticle excitations at zero energy-known as MZMs-can be realized at defects such as vortices or dislocations. MZMs are non-Abelian anyons, whose braiding statistics offer intrinsic topological protection against local decoherence, presenting a promising route towards fault-tolerant quantum computing.

Recent experimental advances have provided evidence for topological superconductivity and MZMs in several FeSC compounds. STM studies on FeTe$_{0.55}$Se$_{0.45}$ first revealed robust zero-bias conductance peaks in vortices \cite{wang2018evidence,kong2019halfinteger,wang2020evidence}, interpreted as signatures of MZMs. Subsequent quantized conductance measurements 
were interpreted as further evidence consistent with a Majorana scenario \cite{zhu2020nearly}. (Li,Fe)OHFeSe with much higher $T_c$ exhibited even clearer spectral isolation of MZMs due to their enhanced superconducting gap and exceptionally small Fermi energies \cite{liu2018robust,zhang2021observation}. Notably, LiFeAs, a stoichiometric pnictide superconductor, demonstrated impurity-assisted vortices and naturally strained terraces leading to ordered arrays of MZMs, thus offering tunable platforms suitable for systematic braiding experiments \cite{liu2022tunable}. Furthermore, CaKFe$_4$As$_4$, with its intrinsic bilayer structure and clear Dirac surface states, expands the FeSC family known to support robust MZM states \cite{liu2020new,cao2024observation}. 

However, identifying MZMs unambiguously remains debated, as an apparent zero-energy vortex bound state is a necessary but insufficient condition \cite{machida2024searching}. Distinguishing MZMs from trivial Caroli-de Gennes-Matricon (CdGM) states is notoriously difficult: their tiny energy spacing ($\delta E \sim \Delta^2/\epsilon_F$) often merges low-lying states into a single zero-bias peak at finite instrumental resolution. Although the small $\epsilon_F$ in Fe(Se,Te) mitigates this, definitive proof demands qualitative signatures unique to MZMs, such as integer level sequences and quantized conductance plateaus. Currently, verifying these is hampered by sample inhomogeneities, multiband effects, and contact-regime challenges \cite{machida2024searching}. Furthermore, STM measurements of FeTe$_{0.55}$Se$_{0.45}$ have also uncovered Yu--Shiba--Rusinov bound states whose zero-bias peaks can mimic several spectroscopic signatures previously ascribed to MZMs \cite{chatzopoulos2021spatially}. Consequently, while the evidence for MZMs in FeSCs is highly suggestive, it is not yet definitive.

Moving forward, experiments should first secure definitive proof of these MZMs via novel, unambiguous signatures \cite{machida2024searching}. Once confirmed, the next frontier is demonstrating their non-Abelian statistics through direct braiding or fusion experiments. Achieving these goals requires advancing scanning probe and nanofabrication techniques for the precise spatial manipulation of individual vortices or engineered defects \cite{liu2022tunable}.

\subsubsection{The Fulde--Ferrell--Larkin--Ovchinnikov (FFLO) state}\label{sec_FFLO}

The Fulde--Ferrell--Larkin--Ovchinnikov (FFLO) state is an exotic superconducting phase in which Cooper pairs possess a finite center-of-mass momentum $\mathbf{q}$, compensating the Zeeman splitting between spin-polarized Fermi surfaces and thereby preventing the collapse of superconductivity when magnetic fields exceed the Pauli paramagnetic limit \cite{fulde1964superconductivity,larkin1964nonuniform}. In the Fulde--Ferrell (FF) variant, the order parameter retains a uniform amplitude but carries a spatially varying phase factor e$^{i \mathbf{q}\cdot\mathbf{r}}$, whereas in the Larkin--Ovchinnikov (LO) variant two counter-propagating components $(\pm\mathbf{q})$ interfere to yield a real order parameter whose amplitude oscillates periodically, producing nodes and sign reversals. Hallmark experimental signatures of an FFLO phase therefore include superconductivity surviving beyond the Pauli limit, strong anisotropy with respect to field orientation, and thermodynamic anomalies such as first-order or double phase transitions below the upper critical field $H_{c2}$.

In FeSCs, experimental evidence supporting FFLO behavior has accumulated, particularly in FeSe and {\KFA}. High-field magnetotransport and thermal-conductivity measurements on FeSe revealed a first-order transition and an upturn of $H_{c2}$ for magnetic fields strictly parallel to the conducting planes \cite{kasahara2020evidence,kasahara2014field}, while scanning-tunnelling spectroscopy identified a quasiparticle nodal-plane structure consistent with a modulated order parameter \cite{kasahara2021quasiparticle}. Intriguingly, this high-field phase persists in samples with controlled disorder, a robustness that challenges conventional FFLO expectations and has stimulated alternative interpretations invoking SDW instabilities or intertwined electronic orders \cite{zhou2021disorder}.

Hole-doped {\KFA} provides one of the clearest thermodynamic cases for an FFLO state among FeSCs \cite{burger2013strong,cho2017thermodynamic}. Specific-heat and magnetic-torque experiments have uncovered a pronounced double superconducting transition that disappears when the field is tilted by more than $\sim1^\circ$ out of the $ab$ plane \cite{cho2017thermodynamic}, mirroring classic FFLO angular sensitivity. Microscopic $^{75}$As NMR measurements further resolve spin-smectic order and Andreev-bound-state spectra that delineate a sharp homogeneous-to-FFLO boundary at $T^* \approx 0.2 T_c$ \cite{liu2025microscopic}, underscoring the decisive role of the multiband electronic structure in stabilizing finite-momentum pairing in this exceptionally clean pnictide. The relation of this high-field phase to the unusual zero-field Fermi-surface, gap, and spin-resonance phenomenology of KFe$_2$As$_2$ is discussed in Sec. \ref{sec_swave_pn}.

Theoretically, multiband models incorporating strong Pauli limitation demonstrate wide parameter windows in which finite-$\mathbf{q}$ pairing outcompetes the homogeneous superconducting state under intense in-plane fields. Real-space Bogoliubov--de Gennes calculations predict observable amplitude-modulation patterns and bound-state ladders that can, in principle, be imaged by high-field scanning probes, offering a route to unite macroscopic thermodynamic evidence with microscopic order-parameter texture \cite{takahashi2014multiband,ptok2013the,gurevich2010upper}.

\subsubsection{Pair-density wave/modulation state}\label{sec_pdw}


Cooper-pair density-wave/modulation (PDW/PDM) states, manifesting as spatial modulations of the superconducting gap or pair amplitude, have recently been reported in several FeSCs \cite{liu2023pair,kong2025cooper,zhao2023smectic,wei2025observation}. These states are conceptually related to nonuniform superconductivity, but they need not all correspond to the same type of finite-momentum pairing. In conventional PDW or LO-like states, the superconducting order parameter varies over multiple lattice constants and breaks translational symmetry. By contrast, an intra-unit-cell PDM can modulate the pair amplitude within the crystallographic unit cell while preserving long-range lattice translation. Unlike conventional LO states, which are induced by strong Pauli-limiting fields, the PDW/PDM states reported in FeSCs arise intrinsically at zero or very low magnetic fields due to internal symmetry-breaking mechanisms.

STM studies have identified notable examples of such modulated superconductivity. In EuRbFe$_{4}$As$_{4}$, STM observations revealed smectic PDW ordering, characterized by an incommensurate superconducting gap modulation over approximately eight lattice constants, reflecting the complex interplay between superconductivity and magnetism \cite{zhao2023smectic}. Another STM study on monolayer Fe(Te,Se) films reported pronounced Cooper-pair density modulation driven by the coexistence of superconductivity and electronic nematicity, further highlighting the critical role of internal symmetry breaking in these phenomena \cite{liu2023pair}.


More recently, measurements on exfoliated FeTe$_{0.55}$Se$_{0.45}$ flakes were interpreted as evidence for a distinct intra-unit-cell PDM state. In this case, the superconducting gap modulation has the periodicity of the underlying lattice and therefore differs from long-wavelength PDW states that break translational symmetry. The observed modulation reflects a large difference in the superconducting gap amplitude on two nominally equivalent Fe sublattices, with a sublattice gap contrast exceeding 30\% in some regions \cite{kong2025cooper}. This lattice-periodic PDM was attributed to glide-mirror symmetry breaking combined with a peculiar nematic distortion in thin flakes, which distinguishes the two Fe sublattices while preserving long-range lattice translation.

Complementary theoretical analysis provided a microscopic basis for these observations, showing that sublattice symmetry breaking combined with a mixed $s_\pm+d$-wave pairing symmetry can naturally lead to such significant intra-unit-cell modulation without requiring a long-wavelength density-wave modulation \cite{papaj2025pair}. Collectively, these recent reports have expanded the experimental landscape for spatially modulated superconducting states in FeSCs, ranging from smectic PDW order to lattice-periodic intra-unit-cell PDM.

\subsubsection{Charge-density wave and charge order}\label{sec_cdw}

Charge-density wave (CDW) or charge order (CO), defined as itinerant (or valence) electrons or holes organizing into periodic spatially modulated patterns with reduced translational symmetry, is recognized as a universal feature in cuprate high-$T_{c}$ superconductors \cite{comin2016resonant}. In FeSCs, CDW phenomena have been observed across multiple systems using diverse experimental techniques \cite{civardi2016superconductivity,martinelli2017experimental,li2017stripes,yuan2021incommensurate,yim2018discovery,walker2023electronic,liu2022tunable,liu2023nematic,chen2023charge}. These studies predominantly reveal short-range, defect-induced, or strain-stabilized electronic modulations rather than long-range order driven by a thermodynamic phase transition \cite{li2017stripes}. Examples include stripe patterns in highly nematic FeSe films \cite{li2017stripes}, incommensurate smectic phases in thin FeSe layers and LiFeAs \cite{yuan2021incommensurate,yim2018discovery}, and stripe-like modulations observed near the nematic quantum critical point in {\FSS} \cite{walker2023electronic}. However, the diverse manifestations of CDWs in FeSCs suggest the absence of a unified theoretical framework to fully capture their complexity.

Theoretically, recent work highlights the close association between CDW formation and electronic nematicity in FeSCs \cite{lahiri2022defect}. A Ginzburg-Landau analysis proposes that localized defects or random strain fields can shift the nematic instability from zero to finite momentum, inducing electronic smectic states confined predominantly to surfaces or near-surface regions. Within this theoretical scenario, CDW patterns observed in STM studies emerge naturally as localized realizations of nematic fluctuations amplified by defects and surface strain. 
Most observed CDW phenomena in FeSCs thus appear intricately connected to the nematic regime. Nonetheless, the exact nature of the interplay between these CDW phenomena and superconductivity remains to be systematically explored through future experimental and theoretical investigations.

\subsubsection{Electronically driven spin-density-wave order}\label{sdw}

The discovery of a double-$\mathbf{Q}$ SDW with tetragonal $C_4$ symmetry in Ba$_{1-x}A_x$Fe$_2$As$_2$ ($A$ = Na, K) and Sr$_{1-x}$Na$_x$Fe$_2$As$_2$ expands our understanding of electronically driven magnetism in FeSCs, by showing that it is not limited to the conventional single-$Q$, orthorhombic stripe state, but can also stabilize a tetragonal magnetic phase when the two stripe ordering vectors are nearly degenerate \cite{avci2013structural,avci2014magnetically,allred2016double,
wang2016complex,bohmer2015superconductivity,yi2018spectral,
guo2019preferred}. This behavior can be understood phenomenologically as a competition between magnetic order parameters associated with the two symmetry-related stripe wave vectors, where near-degeneracy of the two components allows either a single-$Q$, $C_2$ stripe state or a double-$Q$, $C_4$ tetragonal state \cite{allred2016double,fernandes2019intertwined}.

Neutron and x-ray diffraction studies provide definitive evidence for this novel magnetic configuration, characterized by two orthogonal SDW propagation vectors, $\mathbf{Q}=(1, 0)$ and $(0, 1)$ \cite{avci2014magnetically,allred2016double}.
In particular, observations in Sr$_{1-x}$Na$_x$Fe$_2$As$_2$ indicate that magnetic moments develop exclusively on half of the Fe sites and simultaneously realign along the crystallographic $c$-axis, strongly supporting a double-$\mathbf{Q}$ collinear arrangement. Similar diffraction patterns and spin orientations have been consistently identified in Ba$_{1-x}$Na$_x$Fe$_2$As$_2$, confirming that this unique magnetic phase is a ubiquitous electronic instability within hole-doped 122 iron arsenides. These results are different from transverse incommensurate magnetic order in electron-doped BaFe$_2$As$_2$ \cite{PhysRevLett.106.257001}, which is argued to be a spin-glass state unrelated to itinerant electrons in the system \cite{PhysRevB.90.024509,zou2021competitive}.

Detailed explorations of the phase diagrams in Ba$_{1-x}$Na$_x$Fe$_2$As$_2$ and {\BKFA} highlight complex interplay among the 
double-$\mathbf{Q}$
$C_4$ phase, the single-$\mathbf{Q}$ nematic ($C_2$) state, and superconductivity \cite{wang2016complex,bohmer2015superconductivity}. High-resolution thermal expansion and specific heat measurements reveal a rich phase structure, in which the $C_4$ SDW phase emerges distinctly at intermediate doping levels (approximately $0.22 \leq x \leq 0.29$), fully eliminating the orthorhombic distortion \cite{wang2016complex}. Intriguingly, the coexistence region of superconductivity and the $C_4$ phase shows substantial suppression of $T_{c}$, reflecting pronounced competition between these electronic states. Complementary experiments in the {\BKFA} system further emphasize this competition, as superconductivity explicitly drives a re-entrance from the $C_4$ tetragonal phase back into the lower-symmetry $C_2$ nematic phase, suggesting superconductivity's intrinsic affinity for stripe-like magnetic correlations \cite{bohmer2015superconductivity}.

The spin dynamics of the double-$\mathbf{Q}$ $C_4$ phase,
including its polarization-dependent magnetic excitations and weak
superconducting resonance, are discussed together with other
antiferromagnetic spin excitations in Sec.~\ref{sec_highE_spin}.

\section{Pairing symmetry and mechanism}

\subsection{Introduction}

Iron-based superconductivity arises in close proximity to AFM and nematic phases \cite{fernandes2022iron,dai2015antiferromagnetic,si2023iron}. FeSCs exhibit a multiband electronic structure, in which multiple orbitals contribute to the formation of distinct Fermi surface pockets. Despite considerable structural and electronic diversity-ranging from the quasi-nested electron- and hole-like Fermi surfaces characteristic of iron pnictides, to the more complex and anisotropic topologies observed in iron chalcogenides---no single microscopic framework accounts for all FeSCs. A widely discussed framework describes superconductivity in many FeSCs in terms of an even-parity, spin-singlet pairing state commonly referred to as $s_{\pm}$, in which the superconducting order parameter changes sign between different Fermi-surface sheets \cite{wu2024nodal}. Such a state arises from repulsive interactions mediated by spin fluctuations and is consistent with the pronounced AFM correlations in many FeSCs \cite{dai2015antiferromagnetic}. However, the relative importance of spin, orbital, nematic, lattice, and other pairing channels remains material dependent and is still under active discussion.

More broadly, comparison with the cuprates illustrates both a shared phenomenology and essential differences. In both material classes, superconductivity develops in close proximity to AFM correlations, and repulsive interactions at large momentum transfer can favor sign-changing superconducting states. This has motivated spin-fluctuation-mediated pairing as a common conceptual framework for unconventional superconductivity \cite{scalapino2012}. However, the cuprates are often described by a predominantly single-band, strongly correlated CuO$_2$ system with a $d$-wave gap, whereas FeSCs require a multiorbital description in which several Fermi-surface sheets, orbital matrix elements, and Hund's coupling shape the pairing interaction. The resulting gap structures can include $s_{\pm}$, anisotropic or nodal $s$-wave, and competing $d$-wave states. Thus, the comparison is physically instructive but does not imply that a single microscopic balance of interactions applies universally across all FeSCs.

Extensive experimental investigations have provided substantial, although material-dependent, support for the $s_{\pm}$ pairing symmetry in many FeSCs. In particular, the observation of spin-resonance modes via INS at wave vectors corresponding to interband transitions has been widely interpreted as evidence consistent with sign-changing superconducting gaps (Sec. \ref{resonance}). Furthermore, phase-sensitive techniques such as QPI imaging \cite{hirschfeld2015robust,sprau2017discovery,chen2022evidence,du2018sign}, impurity-induced bound state spectroscopy \cite{yin2015observation}, and systematic studies of impurity effects \cite{prozorov2014effect} provide complementary constraints on the internal phase structure of the superconducting order parameter. Complementary measurements by ARPES have elucidated the detailed momentum-space structure of the superconducting gaps across different FeSC families, revealing both nodeless \cite{liu2018orbital,nag2025highly} and nodal superconductivity \cite{wu2024nodal} depending sensitively upon doping, structural modifications, and orbital contributions. Notably, this rich phenomenology has also motivated orbital-selective pairing descriptions in selected compounds, wherein gaps exhibit pronounced differences based on the orbital character of the associated electronic bands.

Recent theoretical and experimental developments have further enriched this picture, with reports of signatures consistent with superconducting states that spontaneously break time-reversal symmetry (TRS) in specific iron chalcogenides, such as FeSe$_{1-x}$S$_x$ \cite{matsuura2023two} and FeSe$_{1-x}$Te$_x$ \cite{roppongi2025topology}. These reported TRS-breaking (TRSB) signatures have been interpreted as complex, multicomponent superconducting states (e.g., $s+id$ or $s+is'$), indicative of the delicate competition and interplay between distinct pairing instabilities. In some cases, these unconventional superconducting phases are accompanied by topologically nontrivial band structures \cite{zhang2018observation,zhang2019multiple}, providing a promising setting for investigating chiral superconductivity and Majorana fermion physics within the FeSC platform. 

Although spin-resonance modes have mostly been interpreted as evidence for
the $s_{\pm}$ pairing symmetry as discussed above, a sign-preserving $s_{++}$ state mediated by orbital fluctuations may also produce spin-resonance-like features in the superconducting state \cite{Kontani2010,Kontani2018}. In particular, self-consistent calculations including the superconductivity-induced reduction of quasiparticle damping show that a resonance-like enhancement near $2\Delta$ can appear even without a sign change \cite{Kontani2018}. Disorder and electron-irradiation studies further complicate the interpretation because the suppression of $T_c$, changes in $H_{c2}$, and impurity-induced bound states depend on both the gap sign and the multiorbital scattering structure \cite{prozorov2024slope}. Thus, neutron resonance, impurity effects, and gap spectroscopy should be evaluated together rather than treated as individually decisive phase-sensitive probes.

In this section, we provide a comprehensive overview of recent progress on pairing symmetry and mechanisms in FeSCs. We begin by examining orbital-selective Cooper pairing and its impact on superconductivity, followed by detailed discussions on nodal and anisotropic $s_{\pm}$ pairing, multicomponent superconductivity, and superconducting states exhibiting TRSB.

\subsection{Orbital-selective Cooper pairing}

\subsubsection{Hund's coupling and orbital selectivity}\label{sec_4a}

As discussed previously, the common structural unit among FeSCs is the FeX$_4$ (X = As, P, Se, etc.) tetrahedron, which splits the Fe $3d$ orbitals into a higher-energy $t_{2g}$ set ($d_{xz}$, $d_{yz}$, $d_{xy}$) and a lower-energy $e_g$ set ($d_{z^2}$, $d_{x^2 - y^2}$). The $t_{2g}$ orbitals lie closer to the Fermi energy, predominantly contributing to the multi-sheet Fermi surface consisting of hole pockets around the BZ center ($\Gamma$) and electron pockets at the zone edges ($M$ or $X/Y$ points) \cite{fernandes2022iron, yi2017role, sobota2021angle,si2016high}. Structural parameters, especially the anion height ($h_{\rm FeX}$), are critically important in tuning electron correlations and spin dynamics in these materials \cite{yin2011kinetic, zhang2014effect}.

It has been widely recognized that both iron pnictides and iron chalcogenides, as multiorbital systems, are correlated Hund's metals or metals with orbital-selective Mott correlations. This complementary perspective captures the interplay of intermediate on-site Coulomb repulsion $U$ and significant intra-atomic Hund's coupling $J_H$ in governing the emergent electronic properties \cite{haule2009coherence,yin2011kinetic,lanata2013orbital,medici2014selective,yu2021orbital,fanfarillo2020synergy,yu2011mott}. In these systems, $U$ (typically $\sim3-5$ eV) discourages double occupancy within the same orbital, while $J_H$ (around $0.5-1$ eV) favors high-spin configurations by aligning spins across different orbitals on the same Fe site, effectively stabilizing local magnetic moments without driving a full Mott insulating state \cite{medici2011hund,yu2011mott}.

Hund's coupling renders inter-orbital hopping incoherent \cite{haule2009coherence,yu2011mott}. The energy cost (or gain) of an electron hopping between different orbitals depends acutely on the specific spin configurations and orbital occupancies involved, introducing strong, configuration-dependent variability into the hopping matrix elements. This ``noisy'' energy landscape destroys the phase coherence necessary for electron (wave) propagation across orbital channels. In contrast, intra-orbital hopping into empty states can maintain kinetic energy and preserve
quasiparticle coherence along individual orbital subbands. This dichotomy leads to a profound \textit{decoupling} between the orbital and electronic structure: the electronic properties (conductivity, Fermi surface topology, quasiparticle coherence) become highly \textit{orbital-selective} \cite{yu2012u,yu2013orbital}. 

The suppressed inter-orbital hybridization prevents the formation of unified, orbitally-mixed bands, resulting in distinct, orbitally-resolved electronic states exhibiting \textit{orbital-dependent correlation} and \textit{coherence} \cite{yin2011kinetic,WOS:000407421200001,yu2011mott}.
For instance, ARPES and optical results of mass enhancement consistently found that the $d_{xy}$ orbital is much more strongly renormalized than $d_{xz}$ and $d_{yz}$ in iron selenides \cite{yin2011kinetic,yin2014spin,yi2017role,huang2022correlationdriven}; STM studies revealed that the $d_{xy}$ orbital is much more incoherent than $d_{xz}$ and $d_{yz}$ orbitals in FeSe \cite{kostin2018imaging}.

This orbital selectivity also profoundly influences magnetic excitations \cite{li2016orbital,wang2017orbital,chen2019anisotropic,tam2020orbital}.
Hund's coupling enforces robust intra-atomic high-spin configurations by energetically penalizing ($\sim J_H$) any deviation from parallel spin alignment across different orbitals on the same Fe site. This strongly suppresses inter-orbital spin fluctuations---collective processes requiring correlated spin changes between orbitals---as they inherently reduce the local total spin moment $S_{\rm total}$. In contrast, intra-orbital spin fluctuations (spin dynamics within a single orbital channel) become the dominant low-energy magnetic mode. While flipping a single spin locally costs $J_{H}$, the propagation of these fluctuations via superexchange ($J_{\rm AFM}\sim t^2/U\ll J_H$) occurs coherently along well-defined orbital-specific bands.

In the itinerant description, the inter-orbital hopping incoherence strongly suppresses inter-orbital spin fluctuations arising from the Fermi surface nesting,  which relies on \textit{coherent} particle-hole excitations across orbitals. In comparison, intra-orbital coherent propagation allows for strong Fermi surface nesting within the same orbital band.

This intra-orbital preference is fundamentally important in driving the orbital-selective or orbital-dependent phenomena, such as the orbital-dependent coherence and electron correlations as discussed above. The orbital dependence (differentiation) is correlated with structural parameters and orbital configurations that determine the intra-orbital electron hopping integrals and coherence \cite{yin2011kinetic,WOS:000407421200001,yu2011mott}. 
The nearest-neighbor Fe--Fe hopping $t_{\alpha, \alpha}$ has two terms---direct Fe--Fe overlap and indirect hopping via the anion---with opposite signs, leading to destructive interference. For $d_{xz/yz}$, the indirect path exceeds the direct one. For $d_{xy}$, the two are comparable; when the anion height is large (e.g., in FeTe), the indirect term is reduced and the two nearly cancel, yielding $t_{xy,xy}\!\approx\!0$. This kinetic frustration accounts for the pronounced mass enhancement of the $d_{xy}$ quasiparticles in iron chalcogenides.
As the anion height increases (monotonically from LaFeAsO to BaFe$_2$As$_2$, LiFeAs, NaFeAs, FeSe and FeTe), hopping integrals involving the $d_{xy}$ orbital reduce significantly, enhancing correlation effects and pushing this orbital toward a more localized state \cite{yi2013observation,yi2015observation,zhang2014effect,yu2013orbital}. In alkali-intercalated FeSe systems, such as $A_x$Fe$_{2-y}$Se$_2$ ($A$=K, Rb, Cs), the $d_{xy}$ orbital was demonstrated to approach a Mott-insulating regime while the other orbitals ($d_{xz}$, $d_{yz}$) remain metallic, giving rise to an orbital-selective Mott phase \cite{yi2015observation,wang2014orbitalselective,yu2013orbital,yi2017role}.

Thus, the intricate interplay between Hund's coupling, orbital-selective electron correlations, and structural parameters sets the stage for the rich and orbital-selective electronic and magnetic behaviors widely observed in FeSCs.

\subsubsection{Orbital-selective Cooper pairing}

\begin{figure*}[htbp!]
\centering
\includegraphics[width=16cm]{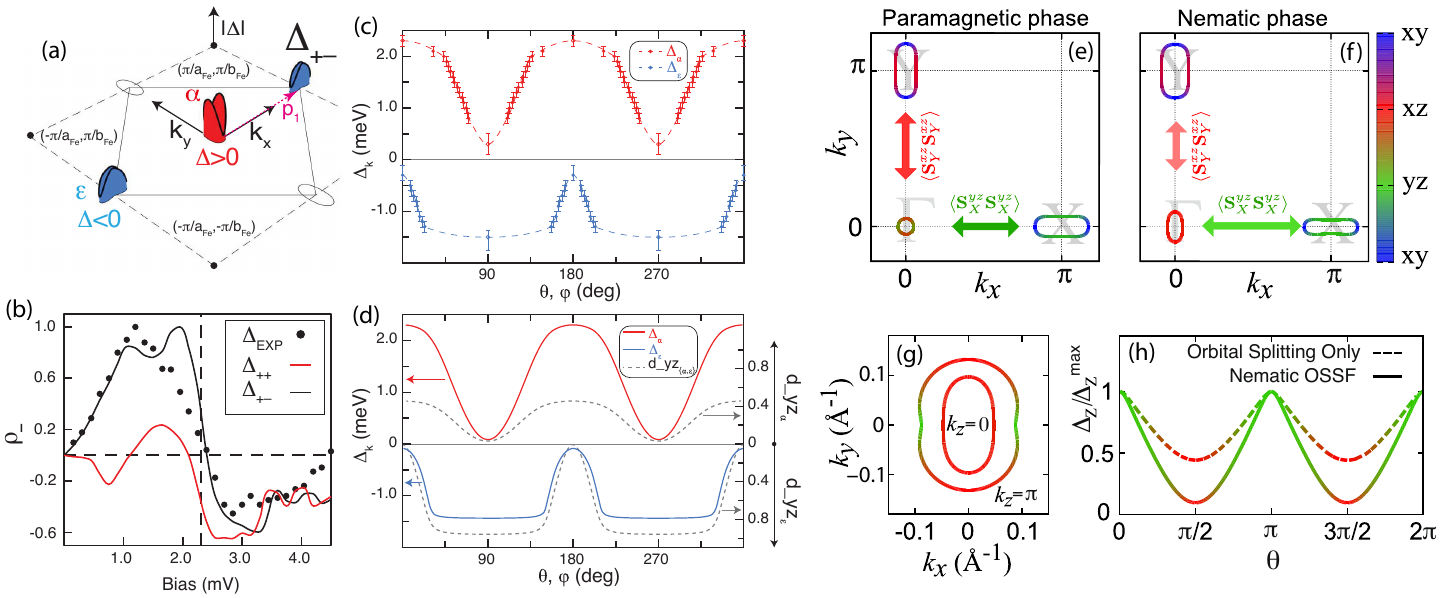}
\caption[]{Orbital-selective Cooper pairing in FeSe \cite{sprau2017discovery,benfatto2018nematic}. (a) Measured $k$-space structure of anisotropic energy gaps of FeSe. The red and blue colors indicate the different signs of the two gap functions. (b) The predicted $\rho_{-}(E)$ for $\pm$ pairing symmetry using the band-gap structure of FeSe is shown as a solid black curve. Here $\rho_-(E)$ denotes the antisymmetrized QPI response obtained by integrating $\mathrm{Re}[g(\mathbf q,+E)]-\mathrm{Re}[g(\mathbf q,-E)]$ over the interband scattering region centered at
$p_1\approx(\pi/a_{\rm Fe},0)$, where $g(\mathbf q,E)$ is the
Fourier-transformed differential conductance.
The predicted $\rho_{-}(E)$ for no gap sign change in FeSe is shown as a solid red curve. The vertical dashed black line marks the energy of the maximum superconducting gap. (c) Angular dependence of FeSe superconducting energy gaps $\Delta_{\alpha}(\mathbf{k})$ about $\Gamma = (0,0)$ and $\Delta_{\varepsilon}(\mathbf{k})$ about $X = (\pi/a_{\rm Fe}, 0)$. (d) Predicted angular dependence of $\Delta_{\alpha}(\mathbf{k})$ and $\Delta_{\varepsilon}(\mathbf{k})$ for an interband pairing interaction that peaks at $(\pi/a_{\rm Fe}, 0)$ and for which pairing is orbital-selective, occurring predominantly for electrons with $d_{yz}$ orbital character. The dashed gray curves show the $d_{yz}$ orbital character of states at the $\alpha$-band and $\varepsilon$-band Fermi surfaces. These energy gap predictions are robust against variations in the quasiparticle orbital weights used in the theoretical calculations, provided that the $d_{yz}$ orbital is kept considerably more coherent than the $d_{xz}$ orbital. (e), (f) FeSe Fermi surfaces at $k_z$= 0 in (e) paramagnetic phase and (f) Nematic phase. The colors represent the main orbital character of the Fermi surface. The green/red arrows denote the orbital-selective spin fluctuations (OSSF), connecting hole and electron pockets at different momenta. The spin fluctuations along $\Gamma X$ and $\Gamma Y$ are equivalent in the paramagnetic phase and become anisotropic in the nematic one. (g) Fermi surface at $k_z= 0$ and $k_z= \pi$ in the nematic phase. At $k_z= \pi$ the hole pocket retains a full $yz$ orbital character at $\theta = 0$. (h) Angular dependence of the SC gap $\Delta_Z$ renormalized to its maximum value. The nematic pairing (solid lines) further enhances the gap anisotropy, leading to larger relative variations on the Z pocket. (a)--(d) are adapted from \cite{sprau2017discovery} and (e)--(h) from \cite{benfatto2018nematic}.
\label{Fig_OSSF}}
\end{figure*}

Considering the presence of orbital selectivity, it naturally follows that superconducting pairing itself could exhibit significant orbital dependence \cite{yu2014orbital,WOS:000407421200001}. In such a scenario, Cooper pairs preferentially form among electrons dominated by a particular orbital character, resulting in superconducting gaps that display pronounced anisotropy, with significant magnitudes primarily on Fermi surface regions associated with that orbital.
The result is orbital-selective pairing, in which the pairing structure is determined by the orbital-dependent quasiparticle weight and the variation of the orbital weight on the Fermi surface \cite{yu2014orbital}.

The phenomenon of such orbital-selective Cooper pairing, first observed in bulk FeSe, represents a substantial breakthrough in the study of FeSCs \cite{sprau2017discovery,kostin2018imaging}. Utilizing sub-Kelvin Bogoliubov QPI (BQPI) imaging, Sprau \textit{et al.} provided the first direct visualization of the superconducting energy gaps in FeSe with unprecedented energy resolution ($\sim75~\mu$eV). As illustrated in Figs.~\ref{Fig_OSSF}(a)--(c), both the hole-like Fermi pocket at the BZ center ($\Gamma$, red) and the electron-like pocket at the zone corner ($X$, blue) exhibit strongly anisotropic, yet fully gapped (nodeless) superconducting states. These findings were subsequently corroborated by ARPES studies \cite{liu2018orbital,kushinirenko2018three}. Notably, the superconducting gap maxima on hole and electron pockets align orthogonally in momentum space, consistent with a scenario involving sign reversal of the superconducting gap between electron and hole Fermi surfaces. 
Crucially, detailed comparisons between the experimental BQPI data and RPA calculations incorporating orbital-dependent quasiparticle coherence \cite{kreisel2017orbital} showed that the observed gap anisotropy and its orientation can be reproduced when pairing is dominated by the Fe $3d_{yz}$ orbital channel [Fig.~\ref{Fig_OSSF}(d)]. Taken together, these results provide strong support for orbital-selective Cooper pairing in FeSe.

A microscopic framework for the observed orbital-selective pairing is provided by OSSF \cite{benfatto2018nematic,kreisel2017orbital,yu2021orbital}. Theoretical studies have elaborated this phenomenon by incorporating orbital-dependent quasiparticle coherence into spin-fluctuation-mediated pairing models. When this orbital selectivity is explicitly included, pair-scattering processes involving less coherent orbitals, especially the strongly correlated $d_{xy}$ orbital, are substantially suppressed, thereby increasing the relative weight of the more coherent $d_{xz}$ and $d_{yz}$ channels in the pairing interaction. The resulting gap structures are consistent with the anisotropic, nodeless gaps observed
in FeSe \cite{sprau2017discovery,kreisel2017orbital}.

Complementary investigations using QPI imaging in the normal-state nematic phase of FeSe by Kostin \textit{et al.} \cite{kostin2018imaging} were initially interpreted as direct evidence for this extreme orbital differentiation. In their scenario, electrons with predominant $d_{xy}$ character exhibit extreme incoherence compared to the $d_{xz}$ and $d_{yz}$ orbitals. However, an alternative explanation for the highly anisotropic QPI patterns involves nonlocal $d_{xy}$ nematicity \cite{rhodes2021nonlocal}. As proposed by Rhodes \textit{et al.}, a sizable nonlocal nematic ordering can trigger a Lifshitz transition that shifts the $d_{xy}$-dominated electron pocket entirely above the Fermi level. While differing in their microscopic details---extreme orbital decoherence versus a band-shifting Lifshitz transition---both models successfully account for the heavily suppressed $d_{xy}$ spectral weight at the Fermi level. This suppression may provide an important electronic environment for the orbital-selective pairing that emerges upon cooling into the superconducting state.


The nematic transition at $T_s \approx 90$~K in FeSe breaks the $C_4$ rotational symmetry, leading to a splitting and reconstruction of the $d_{xz}$- and $d_{yz}$-derived bands. This reconstruction strongly modifies the Fermi surface and the orbital-dependent quasiparticle coherence weights \cite{fanfarillo2016orbital,yi2019nematic}, thereby amplifying orbital differentiation in the normal state. It can in turn
favor orbital-selective pairing, for example by enhancing the relative contribution of the more coherent $d_{yz}$-dominated channel, and is consistent with anisotropic, nodeless superconducting gaps whose maxima are oriented orthogonally on the hole and electron pockets in momentum space.

\begin{figure}
\centering
\includegraphics[width=8cm]{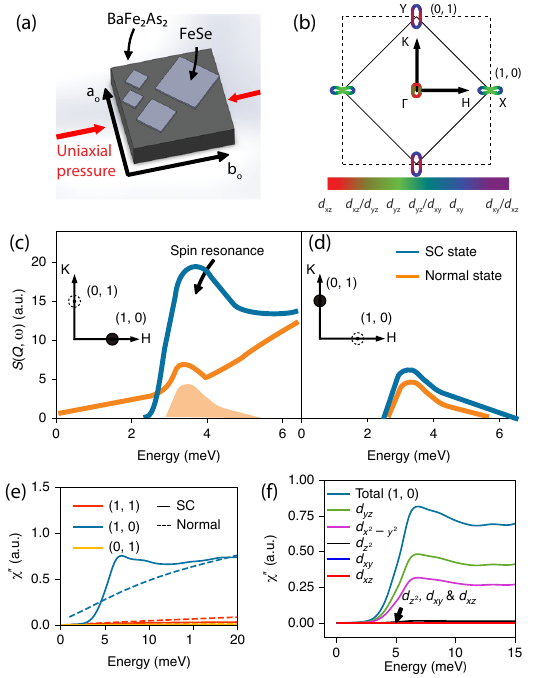}
\caption[]{Low-energy anisotropic spin fluctuations in detwinned FeSe. (a) Schematic diagram of the sample arrangement. FeSe samples are glued on {\BFA} single crystals under a uniaxial pressure. (b) Hole-electron Fermi surfaces of the tight-binding model for FeSe. Fermi surface nesting of $\Gamma\to X$ and $\Gamma\to Y$ corresponds to (1, 0) and (0, 1) in reciprocal lattice units. (c), (d) Schematic illustrations of the magnetic scattering at (1, 0) (c) and (0, 1) (d) in the normal and SC states estimated from the twinned and detwinned samples. (e) The orbital-selective model yields spin fluctuations at low energies that are dominated by peaks at $(\pm1, 0)$ in the low energy range shown. The enhancement of the spin fluctuations at $(\pm1, 0)$ in the superconducting state is clearly seen when plotted as a function of energy. (f) The spin fluctuations at (1, 0) are dominated by the contributions of the $d_{yz}$ orbital. Adapted from \cite{chen2019anisotropic}.
\label{Fig_INS_dFeSe_low}}
\end{figure}

Further theoretical insights link nematicity and OSSF \cite{benfatto2018nematic,hu2018orbital}. Figures \ref{Fig_OSSF}(e) and \ref{Fig_OSSF}(f) display the calculated Fermi surfaces of FeSe in the tetragonal and orthorhombic (nematic) phases, respectively. The red and green bars with arrowheads represent the intra-orbital spin-scattering processes involving the $d_{xz}$ ($\langle S_Y^{xz}S_Y^{xz}\rangle$) and $d_{yz}$ ($\langle S_X^{yz}S_X^{yz}\rangle$) orbitals. As the system transitions into the nematic phase, spin fluctuations associated with the $d_{yz}$ orbital are significantly enhanced, while those involving the $d_{xz}$ orbital are suppressed [length and color change of the bars in Figs.~\ref{Fig_OSSF}(e),(f)]. This indicates that nematic order primarily amplifies anisotropic spin fluctuations within the $d_{yz}$ orbital channel. These anisotropic OSSF, in turn, create conditions highly favorable for Cooper pairing between electron and hole Fermi surface regions dominated by the $d_{yz}$ orbital character. Indeed, as shown in Figs.~\ref{Fig_OSSF}(g), (h), calculations explicitly incorporating nematic pairing mediated by OSSF quantitatively reproduce the experimentally observed gap anisotropy ($\Delta Z/\Delta Z^{\mathrm{max}}$) at the $\Gamma$-point even more accurately [compared to Fig.~\ref{Fig_OSSF}(d)]. Thus, the observed superconducting gap anisotropy in FeSe naturally emerges from this nematic and orbital-selective pairing scenario, highlighting that superconductivity in FeSe is not only orbital-selective but fundamentally intertwined with the nematic state.

This intricate interplay between nematicity, orbital selectivity, and superconductivity clarifies several previously puzzling characteristics of FeSCs. Orbital-dependent quasiparticle coherence, originating fundamentally from Hund's coupling and orbital-dependent electronic correlations, profoundly influences the spin fluctuation spectrum and associated pairing interactions. Kreisel \textit{et al.} \cite{kreisel2017orbital} argued that incorporating orbital selectivity explicitly within itinerant spin-fluctuation theories is crucial for accurately capturing detailed gap anisotropies observed in several strongly correlated FeSCs beyond FeSe, including monolayer FeSe and LiFeAs. These findings suggest that orbital selectivity can play an important role in shaping the pairing interaction and gap anisotropy in several FeSCs. Its quantitative importance, however, is material dependent and should be assessed together with Fermi-surface topology, nematicity, and the strength of spin fluctuations.


\subsubsection{Orbital-selective spin fluctuations}\label{sec_ossf}

The orbital-selective nature of spin fluctuations requires measurements on detwinned samples to reveal intrinsic anisotropies that are otherwise obscured by twinning \cite{chu2010,lu2014nematic,yi2011symmetry,yi2019nematic}. Motivated by the discovery of orbital-selective Cooper pairing in FeSe, INS experiments were subsequently performed on uniaxial pressure/strain-detwinned FeSCs, including NaFe$_{1-x}$Co$_x$As \cite{wang2017orbital}, FeSe \cite{chen2019anisotropic}, and Ba(Fe$_{1-x}$Co$_x$)$_2$As$_2$ \cite{tian2019spin}. A consistent feature observed across these studies is that the low-energy spin fluctuations, especially the superconductivity-induced spin resonance when it is present, are strongly concentrated at the AFM wave vector $(1,0)$ rather than at the orthogonal $(0,1)$ point. This wave-vector is consistent with predominantly intra-orbital spin fluctuations in the $d_{yz}$ orbital.

In this section, we focus on the anisotropic spin fluctuations measured on detwinned FeSe to discuss the identification of the orbital-selectivity of the spin fluctuations \cite{chen2019anisotropic}.
Figure \ref{Fig_INS_dFeSe_low}(a) shows a schematic of the detwinning method for FeSe, where a large number of thin FeSe crystals are affixed to uniaxially pressurized BaFe$_2$As$_2$ crystals. Upon cooling through its structural transition at $T_s \approx 137$~K, {\BFA} crystals are detwinned and generate an orthorhombic distortion that detwins the overlying FeSe crystals for INS experiment. The Fermi surface of superconducting FeSe [Fig.~\ref{Fig_INS_dFeSe_low}(b)] exhibits orbital differentiation, with the $d_{yz}$ character concentrated along the AFM wave vector $\mathbf{Q}_{\mathrm{AFM}} = (1,0)$. As illustrated in Figs.~\ref{Fig_INS_dFeSe_low}(c)--(d), the spin resonance is detected exclusively at this momentum, with no corresponding resonance at $(0,1)$. The pronounced spin-fluctuation anisotropy correlates with the orbital texture of the Fermi surface and provides a stringent experimental constraint on descriptions based on OSSF-induced superconductivity \cite{chen2019anisotropic}.

The experimental findings align naturally with theoretical models that incorporate orbital-dependent quasiparticle coherence. In particular, RPA calculations that account for reduced coherence in the $d_{xy}$ orbital successfully reproduce the observed anisotropic spin resonance [Fig.~\ref{Fig_INS_dFeSe_low}(e)]. Within this calculation, the dynamical susceptibility $\chi''$ is dominated by the $d_{yz}$ orbital channel, as illustrated by orbital decomposition of the magnetic excitations [Fig.\ref{Fig_INS_dFeSe_low}(f)]. The agreement between the measured and calculated momentum dependence strongly supports the view that anisotropic OSSF, enhanced by nematicity, provide an important pairing channel in FeSe \cite{chen2019anisotropic,benfatto2018nematic}.

Together, these neutron scattering experiments and accompanying theoretical analysis provide a coherent link between nematic electronic correlations, orbital-selective spin excitations, and Cooper pairing in FeSe. They substantially strengthen the orbital-selective pairing framework associated with STM studies \cite{sprau2017discovery,kreisel2017orbital,kostin2018imaging}.

In conclusion, detwinned INS, RPA modeling, and spectroscopic
probes strongly support a scenario in which $d_{yz}$-dominated spin fluctuations provide a dominant pairing channel in FeSe. 
Such studies thus provide evidence that
orbital selectivity offers a coherent framework for its gap anisotropy and spin-response anisotropy, with a material-dependent role across FeSCs.

\subsection{Pairing symmetry and superconducting gap structure}

\begin{figure}
\centering
\includegraphics[width=8cm]{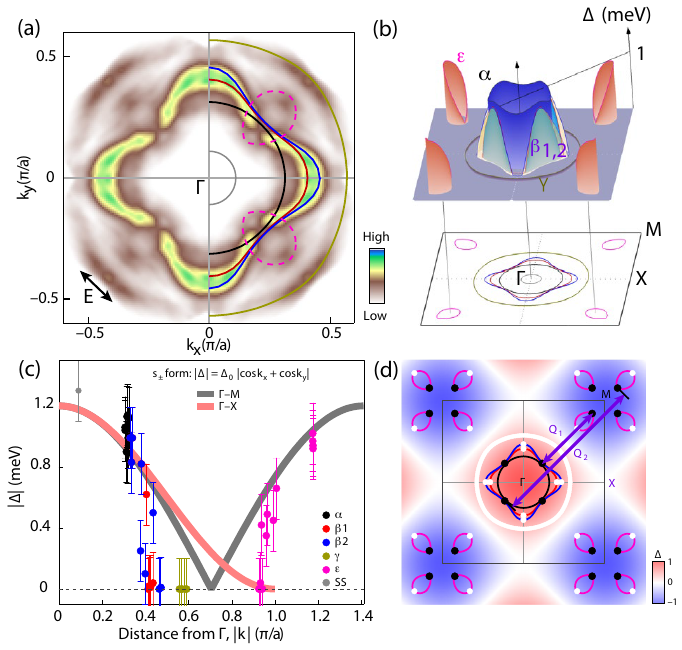}
\caption[]{Nodal $s_\pm$ pairing symmetry in KFe$_2$As$_2$. (a) ARPES mapping of the Fermi surface around the zone center $\Gamma$ point, measured at $0.9$ K. The spectral intensity is integrated within $E_{F}\pm1$ meV and symmetrized assuming $C_{4}$ symmetry. (b) The magnitudes of superconducting gaps in KFe$_2$As$_2$. The corresponding Fermi surface is shown at the bottom. (c) Fermi momentum dependence of superconducting gaps on all the Fermi surface sheets. The thick black line ($\Gamma-M$) and the thick red line ($\Gamma-X$) are expected gap size variations of the $s_\pm$ symmetry ($|\Delta|=|{\rm cos}k_x + {\rm cos}k_y|$). (d) Gap nodes and local gap maxima on the Fermi surface of KFe$_2$As$_2$. The color image shows the gap function cos$k_x$+cos$k_y$ in the $s_\pm$ pairing symmetry, where the positive gap sign, negative gap sign and the gap nodes are represented by red, blue and white colors, respectively. The measured gap nodes are marked by white circles and the gap maxima are marked by black circles. Two interband scattering wave vectors ($\mathbf{Q}_1$ and $\mathbf{Q}_2$) connect the local gap maxima between the $\varepsilon$ band and the ($\alpha$, $\beta$) bands. Adapted from \cite{wu2024nodal}.
\label{Fig_KFA}}
\end{figure}

\begin{figure}
\centering
\includegraphics[width=8cm]{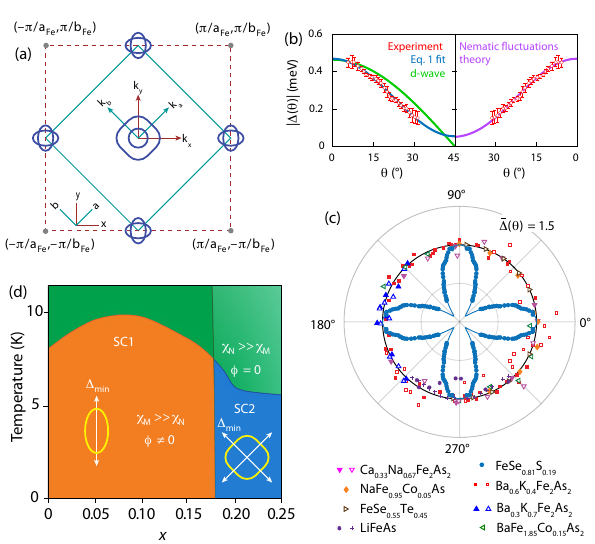}
\caption[]{Highly anisotropic superconducting gap near the nematic quantum critical point of {\FSS}. (a) Schematic of the Fermi surface of tetragonal {\FSS}. The pockets were enlarged by a factor of two for clarity. (b) Gap magnitude $|\Delta(\theta)|$ with 95\% confidence intervals from Gaussian fits. Left panel: fits using $\Delta(\theta) = \Delta_s+ \Delta_s'{\rm cos}^2(2\theta)$ and the $d$-wave form $\Delta(\theta) = \Delta_d {\rm cos}(2\theta)$. Right panel: theoretical calculation considering pairing mediated by nematic quantum critical fluctuations (purple). (c) Polar plot comparing normalized superconducting gaps $|\overline{\Delta}(\theta)|$ for FeSe$_{0.81}$S$_{0.19}$ (outer hole pocket, solid circles) and other tetragonal FeSCs. For LiFeAs, gaps on largest and middle pockets around $\Gamma$ are shown; open (filled) symbols denote outer (inner) hole pockets for other materials. (d) Phase diagram of {\FSS}, indicating the direction of gap minima ($\Delta_{\rm min}$, white arrows) on hole pockets in SC1 and SC2 phases, alongside magnetic fluctuations ($\chi_{\rm M}$), nematic order ($\phi$), and nematic fluctuations ($\chi_{\rm N}$). Adapted from \cite{nag2025highly}.
\label{Fig_FSS19}}
\end{figure}

In the multiorbital FeSCs, Fermi surface sheets at the BZ center and corner (hole and electron pockets, respectively) are typically connected by AFM spin fluctuations, which favor a superconducting order parameter that reverses sign between those Fermi pockets \cite{maier2009neutron,dai2015antiferromagnetic}. This $s_{\pm}$ pairing symmetry was proposed early on as a natural explanation for nodeless gaps in many FeSCs, and it has remained the prevailing candidate. In what follows, we review evidence for sign-changing pairing and discuss how Fermi-surface topology and spin, orbital, and nematic fluctuations generate material-dependent gap structures.

\subsubsection{$s_\pm$-wave pairing in iron pnictides}\label{sec_swave_pn}

Despite this, there have been notable exceptions and debates - for instance, {\KFA} (with only hole pockets) posed a longstanding puzzle and was even speculated to host a $d$-wave state due to its nodal superconductivity \cite{tafti2013sudden,dong2010quantum}. Okazaki \textit{et al.} \cite{okazaki2012} utilized ultrahigh-resolution laser ARPES to map the gap structure and reported a highly unusual Fermi-surface-selective multi-gap structure, including ``octet-line nodes'' on the middle hole pocket, an almost-zero gap on the outer pocket, and a nodeless gap on the inner pocket. A recent high-resolution laser ARPES study revisited this issue \cite{wu2024nodal}.
Figure \ref{Fig_KFA} summarizes the ARPES measurements of the superconducting gap in {\KFA} and the extended $s_{\pm}$ interpretation proposed by Wu \textit{et al.} \cite{wu2024nodal}. In Fig.~\ref{Fig_KFA}(a), the measured Fermi surface at a temperature ($T\approx0.9$~K) much lower than $T_c\approx3.7$~K consists of hole pockets at $\Gamma$ and folded pockets originally located at the zone corner ($M$). Figure \ref{Fig_KFA}(b) schematically illustrates the anisotropic superconducting gap structure, which exhibits pronounced gap nodes on the $\beta$ hole pockets and notable anisotropy, though nodeless, on the $\alpha$ and $\varepsilon$ pockets. 

Wu \textit{et al.} compared the momentum dependence of the measured gap magnitudes $|\Delta(\mathbf{k})|$ with an extended $s$-wave form, $\Delta(\mathbf{k})\propto \cos k_x+\cos k_y$, as shown in Fig.~\ref{Fig_KFA}(c). They interpreted the overall correspondence as supporting an extended $s_{\pm}$-like gap structure, although the measured gap magnitude alone does not determine the relative gap sign.
Figure~\ref{Fig_KFA}(d) superimposes the measured Fermi-surface contours and inferred near-zero-gap locations on a calculated map of the $\cos k_x+\cos k_y$ form factor. The red and blue colors denote the positive and negative values of the model gap function, rather than phases directly measured by ARPES. The authors argued that the observed nodal topology is compatible with an extended $s_{\pm}$ interpretation, while an exact correspondence with the calculated nodal lines is not established.

These ARPES results were interpreted as evidence for an extended $s_{\pm}$-like gap structure in KFe$_2$As$_2$. This interpretation has also been discussed in connection with the unusual incommensurate neutron spin resonance, occurring between the $\Gamma$-centered hole pockets and the unfolded $M$-centered hole pockets in the superconducting state \cite{shen2020neutron}. However, because ARPES directly measures the gap magnitude rather than the phase of the order parameter, the sign structure remains model-dependent. The pairing symmetry in this heavily hole-doped limit should therefore be discussed as an important example of compound-dependent pairing \cite{liu2019evidence} rather than as definitive proof of universal $s_{\pm}$ superconductivity.

Viewed together, KFe$_2$As$_2$ and the nearby overdoped
Ba$_{1-x}$K$_x$Fe$_2$As$_2$ regime illustrate how the strongly
reconstructed, electron-pocket-depleted Fermi surface reshapes the
multiband pairing problem. In KFe$_2$As$_2$, the incommensurate
resonance connecting the $\Gamma$-centered and unfolded $M$-centered
pockets, together with the strongly anisotropic and nodal gap structure, indicates that the relevant interpocket scattering processes differ substantially from those in optimally doped 122 compounds. These
observations constrain, but do not uniquely determine, the relative gap
signs and the leading pairing channel. In the nearby overdoped
Ba$_{1-x}$K$_x$Fe$_2$As$_2$ regime near $x\approx0.8$, the reported
TRSB precursor phase and proposed $s+is'$ state have been interpreted
as consequences of frustrated interband phase coupling between
near-degenerate $s$-wave channels (Sec. \ref{sec_trsb_pn})
\cite{shen2020neutron,wu2024nodal,grinenko2021state,
shipulin2023calorimetric}. The high-field FFLO phase of KFe$_2$As$_2$ (Sec. \ref{sec_FFLO})
is a distinct field-induced instability rather than direct evidence for
$s+is'$ pairing; nevertheless, its sensitivity to Pauli limitation and
multiband electronic structure further underscores the unusual pairing
environment at the heavily hole-doped end of the 122 phase diagram.

\subsubsection{{\FSS}}

{\FSS} has attracted substantial research attention because superconductivity develops in a pristine nematic paramagnetic normal state and undergoes a dramatic change in gap structure across the nematic QCP at $x\approx0.17$. A sharp change in the superconducting gap structure across the NQCP was reported by high-resolution STM \cite{hanaguri2018two}, specific heat and thermal conductivity measurements \cite{sato2018abrupt}, and subsequent studies \cite{matsuura2023two}.
In the nematic ordering regime, the superconducting gap of FeSe and {\FSS} exhibited pronounced $C_2$ anisotropy, consistent with a nematic superconducting state within a $d_{yz}$-dominated OSSF $s_\pm$ framework. What is less clear is how this gap structure evolves once static nematic order collapses and only critical nematic fluctuations survive on crossing the nematic QCP \cite{matsuura2023two,hanaguri2018two}.

Figure \ref{Fig_FSS19} summarizes recent experimental evidence for a highly anisotropic superconducting gap near the nematic QCP of FeSe$_{1-x}$S$_x$ \cite{nag2025highly}. Unlike typical FeSCs, which usually exhibit nearly isotropic or modestly anisotropic gaps consistent with spin-fluctuation-driven $s_{\pm}$ pairing, 
FeSe$_{0.81}$S$_{0.19}$, which lies near or beyond the nematic end point, exhibits a highly anisotropic gap with near-nodal minima on the hole pocket around $\Gamma$ [Figure \ref{Fig_FSS19}(c)]. In the tetragonal regime, the measured pattern has an overall fourfold character, although it may be understood as reflecting the underlying twofold anisotropic tendencies associated with nearby nematic order or nematic fluctuations.
The distinct gap minima, as measured by BQPI, are oriented along a direction $45^\circ$ with respect to the Fe--Fe bond directions (as in KFe$_2$As$_2$). This orientation was argued to be difficult to reconcile with a simple pure $d$-wave scenario [Fig.~\ref{Fig_FSS19}(d)]. Instead, the highly anisotropic gap structure was interpreted as qualitatively consistent with theoretical calculations in which critical nematic fluctuations enhance pairing \cite{klein2018superconductivity}. In this model, the fluctuations selectively enhance pairing interactions along particular directions, yielding deep but nonzero gap minima.

As schematically illustrated in Fig.~\ref{Fig_FSS19}(d), these observations support an important influence of nematic order and fluctuations on the superconducting gap structure. For $x\lesssim0.17$ (SC1 phase), the observed $C_2$ gap anisotropy has been
discussed in terms of nematic order pinning the gap minima along a single crystallographic direction \cite{liu2018orbital,xu2016highly}. Near and beyond the nematic QCP (SC2 phase), as exemplified by FeSe$_{0.81}$S$_{0.19}$, the absence of static nematic order has been proposed to enhance the relative role of nematic fluctuations.

While the gap anisotropy in FeSe$_{0.81}$S$_{0.19}$ is consistent with calculations invoking pairing enhanced by critical nematic fluctuations, two key issues remain unsettled. First, the precise character of the nematic fluctuations and their relationship to spin fluctuations has yet to be established. Second, the full momentum-dependence of the gap, especially on the electron pockets, is still unknown. Momentum-resolved probes such as ARPES and INS are therefore needed to map the gap on all Fermi-surface sheets and to identify the symmetry of the underlying spin fluctuations, which are necessary to clarify the specific pairing symmetry in the SC2 phase of {\FSS}.

\subsubsection{Heavily electron-doped FeSe-based superconductors}

Although these systems differ substantially in dimensionality, disorder,
and interface coupling, electron-only FeSe-based superconductors share
a common Fermi-surface topology. Spectroscopic measurements
generally find fully gapped superconductivity on the electron pockets,
with material-dependent gap anisotropy. Together with the
absence of hole pockets around $\Gamma$, this gap phenomenology places
strong constraints on the possible pairing states and renders a pairing
description based solely on conventional hole--electron nesting
inadequate. Proposed candidates include sign-preserving plain
$s_{++}$ and anisotropic $s_{++}$ states, as well as sign-reversing
incipient or extended $s_{\pm}$ states and nodeless $d$-wave states
\cite{Yang2013,Coh2015,Rademaker2016,Hirschfeld2011,Mazin2011,
Khodas2012,Lin2011,Yamakawa2017,kang2016superconductivity,Chen2015,
linscheid2016high,Gao2016,gao2018possible,hu2013ironbased,
agterberg2017resilient,yin2014spin,Huang2017,yu2013superconductivity}. The absence of nodes
alone does not determine the relative gap sign; spin resonance,
phase-sensitive QPI, impurity-bound-state, and edge-state measurements
provide essential complementary constraints. 

In alkali-intercalated $A_x$Fe$_{2-y}$Se$_2$ ($A$ = K, Rb, Tl, etc., $T_c\sim30$~K), INS reveals a spin resonance at the inter-pocket wave-vector 
$(1, 1)$ in the orthorhombic reciprocal-lattice notation \cite{dai2015antiferromagnetic}, which has commonly been interpreted as favoring a nodeless sign-changing order parameter. Whether the state is best described as a nodeless $d$-wave or a bonding-antibonding $s_{\pm}$ remains unresolved, largely because Fe-vacancy disorder obscures phase-sensitive probes, but the available experimental results are most naturally reconciled with a fully-gapped sign-changing pairing symmetry \cite{kreisel2020on}. The hydroxide-spaced compound (Li,Fe)OHFeSe (optimal $T_c \approx 41$~K) provides much cleaner crystals \cite{lu2015coexistence,dong2015li}.  ARPES revealed an isotropic superconducting gap \cite{zhao2016common}, while phase-sensitive STM/QPI was interpreted as indicating opposite gap signs on the two electron pockets \cite{du2018sign}; a spin resonance near $(1/2, 1/2)$ in the tetragonal structure notation is again observed \cite{pan2017structure,ma2017lowenergy}. Taken together, these results strongly favor a nodeless, sign-reversing state---an ``electron-pocket $s_{\pm}$'' or its nearly degenerate nodeless $d$-wave analogue---consistent with spin-fluctuation-mediated pairing. Concerning these systems, Ref.~\cite{kreisel2020on} provided in-depth theoretical discussions.

In monolayer FeSe/SrTiO$_3$, ARPES and STM studies consistently map a nodeless gap on the two electron pockets at the $M$ points \cite{zhang2016superconducting,zhao2016common,fan2015plain,wei2023identifying}. Early STM/QPI and impurity-scattering experiments, together with step-edge spectroscopy, found no in-gap states for non-magnetic adatoms and observed only topologically trivial edge/corner spectra. These observations were interpreted as favoring a plain, sign-preserving $s$-wave order parameter \cite{fan2015plain}. More recently, atomic-resolution edge-state imaging was also interpreted as favoring plain $s$-wave pairing, based on the absence of the Dirac-like modes expected for sign-reversing gaps \cite{wei2023identifying}. 
By contrast, bound-state STM on intrinsic Fe-vacancy defects reported robust in-gap resonance peaks that do not Zeeman-split, and defect-bound-state QPI reported a $\pi$-phase shift characteristic of sign-changing pairing \cite{zhang2020sign}, consistent with an extended $s_\pm$ or nodeless $d$-wave state.

The momentum dependence of the gap provides an additional constraint.
High-resolution ARPES resolves four deep gap minima locked to the intersections of the two elliptic electron pockets at $M$; the maxima reside on the $d_{xy}$-dominated tips of the ellipses \cite{zhang2016superconducting}.  Such momentum selectivity was argued to be difficult to reproduce using simple $d$- or extended-$s$ harmonics. Within a spin-fluctuation framework incorporating orbital-selective quasiparticle coherence, the calculated gap can reproduce both the ARPES anisotropy and its orientation after a single, orbital-dependent normalization of the interaction kernel \cite{kreisel2017orbital}. These results illustrate that the gap anisotropy alone does not settle the relative gap sign: depending on the degree of orbital-selective $d_{xy}$ decoherence, it can be described by either an orbital-selective sign-changing $s_{\pm}$ state or a plain $s$ state \cite{kreisel2020on}.

\subsubsection{BCS--BEC crossover in FeSe-based superconductors}

FeSe and its variants provide a rare solid-state platform to explore the Bardeen-Cooper-Schrieffer (BCS) to Bose-Einstein condensation (BEC) crossover \cite{kasahara2014field, rinott2017tuning,BCS-BEC}.
In bulk FeSe, exceptionally small Fermi energies ($\epsilon_F \sim 10$ meV) yield a large $\Delta/\epsilon_F \sim 1$ ratio, firmly placing the material in the crossover regime \cite{kasahara2014field}.
Consequently, Zeeman energies comparable to $\epsilon_F$ and $\Delta$ induce a distinct high-field superconducting phase \cite{kasahara2014field}, while the small $k_F\xi$ modifies vortex core states into discrete Friedel-like oscillations rather than a broad zero-energy peak \cite{hanaguri2019quantum}.

This crossover is systematically tunable. In FeSe$_{1-x}$S$_x$, entering the non-nematic tetragonal phase has been interpreted as moving the system toward the BEC side of the crossover, as suggested by giant superconducting fluctuations, non-mean-field thermodynamics \cite{mizukami2023unusual}, and an upward-convex BEC-like Bogoliubov quasiparticle dispersion \cite{hashimoto2020bose}.
Similar BCS--BEC evolutions occur in Fe$_{1+y}$Se$_x$Te$_{1-x}$ tuned by excess iron \cite{rinott2017tuning}, and in FeSe monolayers via substrate-modulated band shifts \cite{lin2023real}.
Unlike ultracold atoms, the crossover in these multiband iron chalcogenides is deeply intertwined with nematic instabilities, which uniquely govern interband couplings and orbital-dependent pseudogap formations above $T_c$ \cite{hanaguri2019quantum, hashimoto2020bose, mizukami2023unusual}.

\subsection{Multicomponent superconductivity and time-reversal-symmetry-breaking}\label{trsb}

\subsubsection{Introduction}
TRSB superconductivity arises when the condensate contains at least two components whose relative phase is neither 0 nor $\pi$, producing a complex superconducting order parameter and tiny spontaneous internal fields. In multiband FeSCs, near-degeneracy between $s_{\pm}$ and $d$-wave pairing instabilities, or two inequivalent $s$ states across bands, provides a natural route to such complex states ($s+id$, $s+is'$) \cite{kreisel2020on}. 
Electronic nematicity could mix symmetries and permit a secondary symmetry-breaking transition below $T_c$ into a TRSB phase \cite{ghosh2025elastocaloric,kang2018time}. Detecting these states requires highly sensitive probes (ZF-$\mu$SR, polar Kerr, elastocaloric, and precision calorimetry). For FeSe, specifically, a microscopic theory shows how near-degenerate $s$ and $d$ channels in a sign-changing nematic background naturally enable an $s{+}e^{i\alpha}d$ transition at $T^{\ast}<T_c$ \cite{kang2018time}.

\subsubsection{Iron chalcogenides}
In {\FSS}, the nematic quantum criticality near $x_{c}\approx0.17$ was suggested to divide the superconducting dome into SC1 and SC2 phases with distinct superconducting gap structures, which was supported by a recent STM study as shown in Fig.~\ref{Fig_FSS19} \cite{nag2025highly}.
In the parent FeSe, local spectroscopic evidence for TRSB at the twin boundaries in FeSe-consistent with a complex $s+id$ order nucleated by the nematic texture-was earlier reported by STM \cite{watashige2015evidence}.
Subsequently, zero-field $\mu$SR on {\FSS} reported signatures consistent with TRSB below $T_c$ on \emph{both} sides of the nematic end point ($x_{c}\approx0.17$), with enhanced muon relaxation (internal fields $\sim10^{-1}$ G) and a markedly reduced superfluid density on the tetragonal side-consistent with an ultranodal state with Bogoliubov Fermi surfaces (BFSs) \cite{matsuura2023two}. In this study, the residual zero-energy density of states on the tetragonal side aligned with theoretical proposals for ultranodal $s$-wave states with BFSs in multiband systems \cite{setty2020topological}. 

In {\FST}, combined polar Kerr and ZF-$\mu$SR measurements locate a \emph{bulk} TRSB phase at $x\approx0.64$ (near maximal $T_c\sim14.5$~K), with onset slightly below $T_c$ and no static magnetism - placing TRSB in a tetragonal, topological regime in which the superconducting order gaps the topological surface state \cite{roppongi2025topology}. Smaller TRSB signals also appear in the nematic regime ($x\approx0.35$). Taken together with {\FSS}, these results show that TRSB is not confined to a single compositional corridor: it emerges in a nematic-hosted SC (SC1), and again in a tetragonal, topological SC (SC2) where multicomponent pairing and spin--orbit coupling can stabilize ultranodal behavior \cite{matsuura2023two,setty2020topological}.
On the theory side, FeSe's sign-changing nematic order linearly mixes $s$ and $d$ at $T_c$, and a weaker bilinear coupling permits a low-$T$ transition to $s{+}e^{i\alpha}d$ (TRSB) \cite{kang2018time}, providing a unified lens on SC1 and on multicomponent TRSB states compatible with ultranodal responses in the tetragonal regime.

\subsubsection{Iron pnictides}\label{sec_trsb_pn}
In overdoped {\BKFA} ($x\approx0.8$), a \emph{precursor} phase with broken TRS appears \emph{above} the superconducting transition: ZF-$\mu$SR and a spontaneous Nernst signal mark a distinct $Z_2$ transition at $T_c^{Z_2}>T_c$, consistent with a bosonic/``quartic-metal'' state or a frustrated $s+is'$ scenario characterized by interband phase ordering without global $U(1)$ coherence \cite{grinenko2021state}. High-resolution calorimetry subsequently resolved two distinct anomalies that track the $Z_2$ (higher $T$) and $U(1)$ (lower $T$) transitions at zero field, corroborating the separation of TRSB and superconducting onsets in a narrow overdoped dome \cite{shipulin2023calorimetric}. These observations provide direct thermodynamic evidence for multicomponent order in {\BKFA} and establish a clean materials platform where $s{+}is'$-type TRSB could be stabilized by near-degenerate interband $s$-wave channels. This overdoped regime is close to the KFe$_2$As$_2$ end member, where the reconstructed Fermi surface, incommensurate spin resonance, and anisotropic gap structure provide complementary constraints on the competing pairing channels discussed in Sec.~\ref{sec_swave_pn}.

For electron-doped {\BFCA}, as discussed in Sec. \ref{sec_strain_tc}, elastocaloric-effect (ECE) measurements uncovered a \emph{second} thermodynamic transition just below $T_c$ when superconductivity condenses \emph{inside} the nematic phase, while only a single transition is observed on the overdoped, tetragonal side. Magnetism and re-entrant tetragonality were excluded by x-ray and ultrasound, pointing to a multicomponent superconducting state stabilized by nematicity-consistent with an $s{+}e^{i\phi}d$ TRSB scenario awaiting direct ZF-$\mu$SR/Kerr confirmation \cite{ghosh2025elastocaloric}. The ECE line shape, its strain derivatives, and an Ehrenfest analysis collectively support a continuous thermodynamic transition within the superconducting state, with a very small associated entropy change (explaining the absence of a clear specific-heat anomaly at $T^{\ast}$).

\subsubsection{Summary and perspective}
Across chalcogenides and pnictides, TRSB emerges where competing pairing channels are delicately balanced-by nematic order [FeSe, {\BFCA}] or by band-structure evolution and Lifshitz physics ({\BKFA}, {\FST}). The spontaneous internal fields and entropic signatures are intrinsically small (sub-gauss, few-percent heat-capacity fractions), underscoring the need for combined ZF-$\mu$SR/Kerr, elastocaloric, and high-resolution calorimetry, ideally under tunable uniaxial stress/strain to control nematicity and train domains. Priority questions now include: (i) phase-sensitive discrimination between $s{+}id$ and $s{+}is'$; (ii) direct imaging and control of TRSB domains and associated edge currents; (iii) quantitative tests of ultranodal/Bogoliubov-Fermi-surface phenomenology in {\FSS} and {\FST} \cite{setty2020topological}; and (iv) exploiting the tetragonal, bulk-TRSB phase in {\FST} to gap topological surface states and engineer chiral Majorana modes \cite{roppongi2025topology}. Establishing the symmetry and topology of these TRSB condensates with phase-sensitive probes and surface spectroscopy will be central to integrating multicomponent FeSCs into the broader landscape of topological and chiral superconductivity.

\section{Antiferromagnetic spin excitations}\label{sec_afs}

\subsection{General introduction}

AFM spin excitations are widely discussed as a leading
candidate for the pairing interaction. Neutron scattering has been central to establishing how magnetic correlations evolve across the FeSC families and how they correlate with superconductivity. The magnetic spectral function is quantified by the $\mathbf{Q}$- and $E(=\hbar\omega)$-dependent imaginary part of the
dynamical susceptibility $\chi''(\mathbf{Q},\omega)$, related to the measured magnetic scattering function by
\begin{equation}
	S(\mathbf{Q},\omega)=\frac{1}{1-\mathrm{e}^{-\hbar\omega/k_B T}}\;\chi''(\mathbf{Q},\omega).
\end{equation}
Parent compounds of iron pnictides typically host stripe-type AFM order at $\mathbf{Q}_{\mathrm{AFM}}=(1,0)$, with spin-wave bandwidths extending to a few hundred meV. Upon carrier doping or isovalent substitution the static order is suppressed, but sizeable paramagnons persist, redistributing spectral weight from coherent magnons into broader continua while retaining strong intensity near $\mathbf{Q}_{\mathrm{AFM}}$. A recurring theme is the intertwining of spin, orbital, and lattice degrees of freedom: orthorhombic/nematic distortions lift the $d_{xz}/d_{yz}$ degeneracy and feed back on the spin anisotropy and wave-vector selection, while Hund's-coupling-driven correlations yield orbital-selective coherence that sculpts $\chi''$ in both energy and momentum. The $\mathbf{Q},\omega$-integrated fluctuating moment, $\langle m^2\rangle \propto \int d\omega\,d\mathbf{Q}\,\chi''$, remains large across wide doping ranges, underscoring the robustness of local-moment physics even in an itinerant setting~\cite{dai2015antiferromagnetic}. The situation becomes more complex in iron chalcogenides and germanides, where stripe magnetism competes with other instabilities such as double stripe, checkerboard, block-type, or ferromagnetic correlations. These competing tendencies, often tuned by subtle changes in lattice structure, chemical composition, or external parameters, generate a rich landscape of magnetic excitations. Understanding how these intertwined spin, orbital, and lattice interactions evolve across material families remains essential for constructing a comprehensive picture of iron-based superconductivity.

\subsection{Superconducting resonance}\label{resonance}

A hallmark of unconventional pairing in the FeSCs is the superconductivity-induced enhancement of $\chi''$ at $\mathbf{Q}_{\mathrm{AFM}}$ and energy $E_r<2\Delta$, the spin resonance detected by INS. Within a spin-exciton picture, a sign change of the superconducting gap between Fermi-surface regions connected by $\mathbf{Q}_{\mathrm{AFM}}$ produces a bound state below the particle--hole continuum, yielding a sharp peak that turns on below $T_c$. Empirically, $E_r$ scales with $T_c$ and/or the superconducting gap, $E_r \sim 5\,k_B T_c$ or $E_r \sim 0.6\times 2\Delta$. The resonance often disperses along $L$, reflecting three-dimensional exchange pathways, and may split into multiple components in systems with multigap superconductivity or strong spin-space anisotropy~\cite{dai2015antiferromagnetic}.

Over the past decade, neutron scattering has mapped the superconducting resonance across a broad landscape beyond the 122 pnictides, encompassing new iron-pnictide structure types as well as (intercalated) iron selenides~\cite{xie2018neutron,xie2018odd,hong2020neutron,wang2016strong,pan2017structure,ma2017prominent,ma2017lowenergy,hu2021polarized,hu2016spin,iida2019coexisting,zhang2016electron,song2016spin,wang2017orbital,iida2017spin,guo2019preferred,adroja2020observation,hong2023interlayer,tan2017phase,zhang2018neutron,li2025neutron}. Across these systems the resonant response exhibits a rich dynamical structure factor $S(\mathbf{Q},\omega)$, including incommensurability, orthogonal in-plane splittings, pronounced $L$-dispersion with odd/even selection linked to  interlayer coupling, and multiple resonance modes, while polarized neutron scattering
measurements resolve strong spin-space anisotropy between in-plane and $c$-axis channels. Symmetry analysis of detwinned superconductors further shows that the resonance inherits  the nematic $C_2$ anisotropy and couples selectively to bands of dominant $d_{xz}/d_{yz}$ or $d_{xy}$ character, supporting a scenario of spin-fluctuation-mediated, orbital-selective Cooper pairing~\cite{chen2019anisotropic,tian2019spin,liu2024low}.

Beyond the originally proposed resonance arising from scattering between nested electron and hole Fermi-surface sheets at the stripe wave vector $\mathbf{Q}_{\mathrm{AFM}}$, resonant modes have also been observed in extremely  electron- or hole-doped systems where one set of pockets is absent or incipient. In HED materials dominated by electron pockets, the resonance shifts to wave vectors connecting electron pockets (often near orthorhombic $(1, 1)$), and can split or become incommensurate due to pocket ellipticity and hybridization. Conversely, in heavily hole-doped compounds lacking electron pockets, a resonance emerges at wave vectors connecting hole sheets (near $(1, 0)$), frequently with marked incommensurability.  These observations suggest that the resonance is a robust fingerprint of sign-changing superconductivity whose characteristic wave vector is set by the actual low energy scattering channels provided by the reconstructed or doped Fermi surface, rather than by a single universal nesting condition.

\begin{figure}[t]
	\includegraphics[width=8cm]{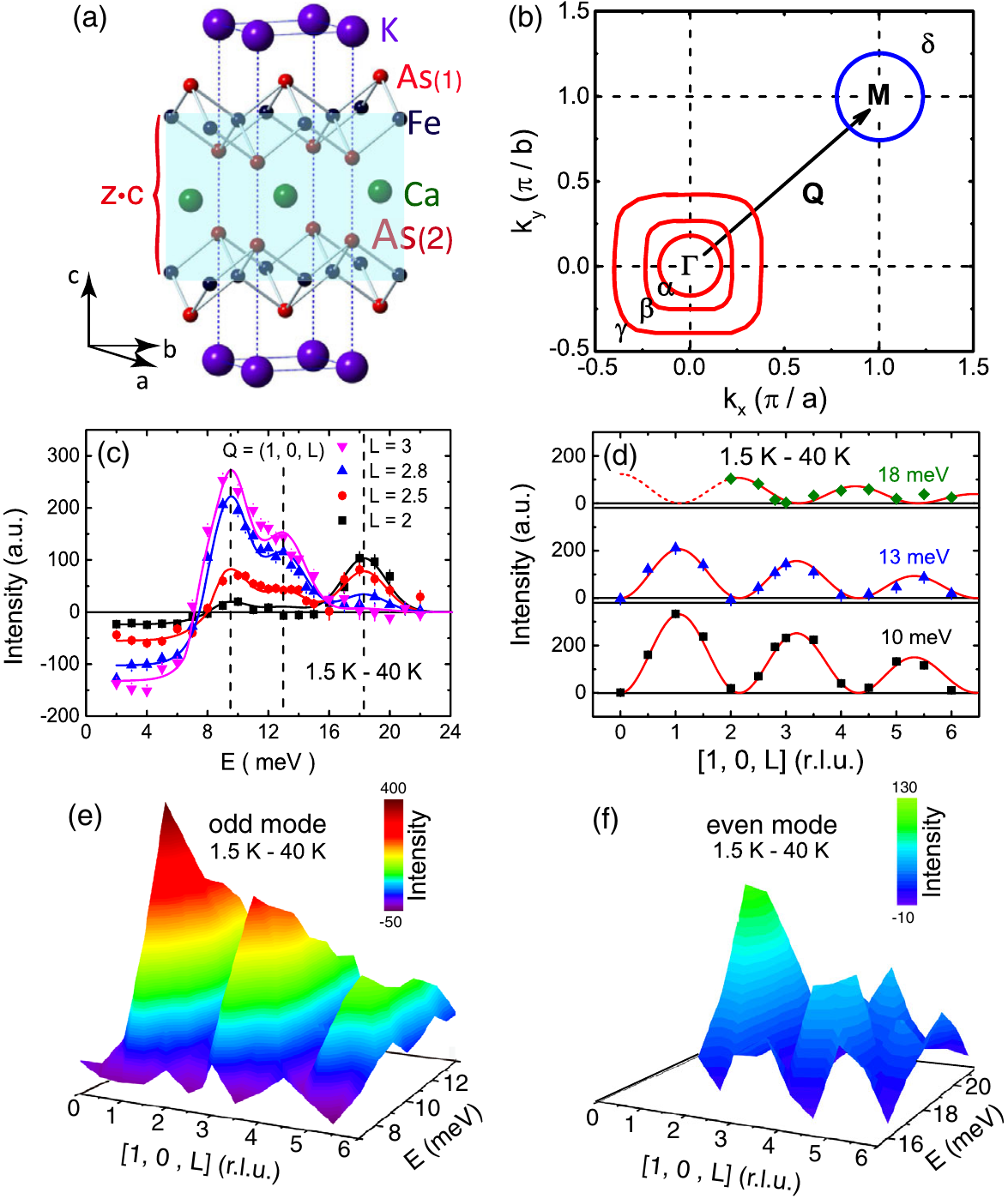}
	\caption{
		(Color online)
		Odd and even spin-resonance modes in bilayer compound CaKFe$_4$As$_4$.
		(a) Crystal structure of CaKFe$_4$As$_4$.
		(b) Schematic two-dimensional Fermi surfaces; the nesting vector $\mathbf{Q}$ connects the $\Gamma$ and $M$ points.
		(c) Energy dependence of the spin-resonance intensity at $\mathbf{Q}=(1,0,L)$; solid curves are guides to the eye.
		(d) $L$ dependence of the three resonance modes at $E=10$, 13, and 18~meV. Red solid and dashed curves are fits to
		$\lvert F(\mathbf{Q})\rvert^{2}\sin^{2}(z\pi L)$ or $\lvert F(\mathbf{Q})\rvert^{2}\cos^{2}(z\pi L)$, respectively.
		(e), (f) Modulation along $L$ of the odd-resonance branches and the even-resonance branch in different energy windows. Adapted from \cite{xie2018odd}.
		\label{fig:oddevenmode}
	}
\end{figure}

In the FePn-112 superconductor Ca$_{0.82}$La$_{0.18}$Fe$_{0.96}$Ni$_{0.04}$As$_2$ ($T_c$ = 33.5~K) with a unique monoclinic structure and zigzag arsenic chains, INS measurements revealed a quasi-two-dimensional spin resonance that is isotropic in spin space and shows negligible dependence on the out-of-plane momentum transfer $L$. These characteristics are consistent with a spin-exciton excitation and support an $s_\pm$ pairing symmetry, despite the material's three-dimensional electronic structure \cite{xie2018neutron}. 

By contrast, bilayer compound CaKFe$_4$As$_4$ ($T_c$ = 35~K) [Fig.~\ref{fig:oddevenmode}(a)] displays three distinct resonance modes with strong $L$-modulation, classified into odd and even symmetries similar to bilayer cuprates [Fig.~\ref{fig:oddevenmode}(c)--(f)]. The resonance energies scale with the total superconducting gap on different Fermi surface sheets [Fig.~\ref{fig:oddevenmode}(b)], indicating multiband superconductivity within the $s_\pm$ framework \cite{xie2018odd}. More recently, it has been proposed that in iron pnictides, the $L$ dependence of both the resonance energy and its intensity universally scales with the interlayer spacing between adjacent FeAs planes, and that the $L$ modulation of the superconducting gap is anticorrelated with this spacing~\cite{hong2023interlayer}.

Polarized INS on CaKFe$_4$As$_4$ further reveals that the high-energy even mode ($\sim$18 meV) is isotropic in spin space, whereas the low-energy odd modes consist of a $c$-axis-polarized branch near 9 meV and an in-plane branch near 12 meV~\cite{xie2020spinexcitation}. This spin anisotropy is attributed to spin--orbit coupling. A similar $c$-axis-polarized low-energy resonance has also been observed in the Ni-doped compound CaK(Fe$_{0.96}$Ni$_{0.04}$)$_4$As$_4$, in which superconductivity coexists with spin-vortex crystal magnetic order \cite{liu2022preferred}. Another related bilayer material, EuRbFe$_4$As$_4$ ($T_c$ = 36.5~K), shows the coexistence of spin resonance and static magnetic order from the Eu sublattice~\cite{iida2019coexisting}; however, the coupling between superconductivity and magnetic order in this system appears to be relatively weak.

\begin{figure}[t]
	\includegraphics[width=8cm]{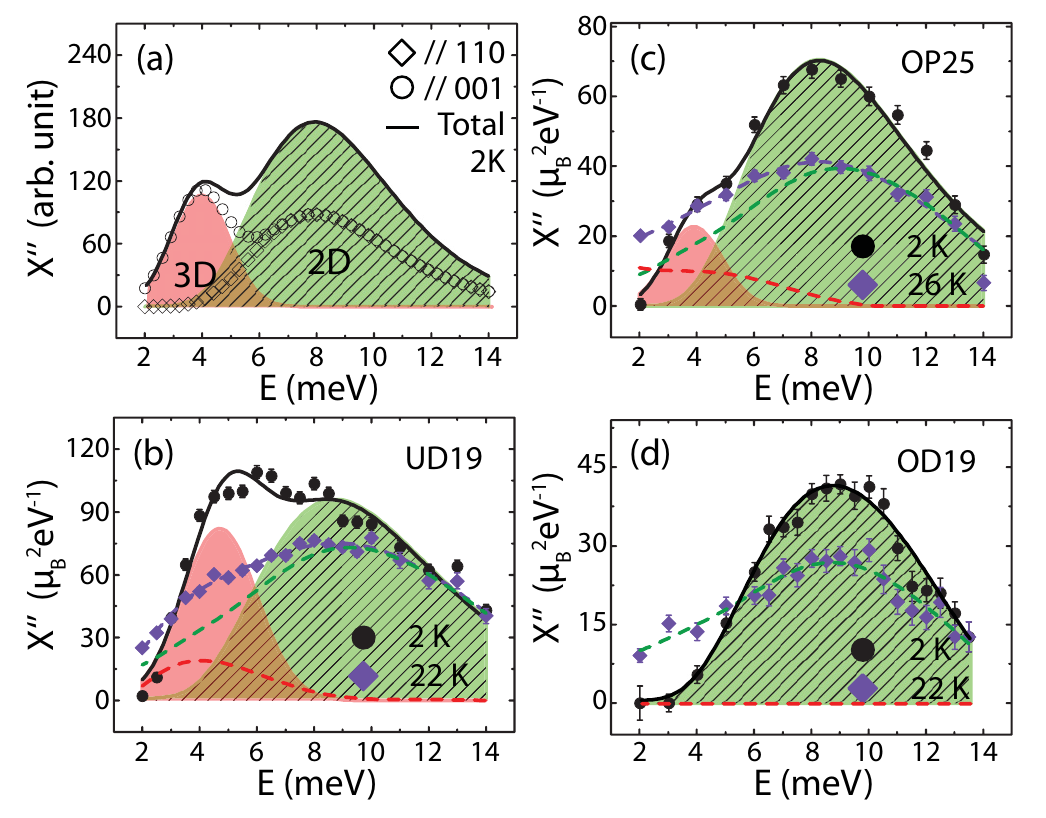}
	\caption{
		(Color online) Doping evolution of the double spin-resonance modes in Ba(Fe$_{1-x}$Co$_x$)$_2$As$_2$.
		(a) Polarized INS result for $\chi^{\prime\prime}(\omega,\mathbf{Q})$ at $\mathbf{Q}=(0.5,0.5,1)$ in the $x=0.06$ ($T_c=24$~K) crystal, reproduced from Ref.~\onlinecite{steffens2013splitting}. Red shading marks the $c$-axis-polarized (out-of-plane) component (anisotropic part); hatched green shading indicates the component present in both in-plane and out-of-plane channels (isotropic part).
		(b)--(d) Normalized $\chi^{\prime\prime}(\omega,\mathbf{Q})$ at $\mathbf{Q}=(0.5,0.5,1)$ for underdoped UD19 ($T_c=19$~K), optimally doped OP25 ($T_c=25$~K), and overdoped OD19 ($T_c=19$~K), measured by unpolarized INS in the superconducting state at 2~K (black) and in the normal state just above $T_c$ (violet). In the superconducting state, red and hatched green shaded areas denote fits to the two resonance components as in panel (a); the black solid curve is the sum of the two fitted peaks. In the normal state, red and olive dashed curves represent fits to the 3D and 2D components of the spin spectrum, respectively, and the violet dashed curve is their sum. Adapted from \cite{wang2016experimental}.
		\label{fig:doublemode}
	}
\end{figure}

Double spin-resonance features have also been reported in iron pnictides without bilayer structures, particularly in the underdoped and optimally doped regimes. Notable examples include electron-doped  Ba[Fe$_{1-x}$(Ni,Co)$_x$]$_2$As$_2$~\cite{lipscombe2010anisotropic,steffens2013splitting,wasser2017anisotropic}, NaFe$_{1-x}$Co$_x$As~\cite{zhang2013measurement,zhang2014anisotropic,zhang2015neutron,zhang2016electron,song2017temperature}, and hole-doped Ba$_{1-x}$K$_x$Fe$_2$As$_2$ ~\cite{song2016spin} and Ba$_{1-x}$Na$_x$Fe$_2$As$_2$~\cite{Waer2019strong}. A comprehensive study by Wang \textit{et al.}~\cite{wang2016experimental} combined INS and ARPES across a broad range of doping levels in Ba(Fe$_{1-x}$Co$_x$)$_2$As$_2$. In the underdoped and optimally doped regimes, two distinct energy scales were resolved: a three-dimensional, magnon-like mode and a two-dimensional mode [Fig.~\ref{fig:doublemode}]. With increasing doping, the three-dimensional mode is progressively weakened, while the two-dimensional mode remains robust and persists into the overdoped regime. The energy of the two-dimensional resonance scales with the superconducting gap and is isotropic in spin space, whereas the low-energy three-dimensional mode is anisotropic and preferentially polarized along the $c$-axis~\cite{steffens2013splitting}. These observations were interpreted as evidence that the double resonance arises from a coexistence of a residual magnon mode and an emergent spin exciton. 

A similar doping evolution of the double resonance has been reported in NaFe$_{1-x}$Co$_x$As~\cite{zhang2013measurement,zhang2014anisotropic,zhang2015neutron,zhang2016electron,song2017temperature}. Under an in-plane magnetic field of $H=12$~T, where superconductivity is only modestly suppressed, the two resonance modes exhibit different responses: the higher-energy mode shows a slight reduction in intensity, whereas the lower-energy mode is nearly fully suppressed~\cite{song2018unusual}. These contrasting behaviors may reflect differences in spin polarization (in-plane versus $c$-axis) and/or in the interband scattering channels shaped by orbital character and gap anisotropy.

Viewed collectively, these pnictide studies establish two recurring motifs: (i) a coexistence of a quasi-two-dimensional spin-exciton with remnants of modes inherited from proximate antiferromagnetism, and (ii) pronounced spin-space anisotropy shaped by spin--orbit coupling and orbital texture.
Compared with the iron pnictides, the iron chalcogenides display distinct parent-state properties. FeSe, for instance, does not exhibit long-range AFM order but instead develops pronounced nematicity and becomes superconducting below $T_c \approx 8.7$~K. These unusual characteristics make FeSe an important platform for investigating how nematicity and spin--orbit coupling shape spin excitations and their interplay with superconductivity. 

In FeSe, a sharp spin resonance has been detected at $\sim$3.5 meV at the stripe wave vector associated with hole--electron Fermi surface nesting \cite{wang2016strong, hu2022lowenergy,chen2020anisotropic,liu2024low}. Polarized neutron scattering further revealed a strong spin-space anisotropy, with fluctuations predominantly polarized along the $c$-axis. This anisotropy was interpreted as evidence of substantial spin--orbit coupling effects \cite{ma2017prominent}. Further studies on detwinned FeSe crystals have shown that spin fluctuations at the resonance energy exhibit $C_2$ symmetry, closely associated with nematicity and orbital-selective pairing \cite{chen2019anisotropic,liu2024low}. Similar symmetry-breaking phenomena have also been observed in detwinned Ba(Fe$_{1-x}$Co$_x$)$_2$As$_2$ \cite{tian2019spin} and NaFe$_{1-x}$Co$_x$As \cite{wang2017orbital}, underscoring the crucial role of electronic nematicity in shaping spin excitations and their interplay with superconductivity (see Sec. \ref{sec_ossf}).

Resonance modes in FeSCs were initially proposed to appear 
near the nesting wave vector connecting the hole pockets at the BZ center 
and the electron pockets at the zone corners, highlighting the importance of both 
types of Fermi surfaces for the superconducting pairing mechanism. Later, resonance modes have also been observed in FeSCs with extreme Fermi surface reconstructions, such as 
$A_x$Fe$_{2-y}$Se$_2$ ($A$ = K, Rb, ...), which exhibit unusual electronic and 
magnetic properties~\cite{park2011magnetic,dai2015antiferromagnetic,
	taylor2012spin,friemel2012conformity}. 

A notable example is the intercalated iron selenide superconductor 
Li$_{1-x}$Fe$_x$OHFeSe ($T_c = 41$~K), which lacks hole pockets at the 
BZ center and can thus be regarded as a heavily electron-doped system. In contrast to $A_x$Fe$_{2-y}$Se$_2$, which generally exhibits phase separation between superconducting regions and AFM insulating $A_2$Fe$_4$Se$_5$, Li$_{1-x}$Fe$_x$OHFeSe is a phase-pure superconductor. In this material, a resonance mode emerges at an energy of approximately 21~meV at an incommensurate wave vector $(1, 0.62)$, connecting electron Fermi pockets near the BZ corners~\cite{pan2017structure,ma2017lowenergy,davies2016spin}. Polarized neutron scattering measurements further reveal that this mode is nearly isotropic in spin space, consistent with an isotropic spin-exciton origin~\cite{hu2021polarized}. These observations are consistent with a sign-reversing superconducting gap between electron pockets, thereby extending the spin-fluctuation-mediated pairing scenario to the regime of extreme electron doping.

Interestingly, analogous phenomena are also found at the opposite end of the doping spectrum. In the heavily hole-doped iron pnictide KFe$_2$As$_2$, which lacks electron pockets~\cite{shen2020neutron}, a resonance mode emerges at an incommensurate wave vector near $(1\pm0.4, 0)$, suggesting a sign reversal of the superconducting gap between hole pockets~\cite{shen2020neutron,wu2024nodal}. Together, these findings provide evidence that spin-resonance modes can arise from sign-changing gap structures even in the absence of conventional electron--hole nesting, underscoring the robustness of spin-fluctuation-mediated pairing across a wide range of Fermi surface topologies in FeSCs.

It is noteworthy that certain resonance-like excitations exhibit behavior deviating from the conventional picture of a simple spin exciton. In KCa$_2$Fe$_4$As$_4$F$_2$ ($T_c$ = 33.5~K), a downward-dispersing two-dimensional resonance mode persists above the maximum superconducting gap energy, contrary to typical spin-exciton expectations confined below the particle-hole continuum \cite{hong2020neutron}. Likewise, a resonance-like feature observed above the superconducting gap in heavily sulfur-doped K$_x$Fe$_{2-y}$Se$_{2-z}$S$_z$ hints at deviations from the pairing symmetry of K$_x$Fe$_{2-y}$Se$_2$ without sulfur doping, implying modifications to the underlying pairing state \cite{wang2016transition}.

Collectively, the body of work above establishes the spin resonance as a robust marker of  unconventional superconductivity whose detailed fingerprints---energy $E_r$, $\mathbf{Q}$ location, layer parity along $L$, and spin polarization---encode material-specific ingredients such as Fermi-surface topology, interlayer coupling, spin--orbit coupling, and nematic order. Moving forward, several questions remain open. First, a quantitative census of absolute spectral weight across families and dopings is needed to benchmark theories and to disentangle the relative contributions of magnon-like remnants and the superconducting exciton to ``double'' resonances. Second, systematic polarized and detwinned measurements under controlled tuning parameters (magnetic field, uniaxial strain, and pressure) can isolate the roles of spin--orbit coupling and nematicity, test odd/even $L$ parity, and track how $E_r/2\Delta$ and polarization anisotropy evolve with dimensionality (bulk and intercalated systems). Third, on the theory side, multiorbital calculations that incorporate realistic band structures, spin--orbit coupling, interlayer matrix elements, and beyond-RPA vertex corrections are required to address deviations from a simple spin-exciton picture (e.g., magnon-exciton hybridization, coupling to Bardasis-Schrieffer or Leggett modes, and effects of incipient bands). Establishing these systematics will clarify when and how the resonance primarily reflects superconducting coherence factors versus proximate magnetism, and will sharpen its use as a quantitative probe of the pairing kernel in FeSCs.

\subsection{High-energy spin excitations in FeSCs}\label{sec_highE_spin}

Over the past decade, advances in the growth of large, high-quality single crystals have been pivotal for comprehensive INS investigations of FeSCs. Such samples have enabled systematic mapping of spin excitations over wide momentum and energy ranges, thereby uncovering their intricate connections to lattice symmetry, electronic degrees of freedom, and unconventional superconductivity. In iron pnictides, for example, well-defined stripe-type, spin-wave-like excitations are routinely observed~\cite{hu2016spin,carr2016electron,li2016orbital,xie2022spina,kim2015spin,sapkota2018doping,murai2018effect,horigane2016spin}. The interplay between stripe-type AFM correlations and nematicity has been systematically investigated across different iron pnictide families~\cite{sapkota2018doping,zhang2016effect,zhang2015neutron,tan2016electron}, providing a relatively coherent picture of how magnetism and nematicity are intertwined in these materials.  

At the same time, high-energy spin excitations display substantial material dependence across the broader Fe-based family, reflecting differences in crystal structure, orbital occupancy, magnetic frustration, dimensionality, and carrier concentration. To highlight both the common stripe-related phenomenology and the material-specific deviations from it, we organize the following
discussion into iron pnictides, iron chalcogenides, and selected
related Fe-based systems.

\begin{figure}[htbp!]
	\includegraphics[width=7cm]{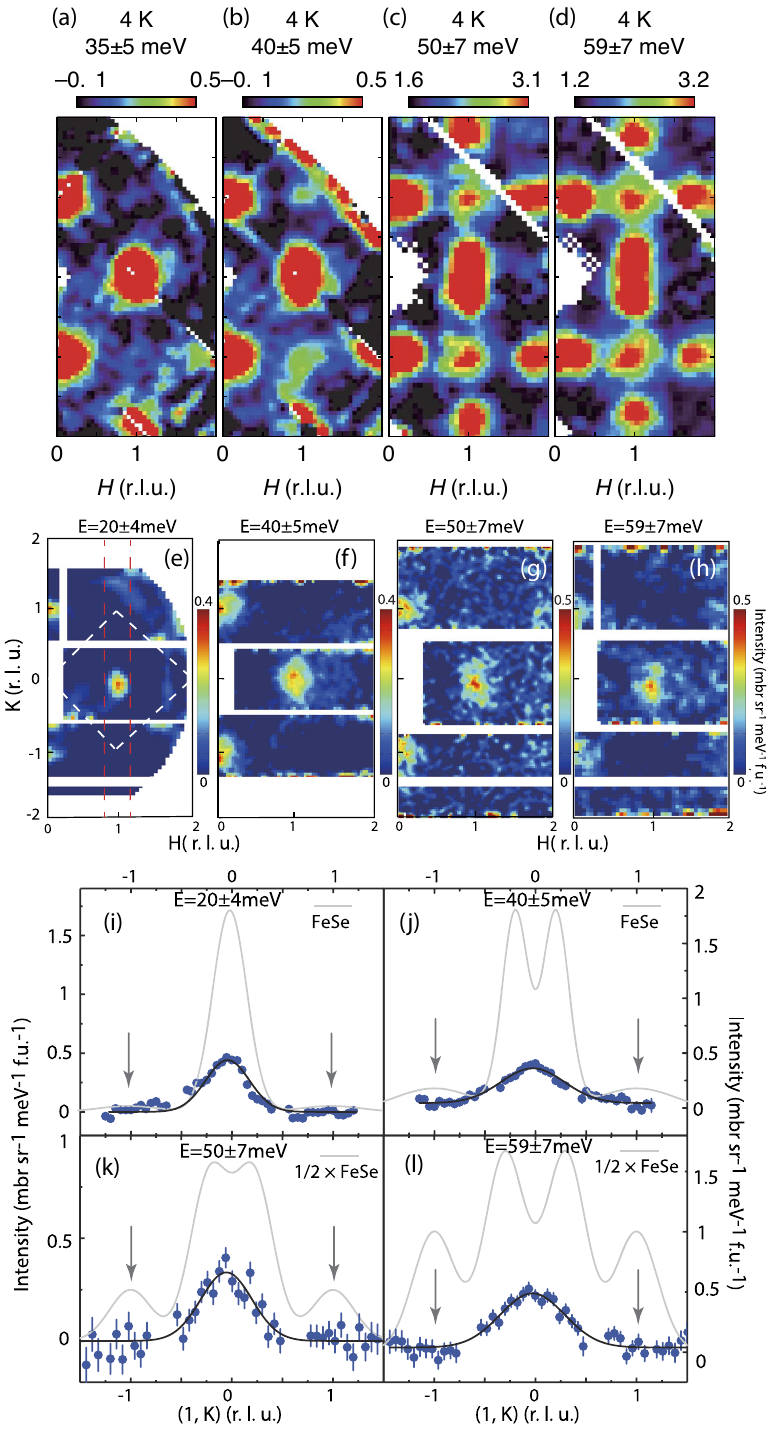}
	\caption{(Color online)
		Spin excitations in FeSe and FeS measured by time-of-flight neutron spectroscopy.
		(a)--(d) Constant-energy maps for FeSe at $T=4$~K; the energy transfer for each panel is indicated at the top.
		(e)--(h) Constant-energy maps for FeS at $T=4$~K with energies labeled similarly.
		In (e), red dashed outlines mark the reciprocal-space region integrated to obtain the one-dimensional cuts shown in (i)--(l).
		The white dashed rectangle denotes the area used to evaluate the local dynamical susceptibility $\chi^{\prime\prime}(E)$.
		(i)--(l) For FeS, constant-energy cuts along the $[1,K]$ direction through the stripe AFM wave vectors, taken at the same energies as in (e)--(h).
		Gray solid curves are fits constrained by the FeSe excitation profiles over the corresponding energy range~\cite{wang2016magnetic}.
		The gray arrow marks the N\'eel spin excitation observed in FeSe but not detected in FeS. Adapted from \cite{wang2016magnetic,man2017spin}.
		\label{fig:feses}
	}
\end{figure}

\subsubsection{Iron pnictides}

Beyond the canonical stripe-type magnetism, several related tetragonal magnetic structures have also been identified in iron pnictides. These include the out-of-plane collinear double-$\mathbf{Q}$ $C_4$ magnetic phase in hole-doped (Ca,Sr,Ba)$_{1-x}$Na$_x$Fe$_2$As$_2$ and Ba$_{1-x}$K$_x$Fe$_2$As$_2$ \cite{taddei2016detailed,taddei2017observation,allred2015tetragonal, allred2016double,avci2014magnetically}, and the spin-vortex-crystal phase in CaKFe$_4$As$_4$ \cite{meier2018hedgehog,kreyssig2018antiferromagnetic, ding2018hedgehog}. Similar magnetic states have also been reported in LaFeAs$_{1-x}$P$_x$O \cite{stadel2022multiple}. These orders are closely related to the stripe instability because they involve one or both of the symmetry-related stripe ordering wave vectors, rather than an unrelated magnetic ordering wave vector.

The relation between these tetragonal magnetic structures and
spin dynamics has been examined by INS in the $C_4$ magnetic phase of
Sr$_{1-x}$Na$_x$Fe$_2$As$_2$ \cite{guo2019preferred}. Across the
transition from the $C_2$ stripe phase to the $C_4$ phase, the
low-energy spin-excitation spectrum and its polarization are modified,
consistent with spin reorientation along the $c$ axis, whereas the
higher-energy magnetic spectrum remains broadly similar to that of the
stripe phase. This behavior supports the view that the double-$Q$ SDW
state remains closely connected to the competing stripe instabilities
at the two symmetry-related wave vectors.

A different route to modifying pnictide spin dynamics is
magnetic dilution. In NaFe$_{1-x}$Cu$_x$As near $x\approx0.5$, Fe--Cu
atomic ordering produces magnetic Fe chains separated by nonmagnetic Cu
chains, yielding an insulating quasi-one-dimensional magnetic system.
Recent INS measurements on NaFe$_{0.53}$Cu$_{0.47}$As resolved highly
anisotropic spin excitations with both stripe- and N\'eel-vector
contributions \cite{wang2026quasi}. The similarity of these excitations
to those in FeSe provides a useful bridge to the frustrated and
quasi-one-dimensional spin correlations discussed below for iron
chalcogenides.

\subsubsection{Iron chalcogenides}

We next turn to iron chalcogenides, where spin excitations display
stronger magnetic frustration and more pronounced material dependence
than in the canonical stripe-ordered iron pnictides. In FeSe-derived
systems, INS experiments reveal coexisting stripe- and N\'eel-type spin
fluctuations, substantial spectral-weight redistribution across the
nematic transition, and, in intercalated compounds, strongly
energy-dependent momentum structures
\cite{wang2016magnetic,man2017spin,pan2017structure,wo2025spin}.
These results highlight the roles of frustration, dimensionality, and
orbital selectivity in shaping the spin dynamics of FeCh-based
superconductors.

A paradigmatic example highlighting the intricate coupling between magnetic frustration, nematicity, and superconductivity is FeSe. Despite the absence of static magnetism, INS experiments reveal strong dynamic spin correlations, characterized by both stripe-type and N\'eel-type fluctuations, over an extended energy range up to $\sim$220 meV [Fig.~\ref{fig:feses}(a)--(d)]~\cite{wang2016magnetic}.

Upon cooling into the nematic phase, a pronounced redistribution of spectral weight occurs:
N\'eel-type spin fluctuations are substantially suppressed, while stripe-type correlations become notably enhanced.
This redistribution demonstrates a strong coupling between spin fluctuations and broken rotational symmetry.
Remarkably, the total fluctuating magnetic moment observed in FeSe is about 60\% larger than that of BaFe$_2$As$_2$, underscoring the significant persistence of magnetic fluctuations despite the lack of long-range magnetic order.

Moreover, replacing Se with S markedly reduces electron correlations, resulting in a pronounced suppression of spin fluctuations and a complete suppression of nematic order [Fig.~\ref{fig:feses}(e)--(l)]. This behavior underscores the strong sensitivity of these intertwined phenomena to subtle electronic tuning achieved through chemical substitution~\cite{man2017spin}.

The momentum-dependent equal-time spin correlation function, $\mathcal{S}(\mathbf{Q})$, in FeSe can be quantitatively captured by self-consistent Gaussian approximation calculations within a frustrated spin model. In this scenario, competing nearest-neighbor ($J_1$) and next-nearest-neighbor ($J_2$) exchange interactions dominate, while further-neighbor couplings remain weak. This strong magnetic frustration suppresses conventional long-range magnetic order, stabilizing instead a quantum-disordered ground state with nematic character~\cite{gu2022frustrated}. More broadly, theoretical analyses have suggested that such quantum-disordered nematic states may be a natural consequence of strong frustration~\cite{wang2015nematicity, yu2015antiferroquadrupolar, glasbrenner2015effect}, offering a unifying perspective on the absence of long-range magnetic order in FeSe.

\begin{figure}[htbp!]
	\includegraphics[width=8.0cm]{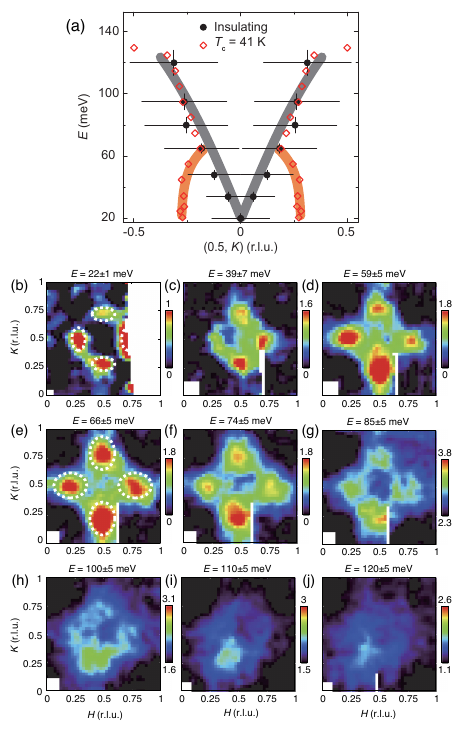}
	\caption{
		(Color online) (a) Dispersion of the spin excitations in superconducting and insulating Li$_{1-x}$Fe$_x$OHFeSe.  (b)--(j), Constant-energy images acquired at 5~K and at indicated energies for superconducting Li$_{1-x}$Fe$_x$OHFeSe ($T_c$ = 41~K). $|\bf{Q}|$-dependent background was subtracted in (b)--(f). For energies $\ge$ 85 meV, raw data are presented (g)--(j).  The measurements in (b) and (c)--(j) were conducted at the incident neutron energies of 49.6 and 191.6 meV, respectively. Symmetry equivalent data were collected and averaged to enhance statistical accuracy. The color bar indicates scattering intensity in units of mbarn sr$^{-1}$ meV$^{-1}$ f. u.$^{-1}$. Adapted from \cite{pan2017structure}.
	}
\label{fig:hourglass}
\end{figure}

A related yet distinct behavior is observed in the high-energy spin excitation spectrum of the intercalated iron selenide superconductor Li$_{1-x}$Fe$_x$OHFeSe, which has $T_c = 41$ K.  In this material, spin excitations exhibit a pronounced energy-dependent momentum structure distinct from FeSe [Fig.~\ref{fig:hourglass}(b)--(k)] \cite{pan2017structure}. Unlike FeSe, which hosts both stripe-type and Néel-type fluctuations, Li$_{1-x}$Fe$_x$OHFeSe shows primarily incommensurate magnetic resonant excitations surrounding the $(0.5, 0.5)$ point in 1-Fe reciprocal-lattice notation at low energies [Fig.~\ref{fig:hourglass}(b)]. As the energy increases, these spin excitations evolve into a diamond-shaped pattern [Fig.~\ref{fig:hourglass}(c)], dispersing outward up to approximately $60$~meV [Fig.~\ref{fig:hourglass}(d)--(e)] and subsequently dispersing inward at higher energies [Fig.~\ref{fig:hourglass}(f)--(k)], thereby forming a characteristic hourglass-shaped momentum-energy dispersion [Fig.~\ref{fig:hourglass}(a)]. This behavior is remarkably similar to spin excitations of cuprates \cite{Tranquada2004quantum,Hayden2004}.

Further insight comes from tuning the carrier concentration. In the insulating Li$_{1-x}$Fe$_x$OHFeSe with reduced electron density, the spin excitation spectrum adopts a simple V-shaped, spin-wave-like dispersion, in sharp contrast to the twisted hourglass pattern of the superconducting counterpart [Fig.~\ref{fig:hourglass}(a)] ~\cite{wo2025spin}. The evolution from a straightforward V-shaped to a more complex hourglass-like spectrum parallels similar trends seen in cuprates as they evolve from AFM insulators to superconductors. This similarity hints at a potentially universal relationship between magnetism and superconductivity that transcends material families, reinforcing the view that spin fluctuations may play an essential role in the pairing mechanism of unconventional superconductors.

\subsubsection{Related Fe-based systems}

\begin{figure}[htbp!]
	\includegraphics[width=8cm]{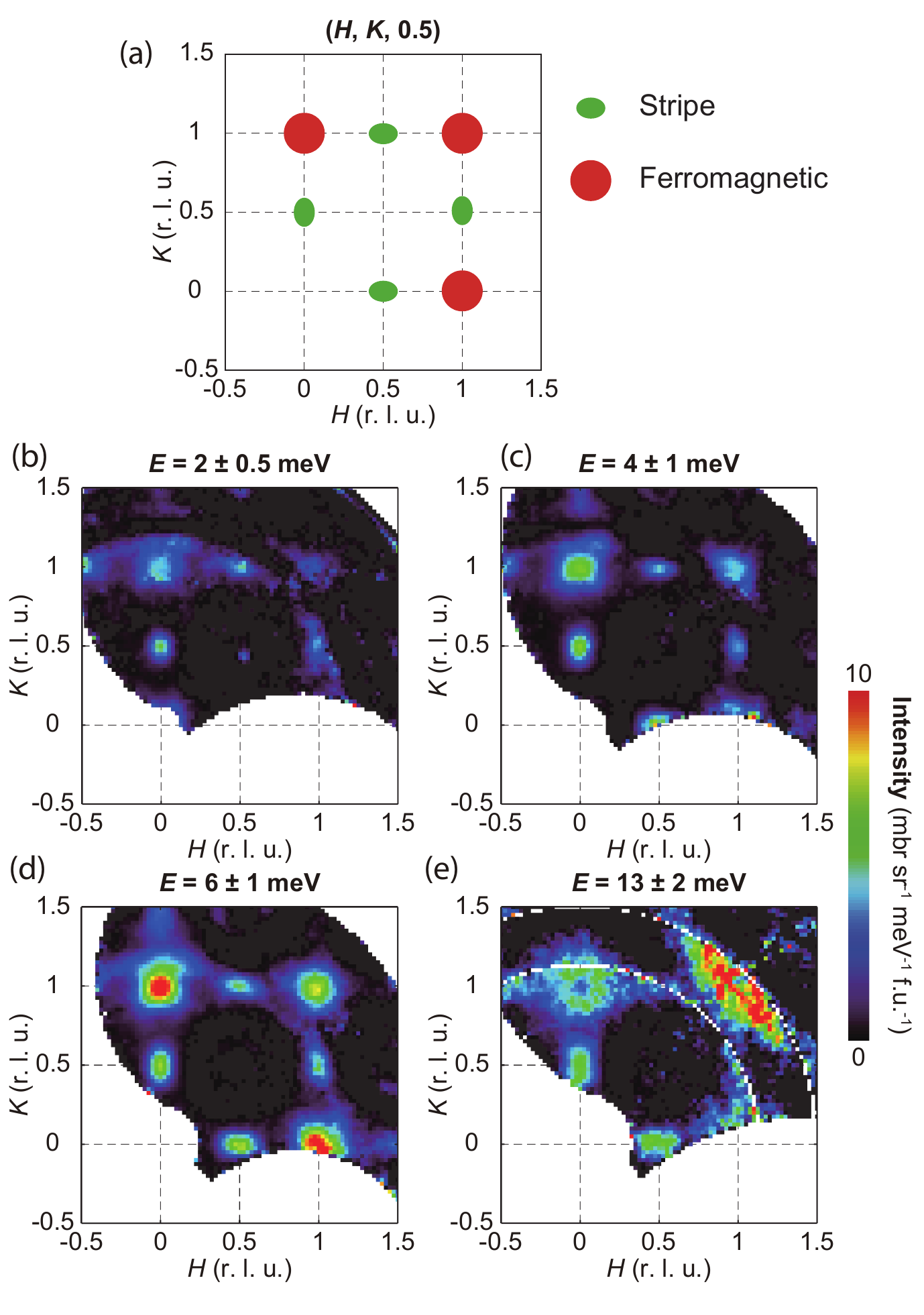}
	\caption{
		(Color online) Momentum dependence of the spin fluctuations in YFe$_2$Ge$_2$ at 4 K. (a) Schematic representation of the stripe and ferromagnetic spin fluctuations in the ($H$, $K$, 0.5) plane. (b)--(e) Constant-energy images at indicated energies. The $|$\textbf{Q}$|$-dependent background has been subtracted using a method similar to that described in Ref.~\onlinecite{wang2016magnetic}. The intensities are normalized to absolute units with acoustic phonons. Adapted from \cite{wo2019coexistence}.
	}
	 \label{fig:yfg}
	
\end{figure}

Beyond FePn and FeCh superconductors, related Fe-based compounds provide
additional examples of how competing magnetic channels shape the spin
response. The emergence of superconductivity in the iron--germanide compound YFe$_2$Ge$_2$ ($T_c \sim 1.8$~K) further broadens the landscape of spin excitations in iron-based materials \cite{zou2014fermi}. Although YFe$_2$Ge$_2$ shares the same crystal structure as the 122 iron pnictides, it lies outside both the pnictide and chalcogenide families. INS studies in the normal state reveal pronounced spin fluctuations at two distinct momentum vectors [Fig.~\ref{fig:yfg}] \cite{wo2019coexistence}. Similar to the iron pnictides, anisotropic stripe-type AFM fluctuations are observed at $(1, 0, 1)$. A key distinguishing feature of YFe$_2$Ge$_2$ is the presence of strong in-plane ferromagnetic fluctuations at ($0$, $0$, $1$), which are nearly isotropic in the ($H$, $K$) plane and, in fact, exceed the intensity of the stripe-type excitations. Both types of fluctuations remain gapless down to the lowest measured energies. The coexistence of robust stripe and ferromagnetic fluctuations provides a natural explanation for the lack of static magnetic order in YFe$_2$Ge$_2$ and highlights the potential importance of ferromagnetic correlations-an aspect often overlooked in the broader context of iron-based superconductors.  Theoretical models have proposed several unconventional pairing 
mechanisms for YFe$_2$Ge$_2$, including singlet pairing mediated by 
AFM fluctuations and triplet pairing linked to in-plane 
ferromagnetic correlations~\cite{singh2014superconductivity,subedi2014unconventional}. 
Future experiments elucidating the interplay between superconductivity 
and these coexisting spin fluctuations, as well as precise determinations 
of the superconducting gap structure, will be crucial for clarifying the 
pairing mechanism in this unique system. 

Related phenomenology has been observed in nonsuperconducting SrCo$_2$As$_2$, where ferromagnetic and stripe AFM spin fluctuations also coexist. In that case, Co substitution in SrFe$_{2-x}$Co$_x$As$_2$ is argued to drive a $t_{2g}$-to-$e_g$ orbital reconfiguration, promoting ferromagnetic fluctuations that are unfavorable for singlet pairing. This contrast underscores how the balance between stripe and ferromagnetic channels, shaped by orbital physics, can either support or suppress superconductivity across the iron-based family \cite{li2019coexistence}.

\subsection{Nematic spin correlations and orbital selectivity}\label{sec_uniaxial}

As discussed in Sec. \ref{sec_ossf}, growing evidence has suggested that spin dynamics in FeSCs are shaped by orbital selectivity. In this section, we discuss more generally orbital-selective and orbital-dependent magnetic excitations, most of which were measured on samples detwinned by uniaxial pressure or strain.
ARPES probes and theory indicate that $d_{xy}$ states tend to be most strongly correlated (heavier mass, smaller coherence scale), while $d_{xz}/d_{yz}$ states couple strongly to nematicity \cite{dai2015antiferromagnetic}. Neutron scattering, in combination with realistic multiorbital calculations, shows that these orbital textures imprint on $\chi''(\mathbf{Q},\omega)$: (i) the relative weight at $(1,0)$ vs.\ $(0,1)$ tracks the imbalance of $d_{xz}$ and $d_{yz}$ spectral weight; (ii) the damping and bandwidth of paramagnons reflect the coherence of the dominant orbital on the nested Fermi-surface sections; and (iii) the polarization anisotropy of the resonance follows the spin--orbit-coupling induced mixing between $d_{xy}$ and $d_{xz}/d_{yz}$.

An interesting example is provided by tetragonal LiFe$_{0.88}$Co$_{0.12}$As, a non-superconducting compound located near the edge of the superconducting dome in LiFe$_{1-x}$Co$_x$As, which exhibits non-Fermi-liquid behavior \cite{li2016orbital}. Neutron scattering measurements on this material reveal a pronounced orbital dependence of the spin dynamics. Unlike most iron pnictides, the spin excitations in LiFe$_{0.88}$Co$_{0.12}$As cannot be captured by an anisotropic Heisenberg Hamiltonian. Comparison with DFT + DMFT calculations indicates that the low-energy, wave-vector-correlated spin excitations are primarily associated with the $d_{xy}$ orbital, while the higher-energy excitations originate mainly from the $d_{yz}$ and $d_{xz}$ orbitals. However, as with most early INS measurements on FeSCs, these experiments were conducted on twinned samples, which inherently average out intrinsic anisotropies in the spin dynamics due to the random orientation of orthorhombic domains.

Recent methodological advances, particularly the application of uniaxial pressure or strain-based detwinning techniques, have enabled measurements on nearly fully detwinned crystals, resolving intrinsic anisotropic spin excitation dispersions over broad energy and momentum ranges in various parent and doped FeSCs~\cite{lu2014nematic,lu2018spin,tam2020orbital,chen2019anisotropic}. These studies have provided crucial insights into the microscopic origin of magnetic ground states, spin excitations, and their coupling to structural and electronic degrees of freedom.

\subsubsection{FePn-122 and FeTe}

\begin{figure}
\centering
\includegraphics[width=6cm]{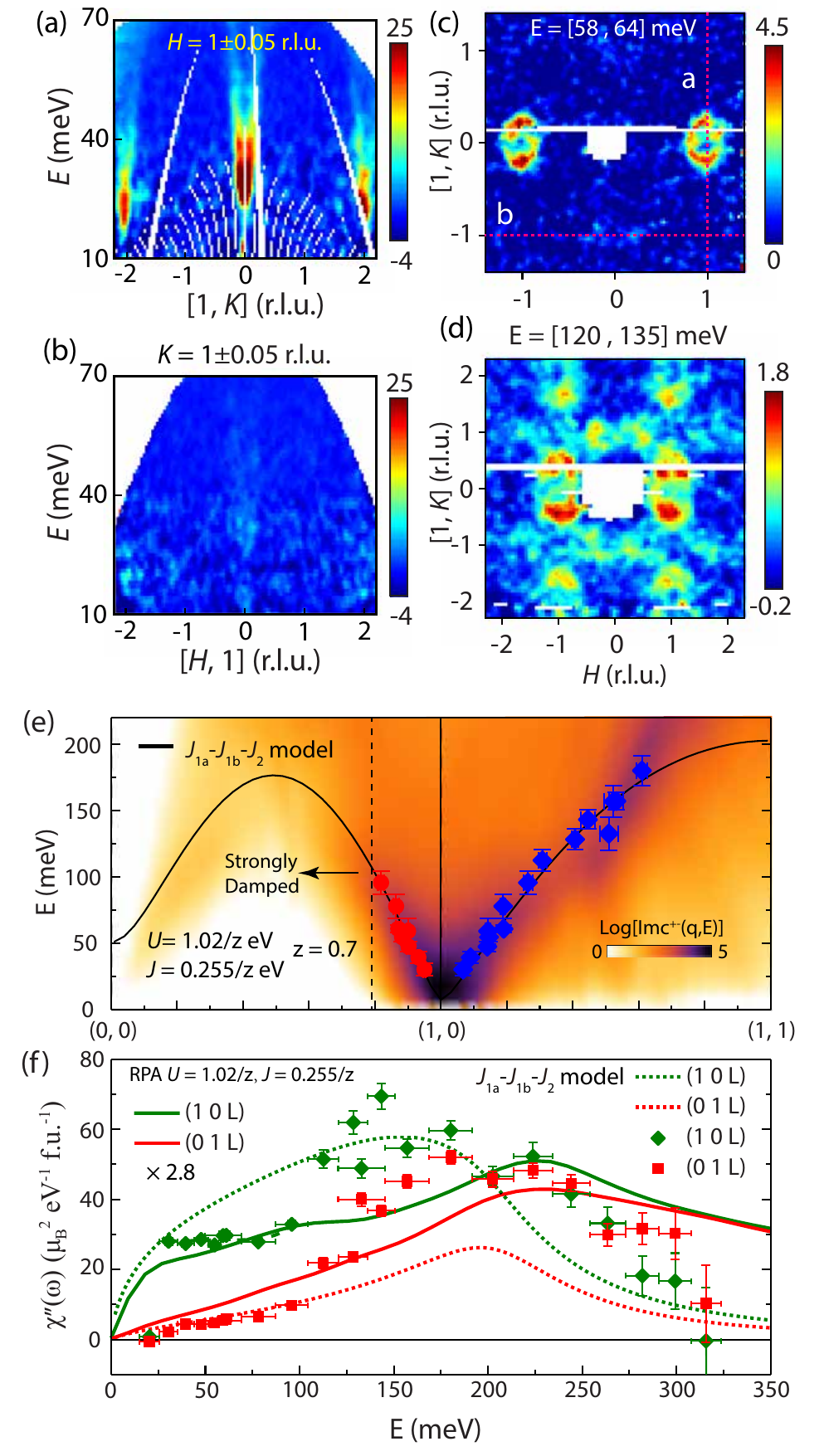}
\caption[]{Spin waves in detwinned {\BFA}.
(a)--(d) Projection of the magnetic scattering intensity onto energy and momentum planes. (a), (b) Magnon dispersions along the (a) $[1, K]$ and (b) $[H, 1]$ directions (red dashed lines in (c)). (c), (d) Constant energy slices in the $[H, K]$ plane. (e) Spin-wave dispersions of a detwinned {\BFA} extracted from constant-energy cuts collected at $T=7$ K. The black curves are obtained from a Heisenberg model ($J_{1a}-J_{1b}-J_2$) fit of twinned {\BFA}. The background shows the spectral weight from the RPA calculation (renormalized with $z=0.7$) with $U=1.02/z$ eV and $J=U/4$. (f) Energy-dependent local susceptibility $\chi''(E)$ for AFM BZs at (1,0) and (0,1). The green and red dashed lines are spin-wave fits from a Heisenberg Hamiltonian obtained from a twinned sample. The green and red solid lines are from MF+RPA calculations, which were multiplied by 2.8 for clear comparison. Adapted from \cite{lu2018spin}.
\label{Fig_INS_dBFA}}
\end{figure}

Comprehensive INS measurements on nearly fully detwinned BaFe$_2$As$_2$ have mapped the spin-excitation spectrum across the BZ up to high energies~\cite{lu2018spin}. As illustrated in Figs.~\ref{Fig_INS_dBFA}(a)--(c) and (e), sharp, dispersive spin waves are clearly observed along the $[1, K]$ direction, whereas excitations along the perpendicular $[H, 1]$ direction are significantly weaker, confirming nearly perfect detwinning. Figures~\ref{Fig_INS_dBFA}(c)--(d) present representative constant-energy slices in the $(H,K)$ plane. Though the intensity at low energies indicates near-perfect detwinning, upon increasing energy [Fig.~\ref{Fig_INS_dBFA}(d)], spectral weight progressively develops near $(0,\pm1)$. At still higher energies the anisotropy is no longer resolved and the response approaches an apparent $C_4$ rotational symmetry [Fig.~\ref{Fig_INS_dBFA}(f)], similar to that of a fully twinned sample. Comparisons between the experimental data and theoretical models suggest that the observed spin dynamics can be qualitatively described within an itinerant-electron picture, provided electronic correlations are adequately accounted for. 
Based on spectral weight considerations, it has been suggested that an effective local-moment model in the presence of biquadratic coupling also provides an effective description \cite{liu2020anisotropic}.
In addition to enabling single-domain conditions, the in-plane uniaxial pressure used for detwinning can itself induce a weak static or quasi-static out-of-plane ($c$-axis) AFM component and associated critical spin fluctuations near $T_s/T_N$~\cite{liu2020inplane}, an effect that should be considered when interpreting strain-dependent measurements.

Similar neutron scattering experiments on optimally doped, uniaxially strained 
BaFe$_{1.9}$Ni$_{0.1}$As$_2$ ($T_c = 20$~K) revealed pronounced anisotropy in spin 
excitations between $\mathbf{Q}_{\mathrm{AFM}} = (\pm 1,0)$ and 
$ (0,\pm 1)$, exhibiting strong energy and temperature 
dependence.  This anisotropy persists up to approximately 60~meV, significantly exceeding the orbital splitting energy between the Fe $d_{xz}$ and $d_{yz}$ orbitals. Such behavior supports a spin-driven Ising-nematic scenario as the principal mechanism underlying the observed resistivity anisotropy and associated electronic nematicity near optimal doping~\cite{song2015energy}. Comparable low-energy spin excitation anisotropies have been observed in underdoped BaFe$_{2-x}$Ni$_x$As$_2$ \cite{zhang2016effect,man2018direct} and BaFe$_{2-x}$Co$_x$As$_2$\cite{tian2019spin}, and SrFe$_2$As$_2$ \cite{tam2019weaker} below the nematic ordering temperature. These results are consistent with transport measurements showing clear evidence of in-plane anisotropy in magnetoresistance \cite{PhysRevB.109.054435} and $H_{c2}$ \cite{WOS:001199902200001}.

Finally, detwinned FeTe exhibits spin dynamics different from iron pnictides, reflecting the unique bicollinear AFM structure characteristic of iron chalcogenides \cite{PhysRevB.79.054503}. Neutron scattering studies demonstrate that the low-energy spin excitations, initially displaying twofold symmetry, evolve into fourfold-symmetric fluctuations at higher energies (above $\sim$ 26 meV) and elevated temperatures~\cite{tam2019plaquette}.  Polarized neutron scattering measurements further show that low-energy spin excitations are anisotropic in spin space, whereas those above $\sim$ 26 meV become isotropic~\cite{tam2019plaquette,song2018spinisotropic}. The observed spin fluctuation patterns align closely with theoretical predictions derived from bilinear-biquadratic spin Hamiltonians. These observations suggest that iron chalcogenides lie near a degeneracy between competing magnetic states, such as bicollinear and plaquette configurations, highlighting the role of magnetic frustration in shaping their exotic magnetic ground states and excitation spectra.

\subsubsection{NaFeAs and FeSe: orbital-dependent spin excitations}\label{sec_odsc}

As discussed in Sec.~\ref{sec_4a}, both the superexchange interactions and coherent propagation of spin waves are inherently orbital-dependent, indicating that high-energy spin waves (excitations) may consist of several components with distinct orbital characters. Earlier itinerant multiorbital calculation \cite{scherer2016collective} showed that spin excitations in FeSCs can be decomposed into several intra-orbital channels, indicating that the spin waves extending to the band top should exhibit clear orbital selectivity \cite{tam2020orbital}.

An essential step in clarifying this issue is to map out the intrinsic spin-excitation spectrum using fully detwinned samples. As described in Fig.~\ref{Fig_INS_dBFA} (Sec. \ref{sec_uniaxial}), INS studies on fully detwinned {\BFA} yielded the intrinsic spin-wave dispersion, which can be effectively captured by an itinerant theoretical model that incorporates moderate electron correlations, with interaction parameters $U \approx 1.46$ eV and Hund's coupling $J_H \approx 0.36$ eV [Fig.~\ref{Fig_INS_dBFA}(e)]. This model also accurately reproduces the observed fact that the high-energy spin excitations do not converge at the $(\pm 1, \pm 1)$ points \cite{lu2018spin}. In an orbital-selective scenario \cite{tam2020orbital}, low-energy spin excitations ($E\lesssim70$ meV in {\BFA}) are predominantly associated with the $d_{yz}$ orbital, while contributions from the $d_{xz}$ orbital remain negligible in this energy range. Notably, at higher energies [Fig.~\ref{Fig_INS_dBFA}(d)], spin excitations begin to emerge at the wave vectors $(0,\pm 1)$, corresponding to contributions from the $d_{xz}$ orbital channel.

Moreover, although the RPA calculation can only reproduce approximately one-third of the measured spectral weight, $\chi''(E)$, from the INS data, it accurately captures the anisotropy of the local dynamical susceptibility, specifically the ratio $\chi_1''(E)/\chi_2''(E)$, where $\chi_1''$ and $\chi_2''$ denote the susceptibilities centered around the magnetic BZ wave vectors $\mathbf{Q}_1=(\pm1,0)$ and $\mathbf{Q}_2=(0,\pm1)$, respectively [Fig.~\ref{Fig_INS_dBFA}(f)].
The considerations on the total spectral weight led to the suggestion that the anisotropy effectively described by the local-moment model in the presence of biquadratic coupling may also provide an effective description of the data \cite{liu2020anisotropic}.

\begin{figure}
\centering
\includegraphics[width=6cm]{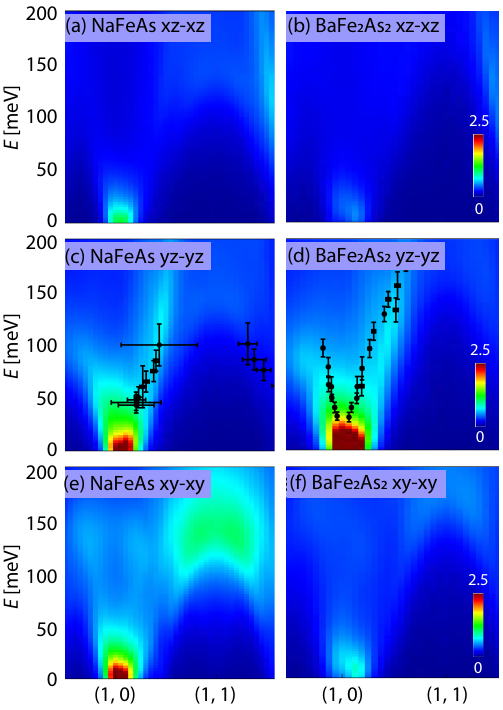}
\caption[]{Orbital selective spin waves in detwinned NaFeAs. DFT + DMFT intensity along $\mathbf{Q}-E$ cuts around the high-symmetry paths in the in-plane orthorhombic unit cell, for NaFeAs [(a), (c), (e)] and {\BFA} [(b), (d), (f)], in (a), (b) the $d_{xz}-d_{xz}$, (c), (d) $d_{yz}-d_{yz}$, and (e), (f) $d_{xy}-d_{xy}$ intraorbital scattering channels. The peak positions from neutron scattering data closely match the DFT + DMFT intensity in the $d_{yz}-d_{yz}$ channel for both (c) NaFeAs and (d) {\BFA}. Above $\sim110$ meV, spin fluctuations are observed at $(1, 1)$ and are consistent with the $d_{xy}-d_{xy}$ scattering channel. Spin fluctuations at $(1, 1)$ are pushed above $E \approx 150$ meV in {\BFA} and persist to the band top of nearly 300 meV, showing that the larger overall bandwidth of {\BFA} is controlled by $d_{xy}-d_{xy}$ fluctuations. Adapted from \cite{tam2020orbital}.
\label{Fig_INS_dNFA}}
\end{figure}

The intensity difference between the intrinsic spin waves at $(\pm1, 0)$ and $(0, \pm1)$ in detwinned NaFeAs closely resembles that observed in {\BFA}. However, previous INS investigations have shown that NaFeAs, characterized by stronger electron correlations in the $d_{xy}$ orbital, exhibits a significantly lower spin-excitation bandwidth ($\sim120$ meV) compared to {\BFA}, along with pronounced damping of the excitations along the longitudinal $[H,1]$ direction \cite{zhang2014effect}, akin to FeSe \cite{liu2025spin}. Figure~\ref{Fig_INS_dNFA} presents DMFT calculations decomposing the intra-orbital spin excitations into distinct $d_{xz}$-$d_{xz}$, $d_{yz}$-$d_{yz}$, and $d_{xy}$-$d_{xy}$ channels for {\BFA} and NaFeAs \cite{tam2020orbital}. These results indicate that in {\BFA}, spin excitations exhibit predominantly $d_{yz}$ character across a large energy range. By contrast, in NaFeAs, excitations below approximately $100$ meV are governed largely by $d_{yz}$-$d_{yz}$ intra-orbital scattering processes, strongly enhanced by the $C_2$-symmetric nematic state. At higher energies, however, spin excitations transition to a dominant $d_{xy}$-$d_{xy}$ character around $(\pm1, \pm1)$, retaining approximate $C_4$ symmetry and exhibiting weaker temperature dependence.

\begin{figure*}[tbp!]
\centering
\includegraphics[width=14cm]{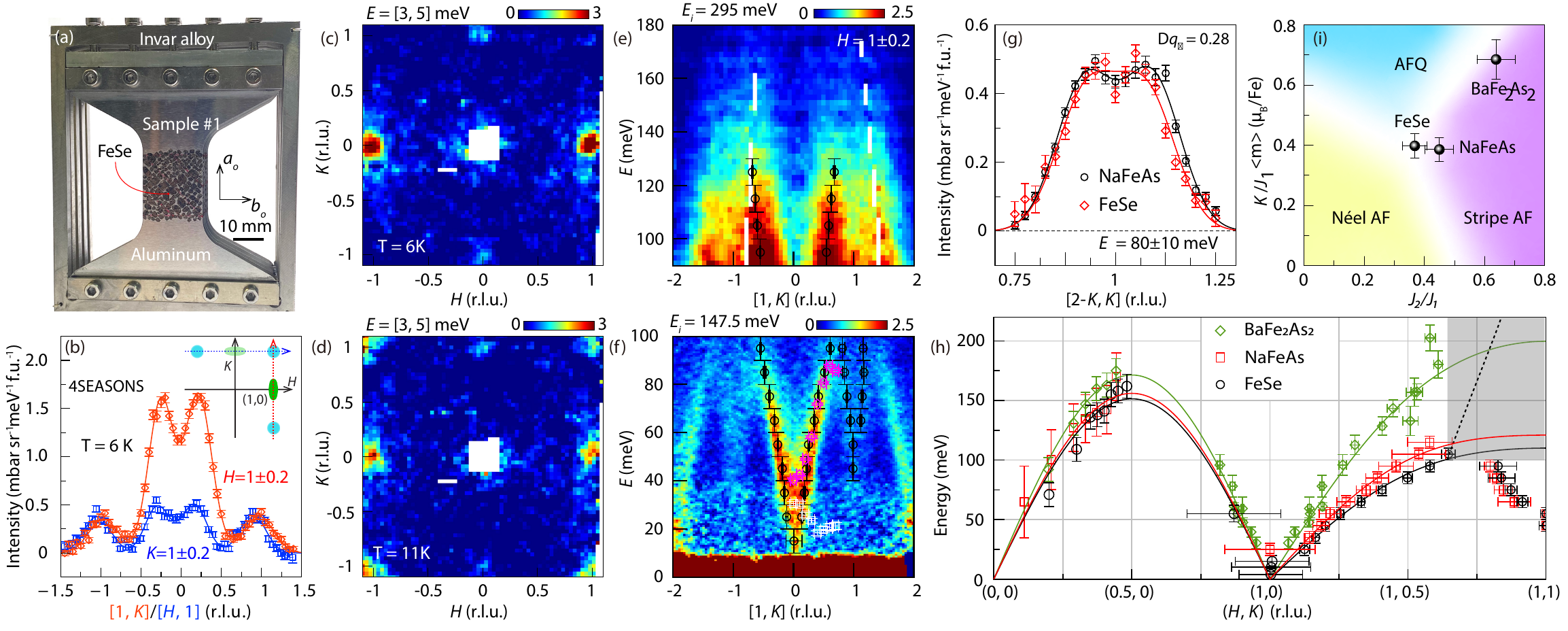}
\caption[]{Spin correlations in the nematic quantum disordered state of detwinned FeSe \cite{liu2025spin,liu2024low}. (a) The uniaxial-strain device based on the differential thermal expansion coefficients between the Invar alloy (Fe$_{0.64}$Ni$_{0.36}$) frame and the aluminum sheet. (b) One-dimensional constant-energy cuts ($E = 55 \pm 10$ meV) of the spin excitations along $[1, K]$ and $[H, 1]$ directions collected at the 4SEASONS time-of-flight spectrometer using incident energy of 80 meV. The inset depicts the positions of the stripe and N\'eel spin excitations in $[H, K]$ space, and the trajectories (blue and red dashed lines with arrowheads) for the 1D cuts. (c),(d) 2D constant-energy slices of the spin excitations in detwinned FeSe with $E=4\pm1$ meV at $T=6$~K and $11$~K, collected with $E_{i}=21$ meV. (e), (f) Magnetic excitation dispersions along $[1, K]$ directions projected onto energy and momentum planes measured at $T \approx 6$ K. The black circle symbols represent the magnetic excitation dispersions at $\mathbf{Q} = (1, 0)$ and $(1, 1)$. The white squares and pink diamonds in (f) correspond to the damping factor $\gamma(q)/2$ and undamped energy $E_0$, respectively. (g) Constant energy cuts along the $[2-K, K]$ direction for NaFeAs (black open circles) and FeSe (red open diamonds) with $E = 80 \pm 10$ meV. The integral interval perpendicular to $[2-K, K]$ is $\Delta q_\perp= 0.28$. (h) Spin-excitation dispersions for {\BFA} (green diamonds), NaFeAs (red squares), and FeSe (black circles). The data points in the range $(0, 0)--(0.5, 0)$ were measured with RIXS. The green, red, and black solid curves are the fittings of the dispersions with the $J_1-K-J_2$ model. The gray-shaded area marks the $(\mathbf{Q}, E)$ region where the spin excitations are heavily damped. (i) The ratios $K/J_1$ and $J_2 /J_1$ for the fittings of the dispersions for {\BFA}, NaFeAs, and FeSe. Adapted from \cite{liu2024low,liu2025spin}.
\label{Fig_INS_dFeSe}}
\end{figure*}

More recently, INS studies of FeSe detwinned via in-plane uniaxial strain have revealed its intrinsic nematic spin dynamics \cite{liu2025spin}.  Figure \ref{Fig_INS_dFeSe}(a) illustrates the low-background strain device, which exploits the differential thermal expansion of Invar and an aluminum alloy to generate uniaxial strain. This strain is effectively transmitted to the FeSe crystals mounted on aluminum sheets, achieving a high single-domain fraction [79\% estimated from Fig.~\ref{Fig_INS_dFeSe}(b)] while enabling accurate background subtraction and full BZ mapping.
The resulting spectra show that stripe excitations at $\mathbf{Q}=(1,0)/(0,1)$ are strongly $C_2$ anisotropic [Figs.~\ref{Fig_INS_dFeSe}(c) and (d)], whereas the N\'eel channel at $\mathbf{Q}=(1,1)$ retains $C_4$ symmetry [Fig.~\ref{Fig_INS_dFeSe}(b)]. The stripe anisotropy persists to $\sim120$ meV, establishing a wide nematic energy window. A distinct N\'eel-like branch appears above $\sim40$ meV and merges with the stripe excitations near $\sim90$ meV [Figs.~\ref{Fig_INS_dFeSe}(e) and (f)]. The dispersion is sharply defined along the transverse $[1, K]$ direction but rapidly overdamped along the longitudinal $[H, 0]$ direction, reflecting highly anisotropic damping. Both the N\'eel excitations and anisotropic damping of the stripe excitations resemble those in NaFeAs [Fig.~\ref{Fig_INS_dFeSe}(g)] but absent in {\BFA}. This similarity has been associated with the large anion height $h_{\rm FeX}$ that controls the electron correlation strength in the $d_{xy}$ orbital.
The momentum-integrated response obeys $\chi''(E)\propto E$ up to $\sim60$ meV and the total fluctuating moment is large, consistent with an $S\!\approx\!1$ local-moment description \cite{liu2025spin}.

A minimal $S=1$ bilinear-biquadratic $J_1-K-J_2$ analysis reproduces the stripe dispersion and places FeSe near a multicritical regime where stripe-AFM, Néel-AFM, and antiferroquadrupolar (spin-nematic) orders compete [Figs.~\ref{Fig_INS_dFeSe}(h), (i)] \cite{liu2025spin,hu2020quantum}. This proximity accounts for the absence of static magnetism and the persistence of nematic spin correlations (which continue under strain above $T_s$), consistent with the proposed nematic quantum-disordered state. 
Within an orbital-selective view, the $C_2$ stripe response, including the spin resonance as shown in Figs.~\ref{Fig_INS_dFeSe}(c) and (d), reflects $d_{xz}/d_{yz}$ channels, while the $C_4$ N\'eel mode [Fig.~\ref{Fig_INS_dFeSe}(b)] is naturally linked to more strongly correlated $d_{xy}$ electrons \cite{kreisel2022theory}.

This orbital-dependent hierarchy indicates that spin excitations in FeSCs are best described as composite collective modes carrying predominant, but not exclusive, orbital characters. Specifically, low-energy excitations near $(1,0)$ are primarily driven by the $d_{yz}$ intra-orbital fluctuations, whereas at higher energies the spectral weight progressively shifts toward the $d_{xy}$ channel, which significantly controls the overall magnetic bandwidth \cite{zhang2014effect}. Within this framework, N\'eel-type spin excitations around the $(1, 1)$ wave vector naturally emerge from the $d_{xy}$ orbital channel, while the strong directional damping along the $[H,0]$ direction---particularly evident in materials with stronger electron correlations, such as FeSe and NaFeAs compared to {\BFA} and CaFe$_2$As$_2$---is explained by the diminished coherence (small quasiparticle weight $Z_{xy}$ and short lifetimes) of the $d_{xy}$ electrons.

An important advantage of framing the spin-excitation spectrum in terms of orbital-selective spin excitations is that it provides a useful framework for interpreting multiple experimental trends, including the restoration from $C_2$ to $C_4$ symmetry at higher energies, systematic variations in excitation bandwidth, and pronounced directional damping. This conceptual approach is broadly consistent with results from itinerant RPA calculations for {\BFA} and orbital decompositions derived from DMFT calculations for NaFeAs \cite{tam2020orbital, lu2018spin, liu2025spin}. Nevertheless, one must acknowledge that spin modes are never purely single-orbital due to spin--orbit coupling, inter-orbital hybridization, and off-diagonal exchange interactions, all of which inevitably mix $d_{xz}$, $d_{yz}$, and $d_{xy}$ contributions. Moreover, local DMFT, employing purely local self-energies and simplified vertex corrections, may underestimate nonlocal and momentum-dependent processes that feed into the N\'eel excitation channel. This limitation partly explains why certain DMFT implementations struggle to accurately reproduce the full zone-boundary intensity and detailed dispersive behavior without incorporating additional itinerant (RPA-like) ingredients.

Taken collectively, these observations emphasize that orbital-dependent spin correlations in FeSCs are not restricted to the low-energy regime governed by Fermi-surface nesting; instead, orbital selectivity extends deeply into the high-energy region dominated by local moment physics. Theoretical frameworks such as RPA and DMFT have underscored how Hund's coupling strongly amplifies orbital differentiation, thereby controlling magnetic exchange anisotropies and spin excitation damping rates. The pervasive orbital selectivity observed throughout the spin-excitation spectrum thus offers a compelling and unified description of both itinerant and localized magnetism in FeSCs. Moreover, it opens new opportunities to further probe the intricate interplay among orbital order, electronic nematicity, and superconductivity.

\subsubsection{Spin fluctuations, nematicity, and superconductivity}

The high-energy bandwidth and exchange parameters quantify the magnetic energy scale and short-range correlations in FeSCs, but cannot by themselves determine their relation to nematicity and superconductivity. This connection is established most directly by the low-energy, momentum-, orbital-, and spin-polarization-dependent response of coherent quasiparticles near the Fermi surface. High- and low-energy spin responses are therefore complementary rather than interchangeable: the former constrains the robustness and frustration of short-range correlations, whereas the latter governs how magnetic correlations couple to nematicity and Cooper pairing.

The detwinned neutron results above directly test the orbital-selective spin-fluctuation framework of Sec.~VI.B.3. Orbital-dependent quasiparticle coherence and the $d_{xz}/d_{yz}$ Fermi-surface texture select interband channels, making the low-energy responses at the two stripe wave vectors inequivalent. In NaFe$_{1-x}$Co$_x$As, Ba(Fe$_{1-x}$Co$_x$)$_2$As$_2$, and FeSe, the resonance and nearby low-energy spin fluctuations concentrate at one wave vector \cite{wang2017orbital,chen2019anisotropic,tian2019spin}. This connection among nematicity, orbital selectivity, and superconductivity cannot be inferred from the high-energy exchange scale alone.

FeSe illustrates this most clearly. Its orbital texture and gap anisotropy favor the $d_{yz}$ hole--electron channel, and detwinned INS detects the resonance only at the corresponding AFM wave vector \cite{sprau2017discovery,kreisel2017orbital,chen2019anisotropic}. Complementary low-energy probes provide the broader context: ARPES resolves the nematic orbital texture, whereas elastoresistivity, Raman, and NMR establish enhanced nematic susceptibility and low-energy spin anisotropy (Sec.~V.A). Together, they show how the low-energy spin spectrum couples nematic, orbital, and superconducting degrees of freedom, without implying universal roles for orbital selectivity or nematic fluctuations. For a broader comparison, see Ref.~\cite{bohmer2022nematicity}. RIXS, discussed next, extends this perspective to high-energy spin dynamics.

\bigskip
\subsection{High-energy spin excitations probed by RIXS}

\begin{figure*}
\centering
\includegraphics[width=16cm]{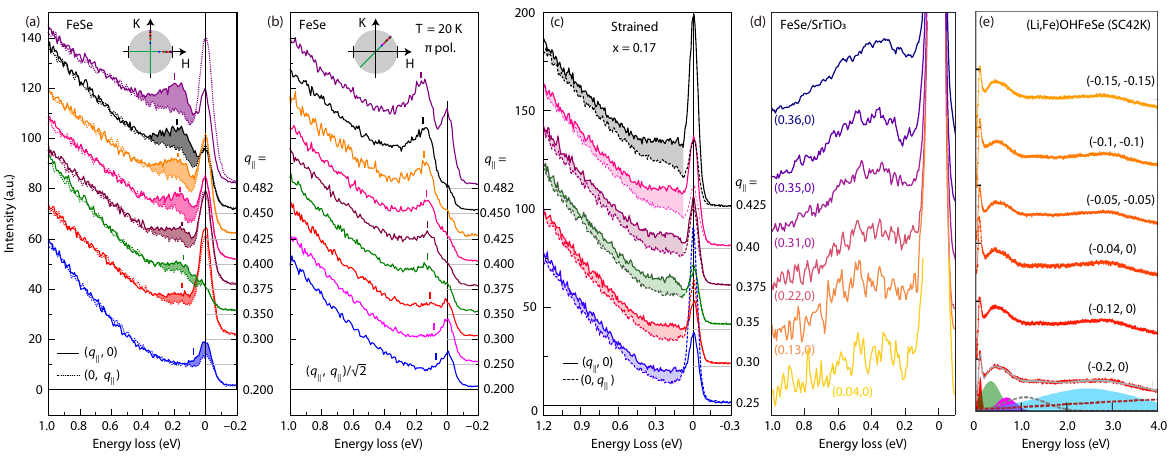}
\caption[]{Representative Fe-$L_3$ RIXS spectra of FeSe-derived superconductors. (a) Momentum-dependent spectra of detwinned FeSe along the $H$ and $K$ directions, and (b) corresponding spectra of twinned FeSe along $[H,H]$ \cite{lu2022spinexcitation}. (c) Strain-dependent spectra of FeSe$_{0.83}$S$_{0.17}$ \cite{liu2024nematic}. (d) Spectra of FeSe/SrTiO$_3$ measured along $[H,0]$ \cite{pelliciari2021evolution}. (e) Selected spectra of (Li,Fe)OHFeSe with $T_c=42$~K along two high-symmetry directions \cite{xiao2022dispersionless}. Adapted from \cite{lu2022spinexcitation,liu2024nematic,pelliciari2021evolution,xiao2022dispersionless}.
\label{Fig_rixs_fig4}}
\end{figure*}

RIXS provides a useful complement to INS for FeSCs, particularly when only small crystals, thin films, heterostructures, or strongly neutron-absorbing materials are available \cite{fujita2012progress,dai2012magnetism,headings2010anomalous,xie2020spinexcitation,zhang2014direct,li2019superconductivity,jin2011link}. Improvements in energy resolution and polarization control at transition-metal $L$ edges have enabled Fe-$L_3$ RIXS studies of low- and high-energy inelastic features in FeSCs \cite{ament2011resonant,degroot2024resonant,mitrano2024exploring, saxes2006,braicovich2009dispersion,brookes2018beamline,zhou2022i21, dvorak2016six,nanoterasu,yamamoto2025status}. In comparison with INS, however, the Fe-$L_3$ spectra contain overlapping magnetic, orbital, charge, lattice, and fluorescence contributions \cite{jia2016using,schlappa2012spin,devereaux2014directly, jia2014persistent,ament2009,haverkort2010theory,gilmore2021description}. Their magnetic interpretation is therefore less direct in multiorbital FeSCs, where several $t_{2g}$ orbitals participate in the intermediate and final states \cite{sobota2021angle}. RIXS thus does not replace the momentum-resolved magnetic information obtained by INS, but extends measurements of high-energy inelastic excitations to materials that are otherwise difficult to study.

Across the FePn-122, 111, and 1111 families, Fe-$L_3$ RIXS has reported broad dispersive inelastic features that are generally compatible with the high-energy magnetic excitations established by INS \cite{zhou2013persistent,peng2018dispersion,rahn2019paramagnon, pelliciari2017local,pelliciari2019reciprocity, garcia2019anisotropic,cantarino2026disorder, pelliciari2016intralayer,song2021spin,pelliciari2016presence}. These studies include hole-doped Ba$_{1-x}$K$_x$Fe$_2$As$_2$, EuFe$_2$As$_2$, P-substituted BaFe$_2$(As,P)$_2$, Mn- and Cr-substituted BaFe$_2$As$_2$, NaFe$_{1-x}$Co$_x$As, NaFe$_{1-x}$Cu$_x$As, and SmFeAsO \cite{wang2013doping,hu2015structural,hu2016spin, pelliciari2019reciprocity,garcia2019anisotropic, cantarino2026disorder,carr2016electron,zhang2014effect, song2016mott,song2021spin,pelliciari2016presence}. The reported dispersion trends support the persistence of substantial high-energy spin correlations across broad doping ranges, while also revealing material-dependent softening, damping, and changes in spectral weight. In systems such as EuFe$_2$As$_2$ and heavily substituted BaFe$_2$As$_2$ derivatives, RIXS is particularly valuable because neutron absorption or limited crystal volume restricts INS measurements. These results support an intermediate-coupling description of FeSC magnetism, although quantitative comparison between RIXS and INS remains complicated by their distinct cross sections, momentum coverage, and treatment of background contributions \cite{pelliciari2017local,pelliciari2019reciprocity,yin2014spin,hancock2010evidence,gilmore2021description}.

RIXS has been especially useful for FeSe-derived materials, where suitable samples for high-energy neutron spectroscopy are often limited. In detwinned FeSe, RIXS reported pronounced $C_2$ anisotropy of the high-energy inelastic response between the two orthogonal in-plane directions, extending to energies of order 200 meV [Fig.~\ref{Fig_rixs_fig4}(a),(b)] \cite{rahn2019paramagnon,lu2022spinexcitation}. The persistence of this anisotropy to high energies shows that spin correlations participate in nematicity over a broad energy range. It does not, by itself, identify a unique primary nematic order parameter, and the relative roles of spin-driven and orbital mechanisms remain under active discussion \cite{chen2019anisotropic,occhialini2023spontaneous}. Strain-dependent RIXS measurements in FeSe$_{1-x}$S$_x$ further reported robust high-energy spin excitations and an enhanced strain-induced anisotropy near the nematic end point [Fig.~\ref{Fig_rixs_fig4}(c)] \cite{liu2024nematic}. These results are consistent with an important role of spin correlations in the nematic response, while their relation to critical nematic fluctuations and superconducting pairing remains to be clarified.

For monolayer FeSe/SrTiO$_3$, RIXS accesses broad inelastic features in a system beyond the reach of conventional INS [Fig.~\ref{Fig_rixs_fig4}(d)] \cite{zhang2014direct,zhou2022i21,pelliciari2021evolution}. Together with Hubbard-model and quantum Monte Carlo calculations, these features were interpreted as reflecting a modified spin-excitation spectrum after the electron-only Fermi-surface reconstruction. Their separation from possible local orbital contributions, and their quantitative connection to the enhanced $T_c$, remain unsettled. RIXS studies of (Li,Fe)OHFeSe likewise reported momentum-independent Raman-like features near 0.3 and 0.7 eV [Fig.~\ref{Fig_rixs_fig4}(e)] \cite{lu2015coexistence,dong2015li,pan2017structure, xiao2022dispersionless}. Atomic-multiplet calculations assigned these features to local transitions within the $e_g$ and $t_{2g}$ manifolds, and their temperature-dependent intensity was interpreted as evidence for strong orbital fluctuations. Overall, RIXS provides a valuable extension of high-energy spectroscopy to small-volume and thin-film FeSCs, while INS remains the primary probe for establishing the momentum-resolved magnetic excitation spectrum.

\section{Summary and Outlook}

Over the past decade or more, FeSCs have matured into a versatile platform for studying unconventional superconductivity in multiband, moderately correlated metals. Chemical substitution, isovalent tuning, pressure, intercalation, and epitaxial/strain engineering have expanded their structural and electronic diversity while enabling increasingly stringent experiments. These tuning parameters reveal a recurring, but strongly material-dependent, landscape in which electronic nematicity and stripe-type antiferromagnetism often occur near superconductivity and evolve with the electronic structure. Structural parameters such as anion height and the Fe--X--Fe bond angle frequently correlate with these changes, although their effects are intertwined with carrier density, orbital reconstruction, disorder, and dimensionality.

No single microscopic framework accounts for all FeSCs. A widely discussed description for many compounds is an even-parity, spin-singlet, sign-changing $s_{\pm}$ state arising from interband interactions and spin fluctuations. The superconductivity-induced spin resonance observed in many materials has been widely interpreted as being consistent with such sign-changing pairing. However, the relative roles of spin, orbital, nematic, lattice, and interface-related interactions vary substantially among compounds. Correspondingly, the gap structure ranges from nearly isotropic to strongly anisotropic or nodal, and can be shaped by orbital selectivity, Fermi-surface reconstruction, and competing pairing channels. In FeSe, BQPI, ARPES, detwinned INS, and orbital-selective calculations provide a coherent framework in which $d_{yz}$-dominated spin fluctuations play an important role in pairing, although its broader applicability remains compound dependent. Reports of TRSB and possible finite-momentum superconductivity further broaden the phenomenology.

In parallel, application-relevant properties---including high $H_{c2}$, moderate anisotropy, and improving critical-current densities---have advanced through progress in coated conductors, thin films, and engineered pinning landscapes. These developments illustrate how the tunability of FeSCs can support both fundamental studies and high-field technologies.

Several frontiers appear particularly promising:

(1) \textit{Quantifying spin-nematic synergy in pairing.} Establishing whether nematic fluctuations provide a cooperative boost to spin-fluctuation-mediated pairing requires phase-sensitive benchmarks that track the superconducting gap structure while independently tuning nematicity (via strain, isoelectronic substitution, or uniaxial pressure) at fixed carrier density.

(2) \textit{Gap sign and structure without hole pockets.} In HED FeSe-derived systems, reconstructions that remove hole pockets test the robustness of sign-changing pairing. Joint ARPES/QPI/INS programs that co-register the Fermi-surface topology, QPI response, and momentum dependence of the resonance in reciprocal space can deliver decisive constraints.

(3) \textit{Interfaces and forward-focused interactions.} Monolayer FeSe/SrTiO$_3$ and related heterostructures continue to illuminate interfacial routes to enhance $T_{c}$. Disentangling intrinsic pairing from interfacial bosons (phonons or polar modes) will benefit from isotope-selective interfaces, momentum-resolved electron-boson coupling maps, and tunable dielectric environments.

(4) \textit{Orbital selectivity and quasiparticle coherence.} The extent to which orbital-dependent correlations sculpt pairing interactions deserves further exploration. Combining orbitally resolved spectroscopies with controlled disorder and DMFT-informed modeling should clarify how coherence recovery (or loss) shifts gap anisotropy and $T_{c}$.

(5) \textit{Topology and superconductivity.} Fe(Te,Se) and related systems provide promising settings for testing topological superconductivity and Majorana bound states. At the same time, reported signatures of time-reversal-symmetry breaking and possible finite-momentum pairing call for bulk-sensitive probes, systematic impurity studies, phase-sensitive measurements, and reproducible vortex-core spectroscopy on high-quality samples.

(6) \textit{Toward deployable conductors.} For applications, priorities include scalable, oxidation-resilient architectures; engineered, anisotropy-appropriate pinning; and long-length uniformity. Strain-tolerant tapes and multilayer films that leverage interface-enhanced pairing represent near-term targets.

By aligning materials design with multi-probe, phase-sensitive experiments and quantitatively constrained theory, the field is positioned to convert qualitative consensus into predictive understanding-and to translate the tunability of FeSCs into both new superconducting phenomena and practical high-field technologies.

\bigskip
\noindent
$\S$ These two authors contributed equally to this work.

\begin{acknowledgments}

We thank Rong Yu, Qimiao Si, and Ming Yi for helpful discussions.
The work at BNU is supported by the Scientific Research Innovation Capability Support Project for Young Faculty (ZYGXQNJSKYCXNLZCXM-M2) (X.L.).
The work at RUC is supported by the National Key R\&D Program of China (Grants No. 2023YFA1406500 and No. 2022YFA1403800) and the National Natural Science Foundation of China (Grant No. 12274459) (H.L.).
The work at Fudan is supported by the Key Program of the National Natural Science Foundation of China (Grant No. 12234006) (J.Z.), the National Key R\&D Program of China (Grant No. 2022YFA1403202) (J.Z.), the Innovation Program for Quantum Science and Technology (Grant No. 2024ZD0300103) (J.Z.), and Large Scientific Facility Open Subject of Songshan Lake Laboratory (Grant No. DG23313511)(J.Z.).
H.H. is supported by The MEXT Elements Strategy Initiative to form the Core Research Center. 
Materials synthesis and neutron scattering research on iron-based superconductors at Rice are supported by the U.S. DOE, BES under Grant Nos. DE-SC0012311 and DE-SC0026179 (P.D.). Part of the materials characterization efforts at Rice is supported by the Robert A. Welch Foundation Grant No. C-1839 (P.D.).  
\end{acknowledgments}


\begin{thebibliography}{686}%
\makeatletter
\providecommand \@ifxundefined [1]{%
 \@ifx{#1\undefined}
}%
\providecommand \@ifnum [1]{%
 \ifnum #1\expandafter \@firstoftwo
 \else \expandafter \@secondoftwo
 \fi
}%
\providecommand \@ifx [1]{%
 \ifx #1\expandafter \@firstoftwo
 \else \expandafter \@secondoftwo
 \fi
}%
\providecommand \natexlab [1]{#1}%
\providecommand \enquote  [1]{``#1''}%
\providecommand \bibnamefont  [1]{#1}%
\providecommand \bibfnamefont [1]{#1}%
\providecommand \citenamefont [1]{#1}%
\providecommand \href@noop [0]{\@secondoftwo}%
\providecommand \href [0]{\begingroup \@sanitize@url \@href}%
\providecommand \@href[1]{\@@startlink{#1}\@@href}%
\providecommand \@@href[1]{\endgroup#1\@@endlink}%
\providecommand \@sanitize@url [0]{\catcode `\\12\catcode `\$12\catcode
  `\&12\catcode `\#12\catcode `\^12\catcode `\_12\catcode `\%12\relax}%
\providecommand \@@startlink[1]{}%
\providecommand \@@endlink[0]{}%
\providecommand \url  [0]{\begingroup\@sanitize@url \@url }%
\providecommand \@url [1]{\endgroup\@href {#1}{\urlprefix }}%
\providecommand \urlprefix  [0]{URL }%
\providecommand \Eprint [0]{\href }%
\providecommand \doibase [0]{https://doi.org/}%
\providecommand \selectlanguage [0]{\@gobble}%
\providecommand \bibinfo  [0]{\@secondoftwo}%
\providecommand \bibfield  [0]{\@secondoftwo}%
\providecommand \translation [1]{[#1]}%
\providecommand \BibitemOpen [0]{}%
\providecommand \bibitemStop [0]{}%
\providecommand \bibitemNoStop [0]{.\EOS\space}%
\providecommand \EOS [0]{\spacefactor3000\relax}%
\providecommand \BibitemShut  [1]{\csname bibitem#1\endcsname}%
\let\auto@bib@innerbib\@empty
\bibitem [{\citenamefont {Adroja}\ \emph {et~al.}(2020)\citenamefont {Adroja},
  \citenamefont {Blundell}, \citenamefont {Lang}, \citenamefont {Luo},
  \citenamefont {Wang},\ and\ \citenamefont {Cao}}]{adroja2020observation}%
  \BibitemOpen
  \bibfield  {author} {\bibinfo {author} {\bibnamefont {Adroja}, \bibfnamefont
  {D.~T.}}, \bibinfo {author} {\bibfnamefont {S.~J.}\ \bibnamefont {Blundell}},
  \bibinfo {author} {\bibfnamefont {F.}~\bibnamefont {Lang}}, \bibinfo {author}
  {\bibfnamefont {H.}~\bibnamefont {Luo}}, \bibinfo {author} {\bibfnamefont
  {Z.-C.}\ \bibnamefont {Wang}}, and\ \bibinfo {author} {\bibfnamefont {G.-H.}\
  \bibnamefont {Cao}}} (\bibinfo {year} {2020}),\ \href
  {https://doi.org/10.1088/1361-648x/aba28f} {\bibfield  {journal} {\bibinfo
  {journal} {J. Phys.: Condens. Matter}\ }\textbf {\bibinfo {volume} {32}},\
  \bibinfo {pages} {435603}}\BibitemShut {NoStop}%
\bibitem [{\citenamefont {Agterberg}\ \emph {et~al.}(2017)\citenamefont
  {Agterberg}, \citenamefont {Shishidou}, \citenamefont {O'Halloran},
  \citenamefont {Brydon},\ and\ \citenamefont
  {Weinert}}]{agterberg2017resilient}%
  \BibitemOpen
  \bibfield  {author} {\bibinfo {author} {\bibnamefont {Agterberg},
  \bibfnamefont {D.~F.}}, \bibinfo {author} {\bibfnamefont {T.}~\bibnamefont
  {Shishidou}}, \bibinfo {author} {\bibfnamefont {J.}~\bibnamefont
  {O'Halloran}}, \bibinfo {author} {\bibfnamefont {P.~M.~R.}\ \bibnamefont
  {Brydon}}, and\ \bibinfo {author} {\bibfnamefont {M.}~\bibnamefont
  {Weinert}}} (\bibinfo {year} {2017}),\ \href
  {https://doi.org/10.1103/PhysRevLett.119.267001} {\bibfield  {journal}
  {\bibinfo  {journal} {Phys. Rev. Lett.}\ }\textbf {\bibinfo {volume} {119}},\
  \bibinfo {pages} {267001}}\BibitemShut {NoStop}%
\bibitem [{\citenamefont {Akbari}\ \emph {et~al.}(2011)\citenamefont {Akbari},
  \citenamefont {Eremin},\ and\ \citenamefont {Thalmeier}}]{Akbari2011}%
  \BibitemOpen
  \bibfield  {author} {\bibinfo {author} {\bibnamefont {Akbari}, \bibfnamefont
  {A.}}, \bibinfo {author} {\bibfnamefont {I.}~\bibnamefont {Eremin}}, and\
  \bibinfo {author} {\bibfnamefont {P.}~\bibnamefont {Thalmeier}}} (\bibinfo
  {year} {2011}),\ \href {https://doi.org/10.1103/PhysRevB.84.134513}
  {\bibfield  {journal} {\bibinfo  {journal} {Phys. Rev. B}\ }\textbf {\bibinfo
  {volume} {84}},\ \bibinfo {pages} {134513}}\BibitemShut {NoStop}%
\bibitem [{\citenamefont {Akbari}\ \emph {et~al.}(2013)\citenamefont {Akbari},
  \citenamefont {Thalmeier},\ and\ \citenamefont {Eremin}}]{Akbari2013}%
  \BibitemOpen
  \bibfield  {author} {\bibinfo {author} {\bibnamefont {Akbari}, \bibfnamefont
  {A.}}, \bibinfo {author} {\bibfnamefont {P.}~\bibnamefont {Thalmeier}}, and\
  \bibinfo {author} {\bibfnamefont {I.}~\bibnamefont {Eremin}}} (\bibinfo
  {year} {2013}),\ \href {https://doi.org/10.1088/1367-2630/15/3/033034}
  {\bibfield  {journal} {\bibinfo  {journal} {New J. Phys.}\ }\textbf {\bibinfo
  {volume} {15}},\ \bibinfo {pages} {033034}}\BibitemShut {NoStop}%
\bibitem [{\citenamefont {Albedah}\ \emph
  {et~al.}(2018{\natexlab{a}})\citenamefont {Albedah}, \citenamefont
  {Nejadsattari}, \citenamefont {Stadnik}, \citenamefont {Liu},\ and\
  \citenamefont {Cao}}]{Albedah2018b}%
  \BibitemOpen
  \bibfield  {author} {\bibinfo {author} {\bibnamefont {Albedah}, \bibfnamefont
  {M.~A.}}, \bibinfo {author} {\bibfnamefont {F.}~\bibnamefont {Nejadsattari}},
  \bibinfo {author} {\bibfnamefont {Z.~M.}\ \bibnamefont {Stadnik}}, \bibinfo
  {author} {\bibfnamefont {Y.}~\bibnamefont {Liu}}, and\ \bibinfo {author}
  {\bibfnamefont {G.-H.}\ \bibnamefont {Cao}}} (\bibinfo {year}
  {2018}{\natexlab{a}}),\ \href {https://doi.org/10.1088/1361-648x/aab4af}
  {\bibfield  {journal} {\bibinfo  {journal} {J. Phys.: Condens. Matter}\
  }\textbf {\bibinfo {volume} {30}},\ \bibinfo {pages} {155803}}\BibitemShut
  {NoStop}%
\bibitem [{\citenamefont {Albedah}\ \emph
  {et~al.}(2018{\natexlab{b}})\citenamefont {Albedah}, \citenamefont
  {Nejadsattari}, \citenamefont {Stadnik}, \citenamefont {Liu},\ and\
  \citenamefont {Cao}}]{Albedah2018a}%
  \BibitemOpen
  \bibfield  {author} {\bibinfo {author} {\bibnamefont {Albedah}, \bibfnamefont
  {M.~A.}}, \bibinfo {author} {\bibfnamefont {F.}~\bibnamefont {Nejadsattari}},
  \bibinfo {author} {\bibfnamefont {Z.~M.}\ \bibnamefont {Stadnik}}, \bibinfo
  {author} {\bibfnamefont {Y.}~\bibnamefont {Liu}}, and\ \bibinfo {author}
  {\bibfnamefont {G.-H.}\ \bibnamefont {Cao}}} (\bibinfo {year}
  {2018}{\natexlab{b}}),\ \href {https://doi.org/10.1103/PhysRevB.97.144426}
  {\bibfield  {journal} {\bibinfo  {journal} {Phys. Rev. B}\ }\textbf {\bibinfo
  {volume} {97}},\ \bibinfo {pages} {144426}}\BibitemShut {NoStop}%
\bibitem [{\citenamefont {Allred}\ \emph {et~al.}(2015)\citenamefont {Allred},
  \citenamefont {Avci}, \citenamefont {Chung}, \citenamefont {Claus},
  \citenamefont {Khalyavin}, \citenamefont {Manuel}, \citenamefont {Taddei},
  \citenamefont {Kanatzidis}, \citenamefont {Rosenkranz}, \citenamefont
  {Osborn},\ and\ \citenamefont {Chmaissem}}]{allred2015tetragonal}%
  \BibitemOpen
  \bibfield  {author} {\bibinfo {author} {\bibnamefont {Allred}, \bibfnamefont
  {J.~M.}}, \bibinfo {author} {\bibfnamefont {S.}~\bibnamefont {Avci}},
  \bibinfo {author} {\bibfnamefont {D.~Y.}\ \bibnamefont {Chung}}, \bibinfo
  {author} {\bibfnamefont {H.}~\bibnamefont {Claus}}, \bibinfo {author}
  {\bibfnamefont {D.~D.}\ \bibnamefont {Khalyavin}}, \bibinfo {author}
  {\bibfnamefont {P.}~\bibnamefont {Manuel}}, \bibinfo {author} {\bibfnamefont
  {K.~M.}\ \bibnamefont {Taddei}}, \bibinfo {author} {\bibfnamefont {M.~G.}\
  \bibnamefont {Kanatzidis}}, \bibinfo {author} {\bibfnamefont
  {S.}~\bibnamefont {Rosenkranz}}, \bibinfo {author} {\bibfnamefont
  {R.}~\bibnamefont {Osborn}}, and\ \bibinfo {author} {\bibfnamefont
  {O.}~\bibnamefont {Chmaissem}}} (\bibinfo {year} {2015}),\ \href
  {https://doi.org/10.1103/PhysRevB.92.094515} {\bibfield  {journal} {\bibinfo
  {journal} {Phys. Rev. B}\ }\textbf {\bibinfo {volume} {92}},\ \bibinfo
  {pages} {094515}}\BibitemShut {NoStop}%
\bibitem [{\citenamefont {Allred}\ \emph {et~al.}(2016)\citenamefont {Allred},
  \citenamefont {Taddei}, \citenamefont {Bugaris}, \citenamefont {Krogstad},
  \citenamefont {Lapidus}, \citenamefont {Chung}, \citenamefont {Claus},
  \citenamefont {Kanatzidis}, \citenamefont {Brown}, \citenamefont {Kang},
  \citenamefont {Fernandes}, \citenamefont {Eremin}, \citenamefont
  {Rosenkranz}, \citenamefont {Chmaissem},\ and\ \citenamefont
  {Osborn}}]{allred2016double}%
  \BibitemOpen
  \bibfield  {author} {\bibinfo {author} {\bibnamefont {Allred}, \bibfnamefont
  {J.~M.}}, \bibinfo {author} {\bibfnamefont {K.~M.}\ \bibnamefont {Taddei}},
  \bibinfo {author} {\bibfnamefont {D.~E.}\ \bibnamefont {Bugaris}}, \bibinfo
  {author} {\bibfnamefont {M.~J.}\ \bibnamefont {Krogstad}}, \bibinfo {author}
  {\bibfnamefont {S.~H.}\ \bibnamefont {Lapidus}}, \bibinfo {author}
  {\bibfnamefont {D.~Y.}\ \bibnamefont {Chung}}, \bibinfo {author}
  {\bibfnamefont {H.}~\bibnamefont {Claus}}, \bibinfo {author} {\bibfnamefont
  {M.~G.}\ \bibnamefont {Kanatzidis}}, \bibinfo {author} {\bibfnamefont
  {D.~E.}\ \bibnamefont {Brown}}, \bibinfo {author} {\bibfnamefont
  {J.}~\bibnamefont {Kang}}, \bibinfo {author} {\bibfnamefont {R.~M.}\
  \bibnamefont {Fernandes}}, \bibinfo {author} {\bibfnamefont {I.}~\bibnamefont
  {Eremin}}, \bibinfo {author} {\bibfnamefont {S.}~\bibnamefont {Rosenkranz}},
  \bibinfo {author} {\bibfnamefont {O.}~\bibnamefont {Chmaissem}}, and\
  \bibinfo {author} {\bibfnamefont {R.}~\bibnamefont {Osborn}}} (\bibinfo
  {year} {2016}),\ \href {https://doi.org/10.1038/nphys3629} {\bibfield
  {journal} {\bibinfo  {journal} {Nat. Phys.}\ }\textbf {\bibinfo {volume}
  {12}},\ \bibinfo {pages} {493}}\BibitemShut {NoStop}%
\bibitem [{\citenamefont {Ament}\ \emph {et~al.}(2009)\citenamefont {Ament},
  \citenamefont {Ghiringhelli}, \citenamefont {Sala}, \citenamefont
  {Braicovich},\ and\ \citenamefont {van~den Brink}}]{ament2009}%
  \BibitemOpen
  \bibfield  {author} {\bibinfo {author} {\bibnamefont {Ament}, \bibfnamefont
  {L.~J.~P.}}, \bibinfo {author} {\bibfnamefont {G.}~\bibnamefont
  {Ghiringhelli}}, \bibinfo {author} {\bibfnamefont {M.~M.}\ \bibnamefont
  {Sala}}, \bibinfo {author} {\bibfnamefont {L.}~\bibnamefont {Braicovich}},
  and\ \bibinfo {author} {\bibfnamefont {J.}~\bibnamefont {van~den Brink}}}
  (\bibinfo {year} {2009}),\ \href
  {https://doi.org/10.1103/PhysRevLett.103.117003} {\bibfield  {journal}
  {\bibinfo  {journal} {Phys. Rev. Lett.}\ }\textbf {\bibinfo {volume} {103}},\
  \bibinfo {pages} {117003}}\BibitemShut {NoStop}%
\bibitem [{\citenamefont {Ament}\ \emph {et~al.}(2011)\citenamefont {Ament},
  \citenamefont {van Veenendaal}, \citenamefont {Devereaux}, \citenamefont
  {Hill},\ and\ \citenamefont {van~den Brink}}]{ament2011resonant}%
  \BibitemOpen
  \bibfield  {author} {\bibinfo {author} {\bibnamefont {Ament}, \bibfnamefont
  {L.~J.~P.}}, \bibinfo {author} {\bibfnamefont {M.}~\bibnamefont {van
  Veenendaal}}, \bibinfo {author} {\bibfnamefont {T.~P.}\ \bibnamefont
  {Devereaux}}, \bibinfo {author} {\bibfnamefont {J.~P.}\ \bibnamefont {Hill}},
  and\ \bibinfo {author} {\bibfnamefont {J.}~\bibnamefont {van~den Brink}}}
  (\bibinfo {year} {2011}),\ \href {https://doi.org/10.1103/RevModPhys.83.705}
  {\bibfield  {journal} {\bibinfo  {journal} {Rev. Mod. Phys.}\ }\textbf
  {\bibinfo {volume} {83}},\ \bibinfo {pages} {705}}\BibitemShut {NoStop}%
\bibitem [{\citenamefont {Asaba}\ \emph {et~al.}(2024)\citenamefont {Asaba},
  \citenamefont {Onishi}, \citenamefont {Kageyama}, \citenamefont {Kiyosue},
  \citenamefont {Ohtsuka}, \citenamefont {Suetsugu}, \citenamefont {Kohsaka},
  \citenamefont {Gaggl}, \citenamefont {Kasahara}, \citenamefont {Murayama},
  \citenamefont {Hashimoto}, \citenamefont {Tazai}, \citenamefont {Kontani},
  \citenamefont {Ortiz}, \citenamefont {Wilson}, \citenamefont {Li},
  \citenamefont {Wen}, \citenamefont {Shibauchi},\ and\ \citenamefont
  {Matsuda}}]{asaba2024evidence}%
  \BibitemOpen
  \bibfield  {author} {\bibinfo {author} {\bibnamefont {Asaba}, \bibfnamefont
  {T.}}, \bibinfo {author} {\bibfnamefont {A.}~\bibnamefont {Onishi}}, \bibinfo
  {author} {\bibfnamefont {Y.}~\bibnamefont {Kageyama}}, \bibinfo {author}
  {\bibfnamefont {T.}~\bibnamefont {Kiyosue}}, \bibinfo {author} {\bibfnamefont
  {K.}~\bibnamefont {Ohtsuka}}, \bibinfo {author} {\bibfnamefont
  {S.}~\bibnamefont {Suetsugu}}, \bibinfo {author} {\bibfnamefont
  {Y.}~\bibnamefont {Kohsaka}}, \bibinfo {author} {\bibfnamefont
  {T.}~\bibnamefont {Gaggl}}, \bibinfo {author} {\bibfnamefont
  {Y.}~\bibnamefont {Kasahara}}, \bibinfo {author} {\bibfnamefont
  {H.}~\bibnamefont {Murayama}}, \bibinfo {author} {\bibfnamefont
  {K.}~\bibnamefont {Hashimoto}}, \bibinfo {author} {\bibfnamefont
  {R.}~\bibnamefont {Tazai}}, \bibinfo {author} {\bibfnamefont
  {H.}~\bibnamefont {Kontani}}, \bibinfo {author} {\bibfnamefont {B.~R.}\
  \bibnamefont {Ortiz}}, \bibinfo {author} {\bibfnamefont {S.~D.}\ \bibnamefont
  {Wilson}}, \bibinfo {author} {\bibfnamefont {Q.}~\bibnamefont {Li}}, \bibinfo
  {author} {\bibfnamefont {H.-H.}\ \bibnamefont {Wen}}, \bibinfo {author}
  {\bibfnamefont {T.}~\bibnamefont {Shibauchi}}, and\ \bibinfo {author}
  {\bibfnamefont {Y.}~\bibnamefont {Matsuda}}} (\bibinfo {year} {2024}),\ \href
  {https://doi.org/10.1038/s41567-023-02272-4} {\bibfield  {journal} {\bibinfo
  {journal} {Nat. Phys.}\ }\textbf {\bibinfo {volume} {20}},\ \bibinfo {pages}
  {40}}\BibitemShut {NoStop}%
\bibitem [{\citenamefont {Avci}\ \emph {et~al.}(2013)\citenamefont {Avci},
  \citenamefont {Allred}, \citenamefont {Chmaissem}, \citenamefont {Chung},
  \citenamefont {Rosenkranz}, \citenamefont {Schlueter}, \citenamefont {Claus},
  \citenamefont {Daoud-Aladine}, \citenamefont {Khalyavin}, \citenamefont
  {Manuel}, \citenamefont {Llobet}, \citenamefont {Suchomel}, \citenamefont
  {Kanatzidis},\ and\ \citenamefont {Osborn}}]{avci2013structural}%
  \BibitemOpen
  \bibfield  {author} {\bibinfo {author} {\bibnamefont {Avci}, \bibfnamefont
  {S.}}, \bibinfo {author} {\bibfnamefont {J.~M.}\ \bibnamefont {Allred}},
  \bibinfo {author} {\bibfnamefont {O.}~\bibnamefont {Chmaissem}}, \bibinfo
  {author} {\bibfnamefont {D.~Y.}\ \bibnamefont {Chung}}, \bibinfo {author}
  {\bibfnamefont {S.}~\bibnamefont {Rosenkranz}}, \bibinfo {author}
  {\bibfnamefont {J.~A.}\ \bibnamefont {Schlueter}}, \bibinfo {author}
  {\bibfnamefont {H.}~\bibnamefont {Claus}}, \bibinfo {author} {\bibfnamefont
  {A.}~\bibnamefont {Daoud-Aladine}}, \bibinfo {author} {\bibfnamefont {D.~D.}\
  \bibnamefont {Khalyavin}}, \bibinfo {author} {\bibfnamefont {P.}~\bibnamefont
  {Manuel}}, \bibinfo {author} {\bibfnamefont {A.}~\bibnamefont {Llobet}},
  \bibinfo {author} {\bibfnamefont {M.~R.}\ \bibnamefont {Suchomel}}, \bibinfo
  {author} {\bibfnamefont {M.~G.}\ \bibnamefont {Kanatzidis}}, and\ \bibinfo
  {author} {\bibfnamefont {R.}~\bibnamefont {Osborn}}} (\bibinfo {year}
  {2013}),\ \href {https://doi.org/10.1103/PhysRevB.88.094510} {\bibfield
  {journal} {\bibinfo  {journal} {Phys. Rev. B}\ }\textbf {\bibinfo {volume}
  {88}},\ \bibinfo {pages} {094510}}\BibitemShut {NoStop}%
\bibitem [{\citenamefont {Avci}\ \emph {et~al.}(2014)\citenamefont {Avci},
  \citenamefont {Chmaissem}, \citenamefont {Allred}, \citenamefont
  {Rosenkranz}, \citenamefont {Eremin}, \citenamefont {Chubukov}, \citenamefont
  {Bugaris}, \citenamefont {Chung}, \citenamefont {Kanatzidis}, \citenamefont
  {Castellan}, \citenamefont {Schlueter}, \citenamefont {Claus}, \citenamefont
  {Khalyavin}, \citenamefont {Manuel}, \citenamefont {Daoud-Aladine},\ and\
  \citenamefont {Osborn}}]{avci2014magnetically}%
  \BibitemOpen
  \bibfield  {author} {\bibinfo {author} {\bibnamefont {Avci}, \bibfnamefont
  {S.}}, \bibinfo {author} {\bibfnamefont {O.}~\bibnamefont {Chmaissem}},
  \bibinfo {author} {\bibfnamefont {J.}~\bibnamefont {Allred}}, \bibinfo
  {author} {\bibfnamefont {S.}~\bibnamefont {Rosenkranz}}, \bibinfo {author}
  {\bibfnamefont {I.}~\bibnamefont {Eremin}}, \bibinfo {author} {\bibfnamefont
  {A.}~\bibnamefont {Chubukov}}, \bibinfo {author} {\bibfnamefont
  {D.}~\bibnamefont {Bugaris}}, \bibinfo {author} {\bibfnamefont
  {D.}~\bibnamefont {Chung}}, \bibinfo {author} {\bibfnamefont
  {M.}~\bibnamefont {Kanatzidis}}, \bibinfo {author} {\bibfnamefont {J.-P.}\
  \bibnamefont {Castellan}}, \bibinfo {author} {\bibfnamefont {J.}~\bibnamefont
  {Schlueter}}, \bibinfo {author} {\bibfnamefont {H.}~\bibnamefont {Claus}},
  \bibinfo {author} {\bibfnamefont {D.}~\bibnamefont {Khalyavin}}, \bibinfo
  {author} {\bibfnamefont {P.}~\bibnamefont {Manuel}}, \bibinfo {author}
  {\bibfnamefont {A.}~\bibnamefont {Daoud-Aladine}}, and\ \bibinfo {author}
  {\bibfnamefont {R.}~\bibnamefont {Osborn}}} (\bibinfo {year} {2014}),\ \href
  {https://doi.org/10.1038/ncomms4845} {\bibfield  {journal} {\bibinfo
  {journal} {Nat. Commun.}\ }\textbf {\bibinfo {volume} {5}},\ \bibinfo {pages}
  {3845}}\BibitemShut {NoStop}%
\bibitem [{\citenamefont {Bao}\ \emph {et~al.}(2011)\citenamefont {Bao},
  \citenamefont {Huang}, \citenamefont {Chen}, \citenamefont {Wang},
  \citenamefont {He},\ and\ \citenamefont {Qiu}}]{Bao2011}%
  \BibitemOpen
  \bibfield  {author} {\bibinfo {author} {\bibnamefont {Bao}, \bibfnamefont
  {W.}}, \bibinfo {author} {\bibfnamefont {Q.-Z.}\ \bibnamefont {Huang}},
  \bibinfo {author} {\bibfnamefont {G.-F.}\ \bibnamefont {Chen}}, \bibinfo
  {author} {\bibfnamefont {D.-M.}\ \bibnamefont {Wang}}, \bibinfo {author}
  {\bibfnamefont {J.-B.}\ \bibnamefont {He}}, and\ \bibinfo {author}
  {\bibfnamefont {Y.-M.}\ \bibnamefont {Qiu}}} (\bibinfo {year} {2011}),\ \href
  {https://doi.org/10.1088/0256-307x/28/8/086104} {\bibfield  {journal}
  {\bibinfo  {journal} {Chin. Phys. Lett.}\ }\textbf {\bibinfo {volume} {28}},\
  \bibinfo {pages} {086104}}\BibitemShut {NoStop}%
\bibitem [{\citenamefont {Bao}\ \emph {et~al.}(2009)\citenamefont {Bao},
  \citenamefont {Qiu}, \citenamefont {Huang}, \citenamefont {Green},
  \citenamefont {Zajdel}, \citenamefont {Fitzsimmons}, \citenamefont
  {Zhernenkov}, \citenamefont {Chang}, \citenamefont {Fang}, \citenamefont
  {Qian}, \citenamefont {Vehstedt}, \citenamefont {Yang}, \citenamefont {Pham},
  \citenamefont {Spinu},\ and\ \citenamefont {Mao}}]{Bao2009}%
  \BibitemOpen
  \bibfield  {author} {\bibinfo {author} {\bibnamefont {Bao}, \bibfnamefont
  {W.}}, \bibinfo {author} {\bibfnamefont {Y.}~\bibnamefont {Qiu}}, \bibinfo
  {author} {\bibfnamefont {Q.}~\bibnamefont {Huang}}, \bibinfo {author}
  {\bibfnamefont {M.~A.}\ \bibnamefont {Green}}, \bibinfo {author}
  {\bibfnamefont {P.}~\bibnamefont {Zajdel}}, \bibinfo {author} {\bibfnamefont
  {M.~R.}\ \bibnamefont {Fitzsimmons}}, \bibinfo {author} {\bibfnamefont
  {M.}~\bibnamefont {Zhernenkov}}, \bibinfo {author} {\bibfnamefont
  {S.}~\bibnamefont {Chang}}, \bibinfo {author} {\bibfnamefont
  {M.}~\bibnamefont {Fang}}, \bibinfo {author} {\bibfnamefont {B.}~\bibnamefont
  {Qian}}, \bibinfo {author} {\bibfnamefont {E.~K.}\ \bibnamefont {Vehstedt}},
  \bibinfo {author} {\bibfnamefont {J.}~\bibnamefont {Yang}}, \bibinfo {author}
  {\bibfnamefont {H.~M.}\ \bibnamefont {Pham}}, \bibinfo {author}
  {\bibfnamefont {L.}~\bibnamefont {Spinu}}, and\ \bibinfo {author}
  {\bibfnamefont {Z.~Q.}\ \bibnamefont {Mao}}} (\bibinfo {year} {2009}),\ \href
  {https://doi.org/10.1103/PhysRevLett.102.247001} {\bibfield  {journal}
  {\bibinfo  {journal} {Phys. Rev. Lett.}\ }\textbf {\bibinfo {volume} {102}},\
  \bibinfo {pages} {247001}}\BibitemShut {NoStop}%
\bibitem [{\citenamefont {Barber}\ \emph {et~al.}(2019)\citenamefont {Barber},
  \citenamefont {Steppke}, \citenamefont {Mackenzie},\ and\ \citenamefont
  {Hicks}}]{barber2019piezoelectric}%
  \BibitemOpen
  \bibfield  {author} {\bibinfo {author} {\bibnamefont {Barber}, \bibfnamefont
  {M.~E.}}, \bibinfo {author} {\bibfnamefont {A.}~\bibnamefont {Steppke}},
  \bibinfo {author} {\bibfnamefont {A.~P.}\ \bibnamefont {Mackenzie}}, and\
  \bibinfo {author} {\bibfnamefont {C.~W.}\ \bibnamefont {Hicks}}} (\bibinfo
  {year} {2019}),\ \href {https://doi.org/10.1063/1.5075485} {\bibfield
  {journal} {\bibinfo  {journal} {Rev. Sci. Instrum.}\ }\textbf {\bibinfo
  {volume} {90}},\ \bibinfo {pages} {023904}}\BibitemShut {NoStop}%
\bibitem [{\citenamefont {Bartlett}\ \emph {et~al.}(2021)\citenamefont
  {Bartlett}, \citenamefont {Steppke}, \citenamefont {Hosoi}, \citenamefont
  {Noad}, \citenamefont {Park}, \citenamefont {Timm}, \citenamefont
  {Shibauchi}, \citenamefont {Mackenzie},\ and\ \citenamefont
  {Hicks}}]{bartlett2021relationship}%
  \BibitemOpen
  \bibfield  {author} {\bibinfo {author} {\bibnamefont {Bartlett},
  \bibfnamefont {J.~M.}}, \bibinfo {author} {\bibfnamefont {A.}~\bibnamefont
  {Steppke}}, \bibinfo {author} {\bibfnamefont {S.}~\bibnamefont {Hosoi}},
  \bibinfo {author} {\bibfnamefont {H.}~\bibnamefont {Noad}}, \bibinfo {author}
  {\bibfnamefont {J.}~\bibnamefont {Park}}, \bibinfo {author} {\bibfnamefont
  {C.}~\bibnamefont {Timm}}, \bibinfo {author} {\bibfnamefont {T.}~\bibnamefont
  {Shibauchi}}, \bibinfo {author} {\bibfnamefont {A.~P.}\ \bibnamefont
  {Mackenzie}}, and\ \bibinfo {author} {\bibfnamefont {C.~W.}\ \bibnamefont
  {Hicks}}} (\bibinfo {year} {2021}),\ \href
  {https://doi.org/10.1103/PhysRevX.11.021038} {\bibfield  {journal} {\bibinfo
  {journal} {Phys. Rev. X}\ }\textbf {\bibinfo {volume} {11}},\ \bibinfo
  {pages} {021038}}\BibitemShut {NoStop}%
\bibitem [{\citenamefont {Benfatto}\ \emph {et~al.}(2018)\citenamefont
  {Benfatto}, \citenamefont {Valenzuela},\ and\ \citenamefont
  {Fanfarillo}}]{benfatto2018nematic}%
  \BibitemOpen
  \bibfield  {author} {\bibinfo {author} {\bibnamefont {Benfatto},
  \bibfnamefont {L.}}, \bibinfo {author} {\bibfnamefont {B.}~\bibnamefont
  {Valenzuela}}, and\ \bibinfo {author} {\bibfnamefont {L.}~\bibnamefont
  {Fanfarillo}}} (\bibinfo {year} {2018}),\ \href
  {https://doi.org/10.1038/s41535-018-0129-9} {\bibfield  {journal} {\bibinfo
  {journal} {npj Quantum Mater.}\ }\textbf {\bibinfo {volume} {3}},\ \bibinfo
  {pages} {56}}\BibitemShut {NoStop}%
\bibitem [{\citenamefont {B{\l}achowski}\ \emph {et~al.}(2011)\citenamefont
  {B{\l}achowski}, \citenamefont {Ruebenbauer}, \citenamefont {{\.Z}ukrowski},
  \citenamefont {Bukowski}, \citenamefont {Rogacki}, \citenamefont {Moll},\
  and\ \citenamefont {Karpinski}}]{Blachowski2011}%
  \BibitemOpen
  \bibfield  {author} {\bibinfo {author} {\bibnamefont {B{\l}achowski},
  \bibfnamefont {A.}}, \bibinfo {author} {\bibfnamefont {K.}~\bibnamefont
  {Ruebenbauer}}, \bibinfo {author} {\bibfnamefont {J.}~\bibnamefont
  {{\.Z}ukrowski}}, \bibinfo {author} {\bibfnamefont {Z.}~\bibnamefont
  {Bukowski}}, \bibinfo {author} {\bibfnamefont {K.}~\bibnamefont {Rogacki}},
  \bibinfo {author} {\bibfnamefont {P.}~\bibnamefont {Moll}}, and\ \bibinfo
  {author} {\bibfnamefont {J.}~\bibnamefont {Karpinski}}} (\bibinfo {year}
  {2011}),\ \href {https://doi.org/10.1103/PhysRevB.84.174503} {\bibfield
  {journal} {\bibinfo  {journal} {Phys. Rev. B}\ }\textbf {\bibinfo {volume}
  {84}},\ \bibinfo {pages} {174503}}\BibitemShut {NoStop}%
\bibitem [{\citenamefont {B{\"o}hmer}\ \emph
  {et~al.}(2015{\natexlab{a}})\citenamefont {B{\"o}hmer}, \citenamefont {Arai},
  \citenamefont {Hardy}, \citenamefont {Hattori}, \citenamefont {Iye},
  \citenamefont {Wolf}, \citenamefont {v.~L{\"o}hneysen}, \citenamefont
  {Ishida},\ and\ \citenamefont {Meingast}}]{bohmer2015origin}%
  \BibitemOpen
  \bibfield  {author} {\bibinfo {author} {\bibnamefont {B{\"o}hmer},
  \bibfnamefont {A.~E.}}, \bibinfo {author} {\bibfnamefont {T.}~\bibnamefont
  {Arai}}, \bibinfo {author} {\bibfnamefont {F.}~\bibnamefont {Hardy}},
  \bibinfo {author} {\bibfnamefont {T.}~\bibnamefont {Hattori}}, \bibinfo
  {author} {\bibfnamefont {T.}~\bibnamefont {Iye}}, \bibinfo {author}
  {\bibfnamefont {T.}~\bibnamefont {Wolf}}, \bibinfo {author} {\bibfnamefont
  {H.}~\bibnamefont {v.~L{\"o}hneysen}}, \bibinfo {author} {\bibfnamefont
  {K.}~\bibnamefont {Ishida}}, and\ \bibinfo {author} {\bibfnamefont
  {C.}~\bibnamefont {Meingast}}} (\bibinfo {year} {2015}{\natexlab{a}}),\ \href
  {https://doi.org/10.1103/PhysRevLett.114.027001} {\bibfield  {journal}
  {\bibinfo  {journal} {Phys. Rev. Lett.}\ }\textbf {\bibinfo {volume} {114}},\
  \bibinfo {pages} {027001}}\BibitemShut {NoStop}%
\bibitem [{\citenamefont {B\"ohmer}\ \emph {et~al.}(2014)\citenamefont
  {B\"ohmer}, \citenamefont {Burger}, \citenamefont {Hardy}, \citenamefont
  {Wolf}, \citenamefont {Schweiss}, \citenamefont {Fromknecht}, \citenamefont
  {Reinecker}, \citenamefont {Schranz},\ and\ \citenamefont
  {Meingast}}]{bohmer2014nematic}%
  \BibitemOpen
  \bibfield  {author} {\bibinfo {author} {\bibnamefont {B\"ohmer},
  \bibfnamefont {A.~E.}}, \bibinfo {author} {\bibfnamefont {P.}~\bibnamefont
  {Burger}}, \bibinfo {author} {\bibfnamefont {F.}~\bibnamefont {Hardy}},
  \bibinfo {author} {\bibfnamefont {T.}~\bibnamefont {Wolf}}, \bibinfo {author}
  {\bibfnamefont {P.}~\bibnamefont {Schweiss}}, \bibinfo {author}
  {\bibfnamefont {R.}~\bibnamefont {Fromknecht}}, \bibinfo {author}
  {\bibfnamefont {M.}~\bibnamefont {Reinecker}}, \bibinfo {author}
  {\bibfnamefont {W.}~\bibnamefont {Schranz}}, and\ \bibinfo {author}
  {\bibfnamefont {C.}~\bibnamefont {Meingast}}} (\bibinfo {year} {2014}),\
  \href {https://doi.org/10.1103/PhysRevLett.112.047001} {\bibfield  {journal}
  {\bibinfo  {journal} {Phys. Rev. Lett.}\ }\textbf {\bibinfo {volume} {112}},\
  \bibinfo {pages} {047001}}\BibitemShut {NoStop}%
\bibitem [{\citenamefont {B{\"o}hmer}\ \emph {et~al.}(2022)\citenamefont
  {B{\"o}hmer}, \citenamefont {Chu}, \citenamefont {Lederer},\ and\
  \citenamefont {Yi}}]{bohmer2022nematicity}%
  \BibitemOpen
  \bibfield  {author} {\bibinfo {author} {\bibnamefont {B{\"o}hmer},
  \bibfnamefont {A.~E.}}, \bibinfo {author} {\bibfnamefont {J.-H.}\
  \bibnamefont {Chu}}, \bibinfo {author} {\bibfnamefont {S.}~\bibnamefont
  {Lederer}}, and\ \bibinfo {author} {\bibfnamefont {M.}~\bibnamefont {Yi}}}
  (\bibinfo {year} {2022}),\ \href {https://doi.org/10.1038/s41567-022-01833-3}
  {\bibfield  {journal} {\bibinfo  {journal} {Nat. Phys.}\ }\textbf {\bibinfo
  {volume} {18}},\ \bibinfo {pages} {1412}}\BibitemShut {NoStop}%
\bibitem [{\citenamefont {B{\"o}hmer}\ \emph
  {et~al.}(2015{\natexlab{b}})\citenamefont {B{\"o}hmer}, \citenamefont
  {Hardy}, \citenamefont {Wang}, \citenamefont {Wolf}, \citenamefont
  {Schweiss},\ and\ \citenamefont {Meingast}}]{bohmer2015superconductivity}%
  \BibitemOpen
  \bibfield  {author} {\bibinfo {author} {\bibnamefont {B{\"o}hmer},
  \bibfnamefont {A.~E.}}, \bibinfo {author} {\bibfnamefont {F.}~\bibnamefont
  {Hardy}}, \bibinfo {author} {\bibfnamefont {L.}~\bibnamefont {Wang}},
  \bibinfo {author} {\bibfnamefont {T.}~\bibnamefont {Wolf}}, \bibinfo {author}
  {\bibfnamefont {P.}~\bibnamefont {Schweiss}}, and\ \bibinfo {author}
  {\bibfnamefont {C.}~\bibnamefont {Meingast}}} (\bibinfo {year}
  {2015}{\natexlab{b}}),\ \href {https://doi.org/10.1038/ncomms8911} {\bibfield
   {journal} {\bibinfo  {journal} {Nat. Commun.}\ }\textbf {\bibinfo {volume}
  {6}},\ \bibinfo {pages} {7911}}\BibitemShut {NoStop}%
\bibitem [{\citenamefont {B{\"o}hmer}\ \emph {et~al.}(2019)\citenamefont
  {B{\"o}hmer}, \citenamefont {Kothapalli}, \citenamefont {Jayasekara},
  \citenamefont {Wilde}, \citenamefont {Li}, \citenamefont {Sapkota},
  \citenamefont {Ueland}, \citenamefont {Das}, \citenamefont {Xiao},
  \citenamefont {Bi}, \citenamefont {Zhao}, \citenamefont {Alp}, \citenamefont
  {Bud'ko}, \citenamefont {Canfield}, \citenamefont {Goldman},\ and\
  \citenamefont {Kreyssig}}]{bohmer2019distinct}%
  \BibitemOpen
  \bibfield  {author} {\bibinfo {author} {\bibnamefont {B{\"o}hmer},
  \bibfnamefont {A.~E.}}, \bibinfo {author} {\bibfnamefont {K.}~\bibnamefont
  {Kothapalli}}, \bibinfo {author} {\bibfnamefont {W.~T.}\ \bibnamefont
  {Jayasekara}}, \bibinfo {author} {\bibfnamefont {J.~M.}\ \bibnamefont
  {Wilde}}, \bibinfo {author} {\bibfnamefont {B.}~\bibnamefont {Li}}, \bibinfo
  {author} {\bibfnamefont {A.}~\bibnamefont {Sapkota}}, \bibinfo {author}
  {\bibfnamefont {B.~G.}\ \bibnamefont {Ueland}}, \bibinfo {author}
  {\bibfnamefont {P.}~\bibnamefont {Das}}, \bibinfo {author} {\bibfnamefont
  {Y.}~\bibnamefont {Xiao}}, \bibinfo {author} {\bibfnamefont {W.}~\bibnamefont
  {Bi}}, \bibinfo {author} {\bibfnamefont {J.}~\bibnamefont {Zhao}}, \bibinfo
  {author} {\bibfnamefont {E.~E.}\ \bibnamefont {Alp}}, \bibinfo {author}
  {\bibfnamefont {S.~L.}\ \bibnamefont {Bud'ko}}, \bibinfo {author}
  {\bibfnamefont {P.~C.}\ \bibnamefont {Canfield}}, \bibinfo {author}
  {\bibfnamefont {A.~I.}\ \bibnamefont {Goldman}}, and\ \bibinfo {author}
  {\bibfnamefont {A.}~\bibnamefont {Kreyssig}}} (\bibinfo {year} {2019}),\
  \href {https://doi.org/10.1103/PhysRevB.100.064515} {\bibfield  {journal}
  {\bibinfo  {journal} {Phys. Rev. B}\ }\textbf {\bibinfo {volume} {100}},\
  \bibinfo {pages} {064515}}\BibitemShut {NoStop}%
\bibitem [{\citenamefont {B\"ohmer}\ and\ \citenamefont
  {Meingast}(2016)}]{bohmer2016electronic}%
  \BibitemOpen
  \bibfield  {author} {\bibinfo {author} {\bibnamefont {B\"ohmer},
  \bibfnamefont {A.~E.}}, and\ \bibinfo {author} {\bibfnamefont
  {C.}~\bibnamefont {Meingast}}} (\bibinfo {year} {2016}),\ \href
  {https://doi.org/10.1016/j.crhy.2015.07.001} {\bibfield  {journal} {\bibinfo
  {journal} {C. R. Physique}\ }\textbf {\bibinfo {volume} {17}},\ \bibinfo
  {pages} {90}}\BibitemShut {NoStop}%
\bibitem [{\citenamefont {Borisov}\ \emph {et~al.}(2018)\citenamefont
  {Borisov}, \citenamefont {Canfield},\ and\ \citenamefont
  {Valent{\'i}}}]{Borisov2018}%
  \BibitemOpen
  \bibfield  {author} {\bibinfo {author} {\bibnamefont {Borisov}, \bibfnamefont
  {V.}}, \bibinfo {author} {\bibfnamefont {P.~C.}\ \bibnamefont {Canfield}},
  and\ \bibinfo {author} {\bibfnamefont {R.}~\bibnamefont {Valent{\'i}}}}
  (\bibinfo {year} {2018}),\ \href {https://doi.org/10.1103/PhysRevB.98.064104}
  {\bibfield  {journal} {\bibinfo  {journal} {Phys. Rev. B}\ }\textbf {\bibinfo
  {volume} {98}},\ \bibinfo {pages} {064104}}\BibitemShut {NoStop}%
\bibitem [{\citenamefont {Braicovich}\ \emph {et~al.}(2009)\citenamefont
  {Braicovich}, \citenamefont {Ament}, \citenamefont {Bisogni}, \citenamefont
  {Forte}, \citenamefont {Aruta}, \citenamefont {Balestrino}, \citenamefont
  {Brookes}, \citenamefont {De~Luca}, \citenamefont {Medaglia}, \citenamefont
  {Granozio}, \citenamefont {Radovic}, \citenamefont {Salluzzo}, \citenamefont
  {van~den Brink},\ and\ \citenamefont
  {Ghiringhelli}}]{braicovich2009dispersion}%
  \BibitemOpen
  \bibfield  {author} {\bibinfo {author} {\bibnamefont {Braicovich},
  \bibfnamefont {L.}}, \bibinfo {author} {\bibfnamefont {L.~J.~P.}\
  \bibnamefont {Ament}}, \bibinfo {author} {\bibfnamefont {V.}~\bibnamefont
  {Bisogni}}, \bibinfo {author} {\bibfnamefont {F.}~\bibnamefont {Forte}},
  \bibinfo {author} {\bibfnamefont {C.}~\bibnamefont {Aruta}}, \bibinfo
  {author} {\bibfnamefont {G.}~\bibnamefont {Balestrino}}, \bibinfo {author}
  {\bibfnamefont {N.~B.}\ \bibnamefont {Brookes}}, \bibinfo {author}
  {\bibfnamefont {G.~M.}\ \bibnamefont {De~Luca}}, \bibinfo {author}
  {\bibfnamefont {P.~G.}\ \bibnamefont {Medaglia}}, \bibinfo {author}
  {\bibfnamefont {F.~M.}\ \bibnamefont {Granozio}}, \bibinfo {author}
  {\bibfnamefont {M.}~\bibnamefont {Radovic}}, \bibinfo {author} {\bibfnamefont
  {M.}~\bibnamefont {Salluzzo}}, \bibinfo {author} {\bibfnamefont
  {J.}~\bibnamefont {van~den Brink}}, and\ \bibinfo {author} {\bibfnamefont
  {G.}~\bibnamefont {Ghiringhelli}}} (\bibinfo {year} {2009}),\ \href
  {https://doi.org/10.1103/PhysRevLett.102.167401} {\bibfield  {journal}
  {\bibinfo  {journal} {Phys. Rev. Lett.}\ }\textbf {\bibinfo {volume} {102}},\
  \bibinfo {pages} {167401}}\BibitemShut {NoStop}%
\bibitem [{\citenamefont {Bristow}\ \emph {et~al.}(2020)\citenamefont
  {Bristow}, \citenamefont {Knafo}, \citenamefont {Reiss}, \citenamefont
  {Meier}, \citenamefont {Canfield}, \citenamefont {Blundell},\ and\
  \citenamefont {Coldea}}]{Bristow2020}%
  \BibitemOpen
  \bibfield  {author} {\bibinfo {author} {\bibnamefont {Bristow}, \bibfnamefont
  {M.}}, \bibinfo {author} {\bibfnamefont {W.}~\bibnamefont {Knafo}}, \bibinfo
  {author} {\bibfnamefont {P.}~\bibnamefont {Reiss}}, \bibinfo {author}
  {\bibfnamefont {W.}~\bibnamefont {Meier}}, \bibinfo {author} {\bibfnamefont
  {P.~C.}\ \bibnamefont {Canfield}}, \bibinfo {author} {\bibfnamefont {S.~J.}\
  \bibnamefont {Blundell}}, and\ \bibinfo {author} {\bibfnamefont {A.~I.}\
  \bibnamefont {Coldea}}} (\bibinfo {year} {2020}),\ \href
  {https://doi.org/10.1103/PhysRevB.101.134502} {\bibfield  {journal} {\bibinfo
   {journal} {Phys. Rev. B}\ }\textbf {\bibinfo {volume} {101}},\ \bibinfo
  {pages} {134502}}\BibitemShut {NoStop}%
\bibitem [{\citenamefont {Brookes}\ \emph {et~al.}(2018)\citenamefont
  {Brookes}, \citenamefont {Yakhou-Harris}, \citenamefont {Kummer},
  \citenamefont {Fondacaro}, \citenamefont {Cezar}, \citenamefont {Betto},
  \citenamefont {Velez-Fort}, \citenamefont {Amorese}, \citenamefont
  {Ghiringhelli}, \citenamefont {Braicovich}, \citenamefont {Barrett},
  \citenamefont {Berruyer}, \citenamefont {Cianciosi}, \citenamefont {Eybert},
  \citenamefont {Marion}, \citenamefont {{van der Linden}},\ and\ \citenamefont
  {Zhang}}]{brookes2018beamline}%
  \BibitemOpen
  \bibfield  {author} {\bibinfo {author} {\bibnamefont {Brookes}, \bibfnamefont
  {N.}}, \bibinfo {author} {\bibfnamefont {F.}~\bibnamefont {Yakhou-Harris}},
  \bibinfo {author} {\bibfnamefont {K.}~\bibnamefont {Kummer}}, \bibinfo
  {author} {\bibfnamefont {A.}~\bibnamefont {Fondacaro}}, \bibinfo {author}
  {\bibfnamefont {J.}~\bibnamefont {Cezar}}, \bibinfo {author} {\bibfnamefont
  {D.}~\bibnamefont {Betto}}, \bibinfo {author} {\bibfnamefont
  {E.}~\bibnamefont {Velez-Fort}}, \bibinfo {author} {\bibfnamefont
  {A.}~\bibnamefont {Amorese}}, \bibinfo {author} {\bibfnamefont
  {G.}~\bibnamefont {Ghiringhelli}}, \bibinfo {author} {\bibfnamefont
  {L.}~\bibnamefont {Braicovich}}, \bibinfo {author} {\bibfnamefont
  {R.}~\bibnamefont {Barrett}}, \bibinfo {author} {\bibfnamefont
  {G.}~\bibnamefont {Berruyer}}, \bibinfo {author} {\bibfnamefont
  {F.}~\bibnamefont {Cianciosi}}, \bibinfo {author} {\bibfnamefont
  {L.}~\bibnamefont {Eybert}}, \bibinfo {author} {\bibfnamefont
  {P.}~\bibnamefont {Marion}}, \bibinfo {author} {\bibfnamefont
  {P.}~\bibnamefont {{van der Linden}}}, and\ \bibinfo {author} {\bibfnamefont
  {L.}~\bibnamefont {Zhang}}} (\bibinfo {year} {2018}),\ \href
  {https://doi.org/10.1016/j.nima.2018.07.001} {\bibfield  {journal} {\bibinfo
  {journal} {Nuclear Inst. and Methods in Physics Research, A}\ }\textbf
  {\bibinfo {volume} {903}},\ \bibinfo {pages} {175}}\BibitemShut {NoStop}%
\bibitem [{\citenamefont {Bukowski}\ \emph {et~al.}(2010)\citenamefont
  {Bukowski}, \citenamefont {Weyeneth}, \citenamefont {Puzniak}, \citenamefont
  {Karpinski},\ and\ \citenamefont {Batlogg}}]{Bukowski2010}%
  \BibitemOpen
  \bibfield  {author} {\bibinfo {author} {\bibnamefont {Bukowski},
  \bibfnamefont {Z.}}, \bibinfo {author} {\bibfnamefont {S.}~\bibnamefont
  {Weyeneth}}, \bibinfo {author} {\bibfnamefont {R.}~\bibnamefont {Puzniak}},
  \bibinfo {author} {\bibfnamefont {J.}~\bibnamefont {Karpinski}}, and\
  \bibinfo {author} {\bibfnamefont {B.}~\bibnamefont {Batlogg}}} (\bibinfo
  {year} {2010}),\ \href {https://doi.org/10.1016/j.physc.2009.11.103}
  {\bibfield  {journal} {\bibinfo  {journal} {Physica C}\ }\textbf {\bibinfo
  {volume} {470}},\ \bibinfo {pages} {S328}}\BibitemShut {NoStop}%
\bibitem [{\citenamefont {Bukowski}\ \emph {et~al.}(2009)\citenamefont
  {Bukowski}, \citenamefont {Weyeneth}, \citenamefont {Puzniak}, \citenamefont
  {Moll}, \citenamefont {Katrych}, \citenamefont {Zhigadlo}, \citenamefont
  {Karpinski}, \citenamefont {Keller},\ and\ \citenamefont
  {Batlogg}}]{Bukowski2009}%
  \BibitemOpen
  \bibfield  {author} {\bibinfo {author} {\bibnamefont {Bukowski},
  \bibfnamefont {Z.}}, \bibinfo {author} {\bibfnamefont {S.}~\bibnamefont
  {Weyeneth}}, \bibinfo {author} {\bibfnamefont {R.}~\bibnamefont {Puzniak}},
  \bibinfo {author} {\bibfnamefont {P.}~\bibnamefont {Moll}}, \bibinfo {author}
  {\bibfnamefont {S.}~\bibnamefont {Katrych}}, \bibinfo {author} {\bibfnamefont
  {N.}~\bibnamefont {Zhigadlo}}, \bibinfo {author} {\bibfnamefont
  {J.}~\bibnamefont {Karpinski}}, \bibinfo {author} {\bibfnamefont
  {H.}~\bibnamefont {Keller}}, and\ \bibinfo {author} {\bibfnamefont
  {B.}~\bibnamefont {Batlogg}}} (\bibinfo {year} {2009}),\ \href
  {https://doi.org/10.1103/PhysRevB.79.104521} {\bibfield  {journal} {\bibinfo
  {journal} {Phys. Rev. B}\ }\textbf {\bibinfo {volume} {79}},\ \bibinfo
  {pages} {104521}}\BibitemShut {NoStop}%
\bibitem [{\citenamefont {Burger}\ \emph {et~al.}(2013)\citenamefont {Burger},
  \citenamefont {Hardy}, \citenamefont {Aoki}, \citenamefont {B\"ohmer},
  \citenamefont {Eder}, \citenamefont {Heid}, \citenamefont {Wolf},
  \citenamefont {Schweiss}, \citenamefont {Fromknecht}, \citenamefont
  {Jackson}, \citenamefont {Paulsen},\ and\ \citenamefont
  {Meingast}}]{burger2013strong}%
  \BibitemOpen
  \bibfield  {author} {\bibinfo {author} {\bibnamefont {Burger}, \bibfnamefont
  {P.}}, \bibinfo {author} {\bibfnamefont {F.}~\bibnamefont {Hardy}}, \bibinfo
  {author} {\bibfnamefont {D.}~\bibnamefont {Aoki}}, \bibinfo {author}
  {\bibfnamefont {A.~E.}\ \bibnamefont {B\"ohmer}}, \bibinfo {author}
  {\bibfnamefont {R.}~\bibnamefont {Eder}}, \bibinfo {author} {\bibfnamefont
  {R.}~\bibnamefont {Heid}}, \bibinfo {author} {\bibfnamefont {T.}~\bibnamefont
  {Wolf}}, \bibinfo {author} {\bibfnamefont {P.}~\bibnamefont {Schweiss}},
  \bibinfo {author} {\bibfnamefont {R.}~\bibnamefont {Fromknecht}}, \bibinfo
  {author} {\bibfnamefont {M.~J.}\ \bibnamefont {Jackson}}, \bibinfo {author}
  {\bibfnamefont {C.}~\bibnamefont {Paulsen}}, and\ \bibinfo {author}
  {\bibfnamefont {C.}~\bibnamefont {Meingast}}} (\bibinfo {year} {2013}),\
  \href {https://doi.org/10.1103/PhysRevB.88.014517} {\bibfield  {journal}
  {\bibinfo  {journal} {Phys. Rev. B}\ }\textbf {\bibinfo {volume} {88}},\
  \bibinfo {pages} {014517}}\BibitemShut {NoStop}%
\bibitem [{\citenamefont {Burrard-Lucas}\ \emph {et~al.}(2013)\citenamefont
  {Burrard-Lucas}, \citenamefont {Free}, \citenamefont {Sedlmaier},
  \citenamefont {Wright}, \citenamefont {Cassidy}, \citenamefont {Hara},
  \citenamefont {Corkett}, \citenamefont {Lancaster}, \citenamefont {Baker},
  \citenamefont {Blundell},\ and\ \citenamefont {Clarke}}]{Burrard-Lucas2013}%
  \BibitemOpen
  \bibfield  {author} {\bibinfo {author} {\bibnamefont {Burrard-Lucas},
  \bibfnamefont {M.}}, \bibinfo {author} {\bibfnamefont {D.~G.}\ \bibnamefont
  {Free}}, \bibinfo {author} {\bibfnamefont {S.~J.}\ \bibnamefont {Sedlmaier}},
  \bibinfo {author} {\bibfnamefont {J.~D.}\ \bibnamefont {Wright}}, \bibinfo
  {author} {\bibfnamefont {S.~J.}\ \bibnamefont {Cassidy}}, \bibinfo {author}
  {\bibfnamefont {Y.}~\bibnamefont {Hara}}, \bibinfo {author} {\bibfnamefont
  {A.~J.}\ \bibnamefont {Corkett}}, \bibinfo {author} {\bibfnamefont
  {T.}~\bibnamefont {Lancaster}}, \bibinfo {author} {\bibfnamefont {P.~J.}\
  \bibnamefont {Baker}}, \bibinfo {author} {\bibfnamefont {S.~J.}\ \bibnamefont
  {Blundell}}, and\ \bibinfo {author} {\bibfnamefont {S.~J.}\ \bibnamefont
  {Clarke}}} (\bibinfo {year} {2013}),\ \href
  {https://doi.org/10.1038/nmat3464} {\bibfield  {journal} {\bibinfo  {journal}
  {Nat. Mater.}\ }\textbf {\bibinfo {volume} {12}},\ \bibinfo {pages}
  {15}}\BibitemShut {NoStop}%
\bibitem [{\citenamefont {Cantarino}\ \emph {et~al.}(2026)\citenamefont
  {Cantarino}, \citenamefont {Teixeira}, \citenamefont {Pakuszewski},
  \citenamefont {Neto}, \citenamefont {de~Abrantes}, \citenamefont
  {Garcia-Fernandez}, \citenamefont {Pagliuso}, \citenamefont {Adriano},
  \citenamefont {Monney}, \citenamefont {Schmitt}, \citenamefont {Andrade},\
  and\ \citenamefont {Garcia}}]{cantarino2026disorder}%
  \BibitemOpen
  \bibfield  {author} {\bibinfo {author} {\bibnamefont {Cantarino},
  \bibfnamefont {M.~R.}}, \bibinfo {author} {\bibfnamefont {R.~M.~P.}\
  \bibnamefont {Teixeira}}, \bibinfo {author} {\bibfnamefont {K.~R.}\
  \bibnamefont {Pakuszewski}}, \bibinfo {author} {\bibfnamefont {W.~R. d.~S.}\
  \bibnamefont {Neto}}, \bibinfo {author} {\bibfnamefont {J.~G.}\ \bibnamefont
  {de~Abrantes}}, \bibinfo {author} {\bibfnamefont {M.}~\bibnamefont
  {Garcia-Fernandez}}, \bibinfo {author} {\bibfnamefont {P.~G.}\ \bibnamefont
  {Pagliuso}}, \bibinfo {author} {\bibfnamefont {C.}~\bibnamefont {Adriano}},
  \bibinfo {author} {\bibfnamefont {C.}~\bibnamefont {Monney}}, \bibinfo
  {author} {\bibfnamefont {T.}~\bibnamefont {Schmitt}}, \bibinfo {author}
  {\bibfnamefont {E.~C.}\ \bibnamefont {Andrade}}, and\ \bibinfo {author}
  {\bibfnamefont {F.~A.}\ \bibnamefont {Garcia}}} (\bibinfo {year} {2026}),\
  \href {https://doi.org/10.1103/rkjn-hf7z} {\bibfield  {journal} {\bibinfo
  {journal} {Phys. Rev. Res.}\ }\textbf {\bibinfo {volume} {8}},\ \bibinfo
  {pages} {L012028}}\BibitemShut {NoStop}%
\bibitem [{\citenamefont {Cao}\ \emph {et~al.}(2011)\citenamefont {Cao},
  \citenamefont {Xu}, \citenamefont {Ren}, \citenamefont {Jiang}, \citenamefont
  {Feng},\ and\ \citenamefont {Xu}}]{Cao2011}%
  \BibitemOpen
  \bibfield  {author} {\bibinfo {author} {\bibnamefont {Cao}, \bibfnamefont
  {G.}}, \bibinfo {author} {\bibfnamefont {S.}~\bibnamefont {Xu}}, \bibinfo
  {author} {\bibfnamefont {Z.}~\bibnamefont {Ren}}, \bibinfo {author}
  {\bibfnamefont {S.}~\bibnamefont {Jiang}}, \bibinfo {author} {\bibfnamefont
  {C.}~\bibnamefont {Feng}}, and\ \bibinfo {author} {\bibfnamefont
  {Z.}~\bibnamefont {Xu}}} (\bibinfo {year} {2011}),\ \href
  {https://doi.org/10.1088/0953-8984/23/46/464204} {\bibfield  {journal}
  {\bibinfo  {journal} {J. Phys.: Condens. Matter}\ }\textbf {\bibinfo {volume}
  {23}},\ \bibinfo {pages} {464204}}\BibitemShut {NoStop}%
\bibitem [{\citenamefont {Cao}\ \emph {et~al.}(2024)\citenamefont {Cao},
  \citenamefont {Li}, \citenamefont {Liu}, \citenamefont {Liu}, \citenamefont
  {Chen}, \citenamefont {Xing}, \citenamefont {Kong}, \citenamefont {Yang},
  \citenamefont {Hu}, \citenamefont {Li}, \citenamefont {Zhou}, \citenamefont
  {Chen}, \citenamefont {Ke}, \citenamefont {Hu}, \citenamefont {Cao},
  \citenamefont {Wu}, \citenamefont {Ding},\ and\ \citenamefont
  {Gao}}]{cao2024observation}%
  \BibitemOpen
  \bibfield  {author} {\bibinfo {author} {\bibnamefont {Cao}, \bibfnamefont
  {L.}}, \bibinfo {author} {\bibfnamefont {G.}~\bibnamefont {Li}}, \bibinfo
  {author} {\bibfnamefont {W.}~\bibnamefont {Liu}}, \bibinfo {author}
  {\bibfnamefont {Y.-B.}\ \bibnamefont {Liu}}, \bibinfo {author} {\bibfnamefont
  {H.}~\bibnamefont {Chen}}, \bibinfo {author} {\bibfnamefont {Y.}~\bibnamefont
  {Xing}}, \bibinfo {author} {\bibfnamefont {L.}~\bibnamefont {Kong}}, \bibinfo
  {author} {\bibfnamefont {F.}~\bibnamefont {Yang}}, \bibinfo {author}
  {\bibfnamefont {Q.}~\bibnamefont {Hu}}, \bibinfo {author} {\bibfnamefont
  {M.}~\bibnamefont {Li}}, \bibinfo {author} {\bibfnamefont {X.}~\bibnamefont
  {Zhou}}, \bibinfo {author} {\bibfnamefont {Z.}~\bibnamefont {Chen}}, \bibinfo
  {author} {\bibfnamefont {C.}~\bibnamefont {Ke}}, \bibinfo {author}
  {\bibfnamefont {L.}~\bibnamefont {Hu}}, \bibinfo {author} {\bibfnamefont
  {G.-H.}\ \bibnamefont {Cao}}, \bibinfo {author} {\bibfnamefont
  {C.}~\bibnamefont {Wu}}, \bibinfo {author} {\bibfnamefont {H.}~\bibnamefont
  {Ding}}, and\ \bibinfo {author} {\bibfnamefont {H.-J.}\ \bibnamefont {Gao}}}
  (\bibinfo {year} {2024}),\ \href
  {https://doi.org/10.1088/0256-307X/41/11/117401} {\bibfield  {journal}
  {\bibinfo  {journal} {Chin. Phys. Lett.}\ }\textbf {\bibinfo {volume} {41}},\
  \bibinfo {pages} {117401}}\BibitemShut {NoStop}%
\bibitem [{\citenamefont {Carr}\ \emph {et~al.}(2016)\citenamefont {Carr},
  \citenamefont {Zhang}, \citenamefont {Song}, \citenamefont {Tan},
  \citenamefont {Li}, \citenamefont {Abernathy}, \citenamefont {Stone},
  \citenamefont {Granroth}, \citenamefont {Perring},\ and\ \citenamefont
  {Dai}}]{carr2016electron}%
  \BibitemOpen
  \bibfield  {author} {\bibinfo {author} {\bibnamefont {Carr}, \bibfnamefont
  {S.~V.}}, \bibinfo {author} {\bibfnamefont {C.}~\bibnamefont {Zhang}},
  \bibinfo {author} {\bibfnamefont {Y.}~\bibnamefont {Song}}, \bibinfo {author}
  {\bibfnamefont {G.}~\bibnamefont {Tan}}, \bibinfo {author} {\bibfnamefont
  {Y.}~\bibnamefont {Li}}, \bibinfo {author} {\bibfnamefont {D.~L.}\
  \bibnamefont {Abernathy}}, \bibinfo {author} {\bibfnamefont {M.~B.}\
  \bibnamefont {Stone}}, \bibinfo {author} {\bibfnamefont {G.~E.}\ \bibnamefont
  {Granroth}}, \bibinfo {author} {\bibfnamefont {T.~G.}\ \bibnamefont
  {Perring}}, and\ \bibinfo {author} {\bibfnamefont {P.}~\bibnamefont {Dai}}}
  (\bibinfo {year} {2016}),\ \href {https://doi.org/10.1103/PhysRevB.93.214506}
  {\bibfield  {journal} {\bibinfo  {journal} {Phys. Rev. B}\ }\textbf {\bibinfo
  {volume} {93}},\ \bibinfo {pages} {214506}}\BibitemShut {NoStop}%
\bibitem [{\citenamefont {Chatzopoulos}\ \emph {et~al.}(2021)\citenamefont
  {Chatzopoulos}, \citenamefont {Cho}, \citenamefont {Bastiaans}, \citenamefont
  {Steffensen}, \citenamefont {Bouwmeester}, \citenamefont {Akbari},
  \citenamefont {Gu}, \citenamefont {Paaske}, \citenamefont {Andersen},\ and\
  \citenamefont {Allan}}]{chatzopoulos2021spatially}%
  \BibitemOpen
  \bibfield  {author} {\bibinfo {author} {\bibnamefont {Chatzopoulos},
  \bibfnamefont {D.}}, \bibinfo {author} {\bibfnamefont {D.}~\bibnamefont
  {Cho}}, \bibinfo {author} {\bibfnamefont {K.~M.}\ \bibnamefont {Bastiaans}},
  \bibinfo {author} {\bibfnamefont {G.~O.}\ \bibnamefont {Steffensen}},
  \bibinfo {author} {\bibfnamefont {D.}~\bibnamefont {Bouwmeester}}, \bibinfo
  {author} {\bibfnamefont {A.}~\bibnamefont {Akbari}}, \bibinfo {author}
  {\bibfnamefont {G.}~\bibnamefont {Gu}}, \bibinfo {author} {\bibfnamefont
  {J.}~\bibnamefont {Paaske}}, \bibinfo {author} {\bibfnamefont {B.~M.}\
  \bibnamefont {Andersen}}, and\ \bibinfo {author} {\bibfnamefont {M.~P.}\
  \bibnamefont {Allan}}} (\bibinfo {year} {2021}),\ \href
  {https://doi.org/10.1038/s41467-020-20529-x} {\bibfield  {journal} {\bibinfo
  {journal} {Nat. Commun.}\ }\textbf {\bibinfo {volume} {12}},\ \bibinfo
  {pages} {298}}\BibitemShut {NoStop}%
\bibitem [{\citenamefont {Chen}\ \emph
  {et~al.}(2020{\natexlab{a}})\citenamefont {Chen}, \citenamefont {Jiang},
  \citenamefont {Zhang}, \citenamefont {Liu}, \citenamefont {Liu},
  \citenamefont {Wang},\ and\ \citenamefont {Wang}}]{chen2020atomic}%
  \BibitemOpen
  \bibfield  {author} {\bibinfo {author} {\bibnamefont {Chen}, \bibfnamefont
  {C.}}, \bibinfo {author} {\bibfnamefont {K.}~\bibnamefont {Jiang}}, \bibinfo
  {author} {\bibfnamefont {Y.}~\bibnamefont {Zhang}}, \bibinfo {author}
  {\bibfnamefont {C.}~\bibnamefont {Liu}}, \bibinfo {author} {\bibfnamefont
  {Y.}~\bibnamefont {Liu}}, \bibinfo {author} {\bibfnamefont {Z.}~\bibnamefont
  {Wang}}, and\ \bibinfo {author} {\bibfnamefont {J.}~\bibnamefont {Wang}}}
  (\bibinfo {year} {2020}{\natexlab{a}}),\ \href
  {https://doi.org/10.1038/s41567-020-0813-0} {\bibfield  {journal} {\bibinfo
  {journal} {Nat. Phys.}\ }\textbf {\bibinfo {volume} {16}},\ \bibinfo {pages}
  {536}}\BibitemShut {NoStop}%
\bibitem [{\citenamefont {Chen}\ \emph {et~al.}(2011)\citenamefont {Chen},
  \citenamefont {Xu}, \citenamefont {Ge}, \citenamefont {Zhang}, \citenamefont
  {Ye}, \citenamefont {Yang}, \citenamefont {Jiang}, \citenamefont {Xie},
  \citenamefont {Che}, \citenamefont {Zhang}, \citenamefont {Wang},
  \citenamefont {Chen}, \citenamefont {Shen}, \citenamefont {Hu},\ and\
  \citenamefont {Feng}}]{chen2011electronic}%
  \BibitemOpen
  \bibfield  {author} {\bibinfo {author} {\bibnamefont {Chen}, \bibfnamefont
  {F.}}, \bibinfo {author} {\bibfnamefont {M.}~\bibnamefont {Xu}}, \bibinfo
  {author} {\bibfnamefont {Q.~Q.}\ \bibnamefont {Ge}}, \bibinfo {author}
  {\bibfnamefont {Y.}~\bibnamefont {Zhang}}, \bibinfo {author} {\bibfnamefont
  {Z.~R.}\ \bibnamefont {Ye}}, \bibinfo {author} {\bibfnamefont {L.~X.}\
  \bibnamefont {Yang}}, \bibinfo {author} {\bibfnamefont {J.}~\bibnamefont
  {Jiang}}, \bibinfo {author} {\bibfnamefont {B.~P.}\ \bibnamefont {Xie}},
  \bibinfo {author} {\bibfnamefont {R.~C.}\ \bibnamefont {Che}}, \bibinfo
  {author} {\bibfnamefont {M.}~\bibnamefont {Zhang}}, \bibinfo {author}
  {\bibfnamefont {A.~F.}\ \bibnamefont {Wang}}, \bibinfo {author}
  {\bibfnamefont {X.~H.}\ \bibnamefont {Chen}}, \bibinfo {author}
  {\bibfnamefont {D.~W.}\ \bibnamefont {Shen}}, \bibinfo {author}
  {\bibfnamefont {J.~P.}\ \bibnamefont {Hu}}, and\ \bibinfo {author}
  {\bibfnamefont {D.~L.}\ \bibnamefont {Feng}}} (\bibinfo {year} {2011}),\
  \href {https://doi.org/10.1103/PhysRevX.1.021020} {\bibfield  {journal}
  {\bibinfo  {journal} {Phys. Rev. X}\ }\textbf {\bibinfo {volume} {1}},\
  \bibinfo {pages} {021020}}\BibitemShut {NoStop}%
\bibitem [{\citenamefont {Chen}\ \emph {et~al.}(2022)\citenamefont {Chen},
  \citenamefont {Aishwarya}, \citenamefont {Hirsbrunner}, \citenamefont
  {Rodriguez}, \citenamefont {Jiao}, \citenamefont {Dong}, \citenamefont
  {Mason}, \citenamefont {Van~Harlingen}, \citenamefont {Harter}, \citenamefont
  {Wilson}, \citenamefont {Hughes},\ and\ \citenamefont
  {Madhavan}}]{chen2022evidence}%
  \BibitemOpen
  \bibfield  {author} {\bibinfo {author} {\bibnamefont {Chen}, \bibfnamefont
  {G.}}, \bibinfo {author} {\bibfnamefont {A.}~\bibnamefont {Aishwarya}},
  \bibinfo {author} {\bibfnamefont {M.~R.}\ \bibnamefont {Hirsbrunner}},
  \bibinfo {author} {\bibfnamefont {J.~O.}\ \bibnamefont {Rodriguez}}, \bibinfo
  {author} {\bibfnamefont {L.}~\bibnamefont {Jiao}}, \bibinfo {author}
  {\bibfnamefont {L.}~\bibnamefont {Dong}}, \bibinfo {author} {\bibfnamefont
  {N.}~\bibnamefont {Mason}}, \bibinfo {author} {\bibfnamefont
  {D.}~\bibnamefont {Van~Harlingen}}, \bibinfo {author} {\bibfnamefont
  {J.}~\bibnamefont {Harter}}, \bibinfo {author} {\bibfnamefont {S.~D.}\
  \bibnamefont {Wilson}}, \bibinfo {author} {\bibfnamefont {T.~L.}\
  \bibnamefont {Hughes}}, and\ \bibinfo {author} {\bibfnamefont
  {V.}~\bibnamefont {Madhavan}}} (\bibinfo {year} {2022}),\ \href
  {https://doi.org/10.1038/s41535-022-00513-y} {\bibfield  {journal} {\bibinfo
  {journal} {npj Quantum Mater.}\ }\textbf {\bibinfo {volume} {7}},\ \bibinfo
  {pages} {110}}\BibitemShut {NoStop}%
\bibitem [{\citenamefont {Chen}\ \emph
  {et~al.}(2008{\natexlab{a}})\citenamefont {Chen}, \citenamefont {Li},
  \citenamefont {Wu}, \citenamefont {Li}, \citenamefont {Hu}, \citenamefont
  {Dong}, \citenamefont {Zheng}, \citenamefont {Luo},\ and\ \citenamefont
  {Wang}}]{Chen2008b}%
  \BibitemOpen
  \bibfield  {author} {\bibinfo {author} {\bibnamefont {Chen}, \bibfnamefont
  {G.}}, \bibinfo {author} {\bibfnamefont {Z.}~\bibnamefont {Li}}, \bibinfo
  {author} {\bibfnamefont {D.}~\bibnamefont {Wu}}, \bibinfo {author}
  {\bibfnamefont {G.}~\bibnamefont {Li}}, \bibinfo {author} {\bibfnamefont
  {W.}~\bibnamefont {Hu}}, \bibinfo {author} {\bibfnamefont {J.}~\bibnamefont
  {Dong}}, \bibinfo {author} {\bibfnamefont {P.}~\bibnamefont {Zheng}},
  \bibinfo {author} {\bibfnamefont {J.}~\bibnamefont {Luo}}, and\ \bibinfo
  {author} {\bibfnamefont {N.}~\bibnamefont {Wang}}} (\bibinfo {year}
  {2008}{\natexlab{a}}),\ \href
  {https://doi.org/10.1103/PhysRevLett.100.247002} {\bibfield  {journal}
  {\bibinfo  {journal} {Phys. Rev. Lett.}\ }\textbf {\bibinfo {volume} {100}},\
  \bibinfo {pages} {247002}}\BibitemShut {NoStop}%
\bibitem [{\citenamefont {Chen}\ \emph {et~al.}(2009)\citenamefont {Chen},
  \citenamefont {Xia}, \citenamefont {Yang}, \citenamefont {Li}, \citenamefont
  {Zheng}, \citenamefont {Luo},\ and\ \citenamefont {Wang}}]{Chen2009}%
  \BibitemOpen
  \bibfield  {author} {\bibinfo {author} {\bibnamefont {Chen}, \bibfnamefont
  {G.}}, \bibinfo {author} {\bibfnamefont {T.}~\bibnamefont {Xia}}, \bibinfo
  {author} {\bibfnamefont {H.}~\bibnamefont {Yang}}, \bibinfo {author}
  {\bibfnamefont {J.}~\bibnamefont {Li}}, \bibinfo {author} {\bibfnamefont
  {P.}~\bibnamefont {Zheng}}, \bibinfo {author} {\bibfnamefont
  {J.}~\bibnamefont {Luo}}, and\ \bibinfo {author} {\bibfnamefont
  {N.}~\bibnamefont {Wang}}} (\bibinfo {year} {2009}),\ \href
  {https://doi.org/10.1088/0953-2048/22/7/072001} {\bibfield  {journal}
  {\bibinfo  {journal} {Supercond. Sci. Technol.}\ }\textbf {\bibinfo {volume}
  {22}},\ \bibinfo {pages} {072001}}\BibitemShut {NoStop}%
\bibitem [{\citenamefont {Chen}\ \emph {et~al.}(2019)\citenamefont {Chen},
  \citenamefont {Chen}, \citenamefont {Kreisel}, \citenamefont {Lu},
  \citenamefont {Schneidewind}, \citenamefont {Qiu}, \citenamefont {Park},
  \citenamefont {Perring}, \citenamefont {Stewart}, \citenamefont {Cao},
  \citenamefont {Zhang}, \citenamefont {Li}, \citenamefont {Rong},
  \citenamefont {Wei}, \citenamefont {Andersen}, \citenamefont {Hirschfeld},
  \citenamefont {Broholm},\ and\ \citenamefont {Dai}}]{chen2019anisotropic}%
  \BibitemOpen
  \bibfield  {author} {\bibinfo {author} {\bibnamefont {Chen}, \bibfnamefont
  {T.}}, \bibinfo {author} {\bibfnamefont {Y.}~\bibnamefont {Chen}}, \bibinfo
  {author} {\bibfnamefont {A.}~\bibnamefont {Kreisel}}, \bibinfo {author}
  {\bibfnamefont {X.}~\bibnamefont {Lu}}, \bibinfo {author} {\bibfnamefont
  {A.}~\bibnamefont {Schneidewind}}, \bibinfo {author} {\bibfnamefont
  {Y.}~\bibnamefont {Qiu}}, \bibinfo {author} {\bibfnamefont {J.~T.}\
  \bibnamefont {Park}}, \bibinfo {author} {\bibfnamefont {T.~G.}\ \bibnamefont
  {Perring}}, \bibinfo {author} {\bibfnamefont {J.~R.}\ \bibnamefont
  {Stewart}}, \bibinfo {author} {\bibfnamefont {H.}~\bibnamefont {Cao}},
  \bibinfo {author} {\bibfnamefont {R.}~\bibnamefont {Zhang}}, \bibinfo
  {author} {\bibfnamefont {Y.}~\bibnamefont {Li}}, \bibinfo {author}
  {\bibfnamefont {Y.}~\bibnamefont {Rong}}, \bibinfo {author} {\bibfnamefont
  {Y.}~\bibnamefont {Wei}}, \bibinfo {author} {\bibfnamefont {B.~M.}\
  \bibnamefont {Andersen}}, \bibinfo {author} {\bibfnamefont {P.~J.}\
  \bibnamefont {Hirschfeld}}, \bibinfo {author} {\bibfnamefont
  {C.}~\bibnamefont {Broholm}}, and\ \bibinfo {author} {\bibfnamefont
  {P.}~\bibnamefont {Dai}}} (\bibinfo {year} {2019}),\ \href
  {https://doi.org/10.1038/s41563-019-0369-5} {\bibfield  {journal} {\bibinfo
  {journal} {Nat. Mater.}\ }\textbf {\bibinfo {volume} {18}},\ \bibinfo {pages}
  {709}}\BibitemShut {NoStop}%
\bibitem [{\citenamefont {Chen}\ \emph
  {et~al.}(2020{\natexlab{b}})\citenamefont {Chen}, \citenamefont {Chen},
  \citenamefont {Tam}, \citenamefont {Gao}, \citenamefont {Qiu}, \citenamefont
  {Schneidewind}, \citenamefont {Radelytskyi}, \citenamefont {Prokes},
  \citenamefont {Chi}, \citenamefont {Matsuda}, \citenamefont {Broholm},\ and\
  \citenamefont {Dai}}]{chen2020anisotropic}%
  \BibitemOpen
  \bibfield  {author} {\bibinfo {author} {\bibnamefont {Chen}, \bibfnamefont
  {T.}}, \bibinfo {author} {\bibfnamefont {Y.}~\bibnamefont {Chen}}, \bibinfo
  {author} {\bibfnamefont {D.~W.}\ \bibnamefont {Tam}}, \bibinfo {author}
  {\bibfnamefont {B.}~\bibnamefont {Gao}}, \bibinfo {author} {\bibfnamefont
  {Y.}~\bibnamefont {Qiu}}, \bibinfo {author} {\bibfnamefont {A.}~\bibnamefont
  {Schneidewind}}, \bibinfo {author} {\bibfnamefont {I.}~\bibnamefont
  {Radelytskyi}}, \bibinfo {author} {\bibfnamefont {K.}~\bibnamefont {Prokes}},
  \bibinfo {author} {\bibfnamefont {S.}~\bibnamefont {Chi}}, \bibinfo {author}
  {\bibfnamefont {M.}~\bibnamefont {Matsuda}}, \bibinfo {author} {\bibfnamefont
  {C.}~\bibnamefont {Broholm}}, and\ \bibinfo {author} {\bibfnamefont
  {P.}~\bibnamefont {Dai}}} (\bibinfo {year} {2020}{\natexlab{b}}),\ \href
  {https://doi.org/10.1103/PhysRevB.101.140504} {\bibfield  {journal} {\bibinfo
   {journal} {Phys. Rev. B}\ }\textbf {\bibinfo {volume} {101}},\ \bibinfo
  {pages} {140504}}\BibitemShut {NoStop}%
\bibitem [{\citenamefont {Chen}\ \emph {et~al.}(2015)\citenamefont {Chen},
  \citenamefont {Maiti}, \citenamefont {Linscheid},\ and\ \citenamefont
  {Hirschfeld}}]{Chen2015}%
  \BibitemOpen
  \bibfield  {author} {\bibinfo {author} {\bibnamefont {Chen}, \bibfnamefont
  {X.}}, \bibinfo {author} {\bibfnamefont {S.}~\bibnamefont {Maiti}}, \bibinfo
  {author} {\bibfnamefont {A.}~\bibnamefont {Linscheid}}, and\ \bibinfo
  {author} {\bibfnamefont {P.}~\bibnamefont {Hirschfeld}}} (\bibinfo {year}
  {2015}),\ \href {https://doi.org/10.1103/PhysRevB.92.224514} {\bibfield
  {journal} {\bibinfo  {journal} {Phys. Rev. B}\ }\textbf {\bibinfo {volume}
  {92}},\ \bibinfo {pages} {224514}}\BibitemShut {NoStop}%
\bibitem [{\citenamefont {Chen}\ \emph
  {et~al.}(2008{\natexlab{b}})\citenamefont {Chen}, \citenamefont {Wu},
  \citenamefont {Wu}, \citenamefont {Liu}, \citenamefont {Chen},\ and\
  \citenamefont {Fang}}]{Chen2008a}%
  \BibitemOpen
  \bibfield  {author} {\bibinfo {author} {\bibnamefont {Chen}, \bibfnamefont
  {X.}}, \bibinfo {author} {\bibfnamefont {T.}~\bibnamefont {Wu}}, \bibinfo
  {author} {\bibfnamefont {G.}~\bibnamefont {Wu}}, \bibinfo {author}
  {\bibfnamefont {R.}~\bibnamefont {Liu}}, \bibinfo {author} {\bibfnamefont
  {H.}~\bibnamefont {Chen}}, and\ \bibinfo {author} {\bibfnamefont
  {D.}~\bibnamefont {Fang}}} (\bibinfo {year} {2008}{\natexlab{b}}),\ \href
  {https://doi.org/10.1038/nature07045} {\bibfield  {journal} {\bibinfo
  {journal} {nature}\ }\textbf {\bibinfo {volume} {453}},\ \bibinfo {pages}
  {761}}\BibitemShut {NoStop}%
\bibitem [{\citenamefont {Chen}\ \emph {et~al.}(2023)\citenamefont {Chen},
  \citenamefont {Li}, \citenamefont {Lu}, \citenamefont {Liu}, \citenamefont
  {Zhang}, \citenamefont {Li}, \citenamefont {Yin}, \citenamefont {Li},
  \citenamefont {Zhang}, \citenamefont {Dong}, \citenamefont {Yan},\ and\
  \citenamefont {Feng}}]{chen2023charge}%
  \BibitemOpen
  \bibfield  {author} {\bibinfo {author} {\bibnamefont {Chen}, \bibfnamefont
  {Z.}}, \bibinfo {author} {\bibfnamefont {D.}~\bibnamefont {Li}}, \bibinfo
  {author} {\bibfnamefont {Z.}~\bibnamefont {Lu}}, \bibinfo {author}
  {\bibfnamefont {Y.}~\bibnamefont {Liu}}, \bibinfo {author} {\bibfnamefont
  {J.}~\bibnamefont {Zhang}}, \bibinfo {author} {\bibfnamefont
  {Y.}~\bibnamefont {Li}}, \bibinfo {author} {\bibfnamefont {R.}~\bibnamefont
  {Yin}}, \bibinfo {author} {\bibfnamefont {M.}~\bibnamefont {Li}}, \bibinfo
  {author} {\bibfnamefont {T.}~\bibnamefont {Zhang}}, \bibinfo {author}
  {\bibfnamefont {X.}~\bibnamefont {Dong}}, \bibinfo {author} {\bibfnamefont
  {Y.-J.}\ \bibnamefont {Yan}}, and\ \bibinfo {author} {\bibfnamefont {D.-L.}\
  \bibnamefont {Feng}}} (\bibinfo {year} {2023}),\ \href
  {https://doi.org/10.1038/s41467-023-37792-3} {\bibfield  {journal} {\bibinfo
  {journal} {Nat. Commun.}\ }\textbf {\bibinfo {volume} {14}},\ \bibinfo
  {pages} {2023}}\BibitemShut {NoStop}%
\bibitem [{\citenamefont {Cheng}\ \emph {et~al.}(2015)\citenamefont {Cheng},
  \citenamefont {Matsubayashi}, \citenamefont {Wu}, \citenamefont {Sun},
  \citenamefont {Lin}, \citenamefont {Luo},\ and\ \citenamefont
  {Uwatoko}}]{cheng2015pressure}%
  \BibitemOpen
  \bibfield  {author} {\bibinfo {author} {\bibnamefont {Cheng}, \bibfnamefont
  {J.-G.}}, \bibinfo {author} {\bibfnamefont {K.}~\bibnamefont {Matsubayashi}},
  \bibinfo {author} {\bibfnamefont {W.}~\bibnamefont {Wu}}, \bibinfo {author}
  {\bibfnamefont {J.~P.}\ \bibnamefont {Sun}}, \bibinfo {author} {\bibfnamefont
  {F.~K.}\ \bibnamefont {Lin}}, \bibinfo {author} {\bibfnamefont {J.~L.}\
  \bibnamefont {Luo}}, and\ \bibinfo {author} {\bibfnamefont {Y.}~\bibnamefont
  {Uwatoko}}} (\bibinfo {year} {2015}),\ \href
  {https://doi.org/10.1103/PhysRevLett.114.117001} {\bibfield  {journal}
  {\bibinfo  {journal} {Phys. Rev. Lett.}\ }\textbf {\bibinfo {volume} {114}},\
  \bibinfo {pages} {117001}}\BibitemShut {NoStop}%
\bibitem [{\citenamefont {Cheng}\ \emph {et~al.}(2025)\citenamefont {Cheng},
  \citenamefont {Wang}, \citenamefont {Ren}, \citenamefont {Deng},
  \citenamefont {Lou}, \citenamefont {Ma}, \citenamefont {Xue},\ and\
  \citenamefont {Song}}]{Cheng2025}%
  \BibitemOpen
  \bibfield  {author} {\bibinfo {author} {\bibnamefont {Cheng}, \bibfnamefont
  {Q.-J.}}, \bibinfo {author} {\bibfnamefont {Y.-W.}\ \bibnamefont {Wang}},
  \bibinfo {author} {\bibfnamefont {M.-Q.}\ \bibnamefont {Ren}}, \bibinfo
  {author} {\bibfnamefont {Z.-X.}\ \bibnamefont {Deng}}, \bibinfo {author}
  {\bibfnamefont {C.-C.}\ \bibnamefont {Lou}}, \bibinfo {author} {\bibfnamefont
  {X.-C.}\ \bibnamefont {Ma}}, \bibinfo {author} {\bibfnamefont {Q.-K.}\
  \bibnamefont {Xue}}, and\ \bibinfo {author} {\bibfnamefont {C.-L.}\
  \bibnamefont {Song}}} (\bibinfo {year} {2025}),\ \href
  {https://doi.org/10.1038/s43246-025-00885-1} {\bibfield  {journal} {\bibinfo
  {journal} {Commun. Mater.}\ }\textbf {\bibinfo {volume} {6}},\ \bibinfo
  {pages} {162}}\BibitemShut {NoStop}%
\bibitem [{\citenamefont {Cheng}\ \emph {et~al.}(2018)\citenamefont {Cheng},
  \citenamefont {Dong}, \citenamefont {Huang}, \citenamefont {Liu},
  \citenamefont {Zhu}, \citenamefont {Wang}, \citenamefont {Vlasko-Vlasov},
  \citenamefont {Welp}, \citenamefont {Kwok},\ and\ \citenamefont
  {Ma}}]{Cheng2019}%
  \BibitemOpen
  \bibfield  {author} {\bibinfo {author} {\bibnamefont {Cheng}, \bibfnamefont
  {Z.}}, \bibinfo {author} {\bibfnamefont {C.}~\bibnamefont {Dong}}, \bibinfo
  {author} {\bibfnamefont {H.}~\bibnamefont {Huang}}, \bibinfo {author}
  {\bibfnamefont {S.}~\bibnamefont {Liu}}, \bibinfo {author} {\bibfnamefont
  {Y.}~\bibnamefont {Zhu}}, \bibinfo {author} {\bibfnamefont {D.}~\bibnamefont
  {Wang}}, \bibinfo {author} {\bibfnamefont {V.}~\bibnamefont {Vlasko-Vlasov}},
  \bibinfo {author} {\bibfnamefont {U.}~\bibnamefont {Welp}}, \bibinfo {author}
  {\bibfnamefont {W.-K.}\ \bibnamefont {Kwok}}, and\ \bibinfo {author}
  {\bibfnamefont {Y.}~\bibnamefont {Ma}}} (\bibinfo {year} {2018}),\ \href
  {https://doi.org/10.1088/1361-6668/aaedff} {\bibfield  {journal} {\bibinfo
  {journal} {Supercond. Sci. Technol.}\ }\textbf {\bibinfo {volume} {32}},\
  \bibinfo {pages} {015008}}\BibitemShut {NoStop}%
\bibitem [{\citenamefont {Chinotti}\ \emph {et~al.}(2017)\citenamefont
  {Chinotti}, \citenamefont {Pal}, \citenamefont {Degiorgi}, \citenamefont
  {B\"ohmer},\ and\ \citenamefont {Canfield}}]{chinotti2017optical}%
  \BibitemOpen
  \bibfield  {author} {\bibinfo {author} {\bibnamefont {Chinotti},
  \bibfnamefont {M.}}, \bibinfo {author} {\bibfnamefont {A.}~\bibnamefont
  {Pal}}, \bibinfo {author} {\bibfnamefont {L.}~\bibnamefont {Degiorgi}},
  \bibinfo {author} {\bibfnamefont {A.~E.}\ \bibnamefont {B\"ohmer}}, and\
  \bibinfo {author} {\bibfnamefont {P.~C.}\ \bibnamefont {Canfield}}} (\bibinfo
  {year} {2017}),\ \href {https://doi.org/10.1103/PhysRevB.96.121112}
  {\bibfield  {journal} {\bibinfo  {journal} {Phys. Rev. B}\ }\textbf {\bibinfo
  {volume} {96}},\ \bibinfo {pages} {121112(R)}}\BibitemShut {NoStop}%
\bibitem [{\citenamefont {Cho}\ \emph {et~al.}(2017)\citenamefont {Cho},
  \citenamefont {Yang}, \citenamefont {Yuan}, \citenamefont {Shen},
  \citenamefont {Wolf},\ and\ \citenamefont {Lortz}}]{cho2017thermodynamic}%
  \BibitemOpen
  \bibfield  {author} {\bibinfo {author} {\bibnamefont {Cho}, \bibfnamefont
  {C.-w.}}, \bibinfo {author} {\bibfnamefont {J.~H.}\ \bibnamefont {Yang}},
  \bibinfo {author} {\bibfnamefont {N.~F.~Q.}\ \bibnamefont {Yuan}}, \bibinfo
  {author} {\bibfnamefont {J.}~\bibnamefont {Shen}}, \bibinfo {author}
  {\bibfnamefont {T.}~\bibnamefont {Wolf}}, and\ \bibinfo {author}
  {\bibfnamefont {R.}~\bibnamefont {Lortz}}} (\bibinfo {year} {2017}),\ \href
  {https://doi.org/10.1103/PhysRevLett.119.217002} {\bibfield  {journal}
  {\bibinfo  {journal} {Phys. Rev. Lett.}\ }\textbf {\bibinfo {volume} {119}},\
  \bibinfo {pages} {217002}}\BibitemShut {NoStop}%
\bibitem [{\citenamefont {Chu}\ \emph {et~al.}(2010)\citenamefont {Chu},
  \citenamefont {Analytis}, \citenamefont {Greve}, \citenamefont {McMahon},
  \citenamefont {Islam}, \citenamefont {Yamamoto},\ and\ \citenamefont
  {Fisher}}]{chu2010}%
  \BibitemOpen
  \bibfield  {author} {\bibinfo {author} {\bibnamefont {Chu}, \bibfnamefont
  {J.-H.}}, \bibinfo {author} {\bibfnamefont {J.~G.}\ \bibnamefont {Analytis}},
  \bibinfo {author} {\bibfnamefont {K.~D.}\ \bibnamefont {Greve}}, \bibinfo
  {author} {\bibfnamefont {P.~L.}\ \bibnamefont {McMahon}}, \bibinfo {author}
  {\bibfnamefont {Z.}~\bibnamefont {Islam}}, \bibinfo {author} {\bibfnamefont
  {Y.}~\bibnamefont {Yamamoto}}, and\ \bibinfo {author} {\bibfnamefont {I.~R.}\
  \bibnamefont {Fisher}}} (\bibinfo {year} {2010}),\ \href
  {https://doi.org/10.1126/science.1190482} {\bibfield  {journal} {\bibinfo
  {journal} {Science}\ }\textbf {\bibinfo {volume} {329}},\ \bibinfo {pages}
  {824}}\BibitemShut {NoStop}%
\bibitem [{\citenamefont {Chu}\ \emph {et~al.}(2012)\citenamefont {Chu},
  \citenamefont {Kuo}, \citenamefont {Analytis},\ and\ \citenamefont
  {Fisher}}]{chu2012}%
  \BibitemOpen
  \bibfield  {author} {\bibinfo {author} {\bibnamefont {Chu}, \bibfnamefont
  {J.-H.}}, \bibinfo {author} {\bibfnamefont {H.-H.}\ \bibnamefont {Kuo}},
  \bibinfo {author} {\bibfnamefont {J.~G.}\ \bibnamefont {Analytis}}, and\
  \bibinfo {author} {\bibfnamefont {I.~R.}\ \bibnamefont {Fisher}}} (\bibinfo
  {year} {2012}),\ \href {https://doi.org/10.1126/science.1221713} {\bibfield
  {journal} {\bibinfo  {journal} {Science}\ }\textbf {\bibinfo {volume}
  {337}},\ \bibinfo {pages} {710}}\BibitemShut {NoStop}%
\bibitem [{\citenamefont {Chubukov}\ \emph {et~al.}(2016)\citenamefont
  {Chubukov}, \citenamefont {Khodas},\ and\ \citenamefont
  {Fernandes}}]{chubukov2016magnetism}%
  \BibitemOpen
  \bibfield  {author} {\bibinfo {author} {\bibnamefont {Chubukov},
  \bibfnamefont {A.~V.}}, \bibinfo {author} {\bibfnamefont {M.}~\bibnamefont
  {Khodas}}, and\ \bibinfo {author} {\bibfnamefont {R.~M.}\ \bibnamefont
  {Fernandes}}} (\bibinfo {year} {2016}),\ \href
  {https://doi.org/10.1103/PhysRevX.6.041045} {\bibfield  {journal} {\bibinfo
  {journal} {Phys. Rev. X}\ }\textbf {\bibinfo {volume} {6}},\ \bibinfo {pages}
  {041045}}\BibitemShut {NoStop}%
\bibitem [{\citenamefont {Civardi}\ \emph {et~al.}(2016)\citenamefont
  {Civardi}, \citenamefont {Moroni}, \citenamefont {Babij}, \citenamefont
  {Bukowski},\ and\ \citenamefont {Carretta}}]{civardi2016superconductivity}%
  \BibitemOpen
  \bibfield  {author} {\bibinfo {author} {\bibnamefont {Civardi}, \bibfnamefont
  {E.}}, \bibinfo {author} {\bibfnamefont {M.}~\bibnamefont {Moroni}}, \bibinfo
  {author} {\bibfnamefont {M.}~\bibnamefont {Babij}}, \bibinfo {author}
  {\bibfnamefont {Z.}~\bibnamefont {Bukowski}}, and\ \bibinfo {author}
  {\bibfnamefont {P.}~\bibnamefont {Carretta}}} (\bibinfo {year} {2016}),\
  \href {https://doi.org/10.1103/PhysRevLett.117.217001} {\bibfield  {journal}
  {\bibinfo  {journal} {Phys. Rev. Lett.}\ }\textbf {\bibinfo {volume} {117}},\
  \bibinfo {pages} {217001}}\BibitemShut {NoStop}%
\bibitem [{\citenamefont {Coh}\ \emph {et~al.}(2015)\citenamefont {Coh},
  \citenamefont {Cohen},\ and\ \citenamefont {Louie}}]{Coh2015}%
  \BibitemOpen
  \bibfield  {author} {\bibinfo {author} {\bibnamefont {Coh}, \bibfnamefont
  {S.}}, \bibinfo {author} {\bibfnamefont {M.~L.}\ \bibnamefont {Cohen}}, and\
  \bibinfo {author} {\bibfnamefont {S.~G.}\ \bibnamefont {Louie}}} (\bibinfo
  {year} {2015}),\ \href {https://doi.org/10.1088/1367-2630/17/7/073027}
  {\bibfield  {journal} {\bibinfo  {journal} {New J. Phys.}\ }\textbf {\bibinfo
  {volume} {17}},\ \bibinfo {pages} {073027}}\BibitemShut {NoStop}%
\bibitem [{\citenamefont {Coldea}(2021)}]{coldea2021electronic}%
  \BibitemOpen
  \bibfield  {author} {\bibinfo {author} {\bibnamefont {Coldea}, \bibfnamefont
  {A.~I.}}} (\bibinfo {year} {2021}),\ \href
  {https://doi.org/10.3389/fphy.2020.594500} {\bibfield  {journal} {\bibinfo
  {journal} {Frontiers in Physics}\ }\textbf {\bibinfo {volume} {8}},\ \bibinfo
  {pages} {594500}}\BibitemShut {NoStop}%
\bibitem [{\citenamefont {Coldea}\ and\ \citenamefont
  {Watson}(2018)}]{coldea2018thekey}%
  \BibitemOpen
  \bibfield  {author} {\bibinfo {author} {\bibnamefont {Coldea}, \bibfnamefont
  {A.~I.}}, and\ \bibinfo {author} {\bibfnamefont {M.~D.}\ \bibnamefont
  {Watson}}} (\bibinfo {year} {2018}),\ \href
  {https://doi.org/10.1146/annurev-conmatphys-033117-054137} {\bibfield
  {journal} {\bibinfo  {journal} {Annu. Rev. Condens. Matter Phys.}\ }\textbf
  {\bibinfo {volume} {9}},\ \bibinfo {pages} {125}}\BibitemShut {NoStop}%
\bibitem [{\citenamefont {Collomb}\ \emph {et~al.}(2021)\citenamefont
  {Collomb}, \citenamefont {Bending}, \citenamefont {Koshelev}, \citenamefont
  {Smylie}, \citenamefont {Farrar}, \citenamefont {Bao}, \citenamefont {Chung},
  \citenamefont {Kanatzidis}, \citenamefont {Kwok},\ and\ \citenamefont
  {Welp}}]{collomb2021observing}%
  \BibitemOpen
  \bibfield  {author} {\bibinfo {author} {\bibnamefont {Collomb}, \bibfnamefont
  {D.}}, \bibinfo {author} {\bibfnamefont {S.~J.}\ \bibnamefont {Bending}},
  \bibinfo {author} {\bibfnamefont {A.~E.}\ \bibnamefont {Koshelev}}, \bibinfo
  {author} {\bibfnamefont {M.~P.}\ \bibnamefont {Smylie}}, \bibinfo {author}
  {\bibfnamefont {L.}~\bibnamefont {Farrar}}, \bibinfo {author} {\bibfnamefont
  {J.-K.}\ \bibnamefont {Bao}}, \bibinfo {author} {\bibfnamefont {D.~Y.}\
  \bibnamefont {Chung}}, \bibinfo {author} {\bibfnamefont {M.~G.}\ \bibnamefont
  {Kanatzidis}}, \bibinfo {author} {\bibfnamefont {W.-K.}\ \bibnamefont
  {Kwok}}, and\ \bibinfo {author} {\bibfnamefont {U.}~\bibnamefont {Welp}}}
  (\bibinfo {year} {2021}),\ \href
  {https://doi.org/10.1103/PhysRevLett.126.157001} {\bibfield  {journal}
  {\bibinfo  {journal} {Phys. Rev. Lett.}\ }\textbf {\bibinfo {volume} {126}},\
  \bibinfo {pages} {157001}}\BibitemShut {NoStop}%
\bibitem [{\citenamefont {Comin}\ and\ \citenamefont
  {Damascelli}(2016)}]{comin2016resonant}%
  \BibitemOpen
  \bibfield  {author} {\bibinfo {author} {\bibnamefont {Comin}, \bibfnamefont
  {R.}}, and\ \bibinfo {author} {\bibfnamefont {A.}~\bibnamefont {Damascelli}}}
  (\bibinfo {year} {2016}),\ \href
  {https://doi.org/10.1146/annurev-conmatphys-031115-011401} {\bibfield
  {journal} {\bibinfo  {journal} {Annu. Rev. Condens. Matter Phys.}\ }\textbf
  {\bibinfo {volume} {7}},\ \bibinfo {pages} {369}}\BibitemShut {NoStop}%
\bibitem [{\citenamefont {Cortes-Gil}\ \emph {et~al.}(2010)\citenamefont
  {Cortes-Gil}, \citenamefont {Parker}, \citenamefont {Pitcher}, \citenamefont
  {Hadermann},\ and\ \citenamefont {Clarke}}]{Cortes-Gil2010}%
  \BibitemOpen
  \bibfield  {author} {\bibinfo {author} {\bibnamefont {Cortes-Gil},
  \bibfnamefont {R.}}, \bibinfo {author} {\bibfnamefont {D.~R.}\ \bibnamefont
  {Parker}}, \bibinfo {author} {\bibfnamefont {M.~J.}\ \bibnamefont {Pitcher}},
  \bibinfo {author} {\bibfnamefont {J.}~\bibnamefont {Hadermann}}, and\
  \bibinfo {author} {\bibfnamefont {S.~J.}\ \bibnamefont {Clarke}}} (\bibinfo
  {year} {2010}),\ \href {https://doi.org/10.1021/cm100956k} {\bibfield
  {journal} {\bibinfo  {journal} {Chemistry of Materials}\ }\textbf {\bibinfo
  {volume} {22}},\ \bibinfo {pages} {4304}}\BibitemShut {NoStop}%
\bibitem [{\citenamefont {Cui}\ \emph {et~al.}(2019)\citenamefont {Cui},
  \citenamefont {Hu}, \citenamefont {Zhang}, \citenamefont {Ma}, \citenamefont
  {Ma}, \citenamefont {Ma}, \citenamefont {Wang}, \citenamefont {Yan},
  \citenamefont {Sun}, \citenamefont {Cheng} \emph {et~al.}}]{Cui2019}%
  \BibitemOpen
  \bibfield  {author} {\bibinfo {author} {\bibnamefont {Cui}, \bibfnamefont
  {Y.}}, \bibinfo {author} {\bibfnamefont {Z.}~\bibnamefont {Hu}}, \bibinfo
  {author} {\bibfnamefont {J.-S.}\ \bibnamefont {Zhang}}, \bibinfo {author}
  {\bibfnamefont {W.-L.}\ \bibnamefont {Ma}}, \bibinfo {author} {\bibfnamefont
  {M.-W.}\ \bibnamefont {Ma}}, \bibinfo {author} {\bibfnamefont
  {Z.}~\bibnamefont {Ma}}, \bibinfo {author} {\bibfnamefont {C.}~\bibnamefont
  {Wang}}, \bibinfo {author} {\bibfnamefont {J.-Q.}\ \bibnamefont {Yan}},
  \bibinfo {author} {\bibfnamefont {J.-P.}\ \bibnamefont {Sun}}, \bibinfo
  {author} {\bibfnamefont {J.-G.}\ \bibnamefont {Cheng}},  \emph {et~al.}}
  (\bibinfo {year} {2019}),\ \href
  {https://doi.org/10.1088/0256-307x/36/7/077401} {\bibfield  {journal}
  {\bibinfo  {journal} {Chin. Phys. Lett.}\ }\textbf {\bibinfo {volume} {36}},\
  \bibinfo {pages} {077401}}\BibitemShut {NoStop}%
\bibitem [{\citenamefont {Cui}\ \emph {et~al.}(2018)\citenamefont {Cui},
  \citenamefont {Zhang}, \citenamefont {Li}, \citenamefont {Lin}, \citenamefont
  {Zhu}, \citenamefont {Wen}, \citenamefont {Wang}, \citenamefont {Sun},
  \citenamefont {Ma}, \citenamefont {Li}, \citenamefont {Gong}, \citenamefont
  {Xie}, \citenamefont {Gu}, \citenamefont {Li}, \citenamefont {Luo},
  \citenamefont {Yu},\ and\ \citenamefont {Yu}}]{Cui2018}%
  \BibitemOpen
  \bibfield  {author} {\bibinfo {author} {\bibnamefont {Cui}, \bibfnamefont
  {Y.}}, \bibinfo {author} {\bibfnamefont {G.}~\bibnamefont {Zhang}}, \bibinfo
  {author} {\bibfnamefont {H.}~\bibnamefont {Li}}, \bibinfo {author}
  {\bibfnamefont {H.}~\bibnamefont {Lin}}, \bibinfo {author} {\bibfnamefont
  {X.}~\bibnamefont {Zhu}}, \bibinfo {author} {\bibfnamefont {H.-H.}\
  \bibnamefont {Wen}}, \bibinfo {author} {\bibfnamefont {G.}~\bibnamefont
  {Wang}}, \bibinfo {author} {\bibfnamefont {J.}~\bibnamefont {Sun}}, \bibinfo
  {author} {\bibfnamefont {M.}~\bibnamefont {Ma}}, \bibinfo {author}
  {\bibfnamefont {Y.}~\bibnamefont {Li}}, \bibinfo {author} {\bibfnamefont
  {D.}~\bibnamefont {Gong}}, \bibinfo {author} {\bibfnamefont {T.}~\bibnamefont
  {Xie}}, \bibinfo {author} {\bibfnamefont {Y.}~\bibnamefont {Gu}}, \bibinfo
  {author} {\bibfnamefont {S.}~\bibnamefont {Li}}, \bibinfo {author}
  {\bibfnamefont {H.}~\bibnamefont {Luo}}, \bibinfo {author} {\bibfnamefont
  {P.}~\bibnamefont {Yu}}, and\ \bibinfo {author} {\bibfnamefont
  {W.}~\bibnamefont {Yu}}} (\bibinfo {year} {2018}),\ \href
  {https://doi.org/10.1016/j.scib.2017.12.009} {\bibfield  {journal} {\bibinfo
  {journal} {Sci. Bull.}\ }\textbf {\bibinfo {volume} {63}},\ \bibinfo {pages}
  {11}}\BibitemShut {NoStop}%
\bibitem [{\citenamefont {Dai}(2015)}]{dai2015antiferromagnetic}%
  \BibitemOpen
  \bibfield  {author} {\bibinfo {author} {\bibnamefont {Dai}, \bibfnamefont
  {P.}}} (\bibinfo {year} {2015}),\ \href
  {https://doi.org/10.1103/RevModPhys.87.855} {\bibfield  {journal} {\bibinfo
  {journal} {Rev. Mod. Phys.}\ }\textbf {\bibinfo {volume} {87}},\ \bibinfo
  {pages} {855}}\BibitemShut {NoStop}%
\bibitem [{\citenamefont {Dai}\ \emph {et~al.}(2012)\citenamefont {Dai},
  \citenamefont {Hu},\ and\ \citenamefont {Dagotto}}]{dai2012magnetism}%
  \BibitemOpen
  \bibfield  {author} {\bibinfo {author} {\bibnamefont {Dai}, \bibfnamefont
  {P.}}, \bibinfo {author} {\bibfnamefont {J.}~\bibnamefont {Hu}}, and\
  \bibinfo {author} {\bibfnamefont {E.}~\bibnamefont {Dagotto}}} (\bibinfo
  {year} {2012}),\ \href {https://doi.org/10.1038/nphys2438} {\bibfield
  {journal} {\bibinfo  {journal} {Nat. Phys.}\ }\textbf {\bibinfo {volume}
  {8}},\ \bibinfo {pages} {709}}\BibitemShut {NoStop}%
\bibitem [{\citenamefont {Das}\ and\ \citenamefont {Balatsky}(2013)}]{Das2013}%
  \BibitemOpen
  \bibfield  {author} {\bibinfo {author} {\bibnamefont {Das}, \bibfnamefont
  {T.}}, and\ \bibinfo {author} {\bibfnamefont {A.~V.}\ \bibnamefont
  {Balatsky}}} (\bibinfo {year} {2013}),\ \href
  {https://doi.org/10.1088/1367-2630/15/9/093045} {\bibfield  {journal}
  {\bibinfo  {journal} {New J. Phys.}\ }\textbf {\bibinfo {volume} {15}},\
  \bibinfo {pages} {093045}}\BibitemShut {NoStop}%
\bibitem [{\citenamefont {Davies}\ \emph {et~al.}(2016)\citenamefont {Davies},
  \citenamefont {Rahn}, \citenamefont {Walker}, \citenamefont {Ewings},
  \citenamefont {Woodruff}, \citenamefont {Clarke},\ and\ \citenamefont
  {Boothroyd}}]{davies2016spin}%
  \BibitemOpen
  \bibfield  {author} {\bibinfo {author} {\bibnamefont {Davies}, \bibfnamefont
  {N.~R.}}, \bibinfo {author} {\bibfnamefont {M.~C.}\ \bibnamefont {Rahn}},
  \bibinfo {author} {\bibfnamefont {H.~C.}\ \bibnamefont {Walker}}, \bibinfo
  {author} {\bibfnamefont {R.~A.}\ \bibnamefont {Ewings}}, \bibinfo {author}
  {\bibfnamefont {D.~N.}\ \bibnamefont {Woodruff}}, \bibinfo {author}
  {\bibfnamefont {S.~J.}\ \bibnamefont {Clarke}}, and\ \bibinfo {author}
  {\bibfnamefont {A.~T.}\ \bibnamefont {Boothroyd}}} (\bibinfo {year} {2016}),\
  \href {https://doi.org/10.1103/PhysRevB.94.144503} {\bibfield  {journal}
  {\bibinfo  {journal} {Phys. Rev. B}\ }\textbf {\bibinfo {volume} {94}},\
  \bibinfo {pages} {144503}}\BibitemShut {NoStop}%
\bibitem [{\citenamefont {De'~Medici}\ \emph {et~al.}(2014)\citenamefont
  {De'~Medici}, \citenamefont {Giovannetti},\ and\ \citenamefont
  {Capone}}]{medici2014selective}%
  \BibitemOpen
  \bibfield  {author} {\bibinfo {author} {\bibnamefont {De'~Medici},
  \bibfnamefont {L.}}, \bibinfo {author} {\bibfnamefont {G.}~\bibnamefont
  {Giovannetti}}, and\ \bibinfo {author} {\bibfnamefont {M.}~\bibnamefont
  {Capone}}} (\bibinfo {year} {2014}),\ \href
  {https://doi.org/10.1103/PhysRevLett.112.177001} {\bibfield  {journal}
  {\bibinfo  {journal} {Phys. Rev. Lett.}\ }\textbf {\bibinfo {volume} {112}},\
  \bibinfo {pages} {177001}}\BibitemShut {NoStop}%
\bibitem [{\citenamefont {Deng}\ \emph {et~al.}(2014)\citenamefont {Deng},
  \citenamefont {Lv}, \citenamefont {Wu}, \citenamefont {Xue}, \citenamefont
  {Zhang}, \citenamefont {Li}, \citenamefont {Wang}, \citenamefont {Ma},
  \citenamefont {Xue},\ and\ \citenamefont {Chu}}]{Deng2014}%
  \BibitemOpen
  \bibfield  {author} {\bibinfo {author} {\bibnamefont {Deng}, \bibfnamefont
  {L.}}, \bibinfo {author} {\bibfnamefont {B.}~\bibnamefont {Lv}}, \bibinfo
  {author} {\bibfnamefont {Z.}~\bibnamefont {Wu}}, \bibinfo {author}
  {\bibfnamefont {Y.}~\bibnamefont {Xue}}, \bibinfo {author} {\bibfnamefont
  {W.}~\bibnamefont {Zhang}}, \bibinfo {author} {\bibfnamefont
  {F.}~\bibnamefont {Li}}, \bibinfo {author} {\bibfnamefont {L.}~\bibnamefont
  {Wang}}, \bibinfo {author} {\bibfnamefont {X.}~\bibnamefont {Ma}}, \bibinfo
  {author} {\bibfnamefont {Q.}~\bibnamefont {Xue}}, and\ \bibinfo {author}
  {\bibfnamefont {C.}~\bibnamefont {Chu}}} (\bibinfo {year} {2014}),\ \href
  {https://doi.org/10.1103/PhysRevB.90.214513} {\bibfield  {journal} {\bibinfo
  {journal} {Phys. Rev. B}\ }\textbf {\bibinfo {volume} {90}},\ \bibinfo
  {pages} {214513}}\BibitemShut {NoStop}%
\bibitem [{\citenamefont {Devereaux}\ \emph {et~al.}(2016)\citenamefont
  {Devereaux}, \citenamefont {Shvaika}, \citenamefont {Wu}, \citenamefont
  {Wohlfeld}, \citenamefont {Jia}, \citenamefont {Wang}, \citenamefont
  {Moritz}, \citenamefont {Chaix}, \citenamefont {Lee}, \citenamefont {Shen},
  \citenamefont {Ghiringhelli},\ and\ \citenamefont
  {Braicovich}}]{devereaux2014directly}%
  \BibitemOpen
  \bibfield  {author} {\bibinfo {author} {\bibnamefont {Devereaux},
  \bibfnamefont {T.~P.}}, \bibinfo {author} {\bibfnamefont {A.~M.}\
  \bibnamefont {Shvaika}}, \bibinfo {author} {\bibfnamefont {K.}~\bibnamefont
  {Wu}}, \bibinfo {author} {\bibfnamefont {K.}~\bibnamefont {Wohlfeld}},
  \bibinfo {author} {\bibfnamefont {C.~J.}\ \bibnamefont {Jia}}, \bibinfo
  {author} {\bibfnamefont {Y.}~\bibnamefont {Wang}}, \bibinfo {author}
  {\bibfnamefont {B.}~\bibnamefont {Moritz}}, \bibinfo {author} {\bibfnamefont
  {L.}~\bibnamefont {Chaix}}, \bibinfo {author} {\bibfnamefont {W.-S.}\
  \bibnamefont {Lee}}, \bibinfo {author} {\bibfnamefont {Z.-X.}\ \bibnamefont
  {Shen}}, \bibinfo {author} {\bibfnamefont {G.}~\bibnamefont {Ghiringhelli}},
  and\ \bibinfo {author} {\bibfnamefont {L.}~\bibnamefont {Braicovich}}}
  (\bibinfo {year} {2016}),\ \href {https://doi.org/10.1103/PhysRevX.6.041019}
  {\bibfield  {journal} {\bibinfo  {journal} {Phys. Rev. X}\ }\textbf {\bibinfo
  {volume} {6}},\ \bibinfo {pages} {041019}}\BibitemShut {NoStop}%
\bibitem [{\citenamefont {Devizorova}\ and\ \citenamefont
  {Buzdin}(2019)}]{Devizorova2019}%
  \BibitemOpen
  \bibfield  {author} {\bibinfo {author} {\bibnamefont {Devizorova},
  \bibfnamefont {Z.}}, and\ \bibinfo {author} {\bibfnamefont {A.}~\bibnamefont
  {Buzdin}}} (\bibinfo {year} {2019}),\ \href
  {https://doi.org/10.1103/PhysRevB.100.104523} {\bibfield  {journal} {\bibinfo
   {journal} {Phys. Rev. B}\ }\textbf {\bibinfo {volume} {100}},\ \bibinfo
  {pages} {104523}}\BibitemShut {NoStop}%
\bibitem [{\citenamefont {Dey}\ \emph {et~al.}(2013)\citenamefont {Dey},
  \citenamefont {Khuntia}, \citenamefont {Mahajan}, \citenamefont {Anupam},\
  and\ \citenamefont {Hossain}}]{Dey2013}%
  \BibitemOpen
  \bibfield  {author} {\bibinfo {author} {\bibnamefont {Dey}, \bibfnamefont
  {T.}}, \bibinfo {author} {\bibfnamefont {P.}~\bibnamefont {Khuntia}},
  \bibinfo {author} {\bibfnamefont {A.~V.}\ \bibnamefont {Mahajan}}, \bibinfo
  {author} {\bibnamefont {Anupam}}, and\ \bibinfo {author} {\bibfnamefont
  {Z.}~\bibnamefont {Hossain}}} (\bibinfo {year} {2013}),\ \href
  {https://doi.org/10.1140/epjb/e2013-40483-6} {\bibfield  {journal} {\bibinfo
  {journal} {Eur. Phys. J. B}\ }\textbf {\bibinfo {volume} {86}},\ \bibinfo
  {pages} {458}}\BibitemShut {NoStop}%
\bibitem [{\citenamefont {Di~Sante}\ \emph {et~al.}(2026)\citenamefont
  {Di~Sante}, \citenamefont {Neupert}, \citenamefont {Sangiovanni},
  \citenamefont {Thomale}, \citenamefont {Comin}, \citenamefont {Checkelsky},
  \citenamefont {Zeljkovic},\ and\ \citenamefont {Wilson}}]{1g9n-wm38}%
  \BibitemOpen
  \bibfield  {author} {\bibinfo {author} {\bibnamefont {Di~Sante},
  \bibfnamefont {D.}}, \bibinfo {author} {\bibfnamefont {T.}~\bibnamefont
  {Neupert}}, \bibinfo {author} {\bibfnamefont {G.}~\bibnamefont
  {Sangiovanni}}, \bibinfo {author} {\bibfnamefont {R.}~\bibnamefont
  {Thomale}}, \bibinfo {author} {\bibfnamefont {R.}~\bibnamefont {Comin}},
  \bibinfo {author} {\bibfnamefont {J.~G.}\ \bibnamefont {Checkelsky}},
  \bibinfo {author} {\bibfnamefont {I.}~\bibnamefont {Zeljkovic}}, and\
  \bibinfo {author} {\bibfnamefont {S.~D.}\ \bibnamefont {Wilson}}} (\bibinfo
  {year} {2026}),\ \href {https://doi.org/10.1103/1g9n-wm38} {\bibfield
  {journal} {\bibinfo  {journal} {Rev. Mod. Phys.}\ }\textbf {\bibinfo {volume}
  {98}},\ \bibinfo {pages} {015002}}\BibitemShut {NoStop}%
\bibitem [{\citenamefont {Ding}\ \emph {et~al.}(2023)\citenamefont {Ding},
  \citenamefont {Zhao}, \citenamefont {Huang}, \citenamefont {Yu},
  \citenamefont {Xu}, \citenamefont {Fang}, \citenamefont {Chen}, \citenamefont
  {Wang}, \citenamefont {Zhang}, \citenamefont {Chen} \emph
  {et~al.}}]{Ding2023}%
  \BibitemOpen
  \bibfield  {author} {\bibinfo {author} {\bibnamefont {Ding}, \bibfnamefont
  {H.}}, \bibinfo {author} {\bibfnamefont {H.}~\bibnamefont {Zhao}}, \bibinfo
  {author} {\bibfnamefont {P.}~\bibnamefont {Huang}}, \bibinfo {author}
  {\bibfnamefont {L.}~\bibnamefont {Yu}}, \bibinfo {author} {\bibfnamefont
  {J.}~\bibnamefont {Xu}}, \bibinfo {author} {\bibfnamefont {Z.}~\bibnamefont
  {Fang}}, \bibinfo {author} {\bibfnamefont {Z.}~\bibnamefont {Chen}}, \bibinfo
  {author} {\bibfnamefont {D.}~\bibnamefont {Wang}}, \bibinfo {author}
  {\bibfnamefont {X.}~\bibnamefont {Zhang}}, \bibinfo {author} {\bibfnamefont
  {W.}~\bibnamefont {Chen}},  \emph {et~al.}} (\bibinfo {year} {2023}),\ \href
  {https://doi.org/10.1088/1361-6668/acfa29} {\bibfield  {journal} {\bibinfo
  {journal} {Supercond. Sci. Technol.}\ }\textbf {\bibinfo {volume} {36}},\
  \bibinfo {pages} {11LT01}}\BibitemShut {NoStop}%
\bibitem [{\citenamefont {Ding}\ \emph {et~al.}(2018)\citenamefont {Ding},
  \citenamefont {Meier}, \citenamefont {Cui}, \citenamefont {Xu}, \citenamefont
  {B\"ohmer}, \citenamefont {Bud'ko}, \citenamefont {Canfield},\ and\
  \citenamefont {Furukawa}}]{ding2018hedgehog}%
  \BibitemOpen
  \bibfield  {author} {\bibinfo {author} {\bibnamefont {Ding}, \bibfnamefont
  {Q.-P.}}, \bibinfo {author} {\bibfnamefont {W.~R.}\ \bibnamefont {Meier}},
  \bibinfo {author} {\bibfnamefont {J.}~\bibnamefont {Cui}}, \bibinfo {author}
  {\bibfnamefont {M.}~\bibnamefont {Xu}}, \bibinfo {author} {\bibfnamefont
  {A.~E.}\ \bibnamefont {B\"ohmer}}, \bibinfo {author} {\bibfnamefont {S.~L.}\
  \bibnamefont {Bud'ko}}, \bibinfo {author} {\bibfnamefont {P.~C.}\
  \bibnamefont {Canfield}}, and\ \bibinfo {author} {\bibfnamefont
  {Y.}~\bibnamefont {Furukawa}}} (\bibinfo {year} {2018}),\ \href
  {https://doi.org/10.1103/PhysRevLett.121.137204} {\bibfield  {journal}
  {\bibinfo  {journal} {Phys. Rev. Lett.}\ }\textbf {\bibinfo {volume} {121}},\
  \bibinfo {pages} {137204}}\BibitemShut {NoStop}%
\bibitem [{\citenamefont {Ding}\ \emph {et~al.}(2014)\citenamefont {Ding},
  \citenamefont {Pan}, \citenamefont {Yang},\ and\ \citenamefont
  {Wen}}]{Ding2014}%
  \BibitemOpen
  \bibfield  {author} {\bibinfo {author} {\bibnamefont {Ding}, \bibfnamefont
  {X.}}, \bibinfo {author} {\bibfnamefont {Y.}~\bibnamefont {Pan}}, \bibinfo
  {author} {\bibfnamefont {H.}~\bibnamefont {Yang}}, and\ \bibinfo {author}
  {\bibfnamefont {H.-H.}\ \bibnamefont {Wen}}} (\bibinfo {year} {2014}),\ \href
  {https://doi.org/10.1103/PhysRevB.89.224515} {\bibfield  {journal} {\bibinfo
  {journal} {Phys. Rev. B}\ }\textbf {\bibinfo {volume} {89}},\ \bibinfo
  {pages} {224515}}\BibitemShut {NoStop}%
\bibitem [{\citenamefont {Dong}\ \emph {et~al.}(2024)\citenamefont {Dong},
  \citenamefont {Xu},\ and\ \citenamefont {Ma}}]{Dong2024}%
  \BibitemOpen
  \bibfield  {author} {\bibinfo {author} {\bibnamefont {Dong}, \bibfnamefont
  {C.}}, \bibinfo {author} {\bibfnamefont {Q.}~\bibnamefont {Xu}}, and\
  \bibinfo {author} {\bibfnamefont {Y.}~\bibnamefont {Ma}}} (\bibinfo {year}
  {2024}),\ \href {https://doi.org/10.1093/nsr/nwae122} {\bibfield  {journal}
  {\bibinfo  {journal} {Natl. Sci. Rev.}\ }\textbf {\bibinfo {volume} {11}},\
  \bibinfo {pages} {nwae122}}\BibitemShut {NoStop}%
\bibitem [{\citenamefont {Dong}\ \emph {et~al.}(2010)\citenamefont {Dong},
  \citenamefont {Zhou}, \citenamefont {Guan}, \citenamefont {Zhang},
  \citenamefont {Dai}, \citenamefont {Qiu}, \citenamefont {Wang}, \citenamefont
  {He}, \citenamefont {Chen},\ and\ \citenamefont {Li}}]{dong2010quantum}%
  \BibitemOpen
  \bibfield  {author} {\bibinfo {author} {\bibnamefont {Dong}, \bibfnamefont
  {J.~K.}}, \bibinfo {author} {\bibfnamefont {S.~Y.}\ \bibnamefont {Zhou}},
  \bibinfo {author} {\bibfnamefont {T.~Y.}\ \bibnamefont {Guan}}, \bibinfo
  {author} {\bibfnamefont {H.}~\bibnamefont {Zhang}}, \bibinfo {author}
  {\bibfnamefont {Y.~F.}\ \bibnamefont {Dai}}, \bibinfo {author} {\bibfnamefont
  {X.}~\bibnamefont {Qiu}}, \bibinfo {author} {\bibfnamefont {X.~F.}\
  \bibnamefont {Wang}}, \bibinfo {author} {\bibfnamefont {Y.}~\bibnamefont
  {He}}, \bibinfo {author} {\bibfnamefont {X.~H.}\ \bibnamefont {Chen}}, and\
  \bibinfo {author} {\bibfnamefont {S.~Y.}\ \bibnamefont {Li}}} (\bibinfo
  {year} {2010}),\ \href {https://doi.org/10.1103/PhysRevLett.104.087005}
  {\bibfield  {journal} {\bibinfo  {journal} {Phys. Rev. Lett.}\ }\textbf
  {\bibinfo {volume} {104}},\ \bibinfo {pages} {087005}}\BibitemShut {NoStop}%
\bibitem [{\citenamefont {Dong}\ \emph
  {et~al.}(2015{\natexlab{a}})\citenamefont {Dong}, \citenamefont {Jin},
  \citenamefont {Yuan}, \citenamefont {Zhou}, \citenamefont {Yuan},
  \citenamefont {Huang}, \citenamefont {Hua}, \citenamefont {Sun},
  \citenamefont {Zheng}, \citenamefont {Hu}, \citenamefont {Mao}, \citenamefont
  {Ma}, \citenamefont {Zhang}, \citenamefont {Zhou},\ and\ \citenamefont
  {Zhao}}]{dong2015li}%
  \BibitemOpen
  \bibfield  {author} {\bibinfo {author} {\bibnamefont {Dong}, \bibfnamefont
  {X.}}, \bibinfo {author} {\bibfnamefont {K.}~\bibnamefont {Jin}}, \bibinfo
  {author} {\bibfnamefont {D.}~\bibnamefont {Yuan}}, \bibinfo {author}
  {\bibfnamefont {H.}~\bibnamefont {Zhou}}, \bibinfo {author} {\bibfnamefont
  {J.}~\bibnamefont {Yuan}}, \bibinfo {author} {\bibfnamefont {Y.}~\bibnamefont
  {Huang}}, \bibinfo {author} {\bibfnamefont {W.}~\bibnamefont {Hua}}, \bibinfo
  {author} {\bibfnamefont {J.}~\bibnamefont {Sun}}, \bibinfo {author}
  {\bibfnamefont {P.}~\bibnamefont {Zheng}}, \bibinfo {author} {\bibfnamefont
  {W.}~\bibnamefont {Hu}}, \bibinfo {author} {\bibfnamefont {Y.}~\bibnamefont
  {Mao}}, \bibinfo {author} {\bibfnamefont {M.}~\bibnamefont {Ma}}, \bibinfo
  {author} {\bibfnamefont {G.}~\bibnamefont {Zhang}}, \bibinfo {author}
  {\bibfnamefont {F.}~\bibnamefont {Zhou}}, and\ \bibinfo {author}
  {\bibfnamefont {Z.}~\bibnamefont {Zhao}}} (\bibinfo {year}
  {2015}{\natexlab{a}}),\ \href {https://doi.org/10.1103/PhysRevB.92.064515}
  {\bibfield  {journal} {\bibinfo  {journal} {Phys. Rev. B}\ }\textbf {\bibinfo
  {volume} {92}},\ \bibinfo {pages} {064515}}\BibitemShut {NoStop}%
\bibitem [{\citenamefont {Dong}\ \emph
  {et~al.}(2015{\natexlab{b}})\citenamefont {Dong}, \citenamefont {Zhou},
  \citenamefont {Yang}, \citenamefont {Yuan}, \citenamefont {Jin},
  \citenamefont {Zhou}, \citenamefont {Yuan}, \citenamefont {Wei},
  \citenamefont {Li}, \citenamefont {Wang}, \citenamefont {Zhang},\ and\
  \citenamefont {Zhao}}]{dong2015phase}%
  \BibitemOpen
  \bibfield  {author} {\bibinfo {author} {\bibnamefont {Dong}, \bibfnamefont
  {X.}}, \bibinfo {author} {\bibfnamefont {H.}~\bibnamefont {Zhou}}, \bibinfo
  {author} {\bibfnamefont {H.}~\bibnamefont {Yang}}, \bibinfo {author}
  {\bibfnamefont {J.}~\bibnamefont {Yuan}}, \bibinfo {author} {\bibfnamefont
  {K.}~\bibnamefont {Jin}}, \bibinfo {author} {\bibfnamefont {F.}~\bibnamefont
  {Zhou}}, \bibinfo {author} {\bibfnamefont {D.}~\bibnamefont {Yuan}}, \bibinfo
  {author} {\bibfnamefont {L.}~\bibnamefont {Wei}}, \bibinfo {author}
  {\bibfnamefont {J.}~\bibnamefont {Li}}, \bibinfo {author} {\bibfnamefont
  {X.}~\bibnamefont {Wang}}, \bibinfo {author} {\bibfnamefont {G.}~\bibnamefont
  {Zhang}}, and\ \bibinfo {author} {\bibfnamefont {Z.}~\bibnamefont {Zhao}}}
  (\bibinfo {year} {2015}{\natexlab{b}}),\ \href
  {https://doi.org/10.1021/ja511292f} {\bibfield  {journal} {\bibinfo
  {journal} {J. Am. Chem. Soc.}\ }\textbf {\bibinfo {volume} {137}},\ \bibinfo
  {pages} {66}}\BibitemShut {NoStop}%
\bibitem [{\citenamefont {Du}\ \emph {et~al.}(2018)\citenamefont {Du},
  \citenamefont {Yang}, \citenamefont {Altenfeld}, \citenamefont {Gu},
  \citenamefont {Yang}, \citenamefont {Eremin}, \citenamefont {Hirschfeld},
  \citenamefont {Mazin}, \citenamefont {Lin}, \citenamefont {Zhu},\ and\
  \citenamefont {Wen}}]{du2018sign}%
  \BibitemOpen
  \bibfield  {author} {\bibinfo {author} {\bibnamefont {Du}, \bibfnamefont
  {Z.}}, \bibinfo {author} {\bibfnamefont {X.}~\bibnamefont {Yang}}, \bibinfo
  {author} {\bibfnamefont {D.}~\bibnamefont {Altenfeld}}, \bibinfo {author}
  {\bibfnamefont {Q.}~\bibnamefont {Gu}}, \bibinfo {author} {\bibfnamefont
  {H.}~\bibnamefont {Yang}}, \bibinfo {author} {\bibfnamefont {I.}~\bibnamefont
  {Eremin}}, \bibinfo {author} {\bibfnamefont {P.~J.}\ \bibnamefont
  {Hirschfeld}}, \bibinfo {author} {\bibfnamefont {I.~I.}\ \bibnamefont
  {Mazin}}, \bibinfo {author} {\bibfnamefont {H.}~\bibnamefont {Lin}}, \bibinfo
  {author} {\bibfnamefont {X.}~\bibnamefont {Zhu}}, and\ \bibinfo {author}
  {\bibfnamefont {H.-H.}\ \bibnamefont {Wen}}} (\bibinfo {year} {2018}),\ \href
  {https://doi.org/10.1038/nphys4299} {\bibfield  {journal} {\bibinfo
  {journal} {Nat. Phys.}\ }\textbf {\bibinfo {volume} {14}},\ \bibinfo {pages}
  {134}}\BibitemShut {NoStop}%
\bibitem [{\citenamefont {Dvorak}\ \emph {et~al.}(2016)\citenamefont {Dvorak},
  \citenamefont {Jarrige}, \citenamefont {Bisogni}, \citenamefont {Coburn},\
  and\ \citenamefont {Leonhardt}}]{dvorak2016six}%
  \BibitemOpen
  \bibfield  {author} {\bibinfo {author} {\bibnamefont {Dvorak}, \bibfnamefont
  {J.}}, \bibinfo {author} {\bibfnamefont {I.}~\bibnamefont {Jarrige}},
  \bibinfo {author} {\bibfnamefont {V.}~\bibnamefont {Bisogni}}, \bibinfo
  {author} {\bibfnamefont {S.}~\bibnamefont {Coburn}}, and\ \bibinfo {author}
  {\bibfnamefont {W.}~\bibnamefont {Leonhardt}}} (\bibinfo {year} {2016}),\
  \href {https://doi.org/10.1063/1.4964847} {\bibfield  {journal} {\bibinfo
  {journal} {Rev. Sci. Instrum.}\ }\textbf {\bibinfo {volume} {87}},\ \bibinfo
  {pages} {115109}}\BibitemShut {NoStop}%
\bibitem [{\citenamefont {Eckberg}\ \emph {et~al.}(2020)\citenamefont
  {Eckberg}, \citenamefont {Campbell}, \citenamefont {Metz}, \citenamefont
  {Collini}, \citenamefont {Hodovanets}, \citenamefont {Drye}, \citenamefont
  {Zavalij}, \citenamefont {Christensen}, \citenamefont {Fernandes},
  \citenamefont {Lee}, \citenamefont {Abbamonte}, \citenamefont {Lynn},\ and\
  \citenamefont {Paglione}}]{eckberg2020}%
  \BibitemOpen
  \bibfield  {author} {\bibinfo {author} {\bibnamefont {Eckberg}, \bibfnamefont
  {C.}}, \bibinfo {author} {\bibfnamefont {D.~J.}\ \bibnamefont {Campbell}},
  \bibinfo {author} {\bibfnamefont {T.}~\bibnamefont {Metz}}, \bibinfo {author}
  {\bibfnamefont {J.}~\bibnamefont {Collini}}, \bibinfo {author} {\bibfnamefont
  {H.}~\bibnamefont {Hodovanets}}, \bibinfo {author} {\bibfnamefont
  {T.}~\bibnamefont {Drye}}, \bibinfo {author} {\bibfnamefont {P.}~\bibnamefont
  {Zavalij}}, \bibinfo {author} {\bibfnamefont {M.~H.}\ \bibnamefont
  {Christensen}}, \bibinfo {author} {\bibfnamefont {R.~M.}\ \bibnamefont
  {Fernandes}}, \bibinfo {author} {\bibfnamefont {S.}~\bibnamefont {Lee}},
  \bibinfo {author} {\bibfnamefont {P.}~\bibnamefont {Abbamonte}}, \bibinfo
  {author} {\bibfnamefont {J.~W.}\ \bibnamefont {Lynn}}, and\ \bibinfo {author}
  {\bibfnamefont {J.}~\bibnamefont {Paglione}}} (\bibinfo {year} {2020}),\
  \href {https://doi.org/10.1038/s41567-019-0736-9} {\bibfield  {journal}
  {\bibinfo  {journal} {Nat. Phys.}\ }\textbf {\bibinfo {volume} {16}},\
  \bibinfo {pages} {346}}\BibitemShut {NoStop}%
\bibitem [{\citenamefont {Eguchi}\ \emph {et~al.}(2013)\citenamefont {Eguchi},
  \citenamefont {Ishikawa}, \citenamefont {Kodama}, \citenamefont
  {Wakabayashi}, \citenamefont {Nakayama}, \citenamefont {Ohmura},\ and\
  \citenamefont {Yamada}}]{Eguchi2013}%
  \BibitemOpen
  \bibfield  {author} {\bibinfo {author} {\bibnamefont {Eguchi}, \bibfnamefont
  {N.}}, \bibinfo {author} {\bibfnamefont {F.}~\bibnamefont {Ishikawa}},
  \bibinfo {author} {\bibfnamefont {M.}~\bibnamefont {Kodama}}, \bibinfo
  {author} {\bibfnamefont {T.}~\bibnamefont {Wakabayashi}}, \bibinfo {author}
  {\bibfnamefont {A.}~\bibnamefont {Nakayama}}, \bibinfo {author}
  {\bibfnamefont {A.}~\bibnamefont {Ohmura}}, and\ \bibinfo {author}
  {\bibfnamefont {Y.}~\bibnamefont {Yamada}}} (\bibinfo {year} {2013}),\ \href
  {https://doi.org/10.7566/jpsj.82.045002} {\bibfield  {journal} {\bibinfo
  {journal} {J. Phys. Soc. Jpn.}\ }\textbf {\bibinfo {volume} {82}},\ \bibinfo
  {pages} {045002}}\BibitemShut {NoStop}%
\bibitem [{\citenamefont {Faeth}\ \emph {et~al.}(2021)\citenamefont {Faeth},
  \citenamefont {Xie}, \citenamefont {Yang}, \citenamefont {Kawasaki},
  \citenamefont {Nelson}, \citenamefont {Zhang}, \citenamefont {Parzyck},
  \citenamefont {Mishra}, \citenamefont {Li}, \citenamefont {Jozwiak},
  \citenamefont {Bostwick}, \citenamefont {Rotenberg}, \citenamefont {Schlom},\
  and\ \citenamefont {Shen}}]{faeth2021interfacial}%
  \BibitemOpen
  \bibfield  {author} {\bibinfo {author} {\bibnamefont {Faeth}, \bibfnamefont
  {B.~D.}}, \bibinfo {author} {\bibfnamefont {S.}~\bibnamefont {Xie}}, \bibinfo
  {author} {\bibfnamefont {S.}~\bibnamefont {Yang}}, \bibinfo {author}
  {\bibfnamefont {J.~K.}\ \bibnamefont {Kawasaki}}, \bibinfo {author}
  {\bibfnamefont {J.~N.}\ \bibnamefont {Nelson}}, \bibinfo {author}
  {\bibfnamefont {S.}~\bibnamefont {Zhang}}, \bibinfo {author} {\bibfnamefont
  {C.}~\bibnamefont {Parzyck}}, \bibinfo {author} {\bibfnamefont
  {P.}~\bibnamefont {Mishra}}, \bibinfo {author} {\bibfnamefont
  {C.}~\bibnamefont {Li}}, \bibinfo {author} {\bibfnamefont {C.}~\bibnamefont
  {Jozwiak}}, \bibinfo {author} {\bibfnamefont {A.}~\bibnamefont {Bostwick}},
  \bibinfo {author} {\bibfnamefont {E.}~\bibnamefont {Rotenberg}}, \bibinfo
  {author} {\bibfnamefont {D.~G.}\ \bibnamefont {Schlom}}, and\ \bibinfo
  {author} {\bibfnamefont {K.~M.}\ \bibnamefont {Shen}}} (\bibinfo {year}
  {2021}),\ \href {https://doi.org/10.1103/PhysRevLett.127.016803} {\bibfield
  {journal} {\bibinfo  {journal} {Phys. Rev. Lett.}\ }\textbf {\bibinfo
  {volume} {127}},\ \bibinfo {pages} {016803}}\BibitemShut {NoStop}%
\bibitem [{\citenamefont {Fan}\ \emph {et~al.}(2015)\citenamefont {Fan},
  \citenamefont {Zhang}, \citenamefont {Liu}, \citenamefont {Yan},
  \citenamefont {Ren}, \citenamefont {Peng}, \citenamefont {Xu}, \citenamefont
  {Xie}, \citenamefont {Hu}, \citenamefont {Zhang},\ and\ \citenamefont
  {Feng}}]{fan2015plain}%
  \BibitemOpen
  \bibfield  {author} {\bibinfo {author} {\bibnamefont {Fan}, \bibfnamefont
  {Q.}}, \bibinfo {author} {\bibfnamefont {W.~H.}\ \bibnamefont {Zhang}},
  \bibinfo {author} {\bibfnamefont {X.}~\bibnamefont {Liu}}, \bibinfo {author}
  {\bibfnamefont {Y.~J.}\ \bibnamefont {Yan}}, \bibinfo {author} {\bibfnamefont
  {M.~Q.}\ \bibnamefont {Ren}}, \bibinfo {author} {\bibfnamefont
  {R.}~\bibnamefont {Peng}}, \bibinfo {author} {\bibfnamefont {H.~C.}\
  \bibnamefont {Xu}}, \bibinfo {author} {\bibfnamefont {B.~P.}\ \bibnamefont
  {Xie}}, \bibinfo {author} {\bibfnamefont {J.~P.}\ \bibnamefont {Hu}},
  \bibinfo {author} {\bibfnamefont {T.}~\bibnamefont {Zhang}}, and\ \bibinfo
  {author} {\bibfnamefont {D.~L.}\ \bibnamefont {Feng}}} (\bibinfo {year}
  {2015}),\ \href {https://doi.org/10.1038/nphys3450} {\bibfield  {journal}
  {\bibinfo  {journal} {Nat. Phys.}\ }\textbf {\bibinfo {volume} {11}},\
  \bibinfo {pages} {946}}\BibitemShut {NoStop}%
\bibitem [{\citenamefont {Fanfarillo}\ \emph {et~al.}(2016)\citenamefont
  {Fanfarillo}, \citenamefont {Mansart}, \citenamefont {Toulemonde},
  \citenamefont {Cercellier}, \citenamefont {Le~F\`evre}, \citenamefont
  {Bertran}, \citenamefont {Valenzuela}, \citenamefont {Benfatto},\ and\
  \citenamefont {Brouet}}]{fanfarillo2016orbital}%
  \BibitemOpen
  \bibfield  {author} {\bibinfo {author} {\bibnamefont {Fanfarillo},
  \bibfnamefont {L.}}, \bibinfo {author} {\bibfnamefont {J.}~\bibnamefont
  {Mansart}}, \bibinfo {author} {\bibfnamefont {P.}~\bibnamefont {Toulemonde}},
  \bibinfo {author} {\bibfnamefont {H.}~\bibnamefont {Cercellier}}, \bibinfo
  {author} {\bibfnamefont {P.}~\bibnamefont {Le~F\`evre}}, \bibinfo {author}
  {\bibfnamefont {F.~m.~c.}\ \bibnamefont {Bertran}}, \bibinfo {author}
  {\bibfnamefont {B.}~\bibnamefont {Valenzuela}}, \bibinfo {author}
  {\bibfnamefont {L.}~\bibnamefont {Benfatto}}, and\ \bibinfo {author}
  {\bibfnamefont {V.}~\bibnamefont {Brouet}}} (\bibinfo {year} {2016}),\ \href
  {https://doi.org/10.1103/PhysRevB.94.155138} {\bibfield  {journal} {\bibinfo
  {journal} {Phys. Rev. B}\ }\textbf {\bibinfo {volume} {94}},\ \bibinfo
  {pages} {155138}}\BibitemShut {NoStop}%
\bibitem [{\citenamefont {Fanfarillo}\ \emph {et~al.}(2020)\citenamefont
  {Fanfarillo}, \citenamefont {Valli},\ and\ \citenamefont
  {Capone}}]{fanfarillo2020synergy}%
  \BibitemOpen
  \bibfield  {author} {\bibinfo {author} {\bibnamefont {Fanfarillo},
  \bibfnamefont {L.}}, \bibinfo {author} {\bibfnamefont {A.}~\bibnamefont
  {Valli}}, and\ \bibinfo {author} {\bibfnamefont {M.}~\bibnamefont {Capone}}}
  (\bibinfo {year} {2020}),\ \href
  {https://doi.org/10.1103/PhysRevLett.125.177001} {\bibfield  {journal}
  {\bibinfo  {journal} {Phys. Rev. Lett.}\ }\textbf {\bibinfo {volume} {125}},\
  \bibinfo {pages} {177001}}\BibitemShut {NoStop}%
\bibitem [{\citenamefont {Fang}\ \emph {et~al.}(2008)\citenamefont {Fang},
  \citenamefont {Pham}, \citenamefont {Qian}, \citenamefont {Liu},
  \citenamefont {Vehstedt}, \citenamefont {Liu}, \citenamefont {Spinu},\ and\
  \citenamefont {Mao}}]{Fang2008}%
  \BibitemOpen
  \bibfield  {author} {\bibinfo {author} {\bibnamefont {Fang}, \bibfnamefont
  {M.}}, \bibinfo {author} {\bibfnamefont {H.}~\bibnamefont {Pham}}, \bibinfo
  {author} {\bibfnamefont {B.}~\bibnamefont {Qian}}, \bibinfo {author}
  {\bibfnamefont {T.}~\bibnamefont {Liu}}, \bibinfo {author} {\bibfnamefont
  {E.}~\bibnamefont {Vehstedt}}, \bibinfo {author} {\bibfnamefont
  {Y.}~\bibnamefont {Liu}}, \bibinfo {author} {\bibfnamefont {L.}~\bibnamefont
  {Spinu}}, and\ \bibinfo {author} {\bibfnamefont {Z.}~\bibnamefont {Mao}}}
  (\bibinfo {year} {2008}),\ \href {https://doi.org/10.1103/PhysRevB.78.224503}
  {\bibfield  {journal} {\bibinfo  {journal} {Phys. Rev. B}\ }\textbf {\bibinfo
  {volume} {78}},\ \bibinfo {pages} {224503}}\BibitemShut {NoStop}%
\bibitem [{\citenamefont {Fernandes}\ \emph {et~al.}(2012)\citenamefont
  {Fernandes}, \citenamefont {Chubukov}, \citenamefont {Knolle}, \citenamefont
  {Eremin},\ and\ \citenamefont {Schmalian}}]{fernandes2012preemptive}%
  \BibitemOpen
  \bibfield  {author} {\bibinfo {author} {\bibnamefont {Fernandes},
  \bibfnamefont {R.~M.}}, \bibinfo {author} {\bibfnamefont {A.~V.}\
  \bibnamefont {Chubukov}}, \bibinfo {author} {\bibfnamefont {J.}~\bibnamefont
  {Knolle}}, \bibinfo {author} {\bibfnamefont {I.}~\bibnamefont {Eremin}}, and\
  \bibinfo {author} {\bibfnamefont {J.}~\bibnamefont {Schmalian}}} (\bibinfo
  {year} {2012}),\ \href {https://doi.org/10.1103/PhysRevB.85.024534}
  {\bibfield  {journal} {\bibinfo  {journal} {Phys. Rev. B}\ }\textbf {\bibinfo
  {volume} {85}},\ \bibinfo {pages} {024534}}\BibitemShut {NoStop}%
\bibitem [{\citenamefont {Fernandes}\ \emph {et~al.}(2014)\citenamefont
  {Fernandes}, \citenamefont {Chubukov},\ and\ \citenamefont
  {Schmalian}}]{fernandes2014what}%
  \BibitemOpen
  \bibfield  {author} {\bibinfo {author} {\bibnamefont {Fernandes},
  \bibfnamefont {R.~M.}}, \bibinfo {author} {\bibfnamefont {A.~V.}\
  \bibnamefont {Chubukov}}, and\ \bibinfo {author} {\bibfnamefont
  {J.}~\bibnamefont {Schmalian}}} (\bibinfo {year} {2014}),\ \href
  {https://doi.org/10.1038/nphys2877} {\bibfield  {journal} {\bibinfo
  {journal} {Nat. Phys.}\ }\textbf {\bibinfo {volume} {10}},\ \bibinfo {pages}
  {97}}\BibitemShut {NoStop}%
\bibitem [{\citenamefont {Fernandes}\ \emph {et~al.}(2022)\citenamefont
  {Fernandes}, \citenamefont {Coldea}, \citenamefont {Ding}, \citenamefont
  {Fisher}, \citenamefont {Hirschfeld},\ and\ \citenamefont
  {Kotliar}}]{fernandes2022iron}%
  \BibitemOpen
  \bibfield  {author} {\bibinfo {author} {\bibnamefont {Fernandes},
  \bibfnamefont {R.~M.}}, \bibinfo {author} {\bibfnamefont {A.~I.}\
  \bibnamefont {Coldea}}, \bibinfo {author} {\bibfnamefont {H.}~\bibnamefont
  {Ding}}, \bibinfo {author} {\bibfnamefont {I.~R.}\ \bibnamefont {Fisher}},
  \bibinfo {author} {\bibfnamefont {P.~J.}\ \bibnamefont {Hirschfeld}}, and\
  \bibinfo {author} {\bibfnamefont {G.}~\bibnamefont {Kotliar}}} (\bibinfo
  {year} {2022}),\ \href {https://doi.org/10.1038/s41586-021-04073-2}
  {\bibfield  {journal} {\bibinfo  {journal} {Nature}\ }\textbf {\bibinfo
  {volume} {601}},\ \bibinfo {pages} {35}}\BibitemShut {NoStop}%
\bibitem [{\citenamefont {Fernandes}\ and\ \citenamefont
  {Millis}(2013)}]{fernandes2013nematicity}%
  \BibitemOpen
  \bibfield  {author} {\bibinfo {author} {\bibnamefont {Fernandes},
  \bibfnamefont {R.~M.}}, and\ \bibinfo {author} {\bibfnamefont {A.~J.}\
  \bibnamefont {Millis}}} (\bibinfo {year} {2013}),\ \href
  {https://doi.org/10.1103/PhysRevLett.111.127001} {\bibfield  {journal}
  {\bibinfo  {journal} {Phys. Rev. Lett.}\ }\textbf {\bibinfo {volume} {111}},\
  \bibinfo {pages} {127001}}\BibitemShut {NoStop}%
\bibitem [{\citenamefont {Fernandes}\ \emph {et~al.}(2019)\citenamefont
  {Fernandes}, \citenamefont {Orth},\ and\ \citenamefont
  {Schmalian}}]{fernandes2019intertwined}%
  \BibitemOpen
  \bibfield  {author} {\bibinfo {author} {\bibnamefont {Fernandes},
  \bibfnamefont {R.~M.}}, \bibinfo {author} {\bibfnamefont {P.~P.}\
  \bibnamefont {Orth}}, and\ \bibinfo {author} {\bibfnamefont {J.}~\bibnamefont
  {Schmalian}}} (\bibinfo {year} {2019}),\ \href
  {https://doi.org/10.1146/annurev-conmatphys-031218-013200} {\bibfield
  {journal} {\bibinfo  {journal} {Annu. Rev. Condens. Matter Phys.}\ }\textbf
  {\bibinfo {volume} {10}},\ \bibinfo {pages} {133}}\BibitemShut {NoStop}%
\bibitem [{\citenamefont {Fernandes}\ and\ \citenamefont
  {Schmalian}(2012)}]{fernandes2012manifestations}%
  \BibitemOpen
  \bibfield  {author} {\bibinfo {author} {\bibnamefont {Fernandes},
  \bibfnamefont {R.~M.}}, and\ \bibinfo {author} {\bibfnamefont
  {J.}~\bibnamefont {Schmalian}}} (\bibinfo {year} {2012}),\ \href
  {https://doi.org/10.1088/0953-2048/25/8/084005} {\bibfield  {journal}
  {\bibinfo  {journal} {Supercond. Sci. Technol.}\ }\textbf {\bibinfo {volume}
  {25}},\ \bibinfo {pages} {084005}}\BibitemShut {NoStop}%
\bibitem [{\citenamefont {Frachet}\ \emph {et~al.}(2024)\citenamefont
  {Frachet}, \citenamefont {Wang}, \citenamefont {Xia}, \citenamefont {Guo},
  \citenamefont {He}, \citenamefont {Maraytta}, \citenamefont {Heid},
  \citenamefont {Haghighirad}, \citenamefont {Merz}, \citenamefont {Meingast},\
  and\ \citenamefont {Hardy}}]{frachet2024colossal}%
  \BibitemOpen
  \bibfield  {author} {\bibinfo {author} {\bibnamefont {Frachet}, \bibfnamefont
  {M.}}, \bibinfo {author} {\bibfnamefont {L.}~\bibnamefont {Wang}}, \bibinfo
  {author} {\bibfnamefont {W.}~\bibnamefont {Xia}}, \bibinfo {author}
  {\bibfnamefont {Y.}~\bibnamefont {Guo}}, \bibinfo {author} {\bibfnamefont
  {M.}~\bibnamefont {He}}, \bibinfo {author} {\bibfnamefont {N.}~\bibnamefont
  {Maraytta}}, \bibinfo {author} {\bibfnamefont {R.}~\bibnamefont {Heid}},
  \bibinfo {author} {\bibfnamefont {A.-A.}\ \bibnamefont {Haghighirad}},
  \bibinfo {author} {\bibfnamefont {M.}~\bibnamefont {Merz}}, \bibinfo {author}
  {\bibfnamefont {C.}~\bibnamefont {Meingast}}, and\ \bibinfo {author}
  {\bibfnamefont {F.}~\bibnamefont {Hardy}}} (\bibinfo {year} {2024}),\ \href
  {https://doi.org/10.1103/PhysRevLett.132.186001} {\bibfield  {journal}
  {\bibinfo  {journal} {Phys. Rev. Lett.}\ }\textbf {\bibinfo {volume} {132}},\
  \bibinfo {pages} {186001}}\BibitemShut {NoStop}%
\bibitem [{\citenamefont {Fradkin}\ \emph {et~al.}(2015)\citenamefont
  {Fradkin}, \citenamefont {Kivelson},\ and\ \citenamefont
  {Tranquada}}]{fradkin2015colloquium}%
  \BibitemOpen
  \bibfield  {author} {\bibinfo {author} {\bibnamefont {Fradkin}, \bibfnamefont
  {E.}}, \bibinfo {author} {\bibfnamefont {S.~A.}\ \bibnamefont {Kivelson}},
  and\ \bibinfo {author} {\bibfnamefont {J.~M.}\ \bibnamefont {Tranquada}}}
  (\bibinfo {year} {2015}),\ \href {https://doi.org/10.1103/RevModPhys.87.457}
  {\bibfield  {journal} {\bibinfo  {journal} {Rev. Mod. Phys.}\ }\textbf
  {\bibinfo {volume} {87}},\ \bibinfo {pages} {457}}\BibitemShut {NoStop}%
\bibitem [{\citenamefont {Frandsen}\ \emph {et~al.}(2017)\citenamefont
  {Frandsen}, \citenamefont {Taddei}, \citenamefont {Yi}, \citenamefont
  {Frano}, \citenamefont {Guguchia}, \citenamefont {Yu}, \citenamefont {Si},
  \citenamefont {Bugaris}, \citenamefont {Stadel}, \citenamefont {Osborn},
  \citenamefont {Rosenkranz}, \citenamefont {Chmaissem},\ and\ \citenamefont
  {Birgeneau}}]{frandsen2017local}%
  \BibitemOpen
  \bibfield  {author} {\bibinfo {author} {\bibnamefont {Frandsen},
  \bibfnamefont {B.~A.}}, \bibinfo {author} {\bibfnamefont {K.~M.}\
  \bibnamefont {Taddei}}, \bibinfo {author} {\bibfnamefont {M.}~\bibnamefont
  {Yi}}, \bibinfo {author} {\bibfnamefont {A.}~\bibnamefont {Frano}}, \bibinfo
  {author} {\bibfnamefont {Z.}~\bibnamefont {Guguchia}}, \bibinfo {author}
  {\bibfnamefont {R.}~\bibnamefont {Yu}}, \bibinfo {author} {\bibfnamefont
  {Q.}~\bibnamefont {Si}}, \bibinfo {author} {\bibfnamefont {D.~E.}\
  \bibnamefont {Bugaris}}, \bibinfo {author} {\bibfnamefont {R.}~\bibnamefont
  {Stadel}}, \bibinfo {author} {\bibfnamefont {R.}~\bibnamefont {Osborn}},
  \bibinfo {author} {\bibfnamefont {S.}~\bibnamefont {Rosenkranz}}, \bibinfo
  {author} {\bibfnamefont {O.}~\bibnamefont {Chmaissem}}, and\ \bibinfo
  {author} {\bibfnamefont {R.~J.}\ \bibnamefont {Birgeneau}}} (\bibinfo {year}
  {2017}),\ \href {https://doi.org/10.1103/PhysRevLett.119.187001} {\bibfield
  {journal} {\bibinfo  {journal} {Phys. Rev. Lett.}\ }\textbf {\bibinfo
  {volume} {119}},\ \bibinfo {pages} {187001}}\BibitemShut {NoStop}%
\bibitem [{\citenamefont {Frandsen}\ \emph {et~al.}(2019)\citenamefont
  {Frandsen}, \citenamefont {Wang}, \citenamefont {Wu}, \citenamefont {Zhao},\
  and\ \citenamefont {Birgeneau}}]{frandsen2019quantitative}%
  \BibitemOpen
  \bibfield  {author} {\bibinfo {author} {\bibnamefont {Frandsen},
  \bibfnamefont {B.~A.}}, \bibinfo {author} {\bibfnamefont {Q.}~\bibnamefont
  {Wang}}, \bibinfo {author} {\bibfnamefont {S.}~\bibnamefont {Wu}}, \bibinfo
  {author} {\bibfnamefont {J.}~\bibnamefont {Zhao}}, and\ \bibinfo {author}
  {\bibfnamefont {R.~J.}\ \bibnamefont {Birgeneau}}} (\bibinfo {year} {2019}),\
  \href {https://doi.org/10.1103/PhysRevB.100.020504} {\bibfield  {journal}
  {\bibinfo  {journal} {Phys. Rev. B}\ }\textbf {\bibinfo {volume} {100}},\
  \bibinfo {pages} {020504(R)}}\BibitemShut {NoStop}%
\bibitem [{\citenamefont {Friemel}\ \emph {et~al.}(2012)\citenamefont
  {Friemel}, \citenamefont {Liu}, \citenamefont {Goremychkin}, \citenamefont
  {Liu}, \citenamefont {Park}, \citenamefont {Sobolev}, \citenamefont {Lin},
  \citenamefont {Keimer},\ and\ \citenamefont
  {Inosov}}]{friemel2012conformity}%
  \BibitemOpen
  \bibfield  {author} {\bibinfo {author} {\bibnamefont {Friemel}, \bibfnamefont
  {G.}}, \bibinfo {author} {\bibfnamefont {W.~P.}\ \bibnamefont {Liu}},
  \bibinfo {author} {\bibfnamefont {E.~A.}\ \bibnamefont {Goremychkin}},
  \bibinfo {author} {\bibfnamefont {Y.}~\bibnamefont {Liu}}, \bibinfo {author}
  {\bibfnamefont {J.~T.}\ \bibnamefont {Park}}, \bibinfo {author}
  {\bibfnamefont {O.}~\bibnamefont {Sobolev}}, \bibinfo {author} {\bibfnamefont
  {C.~T.}\ \bibnamefont {Lin}}, \bibinfo {author} {\bibfnamefont
  {B.}~\bibnamefont {Keimer}}, and\ \bibinfo {author} {\bibfnamefont {D.~S.}\
  \bibnamefont {Inosov}}} (\bibinfo {year} {2012}),\ \href
  {https://doi.org/10.1209/0295-5075/99/67004} {\bibfield  {journal} {\bibinfo
  {journal} {EPL}\ }\textbf {\bibinfo {volume} {99}},\ \bibinfo {pages}
  {67004}}\BibitemShut {NoStop}%
\bibitem [{\citenamefont {Fu}\ and\ \citenamefont
  {Kane}(2008)}]{fu2008superconducting}%
  \BibitemOpen
  \bibfield  {author} {\bibinfo {author} {\bibnamefont {Fu}, \bibfnamefont
  {L.}}, and\ \bibinfo {author} {\bibfnamefont {C.~L.}\ \bibnamefont {Kane}}}
  (\bibinfo {year} {2008}),\ \href
  {https://doi.org/10.1103/PhysRevLett.100.096407} {\bibfield  {journal}
  {\bibinfo  {journal} {Phys. Rev. Lett.}\ }\textbf {\bibinfo {volume} {100}},\
  \bibinfo {pages} {096407}}\BibitemShut {NoStop}%
\bibitem [{\citenamefont {Fujii}\ \emph {et~al.}(2018)\citenamefont {Fujii},
  \citenamefont {Simayi}, \citenamefont {Sakano}, \citenamefont {Sasaki},
  \citenamefont {Nakamura}, \citenamefont {Nakanishi}, \citenamefont {Kihou},
  \citenamefont {Nakajima}, \citenamefont {Lee}, \citenamefont {Iyo},
  \citenamefont {Eisaki}, \citenamefont {Uchida},\ and\ \citenamefont
  {Yoshizawa}}]{fujii2018anisotropic}%
  \BibitemOpen
  \bibfield  {author} {\bibinfo {author} {\bibnamefont {Fujii}, \bibfnamefont
  {C.}}, \bibinfo {author} {\bibfnamefont {S.}~\bibnamefont {Simayi}}, \bibinfo
  {author} {\bibfnamefont {K.}~\bibnamefont {Sakano}}, \bibinfo {author}
  {\bibfnamefont {C.}~\bibnamefont {Sasaki}}, \bibinfo {author} {\bibfnamefont
  {M.}~\bibnamefont {Nakamura}}, \bibinfo {author} {\bibfnamefont
  {Y.}~\bibnamefont {Nakanishi}}, \bibinfo {author} {\bibfnamefont
  {K.}~\bibnamefont {Kihou}}, \bibinfo {author} {\bibfnamefont
  {M.}~\bibnamefont {Nakajima}}, \bibinfo {author} {\bibfnamefont {C.-H.}\
  \bibnamefont {Lee}}, \bibinfo {author} {\bibfnamefont {A.}~\bibnamefont
  {Iyo}}, \bibinfo {author} {\bibfnamefont {H.}~\bibnamefont {Eisaki}},
  \bibinfo {author} {\bibfnamefont {S.-i.}\ \bibnamefont {Uchida}}, and\
  \bibinfo {author} {\bibfnamefont {M.}~\bibnamefont {Yoshizawa}}} (\bibinfo
  {year} {2018}),\ \href {https://doi.org/10.7566/JPSJ.87.074710} {\bibfield
  {journal} {\bibinfo  {journal} {J. Phys. Soc. Jpn.}\ }\textbf {\bibinfo
  {volume} {87}},\ \bibinfo {pages} {074710}}\BibitemShut {NoStop}%
\bibitem [{\citenamefont {Fujioka}\ \emph {et~al.}(2013)\citenamefont
  {Fujioka}, \citenamefont {Denholme}, \citenamefont {Ozaki}, \citenamefont
  {Okazaki}, \citenamefont {Deguchi}, \citenamefont {Demura}, \citenamefont
  {Hara}, \citenamefont {Watanabe}, \citenamefont {Takeya}, \citenamefont
  {Yamaguchi} \emph {et~al.}}]{Fujioka2013}%
  \BibitemOpen
  \bibfield  {author} {\bibinfo {author} {\bibnamefont {Fujioka}, \bibfnamefont
  {M.}}, \bibinfo {author} {\bibfnamefont {S.~J.}\ \bibnamefont {Denholme}},
  \bibinfo {author} {\bibfnamefont {T.}~\bibnamefont {Ozaki}}, \bibinfo
  {author} {\bibfnamefont {H.}~\bibnamefont {Okazaki}}, \bibinfo {author}
  {\bibfnamefont {K.}~\bibnamefont {Deguchi}}, \bibinfo {author} {\bibfnamefont
  {S.}~\bibnamefont {Demura}}, \bibinfo {author} {\bibfnamefont
  {H.}~\bibnamefont {Hara}}, \bibinfo {author} {\bibfnamefont {T.}~\bibnamefont
  {Watanabe}}, \bibinfo {author} {\bibfnamefont {H.}~\bibnamefont {Takeya}},
  \bibinfo {author} {\bibfnamefont {T.}~\bibnamefont {Yamaguchi}},  \emph
  {et~al.}} (\bibinfo {year} {2013}),\ \href
  {https://doi.org/10.1088/0953-2048/26/8/085023} {\bibfield  {journal}
  {\bibinfo  {journal} {Supercond. Sci. Technol.}\ }\textbf {\bibinfo {volume}
  {26}},\ \bibinfo {pages} {085023}}\BibitemShut {NoStop}%
\bibitem [{\citenamefont {Fujita}\ \emph {et~al.}(2012)\citenamefont {Fujita},
  \citenamefont {Hiraka}, \citenamefont {Matsuda}, \citenamefont {Matsuura},
  \citenamefont {M.~Tranquada}, \citenamefont {Wakimoto}, \citenamefont {Xu},\
  and\ \citenamefont {Yamada}}]{fujita2012progress}%
  \BibitemOpen
  \bibfield  {author} {\bibinfo {author} {\bibnamefont {Fujita}, \bibfnamefont
  {M.}}, \bibinfo {author} {\bibfnamefont {H.}~\bibnamefont {Hiraka}}, \bibinfo
  {author} {\bibfnamefont {M.}~\bibnamefont {Matsuda}}, \bibinfo {author}
  {\bibfnamefont {M.}~\bibnamefont {Matsuura}}, \bibinfo {author}
  {\bibfnamefont {J.}~\bibnamefont {M.~Tranquada}}, \bibinfo {author}
  {\bibfnamefont {S.}~\bibnamefont {Wakimoto}}, \bibinfo {author}
  {\bibfnamefont {G.}~\bibnamefont {Xu}}, and\ \bibinfo {author} {\bibfnamefont
  {K.}~\bibnamefont {Yamada}}} (\bibinfo {year} {2012}),\ \href
  {https://doi.org/10.1143/JPSJ.81.011007} {\bibfield  {journal} {\bibinfo
  {journal} {J. Phys. Soc. Jpn.}\ }\textbf {\bibinfo {volume} {81}},\ \bibinfo
  {pages} {011007}}\BibitemShut {NoStop}%
\bibitem [{\citenamefont {Fulde}\ and\ \citenamefont
  {Ferrell}(1964)}]{fulde1964superconductivity}%
  \BibitemOpen
  \bibfield  {author} {\bibinfo {author} {\bibnamefont {Fulde}, \bibfnamefont
  {P.}}, and\ \bibinfo {author} {\bibfnamefont {R.~A.}\ \bibnamefont
  {Ferrell}}} (\bibinfo {year} {1964}),\ \href
  {https://doi.org/10.1103/PhysRev.135.A550} {\bibfield  {journal} {\bibinfo
  {journal} {Phys. Rev.}\ }\textbf {\bibinfo {volume} {135}},\ \bibinfo {pages}
  {A550}}\BibitemShut {NoStop}%
\bibitem [{\citenamefont {Gallais}\ \emph {et~al.}(2013)\citenamefont
  {Gallais}, \citenamefont {Fernandes}, \citenamefont {Paul}, \citenamefont
  {Chauvi\`ere}, \citenamefont {Yang}, \citenamefont {M\'easson}, \citenamefont
  {Cazayous}, \citenamefont {Sacuto}, \citenamefont {Colson},\ and\
  \citenamefont {Forget}}]{gallais2013observation}%
  \BibitemOpen
  \bibfield  {author} {\bibinfo {author} {\bibnamefont {Gallais}, \bibfnamefont
  {Y.}}, \bibinfo {author} {\bibfnamefont {R.~M.}\ \bibnamefont {Fernandes}},
  \bibinfo {author} {\bibfnamefont {I.}~\bibnamefont {Paul}}, \bibinfo {author}
  {\bibfnamefont {L.}~\bibnamefont {Chauvi\`ere}}, \bibinfo {author}
  {\bibfnamefont {Y.-X.}\ \bibnamefont {Yang}}, \bibinfo {author}
  {\bibfnamefont {M.-A.}\ \bibnamefont {M\'easson}}, \bibinfo {author}
  {\bibfnamefont {M.}~\bibnamefont {Cazayous}}, \bibinfo {author}
  {\bibfnamefont {A.}~\bibnamefont {Sacuto}}, \bibinfo {author} {\bibfnamefont
  {D.}~\bibnamefont {Colson}}, and\ \bibinfo {author} {\bibfnamefont
  {A.}~\bibnamefont {Forget}}} (\bibinfo {year} {2013}),\ \href
  {https://doi.org/10.1103/PhysRevLett.111.267001} {\bibfield  {journal}
  {\bibinfo  {journal} {Phys. Rev. Lett.}\ }\textbf {\bibinfo {volume} {111}},\
  \bibinfo {pages} {267001}}\BibitemShut {NoStop}%
\bibitem [{\citenamefont {Gallais}\ and\ \citenamefont
  {Paul}(2016)}]{gallais2016charge}%
  \BibitemOpen
  \bibfield  {author} {\bibinfo {author} {\bibnamefont {Gallais}, \bibfnamefont
  {Y.}}, and\ \bibinfo {author} {\bibfnamefont {I.}~\bibnamefont {Paul}}}
  (\bibinfo {year} {2016}),\ \href {https://doi.org/10.1016/j.crhy.2015.10.001}
  {\bibfield  {journal} {\bibinfo  {journal} {C. R. Physique}\ }\textbf
  {\bibinfo {volume} {17}},\ \bibinfo {pages} {113}}\BibitemShut {NoStop}%
\bibitem [{\citenamefont {Gallais}\ \emph {et~al.}(2016)\citenamefont
  {Gallais}, \citenamefont {Paul}, \citenamefont {Chauvi{\`e}re},\ and\
  \citenamefont {Schmalian}}]{gallais2016nematic}%
  \BibitemOpen
  \bibfield  {author} {\bibinfo {author} {\bibnamefont {Gallais}, \bibfnamefont
  {Y.}}, \bibinfo {author} {\bibfnamefont {I.}~\bibnamefont {Paul}}, \bibinfo
  {author} {\bibfnamefont {L.}~\bibnamefont {Chauvi{\`e}re}}, and\ \bibinfo
  {author} {\bibfnamefont {J.}~\bibnamefont {Schmalian}}} (\bibinfo {year}
  {2016}),\ \href {https://doi.org/10.1103/PhysRevLett.116.017001} {\bibfield
  {journal} {\bibinfo  {journal} {Phys. Rev. Lett.}\ }\textbf {\bibinfo
  {volume} {116}},\ \bibinfo {pages} {017001}}\BibitemShut {NoStop}%
\bibitem [{\citenamefont {Gao}\ \emph {et~al.}(2018)\citenamefont {Gao},
  \citenamefont {Wang}, \citenamefont {Zhou}, \citenamefont {Huang},\ and\
  \citenamefont {Wang}}]{gao2018possible}%
  \BibitemOpen
  \bibfield  {author} {\bibinfo {author} {\bibnamefont {Gao}, \bibfnamefont
  {Y.}}, \bibinfo {author} {\bibfnamefont {Y.}~\bibnamefont {Wang}}, \bibinfo
  {author} {\bibfnamefont {T.}~\bibnamefont {Zhou}}, \bibinfo {author}
  {\bibfnamefont {H.}~\bibnamefont {Huang}}, and\ \bibinfo {author}
  {\bibfnamefont {Q.-H.}\ \bibnamefont {Wang}}} (\bibinfo {year} {2018}),\
  \href {https://doi.org/10.1103/PhysRevLett.121.267005} {\bibfield  {journal}
  {\bibinfo  {journal} {Phys. Rev. Lett.}\ }\textbf {\bibinfo {volume} {121}},\
  \bibinfo {pages} {267005}}\BibitemShut {NoStop}%
\bibitem [{\citenamefont {Gao}\ \emph {et~al.}(2016)\citenamefont {Gao},
  \citenamefont {Yu}, \citenamefont {Zhou}, \citenamefont {Huang},\ and\
  \citenamefont {Wang}}]{Gao2016}%
  \BibitemOpen
  \bibfield  {author} {\bibinfo {author} {\bibnamefont {Gao}, \bibfnamefont
  {Y.}}, \bibinfo {author} {\bibfnamefont {Y.}~\bibnamefont {Yu}}, \bibinfo
  {author} {\bibfnamefont {T.}~\bibnamefont {Zhou}}, \bibinfo {author}
  {\bibfnamefont {H.}~\bibnamefont {Huang}}, and\ \bibinfo {author}
  {\bibfnamefont {Q.-H.}\ \bibnamefont {Wang}}} (\bibinfo {year} {2016}),\
  \href {https://doi.org/10.1103/PhysRevB.94.144512} {\bibfield  {journal}
  {\bibinfo  {journal} {Phys. Rev. B}\ }\textbf {\bibinfo {volume} {94}},\
  \bibinfo {pages} {144512}}\BibitemShut {NoStop}%
\bibitem [{\citenamefont {Gao}\ \emph {et~al.}(2014)\citenamefont {Gao},
  \citenamefont {Togano}, \citenamefont {Matsumoto},\ and\ \citenamefont
  {Kumakura}}]{Gao2015}%
  \BibitemOpen
  \bibfield  {author} {\bibinfo {author} {\bibnamefont {Gao}, \bibfnamefont
  {Z.}}, \bibinfo {author} {\bibfnamefont {K.}~\bibnamefont {Togano}}, \bibinfo
  {author} {\bibfnamefont {A.}~\bibnamefont {Matsumoto}}, and\ \bibinfo
  {author} {\bibfnamefont {H.}~\bibnamefont {Kumakura}}} (\bibinfo {year}
  {2014}),\ \href {https://doi.org/10.1088/0953-2048/28/1/012001} {\bibfield
  {journal} {\bibinfo  {journal} {Supercond. Sci. Technol.}\ }\textbf {\bibinfo
  {volume} {28}},\ \bibinfo {pages} {012001}}\BibitemShut {NoStop}%
\bibitem [{\citenamefont {Garcia}\ \emph {et~al.}(2019)\citenamefont {Garcia},
  \citenamefont {Ivashko}, \citenamefont {McNally}, \citenamefont {Das},
  \citenamefont {Piva}, \citenamefont {Adriano}, \citenamefont {Pagliuso},
  \citenamefont {Chang}, \citenamefont {Schmitt},\ and\ \citenamefont
  {Monney}}]{garcia2019anisotropic}%
  \BibitemOpen
  \bibfield  {author} {\bibinfo {author} {\bibnamefont {Garcia}, \bibfnamefont
  {F.~A.}}, \bibinfo {author} {\bibfnamefont {O.}~\bibnamefont {Ivashko}},
  \bibinfo {author} {\bibfnamefont {D.~E.}\ \bibnamefont {McNally}}, \bibinfo
  {author} {\bibfnamefont {L.}~\bibnamefont {Das}}, \bibinfo {author}
  {\bibfnamefont {M.~M.}\ \bibnamefont {Piva}}, \bibinfo {author}
  {\bibfnamefont {C.}~\bibnamefont {Adriano}}, \bibinfo {author} {\bibfnamefont
  {P.~G.}\ \bibnamefont {Pagliuso}}, \bibinfo {author} {\bibfnamefont
  {J.}~\bibnamefont {Chang}}, \bibinfo {author} {\bibfnamefont
  {T.}~\bibnamefont {Schmitt}}, and\ \bibinfo {author} {\bibfnamefont
  {C.}~\bibnamefont {Monney}}} (\bibinfo {year} {2019}),\ \href
  {https://doi.org/10.1103/PhysRevB.99.115118} {\bibfield  {journal} {\bibinfo
  {journal} {Phys. Rev. B}\ }\textbf {\bibinfo {volume} {99}},\ \bibinfo
  {pages} {115118}}\BibitemShut {NoStop}%
\bibitem [{\citenamefont {Gati}\ \emph {et~al.}(2019)\citenamefont {Gati},
  \citenamefont {B{\"o}hmer}, \citenamefont {Bud'ko},\ and\ \citenamefont
  {Canfield}}]{gati2019bulk}%
  \BibitemOpen
  \bibfield  {author} {\bibinfo {author} {\bibnamefont {Gati}, \bibfnamefont
  {E.}}, \bibinfo {author} {\bibfnamefont {A.~E.}\ \bibnamefont {B{\"o}hmer}},
  \bibinfo {author} {\bibfnamefont {S.~L.}\ \bibnamefont {Bud'ko}}, and\
  \bibinfo {author} {\bibfnamefont {P.~C.}\ \bibnamefont {Canfield}}} (\bibinfo
  {year} {2019}),\ \href {https://doi.org/10.1103/PhysRevLett.123.167002}
  {\bibfield  {journal} {\bibinfo  {journal} {Phys. Rev. Lett.}\ }\textbf
  {\bibinfo {volume} {123}},\ \bibinfo {pages} {167002}}\BibitemShut {NoStop}%
\bibitem [{\citenamefont {Ge}\ \emph {et~al.}(2015)\citenamefont {Ge},
  \citenamefont {Liu}, \citenamefont {Liu}, \citenamefont {Gao}, \citenamefont
  {Qian}, \citenamefont {Xue}, \citenamefont {Liu},\ and\ \citenamefont
  {Jia}}]{ge2015superconductivity}%
  \BibitemOpen
  \bibfield  {author} {\bibinfo {author} {\bibnamefont {Ge}, \bibfnamefont
  {J.-F.}}, \bibinfo {author} {\bibfnamefont {Z.-L.}\ \bibnamefont {Liu}},
  \bibinfo {author} {\bibfnamefont {C.}~\bibnamefont {Liu}}, \bibinfo {author}
  {\bibfnamefont {C.-L.}\ \bibnamefont {Gao}}, \bibinfo {author} {\bibfnamefont
  {D.}~\bibnamefont {Qian}}, \bibinfo {author} {\bibfnamefont {Q.-K.}\
  \bibnamefont {Xue}}, \bibinfo {author} {\bibfnamefont {Y.}~\bibnamefont
  {Liu}}, and\ \bibinfo {author} {\bibfnamefont {J.-F.}\ \bibnamefont {Jia}}}
  (\bibinfo {year} {2015}),\ \href {https://doi.org/10.1038/nmat4153}
  {\bibfield  {journal} {\bibinfo  {journal} {Nat. Mater.}\ }\textbf {\bibinfo
  {volume} {14}},\ \bibinfo {pages} {285}}\BibitemShut {NoStop}%
\bibitem [{\citenamefont {Ghini}\ \emph {et~al.}(2021)\citenamefont {Ghini},
  \citenamefont {Bristow}, \citenamefont {Prentice}, \citenamefont
  {Sutherland}, \citenamefont {Sanna}, \citenamefont {Haghighirad},\ and\
  \citenamefont {Coldea}}]{ghini2021strain}%
  \BibitemOpen
  \bibfield  {author} {\bibinfo {author} {\bibnamefont {Ghini}, \bibfnamefont
  {M.}}, \bibinfo {author} {\bibfnamefont {M.}~\bibnamefont {Bristow}},
  \bibinfo {author} {\bibfnamefont {J.~C.~A.}\ \bibnamefont {Prentice}},
  \bibinfo {author} {\bibfnamefont {S.}~\bibnamefont {Sutherland}}, \bibinfo
  {author} {\bibfnamefont {S.}~\bibnamefont {Sanna}}, \bibinfo {author}
  {\bibfnamefont {A.~A.}\ \bibnamefont {Haghighirad}}, and\ \bibinfo {author}
  {\bibfnamefont {A.~I.}\ \bibnamefont {Coldea}}} (\bibinfo {year} {2021}),\
  \href {https://doi.org/10.1103/PhysRevB.103.205139} {\bibfield  {journal}
  {\bibinfo  {journal} {Phys. Rev. B}\ }\textbf {\bibinfo {volume} {103}},\
  \bibinfo {pages} {205139}}\BibitemShut {NoStop}%
\bibitem [{\citenamefont {Ghiringhelli}\ \emph {et~al.}(2006)\citenamefont
  {Ghiringhelli}, \citenamefont {Piazzalunga}, \citenamefont {Dallera},
  \citenamefont {Trezzi}, \citenamefont {Braicovich}, \citenamefont {Schmitt},
  \citenamefont {Strocov}, \citenamefont {Betemps}, \citenamefont {Patthey},
  \citenamefont {Wang},\ and\ \citenamefont {Grioni}}]{saxes2006}%
  \BibitemOpen
  \bibfield  {author} {\bibinfo {author} {\bibnamefont {Ghiringhelli},
  \bibfnamefont {G.}}, \bibinfo {author} {\bibfnamefont {A.}~\bibnamefont
  {Piazzalunga}}, \bibinfo {author} {\bibfnamefont {C.}~\bibnamefont
  {Dallera}}, \bibinfo {author} {\bibfnamefont {G.}~\bibnamefont {Trezzi}},
  \bibinfo {author} {\bibfnamefont {L.}~\bibnamefont {Braicovich}}, \bibinfo
  {author} {\bibfnamefont {T.}~\bibnamefont {Schmitt}}, \bibinfo {author}
  {\bibfnamefont {V.~N.}\ \bibnamefont {Strocov}}, \bibinfo {author}
  {\bibfnamefont {R.}~\bibnamefont {Betemps}}, \bibinfo {author} {\bibfnamefont
  {L.}~\bibnamefont {Patthey}}, \bibinfo {author} {\bibfnamefont
  {X.}~\bibnamefont {Wang}}, and\ \bibinfo {author} {\bibfnamefont
  {M.}~\bibnamefont {Grioni}}} (\bibinfo {year} {2006}),\ \href
  {https://doi.org/10.1063/1.2372731} {\bibfield  {journal} {\bibinfo
  {journal} {Rev. Sci. Instrum.}\ }\textbf {\bibinfo {volume} {77}},\ \bibinfo
  {pages} {113108}}\BibitemShut {NoStop}%
\bibitem [{\citenamefont {Ghosh}\ \emph {et~al.}(2020)\citenamefont {Ghosh},
  \citenamefont {Brückner}, \citenamefont {Nikitin}, \citenamefont {Grinenko},
  \citenamefont {Elender}, \citenamefont {Mackenzie}, \citenamefont {Luetkens},
  \citenamefont {Klauss},\ and\ \citenamefont
  {Hicks}}]{ghosh2020piezoelectric}%
  \BibitemOpen
  \bibfield  {author} {\bibinfo {author} {\bibnamefont {Ghosh}, \bibfnamefont
  {S.}}, \bibinfo {author} {\bibfnamefont {F.}~\bibnamefont {Brückner}},
  \bibinfo {author} {\bibfnamefont {A.}~\bibnamefont {Nikitin}}, \bibinfo
  {author} {\bibfnamefont {V.}~\bibnamefont {Grinenko}}, \bibinfo {author}
  {\bibfnamefont {M.}~\bibnamefont {Elender}}, \bibinfo {author} {\bibfnamefont
  {A.~P.}\ \bibnamefont {Mackenzie}}, \bibinfo {author} {\bibfnamefont
  {H.}~\bibnamefont {Luetkens}}, \bibinfo {author} {\bibfnamefont {H.-H.}\
  \bibnamefont {Klauss}}, and\ \bibinfo {author} {\bibfnamefont {C.~W.}\
  \bibnamefont {Hicks}}} (\bibinfo {year} {2020}),\ \href
  {https://doi.org/10.1063/5.0025307} {\bibfield  {journal} {\bibinfo
  {journal} {Rev. Sci. Instrum.}\ }\textbf {\bibinfo {volume} {91}},\ \bibinfo
  {pages} {103902}}\BibitemShut {NoStop}%
\bibitem [{\citenamefont {Ghosh}\ \emph {et~al.}(2025)\citenamefont {Ghosh},
  \citenamefont {Ikeda}, \citenamefont {Chakraborty}, \citenamefont
  {Worasaran}, \citenamefont {Theuss}, \citenamefont {Peralta}, \citenamefont
  {Lozano}, \citenamefont {Kim}, \citenamefont {Thompson}, \citenamefont
  {Ryan}, \citenamefont {Ye}, \citenamefont {Kapitulnik}, \citenamefont
  {Kivelson}, \citenamefont {Ramshaw}, \citenamefont {Fernandes},\ and\
  \citenamefont {Fisher}}]{ghosh2025elastocaloric}%
  \BibitemOpen
  \bibfield  {author} {\bibinfo {author} {\bibnamefont {Ghosh}, \bibfnamefont
  {S.}}, \bibinfo {author} {\bibfnamefont {M.~S.}\ \bibnamefont {Ikeda}},
  \bibinfo {author} {\bibfnamefont {A.~R.}\ \bibnamefont {Chakraborty}},
  \bibinfo {author} {\bibfnamefont {T.}~\bibnamefont {Worasaran}}, \bibinfo
  {author} {\bibfnamefont {F.}~\bibnamefont {Theuss}}, \bibinfo {author}
  {\bibfnamefont {L.~B.}\ \bibnamefont {Peralta}}, \bibinfo {author}
  {\bibfnamefont {P.~M.}\ \bibnamefont {Lozano}}, \bibinfo {author}
  {\bibfnamefont {J.-W.}\ \bibnamefont {Kim}}, \bibinfo {author} {\bibfnamefont
  {P.~J.}\ \bibnamefont {Thompson}}, \bibinfo {author} {\bibfnamefont {P.~J.}\
  \bibnamefont {Ryan}}, \bibinfo {author} {\bibfnamefont {L.}~\bibnamefont
  {Ye}}, \bibinfo {author} {\bibfnamefont {A.}~\bibnamefont {Kapitulnik}},
  \bibinfo {author} {\bibfnamefont {S.~A.}\ \bibnamefont {Kivelson}}, \bibinfo
  {author} {\bibfnamefont {B.~J.}\ \bibnamefont {Ramshaw}}, \bibinfo {author}
  {\bibfnamefont {R.~M.}\ \bibnamefont {Fernandes}}, and\ \bibinfo {author}
  {\bibfnamefont {I.~R.}\ \bibnamefont {Fisher}}} (\bibinfo {year} {2025}),\
  \href {https://doi.org/10.1073/pnas.2424833122} {\bibfield  {journal}
  {\bibinfo  {journal} {Proc. Natl. Acad. Sci. U.S.A.}\ }\textbf {\bibinfo
  {volume} {122}},\ \bibinfo {pages} {e2424833122}}\BibitemShut {NoStop}%
\bibitem [{\citenamefont {Gilmore}\ \emph {et~al.}(2021)\citenamefont
  {Gilmore}, \citenamefont {Pelliciari}, \citenamefont {Huang}, \citenamefont
  {Kas}, \citenamefont {Dantz}, \citenamefont {Strocov}, \citenamefont
  {Kasahara}, \citenamefont {Matsuda}, \citenamefont {Das}, \citenamefont
  {Shibauchi},\ and\ \citenamefont {Schmitt}}]{gilmore2021description}%
  \BibitemOpen
  \bibfield  {author} {\bibinfo {author} {\bibnamefont {Gilmore}, \bibfnamefont
  {K.}}, \bibinfo {author} {\bibfnamefont {J.}~\bibnamefont {Pelliciari}},
  \bibinfo {author} {\bibfnamefont {Y.}~\bibnamefont {Huang}}, \bibinfo
  {author} {\bibfnamefont {J.~J.}\ \bibnamefont {Kas}}, \bibinfo {author}
  {\bibfnamefont {M.}~\bibnamefont {Dantz}}, \bibinfo {author} {\bibfnamefont
  {V.~N.}\ \bibnamefont {Strocov}}, \bibinfo {author} {\bibfnamefont
  {S.}~\bibnamefont {Kasahara}}, \bibinfo {author} {\bibfnamefont
  {Y.}~\bibnamefont {Matsuda}}, \bibinfo {author} {\bibfnamefont
  {T.}~\bibnamefont {Das}}, \bibinfo {author} {\bibfnamefont {T.}~\bibnamefont
  {Shibauchi}}, and\ \bibinfo {author} {\bibfnamefont {T.}~\bibnamefont
  {Schmitt}}} (\bibinfo {year} {2021}),\ \href
  {https://doi.org/10.1103/PhysRevX.11.031013} {\bibfield  {journal} {\bibinfo
  {journal} {Phys. Rev. X}\ }\textbf {\bibinfo {volume} {11}},\ \bibinfo
  {pages} {031013}}\BibitemShut {NoStop}%
\bibitem [{\citenamefont {Glasbrenner}\ \emph {et~al.}(2015)\citenamefont
  {Glasbrenner}, \citenamefont {Mazin}, \citenamefont {Jeschke}, \citenamefont
  {Hirschfeld}, \citenamefont {Fernandes},\ and\ \citenamefont
  {Valent{\'i}}}]{glasbrenner2015effect}%
  \BibitemOpen
  \bibfield  {author} {\bibinfo {author} {\bibnamefont {Glasbrenner},
  \bibfnamefont {J.~K.}}, \bibinfo {author} {\bibfnamefont {I.~I.}\
  \bibnamefont {Mazin}}, \bibinfo {author} {\bibfnamefont {H.~O.}\ \bibnamefont
  {Jeschke}}, \bibinfo {author} {\bibfnamefont {P.~J.}\ \bibnamefont
  {Hirschfeld}}, \bibinfo {author} {\bibfnamefont {R.~M.}\ \bibnamefont
  {Fernandes}}, and\ \bibinfo {author} {\bibfnamefont {R.}~\bibnamefont
  {Valent{\'i}}}} (\bibinfo {year} {2015}),\ \href
  {https://doi.org/10.1038/nphys3434} {\bibfield  {journal} {\bibinfo
  {journal} {Nat. Phys.}\ }\textbf {\bibinfo {volume} {11}},\ \bibinfo {pages}
  {953}}\BibitemShut {NoStop}%
\bibitem [{\citenamefont {Graser}\ \emph {et~al.}(2009)\citenamefont {Graser},
  \citenamefont {Maier}, \citenamefont {Hirschfeld},\ and\ \citenamefont
  {Scalapino}}]{Graser2009}%
  \BibitemOpen
  \bibfield  {author} {\bibinfo {author} {\bibnamefont {Graser}, \bibfnamefont
  {S.}}, \bibinfo {author} {\bibfnamefont {T.}~\bibnamefont {Maier}}, \bibinfo
  {author} {\bibfnamefont {P.}~\bibnamefont {Hirschfeld}}, and\ \bibinfo
  {author} {\bibfnamefont {D.}~\bibnamefont {Scalapino}}} (\bibinfo {year}
  {2009}),\ \href {https://doi.org/10.1088/1367-2630/11/2/025016} {\bibfield
  {journal} {\bibinfo  {journal} {New J. Phys.}\ }\textbf {\bibinfo {volume}
  {11}},\ \bibinfo {pages} {025016}}\BibitemShut {NoStop}%
\bibitem [{\citenamefont {Grinenko}\ \emph {et~al.}(2021)\citenamefont
  {Grinenko}, \citenamefont {Weston}, \citenamefont {Caglieris}, \citenamefont
  {Wuttke}, \citenamefont {Hess}, \citenamefont {Gottschall}, \citenamefont
  {Maccari}, \citenamefont {Gorbunov}, \citenamefont {Zherlitsyn},
  \citenamefont {Wosnitza}, \citenamefont {Rydh}, \citenamefont {Kihou},
  \citenamefont {Lee}, \citenamefont {Sarkar}, \citenamefont {Dengre},
  \citenamefont {Garaud}, \citenamefont {Charnukha}, \citenamefont {H{\"u}hne},
  \citenamefont {Nielsch}, \citenamefont {B{\"u}chner}, \citenamefont
  {Klauss},\ and\ \citenamefont {Babaev}}]{grinenko2021state}%
  \BibitemOpen
  \bibfield  {author} {\bibinfo {author} {\bibnamefont {Grinenko},
  \bibfnamefont {V.}}, \bibinfo {author} {\bibfnamefont {D.}~\bibnamefont
  {Weston}}, \bibinfo {author} {\bibfnamefont {F.}~\bibnamefont {Caglieris}},
  \bibinfo {author} {\bibfnamefont {C.}~\bibnamefont {Wuttke}}, \bibinfo
  {author} {\bibfnamefont {C.}~\bibnamefont {Hess}}, \bibinfo {author}
  {\bibfnamefont {T.}~\bibnamefont {Gottschall}}, \bibinfo {author}
  {\bibfnamefont {I.}~\bibnamefont {Maccari}}, \bibinfo {author} {\bibfnamefont
  {D.}~\bibnamefont {Gorbunov}}, \bibinfo {author} {\bibfnamefont
  {S.}~\bibnamefont {Zherlitsyn}}, \bibinfo {author} {\bibfnamefont
  {J.}~\bibnamefont {Wosnitza}}, \bibinfo {author} {\bibfnamefont
  {A.}~\bibnamefont {Rydh}}, \bibinfo {author} {\bibfnamefont {K.}~\bibnamefont
  {Kihou}}, \bibinfo {author} {\bibfnamefont {C.-H.}\ \bibnamefont {Lee}},
  \bibinfo {author} {\bibfnamefont {R.}~\bibnamefont {Sarkar}}, \bibinfo
  {author} {\bibfnamefont {S.}~\bibnamefont {Dengre}}, \bibinfo {author}
  {\bibfnamefont {J.}~\bibnamefont {Garaud}}, \bibinfo {author} {\bibfnamefont
  {A.}~\bibnamefont {Charnukha}}, \bibinfo {author} {\bibfnamefont
  {R.}~\bibnamefont {H{\"u}hne}}, \bibinfo {author} {\bibfnamefont
  {K.}~\bibnamefont {Nielsch}}, \bibinfo {author} {\bibfnamefont
  {B.}~\bibnamefont {B{\"u}chner}}, \bibinfo {author} {\bibfnamefont {H.-H.}\
  \bibnamefont {Klauss}}, and\ \bibinfo {author} {\bibfnamefont
  {E.}~\bibnamefont {Babaev}}} (\bibinfo {year} {2021}),\ \href
  {https://doi.org/10.1038/s41567-021-01350-9} {\bibfield  {journal} {\bibinfo
  {journal} {Nat. Phys.}\ }\textbf {\bibinfo {volume} {17}},\ \bibinfo {pages}
  {1254}}\BibitemShut {NoStop}%
\bibitem [{\citenamefont {de~Groot}\ \emph {et~al.}(2024)\citenamefont
  {de~Groot}, \citenamefont {Haverkort}, \citenamefont {Elnaggar},
  \citenamefont {Juhin}, \citenamefont {Zhou},\ and\ \citenamefont
  {Glatzel}}]{degroot2024resonant}%
  \BibitemOpen
  \bibfield  {author} {\bibinfo {author} {\bibnamefont {de~Groot},
  \bibfnamefont {F.~M.~F.}}, \bibinfo {author} {\bibfnamefont {M.~W.}\
  \bibnamefont {Haverkort}}, \bibinfo {author} {\bibfnamefont {H.}~\bibnamefont
  {Elnaggar}}, \bibinfo {author} {\bibfnamefont {A.}~\bibnamefont {Juhin}},
  \bibinfo {author} {\bibfnamefont {K.-J.}\ \bibnamefont {Zhou}}, and\ \bibinfo
  {author} {\bibfnamefont {P.}~\bibnamefont {Glatzel}}} (\bibinfo {year}
  {2024}),\ \href {https://doi.org/10.1038/s43586-024-00322-6} {\bibfield
  {journal} {\bibinfo  {journal} {Nature Reviews Methods Primers}\ }\textbf
  {\bibinfo {volume} {4}},\ \bibinfo {pages} {45}}\BibitemShut {NoStop}%
\bibitem [{\citenamefont {Gu}\ \emph {et~al.}(2017)\citenamefont {Gu},
  \citenamefont {Liu}, \citenamefont {Xie}, \citenamefont {Zhang},
  \citenamefont {Gong}, \citenamefont {Hu}, \citenamefont {Ma}, \citenamefont
  {Li}, \citenamefont {Zhao}, \citenamefont {Lin}, \citenamefont {Xu},
  \citenamefont {Tan}, \citenamefont {Chen}, \citenamefont {Meng},
  \citenamefont {Yang}, \citenamefont {Luo},\ and\ \citenamefont
  {Li}}]{gu2017unified}%
  \BibitemOpen
  \bibfield  {author} {\bibinfo {author} {\bibnamefont {Gu}, \bibfnamefont
  {Y.}}, \bibinfo {author} {\bibfnamefont {Z.}~\bibnamefont {Liu}}, \bibinfo
  {author} {\bibfnamefont {T.}~\bibnamefont {Xie}}, \bibinfo {author}
  {\bibfnamefont {W.}~\bibnamefont {Zhang}}, \bibinfo {author} {\bibfnamefont
  {D.}~\bibnamefont {Gong}}, \bibinfo {author} {\bibfnamefont {D.}~\bibnamefont
  {Hu}}, \bibinfo {author} {\bibfnamefont {X.}~\bibnamefont {Ma}}, \bibinfo
  {author} {\bibfnamefont {C.}~\bibnamefont {Li}}, \bibinfo {author}
  {\bibfnamefont {L.}~\bibnamefont {Zhao}}, \bibinfo {author} {\bibfnamefont
  {L.}~\bibnamefont {Lin}}, \bibinfo {author} {\bibfnamefont {Z.}~\bibnamefont
  {Xu}}, \bibinfo {author} {\bibfnamefont {G.}~\bibnamefont {Tan}}, \bibinfo
  {author} {\bibfnamefont {G.}~\bibnamefont {Chen}}, \bibinfo {author}
  {\bibfnamefont {Z.~Y.}\ \bibnamefont {Meng}}, \bibinfo {author}
  {\bibfnamefont {Y.-f.}\ \bibnamefont {Yang}}, \bibinfo {author}
  {\bibfnamefont {H.}~\bibnamefont {Luo}}, and\ \bibinfo {author}
  {\bibfnamefont {S.}~\bibnamefont {Li}}} (\bibinfo {year} {2017}),\ \href
  {https://doi.org/10.1103/PhysRevLett.119.157001} {\bibfield  {journal}
  {\bibinfo  {journal} {Phys. Rev. Lett.}\ }\textbf {\bibinfo {volume} {119}},\
  \bibinfo {pages} {157001}}\BibitemShut {NoStop}%
\bibitem [{\citenamefont {Gu}\ \emph {et~al.}(2022)\citenamefont {Gu},
  \citenamefont {Wang}, \citenamefont {Wo}, \citenamefont {He}, \citenamefont
  {Walker}, \citenamefont {Park}, \citenamefont {Enderle}, \citenamefont
  {Christianson}, \citenamefont {Wang},\ and\ \citenamefont
  {Zhao}}]{gu2022frustrated}%
  \BibitemOpen
  \bibfield  {author} {\bibinfo {author} {\bibnamefont {Gu}, \bibfnamefont
  {Y.}}, \bibinfo {author} {\bibfnamefont {Q.}~\bibnamefont {Wang}}, \bibinfo
  {author} {\bibfnamefont {H.}~\bibnamefont {Wo}}, \bibinfo {author}
  {\bibfnamefont {Z.}~\bibnamefont {He}}, \bibinfo {author} {\bibfnamefont
  {H.~C.}\ \bibnamefont {Walker}}, \bibinfo {author} {\bibfnamefont {J.~T.}\
  \bibnamefont {Park}}, \bibinfo {author} {\bibfnamefont {M.}~\bibnamefont
  {Enderle}}, \bibinfo {author} {\bibfnamefont {A.~D.}\ \bibnamefont
  {Christianson}}, \bibinfo {author} {\bibfnamefont {W.}~\bibnamefont {Wang}},
  and\ \bibinfo {author} {\bibfnamefont {J.}~\bibnamefont {Zhao}}} (\bibinfo
  {year} {2022}),\ \href {https://doi.org/10.1103/PhysRevB.106.L060504}
  {\bibfield  {journal} {\bibinfo  {journal} {Phys. Rev. B}\ }\textbf {\bibinfo
  {volume} {106}},\ \bibinfo {pages} {L060504}}\BibitemShut {NoStop}%
\bibitem [{\citenamefont {Guguchia}\ \emph {et~al.}(2011)\citenamefont
  {Guguchia}, \citenamefont {Roos}, \citenamefont {Shengelaya}, \citenamefont
  {Katrych}, \citenamefont {Bukowski}, \citenamefont {Weyeneth}, \citenamefont
  {Mur{\'a}nyi}, \citenamefont {Str{\"a}ssle}, \citenamefont {Maisuradze},
  \citenamefont {Karpinski} \emph {et~al.}}]{Guguchia2011}%
  \BibitemOpen
  \bibfield  {author} {\bibinfo {author} {\bibnamefont {Guguchia},
  \bibfnamefont {Z.}}, \bibinfo {author} {\bibfnamefont {J.}~\bibnamefont
  {Roos}}, \bibinfo {author} {\bibfnamefont {A.}~\bibnamefont {Shengelaya}},
  \bibinfo {author} {\bibfnamefont {S.}~\bibnamefont {Katrych}}, \bibinfo
  {author} {\bibfnamefont {Z.}~\bibnamefont {Bukowski}}, \bibinfo {author}
  {\bibfnamefont {S.}~\bibnamefont {Weyeneth}}, \bibinfo {author}
  {\bibfnamefont {F.}~\bibnamefont {Mur{\'a}nyi}}, \bibinfo {author}
  {\bibfnamefont {S.}~\bibnamefont {Str{\"a}ssle}}, \bibinfo {author}
  {\bibfnamefont {A.}~\bibnamefont {Maisuradze}}, \bibinfo {author}
  {\bibfnamefont {J.}~\bibnamefont {Karpinski}},  \emph {et~al.}} (\bibinfo
  {year} {2011}),\ \href {https://doi.org/10.1103/PhysRevB.83.144516}
  {\bibfield  {journal} {\bibinfo  {journal} {Phys. Rev. B}\ }\textbf {\bibinfo
  {volume} {83}},\ \bibinfo {pages} {144516}}\BibitemShut {NoStop}%
\bibitem [{\citenamefont {Guo}\ \emph {et~al.}(2010)\citenamefont {Guo},
  \citenamefont {Jin}, \citenamefont {Wang}, \citenamefont {Wang},
  \citenamefont {Zhu}, \citenamefont {Zhou}, \citenamefont {He},\ and\
  \citenamefont {Chen}}]{Guo2010}%
  \BibitemOpen
  \bibfield  {author} {\bibinfo {author} {\bibnamefont {Guo}, \bibfnamefont
  {J.}}, \bibinfo {author} {\bibfnamefont {S.}~\bibnamefont {Jin}}, \bibinfo
  {author} {\bibfnamefont {G.}~\bibnamefont {Wang}}, \bibinfo {author}
  {\bibfnamefont {S.}~\bibnamefont {Wang}}, \bibinfo {author} {\bibfnamefont
  {K.}~\bibnamefont {Zhu}}, \bibinfo {author} {\bibfnamefont {T.}~\bibnamefont
  {Zhou}}, \bibinfo {author} {\bibfnamefont {M.}~\bibnamefont {He}}, and\
  \bibinfo {author} {\bibfnamefont {X.}~\bibnamefont {Chen}}} (\bibinfo {year}
  {2010}),\ \href {https://doi.org/10.1103/PhysRevB.82.180520} {\bibfield
  {journal} {\bibinfo  {journal} {Phys. Rev. B}\ }\textbf {\bibinfo {volume}
  {82}},\ \bibinfo {pages} {180520}}\BibitemShut {NoStop}%
\bibitem [{\citenamefont {Guo}\ \emph {et~al.}(2014)\citenamefont {Guo},
  \citenamefont {Lei}, \citenamefont {Hayashi},\ and\ \citenamefont
  {Hosono}}]{guo2014superconductivity}%
  \BibitemOpen
  \bibfield  {author} {\bibinfo {author} {\bibnamefont {Guo}, \bibfnamefont
  {J.}}, \bibinfo {author} {\bibfnamefont {H.}~\bibnamefont {Lei}}, \bibinfo
  {author} {\bibfnamefont {F.}~\bibnamefont {Hayashi}}, and\ \bibinfo {author}
  {\bibfnamefont {H.}~\bibnamefont {Hosono}}} (\bibinfo {year} {2014}),\ \href
  {https://doi.org/10.1038/ncomms5756} {\bibfield  {journal} {\bibinfo
  {journal} {Nat. Commun.}\ }\textbf {\bibinfo {volume} {5}},\ \bibinfo {pages}
  {4756}}\BibitemShut {NoStop}%
\bibitem [{\citenamefont {Guo}\ \emph {et~al.}(2019)\citenamefont {Guo},
  \citenamefont {Yue}, \citenamefont {Iida}, \citenamefont {Kamazawa},
  \citenamefont {Chen}, \citenamefont {Han}, \citenamefont {Zhang},\ and\
  \citenamefont {Li}}]{guo2019preferred}%
  \BibitemOpen
  \bibfield  {author} {\bibinfo {author} {\bibnamefont {Guo}, \bibfnamefont
  {J.}}, \bibinfo {author} {\bibfnamefont {L.}~\bibnamefont {Yue}}, \bibinfo
  {author} {\bibfnamefont {K.}~\bibnamefont {Iida}}, \bibinfo {author}
  {\bibfnamefont {K.}~\bibnamefont {Kamazawa}}, \bibinfo {author}
  {\bibfnamefont {L.}~\bibnamefont {Chen}}, \bibinfo {author} {\bibfnamefont
  {T.}~\bibnamefont {Han}}, \bibinfo {author} {\bibfnamefont {Y.}~\bibnamefont
  {Zhang}}, and\ \bibinfo {author} {\bibfnamefont {Y.}~\bibnamefont {Li}}}
  (\bibinfo {year} {2019}),\ \href
  {https://doi.org/10.1103/PhysRevLett.122.017001} {\bibfield  {journal}
  {\bibinfo  {journal} {Phys. Rev. Lett.}\ }\textbf {\bibinfo {volume} {122}},\
  \bibinfo {pages} {017001}}\BibitemShut {NoStop}%
\bibitem [{\citenamefont {Gurevich}(2010)}]{gurevich2010upper}%
  \BibitemOpen
  \bibfield  {author} {\bibinfo {author} {\bibnamefont {Gurevich},
  \bibfnamefont {A.}}} (\bibinfo {year} {2010}),\ \href
  {https://doi.org/10.1103/PhysRevB.82.184504} {\bibfield  {journal} {\bibinfo
  {journal} {Phys. Rev. B}\ }\textbf {\bibinfo {volume} {82}},\ \bibinfo
  {pages} {184504}}\BibitemShut {NoStop}%
\bibitem [{\citenamefont {Guterding}\ \emph {et~al.}(2015)\citenamefont
  {Guterding}, \citenamefont {Jeschke}, \citenamefont {Hirschfeld},\ and\
  \citenamefont {Valent{\'\i}}}]{Guterding2015}%
  \BibitemOpen
  \bibfield  {author} {\bibinfo {author} {\bibnamefont {Guterding},
  \bibfnamefont {D.}}, \bibinfo {author} {\bibfnamefont {H.~O.}\ \bibnamefont
  {Jeschke}}, \bibinfo {author} {\bibfnamefont {P.}~\bibnamefont {Hirschfeld}},
  and\ \bibinfo {author} {\bibfnamefont {R.}~\bibnamefont {Valent{\'\i}}}}
  (\bibinfo {year} {2015}),\ \href {https://doi.org/10.1103/PhysRevB.91.041112}
  {\bibfield  {journal} {\bibinfo  {journal} {Phys. Rev. B}\ }\textbf {\bibinfo
  {volume} {91}},\ \bibinfo {pages} {041112}}\BibitemShut {NoStop}%
\bibitem [{\citenamefont {Han}\ \emph {et~al.}(2008)\citenamefont {Han},
  \citenamefont {Zhu}, \citenamefont {Mu}, \citenamefont {Cheng},\ and\
  \citenamefont {Wen}}]{Han2008}%
  \BibitemOpen
  \bibfield  {author} {\bibinfo {author} {\bibnamefont {Han}, \bibfnamefont
  {F.}}, \bibinfo {author} {\bibfnamefont {X.}~\bibnamefont {Zhu}}, \bibinfo
  {author} {\bibfnamefont {G.}~\bibnamefont {Mu}}, \bibinfo {author}
  {\bibfnamefont {P.}~\bibnamefont {Cheng}}, and\ \bibinfo {author}
  {\bibfnamefont {H.-H.}\ \bibnamefont {Wen}}} (\bibinfo {year} {2008}),\ \href
  {https://doi.org/10.1103/PhysRevB.78.180503} {\bibfield  {journal} {\bibinfo
  {journal} {Phys. Rev. B}\ }\textbf {\bibinfo {volume} {78}},\ \bibinfo
  {pages} {180503}}\BibitemShut {NoStop}%
\bibitem [{\citenamefont {Han}\ \emph {et~al.}(2025)\citenamefont {Han},
  \citenamefont {Dong}, \citenamefont {Yao}, \citenamefont {Zhang},
  \citenamefont {Zhang}, \citenamefont {Gong}, \citenamefont {Huang},
  \citenamefont {Gong}, \citenamefont {Wang}, \citenamefont {Zhang},
  \citenamefont {Liu}, \citenamefont {Sun}, \citenamefont {Zhu}, \citenamefont
  {Li}, \citenamefont {Luo}, \citenamefont {Awaji}, \citenamefont {Wang},
  \citenamefont {Xie}, \citenamefont {Hosono},\ and\ \citenamefont
  {Ma}}]{Han2025}%
  \BibitemOpen
  \bibfield  {author} {\bibinfo {author} {\bibnamefont {Han}, \bibfnamefont
  {M.}}, \bibinfo {author} {\bibfnamefont {C.}~\bibnamefont {Dong}}, \bibinfo
  {author} {\bibfnamefont {C.}~\bibnamefont {Yao}}, \bibinfo {author}
  {\bibfnamefont {Z.}~\bibnamefont {Zhang}}, \bibinfo {author} {\bibfnamefont
  {Q.}~\bibnamefont {Zhang}}, \bibinfo {author} {\bibfnamefont
  {Y.}~\bibnamefont {Gong}}, \bibinfo {author} {\bibfnamefont {H.}~\bibnamefont
  {Huang}}, \bibinfo {author} {\bibfnamefont {D.}~\bibnamefont {Gong}},
  \bibinfo {author} {\bibfnamefont {D.}~\bibnamefont {Wang}}, \bibinfo {author}
  {\bibfnamefont {X.}~\bibnamefont {Zhang}}, \bibinfo {author} {\bibfnamefont
  {F.}~\bibnamefont {Liu}}, \bibinfo {author} {\bibfnamefont {Y.}~\bibnamefont
  {Sun}}, \bibinfo {author} {\bibfnamefont {Z.}~\bibnamefont {Zhu}}, \bibinfo
  {author} {\bibfnamefont {J.}~\bibnamefont {Li}}, \bibinfo {author}
  {\bibfnamefont {J.}~\bibnamefont {Luo}}, \bibinfo {author} {\bibfnamefont
  {S.}~\bibnamefont {Awaji}}, \bibinfo {author} {\bibfnamefont
  {X.}~\bibnamefont {Wang}}, \bibinfo {author} {\bibfnamefont {J.}~\bibnamefont
  {Xie}}, \bibinfo {author} {\bibfnamefont {H.}~\bibnamefont {Hosono}}, and\
  \bibinfo {author} {\bibfnamefont {Y.}~\bibnamefont {Ma}}} (\bibinfo {year}
  {2025}),\ \href {https://doi.org/10.1002/adma.202513265} {\bibfield
  {journal} {\bibinfo  {journal} {Adv. Mater.}\ }\textbf {\bibinfo {volume}
  {37}},\ \bibinfo {pages} {e13265}}\BibitemShut {NoStop}%
\bibitem [{\citenamefont {Han}\ and\ \citenamefont
  {Savrasov}(2009)}]{Joon2009}%
  \BibitemOpen
  \bibfield  {author} {\bibinfo {author} {\bibnamefont {Han}, \bibfnamefont
  {M.~J.}}, and\ \bibinfo {author} {\bibfnamefont {S.~Y.}\ \bibnamefont
  {Savrasov}}} (\bibinfo {year} {2009}),\ \href
  {https://doi.org/10.1103/PhysRevLett.103.067001} {\bibfield  {journal}
  {\bibinfo  {journal} {Phys. Rev. Lett.}\ }\textbf {\bibinfo {volume} {103}},\
  \bibinfo {pages} {067001}}\BibitemShut {NoStop}%
\bibitem [{\citenamefont {Han}\ \emph {et~al.}(2010)\citenamefont {Han},
  \citenamefont {Li}, \citenamefont {Cao}, \citenamefont {Wang}, \citenamefont
  {Xu}, \citenamefont {Zhao}, \citenamefont {Guo},\ and\ \citenamefont
  {Yang}}]{Han2010}%
  \BibitemOpen
  \bibfield  {author} {\bibinfo {author} {\bibnamefont {Han}, \bibfnamefont
  {Y.}}, \bibinfo {author} {\bibfnamefont {W.~Y.}\ \bibnamefont {Li}}, \bibinfo
  {author} {\bibfnamefont {L.~X.}\ \bibnamefont {Cao}}, \bibinfo {author}
  {\bibfnamefont {X.~Y.}\ \bibnamefont {Wang}}, \bibinfo {author}
  {\bibfnamefont {B.}~\bibnamefont {Xu}}, \bibinfo {author} {\bibfnamefont
  {B.~R.}\ \bibnamefont {Zhao}}, \bibinfo {author} {\bibfnamefont {Y.~Q.}\
  \bibnamefont {Guo}}, and\ \bibinfo {author} {\bibfnamefont {J.~L.}\
  \bibnamefont {Yang}}} (\bibinfo {year} {2010}),\ \href
  {https://doi.org/10.1103/PhysRevLett.104.017003} {\bibfield  {journal}
  {\bibinfo  {journal} {Phys. Rev. Lett.}\ }\textbf {\bibinfo {volume} {104}},\
  \bibinfo {pages} {017003}}\BibitemShut {NoStop}%
\bibitem [{\citenamefont {Hanaguri}\ \emph {et~al.}(2018)\citenamefont
  {Hanaguri}, \citenamefont {Iwaya}, \citenamefont {Kohsaka}, \citenamefont
  {Machida}, \citenamefont {Watashige}, \citenamefont {Kasahara}, \citenamefont
  {Shibauchi},\ and\ \citenamefont {Matsuda}}]{hanaguri2018two}%
  \BibitemOpen
  \bibfield  {author} {\bibinfo {author} {\bibnamefont {Hanaguri},
  \bibfnamefont {T.}}, \bibinfo {author} {\bibfnamefont {K.}~\bibnamefont
  {Iwaya}}, \bibinfo {author} {\bibfnamefont {Y.}~\bibnamefont {Kohsaka}},
  \bibinfo {author} {\bibfnamefont {T.}~\bibnamefont {Machida}}, \bibinfo
  {author} {\bibfnamefont {T.}~\bibnamefont {Watashige}}, \bibinfo {author}
  {\bibfnamefont {S.}~\bibnamefont {Kasahara}}, \bibinfo {author}
  {\bibfnamefont {T.}~\bibnamefont {Shibauchi}}, and\ \bibinfo {author}
  {\bibfnamefont {Y.}~\bibnamefont {Matsuda}}} (\bibinfo {year} {2018}),\ \href
  {https://doi.org/10.1126/sciadv.aar6419} {\bibfield  {journal} {\bibinfo
  {journal} {Sci. Adv.}\ }\textbf {\bibinfo {volume} {4}},\ \bibinfo {pages}
  {eaar6419}}\BibitemShut {NoStop}%
\bibitem [{\citenamefont {Hanaguri}\ \emph {et~al.}(2019)\citenamefont
  {Hanaguri}, \citenamefont {Kasahara}, \citenamefont {B{\"o}ker},
  \citenamefont {Eremin}, \citenamefont {Shibauchi},\ and\ \citenamefont
  {Matsuda}}]{hanaguri2019quantum}%
  \BibitemOpen
  \bibfield  {author} {\bibinfo {author} {\bibnamefont {Hanaguri},
  \bibfnamefont {T.}}, \bibinfo {author} {\bibfnamefont {S.}~\bibnamefont
  {Kasahara}}, \bibinfo {author} {\bibfnamefont {J.}~\bibnamefont {B{\"o}ker}},
  \bibinfo {author} {\bibfnamefont {I.}~\bibnamefont {Eremin}}, \bibinfo
  {author} {\bibfnamefont {T.}~\bibnamefont {Shibauchi}}, and\ \bibinfo
  {author} {\bibfnamefont {Y.}~\bibnamefont {Matsuda}}} (\bibinfo {year}
  {2019}),\ \href {https://doi.org/10.1103/PhysRevLett.122.077001} {\bibfield
  {journal} {\bibinfo  {journal} {Phys. Rev. Lett.}\ }\textbf {\bibinfo
  {volume} {122}},\ \bibinfo {pages} {077001}}\BibitemShut {NoStop}%
\bibitem [{\citenamefont {Hancock}\ \emph {et~al.}(2010)\citenamefont
  {Hancock}, \citenamefont {Viennois}, \citenamefont {van~der Marel},
  \citenamefont {R\o{}nnow}, \citenamefont {Guarise}, \citenamefont {Lin},
  \citenamefont {Grioni}, \citenamefont {Moretti~Sala}, \citenamefont
  {Ghiringhelli}, \citenamefont {Strocov}, \citenamefont {Schlappa},\ and\
  \citenamefont {Schmitt}}]{hancock2010evidence}%
  \BibitemOpen
  \bibfield  {author} {\bibinfo {author} {\bibnamefont {Hancock}, \bibfnamefont
  {J.~N.}}, \bibinfo {author} {\bibfnamefont {R.}~\bibnamefont {Viennois}},
  \bibinfo {author} {\bibfnamefont {D.}~\bibnamefont {van~der Marel}}, \bibinfo
  {author} {\bibfnamefont {H.~M.}\ \bibnamefont {R\o{}nnow}}, \bibinfo {author}
  {\bibfnamefont {M.}~\bibnamefont {Guarise}}, \bibinfo {author} {\bibfnamefont
  {P.-H.}\ \bibnamefont {Lin}}, \bibinfo {author} {\bibfnamefont
  {M.}~\bibnamefont {Grioni}}, \bibinfo {author} {\bibfnamefont
  {M.}~\bibnamefont {Moretti~Sala}}, \bibinfo {author} {\bibfnamefont
  {G.}~\bibnamefont {Ghiringhelli}}, \bibinfo {author} {\bibfnamefont {V.~N.}\
  \bibnamefont {Strocov}}, \bibinfo {author} {\bibfnamefont {J.}~\bibnamefont
  {Schlappa}}, and\ \bibinfo {author} {\bibfnamefont {T.}~\bibnamefont
  {Schmitt}}} (\bibinfo {year} {2010}),\ \href
  {https://doi.org/10.1103/PhysRevB.82.020513} {\bibfield  {journal} {\bibinfo
  {journal} {Phys. Rev. B}\ }\textbf {\bibinfo {volume} {82}},\ \bibinfo
  {pages} {020513}}\BibitemShut {NoStop}%
\bibitem [{\citenamefont {Hanna}\ \emph {et~al.}(2011)\citenamefont {Hanna},
  \citenamefont {Muraba}, \citenamefont {Matsuishi}, \citenamefont {Igawa},
  \citenamefont {Kodama}, \citenamefont {Shamoto},\ and\ \citenamefont
  {Hosono}}]{Hanna2011}%
  \BibitemOpen
  \bibfield  {author} {\bibinfo {author} {\bibnamefont {Hanna}, \bibfnamefont
  {T.}}, \bibinfo {author} {\bibfnamefont {Y.}~\bibnamefont {Muraba}}, \bibinfo
  {author} {\bibfnamefont {S.}~\bibnamefont {Matsuishi}}, \bibinfo {author}
  {\bibfnamefont {N.}~\bibnamefont {Igawa}}, \bibinfo {author} {\bibfnamefont
  {K.}~\bibnamefont {Kodama}}, \bibinfo {author} {\bibfnamefont {S.-i.}\
  \bibnamefont {Shamoto}}, and\ \bibinfo {author} {\bibfnamefont
  {H.}~\bibnamefont {Hosono}}} (\bibinfo {year} {2011}),\ \href
  {https://doi.org/10.1103/PhysRevB.84.024521} {\bibfield  {journal} {\bibinfo
  {journal} {Phys. Rev. B}\ }\textbf {\bibinfo {volume} {84}},\ \bibinfo
  {pages} {024521}}\BibitemShut {NoStop}%
\bibitem [{\citenamefont {Hanzawa}\ \emph {et~al.}(2016)\citenamefont
  {Hanzawa}, \citenamefont {Sato}, \citenamefont {Hiramatsu}, \citenamefont
  {Kamiya},\ and\ \citenamefont {Hosono}}]{Hanzawa2016}%
  \BibitemOpen
  \bibfield  {author} {\bibinfo {author} {\bibnamefont {Hanzawa}, \bibfnamefont
  {K.}}, \bibinfo {author} {\bibfnamefont {H.}~\bibnamefont {Sato}}, \bibinfo
  {author} {\bibfnamefont {H.}~\bibnamefont {Hiramatsu}}, \bibinfo {author}
  {\bibfnamefont {T.}~\bibnamefont {Kamiya}}, and\ \bibinfo {author}
  {\bibfnamefont {H.}~\bibnamefont {Hosono}}} (\bibinfo {year} {2016}),\ \href
  {https://doi.org/10.1073/pnas.1520810113} {\bibfield  {journal} {\bibinfo
  {journal} {Proc. Natl. Acad. Sci. U.S.A.}\ }\textbf {\bibinfo {volume}
  {113}},\ \bibinfo {pages} {3986}}\BibitemShut {NoStop}%
\bibitem [{\citenamefont {Hao}\ and\ \citenamefont
  {Hu}(2014)}]{hao2014topological}%
  \BibitemOpen
  \bibfield  {author} {\bibinfo {author} {\bibnamefont {Hao}, \bibfnamefont
  {N.}}, and\ \bibinfo {author} {\bibfnamefont {J.}~\bibnamefont {Hu}}}
  (\bibinfo {year} {2014}),\ \href {https://doi.org/10.1103/PhysRevX.4.031053}
  {\bibfield  {journal} {\bibinfo  {journal} {Phys. Rev. X}\ }\textbf {\bibinfo
  {volume} {4}},\ \bibinfo {pages} {031053}}\BibitemShut {NoStop}%
\bibitem [{\citenamefont {Hao}\ and\ \citenamefont {Hu}(2019)}]{Hao2019}%
  \BibitemOpen
  \bibfield  {author} {\bibinfo {author} {\bibnamefont {Hao}, \bibfnamefont
  {N.}}, and\ \bibinfo {author} {\bibfnamefont {J.}~\bibnamefont {Hu}}}
  (\bibinfo {year} {2019}),\ \href {https://doi.org/10.1093/nsr/nwy142}
  {\bibfield  {journal} {\bibinfo  {journal} {Natl. Sci. Rev.}\ }\textbf
  {\bibinfo {volume} {6}},\ \bibinfo {pages} {213}}\BibitemShut {NoStop}%
\bibitem [{\citenamefont {Hao}\ and\ \citenamefont {Shen}(2015)}]{Hao2015}%
  \BibitemOpen
  \bibfield  {author} {\bibinfo {author} {\bibnamefont {Hao}, \bibfnamefont
  {N.}}, and\ \bibinfo {author} {\bibfnamefont {S.-Q.}\ \bibnamefont {Shen}}}
  (\bibinfo {year} {2015}),\ \href {https://doi.org/10.1103/PhysRevB.92.165104}
  {\bibfield  {journal} {\bibinfo  {journal} {Phys. Rev. B}\ }\textbf {\bibinfo
  {volume} {92}},\ \bibinfo {pages} {165104}}\BibitemShut {NoStop}%
\bibitem [{\citenamefont {Hashimoto}\ \emph {et~al.}(2020)\citenamefont
  {Hashimoto}, \citenamefont {Ota}, \citenamefont {Tsuzuki}, \citenamefont
  {Nagashima}, \citenamefont {Fukushima}, \citenamefont {Kasahara},
  \citenamefont {Matsuda}, \citenamefont {Matsuura}, \citenamefont {Mizukami},
  \citenamefont {Shibauchi}, \citenamefont {Shin},\ and\ \citenamefont
  {Okazaki}}]{hashimoto2020bose}%
  \BibitemOpen
  \bibfield  {author} {\bibinfo {author} {\bibnamefont {Hashimoto},
  \bibfnamefont {T.}}, \bibinfo {author} {\bibfnamefont {Y.}~\bibnamefont
  {Ota}}, \bibinfo {author} {\bibfnamefont {A.}~\bibnamefont {Tsuzuki}},
  \bibinfo {author} {\bibfnamefont {T.}~\bibnamefont {Nagashima}}, \bibinfo
  {author} {\bibfnamefont {A.}~\bibnamefont {Fukushima}}, \bibinfo {author}
  {\bibfnamefont {S.}~\bibnamefont {Kasahara}}, \bibinfo {author}
  {\bibfnamefont {Y.}~\bibnamefont {Matsuda}}, \bibinfo {author} {\bibfnamefont
  {K.}~\bibnamefont {Matsuura}}, \bibinfo {author} {\bibfnamefont
  {Y.}~\bibnamefont {Mizukami}}, \bibinfo {author} {\bibfnamefont
  {T.}~\bibnamefont {Shibauchi}}, \bibinfo {author} {\bibfnamefont
  {S.}~\bibnamefont {Shin}}, and\ \bibinfo {author} {\bibfnamefont
  {K.}~\bibnamefont {Okazaki}}} (\bibinfo {year} {2020}),\ \href
  {https://doi.org/10.1126/sciadv.abb9052} {\bibfield  {journal} {\bibinfo
  {journal} {Sci. Adv.}\ }\textbf {\bibinfo {volume} {6}},\ \bibinfo {pages}
  {eabb9052}}\BibitemShut {NoStop}%
\bibitem [{\citenamefont {Haule}\ and\ \citenamefont
  {Kotliar}(2009)}]{haule2009coherence}%
  \BibitemOpen
  \bibfield  {author} {\bibinfo {author} {\bibnamefont {Haule}, \bibfnamefont
  {K.}}, and\ \bibinfo {author} {\bibfnamefont {G.}~\bibnamefont {Kotliar}}}
  (\bibinfo {year} {2009}),\ \href
  {https://doi.org/10.1088/1367-2630/11/2/025021} {\bibfield  {journal}
  {\bibinfo  {journal} {New J. Phys.}\ }\textbf {\bibinfo {volume} {11}},\
  \bibinfo {pages} {025021}}\BibitemShut {NoStop}%
\bibitem [{\citenamefont {Haverkort}(2010)}]{haverkort2010theory}%
  \BibitemOpen
  \bibfield  {author} {\bibinfo {author} {\bibnamefont {Haverkort},
  \bibfnamefont {M.~W.}}} (\bibinfo {year} {2010}),\ \href
  {https://doi.org/10.1103/PhysRevLett.105.167404} {\bibfield  {journal}
  {\bibinfo  {journal} {Phys. Rev. Lett.}\ }\textbf {\bibinfo {volume} {105}},\
  \bibinfo {pages} {167404}}\BibitemShut {NoStop}%
\bibitem [{\citenamefont {Hayden}\ \emph {et~al.}(2004)\citenamefont {Hayden},
  \citenamefont {Mook}, \citenamefont {Dai}, \citenamefont {Perring},\ and\
  \citenamefont {Doğan}}]{Hayden2004}%
  \BibitemOpen
  \bibfield  {author} {\bibinfo {author} {\bibnamefont {Hayden}, \bibfnamefont
  {S.~M.}}, \bibinfo {author} {\bibfnamefont {H.~A.}\ \bibnamefont {Mook}},
  \bibinfo {author} {\bibfnamefont {P.}~\bibnamefont {Dai}}, \bibinfo {author}
  {\bibfnamefont {T.~G.}\ \bibnamefont {Perring}}, and\ \bibinfo {author}
  {\bibfnamefont {F.}~\bibnamefont {Doğan}}} (\bibinfo {year} {2004}),\ \href
  {https://doi.org/10.1038/nature02576} {\bibfield  {journal} {\bibinfo
  {journal} {Nature}\ }\textbf {\bibinfo {volume} {429}},\ \bibinfo {pages}
  {531}}\BibitemShut {NoStop}%
\bibitem [{\citenamefont {Hayden}\ and\ \citenamefont
  {Tranquada}(2024)}]{Hayden2024}%
  \BibitemOpen
  \bibfield  {author} {\bibinfo {author} {\bibnamefont {Hayden}, \bibfnamefont
  {S.~M.}}, and\ \bibinfo {author} {\bibfnamefont {J.~M.}\ \bibnamefont
  {Tranquada}}} (\bibinfo {year} {2024}),\ \href
  {https://doi.org/10.1146/annurev-conmatphys-032922-094430} {\bibfield
  {journal} {\bibinfo  {journal} {Annu. Rev. Condens. Matter Phys.}\ }\textbf
  {\bibinfo {volume} {15}},\ \bibinfo {pages} {215}}\BibitemShut {NoStop}%
\bibitem [{\citenamefont {He}\ \emph {et~al.}(2017)\citenamefont {He},
  \citenamefont {Wang}, \citenamefont {Ahn}, \citenamefont {Hardy},
  \citenamefont {Wolf}, \citenamefont {Adelmann}, \citenamefont {Schmalian},
  \citenamefont {Eremin},\ and\ \citenamefont {Meingast}}]{he2017dichotomy}%
  \BibitemOpen
  \bibfield  {author} {\bibinfo {author} {\bibnamefont {He}, \bibfnamefont
  {M.}}, \bibinfo {author} {\bibfnamefont {L.}~\bibnamefont {Wang}}, \bibinfo
  {author} {\bibfnamefont {F.}~\bibnamefont {Ahn}}, \bibinfo {author}
  {\bibfnamefont {F.}~\bibnamefont {Hardy}}, \bibinfo {author} {\bibfnamefont
  {T.}~\bibnamefont {Wolf}}, \bibinfo {author} {\bibfnamefont {P.}~\bibnamefont
  {Adelmann}}, \bibinfo {author} {\bibfnamefont {J.}~\bibnamefont {Schmalian}},
  \bibinfo {author} {\bibfnamefont {I.}~\bibnamefont {Eremin}}, and\ \bibinfo
  {author} {\bibfnamefont {C.}~\bibnamefont {Meingast}}} (\bibinfo {year}
  {2017}),\ \href {https://doi.org/10.1038/s41467-017-00712-3} {\bibfield
  {journal} {\bibinfo  {journal} {Nat. Commun.}\ }\textbf {\bibinfo {volume}
  {8}},\ \bibinfo {pages} {504}}\BibitemShut {NoStop}%
\bibitem [{\citenamefont {He}\ \emph {et~al.}(2014)\citenamefont {He},
  \citenamefont {Liu}, \citenamefont {He}, \citenamefont {Lai}, \citenamefont
  {He}, \citenamefont {Wang}, \citenamefont {Law}, \citenamefont {Lortz},
  \citenamefont {Wang},\ and\ \citenamefont {Sou}}]{He2014}%
  \BibitemOpen
  \bibfield  {author} {\bibinfo {author} {\bibnamefont {He}, \bibfnamefont
  {Q.~L.}}, \bibinfo {author} {\bibfnamefont {H.}~\bibnamefont {Liu}}, \bibinfo
  {author} {\bibfnamefont {M.}~\bibnamefont {He}}, \bibinfo {author}
  {\bibfnamefont {Y.~H.}\ \bibnamefont {Lai}}, \bibinfo {author} {\bibfnamefont
  {H.}~\bibnamefont {He}}, \bibinfo {author} {\bibfnamefont {G.}~\bibnamefont
  {Wang}}, \bibinfo {author} {\bibfnamefont {K.~T.}\ \bibnamefont {Law}},
  \bibinfo {author} {\bibfnamefont {R.}~\bibnamefont {Lortz}}, \bibinfo
  {author} {\bibfnamefont {J.}~\bibnamefont {Wang}}, and\ \bibinfo {author}
  {\bibfnamefont {I.~K.}\ \bibnamefont {Sou}}} (\bibinfo {year} {2014}),\ \href
  {https://doi.org/10.1038/ncomms5247} {\bibfield  {journal} {\bibinfo
  {journal} {Nat. Commun.}\ }\textbf {\bibinfo {volume} {5}},\ \bibinfo {pages}
  {4247}}\BibitemShut {NoStop}%
\bibitem [{\citenamefont {He}\ \emph {et~al.}(2013)\citenamefont {He},
  \citenamefont {He}, \citenamefont {Zhang}, \citenamefont {Zhao},
  \citenamefont {Liu}, \citenamefont {Liu}, \citenamefont {Mou}, \citenamefont
  {Ou}, \citenamefont {Wang}, \citenamefont {Li} \emph {et~al.}}]{He2013}%
  \BibitemOpen
  \bibfield  {author} {\bibinfo {author} {\bibnamefont {He}, \bibfnamefont
  {S.}}, \bibinfo {author} {\bibfnamefont {J.}~\bibnamefont {He}}, \bibinfo
  {author} {\bibfnamefont {W.}~\bibnamefont {Zhang}}, \bibinfo {author}
  {\bibfnamefont {L.}~\bibnamefont {Zhao}}, \bibinfo {author} {\bibfnamefont
  {D.}~\bibnamefont {Liu}}, \bibinfo {author} {\bibfnamefont {X.}~\bibnamefont
  {Liu}}, \bibinfo {author} {\bibfnamefont {D.}~\bibnamefont {Mou}}, \bibinfo
  {author} {\bibfnamefont {Y.-B.}\ \bibnamefont {Ou}}, \bibinfo {author}
  {\bibfnamefont {Q.-Y.}\ \bibnamefont {Wang}}, \bibinfo {author}
  {\bibfnamefont {Z.}~\bibnamefont {Li}},  \emph {et~al.}} (\bibinfo {year}
  {2013}),\ \href {https://doi.org/10.1038/nmat3648} {\bibfield  {journal}
  {\bibinfo  {journal} {Nat. Mater.}\ }\textbf {\bibinfo {volume} {12}},\
  \bibinfo {pages} {605}}\BibitemShut {NoStop}%
\bibitem [{\citenamefont {He}\ \emph {et~al.}(2011)\citenamefont {He},
  \citenamefont {Li}, \citenamefont {Zhang}, \citenamefont {Karki},
  \citenamefont {Jin}, \citenamefont {Sales}, \citenamefont {Sefat},
  \citenamefont {McGuire}, \citenamefont {Mandrus},\ and\ \citenamefont
  {Plummer}}]{he2011nanoscale}%
  \BibitemOpen
  \bibfield  {author} {\bibinfo {author} {\bibnamefont {He}, \bibfnamefont
  {X.}}, \bibinfo {author} {\bibfnamefont {G.}~\bibnamefont {Li}}, \bibinfo
  {author} {\bibfnamefont {J.}~\bibnamefont {Zhang}}, \bibinfo {author}
  {\bibfnamefont {A.~B.}\ \bibnamefont {Karki}}, \bibinfo {author}
  {\bibfnamefont {R.}~\bibnamefont {Jin}}, \bibinfo {author} {\bibfnamefont
  {B.~C.}\ \bibnamefont {Sales}}, \bibinfo {author} {\bibfnamefont {A.~S.}\
  \bibnamefont {Sefat}}, \bibinfo {author} {\bibfnamefont {M.~A.}\ \bibnamefont
  {McGuire}}, \bibinfo {author} {\bibfnamefont {D.}~\bibnamefont {Mandrus}},
  and\ \bibinfo {author} {\bibfnamefont {E.~W.}\ \bibnamefont {Plummer}}}
  (\bibinfo {year} {2011}),\ \href {https://doi.org/10.1103/PhysRevB.83.220502}
  {\bibfield  {journal} {\bibinfo  {journal} {Phys. Rev. B}\ }\textbf {\bibinfo
  {volume} {83}},\ \bibinfo {pages} {220502(R)}}\BibitemShut {NoStop}%
\bibitem [{\citenamefont {Headings}\ \emph {et~al.}(2010)\citenamefont
  {Headings}, \citenamefont {Hayden}, \citenamefont {Coldea},\ and\
  \citenamefont {Perring}}]{headings2010anomalous}%
  \BibitemOpen
  \bibfield  {author} {\bibinfo {author} {\bibnamefont {Headings},
  \bibfnamefont {N.~S.}}, \bibinfo {author} {\bibfnamefont {S.~M.}\
  \bibnamefont {Hayden}}, \bibinfo {author} {\bibfnamefont {R.}~\bibnamefont
  {Coldea}}, and\ \bibinfo {author} {\bibfnamefont {T.~G.}\ \bibnamefont
  {Perring}}} (\bibinfo {year} {2010}),\ \href
  {https://doi.org/10.1103/PhysRevLett.105.247001} {\bibfield  {journal}
  {\bibinfo  {journal} {Phys. Rev. Lett.}\ }\textbf {\bibinfo {volume} {105}},\
  \bibinfo {pages} {247001}}\BibitemShut {NoStop}%
\bibitem [{\citenamefont {Hemmida}\ \emph {et~al.}(2021)\citenamefont
  {Hemmida}, \citenamefont {Winterhalter-Stocker}, \citenamefont {Ehlers},
  \citenamefont {von Nidda}, \citenamefont {Yao}, \citenamefont {Bannies},
  \citenamefont {Rienks}, \citenamefont {Kurleto}, \citenamefont {Felser},
  \citenamefont {B\"uchner}, \citenamefont {Fink}, \citenamefont {Gorol},
  \citenamefont {F\"orster}, \citenamefont {Arsenijevic}, \citenamefont
  {Fritsch},\ and\ \citenamefont {Gegenwart}}]{Hemmida2021}%
  \BibitemOpen
  \bibfield  {author} {\bibinfo {author} {\bibnamefont {Hemmida}, \bibfnamefont
  {M.}}, \bibinfo {author} {\bibfnamefont {N.}~\bibnamefont
  {Winterhalter-Stocker}}, \bibinfo {author} {\bibfnamefont {D.}~\bibnamefont
  {Ehlers}}, \bibinfo {author} {\bibfnamefont {H.-A.~K.}\ \bibnamefont {von
  Nidda}}, \bibinfo {author} {\bibfnamefont {M.}~\bibnamefont {Yao}}, \bibinfo
  {author} {\bibfnamefont {J.}~\bibnamefont {Bannies}}, \bibinfo {author}
  {\bibfnamefont {E.~D.~L.}\ \bibnamefont {Rienks}}, \bibinfo {author}
  {\bibfnamefont {R.}~\bibnamefont {Kurleto}}, \bibinfo {author} {\bibfnamefont
  {C.}~\bibnamefont {Felser}}, \bibinfo {author} {\bibfnamefont
  {B.}~\bibnamefont {B\"uchner}}, \bibinfo {author} {\bibfnamefont
  {J.}~\bibnamefont {Fink}}, \bibinfo {author} {\bibfnamefont {S.}~\bibnamefont
  {Gorol}}, \bibinfo {author} {\bibfnamefont {T.}~\bibnamefont {F\"orster}},
  \bibinfo {author} {\bibfnamefont {S.}~\bibnamefont {Arsenijevic}}, \bibinfo
  {author} {\bibfnamefont {V.}~\bibnamefont {Fritsch}}, and\ \bibinfo {author}
  {\bibfnamefont {P.}~\bibnamefont {Gegenwart}}} (\bibinfo {year} {2021}),\
  \href {https://doi.org/10.1103/PhysRevB.103.195112} {\bibfield  {journal}
  {\bibinfo  {journal} {Phys. Rev. B}\ }\textbf {\bibinfo {volume} {103}},\
  \bibinfo {pages} {195112}}\BibitemShut {NoStop}%
\bibitem [{\citenamefont {Her}\ \emph {et~al.}(2015)\citenamefont {Her},
  \citenamefont {Kohama}, \citenamefont {Matsuda}, \citenamefont {Kindo},
  \citenamefont {Yang}, \citenamefont {Chareev}, \citenamefont {Mitrofanova},
  \citenamefont {Volkova}, \citenamefont {Vasiliev},\ and\ \citenamefont
  {Lin}}]{Her2015}%
  \BibitemOpen
  \bibfield  {author} {\bibinfo {author} {\bibnamefont {Her}, \bibfnamefont
  {J.}}, \bibinfo {author} {\bibfnamefont {Y.}~\bibnamefont {Kohama}}, \bibinfo
  {author} {\bibfnamefont {Y.}~\bibnamefont {Matsuda}}, \bibinfo {author}
  {\bibfnamefont {K.}~\bibnamefont {Kindo}}, \bibinfo {author} {\bibfnamefont
  {W.}~\bibnamefont {Yang}}, \bibinfo {author} {\bibfnamefont {D.}~\bibnamefont
  {Chareev}}, \bibinfo {author} {\bibfnamefont {E.}~\bibnamefont
  {Mitrofanova}}, \bibinfo {author} {\bibfnamefont {O.}~\bibnamefont
  {Volkova}}, \bibinfo {author} {\bibfnamefont {A.}~\bibnamefont {Vasiliev}},
  and\ \bibinfo {author} {\bibfnamefont {J.-Y.}\ \bibnamefont {Lin}}} (\bibinfo
  {year} {2015}),\ \href {https://doi.org/10.1088/0953-2048/28/4/045013}
  {\bibfield  {journal} {\bibinfo  {journal} {Supercond. Sci. Technol.}\
  }\textbf {\bibinfo {volume} {28}},\ \bibinfo {pages} {045013}}\BibitemShut
  {NoStop}%
\bibitem [{\citenamefont {Hicks}\ \emph
  {et~al.}(2014{\natexlab{a}})\citenamefont {Hicks}, \citenamefont {Barber},
  \citenamefont {Edkins}, \citenamefont {Brodsky},\ and\ \citenamefont
  {Mackenzie}}]{hicks2014piezoelectric}%
  \BibitemOpen
  \bibfield  {author} {\bibinfo {author} {\bibnamefont {Hicks}, \bibfnamefont
  {C.~W.}}, \bibinfo {author} {\bibfnamefont {M.~E.}\ \bibnamefont {Barber}},
  \bibinfo {author} {\bibfnamefont {S.~D.}\ \bibnamefont {Edkins}}, \bibinfo
  {author} {\bibfnamefont {D.~O.}\ \bibnamefont {Brodsky}}, and\ \bibinfo
  {author} {\bibfnamefont {A.~P.}\ \bibnamefont {Mackenzie}}} (\bibinfo {year}
  {2014}{\natexlab{a}}),\ \href {https://doi.org/10.1063/1.4881611} {\bibfield
  {journal} {\bibinfo  {journal} {Rev. Sci. Instrum.}\ }\textbf {\bibinfo
  {volume} {85}},\ \bibinfo {pages} {065003}}\BibitemShut {NoStop}%
\bibitem [{\citenamefont {Hicks}\ \emph
  {et~al.}(2014{\natexlab{b}})\citenamefont {Hicks}, \citenamefont {Brodsky},
  \citenamefont {Yelland}, \citenamefont {Gibbs}, \citenamefont {Bruin},
  \citenamefont {Barber}, \citenamefont {Edkins}, \citenamefont {Nishimura},
  \citenamefont {Yonezawa}, \citenamefont {Maeno},\ and\ \citenamefont
  {Mackenzie}}]{hicks2014strong}%
  \BibitemOpen
  \bibfield  {author} {\bibinfo {author} {\bibnamefont {Hicks}, \bibfnamefont
  {C.~W.}}, \bibinfo {author} {\bibfnamefont {D.~O.}\ \bibnamefont {Brodsky}},
  \bibinfo {author} {\bibfnamefont {E.~A.}\ \bibnamefont {Yelland}}, \bibinfo
  {author} {\bibfnamefont {A.~S.}\ \bibnamefont {Gibbs}}, \bibinfo {author}
  {\bibfnamefont {J.~A.~N.}\ \bibnamefont {Bruin}}, \bibinfo {author}
  {\bibfnamefont {M.~E.}\ \bibnamefont {Barber}}, \bibinfo {author}
  {\bibfnamefont {S.~D.}\ \bibnamefont {Edkins}}, \bibinfo {author}
  {\bibfnamefont {K.}~\bibnamefont {Nishimura}}, \bibinfo {author}
  {\bibfnamefont {S.}~\bibnamefont {Yonezawa}}, \bibinfo {author}
  {\bibfnamefont {Y.}~\bibnamefont {Maeno}}, and\ \bibinfo {author}
  {\bibfnamefont {A.~P.}\ \bibnamefont {Mackenzie}}} (\bibinfo {year}
  {2014}{\natexlab{b}}),\ \href {https://doi.org/10.1126/science.1248292}
  {\bibfield  {journal} {\bibinfo  {journal} {Science}\ }\textbf {\bibinfo
  {volume} {344}},\ \bibinfo {pages} {283}}\BibitemShut {NoStop}%
\bibitem [{\citenamefont {Hicks}\ \emph {et~al.}(2025)\citenamefont {Hicks},
  \citenamefont {Jerzembeck}, \citenamefont {Noad}, \citenamefont {Barber},\
  and\ \citenamefont {Mackenzie}}]{hicks2025probing}%
  \BibitemOpen
  \bibfield  {author} {\bibinfo {author} {\bibnamefont {Hicks}, \bibfnamefont
  {C.~W.}}, \bibinfo {author} {\bibfnamefont {F.}~\bibnamefont {Jerzembeck}},
  \bibinfo {author} {\bibfnamefont {H.~M.}\ \bibnamefont {Noad}}, \bibinfo
  {author} {\bibfnamefont {M.~E.}\ \bibnamefont {Barber}}, and\ \bibinfo
  {author} {\bibfnamefont {A.~P.}\ \bibnamefont {Mackenzie}}} (\bibinfo {year}
  {2025}),\ \href {https://doi.org/10.1146/annurev-conmatphys-040521-041041}
  {\bibfield  {journal} {\bibinfo  {journal} {Annu. Rev. Condens. Matter
  Phys.}\ }\textbf {\bibinfo {volume} {16}},\ \bibinfo {pages}
  {417}}\BibitemShut {NoStop}%
\bibitem [{\citenamefont {Hieke}\ \emph {et~al.}(2013)\citenamefont {Hieke},
  \citenamefont {Lippmann}, \citenamefont {St{\"u}rzer}, \citenamefont
  {Friederichs}, \citenamefont {Nitsche}, \citenamefont {Winter}, \citenamefont
  {P{\"o}ttgen},\ and\ \citenamefont {Johrendt}}]{Hieke2013}%
  \BibitemOpen
  \bibfield  {author} {\bibinfo {author} {\bibnamefont {Hieke}, \bibfnamefont
  {C.}}, \bibinfo {author} {\bibfnamefont {J.}~\bibnamefont {Lippmann}},
  \bibinfo {author} {\bibfnamefont {T.}~\bibnamefont {St{\"u}rzer}}, \bibinfo
  {author} {\bibfnamefont {G.}~\bibnamefont {Friederichs}}, \bibinfo {author}
  {\bibfnamefont {F.}~\bibnamefont {Nitsche}}, \bibinfo {author} {\bibfnamefont
  {F.}~\bibnamefont {Winter}}, \bibinfo {author} {\bibfnamefont
  {R.}~\bibnamefont {P{\"o}ttgen}}, and\ \bibinfo {author} {\bibfnamefont
  {D.}~\bibnamefont {Johrendt}}} (\bibinfo {year} {2013}),\ \href
  {https://doi.org/10.1080/14786435.2013.816450} {\bibfield  {journal}
  {\bibinfo  {journal} {Philosophical Magazine}\ }\textbf {\bibinfo {volume}
  {93}},\ \bibinfo {pages} {3680}}\BibitemShut {NoStop}%
\bibitem [{\citenamefont {Hiraishi}\ \emph {et~al.}(2014)\citenamefont
  {Hiraishi}, \citenamefont {Iimura}, \citenamefont {Kojima}, \citenamefont
  {Yamaura}, \citenamefont {Hiraka}, \citenamefont {Ikeda}, \citenamefont
  {Miao}, \citenamefont {Ishikawa}, \citenamefont {Torii}, \citenamefont
  {Miyazaki}, \citenamefont {Yamauchi}, \citenamefont {Koda}, \citenamefont
  {Ishii}, \citenamefont {Yoshida}, \citenamefont {Mizuki}, \citenamefont
  {Kadono}, \citenamefont {Kumai}, \citenamefont {Kamiyama}, \citenamefont
  {Otomo}, \citenamefont {Murakami}, \citenamefont {Matsuishi},\ and\
  \citenamefont {Hosono}}]{hiraishi2014bipartite}%
  \BibitemOpen
  \bibfield  {author} {\bibinfo {author} {\bibnamefont {Hiraishi},
  \bibfnamefont {M.}}, \bibinfo {author} {\bibfnamefont {S.}~\bibnamefont
  {Iimura}}, \bibinfo {author} {\bibfnamefont {K.~M.}\ \bibnamefont {Kojima}},
  \bibinfo {author} {\bibfnamefont {J.}~\bibnamefont {Yamaura}}, \bibinfo
  {author} {\bibfnamefont {H.}~\bibnamefont {Hiraka}}, \bibinfo {author}
  {\bibfnamefont {K.}~\bibnamefont {Ikeda}}, \bibinfo {author} {\bibfnamefont
  {P.}~\bibnamefont {Miao}}, \bibinfo {author} {\bibfnamefont {Y.}~\bibnamefont
  {Ishikawa}}, \bibinfo {author} {\bibfnamefont {S.}~\bibnamefont {Torii}},
  \bibinfo {author} {\bibfnamefont {M.}~\bibnamefont {Miyazaki}}, \bibinfo
  {author} {\bibfnamefont {I.}~\bibnamefont {Yamauchi}}, \bibinfo {author}
  {\bibfnamefont {A.}~\bibnamefont {Koda}}, \bibinfo {author} {\bibfnamefont
  {K.}~\bibnamefont {Ishii}}, \bibinfo {author} {\bibfnamefont
  {M.}~\bibnamefont {Yoshida}}, \bibinfo {author} {\bibfnamefont
  {J.}~\bibnamefont {Mizuki}}, \bibinfo {author} {\bibfnamefont
  {R.}~\bibnamefont {Kadono}}, \bibinfo {author} {\bibfnamefont
  {R.}~\bibnamefont {Kumai}}, \bibinfo {author} {\bibfnamefont
  {T.}~\bibnamefont {Kamiyama}}, \bibinfo {author} {\bibfnamefont
  {T.}~\bibnamefont {Otomo}}, \bibinfo {author} {\bibfnamefont
  {Y.}~\bibnamefont {Murakami}}, \bibinfo {author} {\bibfnamefont
  {S.}~\bibnamefont {Matsuishi}}, and\ \bibinfo {author} {\bibfnamefont
  {H.}~\bibnamefont {Hosono}}} (\bibinfo {year} {2014}),\ \href
  {https://doi.org/10.1038/nphys2906} {\bibfield  {journal} {\bibinfo
  {journal} {Nat. Phys.}\ }\textbf {\bibinfo {volume} {10}},\ \bibinfo {pages}
  {300}}\BibitemShut {NoStop}%
\bibitem [{\citenamefont {Hirschfeld}\ \emph {et~al.}(2011)\citenamefont
  {Hirschfeld}, \citenamefont {Korshunov},\ and\ \citenamefont
  {Mazin}}]{Hirschfeld2011}%
  \BibitemOpen
  \bibfield  {author} {\bibinfo {author} {\bibnamefont {Hirschfeld},
  \bibfnamefont {P.}}, \bibinfo {author} {\bibfnamefont {M.}~\bibnamefont
  {Korshunov}}, and\ \bibinfo {author} {\bibfnamefont {I.}~\bibnamefont
  {Mazin}}} (\bibinfo {year} {2011}),\ \href
  {https://doi.org/10.1088/0034-4885/74/12/124508} {\bibfield  {journal}
  {\bibinfo  {journal} {Reports on Progress in Physics}\ }\textbf {\bibinfo
  {volume} {74}},\ \bibinfo {pages} {124508}}\BibitemShut {NoStop}%
\bibitem [{\citenamefont {Hirschfeld}\ \emph {et~al.}(2015)\citenamefont
  {Hirschfeld}, \citenamefont {Altenfeld}, \citenamefont {Eremin},\ and\
  \citenamefont {Mazin}}]{hirschfeld2015robust}%
  \BibitemOpen
  \bibfield  {author} {\bibinfo {author} {\bibnamefont {Hirschfeld},
  \bibfnamefont {P.~J.}}, \bibinfo {author} {\bibfnamefont {D.}~\bibnamefont
  {Altenfeld}}, \bibinfo {author} {\bibfnamefont {I.}~\bibnamefont {Eremin}},
  and\ \bibinfo {author} {\bibfnamefont {I.~I.}\ \bibnamefont {Mazin}}}
  (\bibinfo {year} {2015}),\ \href {https://doi.org/10.1103/PhysRevB.92.184513}
  {\bibfield  {journal} {\bibinfo  {journal} {Phys. Rev. B}\ }\textbf {\bibinfo
  {volume} {92}},\ \bibinfo {pages} {184513}}\BibitemShut {NoStop}%
\bibitem [{\citenamefont {Holenstein}\ \emph {et~al.}(2019)\citenamefont
  {Holenstein}, \citenamefont {Fischer}, \citenamefont {Liu}, \citenamefont
  {Barbero}, \citenamefont {Simutis}, \citenamefont {Shermadini}, \citenamefont
  {Elender}, \citenamefont {Biswas}, \citenamefont {Khasanov}, \citenamefont
  {Amato} \emph {et~al.}}]{Holenstein2019}%
  \BibitemOpen
  \bibfield  {author} {\bibinfo {author} {\bibnamefont {Holenstein},
  \bibfnamefont {S.}}, \bibinfo {author} {\bibfnamefont {B.}~\bibnamefont
  {Fischer}}, \bibinfo {author} {\bibfnamefont {Y.}~\bibnamefont {Liu}},
  \bibinfo {author} {\bibfnamefont {N.}~\bibnamefont {Barbero}}, \bibinfo
  {author} {\bibfnamefont {G.}~\bibnamefont {Simutis}}, \bibinfo {author}
  {\bibfnamefont {Z.}~\bibnamefont {Shermadini}}, \bibinfo {author}
  {\bibfnamefont {M.}~\bibnamefont {Elender}}, \bibinfo {author} {\bibfnamefont
  {P.}~\bibnamefont {Biswas}}, \bibinfo {author} {\bibfnamefont
  {R.}~\bibnamefont {Khasanov}}, \bibinfo {author} {\bibfnamefont
  {A.}~\bibnamefont {Amato}},  \emph {et~al.}} (\bibinfo {year} {2019}),\ \href
  {https://doi.org/10.48550/arXiv.1911.04325} {\bibfield  {journal} {\bibinfo
  {journal} {arXiv preprint arXiv:1911.04325}\
  }10.48550/arXiv.1911.04325}\BibitemShut {NoStop}%
\bibitem [{\citenamefont {Hong}\ \emph
  {et~al.}(2020{\natexlab{a}})\citenamefont {Hong}, \citenamefont {Song},
  \citenamefont {Liu}, \citenamefont {Li}, \citenamefont {Zeng}, \citenamefont
  {Li}, \citenamefont {Wu}, \citenamefont {Sui}, \citenamefont {Xie},
  \citenamefont {Danilkin}, \citenamefont {Ghosh}, \citenamefont {Ghosh},
  \citenamefont {Hu}, \citenamefont {Zhao}, \citenamefont {Zhou}, \citenamefont
  {Qiu}, \citenamefont {Li},\ and\ \citenamefont {Luo}}]{hong2020neutron}%
  \BibitemOpen
  \bibfield  {author} {\bibinfo {author} {\bibnamefont {Hong}, \bibfnamefont
  {W.}}, \bibinfo {author} {\bibfnamefont {L.}~\bibnamefont {Song}}, \bibinfo
  {author} {\bibfnamefont {B.}~\bibnamefont {Liu}}, \bibinfo {author}
  {\bibfnamefont {Z.}~\bibnamefont {Li}}, \bibinfo {author} {\bibfnamefont
  {Z.}~\bibnamefont {Zeng}}, \bibinfo {author} {\bibfnamefont {Y.}~\bibnamefont
  {Li}}, \bibinfo {author} {\bibfnamefont {D.}~\bibnamefont {Wu}}, \bibinfo
  {author} {\bibfnamefont {Q.}~\bibnamefont {Sui}}, \bibinfo {author}
  {\bibfnamefont {T.}~\bibnamefont {Xie}}, \bibinfo {author} {\bibfnamefont
  {S.}~\bibnamefont {Danilkin}}, \bibinfo {author} {\bibfnamefont
  {H.}~\bibnamefont {Ghosh}}, \bibinfo {author} {\bibfnamefont
  {A.}~\bibnamefont {Ghosh}}, \bibinfo {author} {\bibfnamefont
  {J.}~\bibnamefont {Hu}}, \bibinfo {author} {\bibfnamefont {L.}~\bibnamefont
  {Zhao}}, \bibinfo {author} {\bibfnamefont {X.}~\bibnamefont {Zhou}}, \bibinfo
  {author} {\bibfnamefont {X.}~\bibnamefont {Qiu}}, \bibinfo {author}
  {\bibfnamefont {S.}~\bibnamefont {Li}}, and\ \bibinfo {author} {\bibfnamefont
  {H.}~\bibnamefont {Luo}}} (\bibinfo {year} {2020}{\natexlab{a}}),\ \href
  {https://doi.org/10.1103/PhysRevLett.125.117002} {\bibfield  {journal}
  {\bibinfo  {journal} {Phys. Rev. Lett.}\ }\textbf {\bibinfo {volume} {125}},\
  \bibinfo {pages} {117002}}\BibitemShut {NoStop}%
\bibitem [{\citenamefont {Hong}\ \emph {et~al.}(2023)\citenamefont {Hong},
  \citenamefont {Zhou}, \citenamefont {Li}, \citenamefont {Li}, \citenamefont
  {Stuhr}, \citenamefont {Pokhriyal}, \citenamefont {Ghosh}, \citenamefont
  {Tao}, \citenamefont {Lu}, \citenamefont {Hu}, \citenamefont {Li},\ and\
  \citenamefont {Luo}}]{hong2023interlayer}%
  \BibitemOpen
  \bibfield  {author} {\bibinfo {author} {\bibnamefont {Hong}, \bibfnamefont
  {W.}}, \bibinfo {author} {\bibfnamefont {H.}~\bibnamefont {Zhou}}, \bibinfo
  {author} {\bibfnamefont {Z.}~\bibnamefont {Li}}, \bibinfo {author}
  {\bibfnamefont {Y.}~\bibnamefont {Li}}, \bibinfo {author} {\bibfnamefont
  {U.}~\bibnamefont {Stuhr}}, \bibinfo {author} {\bibfnamefont
  {A.}~\bibnamefont {Pokhriyal}}, \bibinfo {author} {\bibfnamefont
  {H.}~\bibnamefont {Ghosh}}, \bibinfo {author} {\bibfnamefont
  {Z.}~\bibnamefont {Tao}}, \bibinfo {author} {\bibfnamefont {X.}~\bibnamefont
  {Lu}}, \bibinfo {author} {\bibfnamefont {J.}~\bibnamefont {Hu}}, \bibinfo
  {author} {\bibfnamefont {S.}~\bibnamefont {Li}}, and\ \bibinfo {author}
  {\bibfnamefont {H.}~\bibnamefont {Luo}}} (\bibinfo {year} {2023}),\ \href
  {https://doi.org/10.1103/PhysRevB.107.224514} {\bibfield  {journal} {\bibinfo
   {journal} {Phys. Rev. B}\ }\textbf {\bibinfo {volume} {107}},\ \bibinfo
  {pages} {224514}}\BibitemShut {NoStop}%
\bibitem [{\citenamefont {Hong}\ \emph
  {et~al.}(2020{\natexlab{b}})\citenamefont {Hong}, \citenamefont {Caglieris},
  \citenamefont {Kappenberger}, \citenamefont {Wurmehl}, \citenamefont
  {Aswartham}, \citenamefont {Scaravaggi}, \citenamefont {Lepucki},
  \citenamefont {Wolter}, \citenamefont {Grafe}, \citenamefont {B\"uchner},\
  and\ \citenamefont {Hess}}]{hong2020evolution}%
  \BibitemOpen
  \bibfield  {author} {\bibinfo {author} {\bibnamefont {Hong}, \bibfnamefont
  {X.}}, \bibinfo {author} {\bibfnamefont {F.}~\bibnamefont {Caglieris}},
  \bibinfo {author} {\bibfnamefont {R.}~\bibnamefont {Kappenberger}}, \bibinfo
  {author} {\bibfnamefont {S.}~\bibnamefont {Wurmehl}}, \bibinfo {author}
  {\bibfnamefont {S.}~\bibnamefont {Aswartham}}, \bibinfo {author}
  {\bibfnamefont {F.}~\bibnamefont {Scaravaggi}}, \bibinfo {author}
  {\bibfnamefont {P.}~\bibnamefont {Lepucki}}, \bibinfo {author} {\bibfnamefont
  {A.~U.~B.}\ \bibnamefont {Wolter}}, \bibinfo {author} {\bibfnamefont {H.-J.}\
  \bibnamefont {Grafe}}, \bibinfo {author} {\bibfnamefont {B.}~\bibnamefont
  {B\"uchner}}, and\ \bibinfo {author} {\bibfnamefont {C.}~\bibnamefont
  {Hess}}} (\bibinfo {year} {2020}{\natexlab{b}}),\ \href
  {https://doi.org/10.1103/PhysRevLett.125.067001} {\bibfield  {journal}
  {\bibinfo  {journal} {Phys. Rev. Lett.}\ }\textbf {\bibinfo {volume} {125}},\
  \bibinfo {pages} {067001}}\BibitemShut {NoStop}%
\bibitem [{\citenamefont {Horigane}\ \emph {et~al.}(2016)\citenamefont
  {Horigane}, \citenamefont {Kihou}, \citenamefont {Fujita}, \citenamefont
  {Kajimoto}, \citenamefont {Ikeuchi}, \citenamefont {Ji}, \citenamefont
  {Akimitsu},\ and\ \citenamefont {Lee}}]{horigane2016spin}%
  \BibitemOpen
  \bibfield  {author} {\bibinfo {author} {\bibnamefont {Horigane},
  \bibfnamefont {K.}}, \bibinfo {author} {\bibfnamefont {K.}~\bibnamefont
  {Kihou}}, \bibinfo {author} {\bibfnamefont {K.}~\bibnamefont {Fujita}},
  \bibinfo {author} {\bibfnamefont {R.}~\bibnamefont {Kajimoto}}, \bibinfo
  {author} {\bibfnamefont {K.}~\bibnamefont {Ikeuchi}}, \bibinfo {author}
  {\bibfnamefont {S.}~\bibnamefont {Ji}}, \bibinfo {author} {\bibfnamefont
  {J.}~\bibnamefont {Akimitsu}}, and\ \bibinfo {author} {\bibfnamefont {C.~H.}\
  \bibnamefont {Lee}}} (\bibinfo {year} {2016}),\ \href
  {https://doi.org/10.1038/srep33303} {\bibfield  {journal} {\bibinfo
  {journal} {Sci. Rep.}\ }\textbf {\bibinfo {volume} {6}},\ \bibinfo {pages}
  {33303}}\BibitemShut {NoStop}%
\bibitem [{\citenamefont {Hosoi}\ \emph {et~al.}(2016)\citenamefont {Hosoi},
  \citenamefont {Matsuura}, \citenamefont {Ishida}, \citenamefont {Wang},
  \citenamefont {Mizukami}, \citenamefont {Watashige}, \citenamefont
  {Kasahara}, \citenamefont {Matsuda},\ and\ \citenamefont
  {Shibauchi}}]{hosoi2016nematic}%
  \BibitemOpen
  \bibfield  {author} {\bibinfo {author} {\bibnamefont {Hosoi}, \bibfnamefont
  {S.}}, \bibinfo {author} {\bibfnamefont {K.}~\bibnamefont {Matsuura}},
  \bibinfo {author} {\bibfnamefont {K.}~\bibnamefont {Ishida}}, \bibinfo
  {author} {\bibfnamefont {H.}~\bibnamefont {Wang}}, \bibinfo {author}
  {\bibfnamefont {Y.}~\bibnamefont {Mizukami}}, \bibinfo {author}
  {\bibfnamefont {T.}~\bibnamefont {Watashige}}, \bibinfo {author}
  {\bibfnamefont {S.}~\bibnamefont {Kasahara}}, \bibinfo {author}
  {\bibfnamefont {Y.}~\bibnamefont {Matsuda}}, and\ \bibinfo {author}
  {\bibfnamefont {T.}~\bibnamefont {Shibauchi}}} (\bibinfo {year} {2016}),\
  \href {https://doi.org/10.1073/pnas.1605806113} {\bibfield  {journal}
  {\bibinfo  {journal} {Proc. Natl. Acad. Sci. U.S.A.}\ }\textbf {\bibinfo
  {volume} {113}},\ \bibinfo {pages} {8139}}\BibitemShut {NoStop}%
\bibitem [{\citenamefont {Hosono}\ and\ \citenamefont
  {Kuroki}(2015)}]{Hosono2015}%
  \BibitemOpen
  \bibfield  {author} {\bibinfo {author} {\bibnamefont {Hosono}, \bibfnamefont
  {H.}}, and\ \bibinfo {author} {\bibfnamefont {K.}~\bibnamefont {Kuroki}}}
  (\bibinfo {year} {2015}),\ \href
  {https://doi.org/10.1016/j.physc.2015.02.020} {\bibfield  {journal} {\bibinfo
   {journal} {Physica C: Superconductivity and its Applications}\ }\textbf
  {\bibinfo {volume} {514}},\ \bibinfo {pages} {399}}\BibitemShut {NoStop}%
\bibitem [{\citenamefont {Hosono}\ \emph {et~al.}(2018)\citenamefont {Hosono},
  \citenamefont {Yamamoto}, \citenamefont {Hiramatsu},\ and\ \citenamefont
  {Ma}}]{Hosono2018}%
  \BibitemOpen
  \bibfield  {author} {\bibinfo {author} {\bibnamefont {Hosono}, \bibfnamefont
  {H.}}, \bibinfo {author} {\bibfnamefont {A.}~\bibnamefont {Yamamoto}},
  \bibinfo {author} {\bibfnamefont {H.}~\bibnamefont {Hiramatsu}}, and\
  \bibinfo {author} {\bibfnamefont {Y.}~\bibnamefont {Ma}}} (\bibinfo {year}
  {2018}),\ \href {https://doi.org/10.1016/j.mattod.2017.09.006} {\bibfield
  {journal} {\bibinfo  {journal} {Materials today}\ }\textbf {\bibinfo {volume}
  {21}},\ \bibinfo {pages} {278}}\BibitemShut {NoStop}%
\bibitem [{\citenamefont {Hosono}\ \emph {et~al.}(2016)\citenamefont {Hosono},
  \citenamefont {Noji}, \citenamefont {Hatakeda}, \citenamefont {Kawamata},
  \citenamefont {Kato},\ and\ \citenamefont {Koike}}]{Hosono2016}%
  \BibitemOpen
  \bibfield  {author} {\bibinfo {author} {\bibnamefont {Hosono}, \bibfnamefont
  {S.}}, \bibinfo {author} {\bibfnamefont {T.}~\bibnamefont {Noji}}, \bibinfo
  {author} {\bibfnamefont {T.}~\bibnamefont {Hatakeda}}, \bibinfo {author}
  {\bibfnamefont {T.}~\bibnamefont {Kawamata}}, \bibinfo {author}
  {\bibfnamefont {M.}~\bibnamefont {Kato}}, and\ \bibinfo {author}
  {\bibfnamefont {Y.}~\bibnamefont {Koike}}} (\bibinfo {year} {2016}),\ \href
  {https://doi.org/10.7566/jpsj.85.013702} {\bibfield  {journal} {\bibinfo
  {journal} {J. Phys. Soc. Jpn.}\ }\textbf {\bibinfo {volume} {85}},\ \bibinfo
  {pages} {013702}}\BibitemShut {NoStop}%
\bibitem [{\citenamefont {Hsu}\ \emph {et~al.}(2008)\citenamefont {Hsu},
  \citenamefont {Luo}, \citenamefont {Yeh}, \citenamefont {Chen}, \citenamefont
  {Huang}, \citenamefont {Wu}, \citenamefont {Lee}, \citenamefont {Huang},
  \citenamefont {Chu}, \citenamefont {Yan} \emph {et~al.}}]{Hsu2008}%
  \BibitemOpen
  \bibfield  {author} {\bibinfo {author} {\bibnamefont {Hsu}, \bibfnamefont
  {F.-C.}}, \bibinfo {author} {\bibfnamefont {J.-Y.}\ \bibnamefont {Luo}},
  \bibinfo {author} {\bibfnamefont {K.-W.}\ \bibnamefont {Yeh}}, \bibinfo
  {author} {\bibfnamefont {T.-K.}\ \bibnamefont {Chen}}, \bibinfo {author}
  {\bibfnamefont {T.-W.}\ \bibnamefont {Huang}}, \bibinfo {author}
  {\bibfnamefont {P.~M.}\ \bibnamefont {Wu}}, \bibinfo {author} {\bibfnamefont
  {Y.-C.}\ \bibnamefont {Lee}}, \bibinfo {author} {\bibfnamefont {Y.-L.}\
  \bibnamefont {Huang}}, \bibinfo {author} {\bibfnamefont {Y.-Y.}\ \bibnamefont
  {Chu}}, \bibinfo {author} {\bibfnamefont {D.-C.}\ \bibnamefont {Yan}},  \emph
  {et~al.}} (\bibinfo {year} {2008}),\ \href
  {https://doi.org/10.1073/pnas.0807325105} {\bibfield  {journal} {\bibinfo
  {journal} {Proc. Natl. Acad. Sci. U.S.A.}\ }\textbf {\bibinfo {volume}
  {105}},\ \bibinfo {pages} {14262}}\BibitemShut {NoStop}%
\bibitem [{\citenamefont {Hu}\ \emph {et~al.}(2021)\citenamefont {Hu},
  \citenamefont {Feng}, \citenamefont {Park}, \citenamefont {Wo}, \citenamefont
  {Wang}, \citenamefont {Bourdarot}, \citenamefont {Ivanov},\ and\
  \citenamefont {Zhao}}]{hu2021polarized}%
  \BibitemOpen
  \bibfield  {author} {\bibinfo {author} {\bibnamefont {Hu}, \bibfnamefont
  {D.}}, \bibinfo {author} {\bibfnamefont {Y.}~\bibnamefont {Feng}}, \bibinfo
  {author} {\bibfnamefont {J.~T.}\ \bibnamefont {Park}}, \bibinfo {author}
  {\bibfnamefont {H.}~\bibnamefont {Wo}}, \bibinfo {author} {\bibfnamefont
  {Q.}~\bibnamefont {Wang}}, \bibinfo {author} {\bibfnamefont {F.}~\bibnamefont
  {Bourdarot}}, \bibinfo {author} {\bibfnamefont {A.}~\bibnamefont {Ivanov}},
  and\ \bibinfo {author} {\bibfnamefont {J.}~\bibnamefont {Zhao}}} (\bibinfo
  {year} {2021}),\ \href {https://doi.org/10.1088/1361-648x/ac1d16} {\bibfield
  {journal} {\bibinfo  {journal} {J. Phys.: Condens. Matter}\ }\textbf
  {\bibinfo {volume} {33}},\ \bibinfo {pages} {45LT01}}\BibitemShut {NoStop}%
\bibitem [{\citenamefont {Hu}\ \emph {et~al.}(2015)\citenamefont {Hu},
  \citenamefont {Lu}, \citenamefont {Zhang}, \citenamefont {Luo}, \citenamefont
  {Li}, \citenamefont {Wang}, \citenamefont {Chen}, \citenamefont {Han},
  \citenamefont {Banjara}, \citenamefont {Sapkota}, \citenamefont {Kreyssig},
  \citenamefont {Goldman}, \citenamefont {Yamani}, \citenamefont {Niedermayer},
  \citenamefont {Skoulatos}, \citenamefont {Georgii}, \citenamefont {Keller},
  \citenamefont {Wang}, \citenamefont {Yu},\ and\ \citenamefont
  {Dai}}]{hu2015structural}%
  \BibitemOpen
  \bibfield  {author} {\bibinfo {author} {\bibnamefont {Hu}, \bibfnamefont
  {D.}}, \bibinfo {author} {\bibfnamefont {X.}~\bibnamefont {Lu}}, \bibinfo
  {author} {\bibfnamefont {W.}~\bibnamefont {Zhang}}, \bibinfo {author}
  {\bibfnamefont {H.}~\bibnamefont {Luo}}, \bibinfo {author} {\bibfnamefont
  {S.}~\bibnamefont {Li}}, \bibinfo {author} {\bibfnamefont {P.}~\bibnamefont
  {Wang}}, \bibinfo {author} {\bibfnamefont {G.}~\bibnamefont {Chen}}, \bibinfo
  {author} {\bibfnamefont {F.}~\bibnamefont {Han}}, \bibinfo {author}
  {\bibfnamefont {S.~R.}\ \bibnamefont {Banjara}}, \bibinfo {author}
  {\bibfnamefont {A.}~\bibnamefont {Sapkota}}, \bibinfo {author} {\bibfnamefont
  {A.}~\bibnamefont {Kreyssig}}, \bibinfo {author} {\bibfnamefont {A.~I.}\
  \bibnamefont {Goldman}}, \bibinfo {author} {\bibfnamefont {Z.}~\bibnamefont
  {Yamani}}, \bibinfo {author} {\bibfnamefont {C.}~\bibnamefont {Niedermayer}},
  \bibinfo {author} {\bibfnamefont {M.}~\bibnamefont {Skoulatos}}, \bibinfo
  {author} {\bibfnamefont {R.}~\bibnamefont {Georgii}}, \bibinfo {author}
  {\bibfnamefont {T.}~\bibnamefont {Keller}}, \bibinfo {author} {\bibfnamefont
  {P.}~\bibnamefont {Wang}}, \bibinfo {author} {\bibfnamefont {W.}~\bibnamefont
  {Yu}}, and\ \bibinfo {author} {\bibfnamefont {P.}~\bibnamefont {Dai}}}
  (\bibinfo {year} {2015}),\ \href
  {https://doi.org/10.1103/PhysRevLett.114.157002} {\bibfield  {journal}
  {\bibinfo  {journal} {Phys. Rev. Lett.}\ }\textbf {\bibinfo {volume} {114}},\
  \bibinfo {pages} {157002}}\BibitemShut {NoStop}%
\bibitem [{\citenamefont {Hu}\ \emph {et~al.}(2020{\natexlab{a}})\citenamefont
  {Hu}, \citenamefont {Tam}, \citenamefont {Zhang}, \citenamefont {Wei},
  \citenamefont {Georgii}, \citenamefont {Pedersen}, \citenamefont {Roldan},\
  and\ \citenamefont {Dai}}]{hu2020uniaxial}%
  \BibitemOpen
  \bibfield  {author} {\bibinfo {author} {\bibnamefont {Hu}, \bibfnamefont
  {D.}}, \bibinfo {author} {\bibfnamefont {D.~W.}\ \bibnamefont {Tam}},
  \bibinfo {author} {\bibfnamefont {W.}~\bibnamefont {Zhang}}, \bibinfo
  {author} {\bibfnamefont {Y.}~\bibnamefont {Wei}}, \bibinfo {author}
  {\bibfnamefont {R.}~\bibnamefont {Georgii}}, \bibinfo {author} {\bibfnamefont
  {B.}~\bibnamefont {Pedersen}}, \bibinfo {author} {\bibfnamefont {A.~C.}\
  \bibnamefont {Roldan}}, and\ \bibinfo {author} {\bibfnamefont
  {P.}~\bibnamefont {Dai}}} (\bibinfo {year} {2020}{\natexlab{a}}),\ \href
  {https://doi.org/10.1103/PhysRevB.101.020507} {\bibfield  {journal} {\bibinfo
   {journal} {Phys. Rev. B}\ }\textbf {\bibinfo {volume} {101}},\ \bibinfo
  {pages} {020507}}\BibitemShut {NoStop}%
\bibitem [{\citenamefont {Hu}\ \emph {et~al.}(2022)\citenamefont {Hu},
  \citenamefont {Wang}, \citenamefont {Wo}, \citenamefont {Schneidewind},\ and\
  \citenamefont {Zhao}}]{hu2022lowenergy}%
  \BibitemOpen
  \bibfield  {author} {\bibinfo {author} {\bibnamefont {Hu}, \bibfnamefont
  {D.}}, \bibinfo {author} {\bibfnamefont {Q.}~\bibnamefont {Wang}}, \bibinfo
  {author} {\bibfnamefont {H.}~\bibnamefont {Wo}}, \bibinfo {author}
  {\bibfnamefont {A.}~\bibnamefont {Schneidewind}}, and\ \bibinfo {author}
  {\bibfnamefont {J.}~\bibnamefont {Zhao}}} (\bibinfo {year} {2022}),\ \href
  {https://doi.org/10.1103/PhysRevB.106.214522} {\bibfield  {journal} {\bibinfo
   {journal} {Phys. Rev. B}\ }\textbf {\bibinfo {volume} {106}},\ \bibinfo
  {pages} {214522}}\BibitemShut {NoStop}%
\bibitem [{\citenamefont {Hu}\ \emph {et~al.}(2018{\natexlab{a}})\citenamefont
  {Hu}, \citenamefont {Wang}, \citenamefont {Zhang}, \citenamefont {Wei},
  \citenamefont {Gong}, \citenamefont {Tam}, \citenamefont {Zhou},
  \citenamefont {Li}, \citenamefont {Tan}, \citenamefont {Song}, \citenamefont
  {Georgii}, \citenamefont {Pedersen}, \citenamefont {Cao}, \citenamefont
  {Tian}, \citenamefont {Roessli}, \citenamefont {Yin},\ and\ \citenamefont
  {Dai}}]{hu2018caxis}%
  \BibitemOpen
  \bibfield  {author} {\bibinfo {author} {\bibnamefont {Hu}, \bibfnamefont
  {D.}}, \bibinfo {author} {\bibfnamefont {W.}~\bibnamefont {Wang}}, \bibinfo
  {author} {\bibfnamefont {W.}~\bibnamefont {Zhang}}, \bibinfo {author}
  {\bibfnamefont {Y.}~\bibnamefont {Wei}}, \bibinfo {author} {\bibfnamefont
  {D.}~\bibnamefont {Gong}}, \bibinfo {author} {\bibfnamefont {D.~W.}\
  \bibnamefont {Tam}}, \bibinfo {author} {\bibfnamefont {P.}~\bibnamefont
  {Zhou}}, \bibinfo {author} {\bibfnamefont {Y.}~\bibnamefont {Li}}, \bibinfo
  {author} {\bibfnamefont {G.}~\bibnamefont {Tan}}, \bibinfo {author}
  {\bibfnamefont {Y.}~\bibnamefont {Song}}, \bibinfo {author} {\bibfnamefont
  {R.}~\bibnamefont {Georgii}}, \bibinfo {author} {\bibfnamefont
  {B.}~\bibnamefont {Pedersen}}, \bibinfo {author} {\bibfnamefont
  {H.}~\bibnamefont {Cao}}, \bibinfo {author} {\bibfnamefont {W.}~\bibnamefont
  {Tian}}, \bibinfo {author} {\bibfnamefont {B.}~\bibnamefont {Roessli}},
  \bibinfo {author} {\bibfnamefont {Z.}~\bibnamefont {Yin}}, and\ \bibinfo
  {author} {\bibfnamefont {P.}~\bibnamefont {Dai}}} (\bibinfo {year}
  {2018}{\natexlab{a}}),\ \href {https://doi.org/10.1038/s41535-018-0122-3}
  {\bibfield  {journal} {\bibinfo  {journal} {npj Quantum Mater.}\ }\textbf
  {\bibinfo {volume} {3}},\ \bibinfo {pages} {47}}\BibitemShut {NoStop}%
\bibitem [{\citenamefont {Hu}\ \emph {et~al.}(2016)\citenamefont {Hu},
  \citenamefont {Yin}, \citenamefont {Zhang}, \citenamefont {Ewings},
  \citenamefont {Ikeuchi}, \citenamefont {Nakamura}, \citenamefont {Roessli},
  \citenamefont {Wei}, \citenamefont {Zhao}, \citenamefont {Chen},
  \citenamefont {Li}, \citenamefont {Luo}, \citenamefont {Haule}, \citenamefont
  {Kotliar},\ and\ \citenamefont {Dai}}]{hu2016spin}%
  \BibitemOpen
  \bibfield  {author} {\bibinfo {author} {\bibnamefont {Hu}, \bibfnamefont
  {D.}}, \bibinfo {author} {\bibfnamefont {Z.}~\bibnamefont {Yin}}, \bibinfo
  {author} {\bibfnamefont {W.}~\bibnamefont {Zhang}}, \bibinfo {author}
  {\bibfnamefont {R.~A.}\ \bibnamefont {Ewings}}, \bibinfo {author}
  {\bibfnamefont {K.}~\bibnamefont {Ikeuchi}}, \bibinfo {author} {\bibfnamefont
  {M.}~\bibnamefont {Nakamura}}, \bibinfo {author} {\bibfnamefont
  {B.}~\bibnamefont {Roessli}}, \bibinfo {author} {\bibfnamefont
  {Y.}~\bibnamefont {Wei}}, \bibinfo {author} {\bibfnamefont {L.}~\bibnamefont
  {Zhao}}, \bibinfo {author} {\bibfnamefont {G.}~\bibnamefont {Chen}}, \bibinfo
  {author} {\bibfnamefont {S.}~\bibnamefont {Li}}, \bibinfo {author}
  {\bibfnamefont {H.}~\bibnamefont {Luo}}, \bibinfo {author} {\bibfnamefont
  {K.}~\bibnamefont {Haule}}, \bibinfo {author} {\bibfnamefont
  {G.}~\bibnamefont {Kotliar}}, and\ \bibinfo {author} {\bibfnamefont
  {P.}~\bibnamefont {Dai}}} (\bibinfo {year} {2016}),\ \href
  {https://doi.org/10.1103/PhysRevB.94.094504} {\bibfield  {journal} {\bibinfo
  {journal} {Phys. Rev. B}\ }\textbf {\bibinfo {volume} {94}},\ \bibinfo
  {pages} {094504}}\BibitemShut {NoStop}%
\bibitem [{\citenamefont {Hu}\ \emph {et~al.}(2018{\natexlab{b}})\citenamefont
  {Hu}, \citenamefont {Yu}, \citenamefont {Nica}, \citenamefont {Zhu},\ and\
  \citenamefont {Si}}]{hu2018orbital}%
  \BibitemOpen
  \bibfield  {author} {\bibinfo {author} {\bibnamefont {Hu}, \bibfnamefont
  {H.}}, \bibinfo {author} {\bibfnamefont {R.}~\bibnamefont {Yu}}, \bibinfo
  {author} {\bibfnamefont {E.~M.}\ \bibnamefont {Nica}}, \bibinfo {author}
  {\bibfnamefont {J.-X.}\ \bibnamefont {Zhu}}, and\ \bibinfo {author}
  {\bibfnamefont {Q.}~\bibnamefont {Si}}} (\bibinfo {year}
  {2018}{\natexlab{b}}),\ \href {https://doi.org/10.1103/PhysRevB.98.220503}
  {\bibfield  {journal} {\bibinfo  {journal} {Phys. Rev. B}\ }\textbf {\bibinfo
  {volume} {98}},\ \bibinfo {pages} {220503}}\BibitemShut {NoStop}%
\bibitem [{\citenamefont {Hu}\ \emph {et~al.}(2011)\citenamefont {Hu},
  \citenamefont {Zuo}, \citenamefont {Wen}, \citenamefont {Xu}, \citenamefont
  {Lin}, \citenamefont {Li}, \citenamefont {Gu}, \citenamefont {Park},\ and\
  \citenamefont {Greene}}]{hu2011phase}%
  \BibitemOpen
  \bibfield  {author} {\bibinfo {author} {\bibnamefont {Hu}, \bibfnamefont
  {H.}}, \bibinfo {author} {\bibfnamefont {J.-M.}\ \bibnamefont {Zuo}},
  \bibinfo {author} {\bibfnamefont {J.}~\bibnamefont {Wen}}, \bibinfo {author}
  {\bibfnamefont {Z.}~\bibnamefont {Xu}}, \bibinfo {author} {\bibfnamefont
  {Z.}~\bibnamefont {Lin}}, \bibinfo {author} {\bibfnamefont {Q.}~\bibnamefont
  {Li}}, \bibinfo {author} {\bibfnamefont {G.}~\bibnamefont {Gu}}, \bibinfo
  {author} {\bibfnamefont {W.~K.}\ \bibnamefont {Park}}, and\ \bibinfo {author}
  {\bibfnamefont {L.~H.}\ \bibnamefont {Greene}}} (\bibinfo {year} {2011}),\
  \href {https://doi.org/10.1088/1367-2630/13/5/053031} {\bibfield  {journal}
  {\bibinfo  {journal} {New J. Phys.}\ }\textbf {\bibinfo {volume} {13}},\
  \bibinfo {pages} {053031}}\BibitemShut {NoStop}%
\bibitem [{\citenamefont {Hu}(2013)}]{hu2013ironbased}%
  \BibitemOpen
  \bibfield  {author} {\bibinfo {author} {\bibnamefont {Hu}, \bibfnamefont
  {J.}}} (\bibinfo {year} {2013}),\ \href
  {https://doi.org/10.1103/PhysRevX.3.031004} {\bibfield  {journal} {\bibinfo
  {journal} {Phys. Rev. X}\ }\textbf {\bibinfo {volume} {3}},\ \bibinfo {pages}
  {031004}}\BibitemShut {NoStop}%
\bibitem [{\citenamefont {Hu}\ \emph {et~al.}(2020{\natexlab{b}})\citenamefont
  {Hu}, \citenamefont {Lai}, \citenamefont {Gong}, \citenamefont {Yu},
  \citenamefont {Dagotto},\ and\ \citenamefont {Si}}]{hu2020quantum}%
  \BibitemOpen
  \bibfield  {author} {\bibinfo {author} {\bibnamefont {Hu}, \bibfnamefont
  {W.-J.}}, \bibinfo {author} {\bibfnamefont {H.-H.}\ \bibnamefont {Lai}},
  \bibinfo {author} {\bibfnamefont {S.-S.}\ \bibnamefont {Gong}}, \bibinfo
  {author} {\bibfnamefont {R.}~\bibnamefont {Yu}}, \bibinfo {author}
  {\bibfnamefont {E.}~\bibnamefont {Dagotto}}, and\ \bibinfo {author}
  {\bibfnamefont {Q.}~\bibnamefont {Si}}} (\bibinfo {year}
  {2020}{\natexlab{b}}),\ \href
  {https://doi.org/10.1103/physrevresearch.2.023359} {\bibfield  {journal}
  {\bibinfo  {journal} {Phys. Rev. Res.}\ }\textbf {\bibinfo {volume} {2}},\
  \bibinfo {pages} {023359}}\BibitemShut {NoStop}%
\bibitem [{\citenamefont {Huang}\ and\ \citenamefont
  {Hoffman}(2017)}]{Huang2017}%
  \BibitemOpen
  \bibfield  {author} {\bibinfo {author} {\bibnamefont {Huang}, \bibfnamefont
  {D.}}, and\ \bibinfo {author} {\bibfnamefont {J.~E.}\ \bibnamefont
  {Hoffman}}} (\bibinfo {year} {2017}),\ \href
  {https://doi.org/10.1146/annurev-conmatphys-031016-025242} {\bibfield
  {journal} {\bibinfo  {journal} {Annu. Rev. Condens. Matter Phys.}\ }\textbf
  {\bibinfo {volume} {8}},\ \bibinfo {pages} {311}}\BibitemShut {NoStop}%
\bibitem [{\citenamefont {Huang}\ \emph {et~al.}(2022)\citenamefont {Huang},
  \citenamefont {Yu}, \citenamefont {Xu}, \citenamefont {Zhu}, \citenamefont
  {Oh}, \citenamefont {Jiang}, \citenamefont {Wang}, \citenamefont {Wu},
  \citenamefont {Chen}, \citenamefont {Denlinger}, \citenamefont {Mo},
  \citenamefont {Hashimoto}, \citenamefont {Michiardi}, \citenamefont
  {Pedersen}, \citenamefont {Gorovikov}, \citenamefont {Zhdanovich},
  \citenamefont {Damascelli}, \citenamefont {Gu}, \citenamefont {Dai},
  \citenamefont {Chu}, \citenamefont {Lu}, \citenamefont {Si}, \citenamefont
  {Birgeneau},\ and\ \citenamefont {Yi}}]{huang2022correlationdriven}%
  \BibitemOpen
  \bibfield  {author} {\bibinfo {author} {\bibnamefont {Huang}, \bibfnamefont
  {J.}}, \bibinfo {author} {\bibfnamefont {R.}~\bibnamefont {Yu}}, \bibinfo
  {author} {\bibfnamefont {Z.}~\bibnamefont {Xu}}, \bibinfo {author}
  {\bibfnamefont {J.-X.}\ \bibnamefont {Zhu}}, \bibinfo {author} {\bibfnamefont
  {J.~S.}\ \bibnamefont {Oh}}, \bibinfo {author} {\bibfnamefont
  {Q.}~\bibnamefont {Jiang}}, \bibinfo {author} {\bibfnamefont
  {M.}~\bibnamefont {Wang}}, \bibinfo {author} {\bibfnamefont {H.}~\bibnamefont
  {Wu}}, \bibinfo {author} {\bibfnamefont {T.}~\bibnamefont {Chen}}, \bibinfo
  {author} {\bibfnamefont {J.~D.}\ \bibnamefont {Denlinger}}, \bibinfo {author}
  {\bibfnamefont {S.-K.}\ \bibnamefont {Mo}}, \bibinfo {author} {\bibfnamefont
  {M.}~\bibnamefont {Hashimoto}}, \bibinfo {author} {\bibfnamefont
  {M.}~\bibnamefont {Michiardi}}, \bibinfo {author} {\bibfnamefont {T.~M.}\
  \bibnamefont {Pedersen}}, \bibinfo {author} {\bibfnamefont {S.}~\bibnamefont
  {Gorovikov}}, \bibinfo {author} {\bibfnamefont {S.}~\bibnamefont
  {Zhdanovich}}, \bibinfo {author} {\bibfnamefont {A.}~\bibnamefont
  {Damascelli}}, \bibinfo {author} {\bibfnamefont {G.}~\bibnamefont {Gu}},
  \bibinfo {author} {\bibfnamefont {P.}~\bibnamefont {Dai}}, \bibinfo {author}
  {\bibfnamefont {J.-H.}\ \bibnamefont {Chu}}, \bibinfo {author} {\bibfnamefont
  {D.}~\bibnamefont {Lu}}, \bibinfo {author} {\bibfnamefont {Q.}~\bibnamefont
  {Si}}, \bibinfo {author} {\bibfnamefont {R.~J.}\ \bibnamefont {Birgeneau}},
  and\ \bibinfo {author} {\bibfnamefont {M.}~\bibnamefont {Yi}}} (\bibinfo
  {year} {2022}),\ \href {https://doi.org/10.1038/s42005-022-00805-6}
  {\bibfield  {journal} {\bibinfo  {journal} {Commun. Phys.}\ }\textbf
  {\bibinfo {volume} {5}},\ \bibinfo {pages} {29}}\BibitemShut {NoStop}%
\bibitem [{\citenamefont {Iida}(2025)}]{Iida2025}%
  \BibitemOpen
  \bibfield  {author} {\bibinfo {author} {\bibnamefont {Iida}, \bibfnamefont
  {K.}}} (\bibinfo {year} {2025}),\ \href
  {https://doi.org/10.1109/tasc.2025.3530910} {\bibfield  {journal} {\bibinfo
  {journal} {IEEE Transactions on Applied Superconductivity}\ }\textbf
  {\bibinfo {volume} {35}},\ \bibinfo {pages} {7400109}}\BibitemShut {NoStop}%
\bibitem [{\citenamefont {Iida}\ \emph {et~al.}(2017)\citenamefont {Iida},
  \citenamefont {Ishikado}, \citenamefont {Nagai}, \citenamefont {Yoshida},
  \citenamefont {Christianson}, \citenamefont {Murai}, \citenamefont
  {Kawashima}, \citenamefont {Yoshida}, \citenamefont {Eisaki},\ and\
  \citenamefont {Iyo}}]{iida2017spin}%
  \BibitemOpen
  \bibfield  {author} {\bibinfo {author} {\bibnamefont {Iida}, \bibfnamefont
  {K.}}, \bibinfo {author} {\bibfnamefont {M.}~\bibnamefont {Ishikado}},
  \bibinfo {author} {\bibfnamefont {Y.}~\bibnamefont {Nagai}}, \bibinfo
  {author} {\bibfnamefont {H.}~\bibnamefont {Yoshida}}, \bibinfo {author}
  {\bibfnamefont {A.~D.}\ \bibnamefont {Christianson}}, \bibinfo {author}
  {\bibfnamefont {N.}~\bibnamefont {Murai}}, \bibinfo {author} {\bibfnamefont
  {K.}~\bibnamefont {Kawashima}}, \bibinfo {author} {\bibfnamefont
  {Y.}~\bibnamefont {Yoshida}}, \bibinfo {author} {\bibfnamefont
  {H.}~\bibnamefont {Eisaki}}, and\ \bibinfo {author} {\bibfnamefont
  {A.}~\bibnamefont {Iyo}}} (\bibinfo {year} {2017}),\ \href
  {https://doi.org/10.7566/jpsj.86.093703} {\bibfield  {journal} {\bibinfo
  {journal} {J. Phys. Soc. Jpn.}\ }\textbf {\bibinfo {volume} {86}},\ \bibinfo
  {pages} {093703}}\BibitemShut {NoStop}%
\bibitem [{\citenamefont {Iida}\ \emph {et~al.}(2019)\citenamefont {Iida},
  \citenamefont {Nagai}, \citenamefont {Ishida}, \citenamefont {Ishikado},
  \citenamefont {Murai}, \citenamefont {Christianson}, \citenamefont {Yoshida},
  \citenamefont {Inamura}, \citenamefont {Nakamura}, \citenamefont {Nakao},
  \citenamefont {Munakata}, \citenamefont {Kagerbauer}, \citenamefont
  {Eisterer}, \citenamefont {Kawashima}, \citenamefont {Yoshida}, \citenamefont
  {Eisaki},\ and\ \citenamefont {Iyo}}]{iida2019coexisting}%
  \BibitemOpen
  \bibfield  {author} {\bibinfo {author} {\bibnamefont {Iida}, \bibfnamefont
  {K.}}, \bibinfo {author} {\bibfnamefont {Y.}~\bibnamefont {Nagai}}, \bibinfo
  {author} {\bibfnamefont {S.}~\bibnamefont {Ishida}}, \bibinfo {author}
  {\bibfnamefont {M.}~\bibnamefont {Ishikado}}, \bibinfo {author}
  {\bibfnamefont {N.}~\bibnamefont {Murai}}, \bibinfo {author} {\bibfnamefont
  {A.~D.}\ \bibnamefont {Christianson}}, \bibinfo {author} {\bibfnamefont
  {H.}~\bibnamefont {Yoshida}}, \bibinfo {author} {\bibfnamefont
  {Y.}~\bibnamefont {Inamura}}, \bibinfo {author} {\bibfnamefont
  {H.}~\bibnamefont {Nakamura}}, \bibinfo {author} {\bibfnamefont
  {A.}~\bibnamefont {Nakao}}, \bibinfo {author} {\bibfnamefont
  {K.}~\bibnamefont {Munakata}}, \bibinfo {author} {\bibfnamefont
  {D.}~\bibnamefont {Kagerbauer}}, \bibinfo {author} {\bibfnamefont
  {M.}~\bibnamefont {Eisterer}}, \bibinfo {author} {\bibfnamefont
  {K.}~\bibnamefont {Kawashima}}, \bibinfo {author} {\bibfnamefont
  {Y.}~\bibnamefont {Yoshida}}, \bibinfo {author} {\bibfnamefont
  {H.}~\bibnamefont {Eisaki}}, and\ \bibinfo {author} {\bibfnamefont
  {A.}~\bibnamefont {Iyo}}} (\bibinfo {year} {2019}),\ \href
  {https://doi.org/10.1103/PhysRevB.100.014506} {\bibfield  {journal} {\bibinfo
   {journal} {Phys. Rev. B}\ }\textbf {\bibinfo {volume} {100}},\ \bibinfo
  {pages} {014506}}\BibitemShut {NoStop}%
\bibitem [{\citenamefont {Iimura}\ \emph {et~al.}(2016)\citenamefont {Iimura},
  \citenamefont {Matsuishi},\ and\ \citenamefont {Hosono}}]{Iimura2016}%
  \BibitemOpen
  \bibfield  {author} {\bibinfo {author} {\bibnamefont {Iimura}, \bibfnamefont
  {S.}}, \bibinfo {author} {\bibfnamefont {S.}~\bibnamefont {Matsuishi}}, and\
  \bibinfo {author} {\bibfnamefont {H.}~\bibnamefont {Hosono}}} (\bibinfo
  {year} {2016}),\ \href {https://doi.org/10.1103/PhysRevB.94.024512}
  {\bibfield  {journal} {\bibinfo  {journal} {Phys. Rev. B}\ }\textbf {\bibinfo
  {volume} {94}},\ \bibinfo {pages} {024512}}\BibitemShut {NoStop}%
\bibitem [{\citenamefont {Iimura}\ \emph {et~al.}(2012)\citenamefont {Iimura},
  \citenamefont {Matsuishi}, \citenamefont {Sato}, \citenamefont {Hanna},
  \citenamefont {Muraba}, \citenamefont {Kim}, \citenamefont {Kim},
  \citenamefont {Takata},\ and\ \citenamefont {Hosono}}]{Iimura2012}%
  \BibitemOpen
  \bibfield  {author} {\bibinfo {author} {\bibnamefont {Iimura}, \bibfnamefont
  {S.}}, \bibinfo {author} {\bibfnamefont {S.}~\bibnamefont {Matsuishi}},
  \bibinfo {author} {\bibfnamefont {H.}~\bibnamefont {Sato}}, \bibinfo {author}
  {\bibfnamefont {T.}~\bibnamefont {Hanna}}, \bibinfo {author} {\bibfnamefont
  {Y.}~\bibnamefont {Muraba}}, \bibinfo {author} {\bibfnamefont {S.~W.}\
  \bibnamefont {Kim}}, \bibinfo {author} {\bibfnamefont {J.~E.}\ \bibnamefont
  {Kim}}, \bibinfo {author} {\bibfnamefont {M.}~\bibnamefont {Takata}}, and\
  \bibinfo {author} {\bibfnamefont {H.}~\bibnamefont {Hosono}}} (\bibinfo
  {year} {2012}),\ \href {https://doi.org/10.1038/ncomms1913} {\bibfield
  {journal} {\bibinfo  {journal} {Nat. Commun.}\ }\textbf {\bibinfo {volume}
  {3}},\ \bibinfo {pages} {943}}\BibitemShut {NoStop}%
\bibitem [{\citenamefont {Iimura}\ \emph {et~al.}(2017)\citenamefont {Iimura},
  \citenamefont {Okanishi}, \citenamefont {Matsuishi}, \citenamefont {Hiraka},
  \citenamefont {Honda}, \citenamefont {Ikeda}, \citenamefont {Hansen},
  \citenamefont {Otomo},\ and\ \citenamefont {Hosono}}]{Iimura2017}%
  \BibitemOpen
  \bibfield  {author} {\bibinfo {author} {\bibnamefont {Iimura}, \bibfnamefont
  {S.}}, \bibinfo {author} {\bibfnamefont {H.}~\bibnamefont {Okanishi}},
  \bibinfo {author} {\bibfnamefont {S.}~\bibnamefont {Matsuishi}}, \bibinfo
  {author} {\bibfnamefont {H.}~\bibnamefont {Hiraka}}, \bibinfo {author}
  {\bibfnamefont {T.}~\bibnamefont {Honda}}, \bibinfo {author} {\bibfnamefont
  {K.}~\bibnamefont {Ikeda}}, \bibinfo {author} {\bibfnamefont {T.~C.}\
  \bibnamefont {Hansen}}, \bibinfo {author} {\bibfnamefont {T.}~\bibnamefont
  {Otomo}}, and\ \bibinfo {author} {\bibfnamefont {H.}~\bibnamefont {Hosono}}}
  (\bibinfo {year} {2017}),\ \href {https://doi.org/10.1073/pnas.1703295114}
  {\bibfield  {journal} {\bibinfo  {journal} {Proc. Natl. Acad. Sci. U.S.A.}\
  }\textbf {\bibinfo {volume} {114}},\ \bibinfo {pages} {E4354}}\BibitemShut
  {NoStop}%
\bibitem [{\citenamefont {Ikeda}\ \emph {et~al.}(2019)\citenamefont {Ikeda},
  \citenamefont {Straquadine}, \citenamefont {Hristov}, \citenamefont
  {Worasaran}, \citenamefont {Palmstrom}, \citenamefont {Sorensen},
  \citenamefont {Walmsley},\ and\ \citenamefont {Fisher}}]{ikeda2019ac}%
  \BibitemOpen
  \bibfield  {author} {\bibinfo {author} {\bibnamefont {Ikeda}, \bibfnamefont
  {M.~S.}}, \bibinfo {author} {\bibfnamefont {J.~A.~W.}\ \bibnamefont
  {Straquadine}}, \bibinfo {author} {\bibfnamefont {A.~T.}\ \bibnamefont
  {Hristov}}, \bibinfo {author} {\bibfnamefont {T.}~\bibnamefont {Worasaran}},
  \bibinfo {author} {\bibfnamefont {J.~C.}\ \bibnamefont {Palmstrom}}, \bibinfo
  {author} {\bibfnamefont {M.}~\bibnamefont {Sorensen}}, \bibinfo {author}
  {\bibfnamefont {P.}~\bibnamefont {Walmsley}}, and\ \bibinfo {author}
  {\bibfnamefont {I.~R.}\ \bibnamefont {Fisher}}} (\bibinfo {year} {2019}),\
  \href {https://doi.org/10.1063/1.5099924} {\bibfield  {journal} {\bibinfo
  {journal} {Rev. Sci. Instrum.}\ }\textbf {\bibinfo {volume} {90}},\ \bibinfo
  {pages} {083902}}\BibitemShut {NoStop}%
\bibitem [{\citenamefont {Ikeda}\ \emph {et~al.}(2018)\citenamefont {Ikeda},
  \citenamefont {Worasaran}, \citenamefont {Palmstrom}, \citenamefont
  {Straquadine}, \citenamefont {Walmsley},\ and\ \citenamefont
  {Fisher}}]{ikeda2018symmetric}%
  \BibitemOpen
  \bibfield  {author} {\bibinfo {author} {\bibnamefont {Ikeda}, \bibfnamefont
  {M.~S.}}, \bibinfo {author} {\bibfnamefont {T.}~\bibnamefont {Worasaran}},
  \bibinfo {author} {\bibfnamefont {J.~C.}\ \bibnamefont {Palmstrom}}, \bibinfo
  {author} {\bibfnamefont {J.~A.~W.}\ \bibnamefont {Straquadine}}, \bibinfo
  {author} {\bibfnamefont {P.}~\bibnamefont {Walmsley}}, and\ \bibinfo {author}
  {\bibfnamefont {I.~R.}\ \bibnamefont {Fisher}}} (\bibinfo {year} {2018}),\
  \href {https://doi.org/10.1103/PhysRevB.98.245133} {\bibfield  {journal}
  {\bibinfo  {journal} {Phys. Rev. B}\ }\textbf {\bibinfo {volume} {98}},\
  \bibinfo {pages} {245133}}\BibitemShut {NoStop}%
\bibitem [{\citenamefont {Ikeda}\ \emph {et~al.}(2021)\citenamefont {Ikeda},
  \citenamefont {Worasaran}, \citenamefont {Rosenberg}, \citenamefont
  {Palmstrom}, \citenamefont {Kivelson},\ and\ \citenamefont
  {Fisher}}]{ikeda2021elastocaloric}%
  \BibitemOpen
  \bibfield  {author} {\bibinfo {author} {\bibnamefont {Ikeda}, \bibfnamefont
  {M.~S.}}, \bibinfo {author} {\bibfnamefont {T.}~\bibnamefont {Worasaran}},
  \bibinfo {author} {\bibfnamefont {E.~W.}\ \bibnamefont {Rosenberg}}, \bibinfo
  {author} {\bibfnamefont {J.~C.}\ \bibnamefont {Palmstrom}}, \bibinfo {author}
  {\bibfnamefont {S.~A.}\ \bibnamefont {Kivelson}}, and\ \bibinfo {author}
  {\bibfnamefont {I.~R.}\ \bibnamefont {Fisher}}} (\bibinfo {year} {2021}),\
  \href {https://doi.org/10.1073/pnas.2105911118} {\bibfield  {journal}
  {\bibinfo  {journal} {Proc. Natl. Acad. Sci. U.S.A.}\ }\textbf {\bibinfo
  {volume} {118}},\ \bibinfo {pages} {e2105911118}}\BibitemShut {NoStop}%
\bibitem [{\citenamefont {Ishida}\ \emph {et~al.}(2022)\citenamefont {Ishida},
  \citenamefont {Onishi}, \citenamefont {Tsujii}, \citenamefont {Mukasa},
  \citenamefont {Qiu}, \citenamefont {Saito}, \citenamefont {Sugimura},
  \citenamefont {Matsuura}, \citenamefont {Mizukami}, \citenamefont
  {Hashimoto},\ and\ \citenamefont {Shibauchi}}]{ishida2022pure}%
  \BibitemOpen
  \bibfield  {author} {\bibinfo {author} {\bibnamefont {Ishida}, \bibfnamefont
  {K.}}, \bibinfo {author} {\bibfnamefont {Y.}~\bibnamefont {Onishi}}, \bibinfo
  {author} {\bibfnamefont {M.}~\bibnamefont {Tsujii}}, \bibinfo {author}
  {\bibfnamefont {K.}~\bibnamefont {Mukasa}}, \bibinfo {author} {\bibfnamefont
  {M.}~\bibnamefont {Qiu}}, \bibinfo {author} {\bibfnamefont {M.}~\bibnamefont
  {Saito}}, \bibinfo {author} {\bibfnamefont {Y.}~\bibnamefont {Sugimura}},
  \bibinfo {author} {\bibfnamefont {K.}~\bibnamefont {Matsuura}}, \bibinfo
  {author} {\bibfnamefont {Y.}~\bibnamefont {Mizukami}}, \bibinfo {author}
  {\bibfnamefont {K.}~\bibnamefont {Hashimoto}}, and\ \bibinfo {author}
  {\bibfnamefont {T.}~\bibnamefont {Shibauchi}}} (\bibinfo {year} {2022}),\
  \href {https://doi.org/10.1073/pnas.2110501119} {\bibfield  {journal}
  {\bibinfo  {journal} {Proc. Natl. Acad. Sci. U.S.A.}\ }\textbf {\bibinfo
  {volume} {119}},\ \bibinfo {pages} {e2110501119}}\BibitemShut {NoStop}%
\bibitem [{\citenamefont {Ishida}\ \emph {et~al.}(2020)\citenamefont {Ishida},
  \citenamefont {Tsujii}, \citenamefont {Hosoi}, \citenamefont {Mizukami},
  \citenamefont {Ishida}, \citenamefont {Iyo}, \citenamefont {Eisaki},
  \citenamefont {Wolf}, \citenamefont {Grube}, \citenamefont {v.~Löhneysen},
  \citenamefont {Fernandes},\ and\ \citenamefont
  {Shibauchi}}]{ishida2020novel}%
  \BibitemOpen
  \bibfield  {author} {\bibinfo {author} {\bibnamefont {Ishida}, \bibfnamefont
  {K.}}, \bibinfo {author} {\bibfnamefont {M.}~\bibnamefont {Tsujii}}, \bibinfo
  {author} {\bibfnamefont {S.}~\bibnamefont {Hosoi}}, \bibinfo {author}
  {\bibfnamefont {Y.}~\bibnamefont {Mizukami}}, \bibinfo {author}
  {\bibfnamefont {S.}~\bibnamefont {Ishida}}, \bibinfo {author} {\bibfnamefont
  {A.}~\bibnamefont {Iyo}}, \bibinfo {author} {\bibfnamefont {H.}~\bibnamefont
  {Eisaki}}, \bibinfo {author} {\bibfnamefont {T.}~\bibnamefont {Wolf}},
  \bibinfo {author} {\bibfnamefont {K.}~\bibnamefont {Grube}}, \bibinfo
  {author} {\bibfnamefont {H.}~\bibnamefont {v.~Löhneysen}}, \bibinfo {author}
  {\bibfnamefont {R.~M.}\ \bibnamefont {Fernandes}}, and\ \bibinfo {author}
  {\bibfnamefont {T.}~\bibnamefont {Shibauchi}}} (\bibinfo {year} {2020}),\
  \href {https://doi.org/10.1073/pnas.1909172117} {\bibfield  {journal}
  {\bibinfo  {journal} {Proc. Natl. Acad. Sci. U.S.A.}\ }\textbf {\bibinfo
  {volume} {117}},\ \bibinfo {pages} {6424}}\BibitemShut {NoStop}%
\bibitem [{\citenamefont {Iyo}\ \emph {et~al.}(2018)\citenamefont {Iyo},
  \citenamefont {Kawashima}, \citenamefont {Ishida}, \citenamefont {Fujihisa},
  \citenamefont {Gotoh}, \citenamefont {Eisaki},\ and\ \citenamefont
  {Yoshida}}]{Iyo2018}%
  \BibitemOpen
  \bibfield  {author} {\bibinfo {author} {\bibnamefont {Iyo}, \bibfnamefont
  {A.}}, \bibinfo {author} {\bibfnamefont {K.}~\bibnamefont {Kawashima}},
  \bibinfo {author} {\bibfnamefont {S.}~\bibnamefont {Ishida}}, \bibinfo
  {author} {\bibfnamefont {H.}~\bibnamefont {Fujihisa}}, \bibinfo {author}
  {\bibfnamefont {Y.}~\bibnamefont {Gotoh}}, \bibinfo {author} {\bibfnamefont
  {H.}~\bibnamefont {Eisaki}}, and\ \bibinfo {author} {\bibfnamefont
  {Y.}~\bibnamefont {Yoshida}}} (\bibinfo {year} {2018}),\ \href
  {https://doi.org/10.1021/jacs.7b10656} {\bibfield  {journal} {\bibinfo
  {journal} {J. Am. Chem. Soc.}\ }\textbf {\bibinfo {volume} {140}},\ \bibinfo
  {pages} {369}}\BibitemShut {NoStop}%
\bibitem [{\citenamefont {Iyo}\ \emph {et~al.}(2016)\citenamefont {Iyo},
  \citenamefont {Kawashima}, \citenamefont {Kinjo}, \citenamefont {Nishio},
  \citenamefont {Ishida}, \citenamefont {Fujihisa}, \citenamefont {Gotoh},
  \citenamefont {Kihou}, \citenamefont {Eisaki},\ and\ \citenamefont
  {Yoshida}}]{Iyo2016}%
  \BibitemOpen
  \bibfield  {author} {\bibinfo {author} {\bibnamefont {Iyo}, \bibfnamefont
  {A.}}, \bibinfo {author} {\bibfnamefont {K.}~\bibnamefont {Kawashima}},
  \bibinfo {author} {\bibfnamefont {T.}~\bibnamefont {Kinjo}}, \bibinfo
  {author} {\bibfnamefont {T.}~\bibnamefont {Nishio}}, \bibinfo {author}
  {\bibfnamefont {S.}~\bibnamefont {Ishida}}, \bibinfo {author} {\bibfnamefont
  {H.}~\bibnamefont {Fujihisa}}, \bibinfo {author} {\bibfnamefont
  {Y.}~\bibnamefont {Gotoh}}, \bibinfo {author} {\bibfnamefont
  {K.}~\bibnamefont {Kihou}}, \bibinfo {author} {\bibfnamefont
  {H.}~\bibnamefont {Eisaki}}, and\ \bibinfo {author} {\bibfnamefont
  {Y.}~\bibnamefont {Yoshida}}} (\bibinfo {year} {2016}),\ \href
  {https://doi.org/10.1021/jacs.5b12571} {\bibfield  {journal} {\bibinfo
  {journal} {J. Am. Chem. Soc.}\ }\textbf {\bibinfo {volume} {138}},\ \bibinfo
  {pages} {3410}}\BibitemShut {NoStop}%
\bibitem [{\citenamefont {Jackson}\ \emph {et~al.}(2018)\citenamefont
  {Jackson}, \citenamefont {VanGennep}, \citenamefont {Bi}, \citenamefont
  {Zhang}, \citenamefont {Materne}, \citenamefont {Liu}, \citenamefont {Cao},
  \citenamefont {Weir}, \citenamefont {Vohra},\ and\ \citenamefont
  {Hamlin}}]{Jackson2018}%
  \BibitemOpen
  \bibfield  {author} {\bibinfo {author} {\bibnamefont {Jackson}, \bibfnamefont
  {D.~E.}}, \bibinfo {author} {\bibfnamefont {D.}~\bibnamefont {VanGennep}},
  \bibinfo {author} {\bibfnamefont {W.}~\bibnamefont {Bi}}, \bibinfo {author}
  {\bibfnamefont {D.}~\bibnamefont {Zhang}}, \bibinfo {author} {\bibfnamefont
  {P.}~\bibnamefont {Materne}}, \bibinfo {author} {\bibfnamefont
  {Y.}~\bibnamefont {Liu}}, \bibinfo {author} {\bibfnamefont {G.-H.}\
  \bibnamefont {Cao}}, \bibinfo {author} {\bibfnamefont {S.~T.}\ \bibnamefont
  {Weir}}, \bibinfo {author} {\bibfnamefont {Y.~K.}\ \bibnamefont {Vohra}},
  and\ \bibinfo {author} {\bibfnamefont {J.~J.}\ \bibnamefont {Hamlin}}}
  (\bibinfo {year} {2018}),\ \href {https://doi.org/10.1103/PhysRevB.98.014518}
  {\bibfield  {journal} {\bibinfo  {journal} {Phys. Rev. B}\ }\textbf {\bibinfo
  {volume} {98}},\ \bibinfo {pages} {014518}}\BibitemShut {NoStop}%
\bibitem [{\citenamefont {Jeevan}\ \emph {et~al.}(2008)\citenamefont {Jeevan},
  \citenamefont {Hossain}, \citenamefont {Kasinathan}, \citenamefont {Rosner},
  \citenamefont {Geibel},\ and\ \citenamefont {Gegenwart}}]{Jeevan2008}%
  \BibitemOpen
  \bibfield  {author} {\bibinfo {author} {\bibnamefont {Jeevan}, \bibfnamefont
  {H.~S.}}, \bibinfo {author} {\bibfnamefont {Z.}~\bibnamefont {Hossain}},
  \bibinfo {author} {\bibfnamefont {D.}~\bibnamefont {Kasinathan}}, \bibinfo
  {author} {\bibfnamefont {H.}~\bibnamefont {Rosner}}, \bibinfo {author}
  {\bibfnamefont {C.}~\bibnamefont {Geibel}}, and\ \bibinfo {author}
  {\bibfnamefont {P.}~\bibnamefont {Gegenwart}}} (\bibinfo {year} {2008}),\
  \href {https://doi.org/10.1103/PhysRevB.78.092406} {\bibfield  {journal}
  {\bibinfo  {journal} {Phys. Rev. B}\ }\textbf {\bibinfo {volume} {78}},\
  \bibinfo {pages} {092406}}\BibitemShut {NoStop}%
\bibitem [{\citenamefont {Jeevan}\ \emph {et~al.}(2011)\citenamefont {Jeevan},
  \citenamefont {Kasinathan}, \citenamefont {Rosner},\ and\ \citenamefont
  {Gegenwart}}]{Jeevan2011}%
  \BibitemOpen
  \bibfield  {author} {\bibinfo {author} {\bibnamefont {Jeevan}, \bibfnamefont
  {H.~S.}}, \bibinfo {author} {\bibfnamefont {D.}~\bibnamefont {Kasinathan}},
  \bibinfo {author} {\bibfnamefont {H.}~\bibnamefont {Rosner}}, and\ \bibinfo
  {author} {\bibfnamefont {P.}~\bibnamefont {Gegenwart}}} (\bibinfo {year}
  {2011}),\ \href {https://doi.org/10.1103/PhysRevB.83.054511} {\bibfield
  {journal} {\bibinfo  {journal} {Phys. Rev. B}\ }\textbf {\bibinfo {volume}
  {83}},\ \bibinfo {pages} {054511}}\BibitemShut {NoStop}%
\bibitem [{\citenamefont {Jia}\ \emph {et~al.}(2016)\citenamefont {Jia},
  \citenamefont {Wohlfeld}, \citenamefont {Wang}, \citenamefont {Moritz},\ and\
  \citenamefont {Devereaux}}]{jia2016using}%
  \BibitemOpen
  \bibfield  {author} {\bibinfo {author} {\bibnamefont {Jia}, \bibfnamefont
  {C.}}, \bibinfo {author} {\bibfnamefont {K.}~\bibnamefont {Wohlfeld}},
  \bibinfo {author} {\bibfnamefont {Y.}~\bibnamefont {Wang}}, \bibinfo {author}
  {\bibfnamefont {B.}~\bibnamefont {Moritz}}, and\ \bibinfo {author}
  {\bibfnamefont {T.~P.}\ \bibnamefont {Devereaux}}} (\bibinfo {year} {2016}),\
  \href {https://doi.org/10.1103/PhysRevX.6.021020} {\bibfield  {journal}
  {\bibinfo  {journal} {Phys. Rev. X}\ }\textbf {\bibinfo {volume} {6}},\
  \bibinfo {pages} {021020}}\BibitemShut {NoStop}%
\bibitem [{\citenamefont {Jia}\ \emph {et~al.}(2014)\citenamefont {Jia},
  \citenamefont {Nowadnick}, \citenamefont {Wohlfeld}, \citenamefont {Kung},
  \citenamefont {Chen}, \citenamefont {Johnston}, \citenamefont {Tohyama},
  \citenamefont {Moritz},\ and\ \citenamefont {Devereaux}}]{jia2014persistent}%
  \BibitemOpen
  \bibfield  {author} {\bibinfo {author} {\bibnamefont {Jia}, \bibfnamefont
  {C.~J.}}, \bibinfo {author} {\bibfnamefont {E.~A.}\ \bibnamefont
  {Nowadnick}}, \bibinfo {author} {\bibfnamefont {K.}~\bibnamefont {Wohlfeld}},
  \bibinfo {author} {\bibfnamefont {Y.~F.}\ \bibnamefont {Kung}}, \bibinfo
  {author} {\bibfnamefont {C.-C.}\ \bibnamefont {Chen}}, \bibinfo {author}
  {\bibfnamefont {S.}~\bibnamefont {Johnston}}, \bibinfo {author}
  {\bibfnamefont {T.}~\bibnamefont {Tohyama}}, \bibinfo {author} {\bibfnamefont
  {B.}~\bibnamefont {Moritz}}, and\ \bibinfo {author} {\bibfnamefont {T.~P.}\
  \bibnamefont {Devereaux}}} (\bibinfo {year} {2014}),\ \href
  {https://doi.org/10.1038/ncomms4314} {\bibfield  {journal} {\bibinfo
  {journal} {Nat. Commun.}\ }\textbf {\bibinfo {volume} {5}},\ \bibinfo {pages}
  {3314}}\BibitemShut {NoStop}%
\bibitem [{\citenamefont {Jiang}\ \emph {et~al.}(2013)\citenamefont {Jiang},
  \citenamefont {Sun}, \citenamefont {Xu},\ and\ \citenamefont
  {Cao}}]{Jiang2013}%
  \BibitemOpen
  \bibfield  {author} {\bibinfo {author} {\bibnamefont {Jiang}, \bibfnamefont
  {H.}}, \bibinfo {author} {\bibfnamefont {Y.-L.}\ \bibnamefont {Sun}},
  \bibinfo {author} {\bibfnamefont {Z.-A.}\ \bibnamefont {Xu}}, and\ \bibinfo
  {author} {\bibfnamefont {G.-H.}\ \bibnamefont {Cao}}} (\bibinfo {year}
  {2013}),\ \href {https://doi.org/10.1088/1674-1056/22/8/087410} {\bibfield
  {journal} {\bibinfo  {journal} {Chinese Physics B}\ }\textbf {\bibinfo
  {volume} {22}},\ \bibinfo {pages} {087410}}\BibitemShut {NoStop}%
\bibitem [{\citenamefont {Jiang}\ \emph
  {et~al.}(2023{\natexlab{a}})\citenamefont {Jiang}, \citenamefont {Shi},
  \citenamefont {Christensen}, \citenamefont {Sanchez}, \citenamefont {Huang},
  \citenamefont {Lin}, \citenamefont {Liu}, \citenamefont {Malinowski},
  \citenamefont {Xu}, \citenamefont {Fernandes},\ and\ \citenamefont
  {Chu}}]{jiang2023nematic}%
  \BibitemOpen
  \bibfield  {author} {\bibinfo {author} {\bibnamefont {Jiang}, \bibfnamefont
  {Q.}}, \bibinfo {author} {\bibfnamefont {Y.}~\bibnamefont {Shi}}, \bibinfo
  {author} {\bibfnamefont {M.~H.}\ \bibnamefont {Christensen}}, \bibinfo
  {author} {\bibfnamefont {J.~J.}\ \bibnamefont {Sanchez}}, \bibinfo {author}
  {\bibfnamefont {B.}~\bibnamefont {Huang}}, \bibinfo {author} {\bibfnamefont
  {Z.}~\bibnamefont {Lin}}, \bibinfo {author} {\bibfnamefont {Z.}~\bibnamefont
  {Liu}}, \bibinfo {author} {\bibfnamefont {P.}~\bibnamefont {Malinowski}},
  \bibinfo {author} {\bibfnamefont {X.}~\bibnamefont {Xu}}, \bibinfo {author}
  {\bibfnamefont {R.~M.}\ \bibnamefont {Fernandes}}, and\ \bibinfo {author}
  {\bibfnamefont {J.-H.}\ \bibnamefont {Chu}}} (\bibinfo {year}
  {2023}{\natexlab{a}}),\ \href {https://doi.org/10.1038/s42005-023-01154-8}
  {\bibfield  {journal} {\bibinfo  {journal} {Commun. Phys.}\ }\textbf
  {\bibinfo {volume} {6}},\ \bibinfo {pages} {39}}\BibitemShut {NoStop}%
\bibitem [{\citenamefont {Jiang}\ \emph
  {et~al.}(2023{\natexlab{b}})\citenamefont {Jiang}, \citenamefont {Qin},
  \citenamefont {Wei}, \citenamefont {Xu}, \citenamefont {Ke}, \citenamefont
  {Zhu}, \citenamefont {Zhang}, \citenamefont {Zhao}, \citenamefont {Liang},
  \citenamefont {Wei}, \citenamefont {Lin}, \citenamefont {Feng}, \citenamefont
  {Chen}, \citenamefont {Xiong}, \citenamefont {Yuan}, \citenamefont {Zhu},
  \citenamefont {Li}, \citenamefont {Xi}, \citenamefont {Wang}, \citenamefont
  {Yang}, \citenamefont {Wang}, \citenamefont {Xiang}, \citenamefont {Hu},
  \citenamefont {Jiang}, \citenamefont {Chen}, \citenamefont {Jin},\ and\
  \citenamefont {Zhao}}]{jiang2023interplay}%
  \BibitemOpen
  \bibfield  {author} {\bibinfo {author} {\bibnamefont {Jiang}, \bibfnamefont
  {X.}}, \bibinfo {author} {\bibfnamefont {M.}~\bibnamefont {Qin}}, \bibinfo
  {author} {\bibfnamefont {X.}~\bibnamefont {Wei}}, \bibinfo {author}
  {\bibfnamefont {L.}~\bibnamefont {Xu}}, \bibinfo {author} {\bibfnamefont
  {J.}~\bibnamefont {Ke}}, \bibinfo {author} {\bibfnamefont {H.}~\bibnamefont
  {Zhu}}, \bibinfo {author} {\bibfnamefont {R.}~\bibnamefont {Zhang}}, \bibinfo
  {author} {\bibfnamefont {Z.}~\bibnamefont {Zhao}}, \bibinfo {author}
  {\bibfnamefont {Q.}~\bibnamefont {Liang}}, \bibinfo {author} {\bibfnamefont
  {Z.}~\bibnamefont {Wei}}, \bibinfo {author} {\bibfnamefont {Z.}~\bibnamefont
  {Lin}}, \bibinfo {author} {\bibfnamefont {Z.}~\bibnamefont {Feng}}, \bibinfo
  {author} {\bibfnamefont {F.}~\bibnamefont {Chen}}, \bibinfo {author}
  {\bibfnamefont {P.}~\bibnamefont {Xiong}}, \bibinfo {author} {\bibfnamefont
  {J.}~\bibnamefont {Yuan}}, \bibinfo {author} {\bibfnamefont {B.}~\bibnamefont
  {Zhu}}, \bibinfo {author} {\bibfnamefont {Y.}~\bibnamefont {Li}}, \bibinfo
  {author} {\bibfnamefont {C.}~\bibnamefont {Xi}}, \bibinfo {author}
  {\bibfnamefont {Z.}~\bibnamefont {Wang}}, \bibinfo {author} {\bibfnamefont
  {M.}~\bibnamefont {Yang}}, \bibinfo {author} {\bibfnamefont {J.}~\bibnamefont
  {Wang}}, \bibinfo {author} {\bibfnamefont {T.}~\bibnamefont {Xiang}},
  \bibinfo {author} {\bibfnamefont {J.}~\bibnamefont {Hu}}, \bibinfo {author}
  {\bibfnamefont {K.}~\bibnamefont {Jiang}}, \bibinfo {author} {\bibfnamefont
  {Q.}~\bibnamefont {Chen}}, \bibinfo {author} {\bibfnamefont {K.}~\bibnamefont
  {Jin}}, and\ \bibinfo {author} {\bibfnamefont {Z.}~\bibnamefont {Zhao}}}
  (\bibinfo {year} {2023}{\natexlab{b}}),\ \href
  {https://doi.org/10.1038/s41567-022-01894-4} {\bibfield  {journal} {\bibinfo
  {journal} {Nat. Phys.}\ }\textbf {\bibinfo {volume} {19}},\ \bibinfo {pages}
  {365}}\BibitemShut {NoStop}%
\bibitem [{\citenamefont {Jiao}\ \emph {et~al.}(2012)\citenamefont {Jiao},
  \citenamefont {Bao}, \citenamefont {Tao}, \citenamefont {Jiang},
  \citenamefont {Feng}, \citenamefont {Xu},\ and\ \citenamefont
  {Cao}}]{Jiao2012}%
  \BibitemOpen
  \bibfield  {author} {\bibinfo {author} {\bibnamefont {Jiao}, \bibfnamefont
  {W.-H.}}, \bibinfo {author} {\bibfnamefont {J.-K.}\ \bibnamefont {Bao}},
  \bibinfo {author} {\bibfnamefont {Q.}~\bibnamefont {Tao}}, \bibinfo {author}
  {\bibfnamefont {H.}~\bibnamefont {Jiang}}, \bibinfo {author} {\bibfnamefont
  {C.-M.}\ \bibnamefont {Feng}}, \bibinfo {author} {\bibfnamefont {Z.-A.}\
  \bibnamefont {Xu}}, and\ \bibinfo {author} {\bibfnamefont {G.-H.}\
  \bibnamefont {Cao}}} (\bibinfo {year} {2012}),\ \bibfield  {booktitle} {\emph
  {\bibinfo {booktitle} {Journal of Physics: Conference Series}},\ }\href
  {https://doi.org/10.1088/1742-6596/400/2/022038} {\bibfield  {journal}
  {\bibinfo  {journal} {Journal of Physics: Conference Series}\ }\textbf
  {\bibinfo {volume} {400}},\ \bibinfo {pages} {022038}}\BibitemShut {NoStop}%
\bibitem [{\citenamefont {Jiao}\ \emph {et~al.}(2013)\citenamefont {Jiao},
  \citenamefont {Zhai}, \citenamefont {Bao}, \citenamefont {Luo}, \citenamefont
  {Tao}, \citenamefont {Feng}, \citenamefont {Xu},\ and\ \citenamefont
  {Cao}}]{Jiao2013}%
  \BibitemOpen
  \bibfield  {author} {\bibinfo {author} {\bibnamefont {Jiao}, \bibfnamefont
  {W.-H.}}, \bibinfo {author} {\bibfnamefont {H.-F.}\ \bibnamefont {Zhai}},
  \bibinfo {author} {\bibfnamefont {J.-K.}\ \bibnamefont {Bao}}, \bibinfo
  {author} {\bibfnamefont {Y.-K.}\ \bibnamefont {Luo}}, \bibinfo {author}
  {\bibfnamefont {Q.}~\bibnamefont {Tao}}, \bibinfo {author} {\bibfnamefont
  {C.-M.}\ \bibnamefont {Feng}}, \bibinfo {author} {\bibfnamefont {Z.-A.}\
  \bibnamefont {Xu}}, and\ \bibinfo {author} {\bibfnamefont {G.-H.}\
  \bibnamefont {Cao}}} (\bibinfo {year} {2013}),\ \href
  {https://doi.org/10.1088/1367-2630/15/11/113002} {\bibfield  {journal}
  {\bibinfo  {journal} {New J. Phys.}\ }\textbf {\bibinfo {volume} {15}},\
  \bibinfo {pages} {113002}}\BibitemShut {NoStop}%
\bibitem [{\citenamefont {Jin}\ \emph {et~al.}(2011)\citenamefont {Jin},
  \citenamefont {Butch}, \citenamefont {Kirshenbaum}, \citenamefont
  {Paglione},\ and\ \citenamefont {Greene}}]{jin2011link}%
  \BibitemOpen
  \bibfield  {author} {\bibinfo {author} {\bibnamefont {Jin}, \bibfnamefont
  {K.}}, \bibinfo {author} {\bibfnamefont {N.~P.}\ \bibnamefont {Butch}},
  \bibinfo {author} {\bibfnamefont {K.}~\bibnamefont {Kirshenbaum}}, \bibinfo
  {author} {\bibfnamefont {J.}~\bibnamefont {Paglione}}, and\ \bibinfo {author}
  {\bibfnamefont {R.~L.}\ \bibnamefont {Greene}}} (\bibinfo {year} {2011}),\
  \href {https://doi.org/10.1038/nature10308} {\bibfield  {journal} {\bibinfo
  {journal} {Nature}\ }\textbf {\bibinfo {volume} {476}},\ \bibinfo {pages}
  {73}}\BibitemShut {NoStop}%
\bibitem [{\citenamefont {Jin}\ \emph {et~al.}(2016)\citenamefont {Jin},
  \citenamefont {Wu}, \citenamefont {Huang}, \citenamefont {Wu}, \citenamefont
  {Ying}, \citenamefont {Fan}, \citenamefont {Sun}, \citenamefont {Zhao},\ and\
  \citenamefont {Chen}}]{Jin2016}%
  \BibitemOpen
  \bibfield  {author} {\bibinfo {author} {\bibnamefont {Jin}, \bibfnamefont
  {S.}}, \bibinfo {author} {\bibfnamefont {X.}~\bibnamefont {Wu}}, \bibinfo
  {author} {\bibfnamefont {Q.}~\bibnamefont {Huang}}, \bibinfo {author}
  {\bibfnamefont {H.}~\bibnamefont {Wu}}, \bibinfo {author} {\bibfnamefont
  {T.}~\bibnamefont {Ying}}, \bibinfo {author} {\bibfnamefont {X.}~\bibnamefont
  {Fan}}, \bibinfo {author} {\bibfnamefont {R.}~\bibnamefont {Sun}}, \bibinfo
  {author} {\bibfnamefont {L.}~\bibnamefont {Zhao}}, and\ \bibinfo {author}
  {\bibfnamefont {X.}~\bibnamefont {Chen}}} (\bibinfo {year} {2016}),\ \href
  {https://doi.org/10.48550/arXiv.1607.01103} {\bibfield  {journal} {\bibinfo
  {journal} {arXiv preprint arXiv:1607.01103}\
  }10.48550/arXiv.1607.01103}\BibitemShut {NoStop}%
\bibitem [{\citenamefont {Jin}\ \emph {et~al.}(2015)\citenamefont {Jin},
  \citenamefont {Li}, \citenamefont {Su}, \citenamefont {Nandi}, \citenamefont
  {Xiao}, \citenamefont {Jiao}, \citenamefont {Meven}, \citenamefont {Sazonov},
  \citenamefont {Feng}, \citenamefont {Chen} \emph {et~al.}}]{Jin2015}%
  \BibitemOpen
  \bibfield  {author} {\bibinfo {author} {\bibnamefont {Jin}, \bibfnamefont
  {W.}}, \bibinfo {author} {\bibfnamefont {W.}~\bibnamefont {Li}}, \bibinfo
  {author} {\bibfnamefont {Y.}~\bibnamefont {Su}}, \bibinfo {author}
  {\bibfnamefont {S.}~\bibnamefont {Nandi}}, \bibinfo {author} {\bibfnamefont
  {Y.}~\bibnamefont {Xiao}}, \bibinfo {author} {\bibfnamefont {W.}~\bibnamefont
  {Jiao}}, \bibinfo {author} {\bibfnamefont {M.}~\bibnamefont {Meven}},
  \bibinfo {author} {\bibfnamefont {A.}~\bibnamefont {Sazonov}}, \bibinfo
  {author} {\bibfnamefont {E.}~\bibnamefont {Feng}}, \bibinfo {author}
  {\bibfnamefont {Y.}~\bibnamefont {Chen}},  \emph {et~al.}} (\bibinfo {year}
  {2015}),\ \href {https://doi.org/10.1103/PhysRevB.91.064506} {\bibfield
  {journal} {\bibinfo  {journal} {Phys. Rev. B}\ }\textbf {\bibinfo {volume}
  {91}},\ \bibinfo {pages} {064506}}\BibitemShut {NoStop}%
\bibitem [{\citenamefont {Jin}\ \emph {et~al.}(2013)\citenamefont {Jin},
  \citenamefont {Nandi}, \citenamefont {Xiao}, \citenamefont {Su},
  \citenamefont {Zaharko}, \citenamefont {Guguchia}, \citenamefont {Bukowski},
  \citenamefont {Price}, \citenamefont {Jiao}, \citenamefont {Cao} \emph
  {et~al.}}]{Jin2013}%
  \BibitemOpen
  \bibfield  {author} {\bibinfo {author} {\bibnamefont {Jin}, \bibfnamefont
  {W.}}, \bibinfo {author} {\bibfnamefont {S.}~\bibnamefont {Nandi}}, \bibinfo
  {author} {\bibfnamefont {Y.}~\bibnamefont {Xiao}}, \bibinfo {author}
  {\bibfnamefont {Y.}~\bibnamefont {Su}}, \bibinfo {author} {\bibfnamefont
  {O.}~\bibnamefont {Zaharko}}, \bibinfo {author} {\bibfnamefont
  {Z.}~\bibnamefont {Guguchia}}, \bibinfo {author} {\bibfnamefont
  {Z.}~\bibnamefont {Bukowski}}, \bibinfo {author} {\bibfnamefont
  {S.}~\bibnamefont {Price}}, \bibinfo {author} {\bibfnamefont
  {W.}~\bibnamefont {Jiao}}, \bibinfo {author} {\bibfnamefont {G.}~\bibnamefont
  {Cao}},  \emph {et~al.}} (\bibinfo {year} {2013}),\ \href
  {https://doi.org/10.1103/PhysRevB.88.214516} {\bibfield  {journal} {\bibinfo
  {journal} {Phys. Rev. B}\ }\textbf {\bibinfo {volume} {88}},\ \bibinfo
  {pages} {214516}}\BibitemShut {NoStop}%
\bibitem [{\citenamefont {Kakiya}\ \emph {et~al.}(2011)\citenamefont {Kakiya},
  \citenamefont {Kudo}, \citenamefont {Nishikubo}, \citenamefont {Oku},
  \citenamefont {Nishibori}, \citenamefont {Sawa}, \citenamefont {Yamamoto},
  \citenamefont {Nozaka},\ and\ \citenamefont {Nohara}}]{Kakiya2011}%
  \BibitemOpen
  \bibfield  {author} {\bibinfo {author} {\bibnamefont {Kakiya}, \bibfnamefont
  {S.}}, \bibinfo {author} {\bibfnamefont {K.}~\bibnamefont {Kudo}}, \bibinfo
  {author} {\bibfnamefont {Y.}~\bibnamefont {Nishikubo}}, \bibinfo {author}
  {\bibfnamefont {K.}~\bibnamefont {Oku}}, \bibinfo {author} {\bibfnamefont
  {E.}~\bibnamefont {Nishibori}}, \bibinfo {author} {\bibfnamefont
  {H.}~\bibnamefont {Sawa}}, \bibinfo {author} {\bibfnamefont {T.}~\bibnamefont
  {Yamamoto}}, \bibinfo {author} {\bibfnamefont {T.}~\bibnamefont {Nozaka}},
  and\ \bibinfo {author} {\bibfnamefont {M.}~\bibnamefont {Nohara}}} (\bibinfo
  {year} {2011}),\ \href {https://doi.org/10.1143/jpsj.80.093704} {\bibfield
  {journal} {\bibinfo  {journal} {J. Phys. Soc. Jpn.}\ }\textbf {\bibinfo
  {volume} {80}},\ \bibinfo {pages} {093704}}\BibitemShut {NoStop}%
\bibitem [{\citenamefont {Kamihara}\ \emph {et~al.}(2006)\citenamefont
  {Kamihara}, \citenamefont {Hiramatsu}, \citenamefont {Hirano}, \citenamefont
  {Kawamura}, \citenamefont {Yanagi}, \citenamefont {Kamiya},\ and\
  \citenamefont {Hosono}}]{Kamihara2006iron}%
  \BibitemOpen
  \bibfield  {author} {\bibinfo {author} {\bibnamefont {Kamihara},
  \bibfnamefont {Y.}}, \bibinfo {author} {\bibfnamefont {H.}~\bibnamefont
  {Hiramatsu}}, \bibinfo {author} {\bibfnamefont {M.}~\bibnamefont {Hirano}},
  \bibinfo {author} {\bibfnamefont {R.}~\bibnamefont {Kawamura}}, \bibinfo
  {author} {\bibfnamefont {H.}~\bibnamefont {Yanagi}}, \bibinfo {author}
  {\bibfnamefont {T.}~\bibnamefont {Kamiya}}, and\ \bibinfo {author}
  {\bibfnamefont {H.}~\bibnamefont {Hosono}}} (\bibinfo {year} {2006}),\ \href
  {https://doi.org/10.1021/ja063355c} {\bibfield  {journal} {\bibinfo
  {journal} {J. Am. Chem. Soc.}\ }\textbf {\bibinfo {volume} {128}},\ \bibinfo
  {pages} {10012}}\BibitemShut {NoStop}%
\bibitem [{\citenamefont {Kamihara}\ \emph {et~al.}(2008)\citenamefont
  {Kamihara}, \citenamefont {Watanabe}, \citenamefont {Hirano},\ and\
  \citenamefont {Hosono}}]{Kamihara2008}%
  \BibitemOpen
  \bibfield  {author} {\bibinfo {author} {\bibnamefont {Kamihara},
  \bibfnamefont {Y.}}, \bibinfo {author} {\bibfnamefont {T.}~\bibnamefont
  {Watanabe}}, \bibinfo {author} {\bibfnamefont {M.}~\bibnamefont {Hirano}},
  and\ \bibinfo {author} {\bibfnamefont {H.}~\bibnamefont {Hosono}}} (\bibinfo
  {year} {2008}),\ \href {https://doi.org/10.1021/ja800073m} {\bibfield
  {journal} {\bibinfo  {journal} {J. Am. Chem. Soc.}\ }\textbf {\bibinfo
  {volume} {130}},\ \bibinfo {pages} {3296}}\BibitemShut {NoStop}%
\bibitem [{\citenamefont {Kang}\ \emph
  {et~al.}(2018{\natexlab{a}})\citenamefont {Kang}, \citenamefont {Chubukov},\
  and\ \citenamefont {Fernandes}}]{kang2018time}%
  \BibitemOpen
  \bibfield  {author} {\bibinfo {author} {\bibnamefont {Kang}, \bibfnamefont
  {J.}}, \bibinfo {author} {\bibfnamefont {A.~V.}\ \bibnamefont {Chubukov}},
  and\ \bibinfo {author} {\bibfnamefont {R.~M.}\ \bibnamefont {Fernandes}}}
  (\bibinfo {year} {2018}{\natexlab{a}}),\ \href
  {https://doi.org/10.1103/PhysRevB.98.064508} {\bibfield  {journal} {\bibinfo
  {journal} {Phys. Rev. B}\ }\textbf {\bibinfo {volume} {98}},\ \bibinfo
  {pages} {064508}}\BibitemShut {NoStop}%
\bibitem [{\citenamefont {Kang}\ and\ \citenamefont
  {Fernandes}(2016)}]{kang2016superconductivity}%
  \BibitemOpen
  \bibfield  {author} {\bibinfo {author} {\bibnamefont {Kang}, \bibfnamefont
  {J.}}, and\ \bibinfo {author} {\bibfnamefont {R.~M.}\ \bibnamefont
  {Fernandes}}} (\bibinfo {year} {2016}),\ \href
  {https://doi.org/10.1103/PhysRevLett.117.217003} {\bibfield  {journal}
  {\bibinfo  {journal} {Phys. Rev. Lett.}\ }\textbf {\bibinfo {volume} {117}},\
  \bibinfo {pages} {217003}}\BibitemShut {NoStop}%
\bibitem [{\citenamefont {Kang}\ \emph
  {et~al.}(2018{\natexlab{b}})\citenamefont {Kang}, \citenamefont {Fernandes},\
  and\ \citenamefont {Chubukov}}]{kang2018superconductivity}%
  \BibitemOpen
  \bibfield  {author} {\bibinfo {author} {\bibnamefont {Kang}, \bibfnamefont
  {J.}}, \bibinfo {author} {\bibfnamefont {R.~M.}\ \bibnamefont {Fernandes}},
  and\ \bibinfo {author} {\bibfnamefont {A.}~\bibnamefont {Chubukov}}}
  (\bibinfo {year} {2018}{\natexlab{b}}),\ \href
  {https://doi.org/10.1103/PhysRevLett.120.267001} {\bibfield  {journal}
  {\bibinfo  {journal} {Phys. Rev. Lett.}\ }\textbf {\bibinfo {volume} {120}},\
  \bibinfo {pages} {267001}}\BibitemShut {NoStop}%
\bibitem [{\citenamefont {Kasahara}\ \emph {et~al.}(2020)\citenamefont
  {Kasahara}, \citenamefont {Sato}, \citenamefont {Licciardello}, \citenamefont
  {\ifmmode~\check{C}\else \v{C}\fi{}ulo}, \citenamefont
  {Arsenijevi\ifmmode~\acute{c}\else \'{c}\fi{}}, \citenamefont {Ottenbros},
  \citenamefont {Tominaga}, \citenamefont {B\"oker}, \citenamefont {Eremin},
  \citenamefont {Shibauchi}, \citenamefont {Wosnitza}, \citenamefont {Hussey},\
  and\ \citenamefont {Matsuda}}]{kasahara2020evidence}%
  \BibitemOpen
  \bibfield  {author} {\bibinfo {author} {\bibnamefont {Kasahara},
  \bibfnamefont {S.}}, \bibinfo {author} {\bibfnamefont {Y.}~\bibnamefont
  {Sato}}, \bibinfo {author} {\bibfnamefont {S.}~\bibnamefont {Licciardello}},
  \bibinfo {author} {\bibfnamefont {M.}~\bibnamefont {\ifmmode~\check{C}\else
  \v{C}\fi{}ulo}}, \bibinfo {author} {\bibfnamefont {S.}~\bibnamefont
  {Arsenijevi\ifmmode~\acute{c}\else \'{c}\fi{}}}, \bibinfo {author}
  {\bibfnamefont {T.}~\bibnamefont {Ottenbros}}, \bibinfo {author}
  {\bibfnamefont {T.}~\bibnamefont {Tominaga}}, \bibinfo {author}
  {\bibfnamefont {J.}~\bibnamefont {B\"oker}}, \bibinfo {author} {\bibfnamefont
  {I.}~\bibnamefont {Eremin}}, \bibinfo {author} {\bibfnamefont
  {T.}~\bibnamefont {Shibauchi}}, \bibinfo {author} {\bibfnamefont
  {J.}~\bibnamefont {Wosnitza}}, \bibinfo {author} {\bibfnamefont {N.~E.}\
  \bibnamefont {Hussey}}, and\ \bibinfo {author} {\bibfnamefont
  {Y.}~\bibnamefont {Matsuda}}} (\bibinfo {year} {2020}),\ \href
  {https://doi.org/10.1103/PhysRevLett.124.107001} {\bibfield  {journal}
  {\bibinfo  {journal} {Phys. Rev. Lett.}\ }\textbf {\bibinfo {volume} {124}},\
  \bibinfo {pages} {107001}}\BibitemShut {NoStop}%
\bibitem [{\citenamefont {Kasahara}\ \emph {et~al.}(2012)\citenamefont
  {Kasahara}, \citenamefont {Shi}, \citenamefont {Hashimoto}, \citenamefont
  {Tonegawa}, \citenamefont {Mizukami}, \citenamefont {Shibauchi},
  \citenamefont {Sugimoto}, \citenamefont {Fukuda}, \citenamefont {Terashima},
  \citenamefont {Nevidomskyy},\ and\ \citenamefont {Matsuda}}]{Kasahara2012}%
  \BibitemOpen
  \bibfield  {author} {\bibinfo {author} {\bibnamefont {Kasahara},
  \bibfnamefont {S.}}, \bibinfo {author} {\bibfnamefont {H.~J.}\ \bibnamefont
  {Shi}}, \bibinfo {author} {\bibfnamefont {K.}~\bibnamefont {Hashimoto}},
  \bibinfo {author} {\bibfnamefont {S.}~\bibnamefont {Tonegawa}}, \bibinfo
  {author} {\bibfnamefont {Y.}~\bibnamefont {Mizukami}}, \bibinfo {author}
  {\bibfnamefont {T.}~\bibnamefont {Shibauchi}}, \bibinfo {author}
  {\bibfnamefont {K.}~\bibnamefont {Sugimoto}}, \bibinfo {author}
  {\bibfnamefont {T.}~\bibnamefont {Fukuda}}, \bibinfo {author} {\bibfnamefont
  {T.}~\bibnamefont {Terashima}}, \bibinfo {author} {\bibfnamefont {A.~H.}\
  \bibnamefont {Nevidomskyy}}, and\ \bibinfo {author} {\bibfnamefont
  {Y.}~\bibnamefont {Matsuda}}} (\bibinfo {year} {2012}),\ \href
  {https://doi.org/10.1038/nature11178} {\bibfield  {journal} {\bibinfo
  {journal} {Nature}\ }\textbf {\bibinfo {volume} {486}},\ \bibinfo {pages}
  {382}}\BibitemShut {NoStop}%
\bibitem [{\citenamefont {Kasahara}\ \emph {et~al.}(2021)\citenamefont
  {Kasahara}, \citenamefont {Suzuki}, \citenamefont {Machida}, \citenamefont
  {Sato}, \citenamefont {Ukai}, \citenamefont {Murayama}, \citenamefont
  {Suetsugu}, \citenamefont {Kasahara}, \citenamefont {Shibauchi},
  \citenamefont {Hanaguri},\ and\ \citenamefont
  {Matsuda}}]{kasahara2021quasiparticle}%
  \BibitemOpen
  \bibfield  {author} {\bibinfo {author} {\bibnamefont {Kasahara},
  \bibfnamefont {S.}}, \bibinfo {author} {\bibfnamefont {H.}~\bibnamefont
  {Suzuki}}, \bibinfo {author} {\bibfnamefont {T.}~\bibnamefont {Machida}},
  \bibinfo {author} {\bibfnamefont {Y.}~\bibnamefont {Sato}}, \bibinfo {author}
  {\bibfnamefont {Y.}~\bibnamefont {Ukai}}, \bibinfo {author} {\bibfnamefont
  {H.}~\bibnamefont {Murayama}}, \bibinfo {author} {\bibfnamefont
  {S.}~\bibnamefont {Suetsugu}}, \bibinfo {author} {\bibfnamefont
  {Y.}~\bibnamefont {Kasahara}}, \bibinfo {author} {\bibfnamefont
  {T.}~\bibnamefont {Shibauchi}}, \bibinfo {author} {\bibfnamefont
  {T.}~\bibnamefont {Hanaguri}}, and\ \bibinfo {author} {\bibfnamefont
  {Y.}~\bibnamefont {Matsuda}}} (\bibinfo {year} {2021}),\ \href
  {https://doi.org/10.1103/PhysRevLett.127.257001} {\bibfield  {journal}
  {\bibinfo  {journal} {Phys. Rev. Lett.}\ }\textbf {\bibinfo {volume} {127}},\
  \bibinfo {pages} {257001}}\BibitemShut {NoStop}%
\bibitem [{\citenamefont {Kasahara}\ \emph {et~al.}(2014)\citenamefont
  {Kasahara}, \citenamefont {Watashige}, \citenamefont {Hanaguri},
  \citenamefont {Kohsaka}, \citenamefont {Yamashita}, \citenamefont
  {Shimoyama}, \citenamefont {Mizukami}, \citenamefont {Endo}, \citenamefont
  {Ikeda}, \citenamefont {Aoyama}, \citenamefont {Terashima}, \citenamefont
  {Uji}, \citenamefont {Wolf}, \citenamefont {von Löhneysen}, \citenamefont
  {Shibauchi},\ and\ \citenamefont {Matsuda}}]{kasahara2014field}%
  \BibitemOpen
  \bibfield  {author} {\bibinfo {author} {\bibnamefont {Kasahara},
  \bibfnamefont {S.}}, \bibinfo {author} {\bibfnamefont {T.}~\bibnamefont
  {Watashige}}, \bibinfo {author} {\bibfnamefont {T.}~\bibnamefont {Hanaguri}},
  \bibinfo {author} {\bibfnamefont {Y.}~\bibnamefont {Kohsaka}}, \bibinfo
  {author} {\bibfnamefont {T.}~\bibnamefont {Yamashita}}, \bibinfo {author}
  {\bibfnamefont {Y.}~\bibnamefont {Shimoyama}}, \bibinfo {author}
  {\bibfnamefont {Y.}~\bibnamefont {Mizukami}}, \bibinfo {author}
  {\bibfnamefont {R.}~\bibnamefont {Endo}}, \bibinfo {author} {\bibfnamefont
  {H.}~\bibnamefont {Ikeda}}, \bibinfo {author} {\bibfnamefont
  {K.}~\bibnamefont {Aoyama}}, \bibinfo {author} {\bibfnamefont
  {T.}~\bibnamefont {Terashima}}, \bibinfo {author} {\bibfnamefont
  {S.}~\bibnamefont {Uji}}, \bibinfo {author} {\bibfnamefont {T.}~\bibnamefont
  {Wolf}}, \bibinfo {author} {\bibfnamefont {H.}~\bibnamefont {von
  Löhneysen}}, \bibinfo {author} {\bibfnamefont {T.}~\bibnamefont
  {Shibauchi}}, and\ \bibinfo {author} {\bibfnamefont {Y.}~\bibnamefont
  {Matsuda}}} (\bibinfo {year} {2014}),\ \href
  {https://doi.org/10.1073/pnas.1413477111} {\bibfield  {journal} {\bibinfo
  {journal} {Proc. Natl. Acad. Sci. U.S.A.}\ }\textbf {\bibinfo {volume}
  {111}},\ \bibinfo {pages} {16309}}\BibitemShut {NoStop}%
\bibitem [{\citenamefont {Katase}\ \emph {et~al.}(2011)\citenamefont {Katase},
  \citenamefont {Ishimaru}, \citenamefont {Tsukamoto}, \citenamefont
  {Hiramatsu}, \citenamefont {Kamiya}, \citenamefont {Tanabe},\ and\
  \citenamefont {Hosono}}]{Katase2011}%
  \BibitemOpen
  \bibfield  {author} {\bibinfo {author} {\bibnamefont {Katase}, \bibfnamefont
  {T.}}, \bibinfo {author} {\bibfnamefont {Y.}~\bibnamefont {Ishimaru}},
  \bibinfo {author} {\bibfnamefont {A.}~\bibnamefont {Tsukamoto}}, \bibinfo
  {author} {\bibfnamefont {H.}~\bibnamefont {Hiramatsu}}, \bibinfo {author}
  {\bibfnamefont {T.}~\bibnamefont {Kamiya}}, \bibinfo {author} {\bibfnamefont
  {K.}~\bibnamefont {Tanabe}}, and\ \bibinfo {author} {\bibfnamefont
  {H.}~\bibnamefont {Hosono}}} (\bibinfo {year} {2011}),\ \href
  {https://doi.org/10.1038/ncomms1419} {\bibfield  {journal} {\bibinfo
  {journal} {Nat. Commun.}\ }\textbf {\bibinfo {volume} {2}},\ \bibinfo {pages}
  {409}}\BibitemShut {NoStop}%
\bibitem [{\citenamefont {Katayama}\ \emph {et~al.}(2013)\citenamefont
  {Katayama}, \citenamefont {Kudo}, \citenamefont {Onari}, \citenamefont
  {Mizukami}, \citenamefont {Sugawara}, \citenamefont {Sugiyama}, \citenamefont
  {Kitahama}, \citenamefont {Iba}, \citenamefont {Fujimura}, \citenamefont
  {Nishimoto}, \citenamefont {Nohara},\ and\ \citenamefont
  {Sawa}}]{katayama2013superconductivity}%
  \BibitemOpen
  \bibfield  {author} {\bibinfo {author} {\bibnamefont {Katayama},
  \bibfnamefont {N.}}, \bibinfo {author} {\bibfnamefont {K.}~\bibnamefont
  {Kudo}}, \bibinfo {author} {\bibfnamefont {S.}~\bibnamefont {Onari}},
  \bibinfo {author} {\bibfnamefont {T.}~\bibnamefont {Mizukami}}, \bibinfo
  {author} {\bibfnamefont {K.}~\bibnamefont {Sugawara}}, \bibinfo {author}
  {\bibfnamefont {Y.}~\bibnamefont {Sugiyama}}, \bibinfo {author}
  {\bibfnamefont {Y.}~\bibnamefont {Kitahama}}, \bibinfo {author}
  {\bibfnamefont {K.}~\bibnamefont {Iba}}, \bibinfo {author} {\bibfnamefont
  {K.}~\bibnamefont {Fujimura}}, \bibinfo {author} {\bibfnamefont
  {N.}~\bibnamefont {Nishimoto}}, \bibinfo {author} {\bibfnamefont
  {M.}~\bibnamefont {Nohara}}, and\ \bibinfo {author} {\bibfnamefont
  {H.}~\bibnamefont {Sawa}}} (\bibinfo {year} {2013}),\ \href
  {https://doi.org/10.7566/jpsj.82.123702} {\bibfield  {journal} {\bibinfo
  {journal} {J. Phys. Soc. Jpn.}\ }\textbf {\bibinfo {volume} {82}},\ \bibinfo
  {pages} {123702}}\BibitemShut {NoStop}%
\bibitem [{\citenamefont {Kawaguchi}\ \emph {et~al.}(2010)\citenamefont
  {Kawaguchi}, \citenamefont {Ogino}, \citenamefont {Shimizu}, \citenamefont
  {Kishio},\ and\ \citenamefont {Shimoyama}}]{Kawaguchi2010}%
  \BibitemOpen
  \bibfield  {author} {\bibinfo {author} {\bibnamefont {Kawaguchi},
  \bibfnamefont {N.}}, \bibinfo {author} {\bibfnamefont {H.}~\bibnamefont
  {Ogino}}, \bibinfo {author} {\bibfnamefont {Y.}~\bibnamefont {Shimizu}},
  \bibinfo {author} {\bibfnamefont {K.}~\bibnamefont {Kishio}}, and\ \bibinfo
  {author} {\bibfnamefont {J.-i.}\ \bibnamefont {Shimoyama}}} (\bibinfo {year}
  {2010}),\ \href {https://doi.org/10.1143/apex.3.063102} {\bibfield  {journal}
  {\bibinfo  {journal} {Applied physics express}\ }\textbf {\bibinfo {volume}
  {3}},\ \bibinfo {pages} {063102}}\BibitemShut {NoStop}%
\bibitem [{\citenamefont {Kawashima}\ \emph {et~al.}(2018)\citenamefont
  {Kawashima}, \citenamefont {Ishida}, \citenamefont {Oka}, \citenamefont
  {Kito}, \citenamefont {Takeshita}, \citenamefont {Fujihisa}, \citenamefont
  {Gotoh}, \citenamefont {Kihou}, \citenamefont {Eisaki}, \citenamefont
  {Yoshida} \emph {et~al.}}]{Kawashima2018}%
  \BibitemOpen
  \bibfield  {author} {\bibinfo {author} {\bibnamefont {Kawashima},
  \bibfnamefont {K.}}, \bibinfo {author} {\bibfnamefont {S.}~\bibnamefont
  {Ishida}}, \bibinfo {author} {\bibfnamefont {K.}~\bibnamefont {Oka}},
  \bibinfo {author} {\bibfnamefont {H.}~\bibnamefont {Kito}}, \bibinfo {author}
  {\bibfnamefont {N.}~\bibnamefont {Takeshita}}, \bibinfo {author}
  {\bibfnamefont {H.}~\bibnamefont {Fujihisa}}, \bibinfo {author}
  {\bibfnamefont {Y.}~\bibnamefont {Gotoh}}, \bibinfo {author} {\bibfnamefont
  {K.}~\bibnamefont {Kihou}}, \bibinfo {author} {\bibfnamefont
  {H.}~\bibnamefont {Eisaki}}, \bibinfo {author} {\bibfnamefont
  {Y.}~\bibnamefont {Yoshida}},  \emph {et~al.}} (\bibinfo {year} {2018}),\
  \bibfield  {booktitle} {\emph {\bibinfo {booktitle} {Journal of Physics:
  Conference Series}},\ }\href {https://doi.org/10.1088/1742-6596/969/1/012027}
  {\bibfield  {journal} {\bibinfo  {journal} {Journal of Physics: Conference
  Series}\ }\textbf {\bibinfo {volume} {969}},\ \bibinfo {pages}
  {012027}}\BibitemShut {NoStop}%
\bibitem [{\citenamefont {Kawashima}\ \emph {et~al.}(2016)\citenamefont
  {Kawashima}, \citenamefont {Kinjo}, \citenamefont {Nishio}, \citenamefont
  {Ishida}, \citenamefont {Fujihisa}, \citenamefont {Gotoh}, \citenamefont
  {Kihou}, \citenamefont {Eisaki}, \citenamefont {Yoshida},\ and\ \citenamefont
  {Iyo}}]{Kawashima2016}%
  \BibitemOpen
  \bibfield  {author} {\bibinfo {author} {\bibnamefont {Kawashima},
  \bibfnamefont {K.}}, \bibinfo {author} {\bibfnamefont {T.}~\bibnamefont
  {Kinjo}}, \bibinfo {author} {\bibfnamefont {T.}~\bibnamefont {Nishio}},
  \bibinfo {author} {\bibfnamefont {S.}~\bibnamefont {Ishida}}, \bibinfo
  {author} {\bibfnamefont {H.}~\bibnamefont {Fujihisa}}, \bibinfo {author}
  {\bibfnamefont {Y.}~\bibnamefont {Gotoh}}, \bibinfo {author} {\bibfnamefont
  {K.}~\bibnamefont {Kihou}}, \bibinfo {author} {\bibfnamefont
  {H.}~\bibnamefont {Eisaki}}, \bibinfo {author} {\bibfnamefont
  {Y.}~\bibnamefont {Yoshida}}, and\ \bibinfo {author} {\bibfnamefont
  {A.}~\bibnamefont {Iyo}}} (\bibinfo {year} {2016}),\ \href
  {https://doi.org/10.7566/jpsj.85.064710} {\bibfield  {journal} {\bibinfo
  {journal} {J. Phys. Soc. Jpn.}\ }\textbf {\bibinfo {volume} {85}},\ \bibinfo
  {pages} {064710}}\BibitemShut {NoStop}%
\bibitem [{\citenamefont {Keimer}\ \emph {et~al.}(2015)\citenamefont {Keimer},
  \citenamefont {Kivelson}, \citenamefont {Norman}, \citenamefont {Uchida},\
  and\ \citenamefont {Zaanen}}]{WOS:000349190300029}%
  \BibitemOpen
  \bibfield  {author} {\bibinfo {author} {\bibnamefont {Keimer}, \bibfnamefont
  {B.}}, \bibinfo {author} {\bibfnamefont {S.~A.}\ \bibnamefont {Kivelson}},
  \bibinfo {author} {\bibfnamefont {M.~R.}\ \bibnamefont {Norman}}, \bibinfo
  {author} {\bibfnamefont {S.}~\bibnamefont {Uchida}}, and\ \bibinfo {author}
  {\bibfnamefont {J.}~\bibnamefont {Zaanen}}} (\bibinfo {year} {2015}),\ \href
  {https://doi.org/10.1038/nature14165} {\bibfield  {journal} {\bibinfo
  {journal} {Nature}\ }\textbf {\bibinfo {volume} {518}},\ \bibinfo {pages}
  {179}}\BibitemShut {NoStop}%
\bibitem [{\citenamefont {Khodas}\ and\ \citenamefont
  {Chubukov}(2012)}]{Khodas2012}%
  \BibitemOpen
  \bibfield  {author} {\bibinfo {author} {\bibnamefont {Khodas}, \bibfnamefont
  {M.}}, and\ \bibinfo {author} {\bibfnamefont {A.}~\bibnamefont {Chubukov}}}
  (\bibinfo {year} {2012}),\ \href
  {https://doi.org/10.1103/PhysRevLett.108.247003} {\bibfield  {journal}
  {\bibinfo  {journal} {Phys. Rev. Lett.}\ }\textbf {\bibinfo {volume} {108}},\
  \bibinfo {pages} {247003}}\BibitemShut {NoStop}%
\bibitem [{\citenamefont {Kim}\ \emph {et~al.}(2015)\citenamefont {Kim},
  \citenamefont {Wang}, \citenamefont {Tucker}, \citenamefont {Valdivia},
  \citenamefont {Abernathy}, \citenamefont {Chi}, \citenamefont {Christianson},
  \citenamefont {Aczel}, \citenamefont {Hong}, \citenamefont {Heitmann},
  \citenamefont {Ran}, \citenamefont {Canfield}, \citenamefont
  {{Bourret-Courchesne}}, \citenamefont {Kreyssig}, \citenamefont {Lee},
  \citenamefont {Goldman}, \citenamefont {McQueeney},\ and\ \citenamefont
  {Birgeneau}}]{kim2015spin}%
  \BibitemOpen
  \bibfield  {author} {\bibinfo {author} {\bibnamefont {Kim}, \bibfnamefont
  {M.~G.}}, \bibinfo {author} {\bibfnamefont {M.}~\bibnamefont {Wang}},
  \bibinfo {author} {\bibfnamefont {G.~S.}\ \bibnamefont {Tucker}}, \bibinfo
  {author} {\bibfnamefont {P.~N.}\ \bibnamefont {Valdivia}}, \bibinfo {author}
  {\bibfnamefont {D.~L.}\ \bibnamefont {Abernathy}}, \bibinfo {author}
  {\bibfnamefont {S.}~\bibnamefont {Chi}}, \bibinfo {author} {\bibfnamefont
  {A.~D.}\ \bibnamefont {Christianson}}, \bibinfo {author} {\bibfnamefont
  {A.~A.}\ \bibnamefont {Aczel}}, \bibinfo {author} {\bibfnamefont
  {T.}~\bibnamefont {Hong}}, \bibinfo {author} {\bibfnamefont {T.~W.}\
  \bibnamefont {Heitmann}}, \bibinfo {author} {\bibfnamefont {S.}~\bibnamefont
  {Ran}}, \bibinfo {author} {\bibfnamefont {P.~C.}\ \bibnamefont {Canfield}},
  \bibinfo {author} {\bibfnamefont {E.~D.}\ \bibnamefont
  {{Bourret-Courchesne}}}, \bibinfo {author} {\bibfnamefont {A.}~\bibnamefont
  {Kreyssig}}, \bibinfo {author} {\bibfnamefont {D.~H.}\ \bibnamefont {Lee}},
  \bibinfo {author} {\bibfnamefont {A.~I.}\ \bibnamefont {Goldman}}, \bibinfo
  {author} {\bibfnamefont {R.~J.}\ \bibnamefont {McQueeney}}, and\ \bibinfo
  {author} {\bibfnamefont {R.~J.}\ \bibnamefont {Birgeneau}}} (\bibinfo {year}
  {2015}),\ \href {https://doi.org/10.1103/PhysRevB.92.214404} {\bibfield
  {journal} {\bibinfo  {journal} {Phys. Rev. B}\ }\textbf {\bibinfo {volume}
  {92}},\ \bibinfo {pages} {214404}}\BibitemShut {NoStop}%
\bibitem [{\citenamefont {Kim}\ \emph {et~al.}(2021)\citenamefont {Kim},
  \citenamefont {Pervakov}, \citenamefont {Evtushinsky}, \citenamefont {Jung},
  \citenamefont {Poelchen}, \citenamefont {Kummer}, \citenamefont {Vlasenko},
  \citenamefont {Sadakov}, \citenamefont {Usoltsev}, \citenamefont {Pudalov}
  \emph {et~al.}}]{Kim2021}%
  \BibitemOpen
  \bibfield  {author} {\bibinfo {author} {\bibnamefont {Kim}, \bibfnamefont
  {T.~K.}}, \bibinfo {author} {\bibfnamefont {K.}~\bibnamefont {Pervakov}},
  \bibinfo {author} {\bibfnamefont {D.}~\bibnamefont {Evtushinsky}}, \bibinfo
  {author} {\bibfnamefont {S.}~\bibnamefont {Jung}}, \bibinfo {author}
  {\bibfnamefont {G.}~\bibnamefont {Poelchen}}, \bibinfo {author}
  {\bibfnamefont {K.}~\bibnamefont {Kummer}}, \bibinfo {author} {\bibfnamefont
  {V.}~\bibnamefont {Vlasenko}}, \bibinfo {author} {\bibfnamefont {A.~V.}\
  \bibnamefont {Sadakov}}, \bibinfo {author} {\bibfnamefont {A.}~\bibnamefont
  {Usoltsev}}, \bibinfo {author} {\bibfnamefont {V.~M.}\ \bibnamefont
  {Pudalov}},  \emph {et~al.}} (\bibinfo {year} {2021}),\ \href
  {https://doi.org/10.1103/PhysRevB.103.174517} {\bibfield  {journal} {\bibinfo
   {journal} {Phys. Rev. B}\ }\textbf {\bibinfo {volume} {103}},\ \bibinfo
  {pages} {174517}}\BibitemShut {NoStop}%
\bibitem [{\citenamefont {Kissikov}\ \emph {et~al.}(2016)\citenamefont
  {Kissikov}, \citenamefont {Dioguardi}, \citenamefont {Timmons}, \citenamefont
  {Tanatar}, \citenamefont {Prozorov}, \citenamefont {Bud'ko}, \citenamefont
  {Canfield}, \citenamefont {Fernandes},\ and\ \citenamefont
  {Curro}}]{kissikov2016nmr}%
  \BibitemOpen
  \bibfield  {author} {\bibinfo {author} {\bibnamefont {Kissikov},
  \bibfnamefont {T.}}, \bibinfo {author} {\bibfnamefont {A.~P.}\ \bibnamefont
  {Dioguardi}}, \bibinfo {author} {\bibfnamefont {E.~I.}\ \bibnamefont
  {Timmons}}, \bibinfo {author} {\bibfnamefont {M.~A.}\ \bibnamefont
  {Tanatar}}, \bibinfo {author} {\bibfnamefont {R.}~\bibnamefont {Prozorov}},
  \bibinfo {author} {\bibfnamefont {S.~L.}\ \bibnamefont {Bud'ko}}, \bibinfo
  {author} {\bibfnamefont {P.~C.}\ \bibnamefont {Canfield}}, \bibinfo {author}
  {\bibfnamefont {R.~M.}\ \bibnamefont {Fernandes}}, and\ \bibinfo {author}
  {\bibfnamefont {N.~J.}\ \bibnamefont {Curro}}} (\bibinfo {year} {2016}),\
  \href {https://doi.org/10.1103/PhysRevB.94.165123} {\bibfield  {journal}
  {\bibinfo  {journal} {Phys. Rev. B}\ }\textbf {\bibinfo {volume} {94}},\
  \bibinfo {pages} {165123}}\BibitemShut {NoStop}%
\bibitem [{\citenamefont {Kissikov}\ \emph {et~al.}(2018)\citenamefont
  {Kissikov}, \citenamefont {Sarkar}, \citenamefont {Lawson}, \citenamefont
  {Bush}, \citenamefont {Timmons}, \citenamefont {Tanatar}, \citenamefont
  {Prozorov}, \citenamefont {Bud'ko}, \citenamefont {Canfield}, \citenamefont
  {Fernandes},\ and\ \citenamefont {Curro}}]{kissikov2018uniaxial}%
  \BibitemOpen
  \bibfield  {author} {\bibinfo {author} {\bibnamefont {Kissikov},
  \bibfnamefont {T.}}, \bibinfo {author} {\bibfnamefont {R.}~\bibnamefont
  {Sarkar}}, \bibinfo {author} {\bibfnamefont {M.}~\bibnamefont {Lawson}},
  \bibinfo {author} {\bibfnamefont {B.~T.}\ \bibnamefont {Bush}}, \bibinfo
  {author} {\bibfnamefont {E.~I.}\ \bibnamefont {Timmons}}, \bibinfo {author}
  {\bibfnamefont {M.~A.}\ \bibnamefont {Tanatar}}, \bibinfo {author}
  {\bibfnamefont {R.}~\bibnamefont {Prozorov}}, \bibinfo {author}
  {\bibfnamefont {S.~L.}\ \bibnamefont {Bud'ko}}, \bibinfo {author}
  {\bibfnamefont {P.~C.}\ \bibnamefont {Canfield}}, \bibinfo {author}
  {\bibfnamefont {R.~M.}\ \bibnamefont {Fernandes}}, and\ \bibinfo {author}
  {\bibfnamefont {N.~J.}\ \bibnamefont {Curro}}} (\bibinfo {year} {2018}),\
  \href {https://doi.org/10.1038/s41467-018-03377-8} {\bibfield  {journal}
  {\bibinfo  {journal} {Nat. Commun.}\ }\textbf {\bibinfo {volume} {9}},\
  \bibinfo {pages} {1058}}\BibitemShut {NoStop}%
\bibitem [{\citenamefont {Klein}\ and\ \citenamefont
  {Chubukov}(2018)}]{klein2018superconductivity}%
  \BibitemOpen
  \bibfield  {author} {\bibinfo {author} {\bibnamefont {Klein}, \bibfnamefont
  {A.}}, and\ \bibinfo {author} {\bibfnamefont {A.}~\bibnamefont {Chubukov}}}
  (\bibinfo {year} {2018}),\ \href {https://doi.org/10.1103/PhysRevB.98.220501}
  {\bibfield  {journal} {\bibinfo  {journal} {Phys. Rev. B}\ }\textbf {\bibinfo
  {volume} {98}},\ \bibinfo {pages} {220501}}\BibitemShut {NoStop}%
\bibitem [{\citenamefont {Klemm}\ \emph {et~al.}(2024)\citenamefont {Klemm},
  \citenamefont {Mozaffari}, \citenamefont {Zhang}, \citenamefont {Casas},
  \citenamefont {Koshelev}, \citenamefont {Yi}, \citenamefont {Balicas},\ and\
  \citenamefont {Dai}}]{WOS:001199902200001}%
  \BibitemOpen
  \bibfield  {author} {\bibinfo {author} {\bibnamefont {Klemm}, \bibfnamefont
  {M.~L.}}, \bibinfo {author} {\bibfnamefont {S.}~\bibnamefont {Mozaffari}},
  \bibinfo {author} {\bibfnamefont {R.}~\bibnamefont {Zhang}}, \bibinfo
  {author} {\bibfnamefont {B.~W.}\ \bibnamefont {Casas}}, \bibinfo {author}
  {\bibfnamefont {A.~E.}\ \bibnamefont {Koshelev}}, \bibinfo {author}
  {\bibfnamefont {M.}~\bibnamefont {Yi}}, \bibinfo {author} {\bibfnamefont
  {L.}~\bibnamefont {Balicas}}, and\ \bibinfo {author} {\bibfnamefont
  {P.}~\bibnamefont {Dai}}} (\bibinfo {year} {2024}),\ \href
  {https://doi.org/10.1016/j.xcrp.2024.101816} {\bibfield  {journal} {\bibinfo
  {journal} {CELL REPORTS PHYSICAL SCIENCE}\ }\textbf {\bibinfo {volume} {5}},\
  \bibinfo {pages} {101816}}\BibitemShut {NoStop}%
\bibitem [{\citenamefont {Klimczuk}\ \emph {et~al.}(2009)\citenamefont
  {Klimczuk}, \citenamefont {McQueen}, \citenamefont {Williams}, \citenamefont
  {Huang}, \citenamefont {Ronning}, \citenamefont {Bauer}, \citenamefont
  {Thompson}, \citenamefont {Green},\ and\ \citenamefont
  {Cava}}]{Klimczuk2009}%
  \BibitemOpen
  \bibfield  {author} {\bibinfo {author} {\bibnamefont {Klimczuk},
  \bibfnamefont {T.}}, \bibinfo {author} {\bibfnamefont {T.}~\bibnamefont
  {McQueen}}, \bibinfo {author} {\bibfnamefont {A.}~\bibnamefont {Williams}},
  \bibinfo {author} {\bibfnamefont {Q.}~\bibnamefont {Huang}}, \bibinfo
  {author} {\bibfnamefont {F.}~\bibnamefont {Ronning}}, \bibinfo {author}
  {\bibfnamefont {E.~D.}\ \bibnamefont {Bauer}}, \bibinfo {author}
  {\bibfnamefont {J.~D.}\ \bibnamefont {Thompson}}, \bibinfo {author}
  {\bibfnamefont {M.}~\bibnamefont {Green}}, and\ \bibinfo {author}
  {\bibfnamefont {R.~J.}\ \bibnamefont {Cava}}} (\bibinfo {year} {2009}),\
  \href {https://doi.org/10.1103/PhysRevB.79.012505} {\bibfield  {journal}
  {\bibinfo  {journal} {Phys. Rev. B}\ }\textbf {\bibinfo {volume} {79}},\
  \bibinfo {pages} {012505}}\BibitemShut {NoStop}%
\bibitem [{\citenamefont {Kondo}\ \emph {et~al.}(2020)\citenamefont {Kondo},
  \citenamefont {Motoki}, \citenamefont {Hatano}, \citenamefont {Urata},
  \citenamefont {Iida},\ and\ \citenamefont {Ikuta}}]{Kondo2020}%
  \BibitemOpen
  \bibfield  {author} {\bibinfo {author} {\bibnamefont {Kondo}, \bibfnamefont
  {K.}}, \bibinfo {author} {\bibfnamefont {S.}~\bibnamefont {Motoki}}, \bibinfo
  {author} {\bibfnamefont {T.}~\bibnamefont {Hatano}}, \bibinfo {author}
  {\bibfnamefont {T.}~\bibnamefont {Urata}}, \bibinfo {author} {\bibfnamefont
  {K.}~\bibnamefont {Iida}}, and\ \bibinfo {author} {\bibfnamefont
  {H.}~\bibnamefont {Ikuta}}} (\bibinfo {year} {2020}),\ \href
  {https://doi.org/10.1088/1361-6668/aba353} {\bibfield  {journal} {\bibinfo
  {journal} {Supercond. Sci. Technol.}\ }\textbf {\bibinfo {volume} {33}},\
  \bibinfo {pages} {09LT01}}\BibitemShut {NoStop}%
\bibitem [{\citenamefont {Kong}\ \emph {et~al.}(2025)\citenamefont {Kong},
  \citenamefont {Papaj}, \citenamefont {Kim}, \citenamefont {Zhang},
  \citenamefont {Baum}, \citenamefont {Li}, \citenamefont {Watanabe},
  \citenamefont {Taniguchi}, \citenamefont {Gu}, \citenamefont {Lee},\ and\
  \citenamefont {Nadj-Perge}}]{kong2025cooper}%
  \BibitemOpen
  \bibfield  {author} {\bibinfo {author} {\bibnamefont {Kong}, \bibfnamefont
  {L.}}, \bibinfo {author} {\bibfnamefont {M.}~\bibnamefont {Papaj}}, \bibinfo
  {author} {\bibfnamefont {H.}~\bibnamefont {Kim}}, \bibinfo {author}
  {\bibfnamefont {Y.}~\bibnamefont {Zhang}}, \bibinfo {author} {\bibfnamefont
  {E.}~\bibnamefont {Baum}}, \bibinfo {author} {\bibfnamefont {H.}~\bibnamefont
  {Li}}, \bibinfo {author} {\bibfnamefont {K.}~\bibnamefont {Watanabe}},
  \bibinfo {author} {\bibfnamefont {T.}~\bibnamefont {Taniguchi}}, \bibinfo
  {author} {\bibfnamefont {G.}~\bibnamefont {Gu}}, \bibinfo {author}
  {\bibfnamefont {P.~A.}\ \bibnamefont {Lee}}, and\ \bibinfo {author}
  {\bibfnamefont {S.}~\bibnamefont {Nadj-Perge}}} (\bibinfo {year} {2025}),\
  \href {https://doi.org/10.1038/s41586-025-08703-x} {\bibfield  {journal}
  {\bibinfo  {journal} {Nature}\ }\textbf {\bibinfo {volume} {640}},\ \bibinfo
  {pages} {55}}\BibitemShut {NoStop}%
\bibitem [{\citenamefont {Kong}\ \emph {et~al.}(2019)\citenamefont {Kong},
  \citenamefont {Zhu}, \citenamefont {Papaj}, \citenamefont {Chen},
  \citenamefont {Cao}, \citenamefont {Isobe}, \citenamefont {Xing},
  \citenamefont {Liu}, \citenamefont {Wang}, \citenamefont {Fan}, \citenamefont
  {Sun}, \citenamefont {Du}, \citenamefont {Schneeloch}, \citenamefont {Zhong},
  \citenamefont {Gu}, \citenamefont {Fu}, \citenamefont {Gao},\ and\
  \citenamefont {Ding}}]{kong2019halfinteger}%
  \BibitemOpen
  \bibfield  {author} {\bibinfo {author} {\bibnamefont {Kong}, \bibfnamefont
  {L.}}, \bibinfo {author} {\bibfnamefont {S.}~\bibnamefont {Zhu}}, \bibinfo
  {author} {\bibfnamefont {M.}~\bibnamefont {Papaj}}, \bibinfo {author}
  {\bibfnamefont {H.}~\bibnamefont {Chen}}, \bibinfo {author} {\bibfnamefont
  {L.}~\bibnamefont {Cao}}, \bibinfo {author} {\bibfnamefont {H.}~\bibnamefont
  {Isobe}}, \bibinfo {author} {\bibfnamefont {Y.}~\bibnamefont {Xing}},
  \bibinfo {author} {\bibfnamefont {W.}~\bibnamefont {Liu}}, \bibinfo {author}
  {\bibfnamefont {D.}~\bibnamefont {Wang}}, \bibinfo {author} {\bibfnamefont
  {P.}~\bibnamefont {Fan}}, \bibinfo {author} {\bibfnamefont {Y.}~\bibnamefont
  {Sun}}, \bibinfo {author} {\bibfnamefont {S.}~\bibnamefont {Du}}, \bibinfo
  {author} {\bibfnamefont {J.}~\bibnamefont {Schneeloch}}, \bibinfo {author}
  {\bibfnamefont {R.}~\bibnamefont {Zhong}}, \bibinfo {author} {\bibfnamefont
  {G.}~\bibnamefont {Gu}}, \bibinfo {author} {\bibfnamefont {L.}~\bibnamefont
  {Fu}}, \bibinfo {author} {\bibfnamefont {H.-J.}\ \bibnamefont {Gao}}, and\
  \bibinfo {author} {\bibfnamefont {H.}~\bibnamefont {Ding}}} (\bibinfo {year}
  {2019}),\ \href {https://doi.org/10.1038/s41567-019-0630-5} {\bibfield
  {journal} {\bibinfo  {journal} {Nat. Phys.}\ }\textbf {\bibinfo {volume}
  {15}},\ \bibinfo {pages} {1181}}\BibitemShut {NoStop}%
\bibitem [{\citenamefont {Konno}\ \emph {et~al.}(2021)\citenamefont {Konno},
  \citenamefont {Kurokawa}, \citenamefont {Nabeshima}, \citenamefont
  {Sakishita}, \citenamefont {Ogawa}, \citenamefont {Hosako},\ and\
  \citenamefont {Maeda}}]{Konno2021}%
  \BibitemOpen
  \bibfield  {author} {\bibinfo {author} {\bibnamefont {Konno}, \bibfnamefont
  {T.}}, \bibinfo {author} {\bibfnamefont {H.}~\bibnamefont {Kurokawa}},
  \bibinfo {author} {\bibfnamefont {F.}~\bibnamefont {Nabeshima}}, \bibinfo
  {author} {\bibfnamefont {Y.}~\bibnamefont {Sakishita}}, \bibinfo {author}
  {\bibfnamefont {R.}~\bibnamefont {Ogawa}}, \bibinfo {author} {\bibfnamefont
  {I.}~\bibnamefont {Hosako}}, and\ \bibinfo {author} {\bibfnamefont
  {A.}~\bibnamefont {Maeda}}} (\bibinfo {year} {2021}),\ \href
  {https://doi.org/10.1103/PhysRevB.103.014509} {\bibfield  {journal} {\bibinfo
   {journal} {Phys. Rev. B}\ }\textbf {\bibinfo {volume} {103}},\ \bibinfo
  {pages} {014509}}\BibitemShut {NoStop}%
\bibitem [{\citenamefont {Kontani}\ and\ \citenamefont
  {Onari}(2010)}]{Kontani2010}%
  \BibitemOpen
  \bibfield  {author} {\bibinfo {author} {\bibnamefont {Kontani}, \bibfnamefont
  {H.}}, and\ \bibinfo {author} {\bibfnamefont {S.}~\bibnamefont {Onari}}}
  (\bibinfo {year} {2010}),\ \href
  {https://doi.org/10.1103/PhysRevLett.104.157001} {\bibfield  {journal}
  {\bibinfo  {journal} {Phys. Rev. Lett.}\ }\textbf {\bibinfo {volume} {104}},\
  \bibinfo {pages} {157001}}\BibitemShut {NoStop}%
\bibitem [{\citenamefont {Koshelev}(2019)}]{Koshelev2019}%
  \BibitemOpen
  \bibfield  {author} {\bibinfo {author} {\bibnamefont {Koshelev},
  \bibfnamefont {A.}}} (\bibinfo {year} {2019}),\ \href
  {https://doi.org/10.1103/PhysRevB.100.224503} {\bibfield  {journal} {\bibinfo
   {journal} {Phys. Rev. B}\ }\textbf {\bibinfo {volume} {100}},\ \bibinfo
  {pages} {224503}}\BibitemShut {NoStop}%
\bibitem [{\citenamefont {Kostin}\ \emph {et~al.}(2018)\citenamefont {Kostin},
  \citenamefont {Sprau}, \citenamefont {Kreisel}, \citenamefont {Chong},
  \citenamefont {B{\"o}hmer}, \citenamefont {Canfield}, \citenamefont
  {Hirschfeld}, \citenamefont {Andersen},\ and\ \citenamefont
  {Davis}}]{kostin2018imaging}%
  \BibitemOpen
  \bibfield  {author} {\bibinfo {author} {\bibnamefont {Kostin}, \bibfnamefont
  {A.}}, \bibinfo {author} {\bibfnamefont {P.~O.}\ \bibnamefont {Sprau}},
  \bibinfo {author} {\bibfnamefont {A.}~\bibnamefont {Kreisel}}, \bibinfo
  {author} {\bibfnamefont {Y.~X.}\ \bibnamefont {Chong}}, \bibinfo {author}
  {\bibfnamefont {A.~E.}\ \bibnamefont {B{\"o}hmer}}, \bibinfo {author}
  {\bibfnamefont {P.~C.}\ \bibnamefont {Canfield}}, \bibinfo {author}
  {\bibfnamefont {P.~J.}\ \bibnamefont {Hirschfeld}}, \bibinfo {author}
  {\bibfnamefont {B.~M.}\ \bibnamefont {Andersen}}, and\ \bibinfo {author}
  {\bibfnamefont {J.~C.~S.}\ \bibnamefont {Davis}}} (\bibinfo {year} {2018}),\
  \href {https://doi.org/10.1038/s41563-018-0151-0} {\bibfield  {journal}
  {\bibinfo  {journal} {Nat. Mater.}\ }\textbf {\bibinfo {volume} {17}},\
  \bibinfo {pages} {869}}\BibitemShut {NoStop}%
\bibitem [{\citenamefont {Kothapalli}\ \emph {et~al.}(2016)\citenamefont
  {Kothapalli}, \citenamefont {B{\"o}hmer}, \citenamefont {Jayasekara},
  \citenamefont {Ueland}, \citenamefont {Das}, \citenamefont {Sapkota},
  \citenamefont {Taufour}, \citenamefont {Xiao}, \citenamefont {Alp},
  \citenamefont {Bud'ko}, \citenamefont {Canfield}, \citenamefont {Kreyssig},\
  and\ \citenamefont {Goldman}}]{kothapalli2016strong}%
  \BibitemOpen
  \bibfield  {author} {\bibinfo {author} {\bibnamefont {Kothapalli},
  \bibfnamefont {K.}}, \bibinfo {author} {\bibfnamefont {A.~E.}\ \bibnamefont
  {B{\"o}hmer}}, \bibinfo {author} {\bibfnamefont {W.~T.}\ \bibnamefont
  {Jayasekara}}, \bibinfo {author} {\bibfnamefont {B.~G.}\ \bibnamefont
  {Ueland}}, \bibinfo {author} {\bibfnamefont {P.}~\bibnamefont {Das}},
  \bibinfo {author} {\bibfnamefont {A.}~\bibnamefont {Sapkota}}, \bibinfo
  {author} {\bibfnamefont {V.}~\bibnamefont {Taufour}}, \bibinfo {author}
  {\bibfnamefont {Y.}~\bibnamefont {Xiao}}, \bibinfo {author} {\bibfnamefont
  {E.}~\bibnamefont {Alp}}, \bibinfo {author} {\bibfnamefont {S.~L.}\
  \bibnamefont {Bud'ko}}, \bibinfo {author} {\bibfnamefont {P.~C.}\
  \bibnamefont {Canfield}}, \bibinfo {author} {\bibfnamefont {A.}~\bibnamefont
  {Kreyssig}}, and\ \bibinfo {author} {\bibfnamefont {A.~I.}\ \bibnamefont
  {Goldman}}} (\bibinfo {year} {2016}),\ \href
  {https://doi.org/10.1038/ncomms12728} {\bibfield  {journal} {\bibinfo
  {journal} {Nat. Commun.}\ }\textbf {\bibinfo {volume} {7}},\ \bibinfo {pages}
  {12728}}\BibitemShut {NoStop}%
\bibitem [{\citenamefont {Koz}\ \emph {et~al.}(2013)\citenamefont {Koz},
  \citenamefont {R\"o\ss{}ler}, \citenamefont {Tsirlin}, \citenamefont
  {Wirth},\ and\ \citenamefont {Schwarz}}]{Koz2013}%
  \BibitemOpen
  \bibfield  {author} {\bibinfo {author} {\bibnamefont {Koz}, \bibfnamefont
  {C.}}, \bibinfo {author} {\bibfnamefont {S.}~\bibnamefont {R\"o\ss{}ler}},
  \bibinfo {author} {\bibfnamefont {A.~A.}\ \bibnamefont {Tsirlin}}, \bibinfo
  {author} {\bibfnamefont {S.}~\bibnamefont {Wirth}}, and\ \bibinfo {author}
  {\bibfnamefont {U.}~\bibnamefont {Schwarz}}} (\bibinfo {year} {2013}),\ \href
  {https://doi.org/10.1103/PhysRevB.88.094509} {\bibfield  {journal} {\bibinfo
  {journal} {Phys. Rev. B}\ }\textbf {\bibinfo {volume} {88}},\ \bibinfo
  {pages} {094509}}\BibitemShut {NoStop}%
\bibitem [{\citenamefont {Kreisel}\ \emph {et~al.}(2017)\citenamefont
  {Kreisel}, \citenamefont {Andersen}, \citenamefont {Sprau}, \citenamefont
  {Kostin}, \citenamefont {Davis},\ and\ \citenamefont
  {Hirschfeld}}]{kreisel2017orbital}%
  \BibitemOpen
  \bibfield  {author} {\bibinfo {author} {\bibnamefont {Kreisel}, \bibfnamefont
  {A.}}, \bibinfo {author} {\bibfnamefont {B.~M.}\ \bibnamefont {Andersen}},
  \bibinfo {author} {\bibfnamefont {P.~O.}\ \bibnamefont {Sprau}}, \bibinfo
  {author} {\bibfnamefont {A.}~\bibnamefont {Kostin}}, \bibinfo {author}
  {\bibfnamefont {J.~C.~S.}\ \bibnamefont {Davis}}, and\ \bibinfo {author}
  {\bibfnamefont {P.~J.}\ \bibnamefont {Hirschfeld}}} (\bibinfo {year}
  {2017}),\ \href {https://doi.org/10.1103/PhysRevB.95.174504} {\bibfield
  {journal} {\bibinfo  {journal} {Phys. Rev. B}\ }\textbf {\bibinfo {volume}
  {95}},\ \bibinfo {pages} {174504}}\BibitemShut {NoStop}%
\bibitem [{\citenamefont {Kreisel}\ \emph {et~al.}(2020)\citenamefont
  {Kreisel}, \citenamefont {Hirschfeld},\ and\ \citenamefont
  {Andersen}}]{kreisel2020on}%
  \BibitemOpen
  \bibfield  {author} {\bibinfo {author} {\bibnamefont {Kreisel}, \bibfnamefont
  {A.}}, \bibinfo {author} {\bibfnamefont {P.~J.}\ \bibnamefont {Hirschfeld}},
  and\ \bibinfo {author} {\bibfnamefont {B.~M.}\ \bibnamefont {Andersen}}}
  (\bibinfo {year} {2020}),\ \href {https://doi.org/10.3390/sym12091402}
  {\bibfield  {journal} {\bibinfo  {journal} {Symmetry}\ }\textbf {\bibinfo
  {volume} {12}},\ \bibinfo {pages} {1402}}\BibitemShut {NoStop}%
\bibitem [{\citenamefont {Kreisel}\ \emph {et~al.}(2022)\citenamefont
  {Kreisel}, \citenamefont {Hirschfeld},\ and\ \citenamefont
  {Andersen}}]{kreisel2022theory}%
  \BibitemOpen
  \bibfield  {author} {\bibinfo {author} {\bibnamefont {Kreisel}, \bibfnamefont
  {A.}}, \bibinfo {author} {\bibfnamefont {P.~J.}\ \bibnamefont {Hirschfeld}},
  and\ \bibinfo {author} {\bibfnamefont {B.~M.}\ \bibnamefont {Andersen}}}
  (\bibinfo {year} {2022}),\ \href {https://doi.org/10.3389/fphy.2022.859424}
  {\bibfield  {journal} {\bibinfo  {journal} {Frontiers in Physics}\ }\textbf
  {\bibinfo {volume} {10}},\ \bibinfo {pages} {859424}}\BibitemShut {NoStop}%
\bibitem [{\citenamefont {Kreyssig}\ \emph {et~al.}(2018)\citenamefont
  {Kreyssig}, \citenamefont {Wilde}, \citenamefont {B{\"o}hmer}, \citenamefont
  {Tian}, \citenamefont {Meier}, \citenamefont {Li}, \citenamefont {Ueland},
  \citenamefont {Xu}, \citenamefont {Bud'ko}, \citenamefont {Canfield},
  \citenamefont {McQueeney},\ and\ \citenamefont
  {Goldman}}]{kreyssig2018antiferromagnetic}%
  \BibitemOpen
  \bibfield  {author} {\bibinfo {author} {\bibnamefont {Kreyssig},
  \bibfnamefont {A.}}, \bibinfo {author} {\bibfnamefont {J.~M.}\ \bibnamefont
  {Wilde}}, \bibinfo {author} {\bibfnamefont {A.~E.}\ \bibnamefont
  {B{\"o}hmer}}, \bibinfo {author} {\bibfnamefont {W.}~\bibnamefont {Tian}},
  \bibinfo {author} {\bibfnamefont {W.~R.}\ \bibnamefont {Meier}}, \bibinfo
  {author} {\bibfnamefont {B.}~\bibnamefont {Li}}, \bibinfo {author}
  {\bibfnamefont {B.~G.}\ \bibnamefont {Ueland}}, \bibinfo {author}
  {\bibfnamefont {M.}~\bibnamefont {Xu}}, \bibinfo {author} {\bibfnamefont
  {S.~L.}\ \bibnamefont {Bud'ko}}, \bibinfo {author} {\bibfnamefont {P.~C.}\
  \bibnamefont {Canfield}}, \bibinfo {author} {\bibfnamefont {R.~J.}\
  \bibnamefont {McQueeney}}, and\ \bibinfo {author} {\bibfnamefont {A.~I.}\
  \bibnamefont {Goldman}}} (\bibinfo {year} {2018}),\ \href
  {https://doi.org/10.1103/PhysRevB.97.224521} {\bibfield  {journal} {\bibinfo
  {journal} {Phys. Rev. B}\ }\textbf {\bibinfo {volume} {97}},\ \bibinfo
  {pages} {224521}}\BibitemShut {NoStop}%
\bibitem [{\citenamefont {Krzton-Maziopa}\ \emph {et~al.}(2011)\citenamefont
  {Krzton-Maziopa}, \citenamefont {Shermadini}, \citenamefont {Pomjakushina},
  \citenamefont {Pomjakushin}, \citenamefont {Bendele}, \citenamefont {Amato},
  \citenamefont {Khasanov}, \citenamefont {Luetkens},\ and\ \citenamefont
  {Conder}}]{Krzton-Maziopa2011}%
  \BibitemOpen
  \bibfield  {author} {\bibinfo {author} {\bibnamefont {Krzton-Maziopa},
  \bibfnamefont {A.}}, \bibinfo {author} {\bibfnamefont {Z.}~\bibnamefont
  {Shermadini}}, \bibinfo {author} {\bibfnamefont {E.}~\bibnamefont
  {Pomjakushina}}, \bibinfo {author} {\bibfnamefont {V.}~\bibnamefont
  {Pomjakushin}}, \bibinfo {author} {\bibfnamefont {M.}~\bibnamefont
  {Bendele}}, \bibinfo {author} {\bibfnamefont {A.}~\bibnamefont {Amato}},
  \bibinfo {author} {\bibfnamefont {R.}~\bibnamefont {Khasanov}}, \bibinfo
  {author} {\bibfnamefont {H.}~\bibnamefont {Luetkens}}, and\ \bibinfo {author}
  {\bibfnamefont {K.}~\bibnamefont {Conder}}} (\bibinfo {year} {2011}),\ \href
  {https://doi.org/10.1088/0953-8984/23/5/052203} {\bibfield  {journal}
  {\bibinfo  {journal} {J. Phys.: Condens. Matter}\ }\textbf {\bibinfo {volume}
  {23}},\ \bibinfo {pages} {052203}}\BibitemShut {NoStop}%
\bibitem [{\citenamefont {{Krzton-Maziopa}}\ \emph {et~al.}(2016)\citenamefont
  {{Krzton-Maziopa}}, \citenamefont {Svitlyk}, \citenamefont {Pomjakushina},
  \citenamefont {Puzniak},\ and\ \citenamefont
  {Conder}}]{krzton-maziopa2016superconductivity}%
  \BibitemOpen
  \bibfield  {author} {\bibinfo {author} {\bibnamefont {{Krzton-Maziopa}},
  \bibfnamefont {A.}}, \bibinfo {author} {\bibfnamefont {V.}~\bibnamefont
  {Svitlyk}}, \bibinfo {author} {\bibfnamefont {E.}~\bibnamefont
  {Pomjakushina}}, \bibinfo {author} {\bibfnamefont {R.}~\bibnamefont
  {Puzniak}}, and\ \bibinfo {author} {\bibfnamefont {K.}~\bibnamefont
  {Conder}}} (\bibinfo {year} {2016}),\ \href
  {https://doi.org/10.1088/0953-8984/28/29/293002} {\bibfield  {journal}
  {\bibinfo  {journal} {J. Phys.: Condens. Matter}\ }\textbf {\bibinfo {volume}
  {28}},\ \bibinfo {pages} {293002}}\BibitemShut {NoStop}%
\bibitem [{\citenamefont {Kudo}\ \emph
  {et~al.}(2014{\natexlab{a}})\citenamefont {Kudo}, \citenamefont {Kitahama},
  \citenamefont {Fujimura}, \citenamefont {Mizukami}, \citenamefont {Ota},\
  and\ \citenamefont {Nohara}}]{Kudo2014b}%
  \BibitemOpen
  \bibfield  {author} {\bibinfo {author} {\bibnamefont {Kudo}, \bibfnamefont
  {K.}}, \bibinfo {author} {\bibfnamefont {Y.}~\bibnamefont {Kitahama}},
  \bibinfo {author} {\bibfnamefont {K.}~\bibnamefont {Fujimura}}, \bibinfo
  {author} {\bibfnamefont {T.}~\bibnamefont {Mizukami}}, \bibinfo {author}
  {\bibfnamefont {H.}~\bibnamefont {Ota}}, and\ \bibinfo {author}
  {\bibfnamefont {M.}~\bibnamefont {Nohara}}} (\bibinfo {year}
  {2014}{\natexlab{a}}),\ \href {https://doi.org/10.7566/jpsj.83.093705}
  {\bibfield  {journal} {\bibinfo  {journal} {J. Phys. Soc. Jpn.}\ }\textbf
  {\bibinfo {volume} {83}},\ \bibinfo {pages} {093705}}\BibitemShut {NoStop}%
\bibitem [{\citenamefont {Kudo}\ \emph {et~al.}(2013)\citenamefont {Kudo},
  \citenamefont {Mitsuoka}, \citenamefont {Takasuga}, \citenamefont {Sugiyama},
  \citenamefont {Sugawara}, \citenamefont {Katayama}, \citenamefont {Sawa},
  \citenamefont {Kubo}, \citenamefont {Takamori}, \citenamefont {Ichioka} \emph
  {et~al.}}]{Kudo2013}%
  \BibitemOpen
  \bibfield  {author} {\bibinfo {author} {\bibnamefont {Kudo}, \bibfnamefont
  {K.}}, \bibinfo {author} {\bibfnamefont {D.}~\bibnamefont {Mitsuoka}},
  \bibinfo {author} {\bibfnamefont {M.}~\bibnamefont {Takasuga}}, \bibinfo
  {author} {\bibfnamefont {Y.}~\bibnamefont {Sugiyama}}, \bibinfo {author}
  {\bibfnamefont {K.}~\bibnamefont {Sugawara}}, \bibinfo {author}
  {\bibfnamefont {N.}~\bibnamefont {Katayama}}, \bibinfo {author}
  {\bibfnamefont {H.}~\bibnamefont {Sawa}}, \bibinfo {author} {\bibfnamefont
  {H.~S.}\ \bibnamefont {Kubo}}, \bibinfo {author} {\bibfnamefont
  {K.}~\bibnamefont {Takamori}}, \bibinfo {author} {\bibfnamefont
  {M.}~\bibnamefont {Ichioka}},  \emph {et~al.}} (\bibinfo {year} {2013}),\
  \href {https://doi.org/10.1038/srep03101} {\bibfield  {journal} {\bibinfo
  {journal} {Sci. Rep.}\ }\textbf {\bibinfo {volume} {3}},\ \bibinfo {pages}
  {3101}}\BibitemShut {NoStop}%
\bibitem [{\citenamefont {Kudo}\ \emph
  {et~al.}(2014{\natexlab{b}})\citenamefont {Kudo}, \citenamefont {Mizukami},
  \citenamefont {Kitahama}, \citenamefont {Mitsuoka}, \citenamefont {Iba},
  \citenamefont {Fujimura}, \citenamefont {Nishimoto}, \citenamefont
  {Hiraoka},\ and\ \citenamefont {Nohara}}]{Kudo2014a}%
  \BibitemOpen
  \bibfield  {author} {\bibinfo {author} {\bibnamefont {Kudo}, \bibfnamefont
  {K.}}, \bibinfo {author} {\bibfnamefont {T.}~\bibnamefont {Mizukami}},
  \bibinfo {author} {\bibfnamefont {Y.}~\bibnamefont {Kitahama}}, \bibinfo
  {author} {\bibfnamefont {D.}~\bibnamefont {Mitsuoka}}, \bibinfo {author}
  {\bibfnamefont {K.}~\bibnamefont {Iba}}, \bibinfo {author} {\bibfnamefont
  {K.}~\bibnamefont {Fujimura}}, \bibinfo {author} {\bibfnamefont
  {N.}~\bibnamefont {Nishimoto}}, \bibinfo {author} {\bibfnamefont
  {Y.}~\bibnamefont {Hiraoka}}, and\ \bibinfo {author} {\bibfnamefont
  {M.}~\bibnamefont {Nohara}}} (\bibinfo {year} {2014}{\natexlab{b}}),\ \href
  {https://doi.org/10.7566/jpsj.83.025001} {\bibfield  {journal} {\bibinfo
  {journal} {J. Phys. Soc. Jpn.}\ }\textbf {\bibinfo {volume} {83}},\ \bibinfo
  {pages} {025001}}\BibitemShut {NoStop}%
\bibitem [{\citenamefont {Kuo}\ \emph {et~al.}(2016)\citenamefont {Kuo},
  \citenamefont {Chu}, \citenamefont {Palmstrom}, \citenamefont {Kivelson},\
  and\ \citenamefont {Fisher}}]{kuo2016ubiquitous}%
  \BibitemOpen
  \bibfield  {author} {\bibinfo {author} {\bibnamefont {Kuo}, \bibfnamefont
  {H.-H.}}, \bibinfo {author} {\bibfnamefont {J.-H.}\ \bibnamefont {Chu}},
  \bibinfo {author} {\bibfnamefont {J.~C.}\ \bibnamefont {Palmstrom}}, \bibinfo
  {author} {\bibfnamefont {S.~A.}\ \bibnamefont {Kivelson}}, and\ \bibinfo
  {author} {\bibfnamefont {I.~R.}\ \bibnamefont {Fisher}}} (\bibinfo {year}
  {2016}),\ \href {https://doi.org/10.1126/science.aab0103} {\bibfield
  {journal} {\bibinfo  {journal} {Science}\ }\textbf {\bibinfo {volume}
  {352}},\ \bibinfo {pages} {958}}\BibitemShut {NoStop}%
\bibitem [{\citenamefont {Kuo}\ \emph {et~al.}(2013)\citenamefont {Kuo},
  \citenamefont {Shapiro}, \citenamefont {Riggs},\ and\ \citenamefont
  {Fisher}}]{kuo2013measurement}%
  \BibitemOpen
  \bibfield  {author} {\bibinfo {author} {\bibnamefont {Kuo}, \bibfnamefont
  {H.-H.}}, \bibinfo {author} {\bibfnamefont {M.~C.}\ \bibnamefont {Shapiro}},
  \bibinfo {author} {\bibfnamefont {S.~C.}\ \bibnamefont {Riggs}}, and\
  \bibinfo {author} {\bibfnamefont {I.~R.}\ \bibnamefont {Fisher}}} (\bibinfo
  {year} {2013}),\ \href {https://doi.org/10.1103/PhysRevB.88.085113}
  {\bibfield  {journal} {\bibinfo  {journal} {Phys. Rev. B}\ }\textbf {\bibinfo
  {volume} {88}},\ \bibinfo {pages} {085113}}\BibitemShut {NoStop}%
\bibitem [{\citenamefont {Kurita}\ \emph
  {et~al.}(2011{\natexlab{a}})\citenamefont {Kurita}, \citenamefont {Kimata},
  \citenamefont {Kodama}, \citenamefont {Harada}, \citenamefont {Tomita},
  \citenamefont {Suzuki}, \citenamefont {Matsumoto}, \citenamefont {Murata},
  \citenamefont {Uji},\ and\ \citenamefont {Terashima}}]{Kurita2011}%
  \BibitemOpen
  \bibfield  {author} {\bibinfo {author} {\bibnamefont {Kurita}, \bibfnamefont
  {N.}}, \bibinfo {author} {\bibfnamefont {M.}~\bibnamefont {Kimata}}, \bibinfo
  {author} {\bibfnamefont {K.}~\bibnamefont {Kodama}}, \bibinfo {author}
  {\bibfnamefont {A.}~\bibnamefont {Harada}}, \bibinfo {author} {\bibfnamefont
  {M.}~\bibnamefont {Tomita}}, \bibinfo {author} {\bibfnamefont {H.~S.}\
  \bibnamefont {Suzuki}}, \bibinfo {author} {\bibfnamefont {T.}~\bibnamefont
  {Matsumoto}}, \bibinfo {author} {\bibfnamefont {K.}~\bibnamefont {Murata}},
  \bibinfo {author} {\bibfnamefont {S.}~\bibnamefont {Uji}}, and\ \bibinfo
  {author} {\bibfnamefont {T.}~\bibnamefont {Terashima}}} (\bibinfo {year}
  {2011}{\natexlab{a}}),\ \href {https://doi.org/10.1103/PhysRevB.83.214513}
  {\bibfield  {journal} {\bibinfo  {journal} {Phys. Rev. B}\ }\textbf {\bibinfo
  {volume} {83}},\ \bibinfo {pages} {214513}}\BibitemShut {NoStop}%
\bibitem [{\citenamefont {Kurita}\ \emph
  {et~al.}(2011{\natexlab{b}})\citenamefont {Kurita}, \citenamefont {Kimata},
  \citenamefont {Kodama}, \citenamefont {Harada}, \citenamefont {Tomita},
  \citenamefont {Suzuki}, \citenamefont {Matsumoto}, \citenamefont {Murata},
  \citenamefont {Uji},\ and\ \citenamefont {Terashima}}]{Kurita2011b}%
  \BibitemOpen
  \bibfield  {author} {\bibinfo {author} {\bibnamefont {Kurita}, \bibfnamefont
  {N.}}, \bibinfo {author} {\bibfnamefont {M.}~\bibnamefont {Kimata}}, \bibinfo
  {author} {\bibfnamefont {K.}~\bibnamefont {Kodama}}, \bibinfo {author}
  {\bibfnamefont {A.}~\bibnamefont {Harada}}, \bibinfo {author} {\bibfnamefont
  {M.}~\bibnamefont {Tomita}}, \bibinfo {author} {\bibfnamefont {H.~S.}\
  \bibnamefont {Suzuki}}, \bibinfo {author} {\bibfnamefont {T.}~\bibnamefont
  {Matsumoto}}, \bibinfo {author} {\bibfnamefont {K.}~\bibnamefont {Murata}},
  \bibinfo {author} {\bibfnamefont {S.}~\bibnamefont {Uji}}, and\ \bibinfo
  {author} {\bibfnamefont {T.}~\bibnamefont {Terashima}}} (\bibinfo {year}
  {2011}{\natexlab{b}}),\ \href {https://doi.org/10.1103/PhysRevB.83.100501}
  {\bibfield  {journal} {\bibinfo  {journal} {Phys. Rev. B}\ }\textbf {\bibinfo
  {volume} {83}},\ \bibinfo {pages} {100501}}\BibitemShut {NoStop}%
\bibitem [{\citenamefont {Kuroki}\ \emph {et~al.}(2009)\citenamefont {Kuroki},
  \citenamefont {Onari}, \citenamefont {Arita}, \citenamefont {Usui},
  \citenamefont {Tanaka}, \citenamefont {Kontani},\ and\ \citenamefont
  {Aoki}}]{Kuroki2009a}%
  \BibitemOpen
  \bibfield  {author} {\bibinfo {author} {\bibnamefont {Kuroki}, \bibfnamefont
  {K.}}, \bibinfo {author} {\bibfnamefont {S.}~\bibnamefont {Onari}}, \bibinfo
  {author} {\bibfnamefont {R.}~\bibnamefont {Arita}}, \bibinfo {author}
  {\bibfnamefont {H.}~\bibnamefont {Usui}}, \bibinfo {author} {\bibfnamefont
  {Y.}~\bibnamefont {Tanaka}}, \bibinfo {author} {\bibfnamefont
  {H.}~\bibnamefont {Kontani}}, and\ \bibinfo {author} {\bibfnamefont
  {H.}~\bibnamefont {Aoki}}} (\bibinfo {year} {2009}),\ \href
  {https://doi.org/10.1103/PhysRevLett.102.109902} {\bibfield  {journal}
  {\bibinfo  {journal} {Phys. Rev. Lett.}\ }\textbf {\bibinfo {volume} {102}},\
  \bibinfo {pages} {109902}}\BibitemShut {NoStop}%
\bibitem [{\citenamefont {Kushnirenko}\ \emph {et~al.}(2018)\citenamefont
  {Kushnirenko}, \citenamefont {Fedorov}, \citenamefont {Haubold},
  \citenamefont {Thirupathaiah}, \citenamefont {Wolf}, \citenamefont
  {Aswartham}, \citenamefont {Morozov}, \citenamefont {Kim}, \citenamefont
  {B\"uchner},\ and\ \citenamefont {Borisenko}}]{kushinirenko2018three}%
  \BibitemOpen
  \bibfield  {author} {\bibinfo {author} {\bibnamefont {Kushnirenko},
  \bibfnamefont {Y.~S.}}, \bibinfo {author} {\bibfnamefont {A.~V.}\
  \bibnamefont {Fedorov}}, \bibinfo {author} {\bibfnamefont {E.}~\bibnamefont
  {Haubold}}, \bibinfo {author} {\bibfnamefont {S.}~\bibnamefont
  {Thirupathaiah}}, \bibinfo {author} {\bibfnamefont {T.}~\bibnamefont {Wolf}},
  \bibinfo {author} {\bibfnamefont {S.}~\bibnamefont {Aswartham}}, \bibinfo
  {author} {\bibfnamefont {I.}~\bibnamefont {Morozov}}, \bibinfo {author}
  {\bibfnamefont {T.~K.}\ \bibnamefont {Kim}}, \bibinfo {author} {\bibfnamefont
  {B.}~\bibnamefont {B\"uchner}}, and\ \bibinfo {author} {\bibfnamefont
  {S.~V.}\ \bibnamefont {Borisenko}}} (\bibinfo {year} {2018}),\ \href
  {https://doi.org/10.1103/PhysRevB.97.180501} {\bibfield  {journal} {\bibinfo
  {journal} {Phys. Rev. B}\ }\textbf {\bibinfo {volume} {97}},\ \bibinfo
  {pages} {180501}}\BibitemShut {NoStop}%
\bibitem [{\citenamefont {Lahiri}\ \emph {et~al.}(2022)\citenamefont {Lahiri},
  \citenamefont {Klein},\ and\ \citenamefont {Fernandes}}]{lahiri2022defect}%
  \BibitemOpen
  \bibfield  {author} {\bibinfo {author} {\bibnamefont {Lahiri}, \bibfnamefont
  {A.}}, \bibinfo {author} {\bibfnamefont {A.}~\bibnamefont {Klein}}, and\
  \bibinfo {author} {\bibfnamefont {R.~M.}\ \bibnamefont {Fernandes}}}
  (\bibinfo {year} {2022}),\ \href
  {https://doi.org/10.1103/PhysRevB.106.L140503} {\bibfield  {journal}
  {\bibinfo  {journal} {Phys. Rev. B}\ }\textbf {\bibinfo {volume} {106}},\
  \bibinfo {pages} {L140503}}\BibitemShut {NoStop}%
\bibitem [{\citenamefont {Lai}\ \emph {et~al.}(2015)\citenamefont {Lai},
  \citenamefont {Zhang}, \citenamefont {Wang}, \citenamefont {Wang},
  \citenamefont {Zhang}, \citenamefont {Lin},\ and\ \citenamefont
  {Huang}}]{Lai2015}%
  \BibitemOpen
  \bibfield  {author} {\bibinfo {author} {\bibnamefont {Lai}, \bibfnamefont
  {X.}}, \bibinfo {author} {\bibfnamefont {H.}~\bibnamefont {Zhang}}, \bibinfo
  {author} {\bibfnamefont {Y.}~\bibnamefont {Wang}}, \bibinfo {author}
  {\bibfnamefont {X.}~\bibnamefont {Wang}}, \bibinfo {author} {\bibfnamefont
  {X.}~\bibnamefont {Zhang}}, \bibinfo {author} {\bibfnamefont
  {J.}~\bibnamefont {Lin}}, and\ \bibinfo {author} {\bibfnamefont
  {F.}~\bibnamefont {Huang}}} (\bibinfo {year} {2015}),\ \href
  {https://doi.org/10.1021/jacs.5b06687} {\bibfield  {journal} {\bibinfo
  {journal} {J. Am. Chem. Soc.}\ }\textbf {\bibinfo {volume} {137}},\ \bibinfo
  {pages} {10148}}\BibitemShut {NoStop}%
\bibitem [{\citenamefont {Lanat\`a}\ \emph {et~al.}(2013)\citenamefont
  {Lanat\`a}, \citenamefont {Strand}, \citenamefont {Giovannetti},
  \citenamefont {Hellsing}, \citenamefont {de' Medici},\ and\ \citenamefont
  {Capone}}]{lanata2013orbital}%
  \BibitemOpen
  \bibfield  {author} {\bibinfo {author} {\bibnamefont {Lanat\`a},
  \bibfnamefont {N.}}, \bibinfo {author} {\bibfnamefont {H.~U.~R.}\
  \bibnamefont {Strand}}, \bibinfo {author} {\bibfnamefont {G.}~\bibnamefont
  {Giovannetti}}, \bibinfo {author} {\bibfnamefont {B.}~\bibnamefont
  {Hellsing}}, \bibinfo {author} {\bibfnamefont {L.}~\bibnamefont {de'
  Medici}}, and\ \bibinfo {author} {\bibfnamefont {M.}~\bibnamefont {Capone}}}
  (\bibinfo {year} {2013}),\ \href {https://doi.org/10.1103/PhysRevB.87.045122}
  {\bibfield  {journal} {\bibinfo  {journal} {Phys. Rev. B}\ }\textbf {\bibinfo
  {volume} {87}},\ \bibinfo {pages} {045122}}\BibitemShut {NoStop}%
\bibitem [{\citenamefont {Larkin}\ and\ \citenamefont
  {Ovchinnikov}(1964)}]{larkin1964nonuniform}%
  \BibitemOpen
  \bibfield  {author} {\bibinfo {author} {\bibnamefont {Larkin}, \bibfnamefont
  {A.~I.}}, and\ \bibinfo {author} {\bibfnamefont {Y.~N.}\ \bibnamefont
  {Ovchinnikov}}} (\bibinfo {year} {1964}),\ \href@noop {} {\bibfield
  {journal} {\bibinfo  {journal} {Zh. Eksp. Teor. Fiz.}\ }\textbf {\bibinfo
  {volume} {47}},\ \bibinfo {pages} {1136}}\BibitemShut {NoStop}%
\bibitem [{\citenamefont {Lee}\ \emph {et~al.}(2012)\citenamefont {Lee},
  \citenamefont {Kihou}, \citenamefont {Iyo}, \citenamefont {Kito},
  \citenamefont {Shirage},\ and\ \citenamefont {Eisaki}}]{Lee2012}%
  \BibitemOpen
  \bibfield  {author} {\bibinfo {author} {\bibnamefont {Lee}, \bibfnamefont
  {C.}}, \bibinfo {author} {\bibfnamefont {K.}~\bibnamefont {Kihou}}, \bibinfo
  {author} {\bibfnamefont {A.}~\bibnamefont {Iyo}}, \bibinfo {author}
  {\bibfnamefont {H.}~\bibnamefont {Kito}}, \bibinfo {author} {\bibfnamefont
  {P.}~\bibnamefont {Shirage}}, and\ \bibinfo {author} {\bibfnamefont
  {H.}~\bibnamefont {Eisaki}}} (\bibinfo {year} {2012}),\ \href
  {https://doi.org/10.1016/j.ssc.2011.12.012} {\bibfield  {journal} {\bibinfo
  {journal} {Solid State Communications}\ }\textbf {\bibinfo {volume} {152}},\
  \bibinfo {pages} {644}}\BibitemShut {NoStop}%
\bibitem [{\citenamefont {Lee}\ \emph {et~al.}(2008)\citenamefont {Lee},
  \citenamefont {Iyo}, \citenamefont {Eisaki}, \citenamefont {Kito},
  \citenamefont {Teresa Fernandez-Diaz}, \citenamefont {Ito}, \citenamefont
  {Kihou}, \citenamefont {Matsuhata}, \citenamefont {Braden},\ and\
  \citenamefont {Yamada}}]{Lee2008}%
  \BibitemOpen
  \bibfield  {author} {\bibinfo {author} {\bibnamefont {Lee}, \bibfnamefont
  {C.-H.}}, \bibinfo {author} {\bibfnamefont {A.}~\bibnamefont {Iyo}}, \bibinfo
  {author} {\bibfnamefont {H.}~\bibnamefont {Eisaki}}, \bibinfo {author}
  {\bibfnamefont {H.}~\bibnamefont {Kito}}, \bibinfo {author} {\bibfnamefont
  {M.}~\bibnamefont {Teresa Fernandez-Diaz}}, \bibinfo {author} {\bibfnamefont
  {T.}~\bibnamefont {Ito}}, \bibinfo {author} {\bibfnamefont {K.}~\bibnamefont
  {Kihou}}, \bibinfo {author} {\bibfnamefont {H.}~\bibnamefont {Matsuhata}},
  \bibinfo {author} {\bibfnamefont {M.}~\bibnamefont {Braden}}, and\ \bibinfo
  {author} {\bibfnamefont {K.}~\bibnamefont {Yamada}}} (\bibinfo {year}
  {2008}),\ \href {https://doi.org/10.1143/jpsj.77.083704} {\bibfield
  {journal} {\bibinfo  {journal} {J. Phys. Soc. Jpn.}\ }\textbf {\bibinfo
  {volume} {77}},\ \bibinfo {pages} {083704}}\BibitemShut {NoStop}%
\bibitem [{\citenamefont {Lee}(2015)}]{Lee2015}%
  \BibitemOpen
  \bibfield  {author} {\bibinfo {author} {\bibnamefont {Lee}, \bibfnamefont
  {D.-H.}}} (\bibinfo {year} {2015}),\ \href
  {https://doi.org/10.1088/1674-1056/24/11/117405} {\bibfield  {journal}
  {\bibinfo  {journal} {Chinese Physics B}\ }\textbf {\bibinfo {volume} {24}},\
  \bibinfo {pages} {117405}}\BibitemShut {NoStop}%
\bibitem [{\citenamefont {Lee}\ \emph {et~al.}(2014)\citenamefont {Lee},
  \citenamefont {Schmitt}, \citenamefont {Moore}, \citenamefont {Johnston},
  \citenamefont {Cui}, \citenamefont {Li}, \citenamefont {Yi}, \citenamefont
  {Liu}, \citenamefont {Hashimoto}, \citenamefont {Zhang} \emph
  {et~al.}}]{Lee2014}%
  \BibitemOpen
  \bibfield  {author} {\bibinfo {author} {\bibnamefont {Lee}, \bibfnamefont
  {J.}}, \bibinfo {author} {\bibfnamefont {F.}~\bibnamefont {Schmitt}},
  \bibinfo {author} {\bibfnamefont {R.}~\bibnamefont {Moore}}, \bibinfo
  {author} {\bibfnamefont {S.}~\bibnamefont {Johnston}}, \bibinfo {author}
  {\bibfnamefont {Y.-T.}\ \bibnamefont {Cui}}, \bibinfo {author} {\bibfnamefont
  {W.}~\bibnamefont {Li}}, \bibinfo {author} {\bibfnamefont {M.}~\bibnamefont
  {Yi}}, \bibinfo {author} {\bibfnamefont {Z.}~\bibnamefont {Liu}}, \bibinfo
  {author} {\bibfnamefont {M.}~\bibnamefont {Hashimoto}}, \bibinfo {author}
  {\bibfnamefont {Y.}~\bibnamefont {Zhang}},  \emph {et~al.}} (\bibinfo {year}
  {2014}),\ \href {https://doi.org/10.1038/nature13894} {\bibfield  {journal}
  {\bibinfo  {journal} {Nature}\ }\textbf {\bibinfo {volume} {515}},\ \bibinfo
  {pages} {245}}\BibitemShut {NoStop}%
\bibitem [{\citenamefont {Lei}\ \emph {et~al.}(2016)\citenamefont {Lei},
  \citenamefont {Cui}, \citenamefont {Xiang}, \citenamefont {Shang},
  \citenamefont {Wang}, \citenamefont {Ye}, \citenamefont {Luo}, \citenamefont
  {Wu}, \citenamefont {Sun},\ and\ \citenamefont {Chen}}]{lei2016evolution}%
  \BibitemOpen
  \bibfield  {author} {\bibinfo {author} {\bibnamefont {Lei}, \bibfnamefont
  {B.}}, \bibinfo {author} {\bibfnamefont {J.~H.}\ \bibnamefont {Cui}},
  \bibinfo {author} {\bibfnamefont {Z.~J.}\ \bibnamefont {Xiang}}, \bibinfo
  {author} {\bibfnamefont {C.}~\bibnamefont {Shang}}, \bibinfo {author}
  {\bibfnamefont {N.~Z.}\ \bibnamefont {Wang}}, \bibinfo {author}
  {\bibfnamefont {G.~J.}\ \bibnamefont {Ye}}, \bibinfo {author} {\bibfnamefont
  {X.~G.}\ \bibnamefont {Luo}}, \bibinfo {author} {\bibfnamefont
  {T.}~\bibnamefont {Wu}}, \bibinfo {author} {\bibfnamefont {Z.}~\bibnamefont
  {Sun}}, and\ \bibinfo {author} {\bibfnamefont {X.~H.}\ \bibnamefont {Chen}}}
  (\bibinfo {year} {2016}),\ \href
  {https://doi.org/10.1103/PhysRevLett.116.077002} {\bibfield  {journal}
  {\bibinfo  {journal} {Phys. Rev. Lett.}\ }\textbf {\bibinfo {volume} {116}},\
  \bibinfo {pages} {077002}}\BibitemShut {NoStop}%
\bibitem [{\citenamefont {Lei}\ \emph {et~al.}(2017)\citenamefont {Lei},
  \citenamefont {Wang}, \citenamefont {Shang}, \citenamefont {Meng},
  \citenamefont {Ma}, \citenamefont {Luo}, \citenamefont {Wu}, \citenamefont
  {Sun}, \citenamefont {Wang}, \citenamefont {Jiang} \emph {et~al.}}]{Lei2017}%
  \BibitemOpen
  \bibfield  {author} {\bibinfo {author} {\bibnamefont {Lei}, \bibfnamefont
  {B.}}, \bibinfo {author} {\bibfnamefont {N.}~\bibnamefont {Wang}}, \bibinfo
  {author} {\bibfnamefont {C.}~\bibnamefont {Shang}}, \bibinfo {author}
  {\bibfnamefont {F.}~\bibnamefont {Meng}}, \bibinfo {author} {\bibfnamefont
  {L.}~\bibnamefont {Ma}}, \bibinfo {author} {\bibfnamefont {X.}~\bibnamefont
  {Luo}}, \bibinfo {author} {\bibfnamefont {T.}~\bibnamefont {Wu}}, \bibinfo
  {author} {\bibfnamefont {Z.}~\bibnamefont {Sun}}, \bibinfo {author}
  {\bibfnamefont {Y.}~\bibnamefont {Wang}}, \bibinfo {author} {\bibfnamefont
  {Z.}~\bibnamefont {Jiang}},  \emph {et~al.}} (\bibinfo {year} {2017}),\ \href
  {https://doi.org/10.1103/PhysRevB.95.020503} {\bibfield  {journal} {\bibinfo
  {journal} {Phys. Rev. B}\ }\textbf {\bibinfo {volume} {95}},\ \bibinfo
  {pages} {020503}}\BibitemShut {NoStop}%
\bibitem [{\citenamefont {Lei}\ \emph {et~al.}(2011{\natexlab{a}})\citenamefont
  {Lei}, \citenamefont {Abeykoon}, \citenamefont {Bozin},\ and\ \citenamefont
  {Petrovic}}]{Lei2011a}%
  \BibitemOpen
  \bibfield  {author} {\bibinfo {author} {\bibnamefont {Lei}, \bibfnamefont
  {H.}}, \bibinfo {author} {\bibfnamefont {M.}~\bibnamefont {Abeykoon}},
  \bibinfo {author} {\bibfnamefont {E.~S.}\ \bibnamefont {Bozin}}, and\
  \bibinfo {author} {\bibfnamefont {C.}~\bibnamefont {Petrovic}}} (\bibinfo
  {year} {2011}{\natexlab{a}}),\ \href
  {https://doi.org/10.1103/PhysRevB.83.180503} {\bibfield  {journal} {\bibinfo
  {journal} {Phys. Rev. B}\ }\textbf {\bibinfo {volume} {83}},\ \bibinfo
  {pages} {180503}}\BibitemShut {NoStop}%
\bibitem [{\citenamefont {Lei}\ \emph {et~al.}(2011{\natexlab{b}})\citenamefont
  {Lei}, \citenamefont {Bozin}, \citenamefont {Wang},\ and\ \citenamefont
  {Petrovic}}]{Lei2011b}%
  \BibitemOpen
  \bibfield  {author} {\bibinfo {author} {\bibnamefont {Lei}, \bibfnamefont
  {H.}}, \bibinfo {author} {\bibfnamefont {E.~S.}\ \bibnamefont {Bozin}},
  \bibinfo {author} {\bibfnamefont {K.}~\bibnamefont {Wang}}, and\ \bibinfo
  {author} {\bibfnamefont {C.}~\bibnamefont {Petrovic}}} (\bibinfo {year}
  {2011}{\natexlab{b}}),\ \href {https://doi.org/10.1103/PhysRevB.84.060506}
  {\bibfield  {journal} {\bibinfo  {journal} {Phys. Rev. B}\ }\textbf {\bibinfo
  {volume} {84}},\ \bibinfo {pages} {060506}}\BibitemShut {NoStop}%
\bibitem [{\citenamefont {Li}\ \emph {et~al.}(2024)\citenamefont {Li},
  \citenamefont {Mo},\ and\ \citenamefont {Lu}}]{li2024uniaxial}%
  \BibitemOpen
  \bibfield  {author} {\bibinfo {author} {\bibnamefont {Li}, \bibfnamefont
  {C.}}, \bibinfo {author} {\bibfnamefont {Z.}~\bibnamefont {Mo}}, and\
  \bibinfo {author} {\bibfnamefont {X.}~\bibnamefont {Lu}}} (\bibinfo {year}
  {2024}),\ \href {https://doi.org/10.7498/aps.73.20241080} {\bibfield
  {journal} {\bibinfo  {journal} {Acta Phys. Sin.}\ }\textbf {\bibinfo {volume}
  {73}},\ \bibinfo {pages} {197103}}\BibitemShut {NoStop}%
\bibitem [{\citenamefont {Li}\ \emph {et~al.}(2011)\citenamefont {Li},
  \citenamefont {Shen}, \citenamefont {Han}, \citenamefont {Zhu},\ and\
  \citenamefont {Wen}}]{Li2011}%
  \BibitemOpen
  \bibfield  {author} {\bibinfo {author} {\bibnamefont {Li}, \bibfnamefont
  {C.-H.}}, \bibinfo {author} {\bibfnamefont {B.}~\bibnamefont {Shen}},
  \bibinfo {author} {\bibfnamefont {F.}~\bibnamefont {Han}}, \bibinfo {author}
  {\bibfnamefont {X.}~\bibnamefont {Zhu}}, and\ \bibinfo {author}
  {\bibfnamefont {H.-H.}\ \bibnamefont {Wen}}} (\bibinfo {year} {2011}),\ \href
  {https://doi.org/10.1103/PhysRevB.83.184521} {\bibfield  {journal} {\bibinfo
  {journal} {Phys. Rev. B}\ }\textbf {\bibinfo {volume} {83}},\ \bibinfo
  {pages} {184521}}\BibitemShut {NoStop}%
\bibitem [{\citenamefont {Li}\ \emph {et~al.}(2019{\natexlab{a}})\citenamefont
  {Li}, \citenamefont {Lee}, \citenamefont {Wang}, \citenamefont {Osada},
  \citenamefont {Crossley}, \citenamefont {Lee}, \citenamefont {Cui},
  \citenamefont {Hikita},\ and\ \citenamefont
  {Hwang}}]{li2019superconductivity}%
  \BibitemOpen
  \bibfield  {author} {\bibinfo {author} {\bibnamefont {Li}, \bibfnamefont
  {D.}}, \bibinfo {author} {\bibfnamefont {K.}~\bibnamefont {Lee}}, \bibinfo
  {author} {\bibfnamefont {B.~Y.}\ \bibnamefont {Wang}}, \bibinfo {author}
  {\bibfnamefont {M.}~\bibnamefont {Osada}}, \bibinfo {author} {\bibfnamefont
  {S.}~\bibnamefont {Crossley}}, \bibinfo {author} {\bibfnamefont {H.~R.}\
  \bibnamefont {Lee}}, \bibinfo {author} {\bibfnamefont {Y.}~\bibnamefont
  {Cui}}, \bibinfo {author} {\bibfnamefont {Y.}~\bibnamefont {Hikita}}, and\
  \bibinfo {author} {\bibfnamefont {H.~Y.}\ \bibnamefont {Hwang}}} (\bibinfo
  {year} {2019}{\natexlab{a}}),\ \href
  {https://doi.org/10.1038/s41586-019-1496-5} {\bibfield  {journal} {\bibinfo
  {journal} {Nature}\ }\textbf {\bibinfo {volume} {572}},\ \bibinfo {pages}
  {624}}\BibitemShut {NoStop}%
\bibitem [{\citenamefont {Li}\ \emph {et~al.}(2015)\citenamefont {Li},
  \citenamefont {Ding}, \citenamefont {Tang}, \citenamefont {Peng},
  \citenamefont {Zhang}, \citenamefont {Zhang}, \citenamefont {Zhou},
  \citenamefont {Zhang}, \citenamefont {Song}, \citenamefont {He} \emph
  {et~al.}}]{Li2015}%
  \BibitemOpen
  \bibfield  {author} {\bibinfo {author} {\bibnamefont {Li}, \bibfnamefont
  {F.}}, \bibinfo {author} {\bibfnamefont {H.}~\bibnamefont {Ding}}, \bibinfo
  {author} {\bibfnamefont {C.}~\bibnamefont {Tang}}, \bibinfo {author}
  {\bibfnamefont {J.}~\bibnamefont {Peng}}, \bibinfo {author} {\bibfnamefont
  {Q.}~\bibnamefont {Zhang}}, \bibinfo {author} {\bibfnamefont
  {W.}~\bibnamefont {Zhang}}, \bibinfo {author} {\bibfnamefont
  {G.}~\bibnamefont {Zhou}}, \bibinfo {author} {\bibfnamefont {D.}~\bibnamefont
  {Zhang}}, \bibinfo {author} {\bibfnamefont {C.-L.}\ \bibnamefont {Song}},
  \bibinfo {author} {\bibfnamefont {K.}~\bibnamefont {He}},  \emph {et~al.}}
  (\bibinfo {year} {2015}),\ \href {https://doi.org/10.1103/PhysRevB.91.220503}
  {\bibfield  {journal} {\bibinfo  {journal} {Phys. Rev. B}\ }\textbf {\bibinfo
  {volume} {91}},\ \bibinfo {pages} {220503}}\BibitemShut {NoStop}%
\bibitem [{\citenamefont {Li}\ \emph {et~al.}(2020)\citenamefont {Li},
  \citenamefont {Lei}, \citenamefont {Zhao}, \citenamefont {Nie}, \citenamefont
  {Song}, \citenamefont {Zheng}, \citenamefont {Li}, \citenamefont {Kang},
  \citenamefont {Luo}, \citenamefont {Wu},\ and\ \citenamefont
  {Chen}}]{li2020prx}%
  \BibitemOpen
  \bibfield  {author} {\bibinfo {author} {\bibnamefont {Li}, \bibfnamefont
  {J.}}, \bibinfo {author} {\bibfnamefont {B.}~\bibnamefont {Lei}}, \bibinfo
  {author} {\bibfnamefont {D.}~\bibnamefont {Zhao}}, \bibinfo {author}
  {\bibfnamefont {L.~P.}\ \bibnamefont {Nie}}, \bibinfo {author} {\bibfnamefont
  {D.~W.}\ \bibnamefont {Song}}, \bibinfo {author} {\bibfnamefont {L.~X.}\
  \bibnamefont {Zheng}}, \bibinfo {author} {\bibfnamefont {S.~J.}\ \bibnamefont
  {Li}}, \bibinfo {author} {\bibfnamefont {B.~L.}\ \bibnamefont {Kang}},
  \bibinfo {author} {\bibfnamefont {X.~G.}\ \bibnamefont {Luo}}, \bibinfo
  {author} {\bibfnamefont {T.}~\bibnamefont {Wu}}, and\ \bibinfo {author}
  {\bibfnamefont {X.~H.}\ \bibnamefont {Chen}}} (\bibinfo {year} {2020}),\
  \href {https://doi.org/10.1103/PhysRevX.10.011034} {\bibfield  {journal}
  {\bibinfo  {journal} {Phys. Rev. X}\ }\textbf {\bibinfo {volume} {10}},\
  \bibinfo {pages} {011034}}\BibitemShut {NoStop}%
\bibitem [{\citenamefont {Li}\ \emph {et~al.}(2025{\natexlab{a}})\citenamefont
  {Li}, \citenamefont {Li}, \citenamefont {Liu}, \citenamefont {Li},
  \citenamefont {Nie}, \citenamefont {Zhao}, \citenamefont {Shi}, \citenamefont
  {Wu},\ and\ \citenamefont {Chen}}]{li2025npjqm}%
  \BibitemOpen
  \bibfield  {author} {\bibinfo {author} {\bibnamefont {Li}, \bibfnamefont
  {J.}}, \bibinfo {author} {\bibfnamefont {S.}~\bibnamefont {Li}}, \bibinfo
  {author} {\bibfnamefont {K.}~\bibnamefont {Liu}}, \bibinfo {author}
  {\bibfnamefont {X.}~\bibnamefont {Li}}, \bibinfo {author} {\bibfnamefont
  {L.}~\bibnamefont {Nie}}, \bibinfo {author} {\bibfnamefont {D.}~\bibnamefont
  {Zhao}}, \bibinfo {author} {\bibfnamefont {M.}~\bibnamefont {Shi}}, \bibinfo
  {author} {\bibfnamefont {T.}~\bibnamefont {Wu}}, and\ \bibinfo {author}
  {\bibfnamefont {X.}~\bibnamefont {Chen}}} (\bibinfo {year}
  {2025}{\natexlab{a}}),\ \href {https://doi.org/10.1038/s41535-025-00824-w}
  {\bibfield  {journal} {\bibinfo  {journal} {npj Quantum Mater.}\ }\textbf
  {\bibinfo {volume} {10}},\ \bibinfo {pages} {105}}\BibitemShut {NoStop}%
\bibitem [{\citenamefont {Li}\ \emph {et~al.}(2022)\citenamefont {Li},
  \citenamefont {Li}, \citenamefont {Cao}, \citenamefont {Zhou}, \citenamefont
  {Wang}, \citenamefont {Jin}, \citenamefont {Chiu}, \citenamefont {Pennycook},
  \citenamefont {Wang},\ and\ \citenamefont {Gao}}]{li2022ordered}%
  \BibitemOpen
  \bibfield  {author} {\bibinfo {author} {\bibnamefont {Li}, \bibfnamefont
  {M.}}, \bibinfo {author} {\bibfnamefont {G.}~\bibnamefont {Li}}, \bibinfo
  {author} {\bibfnamefont {L.}~\bibnamefont {Cao}}, \bibinfo {author}
  {\bibfnamefont {X.}~\bibnamefont {Zhou}}, \bibinfo {author} {\bibfnamefont
  {X.}~\bibnamefont {Wang}}, \bibinfo {author} {\bibfnamefont {C.}~\bibnamefont
  {Jin}}, \bibinfo {author} {\bibfnamefont {C.-K.}\ \bibnamefont {Chiu}},
  \bibinfo {author} {\bibfnamefont {S.~J.}\ \bibnamefont {Pennycook}}, \bibinfo
  {author} {\bibfnamefont {Z.}~\bibnamefont {Wang}}, and\ \bibinfo {author}
  {\bibfnamefont {H.-J.}\ \bibnamefont {Gao}}} (\bibinfo {year} {2022}),\ \href
  {https://doi.org/10.1038/s41586-022-04744-8} {\bibfield  {journal} {\bibinfo
  {journal} {Nature}\ }\textbf {\bibinfo {volume} {606}},\ \bibinfo {pages}
  {890}}\BibitemShut {NoStop}%
\bibitem [{\citenamefont {Li}\ \emph {et~al.}(2026)\citenamefont {Li},
  \citenamefont {Bansal}, \citenamefont {Nufer}, \citenamefont {Zhang},
  \citenamefont {Haghighirad}, \citenamefont {Sürgers}, \citenamefont
  {Mokrousov},\ and\ \citenamefont {Wulfhekel}}]{Li2026}%
  \BibitemOpen
  \bibfield  {author} {\bibinfo {author} {\bibnamefont {Li}, \bibfnamefont
  {Q.}}, \bibinfo {author} {\bibfnamefont {N.}~\bibnamefont {Bansal}}, \bibinfo
  {author} {\bibfnamefont {P.}~\bibnamefont {Nufer}}, \bibinfo {author}
  {\bibfnamefont {L.}~\bibnamefont {Zhang}}, \bibinfo {author} {\bibfnamefont
  {A.-A.}\ \bibnamefont {Haghighirad}}, \bibinfo {author} {\bibfnamefont
  {C.}~\bibnamefont {Sürgers}}, \bibinfo {author} {\bibfnamefont
  {Y.}~\bibnamefont {Mokrousov}}, and\ \bibinfo {author} {\bibfnamefont
  {W.}~\bibnamefont {Wulfhekel}}} (\bibinfo {year} {2026}),\ \href
  {https://doi.org/10.1038/s41467-026-74465-3} {\bibfield  {journal} {\bibinfo
  {journal} {Nat. Commun.}\ }\textbf {\bibinfo {volume} {17}},\ \bibinfo
  {pages} {5461}}\BibitemShut {NoStop}%
\bibitem [{\citenamefont {Li}\ \emph {et~al.}(2009)\citenamefont {Li},
  \citenamefont {de~la Cruz}, \citenamefont {Huang}, \citenamefont {Chen},
  \citenamefont {Lynn}, \citenamefont {Hu}, \citenamefont {Huang},
  \citenamefont {Hsu}, \citenamefont {Yeh}, \citenamefont {Wu},\ and\
  \citenamefont {Dai}}]{PhysRevB.79.054503}%
  \BibitemOpen
  \bibfield  {author} {\bibinfo {author} {\bibnamefont {Li}, \bibfnamefont
  {S.}}, \bibinfo {author} {\bibfnamefont {C.}~\bibnamefont {de~la Cruz}},
  \bibinfo {author} {\bibfnamefont {Q.}~\bibnamefont {Huang}}, \bibinfo
  {author} {\bibfnamefont {Y.}~\bibnamefont {Chen}}, \bibinfo {author}
  {\bibfnamefont {J.~W.}\ \bibnamefont {Lynn}}, \bibinfo {author}
  {\bibfnamefont {J.}~\bibnamefont {Hu}}, \bibinfo {author} {\bibfnamefont
  {Y.-L.}\ \bibnamefont {Huang}}, \bibinfo {author} {\bibfnamefont {F.-C.}\
  \bibnamefont {Hsu}}, \bibinfo {author} {\bibfnamefont {K.-W.}\ \bibnamefont
  {Yeh}}, \bibinfo {author} {\bibfnamefont {M.-K.}\ \bibnamefont {Wu}}, and\
  \bibinfo {author} {\bibfnamefont {P.}~\bibnamefont {Dai}}} (\bibinfo {year}
  {2009}),\ \href {https://doi.org/10.1103/PhysRevB.79.054503} {\bibfield
  {journal} {\bibinfo  {journal} {Phys. Rev. B}\ }\textbf {\bibinfo {volume}
  {79}},\ \bibinfo {pages} {054503}}\BibitemShut {NoStop}%
\bibitem [{\citenamefont {Li}\ \emph {et~al.}(2017)\citenamefont {Li},
  \citenamefont {Zhang}, \citenamefont {Deng}, \citenamefont {Xu},
  \citenamefont {Mo}, \citenamefont {Yi}, \citenamefont {Ding}, \citenamefont
  {Hashimoto}, \citenamefont {Moore}, \citenamefont {Lu}, \citenamefont {Chen},
  \citenamefont {Shen},\ and\ \citenamefont {Xue}}]{li2017stripes}%
  \BibitemOpen
  \bibfield  {author} {\bibinfo {author} {\bibnamefont {Li}, \bibfnamefont
  {W.}}, \bibinfo {author} {\bibfnamefont {Y.}~\bibnamefont {Zhang}}, \bibinfo
  {author} {\bibfnamefont {P.}~\bibnamefont {Deng}}, \bibinfo {author}
  {\bibfnamefont {Z.}~\bibnamefont {Xu}}, \bibinfo {author} {\bibfnamefont
  {S.-K.}\ \bibnamefont {Mo}}, \bibinfo {author} {\bibfnamefont
  {M.}~\bibnamefont {Yi}}, \bibinfo {author} {\bibfnamefont {H.}~\bibnamefont
  {Ding}}, \bibinfo {author} {\bibfnamefont {M.}~\bibnamefont {Hashimoto}},
  \bibinfo {author} {\bibfnamefont {R.~G.}\ \bibnamefont {Moore}}, \bibinfo
  {author} {\bibfnamefont {D.-H.}\ \bibnamefont {Lu}}, \bibinfo {author}
  {\bibfnamefont {X.}~\bibnamefont {Chen}}, \bibinfo {author} {\bibfnamefont
  {Z.-X.}\ \bibnamefont {Shen}}, and\ \bibinfo {author} {\bibfnamefont {Q.-K.}\
  \bibnamefont {Xue}}} (\bibinfo {year} {2017}),\ \href
  {https://doi.org/10.1038/nphys4186} {\bibfield  {journal} {\bibinfo
  {journal} {Nat. Phys.}\ }\textbf {\bibinfo {volume} {13}},\ \bibinfo {pages}
  {957}}\BibitemShut {NoStop}%
\bibitem [{\citenamefont {Li}\ \emph {et~al.}(2025{\natexlab{b}})\citenamefont
  {Li}, \citenamefont {Wu}, \citenamefont {Shu}, \citenamefont {Liu},
  \citenamefont {Stuhr}, \citenamefont {Deng}, \citenamefont {Stampfl},
  \citenamefont {Zhao}, \citenamefont {Zhou}, \citenamefont {Li}, \citenamefont
  {Pokhriyal}, \citenamefont {Ghosh}, \citenamefont {Hong},\ and\ \citenamefont
  {Luo}}]{li2025neutron}%
  \BibitemOpen
  \bibfield  {author} {\bibinfo {author} {\bibnamefont {Li}, \bibfnamefont
  {Y.}}, \bibinfo {author} {\bibfnamefont {D.}~\bibnamefont {Wu}}, \bibinfo
  {author} {\bibfnamefont {Y.}~\bibnamefont {Shu}}, \bibinfo {author}
  {\bibfnamefont {B.}~\bibnamefont {Liu}}, \bibinfo {author} {\bibfnamefont
  {U.}~\bibnamefont {Stuhr}}, \bibinfo {author} {\bibfnamefont
  {G.}~\bibnamefont {Deng}}, \bibinfo {author} {\bibfnamefont {A.~P.~J.}\
  \bibnamefont {Stampfl}}, \bibinfo {author} {\bibfnamefont {L.}~\bibnamefont
  {Zhao}}, \bibinfo {author} {\bibfnamefont {X.}~\bibnamefont {Zhou}}, \bibinfo
  {author} {\bibfnamefont {S.}~\bibnamefont {Li}}, \bibinfo {author}
  {\bibfnamefont {A.}~\bibnamefont {Pokhriyal}}, \bibinfo {author}
  {\bibfnamefont {H.}~\bibnamefont {Ghosh}}, \bibinfo {author} {\bibfnamefont
  {W.}~\bibnamefont {Hong}}, and\ \bibinfo {author} {\bibfnamefont
  {H.}~\bibnamefont {Luo}}} (\bibinfo {year} {2025}{\natexlab{b}}),\ \href
  {https://doi.org/10.1088/0256-307X/42/6/067405} {\bibfield  {journal}
  {\bibinfo  {journal} {Chin. Phys. Lett.}\ }\textbf {\bibinfo {volume} {42}},\
  \bibinfo {pages} {067405}}\BibitemShut {NoStop}%
\bibitem [{\citenamefont {Li}\ \emph {et~al.}(2018)\citenamefont {Li},
  \citenamefont {Yamani}, \citenamefont {Song}, \citenamefont {Wang},
  \citenamefont {Zhang}, \citenamefont {Tam}, \citenamefont {Chen},
  \citenamefont {Hu}, \citenamefont {Xu}, \citenamefont {Chi}, \citenamefont
  {Xia}, \citenamefont {Zhang}, \citenamefont {Cui}, \citenamefont {Guo},
  \citenamefont {Fang}, \citenamefont {Liu},\ and\ \citenamefont
  {Dai}}]{PhysRevX.8.021056}%
  \BibitemOpen
  \bibfield  {author} {\bibinfo {author} {\bibnamefont {Li}, \bibfnamefont
  {Y.}}, \bibinfo {author} {\bibfnamefont {Z.}~\bibnamefont {Yamani}}, \bibinfo
  {author} {\bibfnamefont {Y.}~\bibnamefont {Song}}, \bibinfo {author}
  {\bibfnamefont {W.}~\bibnamefont {Wang}}, \bibinfo {author} {\bibfnamefont
  {C.}~\bibnamefont {Zhang}}, \bibinfo {author} {\bibfnamefont {D.~W.}\
  \bibnamefont {Tam}}, \bibinfo {author} {\bibfnamefont {T.}~\bibnamefont
  {Chen}}, \bibinfo {author} {\bibfnamefont {D.}~\bibnamefont {Hu}}, \bibinfo
  {author} {\bibfnamefont {Z.}~\bibnamefont {Xu}}, \bibinfo {author}
  {\bibfnamefont {S.}~\bibnamefont {Chi}}, \bibinfo {author} {\bibfnamefont
  {K.}~\bibnamefont {Xia}}, \bibinfo {author} {\bibfnamefont {L.}~\bibnamefont
  {Zhang}}, \bibinfo {author} {\bibfnamefont {S.}~\bibnamefont {Cui}}, \bibinfo
  {author} {\bibfnamefont {W.}~\bibnamefont {Guo}}, \bibinfo {author}
  {\bibfnamefont {Z.}~\bibnamefont {Fang}}, \bibinfo {author} {\bibfnamefont
  {Y.}~\bibnamefont {Liu}}, and\ \bibinfo {author} {\bibfnamefont
  {P.}~\bibnamefont {Dai}}} (\bibinfo {year} {2018}),\ \href
  {https://doi.org/10.1103/PhysRevX.8.021056} {\bibfield  {journal} {\bibinfo
  {journal} {Phys. Rev. X}\ }\textbf {\bibinfo {volume} {8}},\ \bibinfo {pages}
  {021056}}\BibitemShut {NoStop}%
\bibitem [{\citenamefont {Li}\ \emph {et~al.}(2019{\natexlab{b}})\citenamefont
  {Li}, \citenamefont {Yin}, \citenamefont {Liu}, \citenamefont {Wang},
  \citenamefont {Xu}, \citenamefont {Song}, \citenamefont {Tian}, \citenamefont
  {Huang}, \citenamefont {Shen}, \citenamefont {Abernathy}, \citenamefont
  {Niedziela}, \citenamefont {Ewings}, \citenamefont {Perring}, \citenamefont
  {Pajerowski}, \citenamefont {Matsuda}, \citenamefont {Bourges}, \citenamefont
  {Mechthild}, \citenamefont {Su},\ and\ \citenamefont
  {Dai}}]{li2019coexistence}%
  \BibitemOpen
  \bibfield  {author} {\bibinfo {author} {\bibnamefont {Li}, \bibfnamefont
  {Y.}}, \bibinfo {author} {\bibfnamefont {Z.}~\bibnamefont {Yin}}, \bibinfo
  {author} {\bibfnamefont {Z.}~\bibnamefont {Liu}}, \bibinfo {author}
  {\bibfnamefont {W.}~\bibnamefont {Wang}}, \bibinfo {author} {\bibfnamefont
  {Z.}~\bibnamefont {Xu}}, \bibinfo {author} {\bibfnamefont {Y.}~\bibnamefont
  {Song}}, \bibinfo {author} {\bibfnamefont {L.}~\bibnamefont {Tian}}, \bibinfo
  {author} {\bibfnamefont {Y.}~\bibnamefont {Huang}}, \bibinfo {author}
  {\bibfnamefont {D.}~\bibnamefont {Shen}}, \bibinfo {author} {\bibfnamefont
  {D.~L.}\ \bibnamefont {Abernathy}}, \bibinfo {author} {\bibfnamefont {J.~L.}\
  \bibnamefont {Niedziela}}, \bibinfo {author} {\bibfnamefont {R.~A.}\
  \bibnamefont {Ewings}}, \bibinfo {author} {\bibfnamefont {T.~G.}\
  \bibnamefont {Perring}}, \bibinfo {author} {\bibfnamefont {D.~M.}\
  \bibnamefont {Pajerowski}}, \bibinfo {author} {\bibfnamefont
  {M.}~\bibnamefont {Matsuda}}, \bibinfo {author} {\bibfnamefont
  {P.}~\bibnamefont {Bourges}}, \bibinfo {author} {\bibfnamefont
  {E.}~\bibnamefont {Mechthild}}, \bibinfo {author} {\bibfnamefont
  {Y.}~\bibnamefont {Su}}, and\ \bibinfo {author} {\bibfnamefont
  {P.}~\bibnamefont {Dai}}} (\bibinfo {year} {2019}{\natexlab{b}}),\ \href
  {https://doi.org/10.1103/PhysRevLett.122.117204} {\bibfield  {journal}
  {\bibinfo  {journal} {Phys. Rev. Lett.}\ }\textbf {\bibinfo {volume} {122}},\
  \bibinfo {pages} {117204}}\BibitemShut {NoStop}%
\bibitem [{\citenamefont {Li}\ \emph {et~al.}(2016{\natexlab{a}})\citenamefont
  {Li}, \citenamefont {Yin}, \citenamefont {Wang}, \citenamefont {Tam},
  \citenamefont {Abernathy}, \citenamefont {Podlesnyak}, \citenamefont {Zhang},
  \citenamefont {Wang}, \citenamefont {Xing}, \citenamefont {Jin},
  \citenamefont {Haule}, \citenamefont {Kotliar}, \citenamefont {Maier},\ and\
  \citenamefont {Dai}}]{li2016orbital}%
  \BibitemOpen
  \bibfield  {author} {\bibinfo {author} {\bibnamefont {Li}, \bibfnamefont
  {Y.}}, \bibinfo {author} {\bibfnamefont {Z.}~\bibnamefont {Yin}}, \bibinfo
  {author} {\bibfnamefont {X.}~\bibnamefont {Wang}}, \bibinfo {author}
  {\bibfnamefont {D.~W.}\ \bibnamefont {Tam}}, \bibinfo {author} {\bibfnamefont
  {D.~L.}\ \bibnamefont {Abernathy}}, \bibinfo {author} {\bibfnamefont
  {A.}~\bibnamefont {Podlesnyak}}, \bibinfo {author} {\bibfnamefont
  {C.}~\bibnamefont {Zhang}}, \bibinfo {author} {\bibfnamefont
  {M.}~\bibnamefont {Wang}}, \bibinfo {author} {\bibfnamefont {L.}~\bibnamefont
  {Xing}}, \bibinfo {author} {\bibfnamefont {C.}~\bibnamefont {Jin}}, \bibinfo
  {author} {\bibfnamefont {K.}~\bibnamefont {Haule}}, \bibinfo {author}
  {\bibfnamefont {G.}~\bibnamefont {Kotliar}}, \bibinfo {author} {\bibfnamefont
  {T.~A.}\ \bibnamefont {Maier}}, and\ \bibinfo {author} {\bibfnamefont
  {P.}~\bibnamefont {Dai}}} (\bibinfo {year} {2016}{\natexlab{a}}),\ \href
  {https://doi.org/10.1103/PhysRevLett.116.247001} {\bibfield  {journal}
  {\bibinfo  {journal} {Phys. Rev. Lett.}\ }\textbf {\bibinfo {volume} {116}},\
  \bibinfo {pages} {247001}}\BibitemShut {NoStop}%
\bibitem [{\citenamefont {Li}\ \emph {et~al.}(2016{\natexlab{b}})\citenamefont
  {Li}, \citenamefont {Wang}, \citenamefont {Yao},\ and\ \citenamefont
  {Lee}}]{Li2016}%
  \BibitemOpen
  \bibfield  {author} {\bibinfo {author} {\bibnamefont {Li}, \bibfnamefont
  {Z.-X.}}, \bibinfo {author} {\bibfnamefont {F.}~\bibnamefont {Wang}},
  \bibinfo {author} {\bibfnamefont {H.}~\bibnamefont {Yao}}, and\ \bibinfo
  {author} {\bibfnamefont {D.-H.}\ \bibnamefont {Lee}}} (\bibinfo {year}
  {2016}{\natexlab{b}}),\ \href {https://doi.org/10.1007/s11434-016-1087-x}
  {\bibfield  {journal} {\bibinfo  {journal} {Sci. Bull.}\ }\textbf {\bibinfo
  {volume} {61}},\ \bibinfo {pages} {925}}\BibitemShut {NoStop}%
\bibitem [{\citenamefont {Licciardello}\ \emph {et~al.}(2019)\citenamefont
  {Licciardello}, \citenamefont {Buhot}, \citenamefont {Lu}, \citenamefont
  {Ayres}, \citenamefont {Kasahara}, \citenamefont {Matsuda}, \citenamefont
  {Shibauchi},\ and\ \citenamefont {Hussey}}]{licciardello2019electrical}%
  \BibitemOpen
  \bibfield  {author} {\bibinfo {author} {\bibnamefont {Licciardello},
  \bibfnamefont {S.}}, \bibinfo {author} {\bibfnamefont {J.}~\bibnamefont
  {Buhot}}, \bibinfo {author} {\bibfnamefont {J.}~\bibnamefont {Lu}}, \bibinfo
  {author} {\bibfnamefont {J.}~\bibnamefont {Ayres}}, \bibinfo {author}
  {\bibfnamefont {S.}~\bibnamefont {Kasahara}}, \bibinfo {author}
  {\bibfnamefont {Y.}~\bibnamefont {Matsuda}}, \bibinfo {author} {\bibfnamefont
  {T.}~\bibnamefont {Shibauchi}}, and\ \bibinfo {author} {\bibfnamefont
  {N.~E.}\ \bibnamefont {Hussey}}} (\bibinfo {year} {2019}),\ \href
  {https://doi.org/10.1038/s41586-019-0923-y} {\bibfield  {journal} {\bibinfo
  {journal} {Nature}\ }\textbf {\bibinfo {volume} {567}},\ \bibinfo {pages}
  {213}}\BibitemShut {NoStop}%
\bibitem [{\citenamefont {Lin}\ \emph {et~al.}(2011)\citenamefont {Lin},
  \citenamefont {Berlijn}, \citenamefont {Wang}, \citenamefont {Lee},
  \citenamefont {Yin},\ and\ \citenamefont {Ku}}]{Lin2011}%
  \BibitemOpen
  \bibfield  {author} {\bibinfo {author} {\bibnamefont {Lin}, \bibfnamefont
  {C.-H.}}, \bibinfo {author} {\bibfnamefont {T.}~\bibnamefont {Berlijn}},
  \bibinfo {author} {\bibfnamefont {L.}~\bibnamefont {Wang}}, \bibinfo {author}
  {\bibfnamefont {C.-C.}\ \bibnamefont {Lee}}, \bibinfo {author} {\bibfnamefont
  {W.-G.}\ \bibnamefont {Yin}}, and\ \bibinfo {author} {\bibfnamefont
  {W.}~\bibnamefont {Ku}}} (\bibinfo {year} {2011}),\ \href
  {https://doi.org/10.1103/PhysRevLett.107.257001} {\bibfield  {journal}
  {\bibinfo  {journal} {Phys. Rev. Lett.}\ }\textbf {\bibinfo {volume} {107}},\
  \bibinfo {pages} {257001}}\BibitemShut {NoStop}%
\bibitem [{\citenamefont {Lin}\ \emph {et~al.}(2023)\citenamefont {Lin},
  \citenamefont {Huang}, \citenamefont {Rai}, \citenamefont {Yin},
  \citenamefont {He}, \citenamefont {Xue}, \citenamefont {Haas}, \citenamefont
  {Kettemann}, \citenamefont {Chen},\ and\ \citenamefont {Ji}}]{lin2023real}%
  \BibitemOpen
  \bibfield  {author} {\bibinfo {author} {\bibnamefont {Lin}, \bibfnamefont
  {H.}}, \bibinfo {author} {\bibfnamefont {W.}~\bibnamefont {Huang}}, \bibinfo
  {author} {\bibfnamefont {G.}~\bibnamefont {Rai}}, \bibinfo {author}
  {\bibfnamefont {Y.}~\bibnamefont {Yin}}, \bibinfo {author} {\bibfnamefont
  {L.}~\bibnamefont {He}}, \bibinfo {author} {\bibfnamefont {Q.-K.}\
  \bibnamefont {Xue}}, \bibinfo {author} {\bibfnamefont {S.}~\bibnamefont
  {Haas}}, \bibinfo {author} {\bibfnamefont {S.}~\bibnamefont {Kettemann}},
  \bibinfo {author} {\bibfnamefont {X.}~\bibnamefont {Chen}}, and\ \bibinfo
  {author} {\bibfnamefont {S.-H.}\ \bibnamefont {Ji}}} (\bibinfo {year}
  {2023}),\ \href {https://doi.org/10.1103/PhysRevB.107.104517} {\bibfield
  {journal} {\bibinfo  {journal} {Phys. Rev. B}\ }\textbf {\bibinfo {volume}
  {107}},\ \bibinfo {pages} {104517}}\BibitemShut {NoStop}%
\bibitem [{\citenamefont {Linscheid}\ \emph {et~al.}(2016)\citenamefont
  {Linscheid}, \citenamefont {Maiti}, \citenamefont {Wang}, \citenamefont
  {Johnston},\ and\ \citenamefont {Hirschfeld}}]{linscheid2016high}%
  \BibitemOpen
  \bibfield  {author} {\bibinfo {author} {\bibnamefont {Linscheid},
  \bibfnamefont {A.}}, \bibinfo {author} {\bibfnamefont {S.}~\bibnamefont
  {Maiti}}, \bibinfo {author} {\bibfnamefont {Y.}~\bibnamefont {Wang}},
  \bibinfo {author} {\bibfnamefont {S.}~\bibnamefont {Johnston}}, and\ \bibinfo
  {author} {\bibfnamefont {P.~J.}\ \bibnamefont {Hirschfeld}}} (\bibinfo {year}
  {2016}),\ \href {https://doi.org/10.1103/PhysRevLett.117.077003} {\bibfield
  {journal} {\bibinfo  {journal} {Phys. Rev. Lett.}\ }\textbf {\bibinfo
  {volume} {117}},\ \bibinfo {pages} {077003}}\BibitemShut {NoStop}%
\bibitem [{\citenamefont {Lipscombe}\ \emph {et~al.}(2010)\citenamefont
  {Lipscombe}, \citenamefont {Harriger}, \citenamefont {Freeman}, \citenamefont
  {Enderle}, \citenamefont {Zhang}, \citenamefont {Wang}, \citenamefont
  {Egami}, \citenamefont {Hu}, \citenamefont {Xiang}, \citenamefont {Norman},\
  and\ \citenamefont {Dai}}]{lipscombe2010anisotropic}%
  \BibitemOpen
  \bibfield  {author} {\bibinfo {author} {\bibnamefont {Lipscombe},
  \bibfnamefont {O.~J.}}, \bibinfo {author} {\bibfnamefont {L.~W.}\
  \bibnamefont {Harriger}}, \bibinfo {author} {\bibfnamefont {P.~G.}\
  \bibnamefont {Freeman}}, \bibinfo {author} {\bibfnamefont {M.}~\bibnamefont
  {Enderle}}, \bibinfo {author} {\bibfnamefont {C.}~\bibnamefont {Zhang}},
  \bibinfo {author} {\bibfnamefont {M.}~\bibnamefont {Wang}}, \bibinfo {author}
  {\bibfnamefont {T.}~\bibnamefont {Egami}}, \bibinfo {author} {\bibfnamefont
  {J.}~\bibnamefont {Hu}}, \bibinfo {author} {\bibfnamefont {T.}~\bibnamefont
  {Xiang}}, \bibinfo {author} {\bibfnamefont {M.~R.}\ \bibnamefont {Norman}},
  and\ \bibinfo {author} {\bibfnamefont {P.}~\bibnamefont {Dai}}} (\bibinfo
  {year} {2010}),\ \href {https://doi.org/10.1103/PhysRevB.82.064515}
  {\bibfield  {journal} {\bibinfo  {journal} {Phys. Rev. B}\ }\textbf {\bibinfo
  {volume} {82}},\ \bibinfo {pages} {064515}}\BibitemShut {NoStop}%
\bibitem [{\citenamefont {Liu}\ \emph {et~al.}(2022{\natexlab{a}})\citenamefont
  {Liu}, \citenamefont {Bourges}, \citenamefont {Sidis}, \citenamefont {Xie},
  \citenamefont {He}, \citenamefont {Bourdarot}, \citenamefont {Danilkin},
  \citenamefont {Ghosh}, \citenamefont {Ghosh}, \citenamefont {Ma},
  \citenamefont {Li}, \citenamefont {Li},\ and\ \citenamefont
  {Luo}}]{liu2022preferred}%
  \BibitemOpen
  \bibfield  {author} {\bibinfo {author} {\bibnamefont {Liu}, \bibfnamefont
  {C.}}, \bibinfo {author} {\bibfnamefont {P.}~\bibnamefont {Bourges}},
  \bibinfo {author} {\bibfnamefont {Y.}~\bibnamefont {Sidis}}, \bibinfo
  {author} {\bibfnamefont {T.}~\bibnamefont {Xie}}, \bibinfo {author}
  {\bibfnamefont {G.}~\bibnamefont {He}}, \bibinfo {author} {\bibfnamefont
  {F.}~\bibnamefont {Bourdarot}}, \bibinfo {author} {\bibfnamefont
  {S.}~\bibnamefont {Danilkin}}, \bibinfo {author} {\bibfnamefont
  {H.}~\bibnamefont {Ghosh}}, \bibinfo {author} {\bibfnamefont
  {S.}~\bibnamefont {Ghosh}}, \bibinfo {author} {\bibfnamefont
  {X.}~\bibnamefont {Ma}}, \bibinfo {author} {\bibfnamefont {S.}~\bibnamefont
  {Li}}, \bibinfo {author} {\bibfnamefont {Y.}~\bibnamefont {Li}}, and\
  \bibinfo {author} {\bibfnamefont {H.}~\bibnamefont {Luo}}} (\bibinfo {year}
  {2022}{\natexlab{a}}),\ \href
  {https://doi.org/10.1103/PhysRevLett.128.137003} {\bibfield  {journal}
  {\bibinfo  {journal} {Phys. Rev. Lett.}\ }\textbf {\bibinfo {volume} {128}},\
  \bibinfo {pages} {137003}}\BibitemShut {NoStop}%
\bibitem [{\citenamefont {Liu}\ \emph {et~al.}(2021)\citenamefont {Liu},
  \citenamefont {Day}, \citenamefont {Li}, \citenamefont {Roemer},
  \citenamefont {Zhdanovich}, \citenamefont {Gorovikov}, \citenamefont
  {Pedersen}, \citenamefont {Jiang}, \citenamefont {Lee}, \citenamefont
  {Schneider}, \citenamefont {Wong}, \citenamefont {Dosanjh}, \citenamefont
  {Walker}, \citenamefont {Ahn}, \citenamefont {Levy}, \citenamefont
  {Damascelli}, \citenamefont {Sawatzky},\ and\ \citenamefont
  {Zou}}]{liu2021highorder}%
  \BibitemOpen
  \bibfield  {author} {\bibinfo {author} {\bibnamefont {Liu}, \bibfnamefont
  {C.}}, \bibinfo {author} {\bibfnamefont {R.~P.}\ \bibnamefont {Day}},
  \bibinfo {author} {\bibfnamefont {F.}~\bibnamefont {Li}}, \bibinfo {author}
  {\bibfnamefont {R.~L.}\ \bibnamefont {Roemer}}, \bibinfo {author}
  {\bibfnamefont {S.}~\bibnamefont {Zhdanovich}}, \bibinfo {author}
  {\bibfnamefont {S.}~\bibnamefont {Gorovikov}}, \bibinfo {author}
  {\bibfnamefont {T.~M.}\ \bibnamefont {Pedersen}}, \bibinfo {author}
  {\bibfnamefont {J.}~\bibnamefont {Jiang}}, \bibinfo {author} {\bibfnamefont
  {S.}~\bibnamefont {Lee}}, \bibinfo {author} {\bibfnamefont {M.}~\bibnamefont
  {Schneider}}, \bibinfo {author} {\bibfnamefont {D.}~\bibnamefont {Wong}},
  \bibinfo {author} {\bibfnamefont {P.}~\bibnamefont {Dosanjh}}, \bibinfo
  {author} {\bibfnamefont {F.~J.}\ \bibnamefont {Walker}}, \bibinfo {author}
  {\bibfnamefont {C.~H.}\ \bibnamefont {Ahn}}, \bibinfo {author} {\bibfnamefont
  {G.}~\bibnamefont {Levy}}, \bibinfo {author} {\bibfnamefont {A.}~\bibnamefont
  {Damascelli}}, \bibinfo {author} {\bibfnamefont {G.~A.}\ \bibnamefont
  {Sawatzky}}, and\ \bibinfo {author} {\bibfnamefont {K.}~\bibnamefont {Zou}}}
  (\bibinfo {year} {2021}),\ \href {https://doi.org/10.1038/s41467-021-24783-5}
  {\bibfield  {journal} {\bibinfo  {journal} {Nat. Commun.}\ }\textbf {\bibinfo
  {volume} {12}},\ \bibinfo {pages} {4573}}\BibitemShut {NoStop}%
\bibitem [{\citenamefont {Liu}\ \emph {et~al.}(2020{\natexlab{a}})\citenamefont
  {Liu}, \citenamefont {Lu}, \citenamefont {Dai}, \citenamefont {Yu},\ and\
  \citenamefont {Si}}]{liu2020anisotropic}%
  \BibitemOpen
  \bibfield  {author} {\bibinfo {author} {\bibnamefont {Liu}, \bibfnamefont
  {C.}}, \bibinfo {author} {\bibfnamefont {X.}~\bibnamefont {Lu}}, \bibinfo
  {author} {\bibfnamefont {P.}~\bibnamefont {Dai}}, \bibinfo {author}
  {\bibfnamefont {R.}~\bibnamefont {Yu}}, and\ \bibinfo {author} {\bibfnamefont
  {Q.}~\bibnamefont {Si}}} (\bibinfo {year} {2020}{\natexlab{a}}),\ \href
  {https://doi.org/10.1103/PhysRevB.101.024510} {\bibfield  {journal} {\bibinfo
   {journal} {Phys. Rev. B}\ }\textbf {\bibinfo {volume} {101}},\ \bibinfo
  {pages} {024510}}\BibitemShut {NoStop}%
\bibitem [{\citenamefont {Liu}\ \emph {et~al.}(2018{\natexlab{a}})\citenamefont
  {Liu}, \citenamefont {Li}, \citenamefont {Huang}, \citenamefont {Lei},
  \citenamefont {Wang}, \citenamefont {Wu}, \citenamefont {Shen}, \citenamefont
  {Gao}, \citenamefont {Zhang}, \citenamefont {Liu}, \citenamefont {Hu},
  \citenamefont {Xu}, \citenamefont {Liang}, \citenamefont {Liu}, \citenamefont
  {Ai}, \citenamefont {Zhao}, \citenamefont {He}, \citenamefont {Yu},
  \citenamefont {Liu}, \citenamefont {Mao}, \citenamefont {Dong}, \citenamefont
  {Jia}, \citenamefont {Zhang}, \citenamefont {Zhang}, \citenamefont {Yang},
  \citenamefont {Wang}, \citenamefont {Peng}, \citenamefont {Shi},
  \citenamefont {Hu}, \citenamefont {Xiang}, \citenamefont {Chen},
  \citenamefont {Xu}, \citenamefont {Chen},\ and\ \citenamefont
  {Zhou}}]{liu2018orbital}%
  \BibitemOpen
  \bibfield  {author} {\bibinfo {author} {\bibnamefont {Liu}, \bibfnamefont
  {D.}}, \bibinfo {author} {\bibfnamefont {C.}~\bibnamefont {Li}}, \bibinfo
  {author} {\bibfnamefont {J.}~\bibnamefont {Huang}}, \bibinfo {author}
  {\bibfnamefont {B.}~\bibnamefont {Lei}}, \bibinfo {author} {\bibfnamefont
  {L.}~\bibnamefont {Wang}}, \bibinfo {author} {\bibfnamefont {X.}~\bibnamefont
  {Wu}}, \bibinfo {author} {\bibfnamefont {B.}~\bibnamefont {Shen}}, \bibinfo
  {author} {\bibfnamefont {Q.}~\bibnamefont {Gao}}, \bibinfo {author}
  {\bibfnamefont {Y.}~\bibnamefont {Zhang}}, \bibinfo {author} {\bibfnamefont
  {X.}~\bibnamefont {Liu}}, \bibinfo {author} {\bibfnamefont {Y.}~\bibnamefont
  {Hu}}, \bibinfo {author} {\bibfnamefont {Y.}~\bibnamefont {Xu}}, \bibinfo
  {author} {\bibfnamefont {A.}~\bibnamefont {Liang}}, \bibinfo {author}
  {\bibfnamefont {J.}~\bibnamefont {Liu}}, \bibinfo {author} {\bibfnamefont
  {P.}~\bibnamefont {Ai}}, \bibinfo {author} {\bibfnamefont {L.}~\bibnamefont
  {Zhao}}, \bibinfo {author} {\bibfnamefont {S.}~\bibnamefont {He}}, \bibinfo
  {author} {\bibfnamefont {L.}~\bibnamefont {Yu}}, \bibinfo {author}
  {\bibfnamefont {G.}~\bibnamefont {Liu}}, \bibinfo {author} {\bibfnamefont
  {Y.}~\bibnamefont {Mao}}, \bibinfo {author} {\bibfnamefont {X.}~\bibnamefont
  {Dong}}, \bibinfo {author} {\bibfnamefont {X.}~\bibnamefont {Jia}}, \bibinfo
  {author} {\bibfnamefont {F.}~\bibnamefont {Zhang}}, \bibinfo {author}
  {\bibfnamefont {S.}~\bibnamefont {Zhang}}, \bibinfo {author} {\bibfnamefont
  {F.}~\bibnamefont {Yang}}, \bibinfo {author} {\bibfnamefont {Z.}~\bibnamefont
  {Wang}}, \bibinfo {author} {\bibfnamefont {Q.}~\bibnamefont {Peng}}, \bibinfo
  {author} {\bibfnamefont {Y.}~\bibnamefont {Shi}}, \bibinfo {author}
  {\bibfnamefont {J.}~\bibnamefont {Hu}}, \bibinfo {author} {\bibfnamefont
  {T.}~\bibnamefont {Xiang}}, \bibinfo {author} {\bibfnamefont
  {X.}~\bibnamefont {Chen}}, \bibinfo {author} {\bibfnamefont {Z.}~\bibnamefont
  {Xu}}, \bibinfo {author} {\bibfnamefont {C.}~\bibnamefont {Chen}}, and\
  \bibinfo {author} {\bibfnamefont {X.~J.}\ \bibnamefont {Zhou}}} (\bibinfo
  {year} {2018}{\natexlab{a}}),\ \href
  {https://doi.org/10.1103/PhysRevX.8.031033} {\bibfield  {journal} {\bibinfo
  {journal} {Phys. Rev. X}\ }\textbf {\bibinfo {volume} {8}},\ \bibinfo {pages}
  {031033}}\BibitemShut {NoStop}%
\bibitem [{\citenamefont {Liu}\ \emph {et~al.}(2012)\citenamefont {Liu},
  \citenamefont {Zhang}, \citenamefont {Mou}, \citenamefont {He}, \citenamefont
  {Ou}, \citenamefont {Wang}, \citenamefont {Li}, \citenamefont {Wang},
  \citenamefont {Zhao}, \citenamefont {He} \emph {et~al.}}]{Liu2012monoFeSe}%
  \BibitemOpen
  \bibfield  {author} {\bibinfo {author} {\bibnamefont {Liu}, \bibfnamefont
  {D.}}, \bibinfo {author} {\bibfnamefont {W.}~\bibnamefont {Zhang}}, \bibinfo
  {author} {\bibfnamefont {D.}~\bibnamefont {Mou}}, \bibinfo {author}
  {\bibfnamefont {J.}~\bibnamefont {He}}, \bibinfo {author} {\bibfnamefont
  {Y.-B.}\ \bibnamefont {Ou}}, \bibinfo {author} {\bibfnamefont {Q.-Y.}\
  \bibnamefont {Wang}}, \bibinfo {author} {\bibfnamefont {Z.}~\bibnamefont
  {Li}}, \bibinfo {author} {\bibfnamefont {L.}~\bibnamefont {Wang}}, \bibinfo
  {author} {\bibfnamefont {L.}~\bibnamefont {Zhao}}, \bibinfo {author}
  {\bibfnamefont {S.}~\bibnamefont {He}},  \emph {et~al.}} (\bibinfo {year}
  {2012}),\ \href {https://doi.org/10.1038/ncomms1946} {\bibfield  {journal}
  {\bibinfo  {journal} {Nat. Commun.}\ }\textbf {\bibinfo {volume} {3}},\
  \bibinfo {pages} {931}}\BibitemShut {NoStop}%
\bibitem [{\citenamefont {Liu}\ \emph {et~al.}(2023{\natexlab{a}})\citenamefont
  {Liu}, \citenamefont {Ye}, \citenamefont {Mou}, \citenamefont {Zhu},
  \citenamefont {Miao}, \citenamefont {Li}, \citenamefont {Shi}, \citenamefont
  {Qi}, \citenamefont {Ge}, \citenamefont {Liu} \emph {et~al.}}]{Liu2023}%
  \BibitemOpen
  \bibfield  {author} {\bibinfo {author} {\bibnamefont {Liu}, \bibfnamefont
  {L.}}, \bibinfo {author} {\bibfnamefont {J.}~\bibnamefont {Ye}}, \bibinfo
  {author} {\bibfnamefont {S.}~\bibnamefont {Mou}}, \bibinfo {author}
  {\bibfnamefont {R.}~\bibnamefont {Zhu}}, \bibinfo {author} {\bibfnamefont
  {C.}~\bibnamefont {Miao}}, \bibinfo {author} {\bibfnamefont {Y.}~\bibnamefont
  {Li}}, \bibinfo {author} {\bibfnamefont {Z.}~\bibnamefont {Shi}}, \bibinfo
  {author} {\bibfnamefont {Y.}~\bibnamefont {Qi}}, \bibinfo {author}
  {\bibfnamefont {J.}~\bibnamefont {Ge}}, \bibinfo {author} {\bibfnamefont
  {H.}~\bibnamefont {Liu}},  \emph {et~al.}} (\bibinfo {year}
  {2023}{\natexlab{a}}),\ \href {https://doi.org/10.1002/adem.202201536}
  {\bibfield  {journal} {\bibinfo  {journal} {Adv. Eng. Mater.}\ }\textbf
  {\bibinfo {volume} {25}},\ \bibinfo {pages} {2201536}}\BibitemShut {NoStop}%
\bibitem [{\citenamefont {Liu}\ \emph {et~al.}(2020{\natexlab{b}})\citenamefont
  {Liu}, \citenamefont {Klemm}, \citenamefont {Tian}, \citenamefont {Lu},
  \citenamefont {Song}, \citenamefont {Tam}, \citenamefont {Schmalzl},
  \citenamefont {Park}, \citenamefont {Li}, \citenamefont {Tan}, \citenamefont
  {Su}, \citenamefont {Bourdarot}, \citenamefont {Zhao}, \citenamefont {Lynn},
  \citenamefont {Birgeneau},\ and\ \citenamefont {Dai}}]{liu2020inplane}%
  \BibitemOpen
  \bibfield  {author} {\bibinfo {author} {\bibnamefont {Liu}, \bibfnamefont
  {P.}}, \bibinfo {author} {\bibfnamefont {M.~L.}\ \bibnamefont {Klemm}},
  \bibinfo {author} {\bibfnamefont {L.}~\bibnamefont {Tian}}, \bibinfo {author}
  {\bibfnamefont {X.}~\bibnamefont {Lu}}, \bibinfo {author} {\bibfnamefont
  {Y.}~\bibnamefont {Song}}, \bibinfo {author} {\bibfnamefont {D.~W.}\
  \bibnamefont {Tam}}, \bibinfo {author} {\bibfnamefont {K.}~\bibnamefont
  {Schmalzl}}, \bibinfo {author} {\bibfnamefont {J.~T.}\ \bibnamefont {Park}},
  \bibinfo {author} {\bibfnamefont {Y.}~\bibnamefont {Li}}, \bibinfo {author}
  {\bibfnamefont {G.}~\bibnamefont {Tan}}, \bibinfo {author} {\bibfnamefont
  {Y.}~\bibnamefont {Su}}, \bibinfo {author} {\bibfnamefont {F.}~\bibnamefont
  {Bourdarot}}, \bibinfo {author} {\bibfnamefont {Y.}~\bibnamefont {Zhao}},
  \bibinfo {author} {\bibfnamefont {J.~W.}\ \bibnamefont {Lynn}}, \bibinfo
  {author} {\bibfnamefont {R.~J.}\ \bibnamefont {Birgeneau}}, and\ \bibinfo
  {author} {\bibfnamefont {P.}~\bibnamefont {Dai}}} (\bibinfo {year}
  {2020}{\natexlab{b}}),\ \href {https://doi.org/10.1038/s41467-020-19421-5}
  {\bibfield  {journal} {\bibinfo  {journal} {Nat. Commun.}\ }\textbf {\bibinfo
  {volume} {11}},\ \bibinfo {pages} {5728}}\BibitemShut {NoStop}%
\bibitem [{\citenamefont {Liu}\ \emph {et~al.}(2018{\natexlab{b}})\citenamefont
  {Liu}, \citenamefont {Chen}, \citenamefont {Zhang}, \citenamefont {Peng},
  \citenamefont {Yan}, \citenamefont {Wen}, \citenamefont {Lou}, \citenamefont
  {Huang}, \citenamefont {Tian}, \citenamefont {Dong}, \citenamefont {Wang},
  \citenamefont {Bao}, \citenamefont {Wang}, \citenamefont {Yin}, \citenamefont
  {Zhao},\ and\ \citenamefont {Feng}}]{liu2018robust}%
  \BibitemOpen
  \bibfield  {author} {\bibinfo {author} {\bibnamefont {Liu}, \bibfnamefont
  {Q.}}, \bibinfo {author} {\bibfnamefont {C.}~\bibnamefont {Chen}}, \bibinfo
  {author} {\bibfnamefont {T.}~\bibnamefont {Zhang}}, \bibinfo {author}
  {\bibfnamefont {R.}~\bibnamefont {Peng}}, \bibinfo {author} {\bibfnamefont
  {Y.-J.}\ \bibnamefont {Yan}}, \bibinfo {author} {\bibfnamefont {C.-H.-P.}\
  \bibnamefont {Wen}}, \bibinfo {author} {\bibfnamefont {X.}~\bibnamefont
  {Lou}}, \bibinfo {author} {\bibfnamefont {Y.-L.}\ \bibnamefont {Huang}},
  \bibinfo {author} {\bibfnamefont {J.-P.}\ \bibnamefont {Tian}}, \bibinfo
  {author} {\bibfnamefont {X.-L.}\ \bibnamefont {Dong}}, \bibinfo {author}
  {\bibfnamefont {G.-W.}\ \bibnamefont {Wang}}, \bibinfo {author}
  {\bibfnamefont {W.-C.}\ \bibnamefont {Bao}}, \bibinfo {author} {\bibfnamefont
  {Q.-H.}\ \bibnamefont {Wang}}, \bibinfo {author} {\bibfnamefont {Z.-P.}\
  \bibnamefont {Yin}}, \bibinfo {author} {\bibfnamefont {Z.-X.}\ \bibnamefont
  {Zhao}}, and\ \bibinfo {author} {\bibfnamefont {D.-L.}\ \bibnamefont {Feng}}}
  (\bibinfo {year} {2018}{\natexlab{b}}),\ \href
  {https://doi.org/10.1103/PhysRevX.8.041056} {\bibfield  {journal} {\bibinfo
  {journal} {Phys. Rev. X}\ }\textbf {\bibinfo {volume} {8}},\ \bibinfo {pages}
  {041056}}\BibitemShut {NoStop}%
\bibitem [{\citenamefont {Liu}\ \emph {et~al.}(2011)\citenamefont {Liu},
  \citenamefont {Luo}, \citenamefont {Zhang}, \citenamefont {Wang},
  \citenamefont {Ying}, \citenamefont {Wang}, \citenamefont {Yan},
  \citenamefont {Xiang}, \citenamefont {Cheng}, \citenamefont {Ye} \emph
  {et~al.}}]{Liu2011}%
  \BibitemOpen
  \bibfield  {author} {\bibinfo {author} {\bibnamefont {Liu}, \bibfnamefont
  {R.}}, \bibinfo {author} {\bibfnamefont {X.}~\bibnamefont {Luo}}, \bibinfo
  {author} {\bibfnamefont {M.}~\bibnamefont {Zhang}}, \bibinfo {author}
  {\bibfnamefont {A.}~\bibnamefont {Wang}}, \bibinfo {author} {\bibfnamefont
  {J.}~\bibnamefont {Ying}}, \bibinfo {author} {\bibfnamefont {X.}~\bibnamefont
  {Wang}}, \bibinfo {author} {\bibfnamefont {Y.}~\bibnamefont {Yan}}, \bibinfo
  {author} {\bibfnamefont {Z.}~\bibnamefont {Xiang}}, \bibinfo {author}
  {\bibfnamefont {P.}~\bibnamefont {Cheng}}, \bibinfo {author} {\bibfnamefont
  {G.}~\bibnamefont {Ye}},  \emph {et~al.}} (\bibinfo {year} {2011}),\ \href
  {https://doi.org/10.1209/0295-5075/94/27008} {\bibfield  {journal} {\bibinfo
  {journal} {EPL}\ }\textbf {\bibinfo {volume} {94}},\ \bibinfo {pages}
  {27008}}\BibitemShut {NoStop}%
\bibitem [{\citenamefont {Liu}\ \emph {et~al.}(2024{\natexlab{a}})\citenamefont
  {Liu}, \citenamefont {Nakamura}, \citenamefont {Kamazawa},\ and\
  \citenamefont {Lu}}]{liu2024low}%
  \BibitemOpen
  \bibfield  {author} {\bibinfo {author} {\bibnamefont {Liu}, \bibfnamefont
  {R.}}, \bibinfo {author} {\bibfnamefont {M.}~\bibnamefont {Nakamura}},
  \bibinfo {author} {\bibfnamefont {K.}~\bibnamefont {Kamazawa}}, and\ \bibinfo
  {author} {\bibfnamefont {X.}~\bibnamefont {Lu}}} (\bibinfo {year}
  {2024}{\natexlab{a}}),\ \href {https://doi.org/10.1088/0256-307X/41/6/067401}
  {\bibfield  {journal} {\bibinfo  {journal} {Chin. Phys. Lett.}\ }\textbf
  {\bibinfo {volume} {41}},\ \bibinfo {pages} {067401}}\BibitemShut {NoStop}%
\bibitem [{\citenamefont {Liu}\ \emph {et~al.}(2025{\natexlab{a}})\citenamefont
  {Liu}, \citenamefont {Stone}, \citenamefont {Gao}, \citenamefont {Nakamura},
  \citenamefont {Kamazawa}, \citenamefont {Krajewska}, \citenamefont {Walker},
  \citenamefont {Cheng}, \citenamefont {Yu}, \citenamefont {Si}, \citenamefont
  {Dai},\ and\ \citenamefont {Lu}}]{liu2025spin}%
  \BibitemOpen
  \bibfield  {author} {\bibinfo {author} {\bibnamefont {Liu}, \bibfnamefont
  {R.}}, \bibinfo {author} {\bibfnamefont {M.~B.}\ \bibnamefont {Stone}},
  \bibinfo {author} {\bibfnamefont {S.}~\bibnamefont {Gao}}, \bibinfo {author}
  {\bibfnamefont {M.}~\bibnamefont {Nakamura}}, \bibinfo {author}
  {\bibfnamefont {K.}~\bibnamefont {Kamazawa}}, \bibinfo {author}
  {\bibfnamefont {A.}~\bibnamefont {Krajewska}}, \bibinfo {author}
  {\bibfnamefont {H.~C.}\ \bibnamefont {Walker}}, \bibinfo {author}
  {\bibfnamefont {P.}~\bibnamefont {Cheng}}, \bibinfo {author} {\bibfnamefont
  {R.}~\bibnamefont {Yu}}, \bibinfo {author} {\bibfnamefont {Q.}~\bibnamefont
  {Si}}, \bibinfo {author} {\bibfnamefont {P.}~\bibnamefont {Dai}}, and\
  \bibinfo {author} {\bibfnamefont {X.}~\bibnamefont {Lu}}} (\bibinfo {year}
  {2025}{\natexlab{a}}),\ \href {https://doi.org/10.1038/s41467-025-60071-2}
  {\bibfield  {journal} {\bibinfo  {journal} {Nat. Commun.}\ }\textbf {\bibinfo
  {volume} {16}},\ \bibinfo {pages} {5212}}\BibitemShut {NoStop}%
\bibitem [{\citenamefont {Liu}\ \emph {et~al.}(2025{\natexlab{b}})\citenamefont
  {Liu}, \citenamefont {Tang}, \citenamefont {Liu}, \citenamefont {Zhang},
  \citenamefont {Li}, \citenamefont {Zhou}, \citenamefont {Wang},\ and\
  \citenamefont {Lu}}]{liu2024evolution}%
  \BibitemOpen
  \bibfield  {author} {\bibinfo {author} {\bibnamefont {Liu}, \bibfnamefont
  {R.}}, \bibinfo {author} {\bibfnamefont {Q.}~\bibnamefont {Tang}}, \bibinfo
  {author} {\bibfnamefont {C.}~\bibnamefont {Liu}}, \bibinfo {author}
  {\bibfnamefont {W.}~\bibnamefont {Zhang}}, \bibinfo {author} {\bibfnamefont
  {C.}~\bibnamefont {Li}}, \bibinfo {author} {\bibfnamefont {K.}~\bibnamefont
  {Zhou}}, \bibinfo {author} {\bibfnamefont {Q.}~\bibnamefont {Wang}}, and\
  \bibinfo {author} {\bibfnamefont {X.}~\bibnamefont {Lu}}} (\bibinfo {year}
  {2025}{\natexlab{b}}),\ \href {https://doi.org/10.1103/f6jb-yvyj} {\bibfield
  {journal} {\bibinfo  {journal} {Phys. Rev. B}\ }\textbf {\bibinfo {volume}
  {112}},\ \bibinfo {pages} {094514}}\BibitemShut {NoStop}%
\bibitem [{\citenamefont {Liu}\ \emph {et~al.}(2024{\natexlab{b}})\citenamefont
  {Liu}, \citenamefont {Zhang}, \citenamefont {Wei}, \citenamefont {Tao},
  \citenamefont {Asmara}, \citenamefont {Li}, \citenamefont {Strocov},
  \citenamefont {Yu}, \citenamefont {Si}, \citenamefont {Schmitt},\ and\
  \citenamefont {Lu}}]{liu2024nematic}%
  \BibitemOpen
  \bibfield  {author} {\bibinfo {author} {\bibnamefont {Liu}, \bibfnamefont
  {R.}}, \bibinfo {author} {\bibfnamefont {W.}~\bibnamefont {Zhang}}, \bibinfo
  {author} {\bibfnamefont {Y.}~\bibnamefont {Wei}}, \bibinfo {author}
  {\bibfnamefont {Z.}~\bibnamefont {Tao}}, \bibinfo {author} {\bibfnamefont
  {T.~C.}\ \bibnamefont {Asmara}}, \bibinfo {author} {\bibfnamefont
  {Y.}~\bibnamefont {Li}}, \bibinfo {author} {\bibfnamefont {V.~N.}\
  \bibnamefont {Strocov}}, \bibinfo {author} {\bibfnamefont {R.}~\bibnamefont
  {Yu}}, \bibinfo {author} {\bibfnamefont {Q.}~\bibnamefont {Si}}, \bibinfo
  {author} {\bibfnamefont {T.}~\bibnamefont {Schmitt}}, and\ \bibinfo {author}
  {\bibfnamefont {X.}~\bibnamefont {Lu}}} (\bibinfo {year}
  {2024}{\natexlab{b}}),\ \href
  {https://doi.org/10.1103/PhysRevLett.132.016501} {\bibfield  {journal}
  {\bibinfo  {journal} {Phys. Rev. Lett.}\ }\textbf {\bibinfo {volume} {132}},\
  \bibinfo {pages} {016501}}\BibitemShut {NoStop}%
\bibitem [{\citenamefont {Liu}\ \emph {et~al.}(2023{\natexlab{b}})\citenamefont
  {Liu}, \citenamefont {Zhang}, \citenamefont {Wei}, \citenamefont {Tao},
  \citenamefont {Asmara}, \citenamefont {Strocov}, \citenamefont {Schmitt},\
  and\ \citenamefont {Lu}}]{liu2023nematic}%
  \BibitemOpen
  \bibfield  {author} {\bibinfo {author} {\bibnamefont {Liu}, \bibfnamefont
  {R.}}, \bibinfo {author} {\bibfnamefont {W.}~\bibnamefont {Zhang}}, \bibinfo
  {author} {\bibfnamefont {Y.}~\bibnamefont {Wei}}, \bibinfo {author}
  {\bibfnamefont {Z.}~\bibnamefont {Tao}}, \bibinfo {author} {\bibfnamefont
  {T.~C.}\ \bibnamefont {Asmara}}, \bibinfo {author} {\bibfnamefont {V.~N.}\
  \bibnamefont {Strocov}}, \bibinfo {author} {\bibfnamefont {T.}~\bibnamefont
  {Schmitt}}, and\ \bibinfo {author} {\bibfnamefont {X.}~\bibnamefont {Lu}}}
  (\bibinfo {year} {2023}{\natexlab{b}}),\ \href
  {https://doi.org/10.48550/arXiv.2312.12749} {\enquote {\bibinfo {title}
  {{Nematic charge-density-wave correlations in FeSe$_{1-x}$S$_{x}$}},}\
  }\Eprint {https://arxiv.org/abs/2312.12749} {arXiv:2312.12749
  [cond-mat.supr-con]} \BibitemShut {NoStop}%
\bibitem [{\citenamefont {Liu}\ \emph {et~al.}(2010)\citenamefont {Liu},
  \citenamefont {Hu}, \citenamefont {Qian}, \citenamefont {Fobes},
  \citenamefont {Mao}, \citenamefont {Bao}, \citenamefont {Reehuis},
  \citenamefont {Kimber}, \citenamefont {Proke{\v{s}}}, \citenamefont {Matas}
  \emph {et~al.}}]{Liu2010}%
  \BibitemOpen
  \bibfield  {author} {\bibinfo {author} {\bibnamefont {Liu}, \bibfnamefont
  {T.}}, \bibinfo {author} {\bibfnamefont {J.}~\bibnamefont {Hu}}, \bibinfo
  {author} {\bibfnamefont {B.}~\bibnamefont {Qian}}, \bibinfo {author}
  {\bibfnamefont {D.}~\bibnamefont {Fobes}}, \bibinfo {author} {\bibfnamefont
  {Z.~Q.}\ \bibnamefont {Mao}}, \bibinfo {author} {\bibfnamefont
  {W.}~\bibnamefont {Bao}}, \bibinfo {author} {\bibfnamefont {M.}~\bibnamefont
  {Reehuis}}, \bibinfo {author} {\bibfnamefont {S.}~\bibnamefont {Kimber}},
  \bibinfo {author} {\bibfnamefont {K.}~\bibnamefont {Proke{\v{s}}}}, \bibinfo
  {author} {\bibfnamefont {S.}~\bibnamefont {Matas}},  \emph {et~al.}}
  (\bibinfo {year} {2010}),\ \href {https://doi.org/10.1038/nmat2800}
  {\bibfield  {journal} {\bibinfo  {journal} {Nat. Mater.}\ }\textbf {\bibinfo
  {volume} {9}},\ \bibinfo {pages} {718}}\BibitemShut {NoStop}%
\bibitem [{\citenamefont {Liu}\ \emph {et~al.}(2020{\natexlab{c}})\citenamefont
  {Liu}, \citenamefont {Cao}, \citenamefont {Zhu}, \citenamefont {Kong},
  \citenamefont {Wang}, \citenamefont {Papaj}, \citenamefont {Zhang},
  \citenamefont {Liu}, \citenamefont {Chen}, \citenamefont {Li}, \citenamefont
  {Yang}, \citenamefont {Kondo}, \citenamefont {Du}, \citenamefont {Cao},
  \citenamefont {Shin}, \citenamefont {Fu}, \citenamefont {Yin}, \citenamefont
  {Gao},\ and\ \citenamefont {Ding}}]{liu2020new}%
  \BibitemOpen
  \bibfield  {author} {\bibinfo {author} {\bibnamefont {Liu}, \bibfnamefont
  {W.}}, \bibinfo {author} {\bibfnamefont {L.}~\bibnamefont {Cao}}, \bibinfo
  {author} {\bibfnamefont {S.}~\bibnamefont {Zhu}}, \bibinfo {author}
  {\bibfnamefont {L.}~\bibnamefont {Kong}}, \bibinfo {author} {\bibfnamefont
  {G.}~\bibnamefont {Wang}}, \bibinfo {author} {\bibfnamefont {M.}~\bibnamefont
  {Papaj}}, \bibinfo {author} {\bibfnamefont {P.}~\bibnamefont {Zhang}},
  \bibinfo {author} {\bibfnamefont {Y.-B.}\ \bibnamefont {Liu}}, \bibinfo
  {author} {\bibfnamefont {H.}~\bibnamefont {Chen}}, \bibinfo {author}
  {\bibfnamefont {G.}~\bibnamefont {Li}}, \bibinfo {author} {\bibfnamefont
  {F.}~\bibnamefont {Yang}}, \bibinfo {author} {\bibfnamefont {T.}~\bibnamefont
  {Kondo}}, \bibinfo {author} {\bibfnamefont {S.}~\bibnamefont {Du}}, \bibinfo
  {author} {\bibfnamefont {G.-H.}\ \bibnamefont {Cao}}, \bibinfo {author}
  {\bibfnamefont {S.}~\bibnamefont {Shin}}, \bibinfo {author} {\bibfnamefont
  {L.}~\bibnamefont {Fu}}, \bibinfo {author} {\bibfnamefont {Z.}~\bibnamefont
  {Yin}}, \bibinfo {author} {\bibfnamefont {H.-J.}\ \bibnamefont {Gao}}, and\
  \bibinfo {author} {\bibfnamefont {H.}~\bibnamefont {Ding}}} (\bibinfo {year}
  {2020}{\natexlab{c}}),\ \href {https://doi.org/10.1038/s41467-020-19487-1}
  {\bibfield  {journal} {\bibinfo  {journal} {Nat. Commun.}\ }\textbf {\bibinfo
  {volume} {11}},\ \bibinfo {pages} {5688}}\BibitemShut {NoStop}%
\bibitem [{\citenamefont {Liu}\ \emph {et~al.}(2022{\natexlab{b}})\citenamefont
  {Liu}, \citenamefont {Hu}, \citenamefont {Wang}, \citenamefont {Zhong},
  \citenamefont {Yang}, \citenamefont {Kong}, \citenamefont {Cao},
  \citenamefont {Li}, \citenamefont {Peng}, \citenamefont {Okazaki},
  \citenamefont {Kondo}, \citenamefont {Jin}, \citenamefont {Xu}, \citenamefont
  {Gao},\ and\ \citenamefont {Ding}}]{liu2022tunable}%
  \BibitemOpen
  \bibfield  {author} {\bibinfo {author} {\bibnamefont {Liu}, \bibfnamefont
  {W.}}, \bibinfo {author} {\bibfnamefont {Q.}~\bibnamefont {Hu}}, \bibinfo
  {author} {\bibfnamefont {X.}~\bibnamefont {Wang}}, \bibinfo {author}
  {\bibfnamefont {Y.}~\bibnamefont {Zhong}}, \bibinfo {author} {\bibfnamefont
  {F.}~\bibnamefont {Yang}}, \bibinfo {author} {\bibfnamefont {L.}~\bibnamefont
  {Kong}}, \bibinfo {author} {\bibfnamefont {L.}~\bibnamefont {Cao}}, \bibinfo
  {author} {\bibfnamefont {G.}~\bibnamefont {Li}}, \bibinfo {author}
  {\bibfnamefont {Y.}~\bibnamefont {Peng}}, \bibinfo {author} {\bibfnamefont
  {K.}~\bibnamefont {Okazaki}}, \bibinfo {author} {\bibfnamefont
  {T.}~\bibnamefont {Kondo}}, \bibinfo {author} {\bibfnamefont
  {C.}~\bibnamefont {Jin}}, \bibinfo {author} {\bibfnamefont {J.}~\bibnamefont
  {Xu}}, \bibinfo {author} {\bibfnamefont {H.-J.}\ \bibnamefont {Gao}}, and\
  \bibinfo {author} {\bibfnamefont {H.}~\bibnamefont {Ding}}} (\bibinfo {year}
  {2022}{\natexlab{b}}),\ \href {https://doi.org/10.1007/s44214-022-00022-w}
  {\bibfield  {journal} {\bibinfo  {journal} {Quantum Frontiers}\ }\textbf
  {\bibinfo {volume} {1}},\ \bibinfo {pages} {20}}\BibitemShut {NoStop}%
\bibitem [{\citenamefont {Liu}\ \emph {et~al.}(2019{\natexlab{a}})\citenamefont
  {Liu}, \citenamefont {Tao}, \citenamefont {Ren}, \citenamefont {Chen},
  \citenamefont {Yao}, \citenamefont {Wolf}, \citenamefont {Yan}, \citenamefont
  {Zhang},\ and\ \citenamefont {Feng}}]{liu2019evidence}%
  \BibitemOpen
  \bibfield  {author} {\bibinfo {author} {\bibnamefont {Liu}, \bibfnamefont
  {X.}}, \bibinfo {author} {\bibfnamefont {R.}~\bibnamefont {Tao}}, \bibinfo
  {author} {\bibfnamefont {M.}~\bibnamefont {Ren}}, \bibinfo {author}
  {\bibfnamefont {W.}~\bibnamefont {Chen}}, \bibinfo {author} {\bibfnamefont
  {Q.}~\bibnamefont {Yao}}, \bibinfo {author} {\bibfnamefont {T.}~\bibnamefont
  {Wolf}}, \bibinfo {author} {\bibfnamefont {Y.}~\bibnamefont {Yan}}, \bibinfo
  {author} {\bibfnamefont {T.}~\bibnamefont {Zhang}}, and\ \bibinfo {author}
  {\bibfnamefont {D.}~\bibnamefont {Feng}}} (\bibinfo {year}
  {2019}{\natexlab{a}}),\ \href {https://doi.org/10.1038/s41467-019-08962-z}
  {\bibfield  {journal} {\bibinfo  {journal} {Nat. Commun.}\ }\textbf {\bibinfo
  {volume} {10}},\ \bibinfo {pages} {1039}}\BibitemShut {NoStop}%
\bibitem [{\citenamefont {Liu}\ \emph {et~al.}(2025{\natexlab{c}})\citenamefont
  {Liu}, \citenamefont {Kao}, \citenamefont {Luo}, \citenamefont {Yang},
  \citenamefont {Fang}, \citenamefont {Zhao}, \citenamefont {Zhou},\ and\
  \citenamefont {Zheng}}]{liu2025microscopic}%
  \BibitemOpen
  \bibfield  {author} {\bibinfo {author} {\bibnamefont {Liu}, \bibfnamefont
  {X.~Y.}}, \bibinfo {author} {\bibfnamefont {Z.}~\bibnamefont {Kao}}, \bibinfo
  {author} {\bibfnamefont {J.}~\bibnamefont {Luo}}, \bibinfo {author}
  {\bibfnamefont {J.}~\bibnamefont {Yang}}, \bibinfo {author} {\bibfnamefont
  {A.~F.}\ \bibnamefont {Fang}}, \bibinfo {author} {\bibfnamefont
  {J.}~\bibnamefont {Zhao}}, \bibinfo {author} {\bibfnamefont {R.}~\bibnamefont
  {Zhou}}, and\ \bibinfo {author} {\bibfnamefont {G.-q.}\ \bibnamefont
  {Zheng}}} (\bibinfo {year} {2025}{\natexlab{c}}),\ \href
  {https://doi.org/10.1103/7hhv-9fq7} {\bibfield  {journal} {\bibinfo
  {journal} {Phys. Rev. B}\ }\textbf {\bibinfo {volume} {112}},\ \bibinfo
  {pages} {L020505}}\BibitemShut {NoStop}%
\bibitem [{\citenamefont {Liu}\ \emph {et~al.}(2016{\natexlab{a}})\citenamefont
  {Liu}, \citenamefont {Liu}, \citenamefont {Chen}, \citenamefont {Tang},
  \citenamefont {Jiao}, \citenamefont {Tao}, \citenamefont {Xu},\ and\
  \citenamefont {Cao}}]{Liu2016b}%
  \BibitemOpen
  \bibfield  {author} {\bibinfo {author} {\bibnamefont {Liu}, \bibfnamefont
  {Y.}}, \bibinfo {author} {\bibfnamefont {Y.-B.}\ \bibnamefont {Liu}},
  \bibinfo {author} {\bibfnamefont {Q.}~\bibnamefont {Chen}}, \bibinfo {author}
  {\bibfnamefont {Z.-T.}\ \bibnamefont {Tang}}, \bibinfo {author}
  {\bibfnamefont {W.-H.}\ \bibnamefont {Jiao}}, \bibinfo {author}
  {\bibfnamefont {Q.}~\bibnamefont {Tao}}, \bibinfo {author} {\bibfnamefont
  {Z.-A.}\ \bibnamefont {Xu}}, and\ \bibinfo {author} {\bibfnamefont {G.-H.}\
  \bibnamefont {Cao}}} (\bibinfo {year} {2016}{\natexlab{a}}),\ \href
  {https://doi.org/10.1007/s11434-016-1139-2} {\bibfield  {journal} {\bibinfo
  {journal} {Sci. Bull.}\ }\textbf {\bibinfo {volume} {61}},\ \bibinfo {pages}
  {1213}}\BibitemShut {NoStop}%
\bibitem [{\citenamefont {Liu}\ \emph {et~al.}(2016{\natexlab{b}})\citenamefont
  {Liu}, \citenamefont {Liu}, \citenamefont {Tang}, \citenamefont {Jiang},
  \citenamefont {Wang}, \citenamefont {Ablimit}, \citenamefont {Jiao},
  \citenamefont {Tao}, \citenamefont {Feng}, \citenamefont {Xu} \emph
  {et~al.}}]{Liu2016a}%
  \BibitemOpen
  \bibfield  {author} {\bibinfo {author} {\bibnamefont {Liu}, \bibfnamefont
  {Y.}}, \bibinfo {author} {\bibfnamefont {Y.-B.}\ \bibnamefont {Liu}},
  \bibinfo {author} {\bibfnamefont {Z.-T.}\ \bibnamefont {Tang}}, \bibinfo
  {author} {\bibfnamefont {H.}~\bibnamefont {Jiang}}, \bibinfo {author}
  {\bibfnamefont {Z.-C.}\ \bibnamefont {Wang}}, \bibinfo {author}
  {\bibfnamefont {A.}~\bibnamefont {Ablimit}}, \bibinfo {author} {\bibfnamefont
  {W.-H.}\ \bibnamefont {Jiao}}, \bibinfo {author} {\bibfnamefont
  {Q.}~\bibnamefont {Tao}}, \bibinfo {author} {\bibfnamefont {C.-M.}\
  \bibnamefont {Feng}}, \bibinfo {author} {\bibfnamefont {Z.-A.}\ \bibnamefont
  {Xu}},  \emph {et~al.}} (\bibinfo {year} {2016}{\natexlab{b}}),\ \href
  {https://doi.org/10.1103/PhysRevB.93.214503} {\bibfield  {journal} {\bibinfo
  {journal} {Phys. Rev. B}\ }\textbf {\bibinfo {volume} {93}},\ \bibinfo
  {pages} {214503}}\BibitemShut {NoStop}%
\bibitem [{\citenamefont {Liu}\ \emph {et~al.}(2017)\citenamefont {Liu},
  \citenamefont {Liu}, \citenamefont {Yu}, \citenamefont {Tao}, \citenamefont
  {Feng},\ and\ \citenamefont {Cao}}]{Liu2017}%
  \BibitemOpen
  \bibfield  {author} {\bibinfo {author} {\bibnamefont {Liu}, \bibfnamefont
  {Y.}}, \bibinfo {author} {\bibfnamefont {Y.-B.}\ \bibnamefont {Liu}},
  \bibinfo {author} {\bibfnamefont {Y.-L.}\ \bibnamefont {Yu}}, \bibinfo
  {author} {\bibfnamefont {Q.}~\bibnamefont {Tao}}, \bibinfo {author}
  {\bibfnamefont {C.-M.}\ \bibnamefont {Feng}}, and\ \bibinfo {author}
  {\bibfnamefont {G.-H.}\ \bibnamefont {Cao}}} (\bibinfo {year} {2017}),\ \href
  {https://doi.org/10.1103/PhysRevB.96.224510} {\bibfield  {journal} {\bibinfo
  {journal} {Phys. Rev. B}\ }\textbf {\bibinfo {volume} {96}},\ \bibinfo
  {pages} {224510}}\BibitemShut {NoStop}%
\bibitem [{\citenamefont {Liu}\ \emph {et~al.}(2023{\natexlab{c}})\citenamefont
  {Liu}, \citenamefont {Wei}, \citenamefont {He}, \citenamefont {Zhang},
  \citenamefont {Wang},\ and\ \citenamefont {Wang}}]{liu2023pair}%
  \BibitemOpen
  \bibfield  {author} {\bibinfo {author} {\bibnamefont {Liu}, \bibfnamefont
  {Y.}}, \bibinfo {author} {\bibfnamefont {T.}~\bibnamefont {Wei}}, \bibinfo
  {author} {\bibfnamefont {G.}~\bibnamefont {He}}, \bibinfo {author}
  {\bibfnamefont {Y.}~\bibnamefont {Zhang}}, \bibinfo {author} {\bibfnamefont
  {Z.}~\bibnamefont {Wang}}, and\ \bibinfo {author} {\bibfnamefont
  {J.}~\bibnamefont {Wang}}} (\bibinfo {year} {2023}{\natexlab{c}}),\ \href
  {https://doi.org/10.1038/s41586-023-06072-x} {\bibfield  {journal} {\bibinfo
  {journal} {Nature}\ }\textbf {\bibinfo {volume} {618}},\ \bibinfo {pages}
  {934}}\BibitemShut {NoStop}%
\bibitem [{\citenamefont {Liu}\ \emph {et~al.}(2022{\natexlab{c}})\citenamefont
  {Liu}, \citenamefont {Liu},\ and\ \citenamefont {Cao}}]{liu2022ironbased}%
  \BibitemOpen
  \bibfield  {author} {\bibinfo {author} {\bibnamefont {Liu}, \bibfnamefont
  {Y.-B.}}, \bibinfo {author} {\bibfnamefont {Y.}~\bibnamefont {Liu}}, and\
  \bibinfo {author} {\bibfnamefont {G.-H.}\ \bibnamefont {Cao}}} (\bibinfo
  {year} {2022}{\natexlab{c}}),\ \href
  {https://doi.org/10.1088/1361-648x/ac3cf2} {\bibfield  {journal} {\bibinfo
  {journal} {J. Phys.: Condens. Matter}\ }\textbf {\bibinfo {volume} {34}},\
  \bibinfo {pages} {093001}}\BibitemShut {NoStop}%
\bibitem [{\citenamefont {Liu}\ \emph {et~al.}(2020{\natexlab{d}})\citenamefont
  {Liu}, \citenamefont {Liu}, \citenamefont {Cui}, \citenamefont {Ren},\ and\
  \citenamefont {Cao}}]{Liu2020}%
  \BibitemOpen
  \bibfield  {author} {\bibinfo {author} {\bibnamefont {Liu}, \bibfnamefont
  {Y.-B.}}, \bibinfo {author} {\bibfnamefont {Y.}~\bibnamefont {Liu}}, \bibinfo
  {author} {\bibfnamefont {Y.-W.}\ \bibnamefont {Cui}}, \bibinfo {author}
  {\bibfnamefont {Z.}~\bibnamefont {Ren}}, and\ \bibinfo {author}
  {\bibfnamefont {G.-H.}\ \bibnamefont {Cao}}} (\bibinfo {year}
  {2020}{\natexlab{d}}),\ \href {https://doi.org/10.1088/1361-648x/ab68f4}
  {\bibfield  {journal} {\bibinfo  {journal} {J. Phys.: Condens. Matter}\
  }\textbf {\bibinfo {volume} {32}},\ \bibinfo {pages} {175701}}\BibitemShut
  {NoStop}%
\bibitem [{\citenamefont {Liu}\ \emph {et~al.}(2019{\natexlab{b}})\citenamefont
  {Liu}, \citenamefont {Gu}, \citenamefont {Hong}, \citenamefont {Xie},
  \citenamefont {Gong}, \citenamefont {Ma}, \citenamefont {Liu}, \citenamefont
  {Hu}, \citenamefont {Zhao}, \citenamefont {Zhou}, \citenamefont {Fernandes},
  \citenamefont {Yang}, \citenamefont {Luo},\ and\ \citenamefont
  {Li}}]{liu2019nonlinear}%
  \BibitemOpen
  \bibfield  {author} {\bibinfo {author} {\bibnamefont {Liu}, \bibfnamefont
  {Z.}}, \bibinfo {author} {\bibfnamefont {Y.}~\bibnamefont {Gu}}, \bibinfo
  {author} {\bibfnamefont {W.}~\bibnamefont {Hong}}, \bibinfo {author}
  {\bibfnamefont {T.}~\bibnamefont {Xie}}, \bibinfo {author} {\bibfnamefont
  {D.}~\bibnamefont {Gong}}, \bibinfo {author} {\bibfnamefont {X.}~\bibnamefont
  {Ma}}, \bibinfo {author} {\bibfnamefont {J.}~\bibnamefont {Liu}}, \bibinfo
  {author} {\bibfnamefont {C.}~\bibnamefont {Hu}}, \bibinfo {author}
  {\bibfnamefont {L.}~\bibnamefont {Zhao}}, \bibinfo {author} {\bibfnamefont
  {X.}~\bibnamefont {Zhou}}, \bibinfo {author} {\bibfnamefont {R.~M.}\
  \bibnamefont {Fernandes}}, \bibinfo {author} {\bibfnamefont {Y.-f.}\
  \bibnamefont {Yang}}, \bibinfo {author} {\bibfnamefont {H.}~\bibnamefont
  {Luo}}, and\ \bibinfo {author} {\bibfnamefont {S.}~\bibnamefont {Li}}}
  (\bibinfo {year} {2019}{\natexlab{b}}),\ \href
  {https://doi.org/10.1103/PhysRevResearch.1.033154} {\bibfield  {journal}
  {\bibinfo  {journal} {Phys. Rev. Res.}\ }\textbf {\bibinfo {volume} {1}},\
  \bibinfo {pages} {033154}}\BibitemShut {NoStop}%
\bibitem [{\citenamefont {Liu}\ \emph {et~al.}(2016{\natexlab{c}})\citenamefont
  {Liu}, \citenamefont {Gu}, \citenamefont {Zhang}, \citenamefont {Gong},
  \citenamefont {Zhang}, \citenamefont {Xie}, \citenamefont {Lu}, \citenamefont
  {Ma}, \citenamefont {Zhang}, \citenamefont {Zhang}, \citenamefont {Zhu},
  \citenamefont {Ren}, \citenamefont {Shan}, \citenamefont {Qiu}, \citenamefont
  {Dai}, \citenamefont {Yang}, \citenamefont {Luo},\ and\ \citenamefont
  {Li}}]{liu2016nematic}%
  \BibitemOpen
  \bibfield  {author} {\bibinfo {author} {\bibnamefont {Liu}, \bibfnamefont
  {Z.}}, \bibinfo {author} {\bibfnamefont {Y.}~\bibnamefont {Gu}}, \bibinfo
  {author} {\bibfnamefont {W.}~\bibnamefont {Zhang}}, \bibinfo {author}
  {\bibfnamefont {D.}~\bibnamefont {Gong}}, \bibinfo {author} {\bibfnamefont
  {W.}~\bibnamefont {Zhang}}, \bibinfo {author} {\bibfnamefont
  {T.}~\bibnamefont {Xie}}, \bibinfo {author} {\bibfnamefont {X.}~\bibnamefont
  {Lu}}, \bibinfo {author} {\bibfnamefont {X.}~\bibnamefont {Ma}}, \bibinfo
  {author} {\bibfnamefont {X.}~\bibnamefont {Zhang}}, \bibinfo {author}
  {\bibfnamefont {R.}~\bibnamefont {Zhang}}, \bibinfo {author} {\bibfnamefont
  {J.}~\bibnamefont {Zhu}}, \bibinfo {author} {\bibfnamefont {C.}~\bibnamefont
  {Ren}}, \bibinfo {author} {\bibfnamefont {L.}~\bibnamefont {Shan}}, \bibinfo
  {author} {\bibfnamefont {X.}~\bibnamefont {Qiu}}, \bibinfo {author}
  {\bibfnamefont {P.}~\bibnamefont {Dai}}, \bibinfo {author} {\bibfnamefont
  {Y.-f.}\ \bibnamefont {Yang}}, \bibinfo {author} {\bibfnamefont
  {H.}~\bibnamefont {Luo}}, and\ \bibinfo {author} {\bibfnamefont
  {S.}~\bibnamefont {Li}}} (\bibinfo {year} {2016}{\natexlab{c}}),\ \href
  {https://doi.org/10.1103/PhysRevLett.117.157002} {\bibfield  {journal}
  {\bibinfo  {journal} {Phys. Rev. Lett.}\ }\textbf {\bibinfo {volume} {117}},\
  \bibinfo {pages} {157002}}\BibitemShut {NoStop}%
\bibitem [{\citenamefont {Liu}\ \emph {et~al.}(2024{\natexlab{c}})\citenamefont
  {Liu}, \citenamefont {Shi}, \citenamefont {Jiang}, \citenamefont {Rosenberg},
  \citenamefont {DeStefano}, \citenamefont {Liu}, \citenamefont {Hu},
  \citenamefont {Zhao}, \citenamefont {Wang}, \citenamefont {Yao},
  \citenamefont {Graf}, \citenamefont {Dai}, \citenamefont {Yang},
  \citenamefont {Xu},\ and\ \citenamefont {Chu}}]{liu2024absence}%
  \BibitemOpen
  \bibfield  {author} {\bibinfo {author} {\bibnamefont {Liu}, \bibfnamefont
  {Z.}}, \bibinfo {author} {\bibfnamefont {Y.}~\bibnamefont {Shi}}, \bibinfo
  {author} {\bibfnamefont {Q.}~\bibnamefont {Jiang}}, \bibinfo {author}
  {\bibfnamefont {E.~W.}\ \bibnamefont {Rosenberg}}, \bibinfo {author}
  {\bibfnamefont {J.~M.}\ \bibnamefont {DeStefano}}, \bibinfo {author}
  {\bibfnamefont {J.}~\bibnamefont {Liu}}, \bibinfo {author} {\bibfnamefont
  {C.}~\bibnamefont {Hu}}, \bibinfo {author} {\bibfnamefont {Y.}~\bibnamefont
  {Zhao}}, \bibinfo {author} {\bibfnamefont {Z.}~\bibnamefont {Wang}}, \bibinfo
  {author} {\bibfnamefont {Y.}~\bibnamefont {Yao}}, \bibinfo {author}
  {\bibfnamefont {D.}~\bibnamefont {Graf}}, \bibinfo {author} {\bibfnamefont
  {P.}~\bibnamefont {Dai}}, \bibinfo {author} {\bibfnamefont {J.}~\bibnamefont
  {Yang}}, \bibinfo {author} {\bibfnamefont {X.}~\bibnamefont {Xu}}, and\
  \bibinfo {author} {\bibfnamefont {J.-H.}\ \bibnamefont {Chu}}} (\bibinfo
  {year} {2024}{\natexlab{c}}),\ \href
  {https://doi.org/10.1103/PhysRevX.14.031015} {\bibfield  {journal} {\bibinfo
  {journal} {Phys. Rev. X}\ }\textbf {\bibinfo {volume} {14}},\ \bibinfo
  {pages} {031015}}\BibitemShut {NoStop}%
\bibitem [{\citenamefont {L{\"o}hnert}\ \emph {et~al.}(2011)\citenamefont
  {L{\"o}hnert}, \citenamefont {St{\"u}rzer}, \citenamefont {Tegel},
  \citenamefont {Frankovsky}, \citenamefont {Friederichs},\ and\ \citenamefont
  {Johrendt}}]{Lohnert2011}%
  \BibitemOpen
  \bibfield  {author} {\bibinfo {author} {\bibnamefont {L{\"o}hnert},
  \bibfnamefont {C.}}, \bibinfo {author} {\bibfnamefont {T.}~\bibnamefont
  {St{\"u}rzer}}, \bibinfo {author} {\bibfnamefont {M.}~\bibnamefont {Tegel}},
  \bibinfo {author} {\bibfnamefont {R.}~\bibnamefont {Frankovsky}}, \bibinfo
  {author} {\bibfnamefont {G.}~\bibnamefont {Friederichs}}, and\ \bibinfo
  {author} {\bibfnamefont {D.}~\bibnamefont {Johrendt}}} (\bibinfo {year}
  {2011}),\ \href {https://doi.org/10.1002/anie.201104436} {\bibfield
  {journal} {\bibinfo  {journal} {Angewandte Chemie International Edition}\
  }\textbf {\bibinfo {volume} {50}},\ \bibinfo {pages} {9195}}\BibitemShut
  {NoStop}%
\bibitem [{\citenamefont {Louca}\ \emph {et~al.}(2011)\citenamefont {Louca},
  \citenamefont {Yan}, \citenamefont {Llobet},\ and\ \citenamefont
  {Arita}}]{louca2011suppression}%
  \BibitemOpen
  \bibfield  {author} {\bibinfo {author} {\bibnamefont {Louca}, \bibfnamefont
  {D.}}, \bibinfo {author} {\bibfnamefont {J.}~\bibnamefont {Yan}}, \bibinfo
  {author} {\bibfnamefont {A.}~\bibnamefont {Llobet}}, and\ \bibinfo {author}
  {\bibfnamefont {R.}~\bibnamefont {Arita}}} (\bibinfo {year} {2011}),\ \href
  {https://doi.org/10.1103/PhysRevB.84.054522} {\bibfield  {journal} {\bibinfo
  {journal} {Phys. Rev. B}\ }\textbf {\bibinfo {volume} {84}},\ \bibinfo
  {pages} {054522}}\BibitemShut {NoStop}%
\bibitem [{\citenamefont {Lu}\ \emph {et~al.}(2017)\citenamefont {Lu},
  \citenamefont {Zhang}, \citenamefont {Zhang}, \citenamefont {Qiao},
  \citenamefont {He}, \citenamefont {Li}, \citenamefont {Wang}, \citenamefont
  {Guo}, \citenamefont {Zhang}, \citenamefont {Duan} \emph {et~al.}}]{Lu2017}%
  \BibitemOpen
  \bibfield  {author} {\bibinfo {author} {\bibnamefont {Lu}, \bibfnamefont
  {N.}}, \bibinfo {author} {\bibfnamefont {P.}~\bibnamefont {Zhang}}, \bibinfo
  {author} {\bibfnamefont {Q.}~\bibnamefont {Zhang}}, \bibinfo {author}
  {\bibfnamefont {R.}~\bibnamefont {Qiao}}, \bibinfo {author} {\bibfnamefont
  {Q.}~\bibnamefont {He}}, \bibinfo {author} {\bibfnamefont {H.-B.}\
  \bibnamefont {Li}}, \bibinfo {author} {\bibfnamefont {Y.}~\bibnamefont
  {Wang}}, \bibinfo {author} {\bibfnamefont {J.}~\bibnamefont {Guo}}, \bibinfo
  {author} {\bibfnamefont {D.}~\bibnamefont {Zhang}}, \bibinfo {author}
  {\bibfnamefont {Z.}~\bibnamefont {Duan}},  \emph {et~al.}} (\bibinfo {year}
  {2017}),\ \href {https://doi.org/10.1038/nature22389} {\bibfield  {journal}
  {\bibinfo  {journal} {Nature}\ }\textbf {\bibinfo {volume} {546}},\ \bibinfo
  {pages} {124}}\BibitemShut {NoStop}%
\bibitem [{\citenamefont {Lu}\ \emph {et~al.}(2014{\natexlab{a}})\citenamefont
  {Lu}, \citenamefont {Park}, \citenamefont {Zhang}, \citenamefont {Luo},
  \citenamefont {Nevidomskyy}, \citenamefont {Si},\ and\ \citenamefont
  {Dai}}]{lu2014nematic}%
  \BibitemOpen
  \bibfield  {author} {\bibinfo {author} {\bibnamefont {Lu}, \bibfnamefont
  {X.}}, \bibinfo {author} {\bibfnamefont {J.~T.}\ \bibnamefont {Park}},
  \bibinfo {author} {\bibfnamefont {R.}~\bibnamefont {Zhang}}, \bibinfo
  {author} {\bibfnamefont {H.}~\bibnamefont {Luo}}, \bibinfo {author}
  {\bibfnamefont {A.~H.}\ \bibnamefont {Nevidomskyy}}, \bibinfo {author}
  {\bibfnamefont {Q.}~\bibnamefont {Si}}, and\ \bibinfo {author} {\bibfnamefont
  {P.}~\bibnamefont {Dai}}} (\bibinfo {year} {2014}{\natexlab{a}}),\ \href
  {https://doi.org/10.1126/science.1251853} {\bibfield  {journal} {\bibinfo
  {journal} {Science}\ }\textbf {\bibinfo {volume} {345}},\ \bibinfo {pages}
  {657}}\BibitemShut {NoStop}%
\bibitem [{\citenamefont {Lu}\ \emph {et~al.}(2018)\citenamefont {Lu},
  \citenamefont {Scherer}, \citenamefont {Tam}, \citenamefont {Zhang},
  \citenamefont {Zhang}, \citenamefont {Luo}, \citenamefont {Harriger},
  \citenamefont {Walker}, \citenamefont {Adroja}, \citenamefont {Andersen},\
  and\ \citenamefont {Dai}}]{lu2018spin}%
  \BibitemOpen
  \bibfield  {author} {\bibinfo {author} {\bibnamefont {Lu}, \bibfnamefont
  {X.}}, \bibinfo {author} {\bibfnamefont {D.~D.}\ \bibnamefont {Scherer}},
  \bibinfo {author} {\bibfnamefont {D.~W.}\ \bibnamefont {Tam}}, \bibinfo
  {author} {\bibfnamefont {W.}~\bibnamefont {Zhang}}, \bibinfo {author}
  {\bibfnamefont {R.}~\bibnamefont {Zhang}}, \bibinfo {author} {\bibfnamefont
  {H.}~\bibnamefont {Luo}}, \bibinfo {author} {\bibfnamefont {L.~W.}\
  \bibnamefont {Harriger}}, \bibinfo {author} {\bibfnamefont {H.~C.}\
  \bibnamefont {Walker}}, \bibinfo {author} {\bibfnamefont {D.~T.}\
  \bibnamefont {Adroja}}, \bibinfo {author} {\bibfnamefont {B.~M.}\
  \bibnamefont {Andersen}}, and\ \bibinfo {author} {\bibfnamefont
  {P.}~\bibnamefont {Dai}}} (\bibinfo {year} {2018}),\ \href
  {https://doi.org/10.1103/PhysRevLett.121.067002} {\bibfield  {journal}
  {\bibinfo  {journal} {Phys. Rev. Lett.}\ }\textbf {\bibinfo {volume} {121}},\
  \bibinfo {pages} {067002}}\BibitemShut {NoStop}%
\bibitem [{\citenamefont {Lu}\ \emph {et~al.}(2014{\natexlab{b}})\citenamefont
  {Lu}, \citenamefont {Tam}, \citenamefont {Zhang}, \citenamefont {Luo},
  \citenamefont {Wang}, \citenamefont {Zhang}, \citenamefont {Harriger},
  \citenamefont {Keller}, \citenamefont {Keimer}, \citenamefont {Regnault},
  \citenamefont {Maier},\ and\ \citenamefont {Dai}}]{PhysRevB.90.024509}%
  \BibitemOpen
  \bibfield  {author} {\bibinfo {author} {\bibnamefont {Lu}, \bibfnamefont
  {X.}}, \bibinfo {author} {\bibfnamefont {D.~W.}\ \bibnamefont {Tam}},
  \bibinfo {author} {\bibfnamefont {C.}~\bibnamefont {Zhang}}, \bibinfo
  {author} {\bibfnamefont {H.}~\bibnamefont {Luo}}, \bibinfo {author}
  {\bibfnamefont {M.}~\bibnamefont {Wang}}, \bibinfo {author} {\bibfnamefont
  {R.}~\bibnamefont {Zhang}}, \bibinfo {author} {\bibfnamefont {L.~W.}\
  \bibnamefont {Harriger}}, \bibinfo {author} {\bibfnamefont {T.}~\bibnamefont
  {Keller}}, \bibinfo {author} {\bibfnamefont {B.}~\bibnamefont {Keimer}},
  \bibinfo {author} {\bibfnamefont {L.-P.}\ \bibnamefont {Regnault}}, \bibinfo
  {author} {\bibfnamefont {T.~A.}\ \bibnamefont {Maier}}, and\ \bibinfo
  {author} {\bibfnamefont {P.}~\bibnamefont {Dai}}} (\bibinfo {year}
  {2014}{\natexlab{b}}),\ \href {https://doi.org/10.1103/PhysRevB.90.024509}
  {\bibfield  {journal} {\bibinfo  {journal} {Phys. Rev. B}\ }\textbf {\bibinfo
  {volume} {90}},\ \bibinfo {pages} {024509}}\BibitemShut {NoStop}%
\bibitem [{\citenamefont {Lu}\ \emph {et~al.}(2016)\citenamefont {Lu},
  \citenamefont {Tseng}, \citenamefont {Keller}, \citenamefont {Zhang},
  \citenamefont {Hu}, \citenamefont {Song}, \citenamefont {Man}, \citenamefont
  {Park}, \citenamefont {Luo}, \citenamefont {Li}, \citenamefont
  {Nevidomskyy},\ and\ \citenamefont {Dai}}]{lu2016impact}%
  \BibitemOpen
  \bibfield  {author} {\bibinfo {author} {\bibnamefont {Lu}, \bibfnamefont
  {X.}}, \bibinfo {author} {\bibfnamefont {K.-F.}\ \bibnamefont {Tseng}},
  \bibinfo {author} {\bibfnamefont {T.}~\bibnamefont {Keller}}, \bibinfo
  {author} {\bibfnamefont {W.}~\bibnamefont {Zhang}}, \bibinfo {author}
  {\bibfnamefont {D.}~\bibnamefont {Hu}}, \bibinfo {author} {\bibfnamefont
  {Y.}~\bibnamefont {Song}}, \bibinfo {author} {\bibfnamefont {H.}~\bibnamefont
  {Man}}, \bibinfo {author} {\bibfnamefont {J.~T.}\ \bibnamefont {Park}},
  \bibinfo {author} {\bibfnamefont {H.}~\bibnamefont {Luo}}, \bibinfo {author}
  {\bibfnamefont {S.}~\bibnamefont {Li}}, \bibinfo {author} {\bibfnamefont
  {A.~H.}\ \bibnamefont {Nevidomskyy}}, and\ \bibinfo {author} {\bibfnamefont
  {P.}~\bibnamefont {Dai}}} (\bibinfo {year} {2016}),\ \href
  {https://doi.org/10.1103/PhysRevB.93.134519} {\bibfield  {journal} {\bibinfo
  {journal} {Phys. Rev. B}\ }\textbf {\bibinfo {volume} {93}},\ \bibinfo
  {pages} {134519}}\BibitemShut {NoStop}%
\bibitem [{\citenamefont {Lu}\ \emph {et~al.}(2013)\citenamefont {Lu},
  \citenamefont {Wang}, \citenamefont {Zhang}, \citenamefont {Luo},
  \citenamefont {Ma}, \citenamefont {Lei}, \citenamefont {Huang},\ and\
  \citenamefont {Chen}}]{Lu2013}%
  \BibitemOpen
  \bibfield  {author} {\bibinfo {author} {\bibnamefont {Lu}, \bibfnamefont
  {X.}}, \bibinfo {author} {\bibfnamefont {N.}~\bibnamefont {Wang}}, \bibinfo
  {author} {\bibfnamefont {G.}~\bibnamefont {Zhang}}, \bibinfo {author}
  {\bibfnamefont {X.}~\bibnamefont {Luo}}, \bibinfo {author} {\bibfnamefont
  {Z.}~\bibnamefont {Ma}}, \bibinfo {author} {\bibfnamefont {B.}~\bibnamefont
  {Lei}}, \bibinfo {author} {\bibfnamefont {F.}~\bibnamefont {Huang}}, and\
  \bibinfo {author} {\bibfnamefont {X.}~\bibnamefont {Chen}}} (\bibinfo {year}
  {2013}),\ \href {https://doi.org/10.1103/PhysRevB.89.020507} {\bibfield
  {journal} {\bibinfo  {journal} {Phys. Rev. B}\ }\textbf {\bibinfo {volume}
  {89}},\ \bibinfo {pages} {020507(R)}}\BibitemShut {NoStop}%
\bibitem [{\citenamefont {Lu}\ \emph {et~al.}(2022)\citenamefont {Lu},
  \citenamefont {Zhang}, \citenamefont {Tseng}, \citenamefont {Liu},
  \citenamefont {Tao}, \citenamefont {Paris}, \citenamefont {Liu},
  \citenamefont {Chen}, \citenamefont {Strocov}, \citenamefont {Song},
  \citenamefont {Yu}, \citenamefont {Si}, \citenamefont {Dai},\ and\
  \citenamefont {Schmitt}}]{lu2022spinexcitation}%
  \BibitemOpen
  \bibfield  {author} {\bibinfo {author} {\bibnamefont {Lu}, \bibfnamefont
  {X.}}, \bibinfo {author} {\bibfnamefont {W.}~\bibnamefont {Zhang}}, \bibinfo
  {author} {\bibfnamefont {Y.}~\bibnamefont {Tseng}}, \bibinfo {author}
  {\bibfnamefont {R.}~\bibnamefont {Liu}}, \bibinfo {author} {\bibfnamefont
  {Z.}~\bibnamefont {Tao}}, \bibinfo {author} {\bibfnamefont {E.}~\bibnamefont
  {Paris}}, \bibinfo {author} {\bibfnamefont {P.}~\bibnamefont {Liu}}, \bibinfo
  {author} {\bibfnamefont {T.}~\bibnamefont {Chen}}, \bibinfo {author}
  {\bibfnamefont {V.~N.}\ \bibnamefont {Strocov}}, \bibinfo {author}
  {\bibfnamefont {Y.}~\bibnamefont {Song}}, \bibinfo {author} {\bibfnamefont
  {R.}~\bibnamefont {Yu}}, \bibinfo {author} {\bibfnamefont {Q.}~\bibnamefont
  {Si}}, \bibinfo {author} {\bibfnamefont {P.}~\bibnamefont {Dai}}, and\
  \bibinfo {author} {\bibfnamefont {T.}~\bibnamefont {Schmitt}}} (\bibinfo
  {year} {2022}),\ \href {https://doi.org/10.1038/s41567-022-01603-1}
  {\bibfield  {journal} {\bibinfo  {journal} {Nat. Phys.}\ }\textbf {\bibinfo
  {volume} {18}},\ \bibinfo {pages} {806}}\BibitemShut {NoStop}%
\bibitem [{\citenamefont {Lu}\ \emph {et~al.}(2015)\citenamefont {Lu},
  \citenamefont {Wang}, \citenamefont {Wu}, \citenamefont {Wu}, \citenamefont
  {Zhao}, \citenamefont {Zeng}, \citenamefont {Luo}, \citenamefont {Wu},
  \citenamefont {Bao}, \citenamefont {Zhang}, \citenamefont {Huang},
  \citenamefont {Huang},\ and\ \citenamefont {Chen}}]{lu2015coexistence}%
  \BibitemOpen
  \bibfield  {author} {\bibinfo {author} {\bibnamefont {Lu}, \bibfnamefont
  {X.~F.}}, \bibinfo {author} {\bibfnamefont {N.~Z.}\ \bibnamefont {Wang}},
  \bibinfo {author} {\bibfnamefont {H.}~\bibnamefont {Wu}}, \bibinfo {author}
  {\bibfnamefont {Y.~P.}\ \bibnamefont {Wu}}, \bibinfo {author} {\bibfnamefont
  {D.}~\bibnamefont {Zhao}}, \bibinfo {author} {\bibfnamefont {X.~Z.}\
  \bibnamefont {Zeng}}, \bibinfo {author} {\bibfnamefont {X.~G.}\ \bibnamefont
  {Luo}}, \bibinfo {author} {\bibfnamefont {T.}~\bibnamefont {Wu}}, \bibinfo
  {author} {\bibfnamefont {W.}~\bibnamefont {Bao}}, \bibinfo {author}
  {\bibfnamefont {G.~H.}\ \bibnamefont {Zhang}}, \bibinfo {author}
  {\bibfnamefont {F.~Q.}\ \bibnamefont {Huang}}, \bibinfo {author}
  {\bibfnamefont {Q.~Z.}\ \bibnamefont {Huang}}, and\ \bibinfo {author}
  {\bibfnamefont {X.~H.}\ \bibnamefont {Chen}}} (\bibinfo {year} {2015}),\
  \href {https://doi.org/10.1038/nmat4155} {\bibfield  {journal} {\bibinfo
  {journal} {Nat. Mater.}\ }\textbf {\bibinfo {volume} {14}},\ \bibinfo {pages}
  {325}}\BibitemShut {NoStop}%
\bibitem [{\citenamefont {Ma}\ \emph {et~al.}(2009)\citenamefont {Ma},
  \citenamefont {Ji}, \citenamefont {Hu}, \citenamefont {Lu},\ and\
  \citenamefont {Xiang}}]{Ma2009}%
  \BibitemOpen
  \bibfield  {author} {\bibinfo {author} {\bibnamefont {Ma}, \bibfnamefont
  {F.}}, \bibinfo {author} {\bibfnamefont {W.}~\bibnamefont {Ji}}, \bibinfo
  {author} {\bibfnamefont {J.}~\bibnamefont {Hu}}, \bibinfo {author}
  {\bibfnamefont {Z.-Y.}\ \bibnamefont {Lu}}, and\ \bibinfo {author}
  {\bibfnamefont {T.}~\bibnamefont {Xiang}}} (\bibinfo {year} {2009}),\ \href
  {https://doi.org/10.1103/PhysRevLett.102.177003} {\bibfield  {journal}
  {\bibinfo  {journal} {Phys. Rev. Lett.}\ }\textbf {\bibinfo {volume} {102}},\
  \bibinfo {pages} {177003}}\BibitemShut {NoStop}%
\bibitem [{\citenamefont {Ma}\ and\ \citenamefont {Lu}(2008)}]{Ma2008}%
  \BibitemOpen
  \bibfield  {author} {\bibinfo {author} {\bibnamefont {Ma}, \bibfnamefont
  {F.}}, and\ \bibinfo {author} {\bibfnamefont {Z.-Y.}\ \bibnamefont {Lu}}}
  (\bibinfo {year} {2008}),\ \href {https://doi.org/10.1103/PhysRevB.78.033111}
  {\bibfield  {journal} {\bibinfo  {journal} {Phys. Rev. B}\ }\textbf {\bibinfo
  {volume} {78}},\ \bibinfo {pages} {033111}}\BibitemShut {NoStop}%
\bibitem [{\citenamefont {Ma}\ \emph {et~al.}(2017{\natexlab{a}})\citenamefont
  {Ma}, \citenamefont {Bourges}, \citenamefont {Sidis}, \citenamefont {Xu},
  \citenamefont {Li}, \citenamefont {Hu}, \citenamefont {Li}, \citenamefont
  {Wang},\ and\ \citenamefont {Li}}]{ma2017prominent}%
  \BibitemOpen
  \bibfield  {author} {\bibinfo {author} {\bibnamefont {Ma}, \bibfnamefont
  {M.}}, \bibinfo {author} {\bibfnamefont {P.}~\bibnamefont {Bourges}},
  \bibinfo {author} {\bibfnamefont {Y.}~\bibnamefont {Sidis}}, \bibinfo
  {author} {\bibfnamefont {Y.}~\bibnamefont {Xu}}, \bibinfo {author}
  {\bibfnamefont {S.}~\bibnamefont {Li}}, \bibinfo {author} {\bibfnamefont
  {B.}~\bibnamefont {Hu}}, \bibinfo {author} {\bibfnamefont {J.}~\bibnamefont
  {Li}}, \bibinfo {author} {\bibfnamefont {F.}~\bibnamefont {Wang}}, and\
  \bibinfo {author} {\bibfnamefont {Y.}~\bibnamefont {Li}}} (\bibinfo {year}
  {2017}{\natexlab{a}}),\ \href {https://doi.org/10.1103/PhysRevX.7.021025}
  {\bibfield  {journal} {\bibinfo  {journal} {Phys. Rev. X}\ }\textbf {\bibinfo
  {volume} {7}},\ \bibinfo {pages} {021025}}\BibitemShut {NoStop}%
\bibitem [{\citenamefont {Ma}\ \emph {et~al.}(2017{\natexlab{b}})\citenamefont
  {Ma}, \citenamefont {Wang}, \citenamefont {Bourges}, \citenamefont {Sidis},
  \citenamefont {Danilkin},\ and\ \citenamefont {Li}}]{ma2017lowenergy}%
  \BibitemOpen
  \bibfield  {author} {\bibinfo {author} {\bibnamefont {Ma}, \bibfnamefont
  {M.}}, \bibinfo {author} {\bibfnamefont {L.}~\bibnamefont {Wang}}, \bibinfo
  {author} {\bibfnamefont {P.}~\bibnamefont {Bourges}}, \bibinfo {author}
  {\bibfnamefont {Y.}~\bibnamefont {Sidis}}, \bibinfo {author} {\bibfnamefont
  {S.}~\bibnamefont {Danilkin}}, and\ \bibinfo {author} {\bibfnamefont
  {Y.}~\bibnamefont {Li}}} (\bibinfo {year} {2017}{\natexlab{b}}),\ \href
  {https://doi.org/10.1103/PhysRevB.95.100504} {\bibfield  {journal} {\bibinfo
  {journal} {Phys. Rev. B}\ }\textbf {\bibinfo {volume} {95}},\ \bibinfo
  {pages} {100504}}\BibitemShut {NoStop}%
\bibitem [{\citenamefont {Machida}\ and\ \citenamefont
  {Hanaguri}(2024)}]{machida2024searching}%
  \BibitemOpen
  \bibfield  {author} {\bibinfo {author} {\bibnamefont {Machida}, \bibfnamefont
  {T.}}, and\ \bibinfo {author} {\bibfnamefont {T.}~\bibnamefont {Hanaguri}}}
  (\bibinfo {year} {2024}),\ \href {https://doi.org/10.1093/ptep/ptad084}
  {\bibfield  {journal} {\bibinfo  {journal} {Prog. Theor. Exp. Phys.}\
  }\textbf {\bibinfo {volume} {2024}},\ \bibinfo {pages} {08C103}}\BibitemShut
  {NoStop}%
\bibitem [{\citenamefont {Maier}\ \emph {et~al.}(2009)\citenamefont {Maier},
  \citenamefont {Graser}, \citenamefont {Scalapino},\ and\ \citenamefont
  {Hirschfeld}}]{maier2009neutron}%
  \BibitemOpen
  \bibfield  {author} {\bibinfo {author} {\bibnamefont {Maier}, \bibfnamefont
  {T.~A.}}, \bibinfo {author} {\bibfnamefont {S.}~\bibnamefont {Graser}},
  \bibinfo {author} {\bibfnamefont {D.~J.}\ \bibnamefont {Scalapino}}, and\
  \bibinfo {author} {\bibfnamefont {P.}~\bibnamefont {Hirschfeld}}} (\bibinfo
  {year} {2009}),\ \href {https://doi.org/10.1103/PhysRevB.79.134520}
  {\bibfield  {journal} {\bibinfo  {journal} {Phys. Rev. B}\ }\textbf {\bibinfo
  {volume} {79}},\ \bibinfo {pages} {134520}}\BibitemShut {NoStop}%
\bibitem [{\citenamefont {Malinowski}\ \emph {et~al.}(2020)\citenamefont
  {Malinowski}, \citenamefont {Jiang}, \citenamefont {Sanchez}, \citenamefont
  {Mutch}, \citenamefont {Liu}, \citenamefont {Went}, \citenamefont {Liu},
  \citenamefont {Ryan}, \citenamefont {Kim},\ and\ \citenamefont
  {Chu}}]{malinowski2020}%
  \BibitemOpen
  \bibfield  {author} {\bibinfo {author} {\bibnamefont {Malinowski},
  \bibfnamefont {P.}}, \bibinfo {author} {\bibfnamefont {Q.}~\bibnamefont
  {Jiang}}, \bibinfo {author} {\bibfnamefont {J.~J.}\ \bibnamefont {Sanchez}},
  \bibinfo {author} {\bibfnamefont {J.}~\bibnamefont {Mutch}}, \bibinfo
  {author} {\bibfnamefont {Z.}~\bibnamefont {Liu}}, \bibinfo {author}
  {\bibfnamefont {P.}~\bibnamefont {Went}}, \bibinfo {author} {\bibfnamefont
  {J.}~\bibnamefont {Liu}}, \bibinfo {author} {\bibfnamefont {P.~J.}\
  \bibnamefont {Ryan}}, \bibinfo {author} {\bibfnamefont {J.-W.}\ \bibnamefont
  {Kim}}, and\ \bibinfo {author} {\bibfnamefont {J.-H.}\ \bibnamefont {Chu}}}
  (\bibinfo {year} {2020}),\ \href {https://doi.org/10.1038/s41567-020-0983-9}
  {\bibfield  {journal} {\bibinfo  {journal} {Nat. Phys.}\ }\textbf {\bibinfo
  {volume} {16}},\ \bibinfo {pages} {1189}}\BibitemShut {NoStop}%
\bibitem [{\citenamefont {Man}\ \emph {et~al.}(2017)\citenamefont {Man},
  \citenamefont {Guo}, \citenamefont {Zhang}, \citenamefont {Sch{\"o}nemann},
  \citenamefont {Yin}, \citenamefont {Fu}, \citenamefont {Stone}, \citenamefont
  {Huang}, \citenamefont {Song}, \citenamefont {Wang}, \citenamefont {Singh},
  \citenamefont {Lochner}, \citenamefont {Hickel}, \citenamefont {Eremin},
  \citenamefont {Harriger}, \citenamefont {Lynn}, \citenamefont {Broholm},
  \citenamefont {Balicas}, \citenamefont {Si},\ and\ \citenamefont
  {Dai}}]{man2017spin}%
  \BibitemOpen
  \bibfield  {author} {\bibinfo {author} {\bibnamefont {Man}, \bibfnamefont
  {H.}}, \bibinfo {author} {\bibfnamefont {J.}~\bibnamefont {Guo}}, \bibinfo
  {author} {\bibfnamefont {R.}~\bibnamefont {Zhang}}, \bibinfo {author}
  {\bibfnamefont {R.}~\bibnamefont {Sch{\"o}nemann}}, \bibinfo {author}
  {\bibfnamefont {Z.}~\bibnamefont {Yin}}, \bibinfo {author} {\bibfnamefont
  {M.}~\bibnamefont {Fu}}, \bibinfo {author} {\bibfnamefont {M.~B.}\
  \bibnamefont {Stone}}, \bibinfo {author} {\bibfnamefont {Q.}~\bibnamefont
  {Huang}}, \bibinfo {author} {\bibfnamefont {Y.}~\bibnamefont {Song}},
  \bibinfo {author} {\bibfnamefont {W.}~\bibnamefont {Wang}}, \bibinfo {author}
  {\bibfnamefont {D.~J.}\ \bibnamefont {Singh}}, \bibinfo {author}
  {\bibfnamefont {F.}~\bibnamefont {Lochner}}, \bibinfo {author} {\bibfnamefont
  {T.}~\bibnamefont {Hickel}}, \bibinfo {author} {\bibfnamefont
  {I.}~\bibnamefont {Eremin}}, \bibinfo {author} {\bibfnamefont
  {L.}~\bibnamefont {Harriger}}, \bibinfo {author} {\bibfnamefont {J.~W.}\
  \bibnamefont {Lynn}}, \bibinfo {author} {\bibfnamefont {C.}~\bibnamefont
  {Broholm}}, \bibinfo {author} {\bibfnamefont {L.}~\bibnamefont {Balicas}},
  \bibinfo {author} {\bibfnamefont {Q.}~\bibnamefont {Si}}, and\ \bibinfo
  {author} {\bibfnamefont {P.}~\bibnamefont {Dai}}} (\bibinfo {year} {2017}),\
  \href {https://doi.org/10.1038/s41535-017-0019-6} {\bibfield  {journal}
  {\bibinfo  {journal} {npj Quantum Mater.}\ }\textbf {\bibinfo {volume} {2}},\
  \bibinfo {pages} {14}}\BibitemShut {NoStop}%
\bibitem [{\citenamefont {Man}\ \emph {et~al.}(2015)\citenamefont {Man},
  \citenamefont {Lu}, \citenamefont {Chen}, \citenamefont {Zhang},
  \citenamefont {Zhang}, \citenamefont {Luo}, \citenamefont {Kulda},
  \citenamefont {Ivanov}, \citenamefont {Keller}, \citenamefont {Morosan},
  \citenamefont {Si},\ and\ \citenamefont {Dai}}]{man2015electronic}%
  \BibitemOpen
  \bibfield  {author} {\bibinfo {author} {\bibnamefont {Man}, \bibfnamefont
  {H.}}, \bibinfo {author} {\bibfnamefont {X.}~\bibnamefont {Lu}}, \bibinfo
  {author} {\bibfnamefont {J.~S.}\ \bibnamefont {Chen}}, \bibinfo {author}
  {\bibfnamefont {R.}~\bibnamefont {Zhang}}, \bibinfo {author} {\bibfnamefont
  {W.}~\bibnamefont {Zhang}}, \bibinfo {author} {\bibfnamefont
  {H.}~\bibnamefont {Luo}}, \bibinfo {author} {\bibfnamefont {J.}~\bibnamefont
  {Kulda}}, \bibinfo {author} {\bibfnamefont {A.}~\bibnamefont {Ivanov}},
  \bibinfo {author} {\bibfnamefont {T.}~\bibnamefont {Keller}}, \bibinfo
  {author} {\bibfnamefont {E.}~\bibnamefont {Morosan}}, \bibinfo {author}
  {\bibfnamefont {Q.}~\bibnamefont {Si}}, and\ \bibinfo {author} {\bibfnamefont
  {P.}~\bibnamefont {Dai}}} (\bibinfo {year} {2015}),\ \href
  {https://doi.org/10.1103/PhysRevB.92.134521} {\bibfield  {journal} {\bibinfo
  {journal} {Phys. Rev. B}\ }\textbf {\bibinfo {volume} {92}},\ \bibinfo
  {pages} {134521}}\BibitemShut {NoStop}%
\bibitem [{\citenamefont {Man}\ \emph {et~al.}(2018)\citenamefont {Man},
  \citenamefont {Zhang}, \citenamefont {Park}, \citenamefont {Lu},
  \citenamefont {Kulda}, \citenamefont {Ivanov},\ and\ \citenamefont
  {Dai}}]{man2018direct}%
  \BibitemOpen
  \bibfield  {author} {\bibinfo {author} {\bibnamefont {Man}, \bibfnamefont
  {H.}}, \bibinfo {author} {\bibfnamefont {R.}~\bibnamefont {Zhang}}, \bibinfo
  {author} {\bibfnamefont {J.~T.}\ \bibnamefont {Park}}, \bibinfo {author}
  {\bibfnamefont {X.}~\bibnamefont {Lu}}, \bibinfo {author} {\bibfnamefont
  {J.}~\bibnamefont {Kulda}}, \bibinfo {author} {\bibfnamefont
  {A.}~\bibnamefont {Ivanov}}, and\ \bibinfo {author} {\bibfnamefont
  {P.}~\bibnamefont {Dai}}} (\bibinfo {year} {2018}),\ \href
  {https://doi.org/10.1103/PhysRevB.97.060507} {\bibfield  {journal} {\bibinfo
  {journal} {Phys. Rev. B}\ }\textbf {\bibinfo {volume} {97}},\ \bibinfo
  {pages} {060507}}\BibitemShut {NoStop}%
\bibitem [{\citenamefont {Manna}\ \emph {et~al.}(2017)\citenamefont {Manna},
  \citenamefont {Kamlapure}, \citenamefont {Cornils}, \citenamefont
  {H{\"a}nke}, \citenamefont {Hedegaard}, \citenamefont {Bremholm},
  \citenamefont {Iversen}, \citenamefont {Hofmann}, \citenamefont {Wiebe},\
  and\ \citenamefont {Wiesendanger}}]{manna2017interfacial}%
  \BibitemOpen
  \bibfield  {author} {\bibinfo {author} {\bibnamefont {Manna}, \bibfnamefont
  {S.}}, \bibinfo {author} {\bibfnamefont {A.}~\bibnamefont {Kamlapure}},
  \bibinfo {author} {\bibfnamefont {L.}~\bibnamefont {Cornils}}, \bibinfo
  {author} {\bibfnamefont {T.}~\bibnamefont {H{\"a}nke}}, \bibinfo {author}
  {\bibfnamefont {E.~M.~J.}\ \bibnamefont {Hedegaard}}, \bibinfo {author}
  {\bibfnamefont {M.}~\bibnamefont {Bremholm}}, \bibinfo {author}
  {\bibfnamefont {B.~B.}\ \bibnamefont {Iversen}}, \bibinfo {author}
  {\bibfnamefont {{\relax Ph}.}~\bibnamefont {Hofmann}}, \bibinfo {author}
  {\bibfnamefont {J.}~\bibnamefont {Wiebe}}, and\ \bibinfo {author}
  {\bibfnamefont {R.}~\bibnamefont {Wiesendanger}}} (\bibinfo {year} {2017}),\
  \href {https://doi.org/10.1038/ncomms14074} {\bibfield  {journal} {\bibinfo
  {journal} {Nat. Commun.}\ }\textbf {\bibinfo {volume} {8}},\ \bibinfo {pages}
  {14074}}\BibitemShut {NoStop}%
\bibitem [{\citenamefont {Martinelli}\ \emph {et~al.}(2017)\citenamefont
  {Martinelli}, \citenamefont {Manfrinetti}, \citenamefont {Provino},
  \citenamefont {Genovese}, \citenamefont {Caglieris}, \citenamefont {Lamura},
  \citenamefont {Ritter},\ and\ \citenamefont
  {Putti}}]{martinelli2017experimental}%
  \BibitemOpen
  \bibfield  {author} {\bibinfo {author} {\bibnamefont {Martinelli},
  \bibfnamefont {A.}}, \bibinfo {author} {\bibfnamefont {P.}~\bibnamefont
  {Manfrinetti}}, \bibinfo {author} {\bibfnamefont {A.}~\bibnamefont
  {Provino}}, \bibinfo {author} {\bibfnamefont {A.}~\bibnamefont {Genovese}},
  \bibinfo {author} {\bibfnamefont {F.}~\bibnamefont {Caglieris}}, \bibinfo
  {author} {\bibfnamefont {G.}~\bibnamefont {Lamura}}, \bibinfo {author}
  {\bibfnamefont {C.}~\bibnamefont {Ritter}}, and\ \bibinfo {author}
  {\bibfnamefont {M.}~\bibnamefont {Putti}}} (\bibinfo {year} {2017}),\ \href
  {https://doi.org/10.1103/PhysRevLett.118.055701} {\bibfield  {journal}
  {\bibinfo  {journal} {Phys. Rev. Lett.}\ }\textbf {\bibinfo {volume} {118}},\
  \bibinfo {pages} {055701}}\BibitemShut {NoStop}%
\bibitem [{\citenamefont {Massat}\ \emph {et~al.}(2016)\citenamefont {Massat},
  \citenamefont {Farina}, \citenamefont {Paul}, \citenamefont {Karlsson},
  \citenamefont {Strobel}, \citenamefont {Toulemonde}, \citenamefont
  {Méasson}, \citenamefont {Cazayous}, \citenamefont {Sacuto}, \citenamefont
  {Kasahara}, \citenamefont {Shibauchi}, \citenamefont {Matsuda},\ and\
  \citenamefont {Gallais}}]{massat2016charge}%
  \BibitemOpen
  \bibfield  {author} {\bibinfo {author} {\bibnamefont {Massat}, \bibfnamefont
  {P.}}, \bibinfo {author} {\bibfnamefont {D.}~\bibnamefont {Farina}}, \bibinfo
  {author} {\bibfnamefont {I.}~\bibnamefont {Paul}}, \bibinfo {author}
  {\bibfnamefont {S.}~\bibnamefont {Karlsson}}, \bibinfo {author}
  {\bibfnamefont {P.}~\bibnamefont {Strobel}}, \bibinfo {author} {\bibfnamefont
  {P.}~\bibnamefont {Toulemonde}}, \bibinfo {author} {\bibfnamefont {M.-A.}\
  \bibnamefont {Méasson}}, \bibinfo {author} {\bibfnamefont {M.}~\bibnamefont
  {Cazayous}}, \bibinfo {author} {\bibfnamefont {A.}~\bibnamefont {Sacuto}},
  \bibinfo {author} {\bibfnamefont {S.}~\bibnamefont {Kasahara}}, \bibinfo
  {author} {\bibfnamefont {T.}~\bibnamefont {Shibauchi}}, \bibinfo {author}
  {\bibfnamefont {Y.}~\bibnamefont {Matsuda}}, and\ \bibinfo {author}
  {\bibfnamefont {Y.}~\bibnamefont {Gallais}}} (\bibinfo {year} {2016}),\ \href
  {https://doi.org/10.1073/pnas.1606562113} {\bibfield  {journal} {\bibinfo
  {journal} {Proc. Natl. Acad. Sci. U.S.A.}\ }\textbf {\bibinfo {volume}
  {113}},\ \bibinfo {pages} {9177}}\BibitemShut {NoStop}%
\bibitem [{\citenamefont {Massat}\ \emph {et~al.}(2018)\citenamefont {Massat},
  \citenamefont {Quan}, \citenamefont {Grasset}, \citenamefont {M{\'e}asson},
  \citenamefont {Cazayous}, \citenamefont {Sacuto}, \citenamefont {Karlsson},
  \citenamefont {Strobel}, \citenamefont {Toulemonde}, \citenamefont {Yin},\
  and\ \citenamefont {Gallais}}]{massat2018collapse}%
  \BibitemOpen
  \bibfield  {author} {\bibinfo {author} {\bibnamefont {Massat}, \bibfnamefont
  {P.}}, \bibinfo {author} {\bibfnamefont {Y.}~\bibnamefont {Quan}}, \bibinfo
  {author} {\bibfnamefont {R.}~\bibnamefont {Grasset}}, \bibinfo {author}
  {\bibfnamefont {M.-A.}\ \bibnamefont {M{\'e}asson}}, \bibinfo {author}
  {\bibfnamefont {M.}~\bibnamefont {Cazayous}}, \bibinfo {author}
  {\bibfnamefont {A.}~\bibnamefont {Sacuto}}, \bibinfo {author} {\bibfnamefont
  {S.}~\bibnamefont {Karlsson}}, \bibinfo {author} {\bibfnamefont
  {P.}~\bibnamefont {Strobel}}, \bibinfo {author} {\bibfnamefont
  {P.}~\bibnamefont {Toulemonde}}, \bibinfo {author} {\bibfnamefont
  {Z.}~\bibnamefont {Yin}}, and\ \bibinfo {author} {\bibfnamefont
  {Y.}~\bibnamefont {Gallais}}} (\bibinfo {year} {2018}),\ \href
  {https://doi.org/10.1103/PhysRevLett.121.077001} {\bibfield  {journal}
  {\bibinfo  {journal} {Phys. Rev. Lett.}\ }\textbf {\bibinfo {volume} {121}},\
  \bibinfo {pages} {077001}}\BibitemShut {NoStop}%
\bibitem [{\citenamefont {Matsubayashi}\ \emph {et~al.}(2011)\citenamefont
  {Matsubayashi}, \citenamefont {Munakata}, \citenamefont {Isobe},
  \citenamefont {Katayama}, \citenamefont {Ohgushi}, \citenamefont {Ueda},
  \citenamefont {Kawamura}, \citenamefont {Mizumaki}, \citenamefont
  {Ishimatsu}, \citenamefont {Hedo} \emph {et~al.}}]{Matsubayashi2011}%
  \BibitemOpen
  \bibfield  {author} {\bibinfo {author} {\bibnamefont {Matsubayashi},
  \bibfnamefont {K.}}, \bibinfo {author} {\bibfnamefont {K.}~\bibnamefont
  {Munakata}}, \bibinfo {author} {\bibfnamefont {M.}~\bibnamefont {Isobe}},
  \bibinfo {author} {\bibfnamefont {N.}~\bibnamefont {Katayama}}, \bibinfo
  {author} {\bibfnamefont {K.}~\bibnamefont {Ohgushi}}, \bibinfo {author}
  {\bibfnamefont {Y.}~\bibnamefont {Ueda}}, \bibinfo {author} {\bibfnamefont
  {N.}~\bibnamefont {Kawamura}}, \bibinfo {author} {\bibfnamefont
  {M.}~\bibnamefont {Mizumaki}}, \bibinfo {author} {\bibfnamefont
  {N.}~\bibnamefont {Ishimatsu}}, \bibinfo {author} {\bibfnamefont
  {M.}~\bibnamefont {Hedo}},  \emph {et~al.}} (\bibinfo {year} {2011}),\ \href
  {https://doi.org/10.1103/PhysRevB.84.024502} {\bibfield  {journal} {\bibinfo
  {journal} {Phys. Rev. B}\ }\textbf {\bibinfo {volume} {84}},\ \bibinfo
  {pages} {024502}}\BibitemShut {NoStop}%
\bibitem [{\citenamefont {Matsuishi}\ \emph {et~al.}(2012)\citenamefont
  {Matsuishi}, \citenamefont {Hanna}, \citenamefont {Muraba}, \citenamefont
  {Kim}, \citenamefont {Kim}, \citenamefont {Takata}, \citenamefont {Shamoto},
  \citenamefont {Smith},\ and\ \citenamefont {Hosono}}]{Matsuishi2012}%
  \BibitemOpen
  \bibfield  {author} {\bibinfo {author} {\bibnamefont {Matsuishi},
  \bibfnamefont {S.}}, \bibinfo {author} {\bibfnamefont {T.}~\bibnamefont
  {Hanna}}, \bibinfo {author} {\bibfnamefont {Y.}~\bibnamefont {Muraba}},
  \bibinfo {author} {\bibfnamefont {S.~W.}\ \bibnamefont {Kim}}, \bibinfo
  {author} {\bibfnamefont {J.~E.}\ \bibnamefont {Kim}}, \bibinfo {author}
  {\bibfnamefont {M.}~\bibnamefont {Takata}}, \bibinfo {author} {\bibfnamefont
  {S.-i.}\ \bibnamefont {Shamoto}}, \bibinfo {author} {\bibfnamefont {R.~I.}\
  \bibnamefont {Smith}}, and\ \bibinfo {author} {\bibfnamefont
  {H.}~\bibnamefont {Hosono}}} (\bibinfo {year} {2012}),\ \href
  {https://doi.org/10.1103/PhysRevB.85.014514} {\bibfield  {journal} {\bibinfo
  {journal} {Phys. Rev. B}\ }\textbf {\bibinfo {volume} {85}},\ \bibinfo
  {pages} {014514}}\BibitemShut {NoStop}%
\bibitem [{\citenamefont {Matsuishi}\ \emph {et~al.}(2008)\citenamefont
  {Matsuishi}, \citenamefont {Inoue}, \citenamefont {Nomura}, \citenamefont
  {Yanagi}, \citenamefont {Hirano},\ and\ \citenamefont
  {Hosono}}]{Matsuishi2008}%
  \BibitemOpen
  \bibfield  {author} {\bibinfo {author} {\bibnamefont {Matsuishi},
  \bibfnamefont {S.}}, \bibinfo {author} {\bibfnamefont {Y.}~\bibnamefont
  {Inoue}}, \bibinfo {author} {\bibfnamefont {T.}~\bibnamefont {Nomura}},
  \bibinfo {author} {\bibfnamefont {H.}~\bibnamefont {Yanagi}}, \bibinfo
  {author} {\bibfnamefont {M.}~\bibnamefont {Hirano}}, and\ \bibinfo {author}
  {\bibfnamefont {H.}~\bibnamefont {Hosono}}} (\bibinfo {year} {2008}),\ \href
  {https://doi.org/10.1021/ja806357j} {\bibfield  {journal} {\bibinfo
  {journal} {J. Am. Chem. Soc.}\ }\textbf {\bibinfo {volume} {130}},\ \bibinfo
  {pages} {14428}}\BibitemShut {NoStop}%
\bibitem [{\citenamefont {Matsuishi}\ \emph {et~al.}(2014)\citenamefont
  {Matsuishi}, \citenamefont {Maruyama}, \citenamefont {Iimura},\ and\
  \citenamefont {Hosono}}]{Matsuishi2014}%
  \BibitemOpen
  \bibfield  {author} {\bibinfo {author} {\bibnamefont {Matsuishi},
  \bibfnamefont {S.}}, \bibinfo {author} {\bibfnamefont {T.}~\bibnamefont
  {Maruyama}}, \bibinfo {author} {\bibfnamefont {S.}~\bibnamefont {Iimura}},
  and\ \bibinfo {author} {\bibfnamefont {H.}~\bibnamefont {Hosono}}} (\bibinfo
  {year} {2014}),\ \href {https://doi.org/10.1103/PhysRevB.89.094510}
  {\bibfield  {journal} {\bibinfo  {journal} {Phys. Rev. B}\ }\textbf {\bibinfo
  {volume} {89}},\ \bibinfo {pages} {094510}}\BibitemShut {NoStop}%
\bibitem [{\citenamefont {Matsumoto}\ \emph {et~al.}(2019)\citenamefont
  {Matsumoto}, \citenamefont {Hanzawa}, \citenamefont {Sasase}, \citenamefont
  {Haindl}, \citenamefont {Katase}, \citenamefont {Hiramatsu},\ and\
  \citenamefont {Hosono}}]{Matsumoto2019}%
  \BibitemOpen
  \bibfield  {author} {\bibinfo {author} {\bibnamefont {Matsumoto},
  \bibfnamefont {J.}}, \bibinfo {author} {\bibfnamefont {K.}~\bibnamefont
  {Hanzawa}}, \bibinfo {author} {\bibfnamefont {M.}~\bibnamefont {Sasase}},
  \bibinfo {author} {\bibfnamefont {S.}~\bibnamefont {Haindl}}, \bibinfo
  {author} {\bibfnamefont {T.}~\bibnamefont {Katase}}, \bibinfo {author}
  {\bibfnamefont {H.}~\bibnamefont {Hiramatsu}}, and\ \bibinfo {author}
  {\bibfnamefont {H.}~\bibnamefont {Hosono}}} (\bibinfo {year} {2019}),\ \href
  {https://doi.org/10.1103/PhysRevMaterials.3.103401} {\bibfield  {journal}
  {\bibinfo  {journal} {Phys. Rev. Mater.}\ }\textbf {\bibinfo {volume} {3}},\
  \bibinfo {pages} {103401}}\BibitemShut {NoStop}%
\bibitem [{\citenamefont {Matsuura}\ \emph {et~al.}(2017)\citenamefont
  {Matsuura}, \citenamefont {Mizukami}, \citenamefont {Arai}, \citenamefont
  {Sugimura}, \citenamefont {Maejima}, \citenamefont {Machida}, \citenamefont
  {Watanuki}, \citenamefont {Fukuda}, \citenamefont {Yajima}, \citenamefont
  {Hiroi}, \citenamefont {Yip}, \citenamefont {Chan}, \citenamefont {Niu},
  \citenamefont {Hosoi}, \citenamefont {Ishida}, \citenamefont {Mukasa},
  \citenamefont {Kasahara}, \citenamefont {Cheng}, \citenamefont {Goh},
  \citenamefont {Matsuda}, \citenamefont {Uwatoko},\ and\ \citenamefont
  {Shibauchi}}]{matsuura2017maximizing}%
  \BibitemOpen
  \bibfield  {author} {\bibinfo {author} {\bibnamefont {Matsuura},
  \bibfnamefont {K.}}, \bibinfo {author} {\bibfnamefont {Y.}~\bibnamefont
  {Mizukami}}, \bibinfo {author} {\bibfnamefont {Y.}~\bibnamefont {Arai}},
  \bibinfo {author} {\bibfnamefont {Y.}~\bibnamefont {Sugimura}}, \bibinfo
  {author} {\bibfnamefont {N.}~\bibnamefont {Maejima}}, \bibinfo {author}
  {\bibfnamefont {A.}~\bibnamefont {Machida}}, \bibinfo {author} {\bibfnamefont
  {T.}~\bibnamefont {Watanuki}}, \bibinfo {author} {\bibfnamefont
  {T.}~\bibnamefont {Fukuda}}, \bibinfo {author} {\bibfnamefont
  {T.}~\bibnamefont {Yajima}}, \bibinfo {author} {\bibfnamefont
  {Z.}~\bibnamefont {Hiroi}}, \bibinfo {author} {\bibfnamefont {K.~Y.}\
  \bibnamefont {Yip}}, \bibinfo {author} {\bibfnamefont {Y.~C.}\ \bibnamefont
  {Chan}}, \bibinfo {author} {\bibfnamefont {Q.}~\bibnamefont {Niu}}, \bibinfo
  {author} {\bibfnamefont {S.}~\bibnamefont {Hosoi}}, \bibinfo {author}
  {\bibfnamefont {K.}~\bibnamefont {Ishida}}, \bibinfo {author} {\bibfnamefont
  {K.}~\bibnamefont {Mukasa}}, \bibinfo {author} {\bibfnamefont
  {S.}~\bibnamefont {Kasahara}}, \bibinfo {author} {\bibfnamefont {J.-G.}\
  \bibnamefont {Cheng}}, \bibinfo {author} {\bibfnamefont {S.~K.}\ \bibnamefont
  {Goh}}, \bibinfo {author} {\bibfnamefont {Y.}~\bibnamefont {Matsuda}},
  \bibinfo {author} {\bibfnamefont {Y.}~\bibnamefont {Uwatoko}}, and\ \bibinfo
  {author} {\bibfnamefont {T.}~\bibnamefont {Shibauchi}}} (\bibinfo {year}
  {2017}),\ \href {https://doi.org/10.1038/s41467-017-01277-x} {\bibfield
  {journal} {\bibinfo  {journal} {Nat. Commun.}\ }\textbf {\bibinfo {volume}
  {8}},\ \bibinfo {pages} {1143}}\BibitemShut {NoStop}%
\bibitem [{\citenamefont {Matsuura}\ \emph {et~al.}(2023)\citenamefont
  {Matsuura}, \citenamefont {Roppongi}, \citenamefont {Qiu}, \citenamefont
  {Sheng}, \citenamefont {Cai}, \citenamefont {Yamakawa}, \citenamefont
  {Guguchia}, \citenamefont {Day}, \citenamefont {Kojima}, \citenamefont
  {Damascelli}, \citenamefont {Sugimura}, \citenamefont {Saito}, \citenamefont
  {Takenaka}, \citenamefont {Ishihara}, \citenamefont {Mizukami}, \citenamefont
  {Hashimoto}, \citenamefont {Gu}, \citenamefont {Guo}, \citenamefont {Fu},
  \citenamefont {Zhang}, \citenamefont {Ning}, \citenamefont {Zhao},
  \citenamefont {Dai}, \citenamefont {Jin}, \citenamefont {Beare},
  \citenamefont {Luke}, \citenamefont {Uemura},\ and\ \citenamefont
  {Shibauchi}}]{matsuura2023two}%
  \BibitemOpen
  \bibfield  {author} {\bibinfo {author} {\bibnamefont {Matsuura},
  \bibfnamefont {K.}}, \bibinfo {author} {\bibfnamefont {M.}~\bibnamefont
  {Roppongi}}, \bibinfo {author} {\bibfnamefont {M.}~\bibnamefont {Qiu}},
  \bibinfo {author} {\bibfnamefont {Q.}~\bibnamefont {Sheng}}, \bibinfo
  {author} {\bibfnamefont {Y.}~\bibnamefont {Cai}}, \bibinfo {author}
  {\bibfnamefont {K.}~\bibnamefont {Yamakawa}}, \bibinfo {author}
  {\bibfnamefont {Z.}~\bibnamefont {Guguchia}}, \bibinfo {author}
  {\bibfnamefont {R.~P.}\ \bibnamefont {Day}}, \bibinfo {author} {\bibfnamefont
  {K.~M.}\ \bibnamefont {Kojima}}, \bibinfo {author} {\bibfnamefont
  {A.}~\bibnamefont {Damascelli}}, \bibinfo {author} {\bibfnamefont
  {Y.}~\bibnamefont {Sugimura}}, \bibinfo {author} {\bibfnamefont
  {M.}~\bibnamefont {Saito}}, \bibinfo {author} {\bibfnamefont
  {T.}~\bibnamefont {Takenaka}}, \bibinfo {author} {\bibfnamefont
  {K.}~\bibnamefont {Ishihara}}, \bibinfo {author} {\bibfnamefont
  {Y.}~\bibnamefont {Mizukami}}, \bibinfo {author} {\bibfnamefont
  {K.}~\bibnamefont {Hashimoto}}, \bibinfo {author} {\bibfnamefont
  {Y.}~\bibnamefont {Gu}}, \bibinfo {author} {\bibfnamefont {S.}~\bibnamefont
  {Guo}}, \bibinfo {author} {\bibfnamefont {L.}~\bibnamefont {Fu}}, \bibinfo
  {author} {\bibfnamefont {Z.}~\bibnamefont {Zhang}}, \bibinfo {author}
  {\bibfnamefont {F.}~\bibnamefont {Ning}}, \bibinfo {author} {\bibfnamefont
  {G.}~\bibnamefont {Zhao}}, \bibinfo {author} {\bibfnamefont {G.}~\bibnamefont
  {Dai}}, \bibinfo {author} {\bibfnamefont {C.}~\bibnamefont {Jin}}, \bibinfo
  {author} {\bibfnamefont {J.~W.}\ \bibnamefont {Beare}}, \bibinfo {author}
  {\bibfnamefont {G.~M.}\ \bibnamefont {Luke}}, \bibinfo {author}
  {\bibfnamefont {Y.~J.}\ \bibnamefont {Uemura}}, and\ \bibinfo {author}
  {\bibfnamefont {T.}~\bibnamefont {Shibauchi}}} (\bibinfo {year} {2023}),\
  \href {https://doi.org/10.1073/pnas.2208276120} {\bibfield  {journal}
  {\bibinfo  {journal} {Proc. Natl. Acad. Sci. U.S.A.}\ }\textbf {\bibinfo
  {volume} {120}},\ \bibinfo {pages} {e2208276120}}\BibitemShut {NoStop}%
\bibitem [{\citenamefont {Matusiak}\ \emph {et~al.}(2011)\citenamefont
  {Matusiak}, \citenamefont {Bukowski},\ and\ \citenamefont
  {Karpinski}}]{Matusiak2011}%
  \BibitemOpen
  \bibfield  {author} {\bibinfo {author} {\bibnamefont {Matusiak},
  \bibfnamefont {M.}}, \bibinfo {author} {\bibfnamefont {Z.}~\bibnamefont
  {Bukowski}}, and\ \bibinfo {author} {\bibfnamefont {J.}~\bibnamefont
  {Karpinski}}} (\bibinfo {year} {2011}),\ \href
  {https://doi.org/10.1103/PhysRevB.83.224505} {\bibfield  {journal} {\bibinfo
  {journal} {Phys. Rev. B}\ }\textbf {\bibinfo {volume} {83}},\ \bibinfo
  {pages} {224505}}\BibitemShut {NoStop}%
\bibitem [{\citenamefont {Mazin}(2011)}]{Mazin2011}%
  \BibitemOpen
  \bibfield  {author} {\bibinfo {author} {\bibnamefont {Mazin}, \bibfnamefont
  {I.}}} (\bibinfo {year} {2011}),\ \href
  {https://doi.org/10.1103/PhysRevB.84.024529} {\bibfield  {journal} {\bibinfo
  {journal} {Phys. Rev. B}\ }\textbf {\bibinfo {volume} {84}},\ \bibinfo
  {pages} {024529}}\BibitemShut {NoStop}%
\bibitem [{\citenamefont {de' Medici}(2011)}]{medici2011hund}%
  \BibitemOpen
  \bibfield  {author} {\bibinfo {author} {\bibnamefont {de' Medici},
  \bibfnamefont {L.}}} (\bibinfo {year} {2011}),\ \href
  {https://doi.org/10.1103/PhysRevB.83.205112} {\bibfield  {journal} {\bibinfo
  {journal} {Phys. Rev. B}\ }\textbf {\bibinfo {volume} {83}},\ \bibinfo
  {pages} {205112}}\BibitemShut {NoStop}%
\bibitem [{\citenamefont {Medvedev}\ \emph {et~al.}(2009)\citenamefont
  {Medvedev}, \citenamefont {McQueen}, \citenamefont {Troyan}, \citenamefont
  {Palasyuk}, \citenamefont {Eremets}, \citenamefont {Cava}, \citenamefont
  {Naghavi}, \citenamefont {Casper}, \citenamefont {Ksenofontov}, \citenamefont
  {Wortmann} \emph {et~al.}}]{Medvedev2009}%
  \BibitemOpen
  \bibfield  {author} {\bibinfo {author} {\bibnamefont {Medvedev},
  \bibfnamefont {S.}}, \bibinfo {author} {\bibfnamefont {T.}~\bibnamefont
  {McQueen}}, \bibinfo {author} {\bibfnamefont {I.}~\bibnamefont {Troyan}},
  \bibinfo {author} {\bibfnamefont {T.}~\bibnamefont {Palasyuk}}, \bibinfo
  {author} {\bibfnamefont {M.}~\bibnamefont {Eremets}}, \bibinfo {author}
  {\bibfnamefont {R.}~\bibnamefont {Cava}}, \bibinfo {author} {\bibfnamefont
  {S.}~\bibnamefont {Naghavi}}, \bibinfo {author} {\bibfnamefont
  {F.}~\bibnamefont {Casper}}, \bibinfo {author} {\bibfnamefont
  {V.}~\bibnamefont {Ksenofontov}}, \bibinfo {author} {\bibfnamefont
  {G.}~\bibnamefont {Wortmann}},  \emph {et~al.}} (\bibinfo {year} {2009}),\
  \href {https://doi.org/10.1038/nmat2491} {\bibfield  {journal} {\bibinfo
  {journal} {Nat. Mater.}\ }\textbf {\bibinfo {volume} {8}},\ \bibinfo {pages}
  {630}}\BibitemShut {NoStop}%
\bibitem [{\citenamefont {Meier}\ \emph {et~al.}(2018)\citenamefont {Meier},
  \citenamefont {Ding}, \citenamefont {Kreyssig}, \citenamefont {Bud'ko},
  \citenamefont {Sapkota}, \citenamefont {Kothapalli}, \citenamefont {Borisov},
  \citenamefont {Valent{\'i}}, \citenamefont {Batista}, \citenamefont {Orth},
  \citenamefont {Fernandes}, \citenamefont {Goldman}, \citenamefont {Furukawa},
  \citenamefont {B{\"o}hmer},\ and\ \citenamefont
  {Canfield}}]{meier2018hedgehog}%
  \BibitemOpen
  \bibfield  {author} {\bibinfo {author} {\bibnamefont {Meier}, \bibfnamefont
  {W.~R.}}, \bibinfo {author} {\bibfnamefont {Q.-P.}\ \bibnamefont {Ding}},
  \bibinfo {author} {\bibfnamefont {A.}~\bibnamefont {Kreyssig}}, \bibinfo
  {author} {\bibfnamefont {S.~L.}\ \bibnamefont {Bud'ko}}, \bibinfo {author}
  {\bibfnamefont {A.}~\bibnamefont {Sapkota}}, \bibinfo {author} {\bibfnamefont
  {K.}~\bibnamefont {Kothapalli}}, \bibinfo {author} {\bibfnamefont
  {V.}~\bibnamefont {Borisov}}, \bibinfo {author} {\bibfnamefont
  {R.}~\bibnamefont {Valent{\'i}}}, \bibinfo {author} {\bibfnamefont {C.~D.}\
  \bibnamefont {Batista}}, \bibinfo {author} {\bibfnamefont {P.~P.}\
  \bibnamefont {Orth}}, \bibinfo {author} {\bibfnamefont {R.~M.}\ \bibnamefont
  {Fernandes}}, \bibinfo {author} {\bibfnamefont {A.~I.}\ \bibnamefont
  {Goldman}}, \bibinfo {author} {\bibfnamefont {Y.}~\bibnamefont {Furukawa}},
  \bibinfo {author} {\bibfnamefont {A.~E.}\ \bibnamefont {B{\"o}hmer}}, and\
  \bibinfo {author} {\bibfnamefont {P.~C.}\ \bibnamefont {Canfield}}} (\bibinfo
  {year} {2018}),\ \href {https://doi.org/10.1038/s41535-017-0076-x} {\bibfield
   {journal} {\bibinfo  {journal} {npj Quantum Mater.}\ }\textbf {\bibinfo
  {volume} {3}},\ \bibinfo {pages} {5}}\BibitemShut {NoStop}%
\bibitem [{\citenamefont {Meier}\ \emph {et~al.}(2016)\citenamefont {Meier},
  \citenamefont {Kong}, \citenamefont {Kaluarachchi}, \citenamefont {Taufour},
  \citenamefont {Jo}, \citenamefont {Drachuck}, \citenamefont {B{\"o}hmer},
  \citenamefont {Saunders}, \citenamefont {Sapkota}, \citenamefont {Kreyssig}
  \emph {et~al.}}]{Meier2016}%
  \BibitemOpen
  \bibfield  {author} {\bibinfo {author} {\bibnamefont {Meier}, \bibfnamefont
  {W.~R.}}, \bibinfo {author} {\bibfnamefont {T.}~\bibnamefont {Kong}},
  \bibinfo {author} {\bibfnamefont {U.~S.}\ \bibnamefont {Kaluarachchi}},
  \bibinfo {author} {\bibfnamefont {V.}~\bibnamefont {Taufour}}, \bibinfo
  {author} {\bibfnamefont {N.~H.}\ \bibnamefont {Jo}}, \bibinfo {author}
  {\bibfnamefont {G.}~\bibnamefont {Drachuck}}, \bibinfo {author}
  {\bibfnamefont {A.}~\bibnamefont {B{\"o}hmer}}, \bibinfo {author}
  {\bibfnamefont {S.}~\bibnamefont {Saunders}}, \bibinfo {author}
  {\bibfnamefont {A.}~\bibnamefont {Sapkota}}, \bibinfo {author} {\bibfnamefont
  {A.}~\bibnamefont {Kreyssig}},  \emph {et~al.}} (\bibinfo {year} {2016}),\
  \href {https://doi.org/10.1103/PhysRevB.94.064501} {\bibfield  {journal}
  {\bibinfo  {journal} {Phys. Rev. B}\ }\textbf {\bibinfo {volume} {94}},\
  \bibinfo {pages} {064501}}\BibitemShut {NoStop}%
\bibitem [{\citenamefont {de~Melo}\ and\ \citenamefont {Loon}(2024)}]{BCS-BEC}%
  \BibitemOpen
  \bibfield  {author} {\bibinfo {author} {\bibnamefont {de~Melo}, \bibfnamefont
  {C.~A. R.~S.}}, and\ \bibinfo {author} {\bibfnamefont {S.~V.}\ \bibnamefont
  {Loon}}} (\bibinfo {year} {2024}),\ \href
  {https://doi.org/10.1146/annurev-conmatphys-032922-115341} {\bibfield
  {journal} {\bibinfo  {journal} {Annu. Rev. Condens. Matter Phys.}\ }\textbf
  {\bibinfo {volume} {15}},\ \bibinfo {pages} {109}}\BibitemShut {NoStop}%
\bibitem [{\citenamefont {Meng}\ \emph {et~al.}(2022)\citenamefont {Meng},
  \citenamefont {Xing}, \citenamefont {Yi}, \citenamefont {Li}, \citenamefont
  {Zhou}, \citenamefont {Li}, \citenamefont {Zhang}, \citenamefont {Wei},
  \citenamefont {Feng}, \citenamefont {Terashima} \emph {et~al.}}]{Meng2022}%
  \BibitemOpen
  \bibfield  {author} {\bibinfo {author} {\bibnamefont {Meng}, \bibfnamefont
  {Y.}}, \bibinfo {author} {\bibfnamefont {X.}~\bibnamefont {Xing}}, \bibinfo
  {author} {\bibfnamefont {X.}~\bibnamefont {Yi}}, \bibinfo {author}
  {\bibfnamefont {B.}~\bibnamefont {Li}}, \bibinfo {author} {\bibfnamefont
  {N.}~\bibnamefont {Zhou}}, \bibinfo {author} {\bibfnamefont {M.}~\bibnamefont
  {Li}}, \bibinfo {author} {\bibfnamefont {Y.}~\bibnamefont {Zhang}}, \bibinfo
  {author} {\bibfnamefont {W.}~\bibnamefont {Wei}}, \bibinfo {author}
  {\bibfnamefont {J.}~\bibnamefont {Feng}}, \bibinfo {author} {\bibfnamefont
  {K.}~\bibnamefont {Terashima}},  \emph {et~al.}} (\bibinfo {year} {2022}),\
  \href {https://doi.org/10.1103/PhysRevB.105.134506} {\bibfield  {journal}
  {\bibinfo  {journal} {Phys. Rev. B}\ }\textbf {\bibinfo {volume} {105}},\
  \bibinfo {pages} {134506}}\BibitemShut {NoStop}%
\bibitem [{\citenamefont {Merritt}\ \emph {et~al.}(2020)\citenamefont
  {Merritt}, \citenamefont {Weber}, \citenamefont {Castellan}, \citenamefont
  {Wolf}, \citenamefont {Ishikawa}, \citenamefont {Said}, \citenamefont
  {Alatas}, \citenamefont {Fernandes}, \citenamefont {Baron},\ and\
  \citenamefont {Reznik}}]{merritt2020nematic}%
  \BibitemOpen
  \bibfield  {author} {\bibinfo {author} {\bibnamefont {Merritt}, \bibfnamefont
  {A.~M.}}, \bibinfo {author} {\bibfnamefont {F.}~\bibnamefont {Weber}},
  \bibinfo {author} {\bibfnamefont {J.-P.}\ \bibnamefont {Castellan}}, \bibinfo
  {author} {\bibfnamefont {{\relax Th}.}~\bibnamefont {Wolf}}, \bibinfo
  {author} {\bibfnamefont {D.}~\bibnamefont {Ishikawa}}, \bibinfo {author}
  {\bibfnamefont {A.~H.}\ \bibnamefont {Said}}, \bibinfo {author}
  {\bibfnamefont {A.}~\bibnamefont {Alatas}}, \bibinfo {author} {\bibfnamefont
  {R.~M.}\ \bibnamefont {Fernandes}}, \bibinfo {author} {\bibfnamefont
  {A.~Q.~R.}\ \bibnamefont {Baron}}, and\ \bibinfo {author} {\bibfnamefont
  {D.}~\bibnamefont {Reznik}}} (\bibinfo {year} {2020}),\ \href
  {https://doi.org/10.1103/PhysRevLett.124.157001} {\bibfield  {journal}
  {\bibinfo  {journal} {Phys. Rev. Lett.}\ }\textbf {\bibinfo {volume} {124}},\
  \bibinfo {pages} {157001}}\BibitemShut {NoStop}%
\bibitem [{\citenamefont {Miclea}\ \emph {et~al.}(2009)\citenamefont {Miclea},
  \citenamefont {Nicklas}, \citenamefont {Jeevan}, \citenamefont {Kasinathan},
  \citenamefont {Hossain}, \citenamefont {Rosner}, \citenamefont {Gegenwart},
  \citenamefont {Geibel},\ and\ \citenamefont {Steglich}}]{Miclea2009}%
  \BibitemOpen
  \bibfield  {author} {\bibinfo {author} {\bibnamefont {Miclea}, \bibfnamefont
  {C.}}, \bibinfo {author} {\bibfnamefont {M.}~\bibnamefont {Nicklas}},
  \bibinfo {author} {\bibfnamefont {H.~S.}\ \bibnamefont {Jeevan}}, \bibinfo
  {author} {\bibfnamefont {D.}~\bibnamefont {Kasinathan}}, \bibinfo {author}
  {\bibfnamefont {Z.}~\bibnamefont {Hossain}}, \bibinfo {author} {\bibfnamefont
  {H.}~\bibnamefont {Rosner}}, \bibinfo {author} {\bibfnamefont
  {P.}~\bibnamefont {Gegenwart}}, \bibinfo {author} {\bibfnamefont
  {C.}~\bibnamefont {Geibel}}, and\ \bibinfo {author} {\bibfnamefont
  {F.}~\bibnamefont {Steglich}}} (\bibinfo {year} {2009}),\ \href
  {https://doi.org/10.1103/PhysRevB.79.212509} {\bibfield  {journal} {\bibinfo
  {journal} {Phys. Rev. B}\ }\textbf {\bibinfo {volume} {79}},\ \bibinfo
  {pages} {212509}}\BibitemShut {NoStop}%
\bibitem [{\citenamefont {Mirri}\ \emph {et~al.}(2015)\citenamefont {Mirri},
  \citenamefont {Dusza}, \citenamefont {Bastelberger}, \citenamefont
  {Chinotti}, \citenamefont {Degiorgi}, \citenamefont {Chu}, \citenamefont
  {Kuo},\ and\ \citenamefont {Fisher}}]{mirri2015origin}%
  \BibitemOpen
  \bibfield  {author} {\bibinfo {author} {\bibnamefont {Mirri}, \bibfnamefont
  {C.}}, \bibinfo {author} {\bibfnamefont {A.}~\bibnamefont {Dusza}}, \bibinfo
  {author} {\bibfnamefont {S.}~\bibnamefont {Bastelberger}}, \bibinfo {author}
  {\bibfnamefont {M.}~\bibnamefont {Chinotti}}, \bibinfo {author}
  {\bibfnamefont {L.}~\bibnamefont {Degiorgi}}, \bibinfo {author}
  {\bibfnamefont {J.-H.}\ \bibnamefont {Chu}}, \bibinfo {author} {\bibfnamefont
  {H.-H.}\ \bibnamefont {Kuo}}, and\ \bibinfo {author} {\bibfnamefont {I.~R.}\
  \bibnamefont {Fisher}}} (\bibinfo {year} {2015}),\ \href
  {https://doi.org/10.1103/PhysRevLett.115.107001} {\bibfield  {journal}
  {\bibinfo  {journal} {Phys. Rev. Lett.}\ }\textbf {\bibinfo {volume} {115}},\
  \bibinfo {pages} {107001}}\BibitemShut {NoStop}%
\bibitem [{\citenamefont {Misawa}\ and\ \citenamefont
  {Imada}(2014)}]{misawa2014superconductivity}%
  \BibitemOpen
  \bibfield  {author} {\bibinfo {author} {\bibnamefont {Misawa}, \bibfnamefont
  {T.}}, and\ \bibinfo {author} {\bibfnamefont {M.}~\bibnamefont {Imada}}}
  (\bibinfo {year} {2014}),\ \href {https://doi.org/10.1038/ncomms6738}
  {\bibfield  {journal} {\bibinfo  {journal} {Nat. Commun.}\ }\textbf {\bibinfo
  {volume} {5}},\ \bibinfo {pages} {5738}}\BibitemShut {NoStop}%
\bibitem [{\citenamefont {Mishev}\ \emph {et~al.}(2016)\citenamefont {Mishev},
  \citenamefont {Nakajima}, \citenamefont {Eisaki},\ and\ \citenamefont
  {Eisterer}}]{Mishev2016}%
  \BibitemOpen
  \bibfield  {author} {\bibinfo {author} {\bibnamefont {Mishev}, \bibfnamefont
  {V.}}, \bibinfo {author} {\bibfnamefont {M.}~\bibnamefont {Nakajima}},
  \bibinfo {author} {\bibfnamefont {H.}~\bibnamefont {Eisaki}}, and\ \bibinfo
  {author} {\bibfnamefont {M.}~\bibnamefont {Eisterer}}} (\bibinfo {year}
  {2016}),\ \href {https://doi.org/10.1038/srep27783} {\bibfield  {journal}
  {\bibinfo  {journal} {Sci. Rep.}\ }\textbf {\bibinfo {volume} {6}},\ \bibinfo
  {pages} {27783}}\BibitemShut {NoStop}%
\bibitem [{\citenamefont {Mitrano}\ \emph {et~al.}(2024)\citenamefont
  {Mitrano}, \citenamefont {Johnston}, \citenamefont {Kim},\ and\ \citenamefont
  {Dean}}]{mitrano2024exploring}%
  \BibitemOpen
  \bibfield  {author} {\bibinfo {author} {\bibnamefont {Mitrano}, \bibfnamefont
  {M.}}, \bibinfo {author} {\bibfnamefont {S.}~\bibnamefont {Johnston}},
  \bibinfo {author} {\bibfnamefont {Y.-J.}\ \bibnamefont {Kim}}, and\ \bibinfo
  {author} {\bibfnamefont {M.~P.~M.}\ \bibnamefont {Dean}}} (\bibinfo {year}
  {2024}),\ \href {https://doi.org/10.1103/PhysRevX.14.040501} {\bibfield
  {journal} {\bibinfo  {journal} {Phys. Rev. X}\ }\textbf {\bibinfo {volume}
  {14}},\ \bibinfo {pages} {040501}}\BibitemShut {NoStop}%
\bibitem [{\citenamefont {Miura}\ \emph {et~al.}(2024)\citenamefont {Miura},
  \citenamefont {Eley}, \citenamefont {Iida}, \citenamefont {Hanzawa},
  \citenamefont {Matsumoto}, \citenamefont {Hiramatsu}, \citenamefont
  {Ogimoto}, \citenamefont {Suzuki}, \citenamefont {Kobayashi}, \citenamefont
  {Ozaki} \emph {et~al.}}]{Miura2024}%
  \BibitemOpen
  \bibfield  {author} {\bibinfo {author} {\bibnamefont {Miura}, \bibfnamefont
  {M.}}, \bibinfo {author} {\bibfnamefont {S.}~\bibnamefont {Eley}}, \bibinfo
  {author} {\bibfnamefont {K.}~\bibnamefont {Iida}}, \bibinfo {author}
  {\bibfnamefont {K.}~\bibnamefont {Hanzawa}}, \bibinfo {author} {\bibfnamefont
  {J.}~\bibnamefont {Matsumoto}}, \bibinfo {author} {\bibfnamefont
  {H.}~\bibnamefont {Hiramatsu}}, \bibinfo {author} {\bibfnamefont
  {Y.}~\bibnamefont {Ogimoto}}, \bibinfo {author} {\bibfnamefont
  {T.}~\bibnamefont {Suzuki}}, \bibinfo {author} {\bibfnamefont
  {T.}~\bibnamefont {Kobayashi}}, \bibinfo {author} {\bibfnamefont
  {T.}~\bibnamefont {Ozaki}},  \emph {et~al.}} (\bibinfo {year} {2024}),\ \href
  {https://doi.org/10.1038/s41563-024-01952-7} {\bibfield  {journal} {\bibinfo
  {journal} {Nat. Mater.}\ }\textbf {\bibinfo {volume} {23}},\ \bibinfo {pages}
  {1370}}\BibitemShut {NoStop}%
\bibitem [{\citenamefont {Miyata}\ \emph {et~al.}(2015)\citenamefont {Miyata},
  \citenamefont {Nakayama}, \citenamefont {Sugawara}, \citenamefont {Sato},\
  and\ \citenamefont {Takahashi}}]{miyata2015hightemperature}%
  \BibitemOpen
  \bibfield  {author} {\bibinfo {author} {\bibnamefont {Miyata}, \bibfnamefont
  {Y.}}, \bibinfo {author} {\bibfnamefont {K.}~\bibnamefont {Nakayama}},
  \bibinfo {author} {\bibfnamefont {K.}~\bibnamefont {Sugawara}}, \bibinfo
  {author} {\bibfnamefont {T.}~\bibnamefont {Sato}}, and\ \bibinfo {author}
  {\bibfnamefont {T.}~\bibnamefont {Takahashi}}} (\bibinfo {year} {2015}),\
  \href {https://doi.org/10.1038/nmat4302} {\bibfield  {journal} {\bibinfo
  {journal} {Nat. Mater.}\ }\textbf {\bibinfo {volume} {14}},\ \bibinfo {pages}
  {775}}\BibitemShut {NoStop}%
\bibitem [{\citenamefont {Miyawaki}\ \emph {et~al.}(2022)\citenamefont
  {Miyawaki}, \citenamefont {Fujii}, \citenamefont {Imazono}, \citenamefont
  {Horiba}, \citenamefont {Ohtsubo}, \citenamefont {Inami}, \citenamefont
  {Nakatani}, \citenamefont {Inaba}, \citenamefont {Agui}, \citenamefont
  {Kimura},\ and\ \citenamefont {Takahasi}}]{nanoterasu}%
  \BibitemOpen
  \bibfield  {author} {\bibinfo {author} {\bibnamefont {Miyawaki},
  \bibfnamefont {J.}}, \bibinfo {author} {\bibfnamefont {K.}~\bibnamefont
  {Fujii}}, \bibinfo {author} {\bibfnamefont {T.}~\bibnamefont {Imazono}},
  \bibinfo {author} {\bibfnamefont {K.}~\bibnamefont {Horiba}}, \bibinfo
  {author} {\bibfnamefont {Y.}~\bibnamefont {Ohtsubo}}, \bibinfo {author}
  {\bibfnamefont {N.}~\bibnamefont {Inami}}, \bibinfo {author} {\bibfnamefont
  {T.}~\bibnamefont {Nakatani}}, \bibinfo {author} {\bibfnamefont
  {K.}~\bibnamefont {Inaba}}, \bibinfo {author} {\bibfnamefont
  {A.}~\bibnamefont {Agui}}, \bibinfo {author} {\bibfnamefont {H.}~\bibnamefont
  {Kimura}}, and\ \bibinfo {author} {\bibfnamefont {M.}~\bibnamefont
  {Takahasi}}} (\bibinfo {year} {2022}),\ \href
  {https://doi.org/10.1088/1742-6596/2380/1/012030} {\bibfield  {journal}
  {\bibinfo  {journal} {Journal of Physics: Conference Series}\ }\textbf
  {\bibinfo {volume} {2380}},\ \bibinfo {pages} {012030}}\BibitemShut {NoStop}%
\bibitem [{\citenamefont {Mizuguchi}\ \emph {et~al.}(2010)\citenamefont
  {Mizuguchi}, \citenamefont {Hara}, \citenamefont {Deguchi}, \citenamefont
  {Tsuda}, \citenamefont {Yamaguchi}, \citenamefont {Takeda}, \citenamefont
  {Kotegawa}, \citenamefont {Tou},\ and\ \citenamefont
  {Takano}}]{Mizuguchi2010}%
  \BibitemOpen
  \bibfield  {author} {\bibinfo {author} {\bibnamefont {Mizuguchi},
  \bibfnamefont {Y.}}, \bibinfo {author} {\bibfnamefont {Y.}~\bibnamefont
  {Hara}}, \bibinfo {author} {\bibfnamefont {K.}~\bibnamefont {Deguchi}},
  \bibinfo {author} {\bibfnamefont {S.}~\bibnamefont {Tsuda}}, \bibinfo
  {author} {\bibfnamefont {T.}~\bibnamefont {Yamaguchi}}, \bibinfo {author}
  {\bibfnamefont {K.}~\bibnamefont {Takeda}}, \bibinfo {author} {\bibfnamefont
  {H.}~\bibnamefont {Kotegawa}}, \bibinfo {author} {\bibfnamefont
  {H.}~\bibnamefont {Tou}}, and\ \bibinfo {author} {\bibfnamefont
  {Y.}~\bibnamefont {Takano}}} (\bibinfo {year} {2010}),\ \href
  {https://doi.org/10.1088/0953-2048/23/5/054013} {\bibfield  {journal}
  {\bibinfo  {journal} {Supercond. Sci. Technol.}\ }\textbf {\bibinfo {volume}
  {23}},\ \bibinfo {pages} {054013}}\BibitemShut {NoStop}%
\bibitem [{\citenamefont {Mizuguchi}\ \emph
  {et~al.}(2009{\natexlab{a}})\citenamefont {Mizuguchi}, \citenamefont
  {Tomioka}, \citenamefont {Tsuda}, \citenamefont {Yamaguchi},\ and\
  \citenamefont {Takano}}]{Mizuguchi2009a}%
  \BibitemOpen
  \bibfield  {author} {\bibinfo {author} {\bibnamefont {Mizuguchi},
  \bibfnamefont {Y.}}, \bibinfo {author} {\bibfnamefont {F.}~\bibnamefont
  {Tomioka}}, \bibinfo {author} {\bibfnamefont {S.}~\bibnamefont {Tsuda}},
  \bibinfo {author} {\bibfnamefont {T.}~\bibnamefont {Yamaguchi}}, and\
  \bibinfo {author} {\bibfnamefont {Y.}~\bibnamefont {Takano}}} (\bibinfo
  {year} {2009}{\natexlab{a}}),\ \href {https://doi.org/10.1143/jpsj.78.074712}
  {\bibfield  {journal} {\bibinfo  {journal} {J. Phys. Soc. Jpn.}\ }\textbf
  {\bibinfo {volume} {78}},\ \bibinfo {pages} {074712}}\BibitemShut {NoStop}%
\bibitem [{\citenamefont {Mizuguchi}\ \emph
  {et~al.}(2009{\natexlab{b}})\citenamefont {Mizuguchi}, \citenamefont
  {Tomioka}, \citenamefont {Tsuda}, \citenamefont {Yamaguchi},\ and\
  \citenamefont {Takano}}]{Mizuguchi2009b}%
  \BibitemOpen
  \bibfield  {author} {\bibinfo {author} {\bibnamefont {Mizuguchi},
  \bibfnamefont {Y.}}, \bibinfo {author} {\bibfnamefont {F.}~\bibnamefont
  {Tomioka}}, \bibinfo {author} {\bibfnamefont {S.}~\bibnamefont {Tsuda}},
  \bibinfo {author} {\bibfnamefont {T.}~\bibnamefont {Yamaguchi}}, and\
  \bibinfo {author} {\bibfnamefont {Y.}~\bibnamefont {Takano}}} (\bibinfo
  {year} {2009}{\natexlab{b}}),\ \href {https://doi.org/10.1063/1.3058720}
  {\bibfield  {journal} {\bibinfo  {journal} {Appl. Phys. Lett.}\ }\textbf
  {\bibinfo {volume} {94}},\ \bibinfo {pages} {012503}}\BibitemShut {NoStop}%
\bibitem [{\citenamefont {Mizukami}\ \emph {et~al.}(2023)\citenamefont
  {Mizukami}, \citenamefont {Haze}, \citenamefont {Tanaka}, \citenamefont
  {Matsuura}, \citenamefont {Sano}, \citenamefont {B{\"o}ker}, \citenamefont
  {Eremin}, \citenamefont {Kasahara}, \citenamefont {Matsuda},\ and\
  \citenamefont {Shibauchi}}]{mizukami2023unusual}%
  \BibitemOpen
  \bibfield  {author} {\bibinfo {author} {\bibnamefont {Mizukami},
  \bibfnamefont {Y.}}, \bibinfo {author} {\bibfnamefont {M.}~\bibnamefont
  {Haze}}, \bibinfo {author} {\bibfnamefont {O.}~\bibnamefont {Tanaka}},
  \bibinfo {author} {\bibfnamefont {K.}~\bibnamefont {Matsuura}}, \bibinfo
  {author} {\bibfnamefont {D.}~\bibnamefont {Sano}}, \bibinfo {author}
  {\bibfnamefont {J.}~\bibnamefont {B{\"o}ker}}, \bibinfo {author}
  {\bibfnamefont {I.}~\bibnamefont {Eremin}}, \bibinfo {author} {\bibfnamefont
  {S.}~\bibnamefont {Kasahara}}, \bibinfo {author} {\bibfnamefont
  {Y.}~\bibnamefont {Matsuda}}, and\ \bibinfo {author} {\bibfnamefont
  {T.}~\bibnamefont {Shibauchi}}} (\bibinfo {year} {2023}),\ \href
  {https://doi.org/10.1038/s42005-023-01289-8} {\bibfield  {journal} {\bibinfo
  {journal} {Commun. Phys.}\ }\textbf {\bibinfo {volume} {6}},\ \bibinfo
  {pages} {183}}\BibitemShut {NoStop}%
\bibitem [{\citenamefont {Mizukami}\ \emph {et~al.}(2025)\citenamefont
  {Mizukami}, \citenamefont {Tanaka}, \citenamefont {Ishida}, \citenamefont
  {Onishi}, \citenamefont {Kageyama}, \citenamefont {Tsujii}, \citenamefont
  {Ohno}, \citenamefont {Kimura}, \citenamefont {Mitsui}, \citenamefont
  {Kitao}, \citenamefont {Kurokuzu}, \citenamefont {Seto}, \citenamefont
  {Ishida}, \citenamefont {Iyo}, \citenamefont {Eisaki}, \citenamefont
  {Hashimoto},\ and\ \citenamefont {Shibauchi}}]{mizukami2025thermodynamic}%
  \BibitemOpen
  \bibfield  {author} {\bibinfo {author} {\bibnamefont {Mizukami},
  \bibfnamefont {Y.}}, \bibinfo {author} {\bibfnamefont {O.}~\bibnamefont
  {Tanaka}}, \bibinfo {author} {\bibfnamefont {K.}~\bibnamefont {Ishida}},
  \bibinfo {author} {\bibfnamefont {A.}~\bibnamefont {Onishi}}, \bibinfo
  {author} {\bibfnamefont {Y.}~\bibnamefont {Kageyama}}, \bibinfo {author}
  {\bibfnamefont {M.}~\bibnamefont {Tsujii}}, \bibinfo {author} {\bibfnamefont
  {R.}~\bibnamefont {Ohno}}, \bibinfo {author} {\bibfnamefont {N.}~\bibnamefont
  {Kimura}}, \bibinfo {author} {\bibfnamefont {T.}~\bibnamefont {Mitsui}},
  \bibinfo {author} {\bibfnamefont {S.}~\bibnamefont {Kitao}}, \bibinfo
  {author} {\bibfnamefont {M.}~\bibnamefont {Kurokuzu}}, \bibinfo {author}
  {\bibfnamefont {M.}~\bibnamefont {Seto}}, \bibinfo {author} {\bibfnamefont
  {S.}~\bibnamefont {Ishida}}, \bibinfo {author} {\bibfnamefont
  {A.}~\bibnamefont {Iyo}}, \bibinfo {author} {\bibfnamefont {H.}~\bibnamefont
  {Eisaki}}, \bibinfo {author} {\bibfnamefont {K.}~\bibnamefont {Hashimoto}},
  and\ \bibinfo {author} {\bibfnamefont {T.}~\bibnamefont {Shibauchi}}}
  (\bibinfo {year} {2025}),\ \href {https://doi.org/10.1093/pnasnexus/pgaf060}
  {\bibfield  {journal} {\bibinfo  {journal} {PNAS Nexus}\ }\textbf {\bibinfo
  {volume} {4}},\ \bibinfo {pages} {pgaf060}}\BibitemShut {NoStop}%
\bibitem [{\citenamefont {Moodenbaugh}\ \emph {et~al.}(1988)\citenamefont
  {Moodenbaugh}, \citenamefont {Xu}, \citenamefont {Suenaga}, \citenamefont
  {Folkerts},\ and\ \citenamefont {Shelton}}]{Moodenbaugh1988}%
  \BibitemOpen
  \bibfield  {author} {\bibinfo {author} {\bibnamefont {Moodenbaugh},
  \bibfnamefont {A.}}, \bibinfo {author} {\bibfnamefont {Y.}~\bibnamefont
  {Xu}}, \bibinfo {author} {\bibfnamefont {M.}~\bibnamefont {Suenaga}},
  \bibinfo {author} {\bibfnamefont {T.}~\bibnamefont {Folkerts}}, and\ \bibinfo
  {author} {\bibfnamefont {R.}~\bibnamefont {Shelton}}} (\bibinfo {year}
  {1988}),\ \href {https://doi.org/10.1103/PhysRevB.38.4596} {\bibfield
  {journal} {\bibinfo  {journal} {Phys. Rev. B}\ }\textbf {\bibinfo {volume}
  {38}},\ \bibinfo {pages} {4596}}\BibitemShut {NoStop}%
\bibitem [{\citenamefont {Moon}\ \emph {et~al.}(2016)\citenamefont {Moon},
  \citenamefont {Park}, \citenamefont {Haule},\ and\ \citenamefont
  {Shim}}]{Moon2016}%
  \BibitemOpen
  \bibfield  {author} {\bibinfo {author} {\bibnamefont {Moon}, \bibfnamefont
  {C.-Y.}}, \bibinfo {author} {\bibfnamefont {H.}~\bibnamefont {Park}},
  \bibinfo {author} {\bibfnamefont {K.}~\bibnamefont {Haule}}, and\ \bibinfo
  {author} {\bibfnamefont {J.~H.}\ \bibnamefont {Shim}}} (\bibinfo {year}
  {2016}),\ \href {https://doi.org/10.1103/PhysRevB.94.224511} {\bibfield
  {journal} {\bibinfo  {journal} {Phys. Rev. B}\ }\textbf {\bibinfo {volume}
  {94}},\ \bibinfo {pages} {224511}}\BibitemShut {NoStop}%
\bibitem [{\citenamefont {Mou}\ \emph {et~al.}(2011)\citenamefont {Mou},
  \citenamefont {Liu}, \citenamefont {Jia}, \citenamefont {He}, \citenamefont
  {Peng}, \citenamefont {Zhao}, \citenamefont {Yu}, \citenamefont {Liu},
  \citenamefont {He}, \citenamefont {Dong} \emph {et~al.}}]{Mou2011}%
  \BibitemOpen
  \bibfield  {author} {\bibinfo {author} {\bibnamefont {Mou}, \bibfnamefont
  {D.}}, \bibinfo {author} {\bibfnamefont {S.}~\bibnamefont {Liu}}, \bibinfo
  {author} {\bibfnamefont {X.}~\bibnamefont {Jia}}, \bibinfo {author}
  {\bibfnamefont {J.}~\bibnamefont {He}}, \bibinfo {author} {\bibfnamefont
  {Y.}~\bibnamefont {Peng}}, \bibinfo {author} {\bibfnamefont {L.}~\bibnamefont
  {Zhao}}, \bibinfo {author} {\bibfnamefont {L.}~\bibnamefont {Yu}}, \bibinfo
  {author} {\bibfnamefont {G.}~\bibnamefont {Liu}}, \bibinfo {author}
  {\bibfnamefont {S.}~\bibnamefont {He}}, \bibinfo {author} {\bibfnamefont
  {X.}~\bibnamefont {Dong}},  \emph {et~al.}} (\bibinfo {year} {2011}),\ \href
  {https://doi.org/10.1103/PhysRevLett.106.107001} {\bibfield  {journal}
  {\bibinfo  {journal} {Phys. Rev. Lett.}\ }\textbf {\bibinfo {volume} {106}},\
  \bibinfo {pages} {107001}}\BibitemShut {NoStop}%
\bibitem [{\citenamefont {Mueller}\ \emph {et~al.}(2016)\citenamefont
  {Mueller}, \citenamefont {Kusne},\ and\ \citenamefont
  {Ramprasad}}]{Mueller2016}%
  \BibitemOpen
  \bibfield  {author} {\bibinfo {author} {\bibnamefont {Mueller}, \bibfnamefont
  {T.}}, \bibinfo {author} {\bibfnamefont {A.~G.}\ \bibnamefont {Kusne}}, and\
  \bibinfo {author} {\bibfnamefont {R.}~\bibnamefont {Ramprasad}}} (\bibinfo
  {year} {2016}),\ \href {https://doi.org/10.1002/9781119148739.ch4} {\bibfield
   {journal} {\bibinfo  {journal} {Reviews in computational chemistry}\
  }\textbf {\bibinfo {volume} {29}},\ \bibinfo {pages} {186}}\BibitemShut
  {NoStop}%
\bibitem [{\citenamefont {Mukasa}\ \emph {et~al.}(2023)\citenamefont {Mukasa},
  \citenamefont {Ishida}, \citenamefont {Imajo}, \citenamefont {Qiu},
  \citenamefont {Saito}, \citenamefont {Matsuura}, \citenamefont {Sugimura},
  \citenamefont {Liu}, \citenamefont {Uezono}, \citenamefont {Otsuka},
  \citenamefont {{\v C}ulo}, \citenamefont {Kasahara}, \citenamefont {Matsuda},
  \citenamefont {Hussey}, \citenamefont {Watanabe}, \citenamefont {Kindo},\
  and\ \citenamefont {Shibauchi}}]{mukasa2023enhanced}%
  \BibitemOpen
  \bibfield  {author} {\bibinfo {author} {\bibnamefont {Mukasa}, \bibfnamefont
  {K.}}, \bibinfo {author} {\bibfnamefont {K.}~\bibnamefont {Ishida}}, \bibinfo
  {author} {\bibfnamefont {S.}~\bibnamefont {Imajo}}, \bibinfo {author}
  {\bibfnamefont {M.}~\bibnamefont {Qiu}}, \bibinfo {author} {\bibfnamefont
  {M.}~\bibnamefont {Saito}}, \bibinfo {author} {\bibfnamefont
  {K.}~\bibnamefont {Matsuura}}, \bibinfo {author} {\bibfnamefont
  {Y.}~\bibnamefont {Sugimura}}, \bibinfo {author} {\bibfnamefont
  {S.}~\bibnamefont {Liu}}, \bibinfo {author} {\bibfnamefont {Y.}~\bibnamefont
  {Uezono}}, \bibinfo {author} {\bibfnamefont {T.}~\bibnamefont {Otsuka}},
  \bibinfo {author} {\bibfnamefont {M.}~\bibnamefont {{\v C}ulo}}, \bibinfo
  {author} {\bibfnamefont {S.}~\bibnamefont {Kasahara}}, \bibinfo {author}
  {\bibfnamefont {Y.}~\bibnamefont {Matsuda}}, \bibinfo {author} {\bibfnamefont
  {N.~E.}\ \bibnamefont {Hussey}}, \bibinfo {author} {\bibfnamefont
  {T.}~\bibnamefont {Watanabe}}, \bibinfo {author} {\bibfnamefont
  {K.}~\bibnamefont {Kindo}}, and\ \bibinfo {author} {\bibfnamefont
  {T.}~\bibnamefont {Shibauchi}}} (\bibinfo {year} {2023}),\ \href
  {https://doi.org/10.1103/PhysRevX.13.011032} {\bibfield  {journal} {\bibinfo
  {journal} {Phys. Rev. X}\ }\textbf {\bibinfo {volume} {13}},\ \bibinfo
  {pages} {011032}}\BibitemShut {NoStop}%
\bibitem [{\citenamefont {Mukasa}\ \emph {et~al.}(2021)\citenamefont {Mukasa},
  \citenamefont {Matsuura}, \citenamefont {Qiu}, \citenamefont {Saito},
  \citenamefont {Sugimura}, \citenamefont {Ishida}, \citenamefont {Otani},
  \citenamefont {Onishi}, \citenamefont {Mizukami}, \citenamefont {Hashimoto},
  \citenamefont {Gouchi}, \citenamefont {Kumai}, \citenamefont {Uwatoko},\ and\
  \citenamefont {Shibauchi}}]{mukasa2021highpressure}%
  \BibitemOpen
  \bibfield  {author} {\bibinfo {author} {\bibnamefont {Mukasa}, \bibfnamefont
  {K.}}, \bibinfo {author} {\bibfnamefont {K.}~\bibnamefont {Matsuura}},
  \bibinfo {author} {\bibfnamefont {M.}~\bibnamefont {Qiu}}, \bibinfo {author}
  {\bibfnamefont {M.}~\bibnamefont {Saito}}, \bibinfo {author} {\bibfnamefont
  {Y.}~\bibnamefont {Sugimura}}, \bibinfo {author} {\bibfnamefont
  {K.}~\bibnamefont {Ishida}}, \bibinfo {author} {\bibfnamefont
  {M.}~\bibnamefont {Otani}}, \bibinfo {author} {\bibfnamefont
  {Y.}~\bibnamefont {Onishi}}, \bibinfo {author} {\bibfnamefont
  {Y.}~\bibnamefont {Mizukami}}, \bibinfo {author} {\bibfnamefont
  {K.}~\bibnamefont {Hashimoto}}, \bibinfo {author} {\bibfnamefont
  {J.}~\bibnamefont {Gouchi}}, \bibinfo {author} {\bibfnamefont
  {R.}~\bibnamefont {Kumai}}, \bibinfo {author} {\bibfnamefont
  {Y.}~\bibnamefont {Uwatoko}}, and\ \bibinfo {author} {\bibfnamefont
  {T.}~\bibnamefont {Shibauchi}}} (\bibinfo {year} {2021}),\ \href
  {https://doi.org/10.1038/s41467-020-20621-2} {\bibfield  {journal} {\bibinfo
  {journal} {Nat. Commun.}\ }\textbf {\bibinfo {volume} {12}},\ \bibinfo
  {pages} {381}}\BibitemShut {NoStop}%
\bibitem [{\citenamefont {Muraba}\ \emph {et~al.}(2014)\citenamefont {Muraba},
  \citenamefont {Matsuishi},\ and\ \citenamefont {Hosono}}]{Muraba2014}%
  \BibitemOpen
  \bibfield  {author} {\bibinfo {author} {\bibnamefont {Muraba}, \bibfnamefont
  {Y.}}, \bibinfo {author} {\bibfnamefont {S.}~\bibnamefont {Matsuishi}}, and\
  \bibinfo {author} {\bibfnamefont {H.}~\bibnamefont {Hosono}}} (\bibinfo
  {year} {2014}),\ \href {https://doi.org/10.7566/jpsj.83.033705} {\bibfield
  {journal} {\bibinfo  {journal} {J. Phys. Soc. Jpn.}\ }\textbf {\bibinfo
  {volume} {83}},\ \bibinfo {pages} {033705}}\BibitemShut {NoStop}%
\bibitem [{\citenamefont {Murai}\ \emph {et~al.}(2018)\citenamefont {Murai},
  \citenamefont {Suzuki}, \citenamefont {Ideta}, \citenamefont {Nakajima},
  \citenamefont {Tanaka}, \citenamefont {Ikeda},\ and\ \citenamefont
  {Kajimoto}}]{murai2018effect}%
  \BibitemOpen
  \bibfield  {author} {\bibinfo {author} {\bibnamefont {Murai}, \bibfnamefont
  {N.}}, \bibinfo {author} {\bibfnamefont {K.}~\bibnamefont {Suzuki}}, \bibinfo
  {author} {\bibfnamefont {S.-i.}\ \bibnamefont {Ideta}}, \bibinfo {author}
  {\bibfnamefont {M.}~\bibnamefont {Nakajima}}, \bibinfo {author}
  {\bibfnamefont {K.}~\bibnamefont {Tanaka}}, \bibinfo {author} {\bibfnamefont
  {H.}~\bibnamefont {Ikeda}}, and\ \bibinfo {author} {\bibfnamefont
  {R.}~\bibnamefont {Kajimoto}}} (\bibinfo {year} {2018}),\ \href
  {https://doi.org/10.1103/PhysRevB.97.241112} {\bibfield  {journal} {\bibinfo
  {journal} {Phys. Rev. B}\ }\textbf {\bibinfo {volume} {97}},\ \bibinfo
  {pages} {241112}}\BibitemShut {NoStop}%
\bibitem [{\citenamefont {Mutch}\ \emph {et~al.}(2019)\citenamefont {Mutch},
  \citenamefont {Chen}, \citenamefont {Went}, \citenamefont {Qian},
  \citenamefont {Wilson}, \citenamefont {Andreev}, \citenamefont {Chen},\ and\
  \citenamefont {Chu}}]{mutch2019evidence}%
  \BibitemOpen
  \bibfield  {author} {\bibinfo {author} {\bibnamefont {Mutch}, \bibfnamefont
  {J.}}, \bibinfo {author} {\bibfnamefont {W.-C.}\ \bibnamefont {Chen}},
  \bibinfo {author} {\bibfnamefont {P.}~\bibnamefont {Went}}, \bibinfo {author}
  {\bibfnamefont {T.}~\bibnamefont {Qian}}, \bibinfo {author} {\bibfnamefont
  {I.~Z.}\ \bibnamefont {Wilson}}, \bibinfo {author} {\bibfnamefont
  {A.}~\bibnamefont {Andreev}}, \bibinfo {author} {\bibfnamefont {C.-C.}\
  \bibnamefont {Chen}}, and\ \bibinfo {author} {\bibfnamefont {J.-H.}\
  \bibnamefont {Chu}}} (\bibinfo {year} {2019}),\ \href
  {https://doi.org/10.1126/sciadv.aav9771} {\bibfield  {journal} {\bibinfo
  {journal} {Sci. Adv.}\ }\textbf {\bibinfo {volume} {5}},\ \bibinfo {pages}
  {eaav9771}}\BibitemShut {NoStop}%
\bibitem [{\citenamefont {Nag}\ \emph {et~al.}(2025)\citenamefont {Nag},
  \citenamefont {Scott}, \citenamefont {de~Carvalho}, \citenamefont {Byland},
  \citenamefont {Yang}, \citenamefont {Walker}, \citenamefont {Greenberg},
  \citenamefont {Klavins}, \citenamefont {Miranda}, \citenamefont {Gozar},
  \citenamefont {Taufour}, \citenamefont {Fernandes},\ and\ \citenamefont
  {da~Silva~Neto}}]{nag2025highly}%
  \BibitemOpen
  \bibfield  {author} {\bibinfo {author} {\bibnamefont {Nag}, \bibfnamefont
  {P.~K.}}, \bibinfo {author} {\bibfnamefont {K.}~\bibnamefont {Scott}},
  \bibinfo {author} {\bibfnamefont {V.~S.}\ \bibnamefont {de~Carvalho}},
  \bibinfo {author} {\bibfnamefont {J.~K.}\ \bibnamefont {Byland}}, \bibinfo
  {author} {\bibfnamefont {X.}~\bibnamefont {Yang}}, \bibinfo {author}
  {\bibfnamefont {M.}~\bibnamefont {Walker}}, \bibinfo {author} {\bibfnamefont
  {A.~G.}\ \bibnamefont {Greenberg}}, \bibinfo {author} {\bibfnamefont
  {P.}~\bibnamefont {Klavins}}, \bibinfo {author} {\bibfnamefont
  {E.}~\bibnamefont {Miranda}}, \bibinfo {author} {\bibfnamefont
  {A.}~\bibnamefont {Gozar}}, \bibinfo {author} {\bibfnamefont
  {V.}~\bibnamefont {Taufour}}, \bibinfo {author} {\bibfnamefont {R.~M.}\
  \bibnamefont {Fernandes}}, and\ \bibinfo {author} {\bibfnamefont {E.~H.}\
  \bibnamefont {da~Silva~Neto}}} (\bibinfo {year} {2025}),\ \href
  {https://doi.org/10.1038/s41567-024-02683-x} {\bibfield  {journal} {\bibinfo
  {journal} {Nat. Phys.}\ }\textbf {\bibinfo {volume} {21}},\ \bibinfo {pages}
  {89}}\BibitemShut {NoStop}%
\bibitem [{\citenamefont {Nakayama}\ \emph {et~al.}(2014)\citenamefont
  {Nakayama}, \citenamefont {Miyata}, \citenamefont {Phan}, \citenamefont
  {Sato}, \citenamefont {Tanabe}, \citenamefont {Urata}, \citenamefont
  {Tanigaki},\ and\ \citenamefont {Takahashi}}]{Nakayama2014}%
  \BibitemOpen
  \bibfield  {author} {\bibinfo {author} {\bibnamefont {Nakayama},
  \bibfnamefont {K.}}, \bibinfo {author} {\bibfnamefont {Y.}~\bibnamefont
  {Miyata}}, \bibinfo {author} {\bibfnamefont {G.}~\bibnamefont {Phan}},
  \bibinfo {author} {\bibfnamefont {T.}~\bibnamefont {Sato}}, \bibinfo {author}
  {\bibfnamefont {Y.}~\bibnamefont {Tanabe}}, \bibinfo {author} {\bibfnamefont
  {T.}~\bibnamefont {Urata}}, \bibinfo {author} {\bibfnamefont
  {K.}~\bibnamefont {Tanigaki}}, and\ \bibinfo {author} {\bibfnamefont
  {T.}~\bibnamefont {Takahashi}}} (\bibinfo {year} {2014}),\ \href
  {https://doi.org/10.1103/PhysRevLett.113.237001} {\bibfield  {journal}
  {\bibinfo  {journal} {Phys. Rev. Lett.}\ }\textbf {\bibinfo {volume} {113}},\
  \bibinfo {pages} {237001}}\BibitemShut {NoStop}%
\bibitem [{\citenamefont {Neubauer}\ \emph {et~al.}(2024)\citenamefont
  {Neubauer}, \citenamefont {Klemm}, \citenamefont {Mozaffari}, \citenamefont
  {Jiao}, \citenamefont {Koshelev}, \citenamefont {Yaresko}, \citenamefont
  {Yi}, \citenamefont {Balicas},\ and\ \citenamefont
  {Dai}}]{PhysRevB.109.054435}%
  \BibitemOpen
  \bibfield  {author} {\bibinfo {author} {\bibnamefont {Neubauer},
  \bibfnamefont {K.~J.}}, \bibinfo {author} {\bibfnamefont {M.~L.}\
  \bibnamefont {Klemm}}, \bibinfo {author} {\bibfnamefont {S.}~\bibnamefont
  {Mozaffari}}, \bibinfo {author} {\bibfnamefont {L.}~\bibnamefont {Jiao}},
  \bibinfo {author} {\bibfnamefont {A.~E.}\ \bibnamefont {Koshelev}}, \bibinfo
  {author} {\bibfnamefont {A.}~\bibnamefont {Yaresko}}, \bibinfo {author}
  {\bibfnamefont {M.}~\bibnamefont {Yi}}, \bibinfo {author} {\bibfnamefont
  {L.}~\bibnamefont {Balicas}}, and\ \bibinfo {author} {\bibfnamefont
  {P.}~\bibnamefont {Dai}}} (\bibinfo {year} {2024}),\ \href
  {https://doi.org/10.1103/PhysRevB.109.054435} {\bibfield  {journal} {\bibinfo
   {journal} {Phys. Rev. B}\ }\textbf {\bibinfo {volume} {109}},\ \bibinfo
  {pages} {054435}}\BibitemShut {NoStop}%
\bibitem [{\citenamefont {Ni}\ \emph {et~al.}(2011)\citenamefont {Ni},
  \citenamefont {Allred}, \citenamefont {Chan},\ and\ \citenamefont
  {Cava}}]{Ni2011}%
  \BibitemOpen
  \bibfield  {author} {\bibinfo {author} {\bibnamefont {Ni}, \bibfnamefont
  {N.}}, \bibinfo {author} {\bibfnamefont {J.~M.}\ \bibnamefont {Allred}},
  \bibinfo {author} {\bibfnamefont {B.~C.}\ \bibnamefont {Chan}}, and\ \bibinfo
  {author} {\bibfnamefont {R.~J.}\ \bibnamefont {Cava}}} (\bibinfo {year}
  {2011}),\ \href {https://doi.org/10.1073/pnas.1110563108} {\bibfield
  {journal} {\bibinfo  {journal} {Proc. Natl. Acad. Sci. U.S.A.}\ }\textbf
  {\bibinfo {volume} {108}},\ \bibinfo {pages} {E1019}}\BibitemShut {NoStop}%
\bibitem [{\citenamefont {Nica}\ \emph {et~al.}(2017)\citenamefont {Nica},
  \citenamefont {Yu},\ and\ \citenamefont {Si}}]{WOS:000407421200001}%
  \BibitemOpen
  \bibfield  {author} {\bibinfo {author} {\bibnamefont {Nica}, \bibfnamefont
  {E.~M.}}, \bibinfo {author} {\bibfnamefont {R.}~\bibnamefont {Yu}}, and\
  \bibinfo {author} {\bibfnamefont {Q.}~\bibnamefont {Si}}} (\bibinfo {year}
  {2017}),\ \href {https://doi.org/10.1038/s41535-017-0027-6} {\bibfield
  {journal} {\bibinfo  {journal} {npj Quantum Mater.}\ }\textbf {\bibinfo
  {volume} {2}},\ \bibinfo {pages} {24}}\BibitemShut {NoStop}%
\bibitem [{\citenamefont {Nie}\ \emph {et~al.}(2022)\citenamefont {Nie},
  \citenamefont {Sun}, \citenamefont {Ma}, \citenamefont {Song}, \citenamefont
  {Zheng}, \citenamefont {Liang}, \citenamefont {Wu}, \citenamefont {Yu},
  \citenamefont {Li}, \citenamefont {Shan}, \citenamefont {Zhao}, \citenamefont
  {Li}, \citenamefont {Kang}, \citenamefont {Wu}, \citenamefont {Zhou},
  \citenamefont {Liu}, \citenamefont {Xiang}, \citenamefont {Ying},
  \citenamefont {Wang}, \citenamefont {Wu},\ and\ \citenamefont
  {Chen}}]{nie2022charge}%
  \BibitemOpen
  \bibfield  {author} {\bibinfo {author} {\bibnamefont {Nie}, \bibfnamefont
  {L.}}, \bibinfo {author} {\bibfnamefont {K.}~\bibnamefont {Sun}}, \bibinfo
  {author} {\bibfnamefont {W.}~\bibnamefont {Ma}}, \bibinfo {author}
  {\bibfnamefont {D.}~\bibnamefont {Song}}, \bibinfo {author} {\bibfnamefont
  {L.}~\bibnamefont {Zheng}}, \bibinfo {author} {\bibfnamefont
  {Z.}~\bibnamefont {Liang}}, \bibinfo {author} {\bibfnamefont
  {P.}~\bibnamefont {Wu}}, \bibinfo {author} {\bibfnamefont {F.}~\bibnamefont
  {Yu}}, \bibinfo {author} {\bibfnamefont {J.}~\bibnamefont {Li}}, \bibinfo
  {author} {\bibfnamefont {M.}~\bibnamefont {Shan}}, \bibinfo {author}
  {\bibfnamefont {D.}~\bibnamefont {Zhao}}, \bibinfo {author} {\bibfnamefont
  {S.}~\bibnamefont {Li}}, \bibinfo {author} {\bibfnamefont {B.}~\bibnamefont
  {Kang}}, \bibinfo {author} {\bibfnamefont {Z.}~\bibnamefont {Wu}}, \bibinfo
  {author} {\bibfnamefont {Y.}~\bibnamefont {Zhou}}, \bibinfo {author}
  {\bibfnamefont {K.}~\bibnamefont {Liu}}, \bibinfo {author} {\bibfnamefont
  {Z.}~\bibnamefont {Xiang}}, \bibinfo {author} {\bibfnamefont
  {J.}~\bibnamefont {Ying}}, \bibinfo {author} {\bibfnamefont {Z.}~\bibnamefont
  {Wang}}, \bibinfo {author} {\bibfnamefont {T.}~\bibnamefont {Wu}}, and\
  \bibinfo {author} {\bibfnamefont {X.}~\bibnamefont {Chen}}} (\bibinfo {year}
  {2022}),\ \href {https://doi.org/10.1038/s41586-022-04493-8} {\bibfield
  {journal} {\bibinfo  {journal} {Nature}\ }\textbf {\bibinfo {volume} {604}},\
  \bibinfo {pages} {59}}\BibitemShut {NoStop}%
\bibitem [{\citenamefont {Nie}\ \emph {et~al.}(2010)\citenamefont {Nie},
  \citenamefont {Telesca}, \citenamefont {Budnick}, \citenamefont {Sinkovic},\
  and\ \citenamefont {Wells}}]{Nie2010}%
  \BibitemOpen
  \bibfield  {author} {\bibinfo {author} {\bibnamefont {Nie}, \bibfnamefont
  {Y.~F.}}, \bibinfo {author} {\bibfnamefont {D.}~\bibnamefont {Telesca}},
  \bibinfo {author} {\bibfnamefont {J.~I.}\ \bibnamefont {Budnick}}, \bibinfo
  {author} {\bibfnamefont {B.}~\bibnamefont {Sinkovic}}, and\ \bibinfo {author}
  {\bibfnamefont {B.~O.}\ \bibnamefont {Wells}}} (\bibinfo {year} {2010}),\
  \href {https://doi.org/10.1103/PhysRevB.82.020508} {\bibfield  {journal}
  {\bibinfo  {journal} {Phys. Rev. B}\ }\textbf {\bibinfo {volume} {82}},\
  \bibinfo {pages} {020508(R)}}\BibitemShut {NoStop}%
\bibitem [{\citenamefont {Niu}\ \emph {et~al.}(2015)\citenamefont {Niu},
  \citenamefont {Peng}, \citenamefont {Xu}, \citenamefont {Yan}, \citenamefont
  {Jiang}, \citenamefont {Xu}, \citenamefont {Yu}, \citenamefont {Song},
  \citenamefont {Huang}, \citenamefont {Wang} \emph {et~al.}}]{Niu2015}%
  \BibitemOpen
  \bibfield  {author} {\bibinfo {author} {\bibnamefont {Niu}, \bibfnamefont
  {X.}}, \bibinfo {author} {\bibfnamefont {R.}~\bibnamefont {Peng}}, \bibinfo
  {author} {\bibfnamefont {H.}~\bibnamefont {Xu}}, \bibinfo {author}
  {\bibfnamefont {Y.}~\bibnamefont {Yan}}, \bibinfo {author} {\bibfnamefont
  {J.}~\bibnamefont {Jiang}}, \bibinfo {author} {\bibfnamefont
  {D.}~\bibnamefont {Xu}}, \bibinfo {author} {\bibfnamefont {T.}~\bibnamefont
  {Yu}}, \bibinfo {author} {\bibfnamefont {Q.}~\bibnamefont {Song}}, \bibinfo
  {author} {\bibfnamefont {Z.}~\bibnamefont {Huang}}, \bibinfo {author}
  {\bibfnamefont {Y.}~\bibnamefont {Wang}},  \emph {et~al.}} (\bibinfo {year}
  {2015}),\ \href {https://doi.org/10.1103/PhysRevB.92.060504} {\bibfield
  {journal} {\bibinfo  {journal} {Phys. Rev. B}\ }\textbf {\bibinfo {volume}
  {92}},\ \bibinfo {pages} {060504}}\BibitemShut {NoStop}%
\bibitem [{\citenamefont {Noji}\ \emph {et~al.}(2014)\citenamefont {Noji},
  \citenamefont {Hatakeda}, \citenamefont {Hosono}, \citenamefont {Kawamata},
  \citenamefont {Kato},\ and\ \citenamefont {Koike}}]{Noji2014}%
  \BibitemOpen
  \bibfield  {author} {\bibinfo {author} {\bibnamefont {Noji}, \bibfnamefont
  {T.}}, \bibinfo {author} {\bibfnamefont {T.}~\bibnamefont {Hatakeda}},
  \bibinfo {author} {\bibfnamefont {S.}~\bibnamefont {Hosono}}, \bibinfo
  {author} {\bibfnamefont {T.}~\bibnamefont {Kawamata}}, \bibinfo {author}
  {\bibfnamefont {M.}~\bibnamefont {Kato}}, and\ \bibinfo {author}
  {\bibfnamefont {Y.}~\bibnamefont {Koike}}} (\bibinfo {year} {2014}),\ \href
  {https://doi.org/10.1016/j.physc.2014.01.007} {\bibfield  {journal} {\bibinfo
   {journal} {Physica C: Superconductivity and its Applications}\ }\textbf
  {\bibinfo {volume} {504}},\ \bibinfo {pages} {8}}\BibitemShut {NoStop}%
\bibitem [{\citenamefont {Nowik}\ \emph
  {et~al.}(2011{\natexlab{a}})\citenamefont {Nowik}, \citenamefont {Felner},
  \citenamefont {Ren}, \citenamefont {Cao},\ and\ \citenamefont
  {Xu}}]{Nowik2011EuFeCoAs}%
  \BibitemOpen
  \bibfield  {author} {\bibinfo {author} {\bibnamefont {Nowik}, \bibfnamefont
  {I.}}, \bibinfo {author} {\bibfnamefont {I.}~\bibnamefont {Felner}}, \bibinfo
  {author} {\bibfnamefont {Z.}~\bibnamefont {Ren}}, \bibinfo {author}
  {\bibfnamefont {G.}~\bibnamefont {Cao}}, and\ \bibinfo {author}
  {\bibfnamefont {Z.}~\bibnamefont {Xu}}} (\bibinfo {year}
  {2011}{\natexlab{a}}),\ \href {https://doi.org/10.1088/1367-2630/13/2/023033}
  {\bibfield  {journal} {\bibinfo  {journal} {New J. Phys.}\ }\textbf {\bibinfo
  {volume} {13}},\ \bibinfo {pages} {023033}}\BibitemShut {NoStop}%
\bibitem [{\citenamefont {Nowik}\ \emph
  {et~al.}(2011{\natexlab{b}})\citenamefont {Nowik}, \citenamefont {Felner},
  \citenamefont {Ren}, \citenamefont {Cao},\ and\ \citenamefont
  {Xu}}]{Nowik2011EuFeAsP}%
  \BibitemOpen
  \bibfield  {author} {\bibinfo {author} {\bibnamefont {Nowik}, \bibfnamefont
  {I.}}, \bibinfo {author} {\bibfnamefont {I.}~\bibnamefont {Felner}}, \bibinfo
  {author} {\bibfnamefont {Z.}~\bibnamefont {Ren}}, \bibinfo {author}
  {\bibfnamefont {G.}~\bibnamefont {Cao}}, and\ \bibinfo {author}
  {\bibfnamefont {Z.}~\bibnamefont {Xu}}} (\bibinfo {year}
  {2011}{\natexlab{b}}),\ \href {https://doi.org/10.1088/0953-8984/23/6/065701}
  {\bibfield  {journal} {\bibinfo  {journal} {J. Phys.: Condens. Matter}\
  }\textbf {\bibinfo {volume} {23}},\ \bibinfo {pages} {065701}}\BibitemShut
  {NoStop}%
\bibitem [{\citenamefont {Occhialini}\ \emph {et~al.}(2023)\citenamefont
  {Occhialini}, \citenamefont {Sanchez}, \citenamefont {Song}, \citenamefont
  {Fabbris}, \citenamefont {Choi}, \citenamefont {Kim}, \citenamefont {Ryan},\
  and\ \citenamefont {Comin}}]{occhialini2023spontaneous}%
  \BibitemOpen
  \bibfield  {author} {\bibinfo {author} {\bibnamefont {Occhialini},
  \bibfnamefont {C.~A.}}, \bibinfo {author} {\bibfnamefont {J.~J.}\
  \bibnamefont {Sanchez}}, \bibinfo {author} {\bibfnamefont {Q.}~\bibnamefont
  {Song}}, \bibinfo {author} {\bibfnamefont {G.}~\bibnamefont {Fabbris}},
  \bibinfo {author} {\bibfnamefont {Y.}~\bibnamefont {Choi}}, \bibinfo {author}
  {\bibfnamefont {J.-W.}\ \bibnamefont {Kim}}, \bibinfo {author} {\bibfnamefont
  {P.~J.}\ \bibnamefont {Ryan}}, and\ \bibinfo {author} {\bibfnamefont
  {R.}~\bibnamefont {Comin}}} (\bibinfo {year} {2023}),\ \href
  {https://doi.org/10.1038/s41563-023-01585-2} {\bibfield  {journal} {\bibinfo
  {journal} {Nat. Mater.}\ }\textbf {\bibinfo {volume} {22}},\ \bibinfo {pages}
  {985}}\BibitemShut {NoStop}%
\bibitem [{\citenamefont {Ogino}\ \emph {et~al.}(2009)\citenamefont {Ogino},
  \citenamefont {Katsura}, \citenamefont {Horii}, \citenamefont {Kishio},\ and\
  \citenamefont {Shimoyama}}]{Ogino2009}%
  \BibitemOpen
  \bibfield  {author} {\bibinfo {author} {\bibnamefont {Ogino}, \bibfnamefont
  {H.}}, \bibinfo {author} {\bibfnamefont {Y.}~\bibnamefont {Katsura}},
  \bibinfo {author} {\bibfnamefont {S.}~\bibnamefont {Horii}}, \bibinfo
  {author} {\bibfnamefont {K.}~\bibnamefont {Kishio}}, and\ \bibinfo {author}
  {\bibfnamefont {J.-i.}\ \bibnamefont {Shimoyama}}} (\bibinfo {year} {2009}),\
  \href {https://doi.org/10.1088/0953-2048/22/8/085001} {\bibfield  {journal}
  {\bibinfo  {journal} {Supercond. Sci. Technol.}\ }\textbf {\bibinfo {volume}
  {22}},\ \bibinfo {pages} {085001}}\BibitemShut {NoStop}%
\bibitem [{\citenamefont {Ogino}\ \emph
  {et~al.}(2010{\natexlab{a}})\citenamefont {Ogino}, \citenamefont {Machida},
  \citenamefont {Yamamoto}, \citenamefont {Kishio}, \citenamefont {Shimoyama},
  \citenamefont {Tohei},\ and\ \citenamefont {Ikuhara}}]{Ogino2010c}%
  \BibitemOpen
  \bibfield  {author} {\bibinfo {author} {\bibnamefont {Ogino}, \bibfnamefont
  {H.}}, \bibinfo {author} {\bibfnamefont {K.}~\bibnamefont {Machida}},
  \bibinfo {author} {\bibfnamefont {A.}~\bibnamefont {Yamamoto}}, \bibinfo
  {author} {\bibfnamefont {K.}~\bibnamefont {Kishio}}, \bibinfo {author}
  {\bibfnamefont {J.-i.}\ \bibnamefont {Shimoyama}}, \bibinfo {author}
  {\bibfnamefont {T.}~\bibnamefont {Tohei}}, and\ \bibinfo {author}
  {\bibfnamefont {Y.}~\bibnamefont {Ikuhara}}} (\bibinfo {year}
  {2010}{\natexlab{a}}),\ \href
  {https://doi.org/10.1088/0953-2048/23/11/115005} {\bibfield  {journal}
  {\bibinfo  {journal} {Supercond. Sci. Technol.}\ }\textbf {\bibinfo {volume}
  {23}},\ \bibinfo {pages} {115005}}\BibitemShut {NoStop}%
\bibitem [{\citenamefont {Ogino}\ \emph
  {et~al.}(2010{\natexlab{b}})\citenamefont {Ogino}, \citenamefont {Sato},
  \citenamefont {Kishio}, \citenamefont {Shimoyama}, \citenamefont {Tohei},\
  and\ \citenamefont {Ikuhara}}]{Ogino2010b}%
  \BibitemOpen
  \bibfield  {author} {\bibinfo {author} {\bibnamefont {Ogino}, \bibfnamefont
  {H.}}, \bibinfo {author} {\bibfnamefont {S.}~\bibnamefont {Sato}}, \bibinfo
  {author} {\bibfnamefont {K.}~\bibnamefont {Kishio}}, \bibinfo {author}
  {\bibfnamefont {J.-i.}\ \bibnamefont {Shimoyama}}, \bibinfo {author}
  {\bibfnamefont {T.}~\bibnamefont {Tohei}}, and\ \bibinfo {author}
  {\bibfnamefont {Y.}~\bibnamefont {Ikuhara}}} (\bibinfo {year}
  {2010}{\natexlab{b}}),\ \href {https://doi.org/10.1063/1.3478850} {\bibfield
  {journal} {\bibinfo  {journal} {Appl. Phys. Lett.}\ }\textbf {\bibinfo
  {volume} {97}},\ \bibinfo {pages} {072506}}\BibitemShut {NoStop}%
\bibitem [{\citenamefont {Ogino}\ \emph {et~al.}(2011)\citenamefont {Ogino},
  \citenamefont {Shimizu}, \citenamefont {Kawaguchi}, \citenamefont {Kishio},
  \citenamefont {Shimoyama}, \citenamefont {Tohei},\ and\ \citenamefont
  {Ikuhara}}]{Ogino2011}%
  \BibitemOpen
  \bibfield  {author} {\bibinfo {author} {\bibnamefont {Ogino}, \bibfnamefont
  {H.}}, \bibinfo {author} {\bibfnamefont {Y.}~\bibnamefont {Shimizu}},
  \bibinfo {author} {\bibfnamefont {N.}~\bibnamefont {Kawaguchi}}, \bibinfo
  {author} {\bibfnamefont {K.}~\bibnamefont {Kishio}}, \bibinfo {author}
  {\bibfnamefont {J.-i.}\ \bibnamefont {Shimoyama}}, \bibinfo {author}
  {\bibfnamefont {T.}~\bibnamefont {Tohei}}, and\ \bibinfo {author}
  {\bibfnamefont {Y.}~\bibnamefont {Ikuhara}}} (\bibinfo {year} {2011}),\ \href
  {https://doi.org/10.1088/0953-2048/24/8/085020} {\bibfield  {journal}
  {\bibinfo  {journal} {Supercond. Sci. Technol.}\ }\textbf {\bibinfo {volume}
  {24}},\ \bibinfo {pages} {085020}}\BibitemShut {NoStop}%
\bibitem [{\citenamefont {Ogino}\ \emph
  {et~al.}(2010{\natexlab{c}})\citenamefont {Ogino}, \citenamefont {Shimizu},
  \citenamefont {Ushiyama}, \citenamefont {Kawaguchi}, \citenamefont {Kishio},\
  and\ \citenamefont {Shimoyama}}]{Ogino2010a}%
  \BibitemOpen
  \bibfield  {author} {\bibinfo {author} {\bibnamefont {Ogino}, \bibfnamefont
  {H.}}, \bibinfo {author} {\bibfnamefont {Y.}~\bibnamefont {Shimizu}},
  \bibinfo {author} {\bibfnamefont {K.}~\bibnamefont {Ushiyama}}, \bibinfo
  {author} {\bibfnamefont {N.}~\bibnamefont {Kawaguchi}}, \bibinfo {author}
  {\bibfnamefont {K.}~\bibnamefont {Kishio}}, and\ \bibinfo {author}
  {\bibfnamefont {J.-i.}\ \bibnamefont {Shimoyama}}} (\bibinfo {year}
  {2010}{\natexlab{c}}),\ \href {https://doi.org/10.1143/apex.3.063103}
  {\bibfield  {journal} {\bibinfo  {journal} {Applied physics express}\
  }\textbf {\bibinfo {volume} {3}},\ \bibinfo {pages} {063103}}\BibitemShut
  {NoStop}%
\bibitem [{\citenamefont {Okazaki}\ \emph {et~al.}(2012)\citenamefont
  {Okazaki}, \citenamefont {Ota}, \citenamefont {Kotani}, \citenamefont
  {Malaeb}, \citenamefont {Ishida}, \citenamefont {Shimojima}, \citenamefont
  {Kiss}, \citenamefont {Watanabe}, \citenamefont {Chen}, \citenamefont
  {Kihou}, \citenamefont {Lee}, \citenamefont {Iyo}, \citenamefont {Eisaki},
  \citenamefont {Saito}, \citenamefont {Fukazawa}, \citenamefont {Kohori},
  \citenamefont {Hashimoto}, \citenamefont {Shibauchi}, \citenamefont
  {Matsuda}, \citenamefont {Ikeda}, \citenamefont {Miyahara}, \citenamefont
  {Arita}, \citenamefont {Chainani},\ and\ \citenamefont {Shin}}]{okazaki2012}%
  \BibitemOpen
  \bibfield  {author} {\bibinfo {author} {\bibnamefont {Okazaki}, \bibfnamefont
  {K.}}, \bibinfo {author} {\bibfnamefont {Y.}~\bibnamefont {Ota}}, \bibinfo
  {author} {\bibfnamefont {Y.}~\bibnamefont {Kotani}}, \bibinfo {author}
  {\bibfnamefont {W.}~\bibnamefont {Malaeb}}, \bibinfo {author} {\bibfnamefont
  {Y.}~\bibnamefont {Ishida}}, \bibinfo {author} {\bibfnamefont
  {T.}~\bibnamefont {Shimojima}}, \bibinfo {author} {\bibfnamefont
  {T.}~\bibnamefont {Kiss}}, \bibinfo {author} {\bibfnamefont {S.}~\bibnamefont
  {Watanabe}}, \bibinfo {author} {\bibfnamefont {C.-T.}\ \bibnamefont {Chen}},
  \bibinfo {author} {\bibfnamefont {K.}~\bibnamefont {Kihou}}, \bibinfo
  {author} {\bibfnamefont {C.~H.}\ \bibnamefont {Lee}}, \bibinfo {author}
  {\bibfnamefont {A.}~\bibnamefont {Iyo}}, \bibinfo {author} {\bibfnamefont
  {H.}~\bibnamefont {Eisaki}}, \bibinfo {author} {\bibfnamefont
  {T.}~\bibnamefont {Saito}}, \bibinfo {author} {\bibfnamefont
  {H.}~\bibnamefont {Fukazawa}}, \bibinfo {author} {\bibfnamefont
  {Y.}~\bibnamefont {Kohori}}, \bibinfo {author} {\bibfnamefont
  {K.}~\bibnamefont {Hashimoto}}, \bibinfo {author} {\bibfnamefont
  {T.}~\bibnamefont {Shibauchi}}, \bibinfo {author} {\bibfnamefont
  {Y.}~\bibnamefont {Matsuda}}, \bibinfo {author} {\bibfnamefont
  {H.}~\bibnamefont {Ikeda}}, \bibinfo {author} {\bibfnamefont
  {H.}~\bibnamefont {Miyahara}}, \bibinfo {author} {\bibfnamefont
  {R.}~\bibnamefont {Arita}}, \bibinfo {author} {\bibfnamefont
  {A.}~\bibnamefont {Chainani}}, and\ \bibinfo {author} {\bibfnamefont
  {S.}~\bibnamefont {Shin}}} (\bibinfo {year} {2012}),\ \href
  {https://doi.org/10.1126/science.1222793} {\bibfield  {journal} {\bibinfo
  {journal} {Science}\ }\textbf {\bibinfo {volume} {337}},\ \bibinfo {pages}
  {1314}}\BibitemShut {NoStop}%
\bibitem [{\citenamefont {Onishi}\ \emph {et~al.}(2026)\citenamefont {Onishi},
  \citenamefont {Xu}, \citenamefont {Bareille}, \citenamefont {Kageyama},
  \citenamefont {Ishida}, \citenamefont {Eisaki}, \citenamefont {Ishihara},
  \citenamefont {Hashimoto}, \citenamefont {Taniuchi},\ and\ \citenamefont
  {Shibauchi}}]{onishi2026energy}%
  \BibitemOpen
  \bibfield  {author} {\bibinfo {author} {\bibnamefont {Onishi}, \bibfnamefont
  {A.}}, \bibinfo {author} {\bibfnamefont {Z.}~\bibnamefont {Xu}}, \bibinfo
  {author} {\bibfnamefont {C.}~\bibnamefont {Bareille}}, \bibinfo {author}
  {\bibfnamefont {Y.}~\bibnamefont {Kageyama}}, \bibinfo {author}
  {\bibfnamefont {S.}~\bibnamefont {Ishida}}, \bibinfo {author} {\bibfnamefont
  {H.}~\bibnamefont {Eisaki}}, \bibinfo {author} {\bibfnamefont
  {K.}~\bibnamefont {Ishihara}}, \bibinfo {author} {\bibfnamefont
  {K.}~\bibnamefont {Hashimoto}}, \bibinfo {author} {\bibfnamefont
  {T.}~\bibnamefont {Taniuchi}}, and\ \bibinfo {author} {\bibfnamefont
  {T.}~\bibnamefont {Shibauchi}}} (\bibinfo {year} {2026}),\ \href
  {https://doi.org/10.7566/JPSJ.95.033701} {\bibfield  {journal} {\bibinfo
  {journal} {J. Phys. Soc. Jpn.}\ }\textbf {\bibinfo {volume} {95}},\ \bibinfo
  {pages} {033701}}\BibitemShut {NoStop}%
\bibitem [{\citenamefont {{\"O}zer}\ \emph {et~al.}(2006)\citenamefont
  {{\"O}zer}, \citenamefont {Thompson},\ and\ \citenamefont
  {Weitering}}]{Ozer2006}%
  \BibitemOpen
  \bibfield  {author} {\bibinfo {author} {\bibnamefont {{\"O}zer},
  \bibfnamefont {M.~M.}}, \bibinfo {author} {\bibfnamefont {J.~R.}\
  \bibnamefont {Thompson}}, and\ \bibinfo {author} {\bibfnamefont {H.~H.}\
  \bibnamefont {Weitering}}} (\bibinfo {year} {2006}),\ \href
  {https://doi.org/10.1038/nphys244} {\bibfield  {journal} {\bibinfo  {journal}
  {Nat. Phys.}\ }\textbf {\bibinfo {volume} {2}},\ \bibinfo {pages}
  {173}}\BibitemShut {NoStop}%
\bibitem [{\citenamefont {Pachmayr}\ and\ \citenamefont
  {Johrendt}(2015)}]{Pachmayr2015}%
  \BibitemOpen
  \bibfield  {author} {\bibinfo {author} {\bibnamefont {Pachmayr},
  \bibfnamefont {U.}}, and\ \bibinfo {author} {\bibfnamefont {D.}~\bibnamefont
  {Johrendt}}} (\bibinfo {year} {2015}),\ \href
  {https://doi.org/10.1039/C5CC00038F} {\bibfield  {journal} {\bibinfo
  {journal} {Chem. Commun}\ }\textbf {\bibinfo {volume} {51}},\ \bibinfo
  {pages} {4689}}\BibitemShut {NoStop}%
\bibitem [{\citenamefont {Paglione}\ and\ \citenamefont
  {Greene}(2010)}]{paglione2010hightemperature}%
  \BibitemOpen
  \bibfield  {author} {\bibinfo {author} {\bibnamefont {Paglione},
  \bibfnamefont {J.}}, and\ \bibinfo {author} {\bibfnamefont {R.~L.}\
  \bibnamefont {Greene}}} (\bibinfo {year} {2010}),\ \href
  {https://doi.org/10.1038/nphys1759} {\bibfield  {journal} {\bibinfo
  {journal} {Nat. Phys.}\ }\textbf {\bibinfo {volume} {6}},\ \bibinfo {pages}
  {645}}\BibitemShut {NoStop}%
\bibitem [{\citenamefont {Pal}\ \emph {et~al.}(2019)\citenamefont {Pal},
  \citenamefont {Chinotti}, \citenamefont {Chu}, \citenamefont {Kuo},
  \citenamefont {Fisher},\ and\ \citenamefont {Degiorgi}}]{pal2019optical}%
  \BibitemOpen
  \bibfield  {author} {\bibinfo {author} {\bibnamefont {Pal}, \bibfnamefont
  {A.}}, \bibinfo {author} {\bibfnamefont {M.}~\bibnamefont {Chinotti}},
  \bibinfo {author} {\bibfnamefont {J.-H.}\ \bibnamefont {Chu}}, \bibinfo
  {author} {\bibfnamefont {H.-H.}\ \bibnamefont {Kuo}}, \bibinfo {author}
  {\bibfnamefont {I.~R.}\ \bibnamefont {Fisher}}, and\ \bibinfo {author}
  {\bibfnamefont {L.}~\bibnamefont {Degiorgi}}} (\bibinfo {year} {2019}),\
  \href {https://doi.org/10.1038/s41535-018-0140-1} {\bibfield  {journal}
  {\bibinfo  {journal} {npj Quantum Mater.}\ }\textbf {\bibinfo {volume} {4}},\
  \bibinfo {pages} {3}}\BibitemShut {NoStop}%
\bibitem [{\citenamefont {Pan}\ \emph {et~al.}(2017)\citenamefont {Pan},
  \citenamefont {Shen}, \citenamefont {Hu}, \citenamefont {Feng}, \citenamefont
  {Park}, \citenamefont {Christianson}, \citenamefont {Wang}, \citenamefont
  {Hao}, \citenamefont {Wo}, \citenamefont {Yin}, \citenamefont {Maier},\ and\
  \citenamefont {Zhao}}]{pan2017structure}%
  \BibitemOpen
  \bibfield  {author} {\bibinfo {author} {\bibnamefont {Pan}, \bibfnamefont
  {B.}}, \bibinfo {author} {\bibfnamefont {Y.}~\bibnamefont {Shen}}, \bibinfo
  {author} {\bibfnamefont {D.}~\bibnamefont {Hu}}, \bibinfo {author}
  {\bibfnamefont {Y.}~\bibnamefont {Feng}}, \bibinfo {author} {\bibfnamefont
  {J.~T.}\ \bibnamefont {Park}}, \bibinfo {author} {\bibfnamefont {A.~D.}\
  \bibnamefont {Christianson}}, \bibinfo {author} {\bibfnamefont
  {Q.}~\bibnamefont {Wang}}, \bibinfo {author} {\bibfnamefont {Y.}~\bibnamefont
  {Hao}}, \bibinfo {author} {\bibfnamefont {H.}~\bibnamefont {Wo}}, \bibinfo
  {author} {\bibfnamefont {Z.}~\bibnamefont {Yin}}, \bibinfo {author}
  {\bibfnamefont {T.~A.}\ \bibnamefont {Maier}}, and\ \bibinfo {author}
  {\bibfnamefont {J.}~\bibnamefont {Zhao}}} (\bibinfo {year} {2017}),\ \href
  {https://doi.org/10.1038/s41467-017-00162-x} {\bibfield  {journal} {\bibinfo
  {journal} {Nat. Commun.}\ }\textbf {\bibinfo {volume} {8}},\ \bibinfo {pages}
  {123}}\BibitemShut {NoStop}%
\bibitem [{\citenamefont {Papaj}\ \emph {et~al.}(2025)\citenamefont {Papaj},
  \citenamefont {Kong}, \citenamefont {Nadj-Perge},\ and\ \citenamefont
  {Lee}}]{papaj2025pair}%
  \BibitemOpen
  \bibfield  {author} {\bibinfo {author} {\bibnamefont {Papaj}, \bibfnamefont
  {M.}}, \bibinfo {author} {\bibfnamefont {L.}~\bibnamefont {Kong}}, \bibinfo
  {author} {\bibfnamefont {S.}~\bibnamefont {Nadj-Perge}}, and\ \bibinfo
  {author} {\bibfnamefont {P.~A.}\ \bibnamefont {Lee}}} (\bibinfo {year}
  {2025}),\ \href {https://doi.org/10.48550/arXiv.2506.19903} {\enquote
  {\bibinfo {title} {Pair density modulation from glide symmetry breaking and
  nematic superconductivity},}\ }\Eprint {https://arxiv.org/abs/2506.19903}
  {arXiv:2506.19903 [cond-mat.supr-con]} \BibitemShut {NoStop}%
\bibitem [{\citenamefont {Paramanik}\ \emph {et~al.}(2013)\citenamefont
  {Paramanik}, \citenamefont {Das}, \citenamefont {Prasad},\ and\ \citenamefont
  {Hossain}}]{Paramanik2013}%
  \BibitemOpen
  \bibfield  {author} {\bibinfo {author} {\bibnamefont {Paramanik},
  \bibfnamefont {U.}}, \bibinfo {author} {\bibfnamefont {D.}~\bibnamefont
  {Das}}, \bibinfo {author} {\bibfnamefont {R.}~\bibnamefont {Prasad}}, and\
  \bibinfo {author} {\bibfnamefont {Z.}~\bibnamefont {Hossain}}} (\bibinfo
  {year} {2013}),\ \href {https://doi.org/10.1088/0953-8984/25/26/265701}
  {\bibfield  {journal} {\bibinfo  {journal} {J. Phys.: Condens. Matter}\
  }\textbf {\bibinfo {volume} {25}},\ \bibinfo {pages} {265701}}\BibitemShut
  {NoStop}%
\bibitem [{\citenamefont {Park}\ \emph {et~al.}(2020)\citenamefont {Park},
  \citenamefont {Bartlett}, \citenamefont {Noad}, \citenamefont {Stern},
  \citenamefont {Barber}, \citenamefont {König}, \citenamefont {Hosoi},
  \citenamefont {Shibauchi}, \citenamefont {Mackenzie}, \citenamefont
  {Steppke},\ and\ \citenamefont {Hicks}}]{park2020rigid}%
  \BibitemOpen
  \bibfield  {author} {\bibinfo {author} {\bibnamefont {Park}, \bibfnamefont
  {J.}}, \bibinfo {author} {\bibfnamefont {J.~M.}\ \bibnamefont {Bartlett}},
  \bibinfo {author} {\bibfnamefont {H.~M.~L.}\ \bibnamefont {Noad}}, \bibinfo
  {author} {\bibfnamefont {A.~L.}\ \bibnamefont {Stern}}, \bibinfo {author}
  {\bibfnamefont {M.~E.}\ \bibnamefont {Barber}}, \bibinfo {author}
  {\bibfnamefont {M.}~\bibnamefont {König}}, \bibinfo {author} {\bibfnamefont
  {S.}~\bibnamefont {Hosoi}}, \bibinfo {author} {\bibfnamefont
  {T.}~\bibnamefont {Shibauchi}}, \bibinfo {author} {\bibfnamefont {A.~P.}\
  \bibnamefont {Mackenzie}}, \bibinfo {author} {\bibfnamefont {A.}~\bibnamefont
  {Steppke}}, and\ \bibinfo {author} {\bibfnamefont {C.~W.}\ \bibnamefont
  {Hicks}}} (\bibinfo {year} {2020}),\ \href
  {https://doi.org/10.1063/5.0008829} {\bibfield  {journal} {\bibinfo
  {journal} {Rev. Sci. Instrum.}\ }\textbf {\bibinfo {volume} {91}},\ \bibinfo
  {pages} {083902}}\BibitemShut {NoStop}%
\bibitem [{\citenamefont {Park}\ \emph {et~al.}(2011)\citenamefont {Park},
  \citenamefont {Friemel}, \citenamefont {Li}, \citenamefont {Kim},
  \citenamefont {Tsurkan}, \citenamefont {Deisenhofer}, \citenamefont {Krug~von
  Nidda}, \citenamefont {Loidl}, \citenamefont {Ivanov}, \citenamefont
  {Keimer},\ and\ \citenamefont {Inosov}}]{park2011magnetic}%
  \BibitemOpen
  \bibfield  {author} {\bibinfo {author} {\bibnamefont {Park}, \bibfnamefont
  {J.~T.}}, \bibinfo {author} {\bibfnamefont {G.}~\bibnamefont {Friemel}},
  \bibinfo {author} {\bibfnamefont {Y.}~\bibnamefont {Li}}, \bibinfo {author}
  {\bibfnamefont {J.-H.}\ \bibnamefont {Kim}}, \bibinfo {author} {\bibfnamefont
  {V.}~\bibnamefont {Tsurkan}}, \bibinfo {author} {\bibfnamefont
  {J.}~\bibnamefont {Deisenhofer}}, \bibinfo {author} {\bibfnamefont {H.-A.}\
  \bibnamefont {Krug~von Nidda}}, \bibinfo {author} {\bibfnamefont
  {A.}~\bibnamefont {Loidl}}, \bibinfo {author} {\bibfnamefont
  {A.}~\bibnamefont {Ivanov}}, \bibinfo {author} {\bibfnamefont
  {B.}~\bibnamefont {Keimer}}, and\ \bibinfo {author} {\bibfnamefont {D.~S.}\
  \bibnamefont {Inosov}}} (\bibinfo {year} {2011}),\ \href
  {https://doi.org/10.1103/PhysRevLett.107.177005} {\bibfield  {journal}
  {\bibinfo  {journal} {Phys. Rev. Lett.}\ }\textbf {\bibinfo {volume} {107}},\
  \bibinfo {pages} {177005}}\BibitemShut {NoStop}%
\bibitem [{\citenamefont {Parshall}\ \emph {et~al.}(2015)\citenamefont
  {Parshall}, \citenamefont {Pintschovius}, \citenamefont {Niedziela},
  \citenamefont {Castellan}, \citenamefont {Lamago}, \citenamefont {Mittal},
  \citenamefont {Wolf},\ and\ \citenamefont {Reznik}}]{PhysRevB.91.134426}%
  \BibitemOpen
  \bibfield  {author} {\bibinfo {author} {\bibnamefont {Parshall},
  \bibfnamefont {D.}}, \bibinfo {author} {\bibfnamefont {L.}~\bibnamefont
  {Pintschovius}}, \bibinfo {author} {\bibfnamefont {J.~L.}\ \bibnamefont
  {Niedziela}}, \bibinfo {author} {\bibfnamefont {J.-P.}\ \bibnamefont
  {Castellan}}, \bibinfo {author} {\bibfnamefont {D.}~\bibnamefont {Lamago}},
  \bibinfo {author} {\bibfnamefont {R.}~\bibnamefont {Mittal}}, \bibinfo
  {author} {\bibfnamefont {T.}~\bibnamefont {Wolf}}, and\ \bibinfo {author}
  {\bibfnamefont {D.}~\bibnamefont {Reznik}}} (\bibinfo {year} {2015}),\ \href
  {https://doi.org/10.1103/PhysRevB.91.134426} {\bibfield  {journal} {\bibinfo
  {journal} {Phys. Rev. B}\ }\textbf {\bibinfo {volume} {91}},\ \bibinfo
  {pages} {134426}}\BibitemShut {NoStop}%
\bibitem [{\citenamefont {Pelliciari}\ \emph
  {et~al.}(2016{\natexlab{a}})\citenamefont {Pelliciari}, \citenamefont
  {Dantz}, \citenamefont {Huang}, \citenamefont {Strocov}, \citenamefont
  {Xing}, \citenamefont {Wang}, \citenamefont {Jin},\ and\ \citenamefont
  {Schmitt}}]{pelliciari2016presence}%
  \BibitemOpen
  \bibfield  {author} {\bibinfo {author} {\bibnamefont {Pelliciari},
  \bibfnamefont {J.}}, \bibinfo {author} {\bibfnamefont {M.}~\bibnamefont
  {Dantz}}, \bibinfo {author} {\bibfnamefont {Y.}~\bibnamefont {Huang}},
  \bibinfo {author} {\bibfnamefont {V.~N.}\ \bibnamefont {Strocov}}, \bibinfo
  {author} {\bibfnamefont {L.}~\bibnamefont {Xing}}, \bibinfo {author}
  {\bibfnamefont {X.}~\bibnamefont {Wang}}, \bibinfo {author} {\bibfnamefont
  {C.}~\bibnamefont {Jin}}, and\ \bibinfo {author} {\bibfnamefont
  {T.}~\bibnamefont {Schmitt}}} (\bibinfo {year} {2016}{\natexlab{a}}),\ \href
  {https://doi.org/10.1063/1.4962966} {\bibfield  {journal} {\bibinfo
  {journal} {Appl. Phys. Lett.}\ }\textbf {\bibinfo {volume} {109}},\ \bibinfo
  {pages} {122601}}\BibitemShut {NoStop}%
\bibitem [{\citenamefont {Pelliciari}\ \emph
  {et~al.}(2016{\natexlab{b}})\citenamefont {Pelliciari}, \citenamefont
  {Huang}, \citenamefont {Das}, \citenamefont {Dantz}, \citenamefont {Bisogni},
  \citenamefont {Velasco}, \citenamefont {Strocov}, \citenamefont {Xing},
  \citenamefont {Wang}, \citenamefont {Jin},\ and\ \citenamefont
  {Schmitt}}]{pelliciari2016intralayer}%
  \BibitemOpen
  \bibfield  {author} {\bibinfo {author} {\bibnamefont {Pelliciari},
  \bibfnamefont {J.}}, \bibinfo {author} {\bibfnamefont {Y.}~\bibnamefont
  {Huang}}, \bibinfo {author} {\bibfnamefont {T.}~\bibnamefont {Das}}, \bibinfo
  {author} {\bibfnamefont {M.}~\bibnamefont {Dantz}}, \bibinfo {author}
  {\bibfnamefont {V.}~\bibnamefont {Bisogni}}, \bibinfo {author} {\bibfnamefont
  {P.~O.}\ \bibnamefont {Velasco}}, \bibinfo {author} {\bibfnamefont {V.~N.}\
  \bibnamefont {Strocov}}, \bibinfo {author} {\bibfnamefont {L.}~\bibnamefont
  {Xing}}, \bibinfo {author} {\bibfnamefont {X.}~\bibnamefont {Wang}}, \bibinfo
  {author} {\bibfnamefont {C.}~\bibnamefont {Jin}}, and\ \bibinfo {author}
  {\bibfnamefont {T.}~\bibnamefont {Schmitt}}} (\bibinfo {year}
  {2016}{\natexlab{b}}),\ \href {https://doi.org/10.1103/PhysRevB.93.134515}
  {\bibfield  {journal} {\bibinfo  {journal} {Phys. Rev. B}\ }\textbf {\bibinfo
  {volume} {93}},\ \bibinfo {pages} {134515}}\BibitemShut {NoStop}%
\bibitem [{\citenamefont {Pelliciari}\ \emph {et~al.}(2017)\citenamefont
  {Pelliciari}, \citenamefont {Ishii}, \citenamefont {Dantz}, \citenamefont
  {Lu}, \citenamefont {McNally}, \citenamefont {Strocov}, \citenamefont {Xing},
  \citenamefont {Wang}, \citenamefont {Jin}, \citenamefont {Jeevan},
  \citenamefont {Gegenwart},\ and\ \citenamefont
  {Schmitt}}]{pelliciari2017local}%
  \BibitemOpen
  \bibfield  {author} {\bibinfo {author} {\bibnamefont {Pelliciari},
  \bibfnamefont {J.}}, \bibinfo {author} {\bibfnamefont {K.}~\bibnamefont
  {Ishii}}, \bibinfo {author} {\bibfnamefont {M.}~\bibnamefont {Dantz}},
  \bibinfo {author} {\bibfnamefont {X.}~\bibnamefont {Lu}}, \bibinfo {author}
  {\bibfnamefont {D.~E.}\ \bibnamefont {McNally}}, \bibinfo {author}
  {\bibfnamefont {V.~N.}\ \bibnamefont {Strocov}}, \bibinfo {author}
  {\bibfnamefont {L.}~\bibnamefont {Xing}}, \bibinfo {author} {\bibfnamefont
  {X.}~\bibnamefont {Wang}}, \bibinfo {author} {\bibfnamefont {C.}~\bibnamefont
  {Jin}}, \bibinfo {author} {\bibfnamefont {H.~S.}\ \bibnamefont {Jeevan}},
  \bibinfo {author} {\bibfnamefont {P.}~\bibnamefont {Gegenwart}}, and\
  \bibinfo {author} {\bibfnamefont {T.}~\bibnamefont {Schmitt}}} (\bibinfo
  {year} {2017}),\ \href {https://doi.org/10.1103/PhysRevB.95.115152}
  {\bibfield  {journal} {\bibinfo  {journal} {Phys. Rev. B}\ }\textbf {\bibinfo
  {volume} {95}},\ \bibinfo {pages} {115152}}\BibitemShut {NoStop}%
\bibitem [{\citenamefont {Pelliciari}\ \emph {et~al.}(2019)\citenamefont
  {Pelliciari}, \citenamefont {Ishii}, \citenamefont {Huang}, \citenamefont
  {Dantz}, \citenamefont {Lu}, \citenamefont {{Olalde-Velasco}}, \citenamefont
  {Strocov}, \citenamefont {Kasahara}, \citenamefont {Xing}, \citenamefont
  {Wang}, \citenamefont {Jin}, \citenamefont {Matsuda}, \citenamefont
  {Shibauchi}, \citenamefont {Das},\ and\ \citenamefont
  {Schmitt}}]{pelliciari2019reciprocity}%
  \BibitemOpen
  \bibfield  {author} {\bibinfo {author} {\bibnamefont {Pelliciari},
  \bibfnamefont {J.}}, \bibinfo {author} {\bibfnamefont {K.}~\bibnamefont
  {Ishii}}, \bibinfo {author} {\bibfnamefont {Y.}~\bibnamefont {Huang}},
  \bibinfo {author} {\bibfnamefont {M.}~\bibnamefont {Dantz}}, \bibinfo
  {author} {\bibfnamefont {X.}~\bibnamefont {Lu}}, \bibinfo {author}
  {\bibfnamefont {P.}~\bibnamefont {{Olalde-Velasco}}}, \bibinfo {author}
  {\bibfnamefont {V.~N.}\ \bibnamefont {Strocov}}, \bibinfo {author}
  {\bibfnamefont {S.}~\bibnamefont {Kasahara}}, \bibinfo {author}
  {\bibfnamefont {L.}~\bibnamefont {Xing}}, \bibinfo {author} {\bibfnamefont
  {X.}~\bibnamefont {Wang}}, \bibinfo {author} {\bibfnamefont {C.}~\bibnamefont
  {Jin}}, \bibinfo {author} {\bibfnamefont {Y.}~\bibnamefont {Matsuda}},
  \bibinfo {author} {\bibfnamefont {T.}~\bibnamefont {Shibauchi}}, \bibinfo
  {author} {\bibfnamefont {T.}~\bibnamefont {Das}}, and\ \bibinfo {author}
  {\bibfnamefont {T.}~\bibnamefont {Schmitt}}} (\bibinfo {year} {2019}),\ \href
  {https://doi.org/10.1038/s42005-019-0236-3} {\bibfield  {journal} {\bibinfo
  {journal} {Commun. Phys.}\ }\textbf {\bibinfo {volume} {2}},\ \bibinfo
  {pages} {139}}\BibitemShut {NoStop}%
\bibitem [{\citenamefont {Pelliciari}\ \emph {et~al.}(2021)\citenamefont
  {Pelliciari}, \citenamefont {Karakuzu}, \citenamefont {Song}, \citenamefont
  {Arpaia}, \citenamefont {Nag}, \citenamefont {Rossi}, \citenamefont {Li},
  \citenamefont {Yu}, \citenamefont {Chen}, \citenamefont {Peng}, \citenamefont
  {{Garc{\'i}a-Fern{\'a}ndez}}, \citenamefont {Walters}, \citenamefont {Wang},
  \citenamefont {Zhao}, \citenamefont {Ghiringhelli}, \citenamefont {Feng},
  \citenamefont {Maier}, \citenamefont {Zhou}, \citenamefont {Johnston},\ and\
  \citenamefont {Comin}}]{pelliciari2021evolution}%
  \BibitemOpen
  \bibfield  {author} {\bibinfo {author} {\bibnamefont {Pelliciari},
  \bibfnamefont {J.}}, \bibinfo {author} {\bibfnamefont {S.}~\bibnamefont
  {Karakuzu}}, \bibinfo {author} {\bibfnamefont {Q.}~\bibnamefont {Song}},
  \bibinfo {author} {\bibfnamefont {R.}~\bibnamefont {Arpaia}}, \bibinfo
  {author} {\bibfnamefont {A.}~\bibnamefont {Nag}}, \bibinfo {author}
  {\bibfnamefont {M.}~\bibnamefont {Rossi}}, \bibinfo {author} {\bibfnamefont
  {J.}~\bibnamefont {Li}}, \bibinfo {author} {\bibfnamefont {T.}~\bibnamefont
  {Yu}}, \bibinfo {author} {\bibfnamefont {X.}~\bibnamefont {Chen}}, \bibinfo
  {author} {\bibfnamefont {R.}~\bibnamefont {Peng}}, \bibinfo {author}
  {\bibfnamefont {M.}~\bibnamefont {{Garc{\'i}a-Fern{\'a}ndez}}}, \bibinfo
  {author} {\bibfnamefont {A.~C.}\ \bibnamefont {Walters}}, \bibinfo {author}
  {\bibfnamefont {Q.}~\bibnamefont {Wang}}, \bibinfo {author} {\bibfnamefont
  {J.}~\bibnamefont {Zhao}}, \bibinfo {author} {\bibfnamefont {G.}~\bibnamefont
  {Ghiringhelli}}, \bibinfo {author} {\bibfnamefont {D.}~\bibnamefont {Feng}},
  \bibinfo {author} {\bibfnamefont {T.~A.}\ \bibnamefont {Maier}}, \bibinfo
  {author} {\bibfnamefont {K.-J.}\ \bibnamefont {Zhou}}, \bibinfo {author}
  {\bibfnamefont {S.}~\bibnamefont {Johnston}}, and\ \bibinfo {author}
  {\bibfnamefont {R.}~\bibnamefont {Comin}}} (\bibinfo {year} {2021}),\ \href
  {https://doi.org/10.1038/s41467-021-23317-3} {\bibfield  {journal} {\bibinfo
  {journal} {Nat. Commun.}\ }\textbf {\bibinfo {volume} {12}},\ \bibinfo
  {pages} {3122}}\BibitemShut {NoStop}%
\bibitem [{\citenamefont {Peng}\ \emph {et~al.}(2014)\citenamefont {Peng},
  \citenamefont {Shen}, \citenamefont {Xie}, \citenamefont {Xu}, \citenamefont
  {Tan}, \citenamefont {Xia}, \citenamefont {Zhang}, \citenamefont {Cao},
  \citenamefont {Gong}, \citenamefont {Hu} \emph {et~al.}}]{Peng2014}%
  \BibitemOpen
  \bibfield  {author} {\bibinfo {author} {\bibnamefont {Peng}, \bibfnamefont
  {R.}}, \bibinfo {author} {\bibfnamefont {X.}~\bibnamefont {Shen}}, \bibinfo
  {author} {\bibfnamefont {X.}~\bibnamefont {Xie}}, \bibinfo {author}
  {\bibfnamefont {H.}~\bibnamefont {Xu}}, \bibinfo {author} {\bibfnamefont
  {S.}~\bibnamefont {Tan}}, \bibinfo {author} {\bibfnamefont {M.}~\bibnamefont
  {Xia}}, \bibinfo {author} {\bibfnamefont {T.}~\bibnamefont {Zhang}}, \bibinfo
  {author} {\bibfnamefont {H.}~\bibnamefont {Cao}}, \bibinfo {author}
  {\bibfnamefont {X.}~\bibnamefont {Gong}}, \bibinfo {author} {\bibfnamefont
  {J.}~\bibnamefont {Hu}},  \emph {et~al.}} (\bibinfo {year} {2014}),\ \href
  {https://doi.org/10.1103/PhysRevLett.112.107001} {\bibfield  {journal}
  {\bibinfo  {journal} {Phys. Rev. Lett.}\ }\textbf {\bibinfo {volume} {112}},\
  \bibinfo {pages} {107001}}\BibitemShut {NoStop}%
\bibitem [{\citenamefont {Peng}\ \emph {et~al.}(2018)\citenamefont {Peng},
  \citenamefont {Huang}, \citenamefont {Fumagalli}, \citenamefont {Minola},
  \citenamefont {Wang}, \citenamefont {Sun}, \citenamefont {Ding},
  \citenamefont {Kummer}, \citenamefont {Zhou}, \citenamefont {Brookes},
  \citenamefont {Moritz}, \citenamefont {Braicovich}, \citenamefont
  {Devereaux},\ and\ \citenamefont {Ghiringhelli}}]{peng2018dispersion}%
  \BibitemOpen
  \bibfield  {author} {\bibinfo {author} {\bibnamefont {Peng}, \bibfnamefont
  {Y.~Y.}}, \bibinfo {author} {\bibfnamefont {E.~W.}\ \bibnamefont {Huang}},
  \bibinfo {author} {\bibfnamefont {R.}~\bibnamefont {Fumagalli}}, \bibinfo
  {author} {\bibfnamefont {M.}~\bibnamefont {Minola}}, \bibinfo {author}
  {\bibfnamefont {Y.}~\bibnamefont {Wang}}, \bibinfo {author} {\bibfnamefont
  {X.}~\bibnamefont {Sun}}, \bibinfo {author} {\bibfnamefont {Y.}~\bibnamefont
  {Ding}}, \bibinfo {author} {\bibfnamefont {K.}~\bibnamefont {Kummer}},
  \bibinfo {author} {\bibfnamefont {X.~J.}\ \bibnamefont {Zhou}}, \bibinfo
  {author} {\bibfnamefont {N.~B.}\ \bibnamefont {Brookes}}, \bibinfo {author}
  {\bibfnamefont {B.}~\bibnamefont {Moritz}}, \bibinfo {author} {\bibfnamefont
  {L.}~\bibnamefont {Braicovich}}, \bibinfo {author} {\bibfnamefont {T.~P.}\
  \bibnamefont {Devereaux}}, and\ \bibinfo {author} {\bibfnamefont
  {G.}~\bibnamefont {Ghiringhelli}}} (\bibinfo {year} {2018}),\ \href
  {https://doi.org/10.1103/PhysRevB.98.144507} {\bibfield  {journal} {\bibinfo
  {journal} {Phys. Rev. B}\ }\textbf {\bibinfo {volume} {98}},\ \bibinfo
  {pages} {144507}}\BibitemShut {NoStop}%
\bibitem [{\citenamefont {Pfau}\ \emph {et~al.}(2019)\citenamefont {Pfau},
  \citenamefont {Chen}, \citenamefont {Yi}, \citenamefont {Hashimoto},
  \citenamefont {Rotundu}, \citenamefont {Palmstrom}, \citenamefont {Chen},
  \citenamefont {Dai}, \citenamefont {Straquadine}, \citenamefont {Hristov},
  \citenamefont {Birgeneau}, \citenamefont {Fisher}, \citenamefont {Lu},\ and\
  \citenamefont {Shen}}]{pfau2019momentum}%
  \BibitemOpen
  \bibfield  {author} {\bibinfo {author} {\bibnamefont {Pfau}, \bibfnamefont
  {H.}}, \bibinfo {author} {\bibfnamefont {S.~D.}\ \bibnamefont {Chen}},
  \bibinfo {author} {\bibfnamefont {M.}~\bibnamefont {Yi}}, \bibinfo {author}
  {\bibfnamefont {M.}~\bibnamefont {Hashimoto}}, \bibinfo {author}
  {\bibfnamefont {C.~R.}\ \bibnamefont {Rotundu}}, \bibinfo {author}
  {\bibfnamefont {J.~C.}\ \bibnamefont {Palmstrom}}, \bibinfo {author}
  {\bibfnamefont {T.}~\bibnamefont {Chen}}, \bibinfo {author} {\bibfnamefont
  {P.-C.}\ \bibnamefont {Dai}}, \bibinfo {author} {\bibfnamefont
  {J.}~\bibnamefont {Straquadine}}, \bibinfo {author} {\bibfnamefont
  {A.}~\bibnamefont {Hristov}}, \bibinfo {author} {\bibfnamefont {R.~J.}\
  \bibnamefont {Birgeneau}}, \bibinfo {author} {\bibfnamefont {I.~R.}\
  \bibnamefont {Fisher}}, \bibinfo {author} {\bibfnamefont {D.}~\bibnamefont
  {Lu}}, and\ \bibinfo {author} {\bibfnamefont {Z.-X.}\ \bibnamefont {Shen}}}
  (\bibinfo {year} {2019}),\ \href
  {https://doi.org/10.1103/PhysRevLett.123.066402} {\bibfield  {journal}
  {\bibinfo  {journal} {Phys. Rev. Lett.}\ }\textbf {\bibinfo {volume} {123}},\
  \bibinfo {pages} {066402}}\BibitemShut {NoStop}%
\bibitem [{\citenamefont {Philippe}\ \emph {et~al.}(2022)\citenamefont
  {Philippe}, \citenamefont {Lespinas}, \citenamefont {Faria}, \citenamefont
  {Forget}, \citenamefont {Colson}, \citenamefont {Houver}, \citenamefont
  {Cazayous}, \citenamefont {Sacuto}, \citenamefont {Paul},\ and\ \citenamefont
  {Gallais}}]{philippe2022nematic}%
  \BibitemOpen
  \bibfield  {author} {\bibinfo {author} {\bibnamefont {Philippe},
  \bibfnamefont {J.-C.}}, \bibinfo {author} {\bibfnamefont {A.}~\bibnamefont
  {Lespinas}}, \bibinfo {author} {\bibfnamefont {J.}~\bibnamefont {Faria}},
  \bibinfo {author} {\bibfnamefont {A.}~\bibnamefont {Forget}}, \bibinfo
  {author} {\bibfnamefont {D.}~\bibnamefont {Colson}}, \bibinfo {author}
  {\bibfnamefont {S.}~\bibnamefont {Houver}}, \bibinfo {author} {\bibfnamefont
  {M.}~\bibnamefont {Cazayous}}, \bibinfo {author} {\bibfnamefont
  {A.}~\bibnamefont {Sacuto}}, \bibinfo {author} {\bibfnamefont
  {I.}~\bibnamefont {Paul}}, and\ \bibinfo {author} {\bibfnamefont
  {Y.}~\bibnamefont {Gallais}}} (\bibinfo {year} {2022}),\ \href
  {https://doi.org/10.1103/PhysRevLett.129.187002} {\bibfield  {journal}
  {\bibinfo  {journal} {Phys. Rev. Lett.}\ }\textbf {\bibinfo {volume} {129}},\
  \bibinfo {pages} {187002}}\BibitemShut {NoStop}%
\bibitem [{\citenamefont {Pratt}\ \emph {et~al.}(2011)\citenamefont {Pratt},
  \citenamefont {Kim}, \citenamefont {Kreyssig}, \citenamefont {Lee},
  \citenamefont {Tucker}, \citenamefont {Thaler}, \citenamefont {Tian},
  \citenamefont {Zarestky}, \citenamefont {Bud'ko}, \citenamefont {Canfield},
  \citenamefont {Harmon}, \citenamefont {Goldman},\ and\ \citenamefont
  {McQueeney}}]{PhysRevLett.106.257001}%
  \BibitemOpen
  \bibfield  {author} {\bibinfo {author} {\bibnamefont {Pratt}, \bibfnamefont
  {D.~K.}}, \bibinfo {author} {\bibfnamefont {M.~G.}\ \bibnamefont {Kim}},
  \bibinfo {author} {\bibfnamefont {A.}~\bibnamefont {Kreyssig}}, \bibinfo
  {author} {\bibfnamefont {Y.~B.}\ \bibnamefont {Lee}}, \bibinfo {author}
  {\bibfnamefont {G.~S.}\ \bibnamefont {Tucker}}, \bibinfo {author}
  {\bibfnamefont {A.}~\bibnamefont {Thaler}}, \bibinfo {author} {\bibfnamefont
  {W.}~\bibnamefont {Tian}}, \bibinfo {author} {\bibfnamefont {J.~L.}\
  \bibnamefont {Zarestky}}, \bibinfo {author} {\bibfnamefont {S.~L.}\
  \bibnamefont {Bud'ko}}, \bibinfo {author} {\bibfnamefont {P.~C.}\
  \bibnamefont {Canfield}}, \bibinfo {author} {\bibfnamefont {B.~N.}\
  \bibnamefont {Harmon}}, \bibinfo {author} {\bibfnamefont {A.~I.}\
  \bibnamefont {Goldman}}, and\ \bibinfo {author} {\bibfnamefont {R.~J.}\
  \bibnamefont {McQueeney}}} (\bibinfo {year} {2011}),\ \href
  {https://doi.org/10.1103/PhysRevLett.106.257001} {\bibfield  {journal}
  {\bibinfo  {journal} {Phys. Rev. Lett.}\ }\textbf {\bibinfo {volume} {106}},\
  \bibinfo {pages} {257001}}\BibitemShut {NoStop}%
\bibitem [{\citenamefont {Prozorov}\ \emph {et~al.}(2024)\citenamefont
  {Prozorov}, \citenamefont {Kogan}, \citenamefont {Ko{\'n}czykowski},\ and\
  \citenamefont {Tanatar}}]{prozorov2024slope}%
  \BibitemOpen
  \bibfield  {author} {\bibinfo {author} {\bibnamefont {Prozorov},
  \bibfnamefont {R.}}, \bibinfo {author} {\bibfnamefont {V.~G.}\ \bibnamefont
  {Kogan}}, \bibinfo {author} {\bibfnamefont {M.}~\bibnamefont
  {Ko{\'n}czykowski}}, and\ \bibinfo {author} {\bibfnamefont {M.~A.}\
  \bibnamefont {Tanatar}}} (\bibinfo {year} {2024}),\ \href
  {https://doi.org/10.1103/PhysRevB.109.024506} {\bibfield  {journal} {\bibinfo
   {journal} {Phys. Rev. B}\ }\textbf {\bibinfo {volume} {109}},\ \bibinfo
  {pages} {024506}}\BibitemShut {NoStop}%
\bibitem [{\citenamefont {Prozorov}\ \emph {et~al.}(2014)\citenamefont
  {Prozorov}, \citenamefont {Ko{\'n}czykowski}, \citenamefont {Tanatar},
  \citenamefont {Thaler}, \citenamefont {Bud'ko}, \citenamefont {Canfield},
  \citenamefont {Mishra},\ and\ \citenamefont
  {Hirschfeld}}]{prozorov2014effect}%
  \BibitemOpen
  \bibfield  {author} {\bibinfo {author} {\bibnamefont {Prozorov},
  \bibfnamefont {R.}}, \bibinfo {author} {\bibfnamefont {M.}~\bibnamefont
  {Ko{\'n}czykowski}}, \bibinfo {author} {\bibfnamefont {M.~A.}\ \bibnamefont
  {Tanatar}}, \bibinfo {author} {\bibfnamefont {A.}~\bibnamefont {Thaler}},
  \bibinfo {author} {\bibfnamefont {S.~L.}\ \bibnamefont {Bud'ko}}, \bibinfo
  {author} {\bibfnamefont {P.~C.}\ \bibnamefont {Canfield}}, \bibinfo {author}
  {\bibfnamefont {V.}~\bibnamefont {Mishra}}, and\ \bibinfo {author}
  {\bibfnamefont {P.~J.}\ \bibnamefont {Hirschfeld}}} (\bibinfo {year}
  {2014}),\ \href {https://doi.org/10.1103/PhysRevX.4.041032} {\bibfield
  {journal} {\bibinfo  {journal} {Phys. Rev. X}\ }\textbf {\bibinfo {volume}
  {4}},\ \bibinfo {pages} {041032}}\BibitemShut {NoStop}%
\bibitem [{\citenamefont {Ptok}\ and\ \citenamefont
  {Crivelli}(2013)}]{ptok2013the}%
  \BibitemOpen
  \bibfield  {author} {\bibinfo {author} {\bibnamefont {Ptok}, \bibfnamefont
  {A.}}, and\ \bibinfo {author} {\bibfnamefont {D.}~\bibnamefont {Crivelli}}}
  (\bibinfo {year} {2013}),\ \href {https://doi.org/10.1007/s10909-013-0871-0}
  {\bibfield  {journal} {\bibinfo  {journal} {Journal of Low Temperature
  Physics}\ }\textbf {\bibinfo {volume} {172}},\ \bibinfo {pages}
  {226}}\BibitemShut {NoStop}%
\bibitem [{\citenamefont {Pustogow}\ \emph {et~al.}(2019)\citenamefont
  {Pustogow}, \citenamefont {Luo}, \citenamefont {Chronister}, \citenamefont
  {Su}, \citenamefont {Sokolov}, \citenamefont {Jerzembeck}, \citenamefont
  {Mackenzie}, \citenamefont {Hicks}, \citenamefont {Kikugawa}, \citenamefont
  {Raghu}, \citenamefont {Bauer},\ and\ \citenamefont
  {Brown}}]{pustogow2019constraints}%
  \BibitemOpen
  \bibfield  {author} {\bibinfo {author} {\bibnamefont {Pustogow},
  \bibfnamefont {A.}}, \bibinfo {author} {\bibfnamefont {Y.}~\bibnamefont
  {Luo}}, \bibinfo {author} {\bibfnamefont {A.}~\bibnamefont {Chronister}},
  \bibinfo {author} {\bibfnamefont {Y.-S.}\ \bibnamefont {Su}}, \bibinfo
  {author} {\bibfnamefont {D.~A.}\ \bibnamefont {Sokolov}}, \bibinfo {author}
  {\bibfnamefont {F.}~\bibnamefont {Jerzembeck}}, \bibinfo {author}
  {\bibfnamefont {A.~P.}\ \bibnamefont {Mackenzie}}, \bibinfo {author}
  {\bibfnamefont {C.~W.}\ \bibnamefont {Hicks}}, \bibinfo {author}
  {\bibfnamefont {N.}~\bibnamefont {Kikugawa}}, \bibinfo {author}
  {\bibfnamefont {S.}~\bibnamefont {Raghu}}, \bibinfo {author} {\bibfnamefont
  {E.~D.}\ \bibnamefont {Bauer}}, and\ \bibinfo {author} {\bibfnamefont
  {S.~E.}\ \bibnamefont {Brown}}} (\bibinfo {year} {2019}),\ \href
  {https://doi.org/10.1038/s41586-019-1596-2} {\bibfield  {journal} {\bibinfo
  {journal} {Nature}\ }\textbf {\bibinfo {volume} {574}},\ \bibinfo {pages}
  {72}}\BibitemShut {NoStop}%
\bibitem [{\citenamefont {Pyon}\ \emph {et~al.}(2023)\citenamefont {Pyon},
  \citenamefont {Ito}, \citenamefont {Sasaki}, \citenamefont {Sakagami},
  \citenamefont {Tamegai}, \citenamefont {Awaji},\ and\ \citenamefont
  {Kajitani}}]{Pyon2023}%
  \BibitemOpen
  \bibfield  {author} {\bibinfo {author} {\bibnamefont {Pyon}, \bibfnamefont
  {S.}}, \bibinfo {author} {\bibfnamefont {T.}~\bibnamefont {Ito}}, \bibinfo
  {author} {\bibfnamefont {T.}~\bibnamefont {Sasaki}}, \bibinfo {author}
  {\bibfnamefont {R.}~\bibnamefont {Sakagami}}, \bibinfo {author}
  {\bibfnamefont {T.}~\bibnamefont {Tamegai}}, \bibinfo {author} {\bibfnamefont
  {S.}~\bibnamefont {Awaji}}, and\ \bibinfo {author} {\bibfnamefont
  {H.}~\bibnamefont {Kajitani}}} (\bibinfo {year} {2023}),\ \href
  {https://doi.org/10.1016/j.physc.2023.1354354} {\bibfield  {journal}
  {\bibinfo  {journal} {Physica C: Superconductivity and its Applications}\
  }\textbf {\bibinfo {volume} {615}},\ \bibinfo {pages} {1354354}}\BibitemShut
  {NoStop}%
\bibitem [{\citenamefont {Pyon}\ \emph {et~al.}(2021)\citenamefont {Pyon},
  \citenamefont {Mori}, \citenamefont {Tamegai}, \citenamefont {Awaji},
  \citenamefont {Kito}, \citenamefont {Ishida}, \citenamefont {Yoshida},
  \citenamefont {Kajitani},\ and\ \citenamefont {Koizumi}}]{Pyon2021}%
  \BibitemOpen
  \bibfield  {author} {\bibinfo {author} {\bibnamefont {Pyon}, \bibfnamefont
  {S.}}, \bibinfo {author} {\bibfnamefont {H.}~\bibnamefont {Mori}}, \bibinfo
  {author} {\bibfnamefont {T.}~\bibnamefont {Tamegai}}, \bibinfo {author}
  {\bibfnamefont {S.}~\bibnamefont {Awaji}}, \bibinfo {author} {\bibfnamefont
  {H.}~\bibnamefont {Kito}}, \bibinfo {author} {\bibfnamefont {S.}~\bibnamefont
  {Ishida}}, \bibinfo {author} {\bibfnamefont {Y.}~\bibnamefont {Yoshida}},
  \bibinfo {author} {\bibfnamefont {H.}~\bibnamefont {Kajitani}}, and\ \bibinfo
  {author} {\bibfnamefont {N.}~\bibnamefont {Koizumi}}} (\bibinfo {year}
  {2021}),\ \href {https://doi.org/10.1088/1361-6668/ac0cca} {\bibfield
  {journal} {\bibinfo  {journal} {Supercond. Sci. Technol.}\ }\textbf {\bibinfo
  {volume} {34}},\ \bibinfo {pages} {105008}}\BibitemShut {NoStop}%
\bibitem [{\citenamefont {Qi}\ \emph {et~al.}(2012)\citenamefont {Qi},
  \citenamefont {Wang}, \citenamefont {Gao}, \citenamefont {Zhang},
  \citenamefont {Wang}, \citenamefont {Yao}, \citenamefont {Wang},
  \citenamefont {Wang},\ and\ \citenamefont {Ma}}]{Qi2012}%
  \BibitemOpen
  \bibfield  {author} {\bibinfo {author} {\bibnamefont {Qi}, \bibfnamefont
  {Y.}}, \bibinfo {author} {\bibfnamefont {L.}~\bibnamefont {Wang}}, \bibinfo
  {author} {\bibfnamefont {Z.}~\bibnamefont {Gao}}, \bibinfo {author}
  {\bibfnamefont {X.}~\bibnamefont {Zhang}}, \bibinfo {author} {\bibfnamefont
  {D.}~\bibnamefont {Wang}}, \bibinfo {author} {\bibfnamefont {C.}~\bibnamefont
  {Yao}}, \bibinfo {author} {\bibfnamefont {C.}~\bibnamefont {Wang}}, \bibinfo
  {author} {\bibfnamefont {C.}~\bibnamefont {Wang}}, and\ \bibinfo {author}
  {\bibfnamefont {Y.}~\bibnamefont {Ma}}} (\bibinfo {year} {2012}),\ \href
  {https://doi.org/10.1088/1367-2630/14/3/033011} {\bibfield  {journal}
  {\bibinfo  {journal} {New J. Phys.}\ }\textbf {\bibinfo {volume} {14}},\
  \bibinfo {pages} {033011}}\BibitemShut {NoStop}%
\bibitem [{\citenamefont {Qi}\ \emph {et~al.}(2010)\citenamefont {Qi},
  \citenamefont {Wang}, \citenamefont {Wang}, \citenamefont {Zhang},
  \citenamefont {Gao}, \citenamefont {Zhang},\ and\ \citenamefont
  {Ma}}]{Qi2010}%
  \BibitemOpen
  \bibfield  {author} {\bibinfo {author} {\bibnamefont {Qi}, \bibfnamefont
  {Y.}}, \bibinfo {author} {\bibfnamefont {L.}~\bibnamefont {Wang}}, \bibinfo
  {author} {\bibfnamefont {D.}~\bibnamefont {Wang}}, \bibinfo {author}
  {\bibfnamefont {Z.}~\bibnamefont {Zhang}}, \bibinfo {author} {\bibfnamefont
  {Z.}~\bibnamefont {Gao}}, \bibinfo {author} {\bibfnamefont {X.}~\bibnamefont
  {Zhang}}, and\ \bibinfo {author} {\bibfnamefont {Y.}~\bibnamefont {Ma}}}
  (\bibinfo {year} {2010}),\ \href
  {https://doi.org/10.1088/0953-2048/23/5/055009} {\bibfield  {journal}
  {\bibinfo  {journal} {Supercond. Sci. Technol.}\ }\textbf {\bibinfo {volume}
  {23}},\ \bibinfo {pages} {055009}}\BibitemShut {NoStop}%
\bibitem [{\citenamefont {Qian}\ \emph {et~al.}(2011)\citenamefont {Qian},
  \citenamefont {Wang}, \citenamefont {Jin}, \citenamefont {Zhang},
  \citenamefont {Richard}, \citenamefont {Xu}, \citenamefont {Dai},
  \citenamefont {Fang}, \citenamefont {Guo}, \citenamefont {Chen} \emph
  {et~al.}}]{Qian2011}%
  \BibitemOpen
  \bibfield  {author} {\bibinfo {author} {\bibnamefont {Qian}, \bibfnamefont
  {T.}}, \bibinfo {author} {\bibfnamefont {X.-P.}\ \bibnamefont {Wang}},
  \bibinfo {author} {\bibfnamefont {W.-C.}\ \bibnamefont {Jin}}, \bibinfo
  {author} {\bibfnamefont {P.}~\bibnamefont {Zhang}}, \bibinfo {author}
  {\bibfnamefont {P.}~\bibnamefont {Richard}}, \bibinfo {author} {\bibfnamefont
  {G.}~\bibnamefont {Xu}}, \bibinfo {author} {\bibfnamefont {X.}~\bibnamefont
  {Dai}}, \bibinfo {author} {\bibfnamefont {Z.}~\bibnamefont {Fang}}, \bibinfo
  {author} {\bibfnamefont {J.-G.}\ \bibnamefont {Guo}}, \bibinfo {author}
  {\bibfnamefont {X.-L.}\ \bibnamefont {Chen}},  \emph {et~al.}} (\bibinfo
  {year} {2011}),\ \href {https://doi.org/10.1103/PhysRevLett.106.187001}
  {\bibfield  {journal} {\bibinfo  {journal} {Phys. Rev. Lett.}\ }\textbf
  {\bibinfo {volume} {106}},\ \bibinfo {pages} {187001}}\BibitemShut {NoStop}%
\bibitem [{\citenamefont {Rademaker}\ \emph {et~al.}(2016)\citenamefont
  {Rademaker}, \citenamefont {Wang}, \citenamefont {Berlijn},\ and\
  \citenamefont {Johnston}}]{Rademaker2016}%
  \BibitemOpen
  \bibfield  {author} {\bibinfo {author} {\bibnamefont {Rademaker},
  \bibfnamefont {L.}}, \bibinfo {author} {\bibfnamefont {Y.}~\bibnamefont
  {Wang}}, \bibinfo {author} {\bibfnamefont {T.}~\bibnamefont {Berlijn}}, and\
  \bibinfo {author} {\bibfnamefont {S.}~\bibnamefont {Johnston}}} (\bibinfo
  {year} {2016}),\ \href {https://doi.org/10.1088/1367-2630/18/2/022001}
  {\bibfield  {journal} {\bibinfo  {journal} {New J. Phys.}\ }\textbf {\bibinfo
  {volume} {18}},\ \bibinfo {pages} {022001}}\BibitemShut {NoStop}%
\bibitem [{\citenamefont {Raffius}\ \emph {et~al.}(1993)\citenamefont
  {Raffius}, \citenamefont {M{\"o}rsen}, \citenamefont {Mosel}, \citenamefont
  {M{\"u}ller-Warmuth}, \citenamefont {Jeitschko}, \citenamefont
  {Terb{\"u}chte},\ and\ \citenamefont {Vomhof}}]{Raffius1993}%
  \BibitemOpen
  \bibfield  {author} {\bibinfo {author} {\bibnamefont {Raffius}, \bibfnamefont
  {H.}}, \bibinfo {author} {\bibfnamefont {E.}~\bibnamefont {M{\"o}rsen}},
  \bibinfo {author} {\bibfnamefont {B.}~\bibnamefont {Mosel}}, \bibinfo
  {author} {\bibfnamefont {W.}~\bibnamefont {M{\"u}ller-Warmuth}}, \bibinfo
  {author} {\bibfnamefont {W.}~\bibnamefont {Jeitschko}}, \bibinfo {author}
  {\bibfnamefont {L.}~\bibnamefont {Terb{\"u}chte}}, and\ \bibinfo {author}
  {\bibfnamefont {T.}~\bibnamefont {Vomhof}}} (\bibinfo {year} {1993}),\ \href
  {https://doi.org/10.1016/0022-3697(93)90301-7} {\bibfield  {journal}
  {\bibinfo  {journal} {Journal of Physics and Chemistry of Solids}\ }\textbf
  {\bibinfo {volume} {54}},\ \bibinfo {pages} {135}}\BibitemShut {NoStop}%
\bibitem [{\citenamefont {Rahn}\ \emph {et~al.}(2019)\citenamefont {Rahn},
  \citenamefont {Kummer}, \citenamefont {Brookes}, \citenamefont {Haghighirad},
  \citenamefont {Gilmore},\ and\ \citenamefont
  {Boothroyd}}]{rahn2019paramagnon}%
  \BibitemOpen
  \bibfield  {author} {\bibinfo {author} {\bibnamefont {Rahn}, \bibfnamefont
  {M.~C.}}, \bibinfo {author} {\bibfnamefont {K.}~\bibnamefont {Kummer}},
  \bibinfo {author} {\bibfnamefont {N.~B.}\ \bibnamefont {Brookes}}, \bibinfo
  {author} {\bibfnamefont {A.~A.}\ \bibnamefont {Haghighirad}}, \bibinfo
  {author} {\bibfnamefont {K.}~\bibnamefont {Gilmore}}, and\ \bibinfo {author}
  {\bibfnamefont {A.~T.}\ \bibnamefont {Boothroyd}}} (\bibinfo {year} {2019}),\
  \href {https://doi.org/10.1103/PhysRevB.99.014505} {\bibfield  {journal}
  {\bibinfo  {journal} {Phys. Rev. B}\ }\textbf {\bibinfo {volume} {99}},\
  \bibinfo {pages} {014505}}\BibitemShut {NoStop}%
\bibitem [{\citenamefont {Rebec}\ \emph {et~al.}(2017)\citenamefont {Rebec},
  \citenamefont {Jia}, \citenamefont {Zhang}, \citenamefont {Hashimoto},
  \citenamefont {Lu}, \citenamefont {Moore},\ and\ \citenamefont
  {Shen}}]{rebec2017coexistence}%
  \BibitemOpen
  \bibfield  {author} {\bibinfo {author} {\bibnamefont {Rebec}, \bibfnamefont
  {S.~N.}}, \bibinfo {author} {\bibfnamefont {T.}~\bibnamefont {Jia}}, \bibinfo
  {author} {\bibfnamefont {C.}~\bibnamefont {Zhang}}, \bibinfo {author}
  {\bibfnamefont {M.}~\bibnamefont {Hashimoto}}, \bibinfo {author}
  {\bibfnamefont {D.-H.}\ \bibnamefont {Lu}}, \bibinfo {author} {\bibfnamefont
  {R.~G.}\ \bibnamefont {Moore}}, and\ \bibinfo {author} {\bibfnamefont
  {Z.-X.}\ \bibnamefont {Shen}}} (\bibinfo {year} {2017}),\ \href
  {https://doi.org/10.1103/PhysRevLett.118.067002} {\bibfield  {journal}
  {\bibinfo  {journal} {Phys. Rev. Lett.}\ }\textbf {\bibinfo {volume} {118}},\
  \bibinfo {pages} {067002}}\BibitemShut {NoStop}%
\bibitem [{\citenamefont {Reiss}\ \emph {et~al.}(2020)\citenamefont {Reiss},
  \citenamefont {Graf}, \citenamefont {Haghighirad}, \citenamefont {Knafo},
  \citenamefont {Drigo}, \citenamefont {Bristow}, \citenamefont {Schofield},\
  and\ \citenamefont {Coldea}}]{reiss2020quenched}%
  \BibitemOpen
  \bibfield  {author} {\bibinfo {author} {\bibnamefont {Reiss}, \bibfnamefont
  {P.}}, \bibinfo {author} {\bibfnamefont {D.}~\bibnamefont {Graf}}, \bibinfo
  {author} {\bibfnamefont {A.~A.}\ \bibnamefont {Haghighirad}}, \bibinfo
  {author} {\bibfnamefont {W.}~\bibnamefont {Knafo}}, \bibinfo {author}
  {\bibfnamefont {L.}~\bibnamefont {Drigo}}, \bibinfo {author} {\bibfnamefont
  {M.}~\bibnamefont {Bristow}}, \bibinfo {author} {\bibfnamefont {A.~J.}\
  \bibnamefont {Schofield}}, and\ \bibinfo {author} {\bibfnamefont {A.~I.}\
  \bibnamefont {Coldea}}} (\bibinfo {year} {2020}),\ \href
  {https://doi.org/10.1038/s41567-019-0694-2} {\bibfield  {journal} {\bibinfo
  {journal} {Nat. Phys.}\ }\textbf {\bibinfo {volume} {16}},\ \bibinfo {pages}
  {89}}\BibitemShut {NoStop}%
\bibitem [{\citenamefont {Reiss}\ \emph {et~al.}(2017)\citenamefont {Reiss},
  \citenamefont {Watson}, \citenamefont {Kim}, \citenamefont {Haghighirad},
  \citenamefont {Woodruff}, \citenamefont {Bruma}, \citenamefont {Clarke},\
  and\ \citenamefont {Coldea}}]{reiss2017suppression}%
  \BibitemOpen
  \bibfield  {author} {\bibinfo {author} {\bibnamefont {Reiss}, \bibfnamefont
  {P.}}, \bibinfo {author} {\bibfnamefont {M.~D.}\ \bibnamefont {Watson}},
  \bibinfo {author} {\bibfnamefont {T.~K.}\ \bibnamefont {Kim}}, \bibinfo
  {author} {\bibfnamefont {A.~A.}\ \bibnamefont {Haghighirad}}, \bibinfo
  {author} {\bibfnamefont {D.~N.}\ \bibnamefont {Woodruff}}, \bibinfo {author}
  {\bibfnamefont {M.}~\bibnamefont {Bruma}}, \bibinfo {author} {\bibfnamefont
  {S.~J.}\ \bibnamefont {Clarke}}, and\ \bibinfo {author} {\bibfnamefont
  {A.~I.}\ \bibnamefont {Coldea}}} (\bibinfo {year} {2017}),\ \href
  {https://doi.org/10.1103/PhysRevB.96.121103} {\bibfield  {journal} {\bibinfo
  {journal} {Phys. Rev. B}\ }\textbf {\bibinfo {volume} {96}},\ \bibinfo
  {pages} {121103}}\BibitemShut {NoStop}%
\bibitem [{\citenamefont {Ren}\ \emph {et~al.}(2015)\citenamefont {Ren},
  \citenamefont {Duan}, \citenamefont {Hu}, \citenamefont {Li}, \citenamefont
  {Zhang}, \citenamefont {Luo}, \citenamefont {Dai},\ and\ \citenamefont
  {Li}}]{ren2015nematic}%
  \BibitemOpen
  \bibfield  {author} {\bibinfo {author} {\bibnamefont {Ren}, \bibfnamefont
  {X.}}, \bibinfo {author} {\bibfnamefont {L.}~\bibnamefont {Duan}}, \bibinfo
  {author} {\bibfnamefont {Y.}~\bibnamefont {Hu}}, \bibinfo {author}
  {\bibfnamefont {J.}~\bibnamefont {Li}}, \bibinfo {author} {\bibfnamefont
  {R.}~\bibnamefont {Zhang}}, \bibinfo {author} {\bibfnamefont
  {H.}~\bibnamefont {Luo}}, \bibinfo {author} {\bibfnamefont {P.}~\bibnamefont
  {Dai}}, and\ \bibinfo {author} {\bibfnamefont {Y.}~\bibnamefont {Li}}}
  (\bibinfo {year} {2015}),\ \href
  {https://doi.org/10.1103/PhysRevLett.115.197002} {\bibfield  {journal}
  {\bibinfo  {journal} {Phys. Rev. Lett.}\ }\textbf {\bibinfo {volume} {115}},\
  \bibinfo {pages} {197002}}\BibitemShut {NoStop}%
\bibitem [{\citenamefont {Ren}\ \emph {et~al.}(2009)\citenamefont {Ren},
  \citenamefont {Tao}, \citenamefont {Jiang}, \citenamefont {Feng},
  \citenamefont {Wang}, \citenamefont {Dai}, \citenamefont {Cao},\ and\
  \citenamefont {Xu}}]{Ren2009}%
  \BibitemOpen
  \bibfield  {author} {\bibinfo {author} {\bibnamefont {Ren}, \bibfnamefont
  {Z.}}, \bibinfo {author} {\bibfnamefont {Q.}~\bibnamefont {Tao}}, \bibinfo
  {author} {\bibfnamefont {S.}~\bibnamefont {Jiang}}, \bibinfo {author}
  {\bibfnamefont {C.}~\bibnamefont {Feng}}, \bibinfo {author} {\bibfnamefont
  {C.}~\bibnamefont {Wang}}, \bibinfo {author} {\bibfnamefont {J.}~\bibnamefont
  {Dai}}, \bibinfo {author} {\bibfnamefont {G.}~\bibnamefont {Cao}}, and\
  \bibinfo {author} {\bibfnamefont {Z.}~\bibnamefont {Xu}}} (\bibinfo {year}
  {2009}),\ \href {https://doi.org/10.1103/PhysRevLett.102.137002} {\bibfield
  {journal} {\bibinfo  {journal} {Phys. Rev. Lett.}\ }\textbf {\bibinfo
  {volume} {102}},\ \bibinfo {pages} {137002}}\BibitemShut {NoStop}%
\bibitem [{\citenamefont {Ren}\ \emph {et~al.}(2008)\citenamefont {Ren},
  \citenamefont {Zhu}, \citenamefont {Jiang}, \citenamefont {Xu}, \citenamefont
  {Tao}, \citenamefont {Wang}, \citenamefont {Feng}, \citenamefont {Cao},\ and\
  \citenamefont {Xu}}]{Ren2008}%
  \BibitemOpen
  \bibfield  {author} {\bibinfo {author} {\bibnamefont {Ren}, \bibfnamefont
  {Z.}}, \bibinfo {author} {\bibfnamefont {Z.}~\bibnamefont {Zhu}}, \bibinfo
  {author} {\bibfnamefont {S.}~\bibnamefont {Jiang}}, \bibinfo {author}
  {\bibfnamefont {X.}~\bibnamefont {Xu}}, \bibinfo {author} {\bibfnamefont
  {Q.}~\bibnamefont {Tao}}, \bibinfo {author} {\bibfnamefont {C.}~\bibnamefont
  {Wang}}, \bibinfo {author} {\bibfnamefont {C.}~\bibnamefont {Feng}}, \bibinfo
  {author} {\bibfnamefont {G.}~\bibnamefont {Cao}}, and\ \bibinfo {author}
  {\bibfnamefont {Z.}~\bibnamefont {Xu}}} (\bibinfo {year} {2008}),\ \href
  {https://doi.org/10.1103/PhysRevB.78.052501} {\bibfield  {journal} {\bibinfo
  {journal} {Phys. Rev. B}\ }\textbf {\bibinfo {volume} {78}},\ \bibinfo
  {pages} {052501}}\BibitemShut {NoStop}%
\bibitem [{\citenamefont {Rendenbach}\ \emph {et~al.}(2021)\citenamefont
  {Rendenbach}, \citenamefont {Hohl}, \citenamefont {Harm}, \citenamefont
  {Hoch},\ and\ \citenamefont {Johrendt}}]{Rendenbach2021}%
  \BibitemOpen
  \bibfield  {author} {\bibinfo {author} {\bibnamefont {Rendenbach},
  \bibfnamefont {B.}}, \bibinfo {author} {\bibfnamefont {T.}~\bibnamefont
  {Hohl}}, \bibinfo {author} {\bibfnamefont {S.}~\bibnamefont {Harm}}, \bibinfo
  {author} {\bibfnamefont {C.}~\bibnamefont {Hoch}}, and\ \bibinfo {author}
  {\bibfnamefont {D.}~\bibnamefont {Johrendt}}} (\bibinfo {year} {2021}),\
  \href {https://doi.org/10.1021/jacs.0c13396} {\bibfield  {journal} {\bibinfo
  {journal} {J. Am. Chem. Soc.}\ }\textbf {\bibinfo {volume} {143}},\ \bibinfo
  {pages} {3043}}\BibitemShut {NoStop}%
\bibitem [{\citenamefont {Rhodes}\ \emph {et~al.}(2021)\citenamefont {Rhodes},
  \citenamefont {B{\"o}ker}, \citenamefont {M{\"u}ller}, \citenamefont
  {Eschrig},\ and\ \citenamefont {Eremin}}]{rhodes2021nonlocal}%
  \BibitemOpen
  \bibfield  {author} {\bibinfo {author} {\bibnamefont {Rhodes}, \bibfnamefont
  {L.~C.}}, \bibinfo {author} {\bibfnamefont {J.}~\bibnamefont {B{\"o}ker}},
  \bibinfo {author} {\bibfnamefont {M.~A.}\ \bibnamefont {M{\"u}ller}},
  \bibinfo {author} {\bibfnamefont {M.}~\bibnamefont {Eschrig}}, and\ \bibinfo
  {author} {\bibfnamefont {I.~M.}\ \bibnamefont {Eremin}}} (\bibinfo {year}
  {2021}),\ \href {https://doi.org/10.1038/s41535-021-00341-6} {\bibfield
  {journal} {\bibinfo  {journal} {npj Quantum Mater.}\ }\textbf {\bibinfo
  {volume} {6}},\ \bibinfo {pages} {45}}\BibitemShut {NoStop}%
\bibitem [{\citenamefont {Rinott}\ \emph {et~al.}(2017)\citenamefont {Rinott},
  \citenamefont {Chashka}, \citenamefont {Ribak}, \citenamefont {Rienks},
  \citenamefont {Taleb-Ibrahimi}, \citenamefont {Fevre}, \citenamefont
  {Bertran}, \citenamefont {Randeria},\ and\ \citenamefont
  {Kanigel}}]{rinott2017tuning}%
  \BibitemOpen
  \bibfield  {author} {\bibinfo {author} {\bibnamefont {Rinott}, \bibfnamefont
  {S.}}, \bibinfo {author} {\bibfnamefont {K.~B.}\ \bibnamefont {Chashka}},
  \bibinfo {author} {\bibfnamefont {A.}~\bibnamefont {Ribak}}, \bibinfo
  {author} {\bibfnamefont {E.~D.~L.}\ \bibnamefont {Rienks}}, \bibinfo {author}
  {\bibfnamefont {A.}~\bibnamefont {Taleb-Ibrahimi}}, \bibinfo {author}
  {\bibfnamefont {P.~L.}\ \bibnamefont {Fevre}}, \bibinfo {author}
  {\bibfnamefont {F.}~\bibnamefont {Bertran}}, \bibinfo {author} {\bibfnamefont
  {M.}~\bibnamefont {Randeria}}, and\ \bibinfo {author} {\bibfnamefont
  {A.}~\bibnamefont {Kanigel}}} (\bibinfo {year} {2017}),\ \href
  {https://doi.org/10.1126/sciadv.1602372} {\bibfield  {journal} {\bibinfo
  {journal} {Sci. Adv.}\ }\textbf {\bibinfo {volume} {3}},\ \bibinfo {pages}
  {e1602372}}\BibitemShut {NoStop}%
\bibitem [{\citenamefont {Rodriguez}\ \emph {et~al.}(2011)\citenamefont
  {Rodriguez}, \citenamefont {Stock}, \citenamefont {Zajdel}, \citenamefont
  {Krycka}, \citenamefont {Majkrzak}, \citenamefont {Zavalij},\ and\
  \citenamefont {Green}}]{Rodriguez2011}%
  \BibitemOpen
  \bibfield  {author} {\bibinfo {author} {\bibnamefont {Rodriguez},
  \bibfnamefont {E.~E.}}, \bibinfo {author} {\bibfnamefont {C.}~\bibnamefont
  {Stock}}, \bibinfo {author} {\bibfnamefont {P.}~\bibnamefont {Zajdel}},
  \bibinfo {author} {\bibfnamefont {K.~L.}\ \bibnamefont {Krycka}}, \bibinfo
  {author} {\bibfnamefont {C.~F.}\ \bibnamefont {Majkrzak}}, \bibinfo {author}
  {\bibfnamefont {P.}~\bibnamefont {Zavalij}}, and\ \bibinfo {author}
  {\bibfnamefont {M.~A.}\ \bibnamefont {Green}}} (\bibinfo {year} {2011}),\
  \href {https://doi.org/10.1103/PhysRevB.84.064403} {\bibfield  {journal}
  {\bibinfo  {journal} {Phys. Rev. B}\ }\textbf {\bibinfo {volume} {84}},\
  \bibinfo {pages} {064403}}\BibitemShut {NoStop}%
\bibitem [{\citenamefont {Roppongi}\ \emph {et~al.}(2025)\citenamefont
  {Roppongi}, \citenamefont {Cai}, \citenamefont {Ogawa}, \citenamefont {Liu},
  \citenamefont {Zhao}, \citenamefont {Oudah}, \citenamefont {Fujii},
  \citenamefont {Imamura}, \citenamefont {Fang}, \citenamefont {Ishihara},
  \citenamefont {Hashimoto}, \citenamefont {Matsuura}, \citenamefont
  {Mizukami}, \citenamefont {Pula}, \citenamefont {Young}, \citenamefont
  {Marković}, \citenamefont {Bonn}, \citenamefont {Watanabe}, \citenamefont
  {Yamashita}, \citenamefont {Mizuguchi}, \citenamefont {Luke}, \citenamefont
  {Kojima}, \citenamefont {Uemura},\ and\ \citenamefont
  {Shibauchi}}]{roppongi2025topology}%
  \BibitemOpen
  \bibfield  {author} {\bibinfo {author} {\bibnamefont {Roppongi},
  \bibfnamefont {M.}}, \bibinfo {author} {\bibfnamefont {Y.}~\bibnamefont
  {Cai}}, \bibinfo {author} {\bibfnamefont {K.}~\bibnamefont {Ogawa}}, \bibinfo
  {author} {\bibfnamefont {S.}~\bibnamefont {Liu}}, \bibinfo {author}
  {\bibfnamefont {G.}~\bibnamefont {Zhao}}, \bibinfo {author} {\bibfnamefont
  {M.}~\bibnamefont {Oudah}}, \bibinfo {author} {\bibfnamefont
  {T.}~\bibnamefont {Fujii}}, \bibinfo {author} {\bibfnamefont
  {K.}~\bibnamefont {Imamura}}, \bibinfo {author} {\bibfnamefont
  {S.}~\bibnamefont {Fang}}, \bibinfo {author} {\bibfnamefont {K.}~\bibnamefont
  {Ishihara}}, \bibinfo {author} {\bibfnamefont {K.}~\bibnamefont {Hashimoto}},
  \bibinfo {author} {\bibfnamefont {K.}~\bibnamefont {Matsuura}}, \bibinfo
  {author} {\bibfnamefont {Y.}~\bibnamefont {Mizukami}}, \bibinfo {author}
  {\bibfnamefont {M.}~\bibnamefont {Pula}}, \bibinfo {author} {\bibfnamefont
  {C.}~\bibnamefont {Young}}, \bibinfo {author} {\bibfnamefont
  {I.}~\bibnamefont {Marković}}, \bibinfo {author} {\bibfnamefont {D.~A.}\
  \bibnamefont {Bonn}}, \bibinfo {author} {\bibfnamefont {T.}~\bibnamefont
  {Watanabe}}, \bibinfo {author} {\bibfnamefont {A.}~\bibnamefont {Yamashita}},
  \bibinfo {author} {\bibfnamefont {Y.}~\bibnamefont {Mizuguchi}}, \bibinfo
  {author} {\bibfnamefont {G.~M.}\ \bibnamefont {Luke}}, \bibinfo {author}
  {\bibfnamefont {K.~M.}\ \bibnamefont {Kojima}}, \bibinfo {author}
  {\bibfnamefont {Y.~J.}\ \bibnamefont {Uemura}}, and\ \bibinfo {author}
  {\bibfnamefont {T.}~\bibnamefont {Shibauchi}}} (\bibinfo {year} {2025}),\
  \href {https://doi.org/10.1038/s41467-025-61651-y} {\bibfield  {journal}
  {\bibinfo  {journal} {Nat. Commun.}\ }\textbf {\bibinfo {volume} {16}},\
  \bibinfo {pages} {6573}}\BibitemShut {NoStop}%
\bibitem [{\citenamefont {Rosenberg}\ \emph {et~al.}(2019)\citenamefont
  {Rosenberg}, \citenamefont {Chu}, \citenamefont {Ruff}, \citenamefont
  {Hristov},\ and\ \citenamefont {Fisher}}]{rosenberg2019divergence}%
  \BibitemOpen
  \bibfield  {author} {\bibinfo {author} {\bibnamefont {Rosenberg},
  \bibfnamefont {E.~W.}}, \bibinfo {author} {\bibfnamefont {J.-H.}\
  \bibnamefont {Chu}}, \bibinfo {author} {\bibfnamefont {J.~P.~C.}\
  \bibnamefont {Ruff}}, \bibinfo {author} {\bibfnamefont {A.~T.}\ \bibnamefont
  {Hristov}}, and\ \bibinfo {author} {\bibfnamefont {I.~R.}\ \bibnamefont
  {Fisher}}} (\bibinfo {year} {2019}),\ \href
  {https://doi.org/10.1073/pnas.1818910116} {\bibfield  {journal} {\bibinfo
  {journal} {Proc. Natl. Acad. Sci. U.S.A.}\ }\textbf {\bibinfo {volume}
  {116}},\ \bibinfo {pages} {7232}}\BibitemShut {NoStop}%
\bibitem [{\citenamefont {Rosenstein}\ \emph {et~al.}(2016)\citenamefont
  {Rosenstein}, \citenamefont {Shapiro}, \citenamefont {Shapiro},\ and\
  \citenamefont {Li}}]{Rosenstein2016}%
  \BibitemOpen
  \bibfield  {author} {\bibinfo {author} {\bibnamefont {Rosenstein},
  \bibfnamefont {B.}}, \bibinfo {author} {\bibfnamefont {B.~Y.}\ \bibnamefont
  {Shapiro}}, \bibinfo {author} {\bibfnamefont {I.}~\bibnamefont {Shapiro}},
  and\ \bibinfo {author} {\bibfnamefont {D.}~\bibnamefont {Li}}} (\bibinfo
  {year} {2016}),\ \href {https://doi.org/10.1103/PhysRevB.94.024505}
  {\bibfield  {journal} {\bibinfo  {journal} {Phys. Rev. B}\ }\textbf {\bibinfo
  {volume} {94}},\ \bibinfo {pages} {024505}}\BibitemShut {NoStop}%
\bibitem [{\citenamefont {Rotter}\ \emph
  {et~al.}(2008{\natexlab{a}})\citenamefont {Rotter}, \citenamefont {Pangerl},
  \citenamefont {Tegel},\ and\ \citenamefont {Johrendt}}]{Rotter2008b}%
  \BibitemOpen
  \bibfield  {author} {\bibinfo {author} {\bibnamefont {Rotter}, \bibfnamefont
  {M.}}, \bibinfo {author} {\bibfnamefont {M.}~\bibnamefont {Pangerl}},
  \bibinfo {author} {\bibfnamefont {M.}~\bibnamefont {Tegel}}, and\ \bibinfo
  {author} {\bibfnamefont {D.}~\bibnamefont {Johrendt}}} (\bibinfo {year}
  {2008}{\natexlab{a}}),\ \href {https://doi.org/10.1002/anie.200803641}
  {\bibfield  {journal} {\bibinfo  {journal} {Angewandte Chemie International
  Edition}\ }\textbf {\bibinfo {volume} {47}},\ \bibinfo {pages}
  {7949}}\BibitemShut {NoStop}%
\bibitem [{\citenamefont {Rotter}\ \emph
  {et~al.}(2008{\natexlab{b}})\citenamefont {Rotter}, \citenamefont {Tegel},\
  and\ \citenamefont {Johrendt}}]{Rotter2008a}%
  \BibitemOpen
  \bibfield  {author} {\bibinfo {author} {\bibnamefont {Rotter}, \bibfnamefont
  {M.}}, \bibinfo {author} {\bibfnamefont {M.}~\bibnamefont {Tegel}}, and\
  \bibinfo {author} {\bibfnamefont {D.}~\bibnamefont {Johrendt}}} (\bibinfo
  {year} {2008}{\natexlab{b}}),\ \href
  {https://doi.org/10.1103/PhysRevLett.101.107006} {\bibfield  {journal}
  {\bibinfo  {journal} {Phys. Rev. Lett.}\ }\textbf {\bibinfo {volume} {101}},\
  \bibinfo {pages} {107006}}\BibitemShut {NoStop}%
\bibitem [{\citenamefont {Sadakov}\ \emph {et~al.}(2015)\citenamefont
  {Sadakov}, \citenamefont {Romanova}, \citenamefont {Knyazev}, \citenamefont
  {Chareev},\ and\ \citenamefont {Martovitsky}}]{Sadakov2015}%
  \BibitemOpen
  \bibfield  {author} {\bibinfo {author} {\bibnamefont {Sadakov}, \bibfnamefont
  {A.}}, \bibinfo {author} {\bibfnamefont {T.}~\bibnamefont {Romanova}},
  \bibinfo {author} {\bibfnamefont {D.}~\bibnamefont {Knyazev}}, \bibinfo
  {author} {\bibfnamefont {D.}~\bibnamefont {Chareev}}, and\ \bibinfo {author}
  {\bibfnamefont {V.}~\bibnamefont {Martovitsky}}} (\bibinfo {year} {2015}),\
  \href {https://doi.org/10.1016/j.phpro.2015.12.043} {\bibfield  {journal}
  {\bibinfo  {journal} {Physics Procedia}\ }\textbf {\bibinfo {volume} {75}},\
  \bibinfo {pages} {364}},\ \bibinfo {note} {20th International Conference on
  Magnetism, ICM 2015}\BibitemShut {NoStop}%
\bibitem [{\citenamefont {Saito}\ \emph {et~al.}(2016)\citenamefont {Saito},
  \citenamefont {Nojima},\ and\ \citenamefont {Iwasa}}]{Saito2016}%
  \BibitemOpen
  \bibfield  {author} {\bibinfo {author} {\bibnamefont {Saito}, \bibfnamefont
  {Y.}}, \bibinfo {author} {\bibfnamefont {T.}~\bibnamefont {Nojima}}, and\
  \bibinfo {author} {\bibfnamefont {Y.}~\bibnamefont {Iwasa}}} (\bibinfo {year}
  {2016}),\ \href {https://doi.org/10.1088/0953-2048/29/9/093001} {\bibfield
  {journal} {\bibinfo  {journal} {Supercond. Sci. Technol.}\ }\textbf {\bibinfo
  {volume} {29}},\ \bibinfo {pages} {093001}}\BibitemShut {NoStop}%
\bibitem [{\citenamefont {Sala}\ \emph {et~al.}(2014)\citenamefont {Sala},
  \citenamefont {Yakita}, \citenamefont {Ogino}, \citenamefont {Okada},
  \citenamefont {Yamamoto}, \citenamefont {Kishio}, \citenamefont {Ishida},
  \citenamefont {Iyo}, \citenamefont {Eisaki}, \citenamefont {Fujioka} \emph
  {et~al.}}]{Sala2014}%
  \BibitemOpen
  \bibfield  {author} {\bibinfo {author} {\bibnamefont {Sala}, \bibfnamefont
  {A.}}, \bibinfo {author} {\bibfnamefont {H.}~\bibnamefont {Yakita}}, \bibinfo
  {author} {\bibfnamefont {H.}~\bibnamefont {Ogino}}, \bibinfo {author}
  {\bibfnamefont {T.}~\bibnamefont {Okada}}, \bibinfo {author} {\bibfnamefont
  {A.}~\bibnamefont {Yamamoto}}, \bibinfo {author} {\bibfnamefont
  {K.}~\bibnamefont {Kishio}}, \bibinfo {author} {\bibfnamefont
  {S.}~\bibnamefont {Ishida}}, \bibinfo {author} {\bibfnamefont
  {A.}~\bibnamefont {Iyo}}, \bibinfo {author} {\bibfnamefont {H.}~\bibnamefont
  {Eisaki}}, \bibinfo {author} {\bibfnamefont {M.}~\bibnamefont {Fujioka}},
  \emph {et~al.}} (\bibinfo {year} {2014}),\ \href
  {https://doi.org/10.7567/apex.7.073102} {\bibfield  {journal} {\bibinfo
  {journal} {Applied Physics Express}\ }\textbf {\bibinfo {volume} {7}},\
  \bibinfo {pages} {073102}}\BibitemShut {NoStop}%
\bibitem [{\citenamefont {Sanchez}\ \emph {et~al.}(2021)\citenamefont
  {Sanchez}, \citenamefont {Malinowski}, \citenamefont {Mutch}, \citenamefont
  {Liu}, \citenamefont {Kim}, \citenamefont {Ryan},\ and\ \citenamefont
  {Chu}}]{sanchez2021transport}%
  \BibitemOpen
  \bibfield  {author} {\bibinfo {author} {\bibnamefont {Sanchez}, \bibfnamefont
  {J.~J.}}, \bibinfo {author} {\bibfnamefont {P.}~\bibnamefont {Malinowski}},
  \bibinfo {author} {\bibfnamefont {J.}~\bibnamefont {Mutch}}, \bibinfo
  {author} {\bibfnamefont {J.}~\bibnamefont {Liu}}, \bibinfo {author}
  {\bibfnamefont {J.-W.}\ \bibnamefont {Kim}}, \bibinfo {author} {\bibfnamefont
  {P.~J.}\ \bibnamefont {Ryan}}, and\ \bibinfo {author} {\bibfnamefont {J.-H.}\
  \bibnamefont {Chu}}} (\bibinfo {year} {2021}),\ \href
  {https://doi.org/10.1038/s41563-021-01082-4} {\bibfield  {journal} {\bibinfo
  {journal} {Nat. Mater.}\ }\textbf {\bibinfo {volume} {20}},\ \bibinfo {pages}
  {1519}}\BibitemShut {NoStop}%
\bibitem [{\citenamefont {Sapkota}\ \emph {et~al.}(2018)\citenamefont
  {Sapkota}, \citenamefont {Das}, \citenamefont {B{\"o}hmer}, \citenamefont
  {Ueland}, \citenamefont {Abernathy}, \citenamefont {Bud'ko}, \citenamefont
  {Canfield}, \citenamefont {Kreyssig}, \citenamefont {Goldman},\ and\
  \citenamefont {McQueeney}}]{sapkota2018doping}%
  \BibitemOpen
  \bibfield  {author} {\bibinfo {author} {\bibnamefont {Sapkota}, \bibfnamefont
  {A.}}, \bibinfo {author} {\bibfnamefont {P.}~\bibnamefont {Das}}, \bibinfo
  {author} {\bibfnamefont {A.~E.}\ \bibnamefont {B{\"o}hmer}}, \bibinfo
  {author} {\bibfnamefont {B.~G.}\ \bibnamefont {Ueland}}, \bibinfo {author}
  {\bibfnamefont {D.~L.}\ \bibnamefont {Abernathy}}, \bibinfo {author}
  {\bibfnamefont {S.~L.}\ \bibnamefont {Bud'ko}}, \bibinfo {author}
  {\bibfnamefont {P.~C.}\ \bibnamefont {Canfield}}, \bibinfo {author}
  {\bibfnamefont {A.}~\bibnamefont {Kreyssig}}, \bibinfo {author}
  {\bibfnamefont {A.~I.}\ \bibnamefont {Goldman}}, and\ \bibinfo {author}
  {\bibfnamefont {R.~J.}\ \bibnamefont {McQueeney}}} (\bibinfo {year} {2018}),\
  \href {https://doi.org/10.1103/PhysRevB.97.174519} {\bibfield  {journal}
  {\bibinfo  {journal} {Phys. Rev. B}\ }\textbf {\bibinfo {volume} {97}},\
  \bibinfo {pages} {174519}}\BibitemShut {NoStop}%
\bibitem [{\citenamefont {Sasmal}\ \emph {et~al.}(2008)\citenamefont {Sasmal},
  \citenamefont {Lv}, \citenamefont {Lorenz}, \citenamefont {Guloy},
  \citenamefont {Chen}, \citenamefont {Xue},\ and\ \citenamefont
  {Chu}}]{Sasmal2008}%
  \BibitemOpen
  \bibfield  {author} {\bibinfo {author} {\bibnamefont {Sasmal}, \bibfnamefont
  {K.}}, \bibinfo {author} {\bibfnamefont {B.}~\bibnamefont {Lv}}, \bibinfo
  {author} {\bibfnamefont {B.}~\bibnamefont {Lorenz}}, \bibinfo {author}
  {\bibfnamefont {A.~M.}\ \bibnamefont {Guloy}}, \bibinfo {author}
  {\bibfnamefont {F.}~\bibnamefont {Chen}}, \bibinfo {author} {\bibfnamefont
  {Y.-Y.}\ \bibnamefont {Xue}}, and\ \bibinfo {author} {\bibfnamefont {C.-W.}\
  \bibnamefont {Chu}}} (\bibinfo {year} {2008}),\ \href
  {https://doi.org/10.1103/PhysRevLett.101.107007} {\bibfield  {journal}
  {\bibinfo  {journal} {Phys. Rev. Lett.}\ }\textbf {\bibinfo {volume} {101}},\
  \bibinfo {pages} {107007}}\BibitemShut {NoStop}%
\bibitem [{\citenamefont {Sato}\ \emph {et~al.}(2010)\citenamefont {Sato},
  \citenamefont {Ogino}, \citenamefont {Kawaguchi}, \citenamefont {Katsura},
  \citenamefont {Kishio}, \citenamefont {Shimoyama}, \citenamefont {Kotegawa},\
  and\ \citenamefont {Tou}}]{Sato2010}%
  \BibitemOpen
  \bibfield  {author} {\bibinfo {author} {\bibnamefont {Sato}, \bibfnamefont
  {S.}}, \bibinfo {author} {\bibfnamefont {H.}~\bibnamefont {Ogino}}, \bibinfo
  {author} {\bibfnamefont {N.}~\bibnamefont {Kawaguchi}}, \bibinfo {author}
  {\bibfnamefont {Y.}~\bibnamefont {Katsura}}, \bibinfo {author} {\bibfnamefont
  {K.}~\bibnamefont {Kishio}}, \bibinfo {author} {\bibfnamefont {J.-i.}\
  \bibnamefont {Shimoyama}}, \bibinfo {author} {\bibfnamefont {H.}~\bibnamefont
  {Kotegawa}}, and\ \bibinfo {author} {\bibfnamefont {H.}~\bibnamefont {Tou}}}
  (\bibinfo {year} {2010}),\ \href
  {https://doi.org/10.1088/0953-2048/23/4/045001} {\bibfield  {journal}
  {\bibinfo  {journal} {Supercond. Sci. Technol.}\ }\textbf {\bibinfo {volume}
  {23}},\ \bibinfo {pages} {045001}}\BibitemShut {NoStop}%
\bibitem [{\citenamefont {Sato}\ \emph {et~al.}(2018)\citenamefont {Sato},
  \citenamefont {Kasahara}, \citenamefont {Taniguchi}, \citenamefont {Xing},
  \citenamefont {Kasahara}, \citenamefont {Tokiwa}, \citenamefont {Yamakawa},
  \citenamefont {Kontani}, \citenamefont {Shibauchi},\ and\ \citenamefont
  {Matsuda}}]{sato2018abrupt}%
  \BibitemOpen
  \bibfield  {author} {\bibinfo {author} {\bibnamefont {Sato}, \bibfnamefont
  {Y.}}, \bibinfo {author} {\bibfnamefont {S.}~\bibnamefont {Kasahara}},
  \bibinfo {author} {\bibfnamefont {T.}~\bibnamefont {Taniguchi}}, \bibinfo
  {author} {\bibfnamefont {X.}~\bibnamefont {Xing}}, \bibinfo {author}
  {\bibfnamefont {Y.}~\bibnamefont {Kasahara}}, \bibinfo {author}
  {\bibfnamefont {Y.}~\bibnamefont {Tokiwa}}, \bibinfo {author} {\bibfnamefont
  {Y.}~\bibnamefont {Yamakawa}}, \bibinfo {author} {\bibfnamefont
  {H.}~\bibnamefont {Kontani}}, \bibinfo {author} {\bibfnamefont
  {T.}~\bibnamefont {Shibauchi}}, and\ \bibinfo {author} {\bibfnamefont
  {Y.}~\bibnamefont {Matsuda}}} (\bibinfo {year} {2018}),\ \href
  {https://doi.org/10.1073/pnas.1717331115} {\bibfield  {journal} {\bibinfo
  {journal} {Proc. Natl. Acad. Sci. U.S.A.}\ }\textbf {\bibinfo {volume}
  {115}},\ \bibinfo {pages} {1227}}\BibitemShut {NoStop}%
\bibitem [{\citenamefont {Sato}\ \emph {et~al.}(2025)\citenamefont {Sato},
  \citenamefont {Nagahama}, \citenamefont {Kitou}, \citenamefont {Sagayama},
  \citenamefont {Belopolski}, \citenamefont {Yoshimi}, \citenamefont
  {Kawamura}, \citenamefont {Tsukazaki}, \citenamefont {Kanazawa},
  \citenamefont {Nomoto}, \citenamefont {Arita}, \citenamefont {Arima},
  \citenamefont {Kawasaki},\ and\ \citenamefont {Tokura}}]{Sato2025}%
  \BibitemOpen
  \bibfield  {author} {\bibinfo {author} {\bibnamefont {Sato}, \bibfnamefont
  {Y.}}, \bibinfo {author} {\bibfnamefont {S.}~\bibnamefont {Nagahama}},
  \bibinfo {author} {\bibfnamefont {S.}~\bibnamefont {Kitou}}, \bibinfo
  {author} {\bibfnamefont {H.}~\bibnamefont {Sagayama}}, \bibinfo {author}
  {\bibfnamefont {I.}~\bibnamefont {Belopolski}}, \bibinfo {author}
  {\bibfnamefont {R.}~\bibnamefont {Yoshimi}}, \bibinfo {author} {\bibfnamefont
  {M.}~\bibnamefont {Kawamura}}, \bibinfo {author} {\bibfnamefont
  {A.}~\bibnamefont {Tsukazaki}}, \bibinfo {author} {\bibfnamefont
  {N.}~\bibnamefont {Kanazawa}}, \bibinfo {author} {\bibfnamefont
  {T.}~\bibnamefont {Nomoto}}, \bibinfo {author} {\bibfnamefont
  {R.}~\bibnamefont {Arita}}, \bibinfo {author} {\bibfnamefont {T.-h.}\
  \bibnamefont {Arima}}, \bibinfo {author} {\bibfnamefont {M.}~\bibnamefont
  {Kawasaki}}, and\ \bibinfo {author} {\bibfnamefont {Y.}~\bibnamefont
  {Tokura}}} (\bibinfo {year} {2025}),\ \href
  {https://doi.org/10.1038/s41467-025-65902-w} {\bibfield  {journal} {\bibinfo
  {journal} {Nat. Commun.}\ }\textbf {\bibinfo {volume} {16}},\ \bibinfo
  {pages} {10913}}\BibitemShut {NoStop}%
\bibitem [{\citenamefont {Scalapino}(2012)}]{scalapino2012}%
  \BibitemOpen
  \bibfield  {author} {\bibinfo {author} {\bibnamefont {Scalapino},
  \bibfnamefont {D.~J.}}} (\bibinfo {year} {2012}),\ \href
  {https://doi.org/10.1103/RevModPhys.84.1383} {\bibfield  {journal} {\bibinfo
  {journal} {Rev. Mod. Phys.}\ }\textbf {\bibinfo {volume} {84}},\ \bibinfo
  {pages} {1383}}\BibitemShut {NoStop}%
\bibitem [{\citenamefont {Scherer}\ \emph {et~al.}(2016)\citenamefont
  {Scherer}, \citenamefont {Eremin},\ and\ \citenamefont
  {Andersen}}]{scherer2016collective}%
  \BibitemOpen
  \bibfield  {author} {\bibinfo {author} {\bibnamefont {Scherer}, \bibfnamefont
  {D.~D.}}, \bibinfo {author} {\bibfnamefont {I.}~\bibnamefont {Eremin}}, and\
  \bibinfo {author} {\bibfnamefont {B.~M.}\ \bibnamefont {Andersen}}} (\bibinfo
  {year} {2016}),\ \href {https://doi.org/10.1103/PhysRevB.94.180405}
  {\bibfield  {journal} {\bibinfo  {journal} {Phys. Rev. B}\ }\textbf {\bibinfo
  {volume} {94}},\ \bibinfo {pages} {180405}}\BibitemShut {NoStop}%
\bibitem [{\citenamefont {Schlappa}\ \emph {et~al.}(2012)\citenamefont
  {Schlappa}, \citenamefont {Wohlfeld}, \citenamefont {Zhou}, \citenamefont
  {Mourigal}, \citenamefont {Haverkort}, \citenamefont {Strocov}, \citenamefont
  {Hozoi}, \citenamefont {Monney}, \citenamefont {Nishimoto}, \citenamefont
  {Singh}, \citenamefont {Revcolevschi}, \citenamefont {Caux}, \citenamefont
  {Patthey}, \citenamefont {R{\o}nnow}, \citenamefont {van~den Brink},\ and\
  \citenamefont {Schmitt}}]{schlappa2012spin}%
  \BibitemOpen
  \bibfield  {author} {\bibinfo {author} {\bibnamefont {Schlappa},
  \bibfnamefont {J.}}, \bibinfo {author} {\bibfnamefont {K.}~\bibnamefont
  {Wohlfeld}}, \bibinfo {author} {\bibfnamefont {K.~J.}\ \bibnamefont {Zhou}},
  \bibinfo {author} {\bibfnamefont {M.}~\bibnamefont {Mourigal}}, \bibinfo
  {author} {\bibfnamefont {M.~W.}\ \bibnamefont {Haverkort}}, \bibinfo {author}
  {\bibfnamefont {V.~N.}\ \bibnamefont {Strocov}}, \bibinfo {author}
  {\bibfnamefont {L.}~\bibnamefont {Hozoi}}, \bibinfo {author} {\bibfnamefont
  {C.}~\bibnamefont {Monney}}, \bibinfo {author} {\bibfnamefont
  {S.}~\bibnamefont {Nishimoto}}, \bibinfo {author} {\bibfnamefont
  {S.}~\bibnamefont {Singh}}, \bibinfo {author} {\bibfnamefont
  {A.}~\bibnamefont {Revcolevschi}}, \bibinfo {author} {\bibfnamefont {J.-S.}\
  \bibnamefont {Caux}}, \bibinfo {author} {\bibfnamefont {L.}~\bibnamefont
  {Patthey}}, \bibinfo {author} {\bibfnamefont {H.~M.}\ \bibnamefont
  {R{\o}nnow}}, \bibinfo {author} {\bibfnamefont {J.}~\bibnamefont {van~den
  Brink}}, and\ \bibinfo {author} {\bibfnamefont {T.}~\bibnamefont {Schmitt}}}
  (\bibinfo {year} {2012}),\ \href {https://doi.org/10.1038/nature10974}
  {\bibfield  {journal} {\bibinfo  {journal} {Nature}\ }\textbf {\bibinfo
  {volume} {485}},\ \bibinfo {pages} {82}}\BibitemShut {NoStop}%
\bibitem [{\citenamefont {Sefat}\ \emph {et~al.}(2008)\citenamefont {Sefat},
  \citenamefont {Jin}, \citenamefont {McGuire}, \citenamefont {Sales},
  \citenamefont {Singh},\ and\ \citenamefont {Mandrus}}]{Sefat2008}%
  \BibitemOpen
  \bibfield  {author} {\bibinfo {author} {\bibnamefont {Sefat}, \bibfnamefont
  {A.~S.}}, \bibinfo {author} {\bibfnamefont {R.}~\bibnamefont {Jin}}, \bibinfo
  {author} {\bibfnamefont {M.~A.}\ \bibnamefont {McGuire}}, \bibinfo {author}
  {\bibfnamefont {B.~C.}\ \bibnamefont {Sales}}, \bibinfo {author}
  {\bibfnamefont {D.~J.}\ \bibnamefont {Singh}}, and\ \bibinfo {author}
  {\bibfnamefont {D.}~\bibnamefont {Mandrus}}} (\bibinfo {year} {2008}),\ \href
  {https://doi.org/10.1103/PhysRevLett.101.117004} {\bibfield  {journal}
  {\bibinfo  {journal} {Phys. Rev. Lett.}\ }\textbf {\bibinfo {volume} {101}},\
  \bibinfo {pages} {117004}}\BibitemShut {NoStop}%
\bibitem [{\citenamefont {Setty}\ \emph {et~al.}(2020)\citenamefont {Setty},
  \citenamefont {Bhattacharyya}, \citenamefont {Cao}, \citenamefont {Kreisel},\
  and\ \citenamefont {Hirschfeld}}]{setty2020topological}%
  \BibitemOpen
  \bibfield  {author} {\bibinfo {author} {\bibnamefont {Setty}, \bibfnamefont
  {C.}}, \bibinfo {author} {\bibfnamefont {S.}~\bibnamefont {Bhattacharyya}},
  \bibinfo {author} {\bibfnamefont {Y.}~\bibnamefont {Cao}}, \bibinfo {author}
  {\bibfnamefont {A.}~\bibnamefont {Kreisel}}, and\ \bibinfo {author}
  {\bibfnamefont {P.~J.}\ \bibnamefont {Hirschfeld}}} (\bibinfo {year}
  {2020}),\ \href {https://doi.org/10.1038/s41467-020-14357-2} {\bibfield
  {journal} {\bibinfo  {journal} {Nat. Commun.}\ }\textbf {\bibinfo {volume}
  {11}},\ \bibinfo {pages} {523}}\BibitemShut {NoStop}%
\bibitem [{\citenamefont {Shahi}\ \emph {et~al.}(2018)\citenamefont {Shahi},
  \citenamefont {Sun}, \citenamefont {Wang}, \citenamefont {Jiao},
  \citenamefont {Chen}, \citenamefont {Sun}, \citenamefont {Lei}, \citenamefont
  {Uwatoko}, \citenamefont {Wang},\ and\ \citenamefont
  {Cheng}}]{shahi2018hightc}%
  \BibitemOpen
  \bibfield  {author} {\bibinfo {author} {\bibnamefont {Shahi}, \bibfnamefont
  {P.}}, \bibinfo {author} {\bibfnamefont {J.~P.}\ \bibnamefont {Sun}},
  \bibinfo {author} {\bibfnamefont {S.~H.}\ \bibnamefont {Wang}}, \bibinfo
  {author} {\bibfnamefont {Y.~Y.}\ \bibnamefont {Jiao}}, \bibinfo {author}
  {\bibfnamefont {K.~Y.}\ \bibnamefont {Chen}}, \bibinfo {author}
  {\bibfnamefont {S.~S.}\ \bibnamefont {Sun}}, \bibinfo {author} {\bibfnamefont
  {H.~C.}\ \bibnamefont {Lei}}, \bibinfo {author} {\bibfnamefont
  {Y.}~\bibnamefont {Uwatoko}}, \bibinfo {author} {\bibfnamefont {B.~S.}\
  \bibnamefont {Wang}}, and\ \bibinfo {author} {\bibfnamefont {J.-G.}\
  \bibnamefont {Cheng}}} (\bibinfo {year} {2018}),\ \href
  {https://doi.org/10.1103/PhysRevB.97.020508} {\bibfield  {journal} {\bibinfo
  {journal} {Phys. Rev. B}\ }\textbf {\bibinfo {volume} {97}},\ \bibinfo
  {pages} {020508}}\BibitemShut {NoStop}%
\bibitem [{\citenamefont {Shao}\ \emph {et~al.}(2019)\citenamefont {Shao},
  \citenamefont {Wang}, \citenamefont {Li}, \citenamefont {Wu}, \citenamefont
  {Wu}, \citenamefont {Ren}, \citenamefont {Qiu}, \citenamefont {Rao},
  \citenamefont {Wang},\ and\ \citenamefont {Cao}}]{Shao2019}%
  \BibitemOpen
  \bibfield  {author} {\bibinfo {author} {\bibnamefont {Shao}, \bibfnamefont
  {Y.-T.}}, \bibinfo {author} {\bibfnamefont {Z.-C.}\ \bibnamefont {Wang}},
  \bibinfo {author} {\bibfnamefont {B.-Z.}\ \bibnamefont {Li}}, \bibinfo
  {author} {\bibfnamefont {S.-Q.}\ \bibnamefont {Wu}}, \bibinfo {author}
  {\bibfnamefont {J.-F.}\ \bibnamefont {Wu}}, \bibinfo {author} {\bibfnamefont
  {Z.}~\bibnamefont {Ren}}, \bibinfo {author} {\bibfnamefont {S.-W.}\
  \bibnamefont {Qiu}}, \bibinfo {author} {\bibfnamefont {C.}~\bibnamefont
  {Rao}}, \bibinfo {author} {\bibfnamefont {C.}~\bibnamefont {Wang}}, and\
  \bibinfo {author} {\bibfnamefont {G.-H.}\ \bibnamefont {Cao}}} (\bibinfo
  {year} {2019}),\ \href {https://doi.org/10.1007/s40843-019-9438-7} {\bibfield
   {journal} {\bibinfo  {journal} {Science China Materials}\ }\textbf {\bibinfo
  {volume} {62}},\ \bibinfo {pages} {1357}}\BibitemShut {NoStop}%
\bibitem [{\citenamefont {Shen}\ \emph {et~al.}(2020)\citenamefont {Shen},
  \citenamefont {Zhang}, \citenamefont {Wo}, \citenamefont {Shen},
  \citenamefont {Feng}, \citenamefont {Schneidewind}, \citenamefont {{\v
  C}erm{\'a}k}, \citenamefont {Wang},\ and\ \citenamefont
  {Zhao}}]{shen2020neutron}%
  \BibitemOpen
  \bibfield  {author} {\bibinfo {author} {\bibnamefont {Shen}, \bibfnamefont
  {S.}}, \bibinfo {author} {\bibfnamefont {X.}~\bibnamefont {Zhang}}, \bibinfo
  {author} {\bibfnamefont {H.}~\bibnamefont {Wo}}, \bibinfo {author}
  {\bibfnamefont {Y.}~\bibnamefont {Shen}}, \bibinfo {author} {\bibfnamefont
  {Y.}~\bibnamefont {Feng}}, \bibinfo {author} {\bibfnamefont {A.}~\bibnamefont
  {Schneidewind}}, \bibinfo {author} {\bibfnamefont {P.}~\bibnamefont {{\v
  C}erm{\'a}k}}, \bibinfo {author} {\bibfnamefont {W.}~\bibnamefont {Wang}},
  and\ \bibinfo {author} {\bibfnamefont {J.}~\bibnamefont {Zhao}}} (\bibinfo
  {year} {2020}),\ \href {https://doi.org/10.1103/PhysRevLett.124.017001}
  {\bibfield  {journal} {\bibinfo  {journal} {Phys. Rev. Lett.}\ }\textbf
  {\bibinfo {volume} {124}},\ \bibinfo {pages} {017001}}\BibitemShut {NoStop}%
\bibitem [{\citenamefont {Shi}\ \emph {et~al.}(2018{\natexlab{a}})\citenamefont
  {Shi}, \citenamefont {Wang}, \citenamefont {Lei}, \citenamefont {Shang},
  \citenamefont {Meng}, \citenamefont {Ma}, \citenamefont {Zhang},
  \citenamefont {Kuang},\ and\ \citenamefont
  {Chen}}]{shi2018organicionintercalated}%
  \BibitemOpen
  \bibfield  {author} {\bibinfo {author} {\bibnamefont {Shi}, \bibfnamefont
  {M.~Z.}}, \bibinfo {author} {\bibfnamefont {N.~Z.}\ \bibnamefont {Wang}},
  \bibinfo {author} {\bibfnamefont {B.}~\bibnamefont {Lei}}, \bibinfo {author}
  {\bibfnamefont {C.}~\bibnamefont {Shang}}, \bibinfo {author} {\bibfnamefont
  {F.~B.}\ \bibnamefont {Meng}}, \bibinfo {author} {\bibfnamefont {L.~K.}\
  \bibnamefont {Ma}}, \bibinfo {author} {\bibfnamefont {F.~X.}\ \bibnamefont
  {Zhang}}, \bibinfo {author} {\bibfnamefont {D.~Z.}\ \bibnamefont {Kuang}},
  and\ \bibinfo {author} {\bibfnamefont {X.~H.}\ \bibnamefont {Chen}}}
  (\bibinfo {year} {2018}{\natexlab{a}}),\ \href
  {https://doi.org/10.1103/PhysRevMaterials.2.074801} {\bibfield  {journal}
  {\bibinfo  {journal} {Phys. Rev. Mater.}\ }\textbf {\bibinfo {volume} {2}},\
  \bibinfo {pages} {074801}}\BibitemShut {NoStop}%
\bibitem [{\citenamefont {Shi}\ \emph {et~al.}(2018{\natexlab{b}})\citenamefont
  {Shi}, \citenamefont {Wang}, \citenamefont {Lei}, \citenamefont {Ying},
  \citenamefont {Zhu}, \citenamefont {Sun}, \citenamefont {Cui}, \citenamefont
  {Meng}, \citenamefont {Shang}, \citenamefont {Ma},\ and\ \citenamefont
  {H.}}]{Shi2018b}%
  \BibitemOpen
  \bibfield  {author} {\bibinfo {author} {\bibnamefont {Shi}, \bibfnamefont
  {M.~Z.}}, \bibinfo {author} {\bibfnamefont {N.~Z.}\ \bibnamefont {Wang}},
  \bibinfo {author} {\bibfnamefont {B.}~\bibnamefont {Lei}}, \bibinfo {author}
  {\bibfnamefont {J.~J.}\ \bibnamefont {Ying}}, \bibinfo {author}
  {\bibfnamefont {C.~S.}\ \bibnamefont {Zhu}}, \bibinfo {author} {\bibfnamefont
  {Z.~L.}\ \bibnamefont {Sun}}, \bibinfo {author} {\bibfnamefont {J.~H.}\
  \bibnamefont {Cui}}, \bibinfo {author} {\bibfnamefont {F.~B.}\ \bibnamefont
  {Meng}}, \bibinfo {author} {\bibfnamefont {C.}~\bibnamefont {Shang}},
  \bibinfo {author} {\bibfnamefont {L.~K.}\ \bibnamefont {Ma}}, and\ \bibinfo
  {author} {\bibfnamefont {C.~X.}\ \bibnamefont {H.}}} (\bibinfo {year}
  {2018}{\natexlab{b}}),\ \href {https://doi.org/10.1088/1367-2630/aaf312}
  {\bibfield  {journal} {\bibinfo  {journal} {New J. Phys.}\ }\textbf {\bibinfo
  {volume} {20}},\ \bibinfo {pages} {123007}}\BibitemShut {NoStop}%
\bibitem [{\citenamefont {Shibauchi}\ \emph {et~al.}(2020)\citenamefont
  {Shibauchi}, \citenamefont {Hanaguri},\ and\ \citenamefont
  {Matsuda}}]{shibauchi2020exotic}%
  \BibitemOpen
  \bibfield  {author} {\bibinfo {author} {\bibnamefont {Shibauchi},
  \bibfnamefont {T.}}, \bibinfo {author} {\bibfnamefont {T.}~\bibnamefont
  {Hanaguri}}, and\ \bibinfo {author} {\bibfnamefont {Y.}~\bibnamefont
  {Matsuda}}} (\bibinfo {year} {2020}),\ \href
  {https://doi.org/10.7566/JPSJ.89.102002} {\bibfield  {journal} {\bibinfo
  {journal} {J. Phys. Soc. Jpn.}\ }\textbf {\bibinfo {volume} {89}},\ \bibinfo
  {pages} {102002}}\BibitemShut {NoStop}%
\bibitem [{\citenamefont {Shimojima}\ \emph {et~al.}(2017)\citenamefont
  {Shimojima}, \citenamefont {Malaeb}, \citenamefont {Nakamura}, \citenamefont
  {Kondo}, \citenamefont {Kihou}, \citenamefont {Lee}, \citenamefont {Iyo},
  \citenamefont {Eisaki}, \citenamefont {Ishida}, \citenamefont {Nakajima},
  \citenamefont {ichi Uchida}, \citenamefont {Ohgushi}, \citenamefont
  {Ishizaka},\ and\ \citenamefont {Shin}}]{shimojima2017}%
  \BibitemOpen
  \bibfield  {author} {\bibinfo {author} {\bibnamefont {Shimojima},
  \bibfnamefont {T.}}, \bibinfo {author} {\bibfnamefont {W.}~\bibnamefont
  {Malaeb}}, \bibinfo {author} {\bibfnamefont {A.}~\bibnamefont {Nakamura}},
  \bibinfo {author} {\bibfnamefont {T.}~\bibnamefont {Kondo}}, \bibinfo
  {author} {\bibfnamefont {K.}~\bibnamefont {Kihou}}, \bibinfo {author}
  {\bibfnamefont {C.-H.}\ \bibnamefont {Lee}}, \bibinfo {author} {\bibfnamefont
  {A.}~\bibnamefont {Iyo}}, \bibinfo {author} {\bibfnamefont {H.}~\bibnamefont
  {Eisaki}}, \bibinfo {author} {\bibfnamefont {S.}~\bibnamefont {Ishida}},
  \bibinfo {author} {\bibfnamefont {M.}~\bibnamefont {Nakajima}}, \bibinfo
  {author} {\bibfnamefont {S.}~\bibnamefont {ichi Uchida}}, \bibinfo {author}
  {\bibfnamefont {K.}~\bibnamefont {Ohgushi}}, \bibinfo {author} {\bibfnamefont
  {K.}~\bibnamefont {Ishizaka}}, and\ \bibinfo {author} {\bibfnamefont
  {S.}~\bibnamefont {Shin}}} (\bibinfo {year} {2017}),\ \href
  {https://doi.org/10.1126/sciadv.1700466} {\bibfield  {journal} {\bibinfo
  {journal} {Sci. Adv.}\ }\textbf {\bibinfo {volume} {3}},\ \bibinfo {pages}
  {e1700466}}\BibitemShut {NoStop}%
\bibitem [{\citenamefont {Shimojima}\ \emph {et~al.}(2021)\citenamefont
  {Shimojima}, \citenamefont {Motoyui}, \citenamefont {Taniuchi}, \citenamefont
  {Bareille}, \citenamefont {Onari}, \citenamefont {Kontani}, \citenamefont
  {Nakajima}, \citenamefont {Kasahara}, \citenamefont {Shibauchi},
  \citenamefont {Matsuda},\ and\ \citenamefont
  {Shin}}]{shimojima2021discovery}%
  \BibitemOpen
  \bibfield  {author} {\bibinfo {author} {\bibnamefont {Shimojima},
  \bibfnamefont {T.}}, \bibinfo {author} {\bibfnamefont {Y.}~\bibnamefont
  {Motoyui}}, \bibinfo {author} {\bibfnamefont {T.}~\bibnamefont {Taniuchi}},
  \bibinfo {author} {\bibfnamefont {C.}~\bibnamefont {Bareille}}, \bibinfo
  {author} {\bibfnamefont {S.}~\bibnamefont {Onari}}, \bibinfo {author}
  {\bibfnamefont {H.}~\bibnamefont {Kontani}}, \bibinfo {author} {\bibfnamefont
  {M.}~\bibnamefont {Nakajima}}, \bibinfo {author} {\bibfnamefont
  {S.}~\bibnamefont {Kasahara}}, \bibinfo {author} {\bibfnamefont
  {T.}~\bibnamefont {Shibauchi}}, \bibinfo {author} {\bibfnamefont
  {Y.}~\bibnamefont {Matsuda}}, and\ \bibinfo {author} {\bibfnamefont
  {S.}~\bibnamefont {Shin}}} (\bibinfo {year} {2021}),\ \href
  {https://doi.org/10.1126/science.abd6701} {\bibfield  {journal} {\bibinfo
  {journal} {Science}\ }\textbf {\bibinfo {volume} {373}},\ \bibinfo {pages}
  {1122}}\BibitemShut {NoStop}%
\bibitem [{\citenamefont {Shimojima}\ \emph {et~al.}(2014)\citenamefont
  {Shimojima}, \citenamefont {Suzuki}, \citenamefont {Sonobe}, \citenamefont
  {Nakamura}, \citenamefont {Sakano}, \citenamefont {Omachi}, \citenamefont
  {Yoshioka}, \citenamefont {Kuwata-Gonokami}, \citenamefont {Ono},
  \citenamefont {Kumigashira}, \citenamefont {B\"ohmer}, \citenamefont {Hardy},
  \citenamefont {Wolf}, \citenamefont {Meingast}, \citenamefont {L\"ohneysen},
  \citenamefont {Ikeda},\ and\ \citenamefont
  {Ishizaka}}]{shimojima2014lifting}%
  \BibitemOpen
  \bibfield  {author} {\bibinfo {author} {\bibnamefont {Shimojima},
  \bibfnamefont {T.}}, \bibinfo {author} {\bibfnamefont {Y.}~\bibnamefont
  {Suzuki}}, \bibinfo {author} {\bibfnamefont {T.}~\bibnamefont {Sonobe}},
  \bibinfo {author} {\bibfnamefont {A.}~\bibnamefont {Nakamura}}, \bibinfo
  {author} {\bibfnamefont {M.}~\bibnamefont {Sakano}}, \bibinfo {author}
  {\bibfnamefont {J.}~\bibnamefont {Omachi}}, \bibinfo {author} {\bibfnamefont
  {K.}~\bibnamefont {Yoshioka}}, \bibinfo {author} {\bibfnamefont
  {M.}~\bibnamefont {Kuwata-Gonokami}}, \bibinfo {author} {\bibfnamefont
  {K.}~\bibnamefont {Ono}}, \bibinfo {author} {\bibfnamefont {H.}~\bibnamefont
  {Kumigashira}}, \bibinfo {author} {\bibfnamefont {A.~E.}\ \bibnamefont
  {B\"ohmer}}, \bibinfo {author} {\bibfnamefont {F.}~\bibnamefont {Hardy}},
  \bibinfo {author} {\bibfnamefont {T.}~\bibnamefont {Wolf}}, \bibinfo {author}
  {\bibfnamefont {C.}~\bibnamefont {Meingast}}, \bibinfo {author}
  {\bibfnamefont {H.~v.}\ \bibnamefont {L\"ohneysen}}, \bibinfo {author}
  {\bibfnamefont {H.}~\bibnamefont {Ikeda}}, and\ \bibinfo {author}
  {\bibfnamefont {K.}~\bibnamefont {Ishizaka}}} (\bibinfo {year} {2014}),\
  \href {https://doi.org/10.1103/PhysRevB.90.121111} {\bibfield  {journal}
  {\bibinfo  {journal} {Phys. Rev. B}\ }\textbf {\bibinfo {volume} {90}},\
  \bibinfo {pages} {121111(R)}}\BibitemShut {NoStop}%
\bibitem [{\citenamefont {Shiogai}\ \emph {et~al.}(2016)\citenamefont
  {Shiogai}, \citenamefont {Ito}, \citenamefont {Mitsuhashi}, \citenamefont
  {Nojima},\ and\ \citenamefont {Tsukazaki}}]{shiogai2016electricfieldinduced}%
  \BibitemOpen
  \bibfield  {author} {\bibinfo {author} {\bibnamefont {Shiogai}, \bibfnamefont
  {J.}}, \bibinfo {author} {\bibfnamefont {Y.}~\bibnamefont {Ito}}, \bibinfo
  {author} {\bibfnamefont {T.}~\bibnamefont {Mitsuhashi}}, \bibinfo {author}
  {\bibfnamefont {T.}~\bibnamefont {Nojima}}, and\ \bibinfo {author}
  {\bibfnamefont {A.}~\bibnamefont {Tsukazaki}}} (\bibinfo {year} {2016}),\
  \href {https://doi.org/10.1038/nphys3530} {\bibfield  {journal} {\bibinfo
  {journal} {Nat. Phys.}\ }\textbf {\bibinfo {volume} {12}},\ \bibinfo {pages}
  {42}}\BibitemShut {NoStop}%
\bibitem [{\citenamefont {Shipulin}\ \emph {et~al.}(2023)\citenamefont
  {Shipulin}, \citenamefont {Stegani}, \citenamefont {Maccari}, \citenamefont
  {Kihou}, \citenamefont {Lee}, \citenamefont {Hu}, \citenamefont {Zheng},
  \citenamefont {Yang}, \citenamefont {Li}, \citenamefont {Yim}, \citenamefont
  {H{\"u}hne}, \citenamefont {Klauss}, \citenamefont {Putti}, \citenamefont
  {Caglieris}, \citenamefont {Babaev},\ and\ \citenamefont
  {Grinenko}}]{shipulin2023calorimetric}%
  \BibitemOpen
  \bibfield  {author} {\bibinfo {author} {\bibnamefont {Shipulin},
  \bibfnamefont {I.}}, \bibinfo {author} {\bibfnamefont {N.}~\bibnamefont
  {Stegani}}, \bibinfo {author} {\bibfnamefont {I.}~\bibnamefont {Maccari}},
  \bibinfo {author} {\bibfnamefont {K.}~\bibnamefont {Kihou}}, \bibinfo
  {author} {\bibfnamefont {C.-H.}\ \bibnamefont {Lee}}, \bibinfo {author}
  {\bibfnamefont {Q.}~\bibnamefont {Hu}}, \bibinfo {author} {\bibfnamefont
  {Y.}~\bibnamefont {Zheng}}, \bibinfo {author} {\bibfnamefont
  {F.}~\bibnamefont {Yang}}, \bibinfo {author} {\bibfnamefont {Y.}~\bibnamefont
  {Li}}, \bibinfo {author} {\bibfnamefont {C.-M.}\ \bibnamefont {Yim}},
  \bibinfo {author} {\bibfnamefont {R.}~\bibnamefont {H{\"u}hne}}, \bibinfo
  {author} {\bibfnamefont {H.-H.}\ \bibnamefont {Klauss}}, \bibinfo {author}
  {\bibfnamefont {M.}~\bibnamefont {Putti}}, \bibinfo {author} {\bibfnamefont
  {F.}~\bibnamefont {Caglieris}}, \bibinfo {author} {\bibfnamefont
  {E.}~\bibnamefont {Babaev}}, and\ \bibinfo {author} {\bibfnamefont
  {V.}~\bibnamefont {Grinenko}}} (\bibinfo {year} {2023}),\ \href
  {https://doi.org/10.1038/s41467-023-42459-0} {\bibfield  {journal} {\bibinfo
  {journal} {Nat. Commun.}\ }\textbf {\bibinfo {volume} {14}},\ \bibinfo
  {pages} {6734}}\BibitemShut {NoStop}%
\bibitem [{\citenamefont {Shirage}\ \emph {et~al.}(2010)\citenamefont
  {Shirage}, \citenamefont {Kihou}, \citenamefont {Lee}, \citenamefont {Kito},
  \citenamefont {Eisaki},\ and\ \citenamefont {Iyo}}]{Shirage2010}%
  \BibitemOpen
  \bibfield  {author} {\bibinfo {author} {\bibnamefont {Shirage}, \bibfnamefont
  {P.~M.}}, \bibinfo {author} {\bibfnamefont {K.}~\bibnamefont {Kihou}},
  \bibinfo {author} {\bibfnamefont {C.-H.}\ \bibnamefont {Lee}}, \bibinfo
  {author} {\bibfnamefont {H.}~\bibnamefont {Kito}}, \bibinfo {author}
  {\bibfnamefont {H.}~\bibnamefont {Eisaki}}, and\ \bibinfo {author}
  {\bibfnamefont {A.}~\bibnamefont {Iyo}}} (\bibinfo {year} {2010}),\ \href
  {https://doi.org/10.1063/1.3508957} {\bibfield  {journal} {\bibinfo
  {journal} {Appl. Phys. Lett.}\ }\textbf {\bibinfo {volume} {97}},\ \bibinfo
  {pages} {172506}}\BibitemShut {NoStop}%
\bibitem [{\citenamefont {Shirage}\ \emph {et~al.}(2011)\citenamefont
  {Shirage}, \citenamefont {Kihou}, \citenamefont {Lee}, \citenamefont {Kito},
  \citenamefont {Eisaki},\ and\ \citenamefont {Iyo}}]{Shirage2011}%
  \BibitemOpen
  \bibfield  {author} {\bibinfo {author} {\bibnamefont {Shirage}, \bibfnamefont
  {P.~M.}}, \bibinfo {author} {\bibfnamefont {K.}~\bibnamefont {Kihou}},
  \bibinfo {author} {\bibfnamefont {C.-H.}\ \bibnamefont {Lee}}, \bibinfo
  {author} {\bibfnamefont {H.}~\bibnamefont {Kito}}, \bibinfo {author}
  {\bibfnamefont {H.}~\bibnamefont {Eisaki}}, and\ \bibinfo {author}
  {\bibfnamefont {A.}~\bibnamefont {Iyo}}} (\bibinfo {year} {2011}),\ \href
  {https://doi.org/10.1021/ja110729m} {\bibfield  {journal} {\bibinfo
  {journal} {J. Am. Chem. Soc.}\ }\textbf {\bibinfo {volume} {133}},\ \bibinfo
  {pages} {9630}}\BibitemShut {NoStop}%
\bibitem [{\citenamefont {Si}\ and\ \citenamefont
  {Abrahams}(2008)}]{si2008strong}%
  \BibitemOpen
  \bibfield  {author} {\bibinfo {author} {\bibnamefont {Si}, \bibfnamefont
  {Q.}}, and\ \bibinfo {author} {\bibfnamefont {E.}~\bibnamefont {Abrahams}}}
  (\bibinfo {year} {2008}),\ \href
  {https://doi.org/10.1103/PhysRevLett.101.076401} {\bibfield  {journal}
  {\bibinfo  {journal} {Phys. Rev. Lett.}\ }\textbf {\bibinfo {volume} {101}},\
  \bibinfo {pages} {076401}}\BibitemShut {NoStop}%
\bibitem [{\citenamefont {Si}\ and\ \citenamefont {Hussey}(2023)}]{si2023iron}%
  \BibitemOpen
  \bibfield  {author} {\bibinfo {author} {\bibnamefont {Si}, \bibfnamefont
  {Q.}}, and\ \bibinfo {author} {\bibfnamefont {N.~E.}\ \bibnamefont {Hussey}}}
  (\bibinfo {year} {2023}),\ \href {https://doi.org/10.1063/pt.3.5235}
  {\bibfield  {journal} {\bibinfo  {journal} {Physics Today}\ }\textbf
  {\bibinfo {volume} {76}},\ \bibinfo {pages} {34}}\BibitemShut {NoStop}%
\bibitem [{\citenamefont {Si}\ \emph {et~al.}(2016)\citenamefont {Si},
  \citenamefont {Yu},\ and\ \citenamefont {Abrahams}}]{si2016high}%
  \BibitemOpen
  \bibfield  {author} {\bibinfo {author} {\bibnamefont {Si}, \bibfnamefont
  {Q.}}, \bibinfo {author} {\bibfnamefont {R.}~\bibnamefont {Yu}}, and\
  \bibinfo {author} {\bibfnamefont {E.}~\bibnamefont {Abrahams}}} (\bibinfo
  {year} {2016}),\ \href {https://doi.org/10.1038/natrevmats.2016.17}
  {\bibfield  {journal} {\bibinfo  {journal} {Nature Reviews Materials}\
  }\textbf {\bibinfo {volume} {1}},\ \bibinfo {pages} {16017}}\BibitemShut
  {NoStop}%
\bibitem [{\citenamefont {Si}\ \emph {et~al.}(2010)\citenamefont {Si},
  \citenamefont {Jie}, \citenamefont {Wu}, \citenamefont {Zhou}, \citenamefont
  {Gu}, \citenamefont {Johnson},\ and\ \citenamefont {Li}}]{Si2010}%
  \BibitemOpen
  \bibfield  {author} {\bibinfo {author} {\bibnamefont {Si}, \bibfnamefont
  {W.}}, \bibinfo {author} {\bibfnamefont {Q.}~\bibnamefont {Jie}}, \bibinfo
  {author} {\bibfnamefont {L.}~\bibnamefont {Wu}}, \bibinfo {author}
  {\bibfnamefont {J.}~\bibnamefont {Zhou}}, \bibinfo {author} {\bibfnamefont
  {G.}~\bibnamefont {Gu}}, \bibinfo {author} {\bibfnamefont {P.~D.}\
  \bibnamefont {Johnson}}, and\ \bibinfo {author} {\bibfnamefont
  {Q.}~\bibnamefont {Li}}} (\bibinfo {year} {2010}),\ \href
  {https://doi.org/10.1103/PhysRevB.81.092506} {\bibfield  {journal} {\bibinfo
  {journal} {Phys. Rev. B}\ }\textbf {\bibinfo {volume} {81}},\ \bibinfo
  {pages} {092506}}\BibitemShut {NoStop}%
\bibitem [{\citenamefont {Simayi}\ \emph {et~al.}(2013)\citenamefont {Simayi},
  \citenamefont {Sakano}, \citenamefont {Takezawa}, \citenamefont {Nakamura},
  \citenamefont {Nakanishi}, \citenamefont {Kihou}, \citenamefont {Nakajima},
  \citenamefont {Lee}, \citenamefont {Iyo}, \citenamefont {Eisaki},
  \citenamefont {Uchida},\ and\ \citenamefont {Yoshizawa}}]{simayi2013strange}%
  \BibitemOpen
  \bibfield  {author} {\bibinfo {author} {\bibnamefont {Simayi}, \bibfnamefont
  {S.}}, \bibinfo {author} {\bibfnamefont {K.}~\bibnamefont {Sakano}}, \bibinfo
  {author} {\bibfnamefont {H.}~\bibnamefont {Takezawa}}, \bibinfo {author}
  {\bibfnamefont {M.}~\bibnamefont {Nakamura}}, \bibinfo {author}
  {\bibfnamefont {Y.}~\bibnamefont {Nakanishi}}, \bibinfo {author}
  {\bibfnamefont {K.}~\bibnamefont {Kihou}}, \bibinfo {author} {\bibfnamefont
  {M.}~\bibnamefont {Nakajima}}, \bibinfo {author} {\bibfnamefont {C.-H.}\
  \bibnamefont {Lee}}, \bibinfo {author} {\bibfnamefont {A.}~\bibnamefont
  {Iyo}}, \bibinfo {author} {\bibfnamefont {H.}~\bibnamefont {Eisaki}},
  \bibinfo {author} {\bibfnamefont {S.-i.}\ \bibnamefont {Uchida}}, and\
  \bibinfo {author} {\bibfnamefont {M.}~\bibnamefont {Yoshizawa}}} (\bibinfo
  {year} {2013}),\ \href {https://doi.org/10.7566/JPSJ.82.114604} {\bibfield
  {journal} {\bibinfo  {journal} {J. Phys. Soc. Jpn.}\ }\textbf {\bibinfo
  {volume} {82}},\ \bibinfo {pages} {114604}}\BibitemShut {NoStop}%
\bibitem [{\citenamefont {Simonin}(1986)}]{simonin1986surface}%
  \BibitemOpen
  \bibfield  {author} {\bibinfo {author} {\bibnamefont {Simonin}, \bibfnamefont
  {J.}}} (\bibinfo {year} {1986}),\ \href
  {https://doi.org/10.1103/PhysRevB.33.7830} {\bibfield  {journal} {\bibinfo
  {journal} {Phys. Rev. B}\ }\textbf {\bibinfo {volume} {33}},\ \bibinfo
  {pages} {7830}}\BibitemShut {NoStop}%
\bibitem [{\citenamefont {Singh}(2014)}]{singh2014superconductivity}%
  \BibitemOpen
  \bibfield  {author} {\bibinfo {author} {\bibnamefont {Singh}, \bibfnamefont
  {D.~J.}}} (\bibinfo {year} {2014}),\ \href
  {https://doi.org/10.1103/PhysRevB.89.024505} {\bibfield  {journal} {\bibinfo
  {journal} {Phys. Rev. B}\ }\textbf {\bibinfo {volume} {89}},\ \bibinfo
  {pages} {024505}}\BibitemShut {NoStop}%
\bibitem [{\citenamefont {Singh}\ \emph {et~al.}(2018)\citenamefont {Singh},
  \citenamefont {Bristow}, \citenamefont {Meier}, \citenamefont {Taylor},
  \citenamefont {Blundell}, \citenamefont {Canfield},\ and\ \citenamefont
  {Coldea}}]{Singh2018}%
  \BibitemOpen
  \bibfield  {author} {\bibinfo {author} {\bibnamefont {Singh}, \bibfnamefont
  {S.~J.}}, \bibinfo {author} {\bibfnamefont {M.}~\bibnamefont {Bristow}},
  \bibinfo {author} {\bibfnamefont {W.~R.}\ \bibnamefont {Meier}}, \bibinfo
  {author} {\bibfnamefont {P.}~\bibnamefont {Taylor}}, \bibinfo {author}
  {\bibfnamefont {S.~J.}\ \bibnamefont {Blundell}}, \bibinfo {author}
  {\bibfnamefont {P.~C.}\ \bibnamefont {Canfield}}, and\ \bibinfo {author}
  {\bibfnamefont {A.~I.}\ \bibnamefont {Coldea}}} (\bibinfo {year} {2018}),\
  \href {https://doi.org/10.1103/PhysRevMaterials.2.074802} {\bibfield
  {journal} {\bibinfo  {journal} {Phys. Rev. Mater.}\ }\textbf {\bibinfo
  {volume} {2}},\ \bibinfo {pages} {074802}}\BibitemShut {NoStop}%
\bibitem [{\citenamefont {Smylie}\ \emph {et~al.}(2019)\citenamefont {Smylie},
  \citenamefont {Koshelev}, \citenamefont {Willa}, \citenamefont {Willa},
  \citenamefont {Kwok}, \citenamefont {Bao}, \citenamefont {Chung},
  \citenamefont {Kanatzidis}, \citenamefont {Singleton}, \citenamefont
  {Balakirev} \emph {et~al.}}]{Smylie2019}%
  \BibitemOpen
  \bibfield  {author} {\bibinfo {author} {\bibnamefont {Smylie}, \bibfnamefont
  {M.}}, \bibinfo {author} {\bibfnamefont {A.}~\bibnamefont {Koshelev}},
  \bibinfo {author} {\bibfnamefont {K.}~\bibnamefont {Willa}}, \bibinfo
  {author} {\bibfnamefont {R.}~\bibnamefont {Willa}}, \bibinfo {author}
  {\bibfnamefont {W.-K.}\ \bibnamefont {Kwok}}, \bibinfo {author}
  {\bibfnamefont {J.-K.}\ \bibnamefont {Bao}}, \bibinfo {author} {\bibfnamefont
  {D.}~\bibnamefont {Chung}}, \bibinfo {author} {\bibfnamefont
  {M.}~\bibnamefont {Kanatzidis}}, \bibinfo {author} {\bibfnamefont
  {J.}~\bibnamefont {Singleton}}, \bibinfo {author} {\bibfnamefont {F.~F.}\
  \bibnamefont {Balakirev}},  \emph {et~al.}} (\bibinfo {year} {2019}),\ \href
  {https://doi.org/10.1103/PhysRevB.100.054507} {\bibfield  {journal} {\bibinfo
   {journal} {Phys. Rev. B}\ }\textbf {\bibinfo {volume} {100}},\ \bibinfo
  {pages} {054507}}\BibitemShut {NoStop}%
\bibitem [{\citenamefont {Smylie}\ \emph {et~al.}(2018)\citenamefont {Smylie},
  \citenamefont {Willa}, \citenamefont {Bao}, \citenamefont {Ryan},
  \citenamefont {Islam}, \citenamefont {Claus}, \citenamefont {Simsek},
  \citenamefont {Diao}, \citenamefont {Rydh}, \citenamefont {Koshelev} \emph
  {et~al.}}]{Smylie2018}%
  \BibitemOpen
  \bibfield  {author} {\bibinfo {author} {\bibnamefont {Smylie}, \bibfnamefont
  {M.}}, \bibinfo {author} {\bibfnamefont {K.}~\bibnamefont {Willa}}, \bibinfo
  {author} {\bibfnamefont {J.-K.}\ \bibnamefont {Bao}}, \bibinfo {author}
  {\bibfnamefont {K.}~\bibnamefont {Ryan}}, \bibinfo {author} {\bibfnamefont
  {Z.}~\bibnamefont {Islam}}, \bibinfo {author} {\bibfnamefont
  {H.}~\bibnamefont {Claus}}, \bibinfo {author} {\bibfnamefont
  {Y.}~\bibnamefont {Simsek}}, \bibinfo {author} {\bibfnamefont
  {Z.}~\bibnamefont {Diao}}, \bibinfo {author} {\bibfnamefont {A.}~\bibnamefont
  {Rydh}}, \bibinfo {author} {\bibfnamefont {A.}~\bibnamefont {Koshelev}},
  \emph {et~al.}} (\bibinfo {year} {2018}),\ \href
  {https://doi.org/10.1103/PhysRevB.98.104503} {\bibfield  {journal} {\bibinfo
  {journal} {Phys. Rev. B}\ }\textbf {\bibinfo {volume} {98}},\ \bibinfo
  {pages} {104503}}\BibitemShut {NoStop}%
\bibitem [{\citenamefont {Sobota}\ \emph {et~al.}(2021)\citenamefont {Sobota},
  \citenamefont {He},\ and\ \citenamefont {Shen}}]{sobota2021angle}%
  \BibitemOpen
  \bibfield  {author} {\bibinfo {author} {\bibnamefont {Sobota}, \bibfnamefont
  {J.~A.}}, \bibinfo {author} {\bibfnamefont {Y.}~\bibnamefont {He}}, and\
  \bibinfo {author} {\bibfnamefont {Z.-X.}\ \bibnamefont {Shen}}} (\bibinfo
  {year} {2021}),\ \href {https://doi.org/10.1103/RevModPhys.93.025006}
  {\bibfield  {journal} {\bibinfo  {journal} {Rev. Mod. Phys.}\ }\textbf
  {\bibinfo {volume} {93}},\ \bibinfo {pages} {025006}}\BibitemShut {NoStop}%
\bibitem [{\citenamefont {Song}\ \emph
  {et~al.}(2018{\natexlab{a}})\citenamefont {Song}, \citenamefont {Nguyen},
  \citenamefont {Wang}, \citenamefont {Canfield},\ and\ \citenamefont
  {Ho}}]{Song2018b}%
  \BibitemOpen
  \bibfield  {author} {\bibinfo {author} {\bibnamefont {Song}, \bibfnamefont
  {B.}}, \bibinfo {author} {\bibfnamefont {M.~C.}\ \bibnamefont {Nguyen}},
  \bibinfo {author} {\bibfnamefont {C.-Z.}\ \bibnamefont {Wang}}, \bibinfo
  {author} {\bibfnamefont {P.}~\bibnamefont {Canfield}}, and\ \bibinfo {author}
  {\bibfnamefont {K.-M.}\ \bibnamefont {Ho}}} (\bibinfo {year}
  {2018}{\natexlab{a}}),\ \href
  {https://doi.org/10.1103/PhysRevMaterials.2.104802} {\bibfield  {journal}
  {\bibinfo  {journal} {Phys. Rev. Mater.}\ }\textbf {\bibinfo {volume} {2}},\
  \bibinfo {pages} {104802}}\BibitemShut {NoStop}%
\bibitem [{\citenamefont {Song}\ \emph
  {et~al.}(2018{\natexlab{b}})\citenamefont {Song}, \citenamefont {Nguyen},
  \citenamefont {Wang},\ and\ \citenamefont {Ho}}]{Song2018a}%
  \BibitemOpen
  \bibfield  {author} {\bibinfo {author} {\bibnamefont {Song}, \bibfnamefont
  {B.}}, \bibinfo {author} {\bibfnamefont {M.~C.}\ \bibnamefont {Nguyen}},
  \bibinfo {author} {\bibfnamefont {C.-Z.}\ \bibnamefont {Wang}}, and\ \bibinfo
  {author} {\bibfnamefont {K.-M.}\ \bibnamefont {Ho}}} (\bibinfo {year}
  {2018}{\natexlab{b}}),\ \href {https://doi.org/10.1103/PhysRevB.97.094105}
  {\bibfield  {journal} {\bibinfo  {journal} {Phys. Rev. B}\ }\textbf {\bibinfo
  {volume} {97}},\ \bibinfo {pages} {094105}}\BibitemShut {NoStop}%
\bibitem [{\citenamefont {Song}\ \emph
  {et~al.}(2011{\natexlab{a}})\citenamefont {Song}, \citenamefont {Wang},
  \citenamefont {Cheng}, \citenamefont {Jiang}, \citenamefont {Li},
  \citenamefont {Zhang}, \citenamefont {Li}, \citenamefont {He}, \citenamefont
  {Wang}, \citenamefont {Jia} \emph {et~al.}}]{Song2011a}%
  \BibitemOpen
  \bibfield  {author} {\bibinfo {author} {\bibnamefont {Song}, \bibfnamefont
  {C.-L.}}, \bibinfo {author} {\bibfnamefont {Y.-L.}\ \bibnamefont {Wang}},
  \bibinfo {author} {\bibfnamefont {P.}~\bibnamefont {Cheng}}, \bibinfo
  {author} {\bibfnamefont {Y.-P.}\ \bibnamefont {Jiang}}, \bibinfo {author}
  {\bibfnamefont {W.}~\bibnamefont {Li}}, \bibinfo {author} {\bibfnamefont
  {T.}~\bibnamefont {Zhang}}, \bibinfo {author} {\bibfnamefont
  {Z.}~\bibnamefont {Li}}, \bibinfo {author} {\bibfnamefont {K.}~\bibnamefont
  {He}}, \bibinfo {author} {\bibfnamefont {L.}~\bibnamefont {Wang}}, \bibinfo
  {author} {\bibfnamefont {J.-F.}\ \bibnamefont {Jia}},  \emph {et~al.}}
  (\bibinfo {year} {2011}{\natexlab{a}}),\ \href
  {https://doi.org/10.1126/science.1202226} {\bibfield  {journal} {\bibinfo
  {journal} {Science}\ }\textbf {\bibinfo {volume} {332}},\ \bibinfo {pages}
  {1410}}\BibitemShut {NoStop}%
\bibitem [{\citenamefont {Song}\ \emph
  {et~al.}(2011{\natexlab{b}})\citenamefont {Song}, \citenamefont {Wang},
  \citenamefont {Jiang}, \citenamefont {Li}, \citenamefont {Wang},
  \citenamefont {He}, \citenamefont {Chen}, \citenamefont {Ma},\ and\
  \citenamefont {Xue}}]{Song2011b}%
  \BibitemOpen
  \bibfield  {author} {\bibinfo {author} {\bibnamefont {Song}, \bibfnamefont
  {C.-L.}}, \bibinfo {author} {\bibfnamefont {Y.-L.}\ \bibnamefont {Wang}},
  \bibinfo {author} {\bibfnamefont {Y.-P.}\ \bibnamefont {Jiang}}, \bibinfo
  {author} {\bibfnamefont {Z.}~\bibnamefont {Li}}, \bibinfo {author}
  {\bibfnamefont {L.}~\bibnamefont {Wang}}, \bibinfo {author} {\bibfnamefont
  {K.}~\bibnamefont {He}}, \bibinfo {author} {\bibfnamefont {X.}~\bibnamefont
  {Chen}}, \bibinfo {author} {\bibfnamefont {X.-C.}\ \bibnamefont {Ma}}, and\
  \bibinfo {author} {\bibfnamefont {Q.-K.}\ \bibnamefont {Xue}}} (\bibinfo
  {year} {2011}{\natexlab{b}}),\ \href
  {https://doi.org/10.1103/PhysRevB.84.020503} {\bibfield  {journal} {\bibinfo
  {journal} {Phys. Rev. B}\ }\textbf {\bibinfo {volume} {84}},\ \bibinfo
  {pages} {020503}}\BibitemShut {NoStop}%
\bibitem [{\citenamefont {Song}\ \emph
  {et~al.}(2019{\natexlab{a}})\citenamefont {Song}, \citenamefont {Yu},
  \citenamefont {Lou}, \citenamefont {Xie}, \citenamefont {Xu}, \citenamefont
  {Wen}, \citenamefont {Yao}, \citenamefont {Zhang}, \citenamefont {Zhu},
  \citenamefont {Guo}, \citenamefont {Peng},\ and\ \citenamefont
  {Feng}}]{song2019evidence}%
  \BibitemOpen
  \bibfield  {author} {\bibinfo {author} {\bibnamefont {Song}, \bibfnamefont
  {Q.}}, \bibinfo {author} {\bibfnamefont {T.~L.}\ \bibnamefont {Yu}}, \bibinfo
  {author} {\bibfnamefont {X.}~\bibnamefont {Lou}}, \bibinfo {author}
  {\bibfnamefont {B.~P.}\ \bibnamefont {Xie}}, \bibinfo {author} {\bibfnamefont
  {H.~C.}\ \bibnamefont {Xu}}, \bibinfo {author} {\bibfnamefont {C.~H.~P.}\
  \bibnamefont {Wen}}, \bibinfo {author} {\bibfnamefont {Q.}~\bibnamefont
  {Yao}}, \bibinfo {author} {\bibfnamefont {S.~Y.}\ \bibnamefont {Zhang}},
  \bibinfo {author} {\bibfnamefont {X.~T.}\ \bibnamefont {Zhu}}, \bibinfo
  {author} {\bibfnamefont {J.~D.}\ \bibnamefont {Guo}}, \bibinfo {author}
  {\bibfnamefont {R.}~\bibnamefont {Peng}}, and\ \bibinfo {author}
  {\bibfnamefont {D.~L.}\ \bibnamefont {Feng}}} (\bibinfo {year}
  {2019}{\natexlab{a}}),\ \href {https://doi.org/10.1038/s41467-019-08560-z}
  {\bibfield  {journal} {\bibinfo  {journal} {Nat. Commun.}\ }\textbf {\bibinfo
  {volume} {10}},\ \bibinfo {pages} {758}}\BibitemShut {NoStop}%
\bibitem [{\citenamefont {Song}\ \emph
  {et~al.}(2019{\natexlab{b}})\citenamefont {Song}, \citenamefont {Cao},
  \citenamefont {Chakoumakos}, \citenamefont {Zhao}, \citenamefont {Wang},
  \citenamefont {Lei}, \citenamefont {Petrovic},\ and\ \citenamefont
  {Birgeneau}}]{song2019intertwined}%
  \BibitemOpen
  \bibfield  {author} {\bibinfo {author} {\bibnamefont {Song}, \bibfnamefont
  {Y.}}, \bibinfo {author} {\bibfnamefont {H.}~\bibnamefont {Cao}}, \bibinfo
  {author} {\bibfnamefont {B.~C.}\ \bibnamefont {Chakoumakos}}, \bibinfo
  {author} {\bibfnamefont {Y.}~\bibnamefont {Zhao}}, \bibinfo {author}
  {\bibfnamefont {A.}~\bibnamefont {Wang}}, \bibinfo {author} {\bibfnamefont
  {H.}~\bibnamefont {Lei}}, \bibinfo {author} {\bibfnamefont {C.}~\bibnamefont
  {Petrovic}}, and\ \bibinfo {author} {\bibfnamefont {R.~J.}\ \bibnamefont
  {Birgeneau}}} (\bibinfo {year} {2019}{\natexlab{b}}),\ \href
  {https://doi.org/10.1103/PhysRevLett.122.087201} {\bibfield  {journal}
  {\bibinfo  {journal} {Phys. Rev. Lett.}\ }\textbf {\bibinfo {volume} {122}},\
  \bibinfo {pages} {087201}}\BibitemShut {NoStop}%
\bibitem [{\citenamefont {Song}\ \emph {et~al.}(2015)\citenamefont {Song},
  \citenamefont {Lu}, \citenamefont {Abernathy}, \citenamefont {Tam},
  \citenamefont {Niedziela}, \citenamefont {Tian}, \citenamefont {Luo},
  \citenamefont {Si},\ and\ \citenamefont {Dai}}]{song2015energy}%
  \BibitemOpen
  \bibfield  {author} {\bibinfo {author} {\bibnamefont {Song}, \bibfnamefont
  {Y.}}, \bibinfo {author} {\bibfnamefont {X.}~\bibnamefont {Lu}}, \bibinfo
  {author} {\bibfnamefont {D.~L.}\ \bibnamefont {Abernathy}}, \bibinfo {author}
  {\bibfnamefont {D.~W.}\ \bibnamefont {Tam}}, \bibinfo {author} {\bibfnamefont
  {J.~L.}\ \bibnamefont {Niedziela}}, \bibinfo {author} {\bibfnamefont
  {W.}~\bibnamefont {Tian}}, \bibinfo {author} {\bibfnamefont {H.}~\bibnamefont
  {Luo}}, \bibinfo {author} {\bibfnamefont {Q.}~\bibnamefont {Si}}, and\
  \bibinfo {author} {\bibfnamefont {P.}~\bibnamefont {Dai}}} (\bibinfo {year}
  {2015}),\ \href {https://doi.org/10.1103/PhysRevB.92.180504} {\bibfield
  {journal} {\bibinfo  {journal} {Phys. Rev. B}\ }\textbf {\bibinfo {volume}
  {92}},\ \bibinfo {pages} {180504}}\BibitemShut {NoStop}%
\bibitem [{\citenamefont {Song}\ \emph
  {et~al.}(2018{\natexlab{c}})\citenamefont {Song}, \citenamefont {Lu},
  \citenamefont {Regnault}, \citenamefont {Su}, \citenamefont {Lai},
  \citenamefont {Hu}, \citenamefont {Si},\ and\ \citenamefont
  {Dai}}]{song2018spinisotropic}%
  \BibitemOpen
  \bibfield  {author} {\bibinfo {author} {\bibnamefont {Song}, \bibfnamefont
  {Y.}}, \bibinfo {author} {\bibfnamefont {X.}~\bibnamefont {Lu}}, \bibinfo
  {author} {\bibfnamefont {L.-P.}\ \bibnamefont {Regnault}}, \bibinfo {author}
  {\bibfnamefont {Y.}~\bibnamefont {Su}}, \bibinfo {author} {\bibfnamefont
  {H.-H.}\ \bibnamefont {Lai}}, \bibinfo {author} {\bibfnamefont {W.-J.}\
  \bibnamefont {Hu}}, \bibinfo {author} {\bibfnamefont {Q.}~\bibnamefont {Si}},
  and\ \bibinfo {author} {\bibfnamefont {P.}~\bibnamefont {Dai}}} (\bibinfo
  {year} {2018}{\natexlab{c}}),\ \href
  {https://doi.org/10.1103/PhysRevB.97.024519} {\bibfield  {journal} {\bibinfo
  {journal} {Phys. Rev. B}\ }\textbf {\bibinfo {volume} {97}},\ \bibinfo
  {pages} {024519}}\BibitemShut {NoStop}%
\bibitem [{\citenamefont {Song}\ \emph
  {et~al.}(2016{\natexlab{a}})\citenamefont {Song}, \citenamefont {Man},
  \citenamefont {Zhang}, \citenamefont {Lu}, \citenamefont {Zhang},
  \citenamefont {Wang}, \citenamefont {Tan}, \citenamefont {Regnault},
  \citenamefont {Su}, \citenamefont {Kang}, \citenamefont {Fernandes},\ and\
  \citenamefont {Dai}}]{song2016spin}%
  \BibitemOpen
  \bibfield  {author} {\bibinfo {author} {\bibnamefont {Song}, \bibfnamefont
  {Y.}}, \bibinfo {author} {\bibfnamefont {H.}~\bibnamefont {Man}}, \bibinfo
  {author} {\bibfnamefont {R.}~\bibnamefont {Zhang}}, \bibinfo {author}
  {\bibfnamefont {X.}~\bibnamefont {Lu}}, \bibinfo {author} {\bibfnamefont
  {C.}~\bibnamefont {Zhang}}, \bibinfo {author} {\bibfnamefont
  {M.}~\bibnamefont {Wang}}, \bibinfo {author} {\bibfnamefont {G.}~\bibnamefont
  {Tan}}, \bibinfo {author} {\bibfnamefont {L.-P.}\ \bibnamefont {Regnault}},
  \bibinfo {author} {\bibfnamefont {Y.}~\bibnamefont {Su}}, \bibinfo {author}
  {\bibfnamefont {J.}~\bibnamefont {Kang}}, \bibinfo {author} {\bibfnamefont
  {R.~M.}\ \bibnamefont {Fernandes}}, and\ \bibinfo {author} {\bibfnamefont
  {P.}~\bibnamefont {Dai}}} (\bibinfo {year} {2016}{\natexlab{a}}),\ \href
  {https://doi.org/10.1103/PhysRevB.94.214516} {\bibfield  {journal} {\bibinfo
  {journal} {Phys. Rev. B}\ }\textbf {\bibinfo {volume} {94}},\ \bibinfo
  {pages} {214516}}\BibitemShut {NoStop}%
\bibitem [{\citenamefont {Song}\ \emph
  {et~al.}(2018{\natexlab{d}})\citenamefont {Song}, \citenamefont {Tan},
  \citenamefont {Zhang}, \citenamefont {{Toft-Petersen}}, \citenamefont {Yu},\
  and\ \citenamefont {Dai}}]{song2018unusual}%
  \BibitemOpen
  \bibfield  {author} {\bibinfo {author} {\bibnamefont {Song}, \bibfnamefont
  {Y.}}, \bibinfo {author} {\bibfnamefont {G.}~\bibnamefont {Tan}}, \bibinfo
  {author} {\bibfnamefont {C.}~\bibnamefont {Zhang}}, \bibinfo {author}
  {\bibfnamefont {R.}~\bibnamefont {{Toft-Petersen}}}, \bibinfo {author}
  {\bibfnamefont {R.}~\bibnamefont {Yu}}, and\ \bibinfo {author} {\bibfnamefont
  {P.}~\bibnamefont {Dai}}} (\bibinfo {year} {2018}{\natexlab{d}}),\ \href
  {https://doi.org/10.1103/PhysRevB.98.064507} {\bibfield  {journal} {\bibinfo
  {journal} {Phys. Rev. B}\ }\textbf {\bibinfo {volume} {98}},\ \bibinfo
  {pages} {064507}}\BibitemShut {NoStop}%
\bibitem [{\citenamefont {Song}\ \emph {et~al.}(2021)\citenamefont {Song},
  \citenamefont {Wang}, \citenamefont {Paris}, \citenamefont {Lu},
  \citenamefont {Pelliciari}, \citenamefont {Tseng}, \citenamefont {Huang},
  \citenamefont {McNally}, \citenamefont {Dantz}, \citenamefont {Cao},
  \citenamefont {Yu}, \citenamefont {Birgeneau}, \citenamefont {Schmitt},\ and\
  \citenamefont {Dai}}]{song2021spin}%
  \BibitemOpen
  \bibfield  {author} {\bibinfo {author} {\bibnamefont {Song}, \bibfnamefont
  {Y.}}, \bibinfo {author} {\bibfnamefont {W.}~\bibnamefont {Wang}}, \bibinfo
  {author} {\bibfnamefont {E.}~\bibnamefont {Paris}}, \bibinfo {author}
  {\bibfnamefont {X.}~\bibnamefont {Lu}}, \bibinfo {author} {\bibfnamefont
  {J.}~\bibnamefont {Pelliciari}}, \bibinfo {author} {\bibfnamefont
  {Y.}~\bibnamefont {Tseng}}, \bibinfo {author} {\bibfnamefont
  {Y.}~\bibnamefont {Huang}}, \bibinfo {author} {\bibfnamefont
  {D.}~\bibnamefont {McNally}}, \bibinfo {author} {\bibfnamefont
  {M.}~\bibnamefont {Dantz}}, \bibinfo {author} {\bibfnamefont
  {C.}~\bibnamefont {Cao}}, \bibinfo {author} {\bibfnamefont {R.}~\bibnamefont
  {Yu}}, \bibinfo {author} {\bibfnamefont {R.~J.}\ \bibnamefont {Birgeneau}},
  \bibinfo {author} {\bibfnamefont {T.}~\bibnamefont {Schmitt}}, and\ \bibinfo
  {author} {\bibfnamefont {P.}~\bibnamefont {Dai}}} (\bibinfo {year} {2021}),\
  \href {https://doi.org/10.1103/PhysRevB.103.075112} {\bibfield  {journal}
  {\bibinfo  {journal} {Phys. Rev. B}\ }\textbf {\bibinfo {volume} {103}},\
  \bibinfo {pages} {075112}}\BibitemShut {NoStop}%
\bibitem [{\citenamefont {Song}\ \emph {et~al.}(2017)\citenamefont {Song},
  \citenamefont {Wang}, \citenamefont {Zhang}, \citenamefont {Gu},
  \citenamefont {Lu}, \citenamefont {Tan}, \citenamefont {Su}, \citenamefont
  {Bourdarot}, \citenamefont {Christianson}, \citenamefont {Li},\ and\
  \citenamefont {Dai}}]{song2017temperature}%
  \BibitemOpen
  \bibfield  {author} {\bibinfo {author} {\bibnamefont {Song}, \bibfnamefont
  {Y.}}, \bibinfo {author} {\bibfnamefont {W.}~\bibnamefont {Wang}}, \bibinfo
  {author} {\bibfnamefont {C.}~\bibnamefont {Zhang}}, \bibinfo {author}
  {\bibfnamefont {Y.}~\bibnamefont {Gu}}, \bibinfo {author} {\bibfnamefont
  {X.}~\bibnamefont {Lu}}, \bibinfo {author} {\bibfnamefont {G.}~\bibnamefont
  {Tan}}, \bibinfo {author} {\bibfnamefont {Y.}~\bibnamefont {Su}}, \bibinfo
  {author} {\bibfnamefont {F.}~\bibnamefont {Bourdarot}}, \bibinfo {author}
  {\bibfnamefont {A.~D.}\ \bibnamefont {Christianson}}, \bibinfo {author}
  {\bibfnamefont {S.}~\bibnamefont {Li}}, and\ \bibinfo {author} {\bibfnamefont
  {P.}~\bibnamefont {Dai}}} (\bibinfo {year} {2017}),\ \href
  {https://doi.org/10.1103/PhysRevB.96.184512} {\bibfield  {journal} {\bibinfo
  {journal} {Phys. Rev. B}\ }\textbf {\bibinfo {volume} {96}},\ \bibinfo
  {pages} {184512}}\BibitemShut {NoStop}%
\bibitem [{\citenamefont {Song}\ \emph
  {et~al.}(2016{\natexlab{b}})\citenamefont {Song}, \citenamefont {Yamani},
  \citenamefont {Cao}, \citenamefont {Li}, \citenamefont {Zhang}, \citenamefont
  {Chen}, \citenamefont {Huang}, \citenamefont {Wu}, \citenamefont {Tao},
  \citenamefont {Zhu}, \citenamefont {Tian}, \citenamefont {Chi}, \citenamefont
  {Cao}, \citenamefont {Huang}, \citenamefont {Dantz}, \citenamefont {Schmitt},
  \citenamefont {Yu}, \citenamefont {Nevidomskyy}, \citenamefont {Morosan},
  \citenamefont {Si},\ and\ \citenamefont {Dai}}]{song2016mott}%
  \BibitemOpen
  \bibfield  {author} {\bibinfo {author} {\bibnamefont {Song}, \bibfnamefont
  {Y.}}, \bibinfo {author} {\bibfnamefont {Z.}~\bibnamefont {Yamani}}, \bibinfo
  {author} {\bibfnamefont {C.}~\bibnamefont {Cao}}, \bibinfo {author}
  {\bibfnamefont {Y.}~\bibnamefont {Li}}, \bibinfo {author} {\bibfnamefont
  {C.}~\bibnamefont {Zhang}}, \bibinfo {author} {\bibfnamefont {J.~S.}\
  \bibnamefont {Chen}}, \bibinfo {author} {\bibfnamefont {Q.}~\bibnamefont
  {Huang}}, \bibinfo {author} {\bibfnamefont {H.}~\bibnamefont {Wu}}, \bibinfo
  {author} {\bibfnamefont {J.}~\bibnamefont {Tao}}, \bibinfo {author}
  {\bibfnamefont {Y.}~\bibnamefont {Zhu}}, \bibinfo {author} {\bibfnamefont
  {W.}~\bibnamefont {Tian}}, \bibinfo {author} {\bibfnamefont {S.}~\bibnamefont
  {Chi}}, \bibinfo {author} {\bibfnamefont {H.}~\bibnamefont {Cao}}, \bibinfo
  {author} {\bibfnamefont {Y.-B.}\ \bibnamefont {Huang}}, \bibinfo {author}
  {\bibfnamefont {M.}~\bibnamefont {Dantz}}, \bibinfo {author} {\bibfnamefont
  {T.}~\bibnamefont {Schmitt}}, \bibinfo {author} {\bibfnamefont
  {R.}~\bibnamefont {Yu}}, \bibinfo {author} {\bibfnamefont {A.~H.}\
  \bibnamefont {Nevidomskyy}}, \bibinfo {author} {\bibfnamefont
  {E.}~\bibnamefont {Morosan}}, \bibinfo {author} {\bibfnamefont
  {Q.}~\bibnamefont {Si}}, and\ \bibinfo {author} {\bibfnamefont
  {P.}~\bibnamefont {Dai}}} (\bibinfo {year} {2016}{\natexlab{b}}),\ \href
  {https://doi.org/10.1038/ncomms13879} {\bibfield  {journal} {\bibinfo
  {journal} {Nat. Commun.}\ }\textbf {\bibinfo {volume} {7}},\ \bibinfo {pages}
  {13879}}\BibitemShut {NoStop}%
\bibitem [{\citenamefont {Sprau}\ \emph {et~al.}(2017)\citenamefont {Sprau},
  \citenamefont {Kostin}, \citenamefont {Kreisel}, \citenamefont {B{\"o}hmer},
  \citenamefont {Taufour}, \citenamefont {Canfield}, \citenamefont {Mukherjee},
  \citenamefont {Hirschfeld}, \citenamefont {Andersen},\ and\ \citenamefont
  {Davis}}]{sprau2017discovery}%
  \BibitemOpen
  \bibfield  {author} {\bibinfo {author} {\bibnamefont {Sprau}, \bibfnamefont
  {P.~O.}}, \bibinfo {author} {\bibfnamefont {A.}~\bibnamefont {Kostin}},
  \bibinfo {author} {\bibfnamefont {A.}~\bibnamefont {Kreisel}}, \bibinfo
  {author} {\bibfnamefont {A.~E.}\ \bibnamefont {B{\"o}hmer}}, \bibinfo
  {author} {\bibfnamefont {V.}~\bibnamefont {Taufour}}, \bibinfo {author}
  {\bibfnamefont {P.~C.}\ \bibnamefont {Canfield}}, \bibinfo {author}
  {\bibfnamefont {S.}~\bibnamefont {Mukherjee}}, \bibinfo {author}
  {\bibfnamefont {P.~J.}\ \bibnamefont {Hirschfeld}}, \bibinfo {author}
  {\bibfnamefont {B.~M.}\ \bibnamefont {Andersen}}, and\ \bibinfo {author}
  {\bibfnamefont {J.~C.~S.}\ \bibnamefont {Davis}}} (\bibinfo {year} {2017}),\
  \href {https://doi.org/10.1126/science.aal1575} {\bibfield  {journal}
  {\bibinfo  {journal} {Science}\ }\textbf {\bibinfo {volume} {357}},\ \bibinfo
  {pages} {75}}\BibitemShut {NoStop}%
\bibitem [{\citenamefont {Stadel}\ \emph {et~al.}(2022)\citenamefont {Stadel},
  \citenamefont {Khalyavin}, \citenamefont {Manuel}, \citenamefont {Yokoyama},
  \citenamefont {Lapidus}, \citenamefont {Christensen}, \citenamefont
  {Fernandes}, \citenamefont {Phelan}, \citenamefont {Chung}, \citenamefont
  {Osborn}, \citenamefont {Rosenkranz},\ and\ \citenamefont
  {Chmaissem}}]{stadel2022multiple}%
  \BibitemOpen
  \bibfield  {author} {\bibinfo {author} {\bibnamefont {Stadel}, \bibfnamefont
  {R.}}, \bibinfo {author} {\bibfnamefont {D.~D.}\ \bibnamefont {Khalyavin}},
  \bibinfo {author} {\bibfnamefont {P.}~\bibnamefont {Manuel}}, \bibinfo
  {author} {\bibfnamefont {K.}~\bibnamefont {Yokoyama}}, \bibinfo {author}
  {\bibfnamefont {S.}~\bibnamefont {Lapidus}}, \bibinfo {author} {\bibfnamefont
  {M.~H.}\ \bibnamefont {Christensen}}, \bibinfo {author} {\bibfnamefont
  {R.~M.}\ \bibnamefont {Fernandes}}, \bibinfo {author} {\bibfnamefont
  {D.}~\bibnamefont {Phelan}}, \bibinfo {author} {\bibfnamefont {D.~Y.}\
  \bibnamefont {Chung}}, \bibinfo {author} {\bibfnamefont {R.}~\bibnamefont
  {Osborn}}, \bibinfo {author} {\bibfnamefont {S.}~\bibnamefont {Rosenkranz}},
  and\ \bibinfo {author} {\bibfnamefont {O.}~\bibnamefont {Chmaissem}}}
  (\bibinfo {year} {2022}),\ \href {https://doi.org/10.1038/s42005-022-00911-5}
  {\bibfield  {journal} {\bibinfo  {journal} {Commun. Phys.}\ }\textbf
  {\bibinfo {volume} {5}},\ \bibinfo {pages} {146}}\BibitemShut {NoStop}%
\bibitem [{\citenamefont {Stahl}\ and\ \citenamefont
  {Johrendt}(2017)}]{Stahl2017}%
  \BibitemOpen
  \bibfield  {author} {\bibinfo {author} {\bibnamefont {Stahl}, \bibfnamefont
  {J.}}, and\ \bibinfo {author} {\bibfnamefont {D.}~\bibnamefont {Johrendt}}}
  (\bibinfo {year} {2017}),\ \href {https://doi.org/10.48550/arXiv.1706.00314}
  {\bibfield  {journal} {\bibinfo  {journal} {arXiv preprint arXiv:1706.00314}\
  }10.48550/arXiv.1706.00314}\BibitemShut {NoStop}%
\bibitem [{\citenamefont {Steffens}\ \emph {et~al.}(2013)\citenamefont
  {Steffens}, \citenamefont {Lee}, \citenamefont {Qureshi}, \citenamefont
  {Kihou}, \citenamefont {Iyo}, \citenamefont {Eisaki},\ and\ \citenamefont
  {Braden}}]{steffens2013splitting}%
  \BibitemOpen
  \bibfield  {author} {\bibinfo {author} {\bibnamefont {Steffens},
  \bibfnamefont {P.}}, \bibinfo {author} {\bibfnamefont {C.~H.}\ \bibnamefont
  {Lee}}, \bibinfo {author} {\bibfnamefont {N.}~\bibnamefont {Qureshi}},
  \bibinfo {author} {\bibfnamefont {K.}~\bibnamefont {Kihou}}, \bibinfo
  {author} {\bibfnamefont {A.}~\bibnamefont {Iyo}}, \bibinfo {author}
  {\bibfnamefont {H.}~\bibnamefont {Eisaki}}, and\ \bibinfo {author}
  {\bibfnamefont {M.}~\bibnamefont {Braden}}} (\bibinfo {year} {2013}),\ \href
  {https://doi.org/10.1103/PhysRevLett.110.137001} {\bibfield  {journal}
  {\bibinfo  {journal} {Phys. Rev. Lett.}\ }\textbf {\bibinfo {volume} {110}},\
  \bibinfo {pages} {137001}}\BibitemShut {NoStop}%
\bibitem [{\citenamefont {Stewart}(2011)}]{stewart2011superconductivity}%
  \BibitemOpen
  \bibfield  {author} {\bibinfo {author} {\bibnamefont {Stewart}, \bibfnamefont
  {G.~R.}}} (\bibinfo {year} {2011}),\ \href
  {https://doi.org/10.1103/RevModPhys.83.1589} {\bibfield  {journal} {\bibinfo
  {journal} {Rev. Mod. Phys.}\ }\textbf {\bibinfo {volume} {83}},\ \bibinfo
  {pages} {1589}}\BibitemShut {NoStop}%
\bibitem [{\citenamefont {Stolyarov}\ \emph {et~al.}(2018)\citenamefont
  {Stolyarov}, \citenamefont {Casano}, \citenamefont {Belyanchikov},
  \citenamefont {Astrakhantseva}, \citenamefont {Grebenchuk}, \citenamefont
  {Baranov}, \citenamefont {Golovchanskiy}, \citenamefont {Voloshenko},
  \citenamefont {Zhukova}, \citenamefont {Gorshunov} \emph
  {et~al.}}]{Stolyarov2018}%
  \BibitemOpen
  \bibfield  {author} {\bibinfo {author} {\bibnamefont {Stolyarov},
  \bibfnamefont {V.}}, \bibinfo {author} {\bibfnamefont {A.}~\bibnamefont
  {Casano}}, \bibinfo {author} {\bibfnamefont {M.}~\bibnamefont
  {Belyanchikov}}, \bibinfo {author} {\bibfnamefont {A.}~\bibnamefont
  {Astrakhantseva}}, \bibinfo {author} {\bibfnamefont {S.~Y.}\ \bibnamefont
  {Grebenchuk}}, \bibinfo {author} {\bibfnamefont {D.}~\bibnamefont {Baranov}},
  \bibinfo {author} {\bibfnamefont {I.}~\bibnamefont {Golovchanskiy}}, \bibinfo
  {author} {\bibfnamefont {I.}~\bibnamefont {Voloshenko}}, \bibinfo {author}
  {\bibfnamefont {E.}~\bibnamefont {Zhukova}}, \bibinfo {author} {\bibfnamefont
  {B.}~\bibnamefont {Gorshunov}},  \emph {et~al.}} (\bibinfo {year} {2018}),\
  \href {https://doi.org/10.1103/PhysRevB.98.140506} {\bibfield  {journal}
  {\bibinfo  {journal} {Phys. Rev. B}\ }\textbf {\bibinfo {volume} {98}},\
  \bibinfo {pages} {140506}}\BibitemShut {NoStop}%
\bibitem [{\citenamefont {Stolyarov}\ \emph {et~al.}(2020)\citenamefont
  {Stolyarov}, \citenamefont {Pervakov}, \citenamefont {Astrakhantseva},
  \citenamefont {Golovchanskiy}, \citenamefont {Vyalikh}, \citenamefont {Kim},
  \citenamefont {Eremeev}, \citenamefont {Vlasenko}, \citenamefont {Pudalov},
  \citenamefont {Golubov} \emph {et~al.}}]{Stolyarov2020}%
  \BibitemOpen
  \bibfield  {author} {\bibinfo {author} {\bibnamefont {Stolyarov},
  \bibfnamefont {V.~S.}}, \bibinfo {author} {\bibfnamefont {K.~S.}\
  \bibnamefont {Pervakov}}, \bibinfo {author} {\bibfnamefont {A.~S.}\
  \bibnamefont {Astrakhantseva}}, \bibinfo {author} {\bibfnamefont {I.~A.}\
  \bibnamefont {Golovchanskiy}}, \bibinfo {author} {\bibfnamefont {D.~V.}\
  \bibnamefont {Vyalikh}}, \bibinfo {author} {\bibfnamefont {T.~K.}\
  \bibnamefont {Kim}}, \bibinfo {author} {\bibfnamefont {S.~V.}\ \bibnamefont
  {Eremeev}}, \bibinfo {author} {\bibfnamefont {V.~A.}\ \bibnamefont
  {Vlasenko}}, \bibinfo {author} {\bibfnamefont {V.~M.}\ \bibnamefont
  {Pudalov}}, \bibinfo {author} {\bibfnamefont {A.~A.}\ \bibnamefont
  {Golubov}},  \emph {et~al.}} (\bibinfo {year} {2020}),\ \href
  {https://doi.org/10.1021/acs.jpclett.0c02711} {\bibfield  {journal} {\bibinfo
   {journal} {The journal of physical chemistry letters}\ }\textbf {\bibinfo
  {volume} {11}},\ \bibinfo {pages} {9393}}\BibitemShut {NoStop}%
\bibitem [{\citenamefont {Subedi}(2014)}]{subedi2014unconventional}%
  \BibitemOpen
  \bibfield  {author} {\bibinfo {author} {\bibnamefont {Subedi}, \bibfnamefont
  {A.}}} (\bibinfo {year} {2014}),\ \href
  {https://doi.org/10.1103/PhysRevB.89.024504} {\bibfield  {journal} {\bibinfo
  {journal} {Phys. Rev. B}\ }\textbf {\bibinfo {volume} {89}},\ \bibinfo
  {pages} {024504}}\BibitemShut {NoStop}%
\bibitem [{\citenamefont {Sun}\ \emph {et~al.}(2023)\citenamefont {Sun},
  \citenamefont {Huo}, \citenamefont {Hu}, \citenamefont {Li}, \citenamefont
  {Liu}, \citenamefont {Han}, \citenamefont {Tang}, \citenamefont {Mao},
  \citenamefont {Yang}, \citenamefont {Wang}, \citenamefont {Cheng},
  \citenamefont {Yao}, \citenamefont {Zhang},\ and\ \citenamefont
  {Wang}}]{sun2023signatures}%
  \BibitemOpen
  \bibfield  {author} {\bibinfo {author} {\bibnamefont {Sun}, \bibfnamefont
  {H.}}, \bibinfo {author} {\bibfnamefont {M.}~\bibnamefont {Huo}}, \bibinfo
  {author} {\bibfnamefont {X.}~\bibnamefont {Hu}}, \bibinfo {author}
  {\bibfnamefont {J.}~\bibnamefont {Li}}, \bibinfo {author} {\bibfnamefont
  {Z.}~\bibnamefont {Liu}}, \bibinfo {author} {\bibfnamefont {Y.}~\bibnamefont
  {Han}}, \bibinfo {author} {\bibfnamefont {L.}~\bibnamefont {Tang}}, \bibinfo
  {author} {\bibfnamefont {Z.}~\bibnamefont {Mao}}, \bibinfo {author}
  {\bibfnamefont {P.}~\bibnamefont {Yang}}, \bibinfo {author} {\bibfnamefont
  {B.}~\bibnamefont {Wang}}, \bibinfo {author} {\bibfnamefont {J.}~\bibnamefont
  {Cheng}}, \bibinfo {author} {\bibfnamefont {D.-X.}\ \bibnamefont {Yao}},
  \bibinfo {author} {\bibfnamefont {G.-M.}\ \bibnamefont {Zhang}}, and\
  \bibinfo {author} {\bibfnamefont {M.}~\bibnamefont {Wang}}} (\bibinfo {year}
  {2023}),\ \href {https://doi.org/10.1038/s41586-023-06408-7} {\bibfield
  {journal} {\bibinfo  {journal} {Nature}\ }\textbf {\bibinfo {volume} {621}},\
  \bibinfo {pages} {493}}\BibitemShut {NoStop}%
\bibitem [{\citenamefont {Sun}\ \emph {et~al.}(2017{\natexlab{a}})\citenamefont
  {Sun}, \citenamefont {Wang},\ and\ \citenamefont {Cheng}}]{sun2017recent}%
  \BibitemOpen
  \bibfield  {author} {\bibinfo {author} {\bibnamefont {Sun}, \bibfnamefont
  {J.}}, \bibinfo {author} {\bibfnamefont {B.}~\bibnamefont {Wang}}, and\
  \bibinfo {author} {\bibfnamefont {J.}~\bibnamefont {Cheng}}} (\bibinfo {year}
  {2017}{\natexlab{a}}),\ \href {https://doi.org/10.1360/n972017-00713}
  {\bibfield  {journal} {\bibinfo  {journal} {Chinese Science Bulletin}\
  }\textbf {\bibinfo {volume} {62}},\ \bibinfo {pages} {3925}}\BibitemShut
  {NoStop}%
\bibitem [{\citenamefont {Sun}\ \emph {et~al.}(2016)\citenamefont {Sun},
  \citenamefont {Matsuura}, \citenamefont {Ye}, \citenamefont {Mizukami},
  \citenamefont {Shimozawa}, \citenamefont {Matsubayashi}, \citenamefont
  {Yamashita}, \citenamefont {Watashige}, \citenamefont {Kasahara},
  \citenamefont {Matsuda}, \citenamefont {Yan}, \citenamefont {Sales},
  \citenamefont {Uwatoko}, \citenamefont {Cheng},\ and\ \citenamefont
  {Shibauchi}}]{sun2016dome}%
  \BibitemOpen
  \bibfield  {author} {\bibinfo {author} {\bibnamefont {Sun}, \bibfnamefont
  {J.~P.}}, \bibinfo {author} {\bibfnamefont {K.}~\bibnamefont {Matsuura}},
  \bibinfo {author} {\bibfnamefont {G.~Z.}\ \bibnamefont {Ye}}, \bibinfo
  {author} {\bibfnamefont {Y.}~\bibnamefont {Mizukami}}, \bibinfo {author}
  {\bibfnamefont {M.}~\bibnamefont {Shimozawa}}, \bibinfo {author}
  {\bibfnamefont {K.}~\bibnamefont {Matsubayashi}}, \bibinfo {author}
  {\bibfnamefont {M.}~\bibnamefont {Yamashita}}, \bibinfo {author}
  {\bibfnamefont {T.}~\bibnamefont {Watashige}}, \bibinfo {author}
  {\bibfnamefont {S.}~\bibnamefont {Kasahara}}, \bibinfo {author}
  {\bibfnamefont {Y.}~\bibnamefont {Matsuda}}, \bibinfo {author} {\bibfnamefont
  {J.-Q.}\ \bibnamefont {Yan}}, \bibinfo {author} {\bibfnamefont {B.~C.}\
  \bibnamefont {Sales}}, \bibinfo {author} {\bibfnamefont {Y.}~\bibnamefont
  {Uwatoko}}, \bibinfo {author} {\bibfnamefont {J.-G.}\ \bibnamefont {Cheng}},
  and\ \bibinfo {author} {\bibfnamefont {T.}~\bibnamefont {Shibauchi}}}
  (\bibinfo {year} {2016}),\ \href {https://doi.org/10.1038/ncomms12146}
  {\bibfield  {journal} {\bibinfo  {journal} {Nat. Commun.}\ }\textbf {\bibinfo
  {volume} {7}},\ \bibinfo {pages} {12146}}\BibitemShut {NoStop}%
\bibitem [{\citenamefont {Sun}\ \emph {et~al.}(2018)\citenamefont {Sun},
  \citenamefont {Shahi}, \citenamefont {Zhou}, \citenamefont {Huang},
  \citenamefont {Chen}, \citenamefont {Wang}, \citenamefont {Ni}, \citenamefont
  {Li}, \citenamefont {Zhang}, \citenamefont {Yang}, \citenamefont {Uwatoko},
  \citenamefont {Xing}, \citenamefont {Sun}, \citenamefont {Singh},
  \citenamefont {Jin}, \citenamefont {Zhou}, \citenamefont {Zhang},
  \citenamefont {Dong}, \citenamefont {Zhao},\ and\ \citenamefont
  {Cheng}}]{sun2018reemergence}%
  \BibitemOpen
  \bibfield  {author} {\bibinfo {author} {\bibnamefont {Sun}, \bibfnamefont
  {J.~P.}}, \bibinfo {author} {\bibfnamefont {P.}~\bibnamefont {Shahi}},
  \bibinfo {author} {\bibfnamefont {H.~X.}\ \bibnamefont {Zhou}}, \bibinfo
  {author} {\bibfnamefont {Y.~L.}\ \bibnamefont {Huang}}, \bibinfo {author}
  {\bibfnamefont {K.~Y.}\ \bibnamefont {Chen}}, \bibinfo {author}
  {\bibfnamefont {B.~S.}\ \bibnamefont {Wang}}, \bibinfo {author}
  {\bibfnamefont {S.~L.}\ \bibnamefont {Ni}}, \bibinfo {author} {\bibfnamefont
  {N.~N.}\ \bibnamefont {Li}}, \bibinfo {author} {\bibfnamefont
  {K.}~\bibnamefont {Zhang}}, \bibinfo {author} {\bibfnamefont {W.~G.}\
  \bibnamefont {Yang}}, \bibinfo {author} {\bibfnamefont {Y.}~\bibnamefont
  {Uwatoko}}, \bibinfo {author} {\bibfnamefont {G.}~\bibnamefont {Xing}},
  \bibinfo {author} {\bibfnamefont {J.}~\bibnamefont {Sun}}, \bibinfo {author}
  {\bibfnamefont {D.~J.}\ \bibnamefont {Singh}}, \bibinfo {author}
  {\bibfnamefont {K.}~\bibnamefont {Jin}}, \bibinfo {author} {\bibfnamefont
  {F.}~\bibnamefont {Zhou}}, \bibinfo {author} {\bibfnamefont {G.~M.}\
  \bibnamefont {Zhang}}, \bibinfo {author} {\bibfnamefont {X.~L.}\ \bibnamefont
  {Dong}}, \bibinfo {author} {\bibfnamefont {Z.~X.}\ \bibnamefont {Zhao}}, and\
  \bibinfo {author} {\bibfnamefont {J.-G.}\ \bibnamefont {Cheng}}} (\bibinfo
  {year} {2018}),\ \href {https://doi.org/10.1038/s41467-018-02843-7}
  {\bibfield  {journal} {\bibinfo  {journal} {Nat. Commun.}\ }\textbf {\bibinfo
  {volume} {9}},\ \bibinfo {pages} {380}}\BibitemShut {NoStop}%
\bibitem [{\citenamefont {Sun}\ \emph {et~al.}(2012)\citenamefont {Sun},
  \citenamefont {Chen}, \citenamefont {Guo}, \citenamefont {Gao}, \citenamefont
  {Huang}, \citenamefont {Wang}, \citenamefont {Fang}, \citenamefont {Chen},
  \citenamefont {Chen}, \citenamefont {Wu} \emph {et~al.}}]{Sun2012}%
  \BibitemOpen
  \bibfield  {author} {\bibinfo {author} {\bibnamefont {Sun}, \bibfnamefont
  {L.}}, \bibinfo {author} {\bibfnamefont {X.-J.}\ \bibnamefont {Chen}},
  \bibinfo {author} {\bibfnamefont {J.}~\bibnamefont {Guo}}, \bibinfo {author}
  {\bibfnamefont {P.}~\bibnamefont {Gao}}, \bibinfo {author} {\bibfnamefont
  {Q.-Z.}\ \bibnamefont {Huang}}, \bibinfo {author} {\bibfnamefont
  {H.}~\bibnamefont {Wang}}, \bibinfo {author} {\bibfnamefont {M.}~\bibnamefont
  {Fang}}, \bibinfo {author} {\bibfnamefont {X.}~\bibnamefont {Chen}}, \bibinfo
  {author} {\bibfnamefont {G.}~\bibnamefont {Chen}}, \bibinfo {author}
  {\bibfnamefont {Q.}~\bibnamefont {Wu}},  \emph {et~al.}} (\bibinfo {year}
  {2012}),\ \href {https://doi.org/10.1038/nature10813} {\bibfield  {journal}
  {\bibinfo  {journal} {Nature}\ }\textbf {\bibinfo {volume} {483}},\ \bibinfo
  {pages} {67}}\BibitemShut {NoStop}%
\bibitem [{\citenamefont {Sun}\ \emph {et~al.}(2019)\citenamefont {Sun},
  \citenamefont {Jin}, \citenamefont {Gu}, \citenamefont {Zhang}, \citenamefont
  {Huang}, \citenamefont {Ying}, \citenamefont {Peng}, \citenamefont {Deng},
  \citenamefont {Yin},\ and\ \citenamefont {Chen}}]{Sun2019}%
  \BibitemOpen
  \bibfield  {author} {\bibinfo {author} {\bibnamefont {Sun}, \bibfnamefont
  {R.}}, \bibinfo {author} {\bibfnamefont {S.}~\bibnamefont {Jin}}, \bibinfo
  {author} {\bibfnamefont {L.}~\bibnamefont {Gu}}, \bibinfo {author}
  {\bibfnamefont {Q.}~\bibnamefont {Zhang}}, \bibinfo {author} {\bibfnamefont
  {Q.}~\bibnamefont {Huang}}, \bibinfo {author} {\bibfnamefont
  {T.}~\bibnamefont {Ying}}, \bibinfo {author} {\bibfnamefont {Y.}~\bibnamefont
  {Peng}}, \bibinfo {author} {\bibfnamefont {J.}~\bibnamefont {Deng}}, \bibinfo
  {author} {\bibfnamefont {Z.}~\bibnamefont {Yin}}, and\ \bibinfo {author}
  {\bibfnamefont {X.}~\bibnamefont {Chen}}} (\bibinfo {year} {2019}),\ \href
  {https://doi.org/10.1021/jacs.9b05899} {\bibfield  {journal} {\bibinfo
  {journal} {J. Am. Chem. Soc.}\ }\textbf {\bibinfo {volume} {141}},\ \bibinfo
  {pages} {13849}}\BibitemShut {NoStop}%
\bibitem [{\citenamefont {Sun}\ \emph {et~al.}(2017{\natexlab{b}})\citenamefont
  {Sun}, \citenamefont {Wang}, \citenamefont {Yu},\ and\ \citenamefont
  {Lei}}]{sun2017extreme}%
  \BibitemOpen
  \bibfield  {author} {\bibinfo {author} {\bibnamefont {Sun}, \bibfnamefont
  {S.}}, \bibinfo {author} {\bibfnamefont {S.}~\bibnamefont {Wang}}, \bibinfo
  {author} {\bibfnamefont {R.}~\bibnamefont {Yu}}, and\ \bibinfo {author}
  {\bibfnamefont {H.}~\bibnamefont {Lei}}} (\bibinfo {year}
  {2017}{\natexlab{b}}),\ \href {https://doi.org/10.1103/PhysRevB.96.064512}
  {\bibfield  {journal} {\bibinfo  {journal} {Phys. Rev. B}\ }\textbf {\bibinfo
  {volume} {96}},\ \bibinfo {pages} {064512}}\BibitemShut {NoStop}%
\bibitem [{\citenamefont {Sun}\ \emph {et~al.}(2014)\citenamefont {Sun},
  \citenamefont {Zhang}, \citenamefont {Xing}, \citenamefont {Li},
  \citenamefont {Zhao}, \citenamefont {Xia}, \citenamefont {Wang},
  \citenamefont {Ma}, \citenamefont {Xue},\ and\ \citenamefont
  {Wang}}]{Sun2014HighTemperature}%
  \BibitemOpen
  \bibfield  {author} {\bibinfo {author} {\bibnamefont {Sun}, \bibfnamefont
  {Y.}}, \bibinfo {author} {\bibfnamefont {W.}~\bibnamefont {Zhang}}, \bibinfo
  {author} {\bibfnamefont {Y.}~\bibnamefont {Xing}}, \bibinfo {author}
  {\bibfnamefont {F.}~\bibnamefont {Li}}, \bibinfo {author} {\bibfnamefont
  {Y.}~\bibnamefont {Zhao}}, \bibinfo {author} {\bibfnamefont {Z.}~\bibnamefont
  {Xia}}, \bibinfo {author} {\bibfnamefont {L.}~\bibnamefont {Wang}}, \bibinfo
  {author} {\bibfnamefont {X.}~\bibnamefont {Ma}}, \bibinfo {author}
  {\bibfnamefont {Q.-K.}\ \bibnamefont {Xue}}, and\ \bibinfo {author}
  {\bibfnamefont {J.}~\bibnamefont {Wang}}} (\bibinfo {year} {2014}),\ \href
  {https://doi.org/10.1038/srep06040} {\bibfield  {journal} {\bibinfo
  {journal} {Sci. Rep.}\ }\textbf {\bibinfo {volume} {4}},\ \bibinfo {pages}
  {6040}}\BibitemShut {NoStop}%
\bibitem [{\citenamefont {Suzuki}\ \emph {et~al.}(2014)\citenamefont {Suzuki},
  \citenamefont {Usui}, \citenamefont {Iimura}, \citenamefont {Sato},
  \citenamefont {Matsuishi}, \citenamefont {Hosono},\ and\ \citenamefont
  {Kuroki}}]{Suzuki2014}%
  \BibitemOpen
  \bibfield  {author} {\bibinfo {author} {\bibnamefont {Suzuki}, \bibfnamefont
  {K.}}, \bibinfo {author} {\bibfnamefont {H.}~\bibnamefont {Usui}}, \bibinfo
  {author} {\bibfnamefont {S.}~\bibnamefont {Iimura}}, \bibinfo {author}
  {\bibfnamefont {Y.}~\bibnamefont {Sato}}, \bibinfo {author} {\bibfnamefont
  {S.}~\bibnamefont {Matsuishi}}, \bibinfo {author} {\bibfnamefont
  {H.}~\bibnamefont {Hosono}}, and\ \bibinfo {author} {\bibfnamefont
  {K.}~\bibnamefont {Kuroki}}} (\bibinfo {year} {2014}),\ \href
  {https://doi.org/10.1103/PhysRevLett.113.027002} {\bibfield  {journal}
  {\bibinfo  {journal} {Phys. Rev. Lett.}\ }\textbf {\bibinfo {volume} {113}},\
  \bibinfo {pages} {027002}}\BibitemShut {NoStop}%
\bibitem [{\citenamefont {Suzuki}\ \emph {et~al.}(2015)\citenamefont {Suzuki},
  \citenamefont {Shimojima}, \citenamefont {Sonobe}, \citenamefont {Nakamura},
  \citenamefont {Sakano}, \citenamefont {Tsuji}, \citenamefont {Omachi},
  \citenamefont {Yoshioka}, \citenamefont {Kuwata-Gonokami}, \citenamefont
  {Watashige}, \citenamefont {Kobayashi}, \citenamefont {Kasahara},
  \citenamefont {Shibauchi}, \citenamefont {Matsuda}, \citenamefont {Yamakawa},
  \citenamefont {Kontani},\ and\ \citenamefont
  {Ishizaka}}]{suzuki2015momentum}%
  \BibitemOpen
  \bibfield  {author} {\bibinfo {author} {\bibnamefont {Suzuki}, \bibfnamefont
  {Y.}}, \bibinfo {author} {\bibfnamefont {T.}~\bibnamefont {Shimojima}},
  \bibinfo {author} {\bibfnamefont {T.}~\bibnamefont {Sonobe}}, \bibinfo
  {author} {\bibfnamefont {A.}~\bibnamefont {Nakamura}}, \bibinfo {author}
  {\bibfnamefont {M.}~\bibnamefont {Sakano}}, \bibinfo {author} {\bibfnamefont
  {H.}~\bibnamefont {Tsuji}}, \bibinfo {author} {\bibfnamefont
  {J.}~\bibnamefont {Omachi}}, \bibinfo {author} {\bibfnamefont
  {K.}~\bibnamefont {Yoshioka}}, \bibinfo {author} {\bibfnamefont
  {M.}~\bibnamefont {Kuwata-Gonokami}}, \bibinfo {author} {\bibfnamefont
  {T.}~\bibnamefont {Watashige}}, \bibinfo {author} {\bibfnamefont
  {R.}~\bibnamefont {Kobayashi}}, \bibinfo {author} {\bibfnamefont
  {S.}~\bibnamefont {Kasahara}}, \bibinfo {author} {\bibfnamefont
  {T.}~\bibnamefont {Shibauchi}}, \bibinfo {author} {\bibfnamefont
  {Y.}~\bibnamefont {Matsuda}}, \bibinfo {author} {\bibfnamefont
  {Y.}~\bibnamefont {Yamakawa}}, \bibinfo {author} {\bibfnamefont
  {H.}~\bibnamefont {Kontani}}, and\ \bibinfo {author} {\bibfnamefont
  {K.}~\bibnamefont {Ishizaka}}} (\bibinfo {year} {2015}),\ \href
  {https://doi.org/10.1103/PhysRevB.92.205117} {\bibfield  {journal} {\bibinfo
  {journal} {Phys. Rev. B}\ }\textbf {\bibinfo {volume} {92}},\ \bibinfo
  {pages} {205117}}\BibitemShut {NoStop}%
\bibitem [{\citenamefont {Taddei}\ \emph {et~al.}(2016)\citenamefont {Taddei},
  \citenamefont {Allred}, \citenamefont {Bugaris}, \citenamefont {Lapidus},
  \citenamefont {Krogstad}, \citenamefont {Stadel}, \citenamefont {Claus},
  \citenamefont {Chung}, \citenamefont {Kanatzidis}, \citenamefont
  {Rosenkranz}, \citenamefont {Osborn},\ and\ \citenamefont
  {Chmaissem}}]{taddei2016detailed}%
  \BibitemOpen
  \bibfield  {author} {\bibinfo {author} {\bibnamefont {Taddei}, \bibfnamefont
  {K.~M.}}, \bibinfo {author} {\bibfnamefont {J.~M.}\ \bibnamefont {Allred}},
  \bibinfo {author} {\bibfnamefont {D.~E.}\ \bibnamefont {Bugaris}}, \bibinfo
  {author} {\bibfnamefont {S.}~\bibnamefont {Lapidus}}, \bibinfo {author}
  {\bibfnamefont {M.~J.}\ \bibnamefont {Krogstad}}, \bibinfo {author}
  {\bibfnamefont {R.}~\bibnamefont {Stadel}}, \bibinfo {author} {\bibfnamefont
  {H.}~\bibnamefont {Claus}}, \bibinfo {author} {\bibfnamefont {D.~Y.}\
  \bibnamefont {Chung}}, \bibinfo {author} {\bibfnamefont {M.~G.}\ \bibnamefont
  {Kanatzidis}}, \bibinfo {author} {\bibfnamefont {S.}~\bibnamefont
  {Rosenkranz}}, \bibinfo {author} {\bibfnamefont {R.}~\bibnamefont {Osborn}},
  and\ \bibinfo {author} {\bibfnamefont {O.}~\bibnamefont {Chmaissem}}}
  (\bibinfo {year} {2016}),\ \href {https://doi.org/10.1103/PhysRevB.93.134510}
  {\bibfield  {journal} {\bibinfo  {journal} {Phys. Rev. B}\ }\textbf {\bibinfo
  {volume} {93}},\ \bibinfo {pages} {134510}}\BibitemShut {NoStop}%
\bibitem [{\citenamefont {Taddei}\ \emph {et~al.}(2017)\citenamefont {Taddei},
  \citenamefont {Allred}, \citenamefont {Bugaris}, \citenamefont {Lapidus},
  \citenamefont {Krogstad}, \citenamefont {Claus}, \citenamefont {Chung},
  \citenamefont {Kanatzidis}, \citenamefont {Osborn}, \citenamefont
  {Rosenkranz},\ and\ \citenamefont {Chmaissem}}]{taddei2017observation}%
  \BibitemOpen
  \bibfield  {author} {\bibinfo {author} {\bibnamefont {Taddei}, \bibfnamefont
  {K.~M.}}, \bibinfo {author} {\bibfnamefont {J.~M.}\ \bibnamefont {Allred}},
  \bibinfo {author} {\bibfnamefont {D.~E.}\ \bibnamefont {Bugaris}}, \bibinfo
  {author} {\bibfnamefont {S.~H.}\ \bibnamefont {Lapidus}}, \bibinfo {author}
  {\bibfnamefont {M.~J.}\ \bibnamefont {Krogstad}}, \bibinfo {author}
  {\bibfnamefont {H.}~\bibnamefont {Claus}}, \bibinfo {author} {\bibfnamefont
  {D.~Y.}\ \bibnamefont {Chung}}, \bibinfo {author} {\bibfnamefont {M.~G.}\
  \bibnamefont {Kanatzidis}}, \bibinfo {author} {\bibfnamefont
  {R.}~\bibnamefont {Osborn}}, \bibinfo {author} {\bibfnamefont
  {S.}~\bibnamefont {Rosenkranz}}, and\ \bibinfo {author} {\bibfnamefont
  {O.}~\bibnamefont {Chmaissem}}} (\bibinfo {year} {2017}),\ \href
  {https://doi.org/10.1103/PhysRevB.95.064508} {\bibfield  {journal} {\bibinfo
  {journal} {Phys. Rev. B}\ }\textbf {\bibinfo {volume} {95}},\ \bibinfo
  {pages} {064508}}\BibitemShut {NoStop}%
\bibitem [{\citenamefont {Tafti}\ \emph {et~al.}(2013)\citenamefont {Tafti},
  \citenamefont {Juneau-Fecteau}, \citenamefont {Delage}, \citenamefont
  {René~de Cotret}, \citenamefont {Reid}, \citenamefont {Wang}, \citenamefont
  {Luo}, \citenamefont {Chen}, \citenamefont {Doiron-Leyraud},\ and\
  \citenamefont {Taillefer}}]{tafti2013sudden}%
  \BibitemOpen
  \bibfield  {author} {\bibinfo {author} {\bibnamefont {Tafti}, \bibfnamefont
  {F.~F.}}, \bibinfo {author} {\bibfnamefont {A.}~\bibnamefont
  {Juneau-Fecteau}}, \bibinfo {author} {\bibfnamefont {M.-E.}\ \bibnamefont
  {Delage}}, \bibinfo {author} {\bibfnamefont {S.}~\bibnamefont {René~de
  Cotret}}, \bibinfo {author} {\bibfnamefont {J.-P.}\ \bibnamefont {Reid}},
  \bibinfo {author} {\bibfnamefont {A.~F.}\ \bibnamefont {Wang}}, \bibinfo
  {author} {\bibfnamefont {X.-G.}\ \bibnamefont {Luo}}, \bibinfo {author}
  {\bibfnamefont {X.~H.}\ \bibnamefont {Chen}}, \bibinfo {author}
  {\bibfnamefont {N.}~\bibnamefont {Doiron-Leyraud}}, and\ \bibinfo {author}
  {\bibfnamefont {L.}~\bibnamefont {Taillefer}}} (\bibinfo {year} {2013}),\
  \href {https://doi.org/10.1038/nphys2617} {\bibfield  {journal} {\bibinfo
  {journal} {Nat. Phys.}\ }\textbf {\bibinfo {volume} {9}},\ \bibinfo {pages}
  {349}}\BibitemShut {NoStop}%
\bibitem [{\citenamefont {Takahashi}\ \emph {et~al.}(2015)\citenamefont
  {Takahashi}, \citenamefont {Soeda}, \citenamefont {Nukii}, \citenamefont
  {Kawashima}, \citenamefont {Nakanishi}, \citenamefont {Iimura}, \citenamefont
  {Muraba}, \citenamefont {Matsuishi},\ and\ \citenamefont
  {Hosono}}]{Takahashi2015}%
  \BibitemOpen
  \bibfield  {author} {\bibinfo {author} {\bibnamefont {Takahashi},
  \bibfnamefont {H.}}, \bibinfo {author} {\bibfnamefont {H.}~\bibnamefont
  {Soeda}}, \bibinfo {author} {\bibfnamefont {M.}~\bibnamefont {Nukii}},
  \bibinfo {author} {\bibfnamefont {C.}~\bibnamefont {Kawashima}}, \bibinfo
  {author} {\bibfnamefont {T.}~\bibnamefont {Nakanishi}}, \bibinfo {author}
  {\bibfnamefont {S.}~\bibnamefont {Iimura}}, \bibinfo {author} {\bibfnamefont
  {Y.}~\bibnamefont {Muraba}}, \bibinfo {author} {\bibfnamefont
  {S.}~\bibnamefont {Matsuishi}}, and\ \bibinfo {author} {\bibfnamefont
  {H.}~\bibnamefont {Hosono}}} (\bibinfo {year} {2015}),\ \href
  {https://doi.org/10.1038/srep07829} {\bibfield  {journal} {\bibinfo
  {journal} {Sci. Rep.}\ }\textbf {\bibinfo {volume} {5}},\ \bibinfo {pages}
  {7829}}\BibitemShut {NoStop}%
\bibitem [{\citenamefont {Takahashi}\ \emph {et~al.}(2014)\citenamefont
  {Takahashi}, \citenamefont {Mizushima},\ and\ \citenamefont
  {Machida}}]{takahashi2014multiband}%
  \BibitemOpen
  \bibfield  {author} {\bibinfo {author} {\bibnamefont {Takahashi},
  \bibfnamefont {M.}}, \bibinfo {author} {\bibfnamefont {T.}~\bibnamefont
  {Mizushima}}, and\ \bibinfo {author} {\bibfnamefont {K.}~\bibnamefont
  {Machida}}} (\bibinfo {year} {2014}),\ \href
  {https://doi.org/10.1103/PhysRevB.89.064505} {\bibfield  {journal} {\bibinfo
  {journal} {Phys. Rev. B}\ }\textbf {\bibinfo {volume} {89}},\ \bibinfo
  {pages} {064505}}\BibitemShut {NoStop}%
\bibitem [{\citenamefont {Takeuchi}\ \emph {et~al.}(2018)\citenamefont
  {Takeuchi}, \citenamefont {Yamakawa},\ and\ \citenamefont
  {Kontani}}]{Kontani2018}%
  \BibitemOpen
  \bibfield  {author} {\bibinfo {author} {\bibnamefont {Takeuchi},
  \bibfnamefont {L.}}, \bibinfo {author} {\bibfnamefont {Y.}~\bibnamefont
  {Yamakawa}}, and\ \bibinfo {author} {\bibfnamefont {H.}~\bibnamefont
  {Kontani}}} (\bibinfo {year} {2018}),\ \href
  {https://doi.org/10.1103/PhysRevB.98.165143} {\bibfield  {journal} {\bibinfo
  {journal} {Phys. Rev. B}\ }\textbf {\bibinfo {volume} {98}},\ \bibinfo
  {pages} {165143}}\BibitemShut {NoStop}%
\bibitem [{\citenamefont {Tam}\ \emph {et~al.}(2019{\natexlab{a}})\citenamefont
  {Tam}, \citenamefont {Lai}, \citenamefont {Hu}, \citenamefont {Lu},
  \citenamefont {Walker}, \citenamefont {Abernathy}, \citenamefont {Niedziela},
  \citenamefont {Weber}, \citenamefont {Enderle}, \citenamefont {Su},
  \citenamefont {Mao}, \citenamefont {Si},\ and\ \citenamefont
  {Dai}}]{tam2019plaquette}%
  \BibitemOpen
  \bibfield  {author} {\bibinfo {author} {\bibnamefont {Tam}, \bibfnamefont
  {D.~W.}}, \bibinfo {author} {\bibfnamefont {H.-H.}\ \bibnamefont {Lai}},
  \bibinfo {author} {\bibfnamefont {J.}~\bibnamefont {Hu}}, \bibinfo {author}
  {\bibfnamefont {X.}~\bibnamefont {Lu}}, \bibinfo {author} {\bibfnamefont
  {H.~C.}\ \bibnamefont {Walker}}, \bibinfo {author} {\bibfnamefont {D.~L.}\
  \bibnamefont {Abernathy}}, \bibinfo {author} {\bibfnamefont {J.~L.}\
  \bibnamefont {Niedziela}}, \bibinfo {author} {\bibfnamefont {T.}~\bibnamefont
  {Weber}}, \bibinfo {author} {\bibfnamefont {M.}~\bibnamefont {Enderle}},
  \bibinfo {author} {\bibfnamefont {Y.}~\bibnamefont {Su}}, \bibinfo {author}
  {\bibfnamefont {Z.~Q.}\ \bibnamefont {Mao}}, \bibinfo {author} {\bibfnamefont
  {Q.}~\bibnamefont {Si}}, and\ \bibinfo {author} {\bibfnamefont
  {P.}~\bibnamefont {Dai}}} (\bibinfo {year} {2019}{\natexlab{a}}),\ \href
  {https://doi.org/10.1103/PhysRevB.100.054405} {\bibfield  {journal} {\bibinfo
   {journal} {Phys. Rev. B}\ }\textbf {\bibinfo {volume} {100}},\ \bibinfo
  {pages} {054405}}\BibitemShut {NoStop}%
\bibitem [{\citenamefont {Tam}\ \emph {et~al.}(2017)\citenamefont {Tam},
  \citenamefont {Song}, \citenamefont {Man}, \citenamefont {Cheung},
  \citenamefont {Yin}, \citenamefont {Lu}, \citenamefont {Wang}, \citenamefont
  {Frandsen}, \citenamefont {Liu}, \citenamefont {Gong}, \citenamefont {Ito},
  \citenamefont {Cai}, \citenamefont {Wilson}, \citenamefont {Guo},
  \citenamefont {Koshiishi}, \citenamefont {Tian}, \citenamefont {Hitti},
  \citenamefont {Ivanov}, \citenamefont {Zhao}, \citenamefont {Lynn},
  \citenamefont {Luke}, \citenamefont {Berlijn}, \citenamefont {Maier},
  \citenamefont {Uemura},\ and\ \citenamefont {Dai}}]{tam2017uniaxial}%
  \BibitemOpen
  \bibfield  {author} {\bibinfo {author} {\bibnamefont {Tam}, \bibfnamefont
  {D.~W.}}, \bibinfo {author} {\bibfnamefont {Y.}~\bibnamefont {Song}},
  \bibinfo {author} {\bibfnamefont {H.}~\bibnamefont {Man}}, \bibinfo {author}
  {\bibfnamefont {S.~C.}\ \bibnamefont {Cheung}}, \bibinfo {author}
  {\bibfnamefont {Z.}~\bibnamefont {Yin}}, \bibinfo {author} {\bibfnamefont
  {X.}~\bibnamefont {Lu}}, \bibinfo {author} {\bibfnamefont {W.}~\bibnamefont
  {Wang}}, \bibinfo {author} {\bibfnamefont {B.~A.}\ \bibnamefont {Frandsen}},
  \bibinfo {author} {\bibfnamefont {L.}~\bibnamefont {Liu}}, \bibinfo {author}
  {\bibfnamefont {Z.}~\bibnamefont {Gong}}, \bibinfo {author} {\bibfnamefont
  {T.~U.}\ \bibnamefont {Ito}}, \bibinfo {author} {\bibfnamefont
  {Y.}~\bibnamefont {Cai}}, \bibinfo {author} {\bibfnamefont {M.~N.}\
  \bibnamefont {Wilson}}, \bibinfo {author} {\bibfnamefont {S.}~\bibnamefont
  {Guo}}, \bibinfo {author} {\bibfnamefont {K.}~\bibnamefont {Koshiishi}},
  \bibinfo {author} {\bibfnamefont {W.}~\bibnamefont {Tian}}, \bibinfo {author}
  {\bibfnamefont {B.}~\bibnamefont {Hitti}}, \bibinfo {author} {\bibfnamefont
  {A.}~\bibnamefont {Ivanov}}, \bibinfo {author} {\bibfnamefont
  {Y.}~\bibnamefont {Zhao}}, \bibinfo {author} {\bibfnamefont {J.~W.}\
  \bibnamefont {Lynn}}, \bibinfo {author} {\bibfnamefont {G.~M.}\ \bibnamefont
  {Luke}}, \bibinfo {author} {\bibfnamefont {T.}~\bibnamefont {Berlijn}},
  \bibinfo {author} {\bibfnamefont {T.~A.}\ \bibnamefont {Maier}}, \bibinfo
  {author} {\bibfnamefont {Y.~J.}\ \bibnamefont {Uemura}}, and\ \bibinfo
  {author} {\bibfnamefont {P.}~\bibnamefont {Dai}}} (\bibinfo {year} {2017}),\
  \href {https://doi.org/10.1103/PhysRevB.95.060505} {\bibfield  {journal}
  {\bibinfo  {journal} {Phys. Rev. B}\ }\textbf {\bibinfo {volume} {95}},\
  \bibinfo {pages} {060505}}\BibitemShut {NoStop}%
\bibitem [{\citenamefont {Tam}\ \emph {et~al.}(2019{\natexlab{b}})\citenamefont
  {Tam}, \citenamefont {Wang}, \citenamefont {Zhang}, \citenamefont {Song},
  \citenamefont {Zhang}, \citenamefont {Carr}, \citenamefont {Walker},
  \citenamefont {Perring}, \citenamefont {Adroja},\ and\ \citenamefont
  {Dai}}]{tam2019weaker}%
  \BibitemOpen
  \bibfield  {author} {\bibinfo {author} {\bibnamefont {Tam}, \bibfnamefont
  {D.~W.}}, \bibinfo {author} {\bibfnamefont {W.}~\bibnamefont {Wang}},
  \bibinfo {author} {\bibfnamefont {L.}~\bibnamefont {Zhang}}, \bibinfo
  {author} {\bibfnamefont {Y.}~\bibnamefont {Song}}, \bibinfo {author}
  {\bibfnamefont {R.}~\bibnamefont {Zhang}}, \bibinfo {author} {\bibfnamefont
  {S.~V.}\ \bibnamefont {Carr}}, \bibinfo {author} {\bibfnamefont {H.~C.}\
  \bibnamefont {Walker}}, \bibinfo {author} {\bibfnamefont {T.~G.}\
  \bibnamefont {Perring}}, \bibinfo {author} {\bibfnamefont {D.~T.}\
  \bibnamefont {Adroja}}, and\ \bibinfo {author} {\bibfnamefont
  {P.}~\bibnamefont {Dai}}} (\bibinfo {year} {2019}{\natexlab{b}}),\ \href
  {https://doi.org/10.1103/PhysRevB.99.134519} {\bibfield  {journal} {\bibinfo
  {journal} {Phys. Rev. B}\ }\textbf {\bibinfo {volume} {99}},\ \bibinfo
  {pages} {134519}}\BibitemShut {NoStop}%
\bibitem [{\citenamefont {Tam}\ \emph {et~al.}(2020)\citenamefont {Tam},
  \citenamefont {Yin}, \citenamefont {Xie}, \citenamefont {Wang}, \citenamefont
  {Stone}, \citenamefont {Adroja}, \citenamefont {Walker}, \citenamefont {Yi},\
  and\ \citenamefont {Dai}}]{tam2020orbital}%
  \BibitemOpen
  \bibfield  {author} {\bibinfo {author} {\bibnamefont {Tam}, \bibfnamefont
  {D.~W.}}, \bibinfo {author} {\bibfnamefont {Z.}~\bibnamefont {Yin}}, \bibinfo
  {author} {\bibfnamefont {Y.}~\bibnamefont {Xie}}, \bibinfo {author}
  {\bibfnamefont {W.}~\bibnamefont {Wang}}, \bibinfo {author} {\bibfnamefont
  {M.~B.}\ \bibnamefont {Stone}}, \bibinfo {author} {\bibfnamefont {D.~T.}\
  \bibnamefont {Adroja}}, \bibinfo {author} {\bibfnamefont {H.~C.}\
  \bibnamefont {Walker}}, \bibinfo {author} {\bibfnamefont {M.}~\bibnamefont
  {Yi}}, and\ \bibinfo {author} {\bibfnamefont {P.}~\bibnamefont {Dai}}}
  (\bibinfo {year} {2020}),\ \href
  {https://doi.org/10.1103/PhysRevB.102.054430} {\bibfield  {journal} {\bibinfo
   {journal} {Phys. Rev. B}\ }\textbf {\bibinfo {volume} {102}},\ \bibinfo
  {pages} {054430}}\BibitemShut {NoStop}%
\bibitem [{\citenamefont {Tan}\ \emph {et~al.}(2016)\citenamefont {Tan},
  \citenamefont {Song}, \citenamefont {Zhang}, \citenamefont {Lin},
  \citenamefont {Xu}, \citenamefont {Hou}, \citenamefont {Tian}, \citenamefont
  {Cao}, \citenamefont {Li}, \citenamefont {Feng},\ and\ \citenamefont
  {Dai}}]{tan2016electron}%
  \BibitemOpen
  \bibfield  {author} {\bibinfo {author} {\bibnamefont {Tan}, \bibfnamefont
  {G.}}, \bibinfo {author} {\bibfnamefont {Y.}~\bibnamefont {Song}}, \bibinfo
  {author} {\bibfnamefont {C.}~\bibnamefont {Zhang}}, \bibinfo {author}
  {\bibfnamefont {L.}~\bibnamefont {Lin}}, \bibinfo {author} {\bibfnamefont
  {Z.}~\bibnamefont {Xu}}, \bibinfo {author} {\bibfnamefont {T.}~\bibnamefont
  {Hou}}, \bibinfo {author} {\bibfnamefont {W.}~\bibnamefont {Tian}}, \bibinfo
  {author} {\bibfnamefont {H.}~\bibnamefont {Cao}}, \bibinfo {author}
  {\bibfnamefont {S.}~\bibnamefont {Li}}, \bibinfo {author} {\bibfnamefont
  {S.}~\bibnamefont {Feng}}, and\ \bibinfo {author} {\bibfnamefont
  {P.}~\bibnamefont {Dai}}} (\bibinfo {year} {2016}),\ \href
  {https://doi.org/10.1103/PhysRevB.94.014509} {\bibfield  {journal} {\bibinfo
  {journal} {Phys. Rev. B}\ }\textbf {\bibinfo {volume} {94}},\ \bibinfo
  {pages} {014509}}\BibitemShut {NoStop}%
\bibitem [{\citenamefont {Tan}\ \emph {et~al.}(2017)\citenamefont {Tan},
  \citenamefont {Song}, \citenamefont {Zhang}, \citenamefont {Lin},
  \citenamefont {Xu}, \citenamefont {Tian}, \citenamefont {Chi}, \citenamefont
  {{Graves-Brook}}, \citenamefont {Li},\ and\ \citenamefont
  {Dai}}]{tan2017phase}%
  \BibitemOpen
  \bibfield  {author} {\bibinfo {author} {\bibnamefont {Tan}, \bibfnamefont
  {G.}}, \bibinfo {author} {\bibfnamefont {Y.}~\bibnamefont {Song}}, \bibinfo
  {author} {\bibfnamefont {R.}~\bibnamefont {Zhang}}, \bibinfo {author}
  {\bibfnamefont {L.}~\bibnamefont {Lin}}, \bibinfo {author} {\bibfnamefont
  {Z.}~\bibnamefont {Xu}}, \bibinfo {author} {\bibfnamefont {L.}~\bibnamefont
  {Tian}}, \bibinfo {author} {\bibfnamefont {S.}~\bibnamefont {Chi}}, \bibinfo
  {author} {\bibfnamefont {M.~K.}\ \bibnamefont {{Graves-Brook}}}, \bibinfo
  {author} {\bibfnamefont {S.}~\bibnamefont {Li}}, and\ \bibinfo {author}
  {\bibfnamefont {P.}~\bibnamefont {Dai}}} (\bibinfo {year} {2017}),\ \href
  {https://doi.org/10.1103/PhysRevB.95.054501} {\bibfield  {journal} {\bibinfo
  {journal} {Phys. Rev. B}\ }\textbf {\bibinfo {volume} {95}},\ \bibinfo
  {pages} {054501}}\BibitemShut {NoStop}%
\bibitem [{\citenamefont {Tan}\ \emph {et~al.}(2013)\citenamefont {Tan},
  \citenamefont {Zhang}, \citenamefont {Xia}, \citenamefont {Ye}, \citenamefont
  {Chen}, \citenamefont {Xie}, \citenamefont {Peng}, \citenamefont {Xu},
  \citenamefont {Fan}, \citenamefont {Xu} \emph {et~al.}}]{Tan2013}%
  \BibitemOpen
  \bibfield  {author} {\bibinfo {author} {\bibnamefont {Tan}, \bibfnamefont
  {S.}}, \bibinfo {author} {\bibfnamefont {Y.}~\bibnamefont {Zhang}}, \bibinfo
  {author} {\bibfnamefont {M.}~\bibnamefont {Xia}}, \bibinfo {author}
  {\bibfnamefont {Z.}~\bibnamefont {Ye}}, \bibinfo {author} {\bibfnamefont
  {F.}~\bibnamefont {Chen}}, \bibinfo {author} {\bibfnamefont {X.}~\bibnamefont
  {Xie}}, \bibinfo {author} {\bibfnamefont {R.}~\bibnamefont {Peng}}, \bibinfo
  {author} {\bibfnamefont {D.}~\bibnamefont {Xu}}, \bibinfo {author}
  {\bibfnamefont {Q.}~\bibnamefont {Fan}}, \bibinfo {author} {\bibfnamefont
  {H.}~\bibnamefont {Xu}},  \emph {et~al.}} (\bibinfo {year} {2013}),\ \href
  {https://doi.org/10.1038/nmat3654} {\bibfield  {journal} {\bibinfo  {journal}
  {Nat. Mater.}\ }\textbf {\bibinfo {volume} {12}},\ \bibinfo {pages}
  {634}}\BibitemShut {NoStop}%
\bibitem [{\citenamefont {Tanatar}\ \emph {et~al.}(2016)\citenamefont
  {Tanatar}, \citenamefont {B{\"o}hmer}, \citenamefont {Timmons}, \citenamefont
  {Sch{\"u}tt}, \citenamefont {Drachuck}, \citenamefont {Taufour},
  \citenamefont {Kothapalli}, \citenamefont {Kreyssig}, \citenamefont {Bud'ko},
  \citenamefont {Canfield}, \citenamefont {Fernandes},\ and\ \citenamefont
  {Prozorov}}]{tanatar2016origin}%
  \BibitemOpen
  \bibfield  {author} {\bibinfo {author} {\bibnamefont {Tanatar}, \bibfnamefont
  {M.~A.}}, \bibinfo {author} {\bibfnamefont {A.~E.}\ \bibnamefont
  {B{\"o}hmer}}, \bibinfo {author} {\bibfnamefont {E.~I.}\ \bibnamefont
  {Timmons}}, \bibinfo {author} {\bibfnamefont {M.}~\bibnamefont {Sch{\"u}tt}},
  \bibinfo {author} {\bibfnamefont {G.}~\bibnamefont {Drachuck}}, \bibinfo
  {author} {\bibfnamefont {V.}~\bibnamefont {Taufour}}, \bibinfo {author}
  {\bibfnamefont {K.}~\bibnamefont {Kothapalli}}, \bibinfo {author}
  {\bibfnamefont {A.}~\bibnamefont {Kreyssig}}, \bibinfo {author}
  {\bibfnamefont {S.~L.}\ \bibnamefont {Bud'ko}}, \bibinfo {author}
  {\bibfnamefont {P.~C.}\ \bibnamefont {Canfield}}, \bibinfo {author}
  {\bibfnamefont {R.~M.}\ \bibnamefont {Fernandes}}, and\ \bibinfo {author}
  {\bibfnamefont {R.}~\bibnamefont {Prozorov}}} (\bibinfo {year} {2016}),\
  \href {https://doi.org/10.1103/PhysRevLett.117.127001} {\bibfield  {journal}
  {\bibinfo  {journal} {Phys. Rev. Lett.}\ }\textbf {\bibinfo {volume} {117}},\
  \bibinfo {pages} {127001}}\BibitemShut {NoStop}%
\bibitem [{\citenamefont {Tang}\ \emph {et~al.}(2016)\citenamefont {Tang},
  \citenamefont {Liu}, \citenamefont {Zhou}, \citenamefont {Li}, \citenamefont
  {Ding}, \citenamefont {Li}, \citenamefont {Zhang}, \citenamefont {Li},
  \citenamefont {Song}, \citenamefont {Ji} \emph {et~al.}}]{Tang2016}%
  \BibitemOpen
  \bibfield  {author} {\bibinfo {author} {\bibnamefont {Tang}, \bibfnamefont
  {C.}}, \bibinfo {author} {\bibfnamefont {C.}~\bibnamefont {Liu}}, \bibinfo
  {author} {\bibfnamefont {G.}~\bibnamefont {Zhou}}, \bibinfo {author}
  {\bibfnamefont {F.}~\bibnamefont {Li}}, \bibinfo {author} {\bibfnamefont
  {H.}~\bibnamefont {Ding}}, \bibinfo {author} {\bibfnamefont {Z.}~\bibnamefont
  {Li}}, \bibinfo {author} {\bibfnamefont {D.}~\bibnamefont {Zhang}}, \bibinfo
  {author} {\bibfnamefont {Z.}~\bibnamefont {Li}}, \bibinfo {author}
  {\bibfnamefont {C.}~\bibnamefont {Song}}, \bibinfo {author} {\bibfnamefont
  {S.}~\bibnamefont {Ji}},  \emph {et~al.}} (\bibinfo {year} {2016}),\ \href
  {https://doi.org/10.1103/PhysRevB.93.020507} {\bibfield  {journal} {\bibinfo
  {journal} {Phys. Rev. B}\ }\textbf {\bibinfo {volume} {93}},\ \bibinfo
  {pages} {020507}}\BibitemShut {NoStop}%
\bibitem [{\citenamefont {Tang}\ \emph {et~al.}(2000)\citenamefont {Tang},
  \citenamefont {Ng}, \citenamefont {Yau},\ and\ \citenamefont
  {Gao}}]{Tang2000}%
  \BibitemOpen
  \bibfield  {author} {\bibinfo {author} {\bibnamefont {Tang}, \bibfnamefont
  {W.}}, \bibinfo {author} {\bibfnamefont {C.}~\bibnamefont {Ng}}, \bibinfo
  {author} {\bibfnamefont {C.}~\bibnamefont {Yau}}, and\ \bibinfo {author}
  {\bibfnamefont {J.}~\bibnamefont {Gao}}} (\bibinfo {year} {2000}),\ \href
  {https://doi.org/10.1088/0953-2048/13/5/329} {\bibfield  {journal} {\bibinfo
  {journal} {Supercond. Sci. Technol.}\ }\textbf {\bibinfo {volume} {13}},\
  \bibinfo {pages} {580}}\BibitemShut {NoStop}%
\bibitem [{\citenamefont {Tapp}\ \emph {et~al.}(2008)\citenamefont {Tapp},
  \citenamefont {Tang}, \citenamefont {Lv}, \citenamefont {Sasmal},
  \citenamefont {Lorenz}, \citenamefont {Chu},\ and\ \citenamefont
  {Guloy}}]{Tapp2008}%
  \BibitemOpen
  \bibfield  {author} {\bibinfo {author} {\bibnamefont {Tapp}, \bibfnamefont
  {J.~H.}}, \bibinfo {author} {\bibfnamefont {Z.}~\bibnamefont {Tang}},
  \bibinfo {author} {\bibfnamefont {B.}~\bibnamefont {Lv}}, \bibinfo {author}
  {\bibfnamefont {K.}~\bibnamefont {Sasmal}}, \bibinfo {author} {\bibfnamefont
  {B.}~\bibnamefont {Lorenz}}, \bibinfo {author} {\bibfnamefont {P.~C.}\
  \bibnamefont {Chu}}, and\ \bibinfo {author} {\bibfnamefont {A.~M.}\
  \bibnamefont {Guloy}}} (\bibinfo {year} {2008}),\ \href
  {https://doi.org/10.1103/PhysRevB.78.060505} {\bibfield  {journal} {\bibinfo
  {journal} {Phys. Rev. B}\ }\textbf {\bibinfo {volume} {78}},\ \bibinfo
  {pages} {060505}}\BibitemShut {NoStop}%
\bibitem [{\citenamefont {Taylor}\ \emph {et~al.}(2012)\citenamefont {Taylor},
  \citenamefont {Ewings}, \citenamefont {Perring}, \citenamefont {White},
  \citenamefont {Babkevich}, \citenamefont {Krzton-Maziopa}, \citenamefont
  {Pomjakushina}, \citenamefont {Conder},\ and\ \citenamefont
  {Boothroyd}}]{taylor2012spin}%
  \BibitemOpen
  \bibfield  {author} {\bibinfo {author} {\bibnamefont {Taylor}, \bibfnamefont
  {A.~E.}}, \bibinfo {author} {\bibfnamefont {R.~A.}\ \bibnamefont {Ewings}},
  \bibinfo {author} {\bibfnamefont {T.~G.}\ \bibnamefont {Perring}}, \bibinfo
  {author} {\bibfnamefont {J.~S.}\ \bibnamefont {White}}, \bibinfo {author}
  {\bibfnamefont {P.}~\bibnamefont {Babkevich}}, \bibinfo {author}
  {\bibfnamefont {A.}~\bibnamefont {Krzton-Maziopa}}, \bibinfo {author}
  {\bibfnamefont {E.}~\bibnamefont {Pomjakushina}}, \bibinfo {author}
  {\bibfnamefont {K.}~\bibnamefont {Conder}}, and\ \bibinfo {author}
  {\bibfnamefont {A.~T.}\ \bibnamefont {Boothroyd}}} (\bibinfo {year} {2012}),\
  \href {https://doi.org/10.1103/PhysRevB.86.094528} {\bibfield  {journal}
  {\bibinfo  {journal} {Phys. Rev. B}\ }\textbf {\bibinfo {volume} {86}},\
  \bibinfo {pages} {094528}}\BibitemShut {NoStop}%
\bibitem [{\citenamefont {Tegel}\ \emph {et~al.}(2008)\citenamefont {Tegel},
  \citenamefont {Rotter}, \citenamefont {Weiss}, \citenamefont {Schappacher},
  \citenamefont {P{\"o}ttgen},\ and\ \citenamefont {Johrendt}}]{Tegel2008}%
  \BibitemOpen
  \bibfield  {author} {\bibinfo {author} {\bibnamefont {Tegel}, \bibfnamefont
  {M.}}, \bibinfo {author} {\bibfnamefont {M.}~\bibnamefont {Rotter}}, \bibinfo
  {author} {\bibfnamefont {V.}~\bibnamefont {Weiss}}, \bibinfo {author}
  {\bibfnamefont {F.~M.}\ \bibnamefont {Schappacher}}, \bibinfo {author}
  {\bibfnamefont {R.}~\bibnamefont {P{\"o}ttgen}}, and\ \bibinfo {author}
  {\bibfnamefont {D.}~\bibnamefont {Johrendt}}} (\bibinfo {year} {2008}),\
  \href {https://doi.org/10.1088/0953-8984/20/45/452201} {\bibfield  {journal}
  {\bibinfo  {journal} {J. Phys.: Condens. Matter}\ }\textbf {\bibinfo {volume}
  {20}},\ \bibinfo {pages} {452201}}\BibitemShut {NoStop}%
\bibitem [{\citenamefont {Terashima}\ \emph {et~al.}(2015)\citenamefont
  {Terashima}, \citenamefont {Kikugawa}, \citenamefont {Kasahara},
  \citenamefont {Watashige}, \citenamefont {Shibauchi}, \citenamefont
  {Matsuda}, \citenamefont {Wolf}, \citenamefont {B\"{o}hmer}, \citenamefont
  {Hardy}, \citenamefont {Meingast}, \citenamefont {L\"{o}hneysen},\ and\
  \citenamefont {Uji}}]{terashima2015pressure}%
  \BibitemOpen
  \bibfield  {author} {\bibinfo {author} {\bibnamefont {Terashima},
  \bibfnamefont {T.}}, \bibinfo {author} {\bibfnamefont {N.}~\bibnamefont
  {Kikugawa}}, \bibinfo {author} {\bibfnamefont {S.}~\bibnamefont {Kasahara}},
  \bibinfo {author} {\bibfnamefont {T.}~\bibnamefont {Watashige}}, \bibinfo
  {author} {\bibfnamefont {T.}~\bibnamefont {Shibauchi}}, \bibinfo {author}
  {\bibfnamefont {Y.}~\bibnamefont {Matsuda}}, \bibinfo {author} {\bibfnamefont
  {T.}~\bibnamefont {Wolf}}, \bibinfo {author} {\bibfnamefont {A.~E.}\
  \bibnamefont {B\"{o}hmer}}, \bibinfo {author} {\bibfnamefont
  {F.}~\bibnamefont {Hardy}}, \bibinfo {author} {\bibfnamefont
  {C.}~\bibnamefont {Meingast}}, \bibinfo {author} {\bibfnamefont {H.~v.}\
  \bibnamefont {L\"{o}hneysen}}, and\ \bibinfo {author} {\bibfnamefont
  {S.}~\bibnamefont {Uji}}} (\bibinfo {year} {2015}),\ \href
  {https://doi.org/10.7566/JPSJ.84.063701} {\bibfield  {journal} {\bibinfo
  {journal} {J. Phys. Soc. Jpn.}\ }\textbf {\bibinfo {volume} {84}},\ \bibinfo
  {pages} {063701}}\BibitemShut {NoStop}%
\bibitem [{\citenamefont {Terashima}\ \emph {et~al.}(2009)\citenamefont
  {Terashima}, \citenamefont {Kimata}, \citenamefont {Satsukawa}, \citenamefont
  {Harada}, \citenamefont {Hazama}, \citenamefont {Uji}, \citenamefont
  {S.~Suzuki}, \citenamefont {Matsumoto},\ and\ \citenamefont
  {Murata}}]{Terashima2009}%
  \BibitemOpen
  \bibfield  {author} {\bibinfo {author} {\bibnamefont {Terashima},
  \bibfnamefont {T.}}, \bibinfo {author} {\bibfnamefont {M.}~\bibnamefont
  {Kimata}}, \bibinfo {author} {\bibfnamefont {H.}~\bibnamefont {Satsukawa}},
  \bibinfo {author} {\bibfnamefont {A.}~\bibnamefont {Harada}}, \bibinfo
  {author} {\bibfnamefont {K.}~\bibnamefont {Hazama}}, \bibinfo {author}
  {\bibfnamefont {S.}~\bibnamefont {Uji}}, \bibinfo {author} {\bibfnamefont
  {H.}~\bibnamefont {S.~Suzuki}}, \bibinfo {author} {\bibfnamefont
  {T.}~\bibnamefont {Matsumoto}}, and\ \bibinfo {author} {\bibfnamefont
  {K.}~\bibnamefont {Murata}}} (\bibinfo {year} {2009}),\ \href
  {https://doi.org/10.1143/jpsj.78.083701} {\bibfield  {journal} {\bibinfo
  {journal} {J. Phys. Soc. Jpn.}\ }\textbf {\bibinfo {volume} {78}},\ \bibinfo
  {pages} {083701}}\BibitemShut {NoStop}%
\bibitem [{\citenamefont {Thorsm{\o}lle}\ \emph {et~al.}(2016)\citenamefont
  {Thorsm{\o}lle}, \citenamefont {Khodas}, \citenamefont {Yin}, \citenamefont
  {Zhang}, \citenamefont {Carr}, \citenamefont {Dai},\ and\ \citenamefont
  {Blumberg}}]{thorsmolle2016critical}%
  \BibitemOpen
  \bibfield  {author} {\bibinfo {author} {\bibnamefont {Thorsm{\o}lle},
  \bibfnamefont {V.~K.}}, \bibinfo {author} {\bibfnamefont {M.}~\bibnamefont
  {Khodas}}, \bibinfo {author} {\bibfnamefont {Z.~P.}\ \bibnamefont {Yin}},
  \bibinfo {author} {\bibfnamefont {C.}~\bibnamefont {Zhang}}, \bibinfo
  {author} {\bibfnamefont {S.~V.}\ \bibnamefont {Carr}}, \bibinfo {author}
  {\bibfnamefont {P.}~\bibnamefont {Dai}}, and\ \bibinfo {author}
  {\bibfnamefont {G.}~\bibnamefont {Blumberg}}} (\bibinfo {year} {2016}),\
  \href {https://doi.org/10.1103/PhysRevB.93.054515} {\bibfield  {journal}
  {\bibinfo  {journal} {Phys. Rev. B}\ }\textbf {\bibinfo {volume} {93}},\
  \bibinfo {pages} {054515}}\BibitemShut {NoStop}%
\bibitem [{\citenamefont {Tian}\ \emph {et~al.}(2019)\citenamefont {Tian},
  \citenamefont {Liu}, \citenamefont {Xu}, \citenamefont {Li}, \citenamefont
  {Lu}, \citenamefont {Walker}, \citenamefont {Stuhr}, \citenamefont {Tan},
  \citenamefont {Lu},\ and\ \citenamefont {Dai}}]{tian2019spin}%
  \BibitemOpen
  \bibfield  {author} {\bibinfo {author} {\bibnamefont {Tian}, \bibfnamefont
  {L.}}, \bibinfo {author} {\bibfnamefont {P.}~\bibnamefont {Liu}}, \bibinfo
  {author} {\bibfnamefont {Z.}~\bibnamefont {Xu}}, \bibinfo {author}
  {\bibfnamefont {Y.}~\bibnamefont {Li}}, \bibinfo {author} {\bibfnamefont
  {Z.}~\bibnamefont {Lu}}, \bibinfo {author} {\bibfnamefont {H.~C.}\
  \bibnamefont {Walker}}, \bibinfo {author} {\bibfnamefont {U.}~\bibnamefont
  {Stuhr}}, \bibinfo {author} {\bibfnamefont {G.}~\bibnamefont {Tan}}, \bibinfo
  {author} {\bibfnamefont {X.}~\bibnamefont {Lu}}, and\ \bibinfo {author}
  {\bibfnamefont {P.}~\bibnamefont {Dai}}} (\bibinfo {year} {2019}),\ \href
  {https://doi.org/10.1103/PhysRevB.100.134509} {\bibfield  {journal} {\bibinfo
   {journal} {Phys. Rev. B}\ }\textbf {\bibinfo {volume} {100}},\ \bibinfo
  {pages} {134509}}\BibitemShut {NoStop}%
\bibitem [{\citenamefont {Tian}\ \emph {et~al.}(2016)\citenamefont {Tian},
  \citenamefont {Zhang}, \citenamefont {Li}, \citenamefont {Wu}, \citenamefont
  {Wu}, \citenamefont {Sun}, \citenamefont {Zhou}, \citenamefont {Wang},
  \citenamefont {Ma}, \citenamefont {Xue} \emph {et~al.}}]{Tian2016}%
  \BibitemOpen
  \bibfield  {author} {\bibinfo {author} {\bibnamefont {Tian}, \bibfnamefont
  {Y.}}, \bibinfo {author} {\bibfnamefont {W.}~\bibnamefont {Zhang}}, \bibinfo
  {author} {\bibfnamefont {F.}~\bibnamefont {Li}}, \bibinfo {author}
  {\bibfnamefont {Y.}~\bibnamefont {Wu}}, \bibinfo {author} {\bibfnamefont
  {Q.}~\bibnamefont {Wu}}, \bibinfo {author} {\bibfnamefont {F.}~\bibnamefont
  {Sun}}, \bibinfo {author} {\bibfnamefont {G.}~\bibnamefont {Zhou}}, \bibinfo
  {author} {\bibfnamefont {L.}~\bibnamefont {Wang}}, \bibinfo {author}
  {\bibfnamefont {X.}~\bibnamefont {Ma}}, \bibinfo {author} {\bibfnamefont
  {Q.-K.}\ \bibnamefont {Xue}},  \emph {et~al.}} (\bibinfo {year} {2016}),\
  \href {https://doi.org/10.1103/PhysRevLett.116.107001} {\bibfield  {journal}
  {\bibinfo  {journal} {Phys. Rev. Lett.}\ }\textbf {\bibinfo {volume} {116}},\
  \bibinfo {pages} {107001}}\BibitemShut {NoStop}%
\bibitem [{\citenamefont {Tokiwa}\ \emph {et~al.}(2012)\citenamefont {Tokiwa},
  \citenamefont {H{\"u}bner}, \citenamefont {Beck}, \citenamefont {Jeevan},\
  and\ \citenamefont {Gegenwart}}]{Tokiwa2012}%
  \BibitemOpen
  \bibfield  {author} {\bibinfo {author} {\bibnamefont {Tokiwa}, \bibfnamefont
  {Y.}}, \bibinfo {author} {\bibfnamefont {S.-H.}\ \bibnamefont {H{\"u}bner}},
  \bibinfo {author} {\bibfnamefont {O.}~\bibnamefont {Beck}}, \bibinfo {author}
  {\bibfnamefont {H.~S.}\ \bibnamefont {Jeevan}}, and\ \bibinfo {author}
  {\bibfnamefont {P.}~\bibnamefont {Gegenwart}}} (\bibinfo {year} {2012}),\
  \href {https://doi.org/10.1103/PhysRevB.86.220505} {\bibfield  {journal}
  {\bibinfo  {journal} {Phys. Rev. B}\ }\textbf {\bibinfo {volume} {86}},\
  \bibinfo {pages} {220505}}\BibitemShut {NoStop}%
\bibitem [{\citenamefont {Tranquada}(2020)}]{Tranquada01102020}%
  \BibitemOpen
  \bibfield  {author} {\bibinfo {author} {\bibnamefont {Tranquada},
  \bibfnamefont {J.~M.}}} (\bibinfo {year} {2020}),\ \href
  {https://doi.org/10.1080/00018732.2021.1935698} {\bibfield  {journal}
  {\bibinfo  {journal} {Advances in Physics}\ }\textbf {\bibinfo {volume}
  {69}},\ \bibinfo {pages} {437}}\BibitemShut {NoStop}%
\bibitem [{\citenamefont {Tranquada}\ \emph {et~al.}(2004)\citenamefont
  {Tranquada}, \citenamefont {Woo}, \citenamefont {Perring}, \citenamefont
  {Goka}, \citenamefont {Gu}, \citenamefont {Xu}, \citenamefont {Fujita},\ and\
  \citenamefont {Yamada}}]{Tranquada2004quantum}%
  \BibitemOpen
  \bibfield  {author} {\bibinfo {author} {\bibnamefont {Tranquada},
  \bibfnamefont {J.~M.}}, \bibinfo {author} {\bibfnamefont {H.}~\bibnamefont
  {Woo}}, \bibinfo {author} {\bibfnamefont {T.~G.}\ \bibnamefont {Perring}},
  \bibinfo {author} {\bibfnamefont {H.}~\bibnamefont {Goka}}, \bibinfo {author}
  {\bibfnamefont {G.~D.}\ \bibnamefont {Gu}}, \bibinfo {author} {\bibfnamefont
  {G.}~\bibnamefont {Xu}}, \bibinfo {author} {\bibfnamefont {M.}~\bibnamefont
  {Fujita}}, and\ \bibinfo {author} {\bibfnamefont {K.}~\bibnamefont {Yamada}}}
  (\bibinfo {year} {2004}),\ \href {https://doi.org/10.1038/nature02574}
  {\bibfield  {journal} {\bibinfo  {journal} {Nature}\ }\textbf {\bibinfo
  {volume} {429}},\ \bibinfo {pages} {534}}\BibitemShut {NoStop}%
\bibitem [{\citenamefont {Uhoya}\ \emph {et~al.}(2010)\citenamefont {Uhoya},
  \citenamefont {Tsoi}, \citenamefont {Vohra}, \citenamefont {McGuire},
  \citenamefont {Sefat}, \citenamefont {Sales}, \citenamefont {Mandrus},\ and\
  \citenamefont {Weir}}]{Uhoya2010}%
  \BibitemOpen
  \bibfield  {author} {\bibinfo {author} {\bibnamefont {Uhoya}, \bibfnamefont
  {W.}}, \bibinfo {author} {\bibfnamefont {G.}~\bibnamefont {Tsoi}}, \bibinfo
  {author} {\bibfnamefont {Y.~K.}\ \bibnamefont {Vohra}}, \bibinfo {author}
  {\bibfnamefont {M.~A.}\ \bibnamefont {McGuire}}, \bibinfo {author}
  {\bibfnamefont {A.~S.}\ \bibnamefont {Sefat}}, \bibinfo {author}
  {\bibfnamefont {B.~C.}\ \bibnamefont {Sales}}, \bibinfo {author}
  {\bibfnamefont {D.}~\bibnamefont {Mandrus}}, and\ \bibinfo {author}
  {\bibfnamefont {S.~T.}\ \bibnamefont {Weir}}} (\bibinfo {year} {2010}),\
  \href {https://doi.org/10.1088/0953-8984/22/29/292202} {\bibfield  {journal}
  {\bibinfo  {journal} {J. Phys.: Condens. Matter}\ }\textbf {\bibinfo {volume}
  {22}},\ \bibinfo {pages} {292202}}\BibitemShut {NoStop}%
\bibitem [{\citenamefont {Valadkhani}\ \emph {et~al.}(2024)\citenamefont
  {Valadkhani}, \citenamefont {C\'espedes}, \citenamefont {Mandloi},
  \citenamefont {Xu}, \citenamefont {Schmidt}, \citenamefont {Bud'ko},
  \citenamefont {Canfield}, \citenamefont {Valent\'{\i}},\ and\ \citenamefont
  {Gati}}]{strain1144}%
  \BibitemOpen
  \bibfield  {author} {\bibinfo {author} {\bibnamefont {Valadkhani},
  \bibfnamefont {A.}}, \bibinfo {author} {\bibfnamefont {B.~Z.~n.}\
  \bibnamefont {C\'espedes}}, \bibinfo {author} {\bibfnamefont
  {S.}~\bibnamefont {Mandloi}}, \bibinfo {author} {\bibfnamefont
  {M.}~\bibnamefont {Xu}}, \bibinfo {author} {\bibfnamefont {J.}~\bibnamefont
  {Schmidt}}, \bibinfo {author} {\bibfnamefont {S.~L.}\ \bibnamefont {Bud'ko}},
  \bibinfo {author} {\bibfnamefont {P.~C.}\ \bibnamefont {Canfield}}, \bibinfo
  {author} {\bibfnamefont {R.}~\bibnamefont {Valent\'{\i}}}, and\ \bibinfo
  {author} {\bibfnamefont {E.}~\bibnamefont {Gati}}} (\bibinfo {year} {2024}),\
  \href {https://doi.org/10.1103/PhysRevB.109.L180503} {\bibfield  {journal}
  {\bibinfo  {journal} {Phys. Rev. B}\ }\textbf {\bibinfo {volume} {109}},\
  \bibinfo {pages} {L180503}}\BibitemShut {NoStop}%
\bibitem [{\citenamefont {Walker}\ \emph {et~al.}(2023)\citenamefont {Walker},
  \citenamefont {Scott}, \citenamefont {Boyle}, \citenamefont {Byland},
  \citenamefont {B{\"o}tzel}, \citenamefont {Zhao}, \citenamefont {Day},
  \citenamefont {Zhdanovich}, \citenamefont {Gorovikov}, \citenamefont
  {Pedersen}, \citenamefont {Klavins}, \citenamefont {Damascelli},
  \citenamefont {Eremin}, \citenamefont {Gozar}, \citenamefont {Taufour},\ and\
  \citenamefont {Da~Silva~Neto}}]{walker2023electronic}%
  \BibitemOpen
  \bibfield  {author} {\bibinfo {author} {\bibnamefont {Walker}, \bibfnamefont
  {M.}}, \bibinfo {author} {\bibfnamefont {K.}~\bibnamefont {Scott}}, \bibinfo
  {author} {\bibfnamefont {T.~J.}\ \bibnamefont {Boyle}}, \bibinfo {author}
  {\bibfnamefont {J.~K.}\ \bibnamefont {Byland}}, \bibinfo {author}
  {\bibfnamefont {S.}~\bibnamefont {B{\"o}tzel}}, \bibinfo {author}
  {\bibfnamefont {Z.}~\bibnamefont {Zhao}}, \bibinfo {author} {\bibfnamefont
  {R.~P.}\ \bibnamefont {Day}}, \bibinfo {author} {\bibfnamefont
  {S.}~\bibnamefont {Zhdanovich}}, \bibinfo {author} {\bibfnamefont
  {S.}~\bibnamefont {Gorovikov}}, \bibinfo {author} {\bibfnamefont {T.~M.}\
  \bibnamefont {Pedersen}}, \bibinfo {author} {\bibfnamefont {P.}~\bibnamefont
  {Klavins}}, \bibinfo {author} {\bibfnamefont {A.}~\bibnamefont {Damascelli}},
  \bibinfo {author} {\bibfnamefont {I.~M.}\ \bibnamefont {Eremin}}, \bibinfo
  {author} {\bibfnamefont {A.}~\bibnamefont {Gozar}}, \bibinfo {author}
  {\bibfnamefont {V.}~\bibnamefont {Taufour}}, and\ \bibinfo {author}
  {\bibfnamefont {E.~H.}\ \bibnamefont {Da~Silva~Neto}}} (\bibinfo {year}
  {2023}),\ \href {https://doi.org/10.1038/s41535-023-00592-5} {\bibfield
  {journal} {\bibinfo  {journal} {npj Quantum Mater.}\ }\textbf {\bibinfo
  {volume} {8}},\ \bibinfo {pages} {60}}\BibitemShut {NoStop}%
\bibitem [{\citenamefont {Wang}\ \emph
  {et~al.}(2008{\natexlab{a}})\citenamefont {Wang}, \citenamefont {Li},
  \citenamefont {Chi}, \citenamefont {Zhu}, \citenamefont {Ren}, \citenamefont
  {Li}, \citenamefont {Wang}, \citenamefont {Lin}, \citenamefont {Luo},
  \citenamefont {Jiang} \emph {et~al.}}]{Wang2008b}%
  \BibitemOpen
  \bibfield  {author} {\bibinfo {author} {\bibnamefont {Wang}, \bibfnamefont
  {C.}}, \bibinfo {author} {\bibfnamefont {L.}~\bibnamefont {Li}}, \bibinfo
  {author} {\bibfnamefont {S.}~\bibnamefont {Chi}}, \bibinfo {author}
  {\bibfnamefont {Z.}~\bibnamefont {Zhu}}, \bibinfo {author} {\bibfnamefont
  {Z.}~\bibnamefont {Ren}}, \bibinfo {author} {\bibfnamefont {Y.}~\bibnamefont
  {Li}}, \bibinfo {author} {\bibfnamefont {Y.}~\bibnamefont {Wang}}, \bibinfo
  {author} {\bibfnamefont {X.}~\bibnamefont {Lin}}, \bibinfo {author}
  {\bibfnamefont {Y.}~\bibnamefont {Luo}}, \bibinfo {author} {\bibfnamefont
  {S.}~\bibnamefont {Jiang}},  \emph {et~al.}} (\bibinfo {year}
  {2008}{\natexlab{a}}),\ \href {https://doi.org/10.1209/0295-5075/83/67006}
  {\bibfield  {journal} {\bibinfo  {journal} {EPL}\ }\textbf {\bibinfo {volume}
  {83}},\ \bibinfo {pages} {67006}}\BibitemShut {NoStop}%
\bibitem [{\citenamefont {Wang}\ \emph
  {et~al.}(2016{\natexlab{a}})\citenamefont {Wang}, \citenamefont {Wang},
  \citenamefont {Mei}, \citenamefont {Li}, \citenamefont {Li}, \citenamefont
  {Tang}, \citenamefont {Liu}, \citenamefont {Zhang}, \citenamefont {Zhai},
  \citenamefont {Xu} \emph {et~al.}}]{Wang2013}%
  \BibitemOpen
  \bibfield  {author} {\bibinfo {author} {\bibnamefont {Wang}, \bibfnamefont
  {C.}}, \bibinfo {author} {\bibfnamefont {Z.-C.}\ \bibnamefont {Wang}},
  \bibinfo {author} {\bibfnamefont {Y.-X.}\ \bibnamefont {Mei}}, \bibinfo
  {author} {\bibfnamefont {Y.-K.}\ \bibnamefont {Li}}, \bibinfo {author}
  {\bibfnamefont {L.}~\bibnamefont {Li}}, \bibinfo {author} {\bibfnamefont
  {Z.-T.}\ \bibnamefont {Tang}}, \bibinfo {author} {\bibfnamefont
  {Y.}~\bibnamefont {Liu}}, \bibinfo {author} {\bibfnamefont {P.}~\bibnamefont
  {Zhang}}, \bibinfo {author} {\bibfnamefont {H.-F.}\ \bibnamefont {Zhai}},
  \bibinfo {author} {\bibfnamefont {Z.-A.}\ \bibnamefont {Xu}},  \emph
  {et~al.}} (\bibinfo {year} {2016}{\natexlab{a}}),\ \href
  {https://doi.org/10.1021/jacs.6b00236} {\bibfield  {journal} {\bibinfo
  {journal} {J. Am. Chem. Soc.}\ }\textbf {\bibinfo {volume} {138}},\ \bibinfo
  {pages} {2170}}\BibitemShut {NoStop}%
\bibitem [{\citenamefont {Wang}\ \emph
  {et~al.}(2018{\natexlab{a}})\citenamefont {Wang}, \citenamefont {Kong},
  \citenamefont {Fan}, \citenamefont {Chen}, \citenamefont {Zhu}, \citenamefont
  {Liu}, \citenamefont {Cao}, \citenamefont {Sun}, \citenamefont {Du},
  \citenamefont {Schneeloch}, \citenamefont {Zhong}, \citenamefont {Gu},
  \citenamefont {Fu}, \citenamefont {Ding},\ and\ \citenamefont
  {Gao}}]{wang2018evidence}%
  \BibitemOpen
  \bibfield  {author} {\bibinfo {author} {\bibnamefont {Wang}, \bibfnamefont
  {D.}}, \bibinfo {author} {\bibfnamefont {L.}~\bibnamefont {Kong}}, \bibinfo
  {author} {\bibfnamefont {P.}~\bibnamefont {Fan}}, \bibinfo {author}
  {\bibfnamefont {H.}~\bibnamefont {Chen}}, \bibinfo {author} {\bibfnamefont
  {S.}~\bibnamefont {Zhu}}, \bibinfo {author} {\bibfnamefont {W.}~\bibnamefont
  {Liu}}, \bibinfo {author} {\bibfnamefont {L.}~\bibnamefont {Cao}}, \bibinfo
  {author} {\bibfnamefont {Y.}~\bibnamefont {Sun}}, \bibinfo {author}
  {\bibfnamefont {S.}~\bibnamefont {Du}}, \bibinfo {author} {\bibfnamefont
  {J.}~\bibnamefont {Schneeloch}}, \bibinfo {author} {\bibfnamefont
  {R.}~\bibnamefont {Zhong}}, \bibinfo {author} {\bibfnamefont
  {G.}~\bibnamefont {Gu}}, \bibinfo {author} {\bibfnamefont {L.}~\bibnamefont
  {Fu}}, \bibinfo {author} {\bibfnamefont {H.}~\bibnamefont {Ding}}, and\
  \bibinfo {author} {\bibfnamefont {H.-J.}\ \bibnamefont {Gao}}} (\bibinfo
  {year} {2018}{\natexlab{a}}),\ \href
  {https://doi.org/10.1126/science.aao1797} {\bibfield  {journal} {\bibinfo
  {journal} {Science}\ }\textbf {\bibinfo {volume} {362}},\ \bibinfo {pages}
  {333}}\BibitemShut {NoStop}%
\bibitem [{\citenamefont {Wang}\ \emph
  {et~al.}(2015{\natexlab{a}})\citenamefont {Wang}, \citenamefont {Kivelson},\
  and\ \citenamefont {Lee}}]{wang2015nematicity}%
  \BibitemOpen
  \bibfield  {author} {\bibinfo {author} {\bibnamefont {Wang}, \bibfnamefont
  {F.}}, \bibinfo {author} {\bibfnamefont {S.~A.}\ \bibnamefont {Kivelson}},
  and\ \bibinfo {author} {\bibfnamefont {D.-H.}\ \bibnamefont {Lee}}} (\bibinfo
  {year} {2015}{\natexlab{a}}),\ \href {https://doi.org/10.1038/nphys3456}
  {\bibfield  {journal} {\bibinfo  {journal} {Nat. Phys.}\ }\textbf {\bibinfo
  {volume} {11}},\ \bibinfo {pages} {959}}\BibitemShut {NoStop}%
\bibitem [{\citenamefont {Wang}\ \emph {et~al.}(2011)\citenamefont {Wang},
  \citenamefont {Dong}, \citenamefont {Li}, \citenamefont {Mao}, \citenamefont
  {Zhu}, \citenamefont {Feng}, \citenamefont {Yuan},\ and\ \citenamefont
  {Fang}}]{Wang2011}%
  \BibitemOpen
  \bibfield  {author} {\bibinfo {author} {\bibnamefont {Wang}, \bibfnamefont
  {H.-D.}}, \bibinfo {author} {\bibfnamefont {C.-H.}\ \bibnamefont {Dong}},
  \bibinfo {author} {\bibfnamefont {Z.-J.}\ \bibnamefont {Li}}, \bibinfo
  {author} {\bibfnamefont {Q.-H.}\ \bibnamefont {Mao}}, \bibinfo {author}
  {\bibfnamefont {S.-S.}\ \bibnamefont {Zhu}}, \bibinfo {author} {\bibfnamefont
  {C.-M.}\ \bibnamefont {Feng}}, \bibinfo {author} {\bibfnamefont
  {H.}~\bibnamefont {Yuan}}, and\ \bibinfo {author} {\bibfnamefont {M.-H.}\
  \bibnamefont {Fang}}} (\bibinfo {year} {2011}),\ \href
  {https://doi.org/10.1209/0295-5075/93/47004} {\bibfield  {journal} {\bibinfo
  {journal} {EPL}\ }\textbf {\bibinfo {volume} {93}},\ \bibinfo {pages}
  {47004}}\BibitemShut {NoStop}%
\bibitem [{\citenamefont {Wang}\ \emph
  {et~al.}(2016{\natexlab{b}})\citenamefont {Wang}, \citenamefont {Hardy},
  \citenamefont {B\"ohmer}, \citenamefont {Wolf}, \citenamefont {Schweiss},\
  and\ \citenamefont {Meingast}}]{wang2016complex}%
  \BibitemOpen
  \bibfield  {author} {\bibinfo {author} {\bibnamefont {Wang}, \bibfnamefont
  {L.}}, \bibinfo {author} {\bibfnamefont {F.}~\bibnamefont {Hardy}}, \bibinfo
  {author} {\bibfnamefont {A.~E.}\ \bibnamefont {B\"ohmer}}, \bibinfo {author}
  {\bibfnamefont {T.}~\bibnamefont {Wolf}}, \bibinfo {author} {\bibfnamefont
  {P.}~\bibnamefont {Schweiss}}, and\ \bibinfo {author} {\bibfnamefont
  {C.}~\bibnamefont {Meingast}}} (\bibinfo {year} {2016}{\natexlab{b}}),\ \href
  {https://doi.org/10.1103/PhysRevB.93.014514} {\bibfield  {journal} {\bibinfo
  {journal} {Phys. Rev. B}\ }\textbf {\bibinfo {volume} {93}},\ \bibinfo
  {pages} {014514}}\BibitemShut {NoStop}%
\bibitem [{\citenamefont {Wang}\ \emph
  {et~al.}(2016{\natexlab{c}})\citenamefont {Wang}, \citenamefont {Yi},
  \citenamefont {Sun}, \citenamefont {Valdivia}, \citenamefont {Kim},
  \citenamefont {Xu}, \citenamefont {Berlijn}, \citenamefont {Christianson},
  \citenamefont {Chi}, \citenamefont {Hashimoto}, \citenamefont {Lu},
  \citenamefont {Li}, \citenamefont {{Bourret-Courchesne}}, \citenamefont
  {Dai}, \citenamefont {Lee}, \citenamefont {Maier},\ and\ \citenamefont
  {Birgeneau}}]{wang2016experimental}%
  \BibitemOpen
  \bibfield  {author} {\bibinfo {author} {\bibnamefont {Wang}, \bibfnamefont
  {M.}}, \bibinfo {author} {\bibfnamefont {M.}~\bibnamefont {Yi}}, \bibinfo
  {author} {\bibfnamefont {H.~L.}\ \bibnamefont {Sun}}, \bibinfo {author}
  {\bibfnamefont {P.}~\bibnamefont {Valdivia}}, \bibinfo {author}
  {\bibfnamefont {M.~G.}\ \bibnamefont {Kim}}, \bibinfo {author} {\bibfnamefont
  {Z.~J.}\ \bibnamefont {Xu}}, \bibinfo {author} {\bibfnamefont
  {T.}~\bibnamefont {Berlijn}}, \bibinfo {author} {\bibfnamefont {A.~D.}\
  \bibnamefont {Christianson}}, \bibinfo {author} {\bibfnamefont
  {S.}~\bibnamefont {Chi}}, \bibinfo {author} {\bibfnamefont {M.}~\bibnamefont
  {Hashimoto}}, \bibinfo {author} {\bibfnamefont {D.~H.}\ \bibnamefont {Lu}},
  \bibinfo {author} {\bibfnamefont {X.~D.}\ \bibnamefont {Li}}, \bibinfo
  {author} {\bibfnamefont {E.}~\bibnamefont {{Bourret-Courchesne}}}, \bibinfo
  {author} {\bibfnamefont {P.}~\bibnamefont {Dai}}, \bibinfo {author}
  {\bibfnamefont {D.~H.}\ \bibnamefont {Lee}}, \bibinfo {author} {\bibfnamefont
  {T.~A.}\ \bibnamefont {Maier}}, and\ \bibinfo {author} {\bibfnamefont
  {R.~J.}\ \bibnamefont {Birgeneau}}} (\bibinfo {year} {2016}{\natexlab{c}}),\
  \href {https://doi.org/10.1103/PhysRevB.93.205149} {\bibfield  {journal}
  {\bibinfo  {journal} {Phys. Rev. B}\ }\textbf {\bibinfo {volume} {93}},\
  \bibinfo {pages} {205149}}\BibitemShut {NoStop}%
\bibitem [{\citenamefont {Wang}\ \emph {et~al.}(2013)\citenamefont {Wang},
  \citenamefont {Zhang}, \citenamefont {Lu}, \citenamefont {Tan}, \citenamefont
  {Luo}, \citenamefont {Song}, \citenamefont {Wang}, \citenamefont {Zhang},
  \citenamefont {Goremychkin}, \citenamefont {Perring}, \citenamefont {Maier},
  \citenamefont {Yin}, \citenamefont {Haule}, \citenamefont {Kotliar},\ and\
  \citenamefont {Dai}}]{wang2013doping}%
  \BibitemOpen
  \bibfield  {author} {\bibinfo {author} {\bibnamefont {Wang}, \bibfnamefont
  {M.}}, \bibinfo {author} {\bibfnamefont {C.}~\bibnamefont {Zhang}}, \bibinfo
  {author} {\bibfnamefont {X.}~\bibnamefont {Lu}}, \bibinfo {author}
  {\bibfnamefont {G.}~\bibnamefont {Tan}}, \bibinfo {author} {\bibfnamefont
  {H.}~\bibnamefont {Luo}}, \bibinfo {author} {\bibfnamefont {Y.}~\bibnamefont
  {Song}}, \bibinfo {author} {\bibfnamefont {M.}~\bibnamefont {Wang}}, \bibinfo
  {author} {\bibfnamefont {X.}~\bibnamefont {Zhang}}, \bibinfo {author}
  {\bibfnamefont {E.~A.}\ \bibnamefont {Goremychkin}}, \bibinfo {author}
  {\bibfnamefont {T.~G.}\ \bibnamefont {Perring}}, \bibinfo {author}
  {\bibfnamefont {T.~A.}\ \bibnamefont {Maier}}, \bibinfo {author}
  {\bibfnamefont {Z.}~\bibnamefont {Yin}}, \bibinfo {author} {\bibfnamefont
  {K.}~\bibnamefont {Haule}}, \bibinfo {author} {\bibfnamefont
  {G.}~\bibnamefont {Kotliar}}, and\ \bibinfo {author} {\bibfnamefont
  {P.}~\bibnamefont {Dai}}} (\bibinfo {year} {2013}),\ \href
  {https://doi.org/10.1038/ncomms3874} {\bibfield  {journal} {\bibinfo
  {journal} {Nat. Commun.}\ }\textbf {\bibinfo {volume} {4}},\ \bibinfo {pages}
  {2874}}\BibitemShut {NoStop}%
\bibitem [{\citenamefont {Wang}\ \emph
  {et~al.}(2016{\natexlab{d}})\citenamefont {Wang}, \citenamefont {Park},
  \citenamefont {Feng}, \citenamefont {Shen}, \citenamefont {Hao},
  \citenamefont {Pan}, \citenamefont {Lynn}, \citenamefont {Ivanov},
  \citenamefont {Chi}, \citenamefont {Matsuda}, \citenamefont {Cao},
  \citenamefont {Birgeneau}, \citenamefont {Efremov},\ and\ \citenamefont
  {Zhao}}]{wang2016transition}%
  \BibitemOpen
  \bibfield  {author} {\bibinfo {author} {\bibnamefont {Wang}, \bibfnamefont
  {Q.}}, \bibinfo {author} {\bibfnamefont {J.~T.}\ \bibnamefont {Park}},
  \bibinfo {author} {\bibfnamefont {Y.}~\bibnamefont {Feng}}, \bibinfo {author}
  {\bibfnamefont {Y.}~\bibnamefont {Shen}}, \bibinfo {author} {\bibfnamefont
  {Y.}~\bibnamefont {Hao}}, \bibinfo {author} {\bibfnamefont {B.}~\bibnamefont
  {Pan}}, \bibinfo {author} {\bibfnamefont {J.~W.}\ \bibnamefont {Lynn}},
  \bibinfo {author} {\bibfnamefont {A.}~\bibnamefont {Ivanov}}, \bibinfo
  {author} {\bibfnamefont {S.}~\bibnamefont {Chi}}, \bibinfo {author}
  {\bibfnamefont {M.}~\bibnamefont {Matsuda}}, \bibinfo {author} {\bibfnamefont
  {H.}~\bibnamefont {Cao}}, \bibinfo {author} {\bibfnamefont {R.~J.}\
  \bibnamefont {Birgeneau}}, \bibinfo {author} {\bibfnamefont {D.~V.}\
  \bibnamefont {Efremov}}, and\ \bibinfo {author} {\bibfnamefont
  {J.}~\bibnamefont {Zhao}}} (\bibinfo {year} {2016}{\natexlab{d}}),\ \href
  {https://doi.org/10.1103/PhysRevLett.116.197004} {\bibfield  {journal}
  {\bibinfo  {journal} {Phys. Rev. Lett.}\ }\textbf {\bibinfo {volume} {116}},\
  \bibinfo {pages} {197004}}\BibitemShut {NoStop}%
\bibitem [{\citenamefont {Wang}\ \emph
  {et~al.}(2016{\natexlab{e}})\citenamefont {Wang}, \citenamefont {Shen},
  \citenamefont {Pan}, \citenamefont {Hao}, \citenamefont {Ma}, \citenamefont
  {Zhou}, \citenamefont {Steffens}, \citenamefont {Schmalzl}, \citenamefont
  {Forrest}, \citenamefont {{Abdel-Hafiez}}, \citenamefont {Chen},
  \citenamefont {Chareev}, \citenamefont {Vasiliev}, \citenamefont {Bourges},
  \citenamefont {Sidis}, \citenamefont {Cao},\ and\ \citenamefont
  {Zhao}}]{wang2016strong}%
  \BibitemOpen
  \bibfield  {author} {\bibinfo {author} {\bibnamefont {Wang}, \bibfnamefont
  {Q.}}, \bibinfo {author} {\bibfnamefont {Y.}~\bibnamefont {Shen}}, \bibinfo
  {author} {\bibfnamefont {B.}~\bibnamefont {Pan}}, \bibinfo {author}
  {\bibfnamefont {Y.}~\bibnamefont {Hao}}, \bibinfo {author} {\bibfnamefont
  {M.}~\bibnamefont {Ma}}, \bibinfo {author} {\bibfnamefont {F.}~\bibnamefont
  {Zhou}}, \bibinfo {author} {\bibfnamefont {P.}~\bibnamefont {Steffens}},
  \bibinfo {author} {\bibfnamefont {K.}~\bibnamefont {Schmalzl}}, \bibinfo
  {author} {\bibfnamefont {T.~R.}\ \bibnamefont {Forrest}}, \bibinfo {author}
  {\bibfnamefont {M.}~\bibnamefont {{Abdel-Hafiez}}}, \bibinfo {author}
  {\bibfnamefont {X.}~\bibnamefont {Chen}}, \bibinfo {author} {\bibfnamefont
  {D.~A.}\ \bibnamefont {Chareev}}, \bibinfo {author} {\bibfnamefont {A.~N.}\
  \bibnamefont {Vasiliev}}, \bibinfo {author} {\bibfnamefont {P.}~\bibnamefont
  {Bourges}}, \bibinfo {author} {\bibfnamefont {Y.}~\bibnamefont {Sidis}},
  \bibinfo {author} {\bibfnamefont {H.}~\bibnamefont {Cao}}, and\ \bibinfo
  {author} {\bibfnamefont {J.}~\bibnamefont {Zhao}}} (\bibinfo {year}
  {2016}{\natexlab{e}}),\ \href {https://doi.org/10.1038/nmat4492} {\bibfield
  {journal} {\bibinfo  {journal} {Nat. Mater.}\ }\textbf {\bibinfo {volume}
  {15}},\ \bibinfo {pages} {159}}\BibitemShut {NoStop}%
\bibitem [{\citenamefont {Wang}\ \emph
  {et~al.}(2016{\natexlab{f}})\citenamefont {Wang}, \citenamefont {Shen},
  \citenamefont {Pan}, \citenamefont {Zhang}, \citenamefont {Ikeuchi},
  \citenamefont {Iida}, \citenamefont {Christianson}, \citenamefont {Walker},
  \citenamefont {Adroja}, \citenamefont {{Abdel-Hafiez}}, \citenamefont {Chen},
  \citenamefont {Chareev}, \citenamefont {Vasiliev},\ and\ \citenamefont
  {Zhao}}]{wang2016magnetic}%
  \BibitemOpen
  \bibfield  {author} {\bibinfo {author} {\bibnamefont {Wang}, \bibfnamefont
  {Q.}}, \bibinfo {author} {\bibfnamefont {Y.}~\bibnamefont {Shen}}, \bibinfo
  {author} {\bibfnamefont {B.}~\bibnamefont {Pan}}, \bibinfo {author}
  {\bibfnamefont {X.}~\bibnamefont {Zhang}}, \bibinfo {author} {\bibfnamefont
  {K.}~\bibnamefont {Ikeuchi}}, \bibinfo {author} {\bibfnamefont
  {K.}~\bibnamefont {Iida}}, \bibinfo {author} {\bibfnamefont {A.~D.}\
  \bibnamefont {Christianson}}, \bibinfo {author} {\bibfnamefont {H.~C.}\
  \bibnamefont {Walker}}, \bibinfo {author} {\bibfnamefont {D.~T.}\
  \bibnamefont {Adroja}}, \bibinfo {author} {\bibfnamefont {M.}~\bibnamefont
  {{Abdel-Hafiez}}}, \bibinfo {author} {\bibfnamefont {X.}~\bibnamefont
  {Chen}}, \bibinfo {author} {\bibfnamefont {D.~A.}\ \bibnamefont {Chareev}},
  \bibinfo {author} {\bibfnamefont {A.~N.}\ \bibnamefont {Vasiliev}}, and\
  \bibinfo {author} {\bibfnamefont {J.}~\bibnamefont {Zhao}}} (\bibinfo {year}
  {2016}{\natexlab{f}}),\ \href {https://doi.org/10.1038/ncomms12182}
  {\bibfield  {journal} {\bibinfo  {journal} {Nat. Commun.}\ }\textbf {\bibinfo
  {volume} {7}},\ \bibinfo {pages} {12182}}\BibitemShut {NoStop}%
\bibitem [{\citenamefont {Wang}\ \emph {et~al.}(2012)\citenamefont {Wang},
  \citenamefont {Li}, \citenamefont {Zhang}, \citenamefont {Zhang},
  \citenamefont {Zhang}, \citenamefont {Li}, \citenamefont {Ding},
  \citenamefont {Ou}, \citenamefont {Deng}, \citenamefont {Chang} \emph
  {et~al.}}]{Wang2012}%
  \BibitemOpen
  \bibfield  {author} {\bibinfo {author} {\bibnamefont {Wang}, \bibfnamefont
  {Q.-Y.}}, \bibinfo {author} {\bibfnamefont {Z.}~\bibnamefont {Li}}, \bibinfo
  {author} {\bibfnamefont {W.-H.}\ \bibnamefont {Zhang}}, \bibinfo {author}
  {\bibfnamefont {Z.-C.}\ \bibnamefont {Zhang}}, \bibinfo {author}
  {\bibfnamefont {J.-S.}\ \bibnamefont {Zhang}}, \bibinfo {author}
  {\bibfnamefont {W.}~\bibnamefont {Li}}, \bibinfo {author} {\bibfnamefont
  {H.}~\bibnamefont {Ding}}, \bibinfo {author} {\bibfnamefont {Y.-B.}\
  \bibnamefont {Ou}}, \bibinfo {author} {\bibfnamefont {P.}~\bibnamefont
  {Deng}}, \bibinfo {author} {\bibfnamefont {K.}~\bibnamefont {Chang}},  \emph
  {et~al.}} (\bibinfo {year} {2012}),\ \href
  {https://doi.org/10.1088/0256-307x/29/3/037402} {\bibfield  {journal}
  {\bibinfo  {journal} {Chin. Phys. Lett.}\ }\textbf {\bibinfo {volume} {29}},\
  \bibinfo {pages} {037402}}\BibitemShut {NoStop}%
\bibitem [{\citenamefont {Wang}\ \emph
  {et~al.}(2022{\natexlab{a}})\citenamefont {Wang}, \citenamefont {Lei},
  \citenamefont {Zhu}, \citenamefont {Cui}, \citenamefont {Zhuo}, \citenamefont
  {Xiang}, \citenamefont {Luo},\ and\ \citenamefont {Chen}}]{Wang2022}%
  \BibitemOpen
  \bibfield  {author} {\bibinfo {author} {\bibnamefont {Wang}, \bibfnamefont
  {W.}}, \bibinfo {author} {\bibfnamefont {B.}~\bibnamefont {Lei}}, \bibinfo
  {author} {\bibfnamefont {C.}~\bibnamefont {Zhu}}, \bibinfo {author}
  {\bibfnamefont {J.}~\bibnamefont {Cui}}, \bibinfo {author} {\bibfnamefont
  {W.}~\bibnamefont {Zhuo}}, \bibinfo {author} {\bibfnamefont {Z.}~\bibnamefont
  {Xiang}}, \bibinfo {author} {\bibfnamefont {X.}~\bibnamefont {Luo}}, and\
  \bibinfo {author} {\bibfnamefont {X.}~\bibnamefont {Chen}}} (\bibinfo {year}
  {2022}{\natexlab{a}}),\ \href {https://doi.org/10.1103/PhysRevB.106.014509}
  {\bibfield  {journal} {\bibinfo  {journal} {Phys. Rev. B}\ }\textbf {\bibinfo
  {volume} {106}},\ \bibinfo {pages} {014509}}\BibitemShut {NoStop}%
\bibitem [{\citenamefont {Wang}\ \emph
  {et~al.}(2017{\natexlab{a}})\citenamefont {Wang}, \citenamefont {Park},
  \citenamefont {Yu}, \citenamefont {Li}, \citenamefont {Song}, \citenamefont
  {Zhang}, \citenamefont {Ivanov}, \citenamefont {Kulda},\ and\ \citenamefont
  {Dai}}]{wang2017orbital}%
  \BibitemOpen
  \bibfield  {author} {\bibinfo {author} {\bibnamefont {Wang}, \bibfnamefont
  {W.}}, \bibinfo {author} {\bibfnamefont {J.~T.}\ \bibnamefont {Park}},
  \bibinfo {author} {\bibfnamefont {R.}~\bibnamefont {Yu}}, \bibinfo {author}
  {\bibfnamefont {Y.}~\bibnamefont {Li}}, \bibinfo {author} {\bibfnamefont
  {Y.}~\bibnamefont {Song}}, \bibinfo {author} {\bibfnamefont {Z.}~\bibnamefont
  {Zhang}}, \bibinfo {author} {\bibfnamefont {A.}~\bibnamefont {Ivanov}},
  \bibinfo {author} {\bibfnamefont {J.}~\bibnamefont {Kulda}}, and\ \bibinfo
  {author} {\bibfnamefont {P.}~\bibnamefont {Dai}}} (\bibinfo {year}
  {2017}{\natexlab{a}}),\ \href {https://doi.org/10.1103/PhysRevB.95.094519}
  {\bibfield  {journal} {\bibinfo  {journal} {Phys. Rev. B}\ }\textbf {\bibinfo
  {volume} {95}},\ \bibinfo {pages} {094519}}\BibitemShut {NoStop}%
\bibitem [{\citenamefont {Wang}\ \emph
  {et~al.}(2018{\natexlab{b}})\citenamefont {Wang}, \citenamefont {Song},
  \citenamefont {Cao}, \citenamefont {Tseng}, \citenamefont {Keller},
  \citenamefont {Li}, \citenamefont {Harriger}, \citenamefont {Tian},
  \citenamefont {Chi}, \citenamefont {Yu}, \citenamefont {Nevidomskyy},\ and\
  \citenamefont {Dai}}]{wang2018local}%
  \BibitemOpen
  \bibfield  {author} {\bibinfo {author} {\bibnamefont {Wang}, \bibfnamefont
  {W.}}, \bibinfo {author} {\bibfnamefont {Y.}~\bibnamefont {Song}}, \bibinfo
  {author} {\bibfnamefont {C.}~\bibnamefont {Cao}}, \bibinfo {author}
  {\bibfnamefont {K.-F.}\ \bibnamefont {Tseng}}, \bibinfo {author}
  {\bibfnamefont {T.}~\bibnamefont {Keller}}, \bibinfo {author} {\bibfnamefont
  {Y.}~\bibnamefont {Li}}, \bibinfo {author} {\bibfnamefont {L.~W.}\
  \bibnamefont {Harriger}}, \bibinfo {author} {\bibfnamefont {W.}~\bibnamefont
  {Tian}}, \bibinfo {author} {\bibfnamefont {S.}~\bibnamefont {Chi}}, \bibinfo
  {author} {\bibfnamefont {R.}~\bibnamefont {Yu}}, \bibinfo {author}
  {\bibfnamefont {A.~H.}\ \bibnamefont {Nevidomskyy}}, and\ \bibinfo {author}
  {\bibfnamefont {P.}~\bibnamefont {Dai}}} (\bibinfo {year}
  {2018}{\natexlab{b}}),\ \href {https://doi.org/10.1038/s41467-018-05529-2}
  {\bibfield  {journal} {\bibinfo  {journal} {Nat. Commun.}\ }\textbf {\bibinfo
  {volume} {9}},\ \bibinfo {pages} {3128}}\BibitemShut {NoStop}%
\bibitem [{\citenamefont {Wang}\ \emph
  {et~al.}(2022{\natexlab{b}})\citenamefont {Wang}, \citenamefont {Chen},
  \citenamefont {Zhang}, \citenamefont {Zhang},\ and\ \citenamefont
  {Feng}}]{wang2022stm}%
  \BibitemOpen
  \bibfield  {author} {\bibinfo {author} {\bibnamefont {Wang}, \bibfnamefont
  {X.}}, \bibinfo {author} {\bibfnamefont {C.}~\bibnamefont {Chen}}, \bibinfo
  {author} {\bibfnamefont {T.}~\bibnamefont {Zhang}}, \bibinfo {author}
  {\bibfnamefont {T.}~\bibnamefont {Zhang}}, and\ \bibinfo {author}
  {\bibfnamefont {D.}~\bibnamefont {Feng}}} (\bibinfo {year}
  {2022}{\natexlab{b}}),\ \href {https://doi.org/10.1007/s44214-022-00014-w}
  {\bibfield  {journal} {\bibinfo  {journal} {Quantum Frontiers}\ }\textbf
  {\bibinfo {volume} {1}},\ \bibinfo {pages} {12}}\BibitemShut {NoStop}%
\bibitem [{\citenamefont {Wang}\ \emph
  {et~al.}(2008{\natexlab{b}})\citenamefont {Wang}, \citenamefont {Liu},
  \citenamefont {Lv}, \citenamefont {Gao}, \citenamefont {Yang}, \citenamefont
  {Yu}, \citenamefont {Li},\ and\ \citenamefont {Jin}}]{Wang2008LiFeAs}%
  \BibitemOpen
  \bibfield  {author} {\bibinfo {author} {\bibnamefont {Wang}, \bibfnamefont
  {X.}}, \bibinfo {author} {\bibfnamefont {Q.}~\bibnamefont {Liu}}, \bibinfo
  {author} {\bibfnamefont {Y.}~\bibnamefont {Lv}}, \bibinfo {author}
  {\bibfnamefont {W.}~\bibnamefont {Gao}}, \bibinfo {author} {\bibfnamefont
  {L.}~\bibnamefont {Yang}}, \bibinfo {author} {\bibfnamefont {R.}~\bibnamefont
  {Yu}}, \bibinfo {author} {\bibfnamefont {F.}~\bibnamefont {Li}}, and\
  \bibinfo {author} {\bibfnamefont {C.}~\bibnamefont {Jin}}} (\bibinfo {year}
  {2008}{\natexlab{b}}),\ \href {https://doi.org/10.1016/j.ssc.2008.09.057}
  {\bibfield  {journal} {\bibinfo  {journal} {Solid State Communications}\
  }\textbf {\bibinfo {volume} {148}},\ \bibinfo {pages} {538}}\BibitemShut
  {NoStop}%
\bibitem [{\citenamefont {Wang}\ \emph {et~al.}(2010)\citenamefont {Wang},
  \citenamefont {Yan}, \citenamefont {Ying}, \citenamefont {Li}, \citenamefont
  {Zhang}, \citenamefont {Xu},\ and\ \citenamefont {Chen}}]{Wang2010}%
  \BibitemOpen
  \bibfield  {author} {\bibinfo {author} {\bibnamefont {Wang}, \bibfnamefont
  {X.}}, \bibinfo {author} {\bibfnamefont {Y.}~\bibnamefont {Yan}}, \bibinfo
  {author} {\bibfnamefont {J.}~\bibnamefont {Ying}}, \bibinfo {author}
  {\bibfnamefont {Q.}~\bibnamefont {Li}}, \bibinfo {author} {\bibfnamefont
  {M.}~\bibnamefont {Zhang}}, \bibinfo {author} {\bibfnamefont
  {N.}~\bibnamefont {Xu}}, and\ \bibinfo {author} {\bibfnamefont
  {X.}~\bibnamefont {Chen}}} (\bibinfo {year} {2010}),\ \href
  {https://doi.org/10.1088/0953-8984/22/7/075702} {\bibfield  {journal}
  {\bibinfo  {journal} {J. Phys.: Condens. Matter}\ }\textbf {\bibinfo {volume}
  {22}},\ \bibinfo {pages} {075702}}\BibitemShut {NoStop}%
\bibitem [{\citenamefont {Wang}\ \emph {et~al.}(2026)\citenamefont {Wang},
  \citenamefont {Tam}, \citenamefont {Wang}, \citenamefont {Ewings},
  \citenamefont {Stewart}, \citenamefont {Matsuda}, \citenamefont {Cao},
  \citenamefont {Liu}, \citenamefont {Yu}, \citenamefont {Dai},\ and\
  \citenamefont {Song}}]{wang2026quasi}%
  \BibitemOpen
  \bibfield  {author} {\bibinfo {author} {\bibnamefont {Wang}, \bibfnamefont
  {Y.}}, \bibinfo {author} {\bibfnamefont {D.~W.}\ \bibnamefont {Tam}},
  \bibinfo {author} {\bibfnamefont {W.}~\bibnamefont {Wang}}, \bibinfo {author}
  {\bibfnamefont {R.~A.}\ \bibnamefont {Ewings}}, \bibinfo {author}
  {\bibfnamefont {J.~R.}\ \bibnamefont {Stewart}}, \bibinfo {author}
  {\bibfnamefont {M.}~\bibnamefont {Matsuda}}, \bibinfo {author} {\bibfnamefont
  {C.}~\bibnamefont {Cao}}, \bibinfo {author} {\bibfnamefont {C.}~\bibnamefont
  {Liu}}, \bibinfo {author} {\bibfnamefont {R.}~\bibnamefont {Yu}}, \bibinfo
  {author} {\bibfnamefont {P.}~\bibnamefont {Dai}}, and\ \bibinfo {author}
  {\bibfnamefont {Y.}~\bibnamefont {Song}}} (\bibinfo {year} {2026}),\ \href
  {https://doi.org/10.1103/5vdj-9wlt} {\bibfield  {journal} {\bibinfo
  {journal} {Phys. Rev. Lett.}\ }\textbf {\bibinfo {volume} {136}},\ \bibinfo
  {pages} {066503}}\BibitemShut {NoStop}%
\bibitem [{\citenamefont {Wang}\ \emph
  {et~al.}(2017{\natexlab{b}})\citenamefont {Wang}, \citenamefont {He},
  \citenamefont {Tang}, \citenamefont {Wu},\ and\ \citenamefont
  {Cao}}]{Wang2017}%
  \BibitemOpen
  \bibfield  {author} {\bibinfo {author} {\bibnamefont {Wang}, \bibfnamefont
  {Z.}}, \bibinfo {author} {\bibfnamefont {C.}~\bibnamefont {He}}, \bibinfo
  {author} {\bibfnamefont {Z.}~\bibnamefont {Tang}}, \bibinfo {author}
  {\bibfnamefont {S.}~\bibnamefont {Wu}}, and\ \bibinfo {author} {\bibfnamefont
  {G.}~\bibnamefont {Cao}}} (\bibinfo {year} {2017}{\natexlab{b}}),\ \href
  {https://doi.org/10.1007/s40843-016-5150-x} {\bibfield  {journal} {\bibinfo
  {journal} {Science China Materials}\ }\textbf {\bibinfo {volume} {60}},\
  \bibinfo {pages} {83}}\BibitemShut {NoStop}%
\bibitem [{\citenamefont {Wang}\ \emph {et~al.}(2020)\citenamefont {Wang},
  \citenamefont {Rodriguez}, \citenamefont {Jiao}, \citenamefont {Howard},
  \citenamefont {Graham}, \citenamefont {Gu}, \citenamefont {Hughes},
  \citenamefont {Morr},\ and\ \citenamefont {Madhavan}}]{wang2020evidence}%
  \BibitemOpen
  \bibfield  {author} {\bibinfo {author} {\bibnamefont {Wang}, \bibfnamefont
  {Z.}}, \bibinfo {author} {\bibfnamefont {J.~O.}\ \bibnamefont {Rodriguez}},
  \bibinfo {author} {\bibfnamefont {L.}~\bibnamefont {Jiao}}, \bibinfo {author}
  {\bibfnamefont {S.}~\bibnamefont {Howard}}, \bibinfo {author} {\bibfnamefont
  {M.}~\bibnamefont {Graham}}, \bibinfo {author} {\bibfnamefont {G.~D.}\
  \bibnamefont {Gu}}, \bibinfo {author} {\bibfnamefont {T.~L.}\ \bibnamefont
  {Hughes}}, \bibinfo {author} {\bibfnamefont {D.~K.}\ \bibnamefont {Morr}},
  and\ \bibinfo {author} {\bibfnamefont {V.}~\bibnamefont {Madhavan}}}
  (\bibinfo {year} {2020}),\ \href {https://doi.org/10.1126/science.aaw8419}
  {\bibfield  {journal} {\bibinfo  {journal} {Science}\ }\textbf {\bibinfo
  {volume} {367}},\ \bibinfo {pages} {104}}\BibitemShut {NoStop}%
\bibitem [{\citenamefont {Wang}\ \emph {et~al.}(2014)\citenamefont {Wang},
  \citenamefont {Schmidt}, \citenamefont {Fischer}, \citenamefont {Tsurkan},
  \citenamefont {Greger}, \citenamefont {Vollhardt}, \citenamefont {Loidl},\
  and\ \citenamefont {Deisenhofer}}]{wang2014orbitalselective}%
  \BibitemOpen
  \bibfield  {author} {\bibinfo {author} {\bibnamefont {Wang}, \bibfnamefont
  {Z.}}, \bibinfo {author} {\bibfnamefont {M.}~\bibnamefont {Schmidt}},
  \bibinfo {author} {\bibfnamefont {J.}~\bibnamefont {Fischer}}, \bibinfo
  {author} {\bibfnamefont {V.}~\bibnamefont {Tsurkan}}, \bibinfo {author}
  {\bibfnamefont {M.}~\bibnamefont {Greger}}, \bibinfo {author} {\bibfnamefont
  {D.}~\bibnamefont {Vollhardt}}, \bibinfo {author} {\bibfnamefont
  {A.}~\bibnamefont {Loidl}}, and\ \bibinfo {author} {\bibfnamefont
  {J.}~\bibnamefont {Deisenhofer}}} (\bibinfo {year} {2014}),\ \href
  {https://doi.org/10.1038/ncomms4202} {\bibfield  {journal} {\bibinfo
  {journal} {Nat. Commun.}\ }\textbf {\bibinfo {volume} {5}},\ \bibinfo {pages}
  {3202}}\BibitemShut {NoStop}%
\bibitem [{\citenamefont {Wang}\ \emph
  {et~al.}(2015{\natexlab{b}})\citenamefont {Wang}, \citenamefont {Zhang},
  \citenamefont {Xu}, \citenamefont {Zeng}, \citenamefont {Miao}, \citenamefont
  {Xu}, \citenamefont {Qian}, \citenamefont {Weng}, \citenamefont {Richard},
  \citenamefont {Fedorov}, \citenamefont {Ding}, \citenamefont {Dai},\ and\
  \citenamefont {Fang}}]{wang2015topological}%
  \BibitemOpen
  \bibfield  {author} {\bibinfo {author} {\bibnamefont {Wang}, \bibfnamefont
  {Z.}}, \bibinfo {author} {\bibfnamefont {P.}~\bibnamefont {Zhang}}, \bibinfo
  {author} {\bibfnamefont {G.}~\bibnamefont {Xu}}, \bibinfo {author}
  {\bibfnamefont {L.~K.}\ \bibnamefont {Zeng}}, \bibinfo {author}
  {\bibfnamefont {H.}~\bibnamefont {Miao}}, \bibinfo {author} {\bibfnamefont
  {X.}~\bibnamefont {Xu}}, \bibinfo {author} {\bibfnamefont {T.}~\bibnamefont
  {Qian}}, \bibinfo {author} {\bibfnamefont {H.}~\bibnamefont {Weng}}, \bibinfo
  {author} {\bibfnamefont {P.}~\bibnamefont {Richard}}, \bibinfo {author}
  {\bibfnamefont {A.~V.}\ \bibnamefont {Fedorov}}, \bibinfo {author}
  {\bibfnamefont {H.}~\bibnamefont {Ding}}, \bibinfo {author} {\bibfnamefont
  {X.}~\bibnamefont {Dai}}, and\ \bibinfo {author} {\bibfnamefont
  {Z.}~\bibnamefont {Fang}}} (\bibinfo {year} {2015}{\natexlab{b}}),\ \href
  {https://doi.org/10.1103/PhysRevB.92.115119} {\bibfield  {journal} {\bibinfo
  {journal} {Phys. Rev. B}\ }\textbf {\bibinfo {volume} {92}},\ \bibinfo
  {pages} {115119}}\BibitemShut {NoStop}%
\bibitem [{\citenamefont {Wang}\ \emph
  {et~al.}(2016{\natexlab{g}})\citenamefont {Wang}, \citenamefont {He},
  \citenamefont {Wu}, \citenamefont {Tang}, \citenamefont {Liu}, \citenamefont
  {Ablimit}, \citenamefont {Feng},\ and\ \citenamefont {Cao}}]{Wang2016}%
  \BibitemOpen
  \bibfield  {author} {\bibinfo {author} {\bibnamefont {Wang}, \bibfnamefont
  {Z.-C.}}, \bibinfo {author} {\bibfnamefont {C.-Y.}\ \bibnamefont {He}},
  \bibinfo {author} {\bibfnamefont {S.-Q.}\ \bibnamefont {Wu}}, \bibinfo
  {author} {\bibfnamefont {Z.-T.}\ \bibnamefont {Tang}}, \bibinfo {author}
  {\bibfnamefont {Y.}~\bibnamefont {Liu}}, \bibinfo {author} {\bibfnamefont
  {A.}~\bibnamefont {Ablimit}}, \bibinfo {author} {\bibfnamefont {C.-M.}\
  \bibnamefont {Feng}}, and\ \bibinfo {author} {\bibfnamefont {G.-H.}\
  \bibnamefont {Cao}}} (\bibinfo {year} {2016}{\natexlab{g}}),\ \href
  {https://doi.org/10.1021/jacs.6b04538} {\bibfield  {journal} {\bibinfo
  {journal} {J. Am. Chem. Soc.}\ }\textbf {\bibinfo {volume} {138}},\ \bibinfo
  {pages} {7856}}\BibitemShut {NoStop}%
\bibitem [{\citenamefont {Wang}\ \emph
  {et~al.}(2017{\natexlab{c}})\citenamefont {Wang}, \citenamefont {He},
  \citenamefont {Wu}, \citenamefont {Tang}, \citenamefont {Liu}, \citenamefont
  {Ablimit}, \citenamefont {Tao}, \citenamefont {Feng}, \citenamefont {Xu},\
  and\ \citenamefont {Cao}}]{Wu2017a}%
  \BibitemOpen
  \bibfield  {author} {\bibinfo {author} {\bibnamefont {Wang}, \bibfnamefont
  {Z.-C.}}, \bibinfo {author} {\bibfnamefont {C.-Y.}\ \bibnamefont {He}},
  \bibinfo {author} {\bibfnamefont {S.-Q.}\ \bibnamefont {Wu}}, \bibinfo
  {author} {\bibfnamefont {Z.-T.}\ \bibnamefont {Tang}}, \bibinfo {author}
  {\bibfnamefont {Y.}~\bibnamefont {Liu}}, \bibinfo {author} {\bibfnamefont
  {A.}~\bibnamefont {Ablimit}}, \bibinfo {author} {\bibfnamefont
  {Q.}~\bibnamefont {Tao}}, \bibinfo {author} {\bibfnamefont {C.-M.}\
  \bibnamefont {Feng}}, \bibinfo {author} {\bibfnamefont {Z.-A.}\ \bibnamefont
  {Xu}}, and\ \bibinfo {author} {\bibfnamefont {G.-H.}\ \bibnamefont {Cao}}}
  (\bibinfo {year} {2017}{\natexlab{c}}),\ \href
  {https://doi.org/10.1088/1361-648x/aa58d2} {\bibfield  {journal} {\bibinfo
  {journal} {J. Phys.: Condens. Matter}\ }\textbf {\bibinfo {volume} {29}},\
  \bibinfo {pages} {11LT01}}\BibitemShut {NoStop}%
\bibitem [{\citenamefont {Wang}\ \emph
  {et~al.}(2017{\natexlab{d}})\citenamefont {Wang}, \citenamefont {He},
  \citenamefont {Wu}, \citenamefont {Tang}, \citenamefont {Liu},\ and\
  \citenamefont {Cao}}]{Wu2017b}%
  \BibitemOpen
  \bibfield  {author} {\bibinfo {author} {\bibnamefont {Wang}, \bibfnamefont
  {Z.-C.}}, \bibinfo {author} {\bibfnamefont {C.-Y.}\ \bibnamefont {He}},
  \bibinfo {author} {\bibfnamefont {S.-Q.}\ \bibnamefont {Wu}}, \bibinfo
  {author} {\bibfnamefont {Z.-T.}\ \bibnamefont {Tang}}, \bibinfo {author}
  {\bibfnamefont {Y.}~\bibnamefont {Liu}}, and\ \bibinfo {author}
  {\bibfnamefont {G.-H.}\ \bibnamefont {Cao}}} (\bibinfo {year}
  {2017}{\natexlab{d}}),\ \href {https://doi.org/10.1021/acs.chemmater.6b05458}
  {\bibfield  {journal} {\bibinfo  {journal} {Chemistry of Materials}\ }\textbf
  {\bibinfo {volume} {29}},\ \bibinfo {pages} {1805}}\BibitemShut {NoStop}%
\bibitem [{\citenamefont {Wa{\ss}er}\ \emph {et~al.}(2017)\citenamefont
  {Wa{\ss}er}, \citenamefont {Lee}, \citenamefont {Kihou}, \citenamefont
  {Steffens}, \citenamefont {Schmalzl}, \citenamefont {Qureshi},\ and\
  \citenamefont {Braden}}]{wasser2017anisotropic}%
  \BibitemOpen
  \bibfield  {author} {\bibinfo {author} {\bibnamefont {Wa{\ss}er},
  \bibfnamefont {F.}}, \bibinfo {author} {\bibfnamefont {C.~H.}\ \bibnamefont
  {Lee}}, \bibinfo {author} {\bibfnamefont {K.}~\bibnamefont {Kihou}}, \bibinfo
  {author} {\bibfnamefont {P.}~\bibnamefont {Steffens}}, \bibinfo {author}
  {\bibfnamefont {K.}~\bibnamefont {Schmalzl}}, \bibinfo {author}
  {\bibfnamefont {N.}~\bibnamefont {Qureshi}}, and\ \bibinfo {author}
  {\bibfnamefont {M.}~\bibnamefont {Braden}}} (\bibinfo {year} {2017}),\ \href
  {https://doi.org/10.1038/s41598-017-10208-1} {\bibfield  {journal} {\bibinfo
  {journal} {Sci. Rep.}\ }\textbf {\bibinfo {volume} {7}},\ \bibinfo {pages}
  {10307}}\BibitemShut {NoStop}%
\bibitem [{\citenamefont {Watashige}\ \emph {et~al.}(2015)\citenamefont
  {Watashige}, \citenamefont {Tsutsumi}, \citenamefont {Hanaguri},
  \citenamefont {Kohsaka}, \citenamefont {Kasahara}, \citenamefont {Furusaki},
  \citenamefont {Sigrist}, \citenamefont {Meingast}, \citenamefont {Wolf},
  \citenamefont {v.~L{\"o}hneysen}, \citenamefont {Shibauchi},\ and\
  \citenamefont {Matsuda}}]{watashige2015evidence}%
  \BibitemOpen
  \bibfield  {author} {\bibinfo {author} {\bibnamefont {Watashige},
  \bibfnamefont {T.}}, \bibinfo {author} {\bibfnamefont {Y.}~\bibnamefont
  {Tsutsumi}}, \bibinfo {author} {\bibfnamefont {T.}~\bibnamefont {Hanaguri}},
  \bibinfo {author} {\bibfnamefont {Y.}~\bibnamefont {Kohsaka}}, \bibinfo
  {author} {\bibfnamefont {S.}~\bibnamefont {Kasahara}}, \bibinfo {author}
  {\bibfnamefont {A.}~\bibnamefont {Furusaki}}, \bibinfo {author}
  {\bibfnamefont {M.}~\bibnamefont {Sigrist}}, \bibinfo {author} {\bibfnamefont
  {C.}~\bibnamefont {Meingast}}, \bibinfo {author} {\bibfnamefont
  {T.}~\bibnamefont {Wolf}}, \bibinfo {author} {\bibfnamefont {H.}~\bibnamefont
  {v.~L{\"o}hneysen}}, \bibinfo {author} {\bibfnamefont {T.}~\bibnamefont
  {Shibauchi}}, and\ \bibinfo {author} {\bibfnamefont {Y.}~\bibnamefont
  {Matsuda}}} (\bibinfo {year} {2015}),\ \href
  {https://doi.org/10.1103/PhysRevX.5.031022} {\bibfield  {journal} {\bibinfo
  {journal} {Phys. Rev. X}\ }\textbf {\bibinfo {volume} {5}},\ \bibinfo {pages}
  {031022}}\BibitemShut {NoStop}%
\bibitem [{\citenamefont {Watson}\ \emph {et~al.}(2015)\citenamefont {Watson},
  \citenamefont {Yamashita}, \citenamefont {Kasahara}, \citenamefont {Knafo},
  \citenamefont {Nardone}, \citenamefont {B{\'e}ard}, \citenamefont {Hardy},
  \citenamefont {McCollam}, \citenamefont {Narayanan}, \citenamefont {Blake},
  \citenamefont {Wolf}, \citenamefont {Haghighirad}, \citenamefont {Meingast},
  \citenamefont {Schofield}, \citenamefont {V.~L{\"o}hneysen}, \citenamefont
  {Matsuda}, \citenamefont {Coldea},\ and\ \citenamefont
  {Shibauchi}}]{watson2015}%
  \BibitemOpen
  \bibfield  {author} {\bibinfo {author} {\bibnamefont {Watson}, \bibfnamefont
  {M.~D.}}, \bibinfo {author} {\bibfnamefont {T.}~\bibnamefont {Yamashita}},
  \bibinfo {author} {\bibfnamefont {S.}~\bibnamefont {Kasahara}}, \bibinfo
  {author} {\bibfnamefont {W.}~\bibnamefont {Knafo}}, \bibinfo {author}
  {\bibfnamefont {M.}~\bibnamefont {Nardone}}, \bibinfo {author} {\bibfnamefont
  {J.}~\bibnamefont {B{\'e}ard}}, \bibinfo {author} {\bibfnamefont
  {F.}~\bibnamefont {Hardy}}, \bibinfo {author} {\bibfnamefont
  {A.}~\bibnamefont {McCollam}}, \bibinfo {author} {\bibfnamefont
  {A.}~\bibnamefont {Narayanan}}, \bibinfo {author} {\bibfnamefont {S.~F.}\
  \bibnamefont {Blake}}, \bibinfo {author} {\bibfnamefont {T.}~\bibnamefont
  {Wolf}}, \bibinfo {author} {\bibfnamefont {A.~A.}\ \bibnamefont
  {Haghighirad}}, \bibinfo {author} {\bibfnamefont {C.}~\bibnamefont
  {Meingast}}, \bibinfo {author} {\bibfnamefont {A.~J.}\ \bibnamefont
  {Schofield}}, \bibinfo {author} {\bibfnamefont {H.}~\bibnamefont
  {V.~L{\"o}hneysen}}, \bibinfo {author} {\bibfnamefont {Y.}~\bibnamefont
  {Matsuda}}, \bibinfo {author} {\bibfnamefont {A.~I.}\ \bibnamefont {Coldea}},
  and\ \bibinfo {author} {\bibfnamefont {T.}~\bibnamefont {Shibauchi}}}
  (\bibinfo {year} {2015}),\ \href
  {https://doi.org/10.1103/PhysRevLett.115.027006} {\bibfield  {journal}
  {\bibinfo  {journal} {Phys. Rev. Lett.}\ }\textbf {\bibinfo {volume} {115}},\
  \bibinfo {pages} {027006}}\BibitemShut {NoStop}%
\bibitem [{\citenamefont {Waßer}\ \emph {et~al.}(2019)\citenamefont {Waßer},
  \citenamefont {Park}, \citenamefont {Aswartham}, \citenamefont {Wurmehl},
  \citenamefont {Sidis}, \citenamefont {Steffens}, \citenamefont {Schmalzl},
  \citenamefont {B\"{u}chner},\ and\ \citenamefont {Braden}}]{Waer2019strong}%
  \BibitemOpen
  \bibfield  {author} {\bibinfo {author} {\bibnamefont {Waßer}, \bibfnamefont
  {F.}}, \bibinfo {author} {\bibfnamefont {J.~T.}\ \bibnamefont {Park}},
  \bibinfo {author} {\bibfnamefont {S.}~\bibnamefont {Aswartham}}, \bibinfo
  {author} {\bibfnamefont {S.}~\bibnamefont {Wurmehl}}, \bibinfo {author}
  {\bibfnamefont {Y.}~\bibnamefont {Sidis}}, \bibinfo {author} {\bibfnamefont
  {P.}~\bibnamefont {Steffens}}, \bibinfo {author} {\bibfnamefont
  {K.}~\bibnamefont {Schmalzl}}, \bibinfo {author} {\bibfnamefont
  {B.}~\bibnamefont {B\"{u}chner}}, and\ \bibinfo {author} {\bibfnamefont
  {M.}~\bibnamefont {Braden}}} (\bibinfo {year} {2019}),\ \href
  {https://doi.org/10.1038/s41535-019-0198-4} {\bibfield  {journal} {\bibinfo
  {journal} {npj Quantum Mater.}\ }\textbf {\bibinfo {volume} {4}},\ \bibinfo
  {pages} {59}}\BibitemShut {NoStop}%
\bibitem [{\citenamefont {Weber}\ \emph {et~al.}(2018)\citenamefont {Weber},
  \citenamefont {Parshall}, \citenamefont {Pintschovius}, \citenamefont
  {Castellan}, \citenamefont {Kauth}, \citenamefont {Merz}, \citenamefont
  {Wolf}, \citenamefont {Sch\"utt}, \citenamefont {Schmalian}, \citenamefont
  {Fernandes},\ and\ \citenamefont {Reznik}}]{weber2018soft}%
  \BibitemOpen
  \bibfield  {author} {\bibinfo {author} {\bibnamefont {Weber}, \bibfnamefont
  {F.}}, \bibinfo {author} {\bibfnamefont {D.}~\bibnamefont {Parshall}},
  \bibinfo {author} {\bibfnamefont {L.}~\bibnamefont {Pintschovius}}, \bibinfo
  {author} {\bibfnamefont {J.-P.}\ \bibnamefont {Castellan}}, \bibinfo {author}
  {\bibfnamefont {M.}~\bibnamefont {Kauth}}, \bibinfo {author} {\bibfnamefont
  {M.}~\bibnamefont {Merz}}, \bibinfo {author} {\bibfnamefont {T.}~\bibnamefont
  {Wolf}}, \bibinfo {author} {\bibfnamefont {M.}~\bibnamefont {Sch\"utt}},
  \bibinfo {author} {\bibfnamefont {J.}~\bibnamefont {Schmalian}}, \bibinfo
  {author} {\bibfnamefont {R.~M.}\ \bibnamefont {Fernandes}}, and\ \bibinfo
  {author} {\bibfnamefont {D.}~\bibnamefont {Reznik}}} (\bibinfo {year}
  {2018}),\ \href {https://doi.org/10.1103/PhysRevB.98.014516} {\bibfield
  {journal} {\bibinfo  {journal} {Phys. Rev. B}\ }\textbf {\bibinfo {volume}
  {98}},\ \bibinfo {pages} {014516}}\BibitemShut {NoStop}%
\bibitem [{\citenamefont {Wei}\ \emph {et~al.}(2025)\citenamefont {Wei},
  \citenamefont {Liu}, \citenamefont {Ren}, \citenamefont {Liang},
  \citenamefont {Wang},\ and\ \citenamefont {Wang}}]{wei2025observation}%
  \BibitemOpen
  \bibfield  {author} {\bibinfo {author} {\bibnamefont {Wei}, \bibfnamefont
  {T.}}, \bibinfo {author} {\bibfnamefont {Y.}~\bibnamefont {Liu}}, \bibinfo
  {author} {\bibfnamefont {W.}~\bibnamefont {Ren}}, \bibinfo {author}
  {\bibfnamefont {Z.}~\bibnamefont {Liang}}, \bibinfo {author} {\bibfnamefont
  {Z.}~\bibnamefont {Wang}}, and\ \bibinfo {author} {\bibfnamefont
  {J.}~\bibnamefont {Wang}}} (\bibinfo {year} {2025}),\ \href
  {https://doi.org/10.1088/0256-307X/42/2/027404} {\bibfield  {journal}
  {\bibinfo  {journal} {Chin. Phys. Lett.}\ }\textbf {\bibinfo {volume} {42}},\
  \bibinfo {pages} {027404}}\BibitemShut {NoStop}%
\bibitem [{\citenamefont {Wei}\ \emph {et~al.}(2023)\citenamefont {Wei},
  \citenamefont {Qin}, \citenamefont {Ding}, \citenamefont {Wu}, \citenamefont
  {Hu}, \citenamefont {Sun}, \citenamefont {Wang},\ and\ \citenamefont
  {Xue}}]{wei2023identifying}%
  \BibitemOpen
  \bibfield  {author} {\bibinfo {author} {\bibnamefont {Wei}, \bibfnamefont
  {Z.}}, \bibinfo {author} {\bibfnamefont {S.}~\bibnamefont {Qin}}, \bibinfo
  {author} {\bibfnamefont {C.}~\bibnamefont {Ding}}, \bibinfo {author}
  {\bibfnamefont {X.}~\bibnamefont {Wu}}, \bibinfo {author} {\bibfnamefont
  {J.}~\bibnamefont {Hu}}, \bibinfo {author} {\bibfnamefont {Y.-J.}\
  \bibnamefont {Sun}}, \bibinfo {author} {\bibfnamefont {L.}~\bibnamefont
  {Wang}}, and\ \bibinfo {author} {\bibfnamefont {Q.-K.}\ \bibnamefont {Xue}}}
  (\bibinfo {year} {2023}),\ \href {https://doi.org/10.1038/s41467-023-40931-5}
  {\bibfield  {journal} {\bibinfo  {journal} {Nat. Commun.}\ }\textbf {\bibinfo
  {volume} {14}},\ \bibinfo {pages} {5302}}\BibitemShut {NoStop}%
\bibitem [{\citenamefont {Weiss}\ \emph {et~al.}(2012)\citenamefont {Weiss},
  \citenamefont {Tarantini}, \citenamefont {Jiang}, \citenamefont {Kametani},
  \citenamefont {Polyanskii}, \citenamefont {Larbalestier},\ and\ \citenamefont
  {Hellstrom}}]{Weiss2012}%
  \BibitemOpen
  \bibfield  {author} {\bibinfo {author} {\bibnamefont {Weiss}, \bibfnamefont
  {J.}}, \bibinfo {author} {\bibfnamefont {C.}~\bibnamefont {Tarantini}},
  \bibinfo {author} {\bibfnamefont {J.}~\bibnamefont {Jiang}}, \bibinfo
  {author} {\bibfnamefont {F.}~\bibnamefont {Kametani}}, \bibinfo {author}
  {\bibfnamefont {A.}~\bibnamefont {Polyanskii}}, \bibinfo {author}
  {\bibfnamefont {D.}~\bibnamefont {Larbalestier}}, and\ \bibinfo {author}
  {\bibfnamefont {E.}~\bibnamefont {Hellstrom}}} (\bibinfo {year} {2012}),\
  \href {https://doi.org/10.1038/nmat3333} {\bibfield  {journal} {\bibinfo
  {journal} {Nat. Mater.}\ }\textbf {\bibinfo {volume} {11}},\ \bibinfo {pages}
  {682}}\BibitemShut {NoStop}%
\bibitem [{\citenamefont {Wen}\ \emph {et~al.}(2016)\citenamefont {Wen},
  \citenamefont {Xu}, \citenamefont {Chen}, \citenamefont {Huang},
  \citenamefont {Lou}, \citenamefont {Pu}, \citenamefont {Song}, \citenamefont
  {Xie}, \citenamefont {{Abdel-Hafiez}}, \citenamefont {Chareev}, \citenamefont
  {Vasiliev}, \citenamefont {Peng},\ and\ \citenamefont
  {Feng}}]{wen2016anomalous}%
  \BibitemOpen
  \bibfield  {author} {\bibinfo {author} {\bibnamefont {Wen}, \bibfnamefont
  {C.~H.~P.}}, \bibinfo {author} {\bibfnamefont {H.~C.}\ \bibnamefont {Xu}},
  \bibinfo {author} {\bibfnamefont {C.}~\bibnamefont {Chen}}, \bibinfo {author}
  {\bibfnamefont {Z.~C.}\ \bibnamefont {Huang}}, \bibinfo {author}
  {\bibfnamefont {X.}~\bibnamefont {Lou}}, \bibinfo {author} {\bibfnamefont
  {Y.~J.}\ \bibnamefont {Pu}}, \bibinfo {author} {\bibfnamefont
  {Q.}~\bibnamefont {Song}}, \bibinfo {author} {\bibfnamefont {B.~P.}\
  \bibnamefont {Xie}}, \bibinfo {author} {\bibfnamefont {M.}~\bibnamefont
  {{Abdel-Hafiez}}}, \bibinfo {author} {\bibfnamefont {D.~A.}\ \bibnamefont
  {Chareev}}, \bibinfo {author} {\bibfnamefont {A.~N.}\ \bibnamefont
  {Vasiliev}}, \bibinfo {author} {\bibfnamefont {R.}~\bibnamefont {Peng}}, and\
  \bibinfo {author} {\bibfnamefont {D.~L.}\ \bibnamefont {Feng}}} (\bibinfo
  {year} {2016}),\ \href {https://doi.org/10.1038/ncomms10840} {\bibfield
  {journal} {\bibinfo  {journal} {Nat. Commun.}\ }\textbf {\bibinfo {volume}
  {7}},\ \bibinfo {pages} {10840}}\BibitemShut {NoStop}%
\bibitem [{\citenamefont {Wen}\ \emph {et~al.}(2008)\citenamefont {Wen},
  \citenamefont {Mu}, \citenamefont {Fang}, \citenamefont {Yang},\ and\
  \citenamefont {Zhu}}]{Wen2008}%
  \BibitemOpen
  \bibfield  {author} {\bibinfo {author} {\bibnamefont {Wen}, \bibfnamefont
  {H.-H.}}, \bibinfo {author} {\bibfnamefont {G.}~\bibnamefont {Mu}}, \bibinfo
  {author} {\bibfnamefont {L.}~\bibnamefont {Fang}}, \bibinfo {author}
  {\bibfnamefont {H.}~\bibnamefont {Yang}}, and\ \bibinfo {author}
  {\bibfnamefont {X.}~\bibnamefont {Zhu}}} (\bibinfo {year} {2008}),\ \href
  {https://doi.org/10.1209/0295-5075/82/17009} {\bibfield  {journal} {\bibinfo
  {journal} {EPL}\ }\textbf {\bibinfo {volume} {82}},\ \bibinfo {pages}
  {17009}}\BibitemShut {NoStop}%
\bibitem [{\citenamefont {Wiecki}\ \emph {et~al.}(2021)\citenamefont {Wiecki},
  \citenamefont {Frachet}, \citenamefont {Haghighirad}, \citenamefont {Wolf},
  \citenamefont {Meingast}, \citenamefont {Heid},\ and\ \citenamefont
  {B{\"o}hmer}}]{wiecki2021emerging}%
  \BibitemOpen
  \bibfield  {author} {\bibinfo {author} {\bibnamefont {Wiecki}, \bibfnamefont
  {P.}}, \bibinfo {author} {\bibfnamefont {M.}~\bibnamefont {Frachet}},
  \bibinfo {author} {\bibfnamefont {A.-A.}\ \bibnamefont {Haghighirad}},
  \bibinfo {author} {\bibfnamefont {T.}~\bibnamefont {Wolf}}, \bibinfo {author}
  {\bibfnamefont {C.}~\bibnamefont {Meingast}}, \bibinfo {author}
  {\bibfnamefont {R.}~\bibnamefont {Heid}}, and\ \bibinfo {author}
  {\bibfnamefont {A.~E.}\ \bibnamefont {B{\"o}hmer}}} (\bibinfo {year}
  {2021}),\ \href {https://doi.org/10.1038/s41467-021-25121-5} {\bibfield
  {journal} {\bibinfo  {journal} {Nat. Commun.}\ }\textbf {\bibinfo {volume}
  {12}},\ \bibinfo {pages} {4824}}\BibitemShut {NoStop}%
\bibitem [{\citenamefont {Wiecki}\ \emph {et~al.}(2020)\citenamefont {Wiecki},
  \citenamefont {Haghighirad}, \citenamefont {Weber}, \citenamefont {Merz},
  \citenamefont {Heid},\ and\ \citenamefont {B\"ohmer}}]{wiecki2020dominant}%
  \BibitemOpen
  \bibfield  {author} {\bibinfo {author} {\bibnamefont {Wiecki}, \bibfnamefont
  {P.}}, \bibinfo {author} {\bibfnamefont {A.-A.}\ \bibnamefont {Haghighirad}},
  \bibinfo {author} {\bibfnamefont {F.}~\bibnamefont {Weber}}, \bibinfo
  {author} {\bibfnamefont {M.}~\bibnamefont {Merz}}, \bibinfo {author}
  {\bibfnamefont {R.}~\bibnamefont {Heid}}, and\ \bibinfo {author}
  {\bibfnamefont {A.~E.}\ \bibnamefont {B\"ohmer}}} (\bibinfo {year} {2020}),\
  \href {https://doi.org/10.1103/PhysRevLett.125.187001} {\bibfield  {journal}
  {\bibinfo  {journal} {Phys. Rev. Lett.}\ }\textbf {\bibinfo {volume} {125}},\
  \bibinfo {pages} {187001}}\BibitemShut {NoStop}%
\bibitem [{\citenamefont {Williams}\ \emph {et~al.}(2009)\citenamefont
  {Williams}, \citenamefont {McQueen}, \citenamefont {Ksenofontov},
  \citenamefont {Felser},\ and\ \citenamefont {Cava}}]{Williams2009}%
  \BibitemOpen
  \bibfield  {author} {\bibinfo {author} {\bibnamefont {Williams},
  \bibfnamefont {A.}}, \bibinfo {author} {\bibfnamefont {T.}~\bibnamefont
  {McQueen}}, \bibinfo {author} {\bibfnamefont {V.}~\bibnamefont
  {Ksenofontov}}, \bibinfo {author} {\bibfnamefont {C.}~\bibnamefont {Felser}},
  and\ \bibinfo {author} {\bibfnamefont {R.}~\bibnamefont {Cava}}} (\bibinfo
  {year} {2009}),\ \href {https://doi.org/10.1088/0953-8984/21/30/305701}
  {\bibfield  {journal} {\bibinfo  {journal} {J. Phys.: Condens. Matter}\
  }\textbf {\bibinfo {volume} {21}},\ \bibinfo {pages} {305701}}\BibitemShut
  {NoStop}%
\bibitem [{\citenamefont {Wilson}\ and\ \citenamefont
  {Ortiz}(2024)}]{Wilson2024AV3Sb5}%
  \BibitemOpen
  \bibfield  {author} {\bibinfo {author} {\bibnamefont {Wilson}, \bibfnamefont
  {S.~D.}}, and\ \bibinfo {author} {\bibfnamefont {B.~R.}\ \bibnamefont
  {Ortiz}}} (\bibinfo {year} {2024}),\ \href
  {https://doi.org/10.1038/s41578-024-00677-y} {\bibfield  {journal} {\bibinfo
  {journal} {Nature Reviews Materials}\ }\textbf {\bibinfo {volume} {9}},\
  \bibinfo {pages} {420}}\BibitemShut {NoStop}%
\bibitem [{\citenamefont {Wo}\ \emph {et~al.}(2025)\citenamefont {Wo},
  \citenamefont {Pan}, \citenamefont {Hu}, \citenamefont {Feng}, \citenamefont
  {Christianson},\ and\ \citenamefont {Zhao}}]{wo2025spin}%
  \BibitemOpen
  \bibfield  {author} {\bibinfo {author} {\bibnamefont {Wo}, \bibfnamefont
  {H.}}, \bibinfo {author} {\bibfnamefont {B.}~\bibnamefont {Pan}}, \bibinfo
  {author} {\bibfnamefont {D.}~\bibnamefont {Hu}}, \bibinfo {author}
  {\bibfnamefont {Y.}~\bibnamefont {Feng}}, \bibinfo {author} {\bibfnamefont
  {A.~D.}\ \bibnamefont {Christianson}}, and\ \bibinfo {author} {\bibfnamefont
  {J.}~\bibnamefont {Zhao}}} (\bibinfo {year} {2025}),\ \href
  {https://doi.org/10.1103/PhysRevLett.134.016501} {\bibfield  {journal}
  {\bibinfo  {journal} {Phys. Rev. Lett.}\ }\textbf {\bibinfo {volume} {134}},\
  \bibinfo {pages} {016501}}\BibitemShut {NoStop}%
\bibitem [{\citenamefont {Wo}\ \emph {et~al.}(2019)\citenamefont {Wo},
  \citenamefont {Wang}, \citenamefont {Shen}, \citenamefont {Zhang},
  \citenamefont {Hao}, \citenamefont {Feng}, \citenamefont {Shen},
  \citenamefont {He}, \citenamefont {Pan}, \citenamefont {Wang}, \citenamefont
  {Nakajima}, \citenamefont {{Ohira-Kawamura}}, \citenamefont {Steffens},
  \citenamefont {Boehm}, \citenamefont {Schmalzl}, \citenamefont {Forrest},
  \citenamefont {Matsuda}, \citenamefont {Zhao}, \citenamefont {Lynn},
  \citenamefont {Yin},\ and\ \citenamefont {Zhao}}]{wo2019coexistence}%
  \BibitemOpen
  \bibfield  {author} {\bibinfo {author} {\bibnamefont {Wo}, \bibfnamefont
  {H.}}, \bibinfo {author} {\bibfnamefont {Q.}~\bibnamefont {Wang}}, \bibinfo
  {author} {\bibfnamefont {Y.}~\bibnamefont {Shen}}, \bibinfo {author}
  {\bibfnamefont {X.}~\bibnamefont {Zhang}}, \bibinfo {author} {\bibfnamefont
  {Y.}~\bibnamefont {Hao}}, \bibinfo {author} {\bibfnamefont {Y.}~\bibnamefont
  {Feng}}, \bibinfo {author} {\bibfnamefont {S.}~\bibnamefont {Shen}}, \bibinfo
  {author} {\bibfnamefont {Z.}~\bibnamefont {He}}, \bibinfo {author}
  {\bibfnamefont {B.}~\bibnamefont {Pan}}, \bibinfo {author} {\bibfnamefont
  {W.}~\bibnamefont {Wang}}, \bibinfo {author} {\bibfnamefont {K.}~\bibnamefont
  {Nakajima}}, \bibinfo {author} {\bibfnamefont {S.}~\bibnamefont
  {{Ohira-Kawamura}}}, \bibinfo {author} {\bibfnamefont {P.}~\bibnamefont
  {Steffens}}, \bibinfo {author} {\bibfnamefont {M.}~\bibnamefont {Boehm}},
  \bibinfo {author} {\bibfnamefont {K.}~\bibnamefont {Schmalzl}}, \bibinfo
  {author} {\bibfnamefont {T.~R.}\ \bibnamefont {Forrest}}, \bibinfo {author}
  {\bibfnamefont {M.}~\bibnamefont {Matsuda}}, \bibinfo {author} {\bibfnamefont
  {Y.}~\bibnamefont {Zhao}}, \bibinfo {author} {\bibfnamefont {J.~W.}\
  \bibnamefont {Lynn}}, \bibinfo {author} {\bibfnamefont {Z.}~\bibnamefont
  {Yin}}, and\ \bibinfo {author} {\bibfnamefont {J.}~\bibnamefont {Zhao}}}
  (\bibinfo {year} {2019}),\ \href
  {https://doi.org/10.1103/PhysRevLett.122.217003} {\bibfield  {journal}
  {\bibinfo  {journal} {Phys. Rev. Lett.}\ }\textbf {\bibinfo {volume} {122}},\
  \bibinfo {pages} {217003}}\BibitemShut {NoStop}%
\bibitem [{\citenamefont {Worasaran}\ \emph {et~al.}(2021)\citenamefont
  {Worasaran}, \citenamefont {Ikeda}, \citenamefont {Palmstrom}, \citenamefont
  {Straquadine}, \citenamefont {Kivelson},\ and\ \citenamefont
  {Fisher}}]{worasaran2021nematic}%
  \BibitemOpen
  \bibfield  {author} {\bibinfo {author} {\bibnamefont {Worasaran},
  \bibfnamefont {T.}}, \bibinfo {author} {\bibfnamefont {M.~S.}\ \bibnamefont
  {Ikeda}}, \bibinfo {author} {\bibfnamefont {J.~C.}\ \bibnamefont
  {Palmstrom}}, \bibinfo {author} {\bibfnamefont {J.~A.~W.}\ \bibnamefont
  {Straquadine}}, \bibinfo {author} {\bibfnamefont {S.~A.}\ \bibnamefont
  {Kivelson}}, and\ \bibinfo {author} {\bibfnamefont {I.~R.}\ \bibnamefont
  {Fisher}}} (\bibinfo {year} {2021}),\ \href
  {https://doi.org/10.1126/science.abb9280} {\bibfield  {journal} {\bibinfo
  {journal} {Science}\ }\textbf {\bibinfo {volume} {372}},\ \bibinfo {pages}
  {973}}\BibitemShut {NoStop}%
\bibitem [{\citenamefont {Wu}\ \emph {et~al.}(2011)\citenamefont {Wu},
  \citenamefont {Chanda}, \citenamefont {Jeevan}, \citenamefont {Gegenwart},\
  and\ \citenamefont {Dressel}}]{Wu2009}%
  \BibitemOpen
  \bibfield  {author} {\bibinfo {author} {\bibnamefont {Wu}, \bibfnamefont
  {D.}}, \bibinfo {author} {\bibfnamefont {G.}~\bibnamefont {Chanda}}, \bibinfo
  {author} {\bibfnamefont {H.~S.}\ \bibnamefont {Jeevan}}, \bibinfo {author}
  {\bibfnamefont {P.}~\bibnamefont {Gegenwart}}, and\ \bibinfo {author}
  {\bibfnamefont {M.}~\bibnamefont {Dressel}}} (\bibinfo {year} {2011}),\ \href
  {https://doi.org/10.1103/PhysRevB.83.100503} {\bibfield  {journal} {\bibinfo
  {journal} {Phys. Rev. B}\ }\textbf {\bibinfo {volume} {83}},\ \bibinfo
  {pages} {100503}}\BibitemShut {NoStop}%
\bibitem [{\citenamefont {Wu}\ \emph {et~al.}(2024)\citenamefont {Wu},
  \citenamefont {Jia}, \citenamefont {Yang}, \citenamefont {Hong},
  \citenamefont {Shu}, \citenamefont {Miao}, \citenamefont {Yan}, \citenamefont
  {Rong}, \citenamefont {Ai}, \citenamefont {Zhang}, \citenamefont {Yin},
  \citenamefont {Liu}, \citenamefont {Chen}, \citenamefont {Yang},
  \citenamefont {Peng}, \citenamefont {Li}, \citenamefont {Zhang},
  \citenamefont {Zhang}, \citenamefont {Yang}, \citenamefont {Wang},
  \citenamefont {Zong}, \citenamefont {Liu}, \citenamefont {Li}, \citenamefont
  {Wang}, \citenamefont {Peng}, \citenamefont {Mao}, \citenamefont {Liu},
  \citenamefont {Li}, \citenamefont {Chen}, \citenamefont {Luo}, \citenamefont
  {Wu}, \citenamefont {Xu}, \citenamefont {Zhao},\ and\ \citenamefont
  {Zhou}}]{wu2024nodal}%
  \BibitemOpen
  \bibfield  {author} {\bibinfo {author} {\bibnamefont {Wu}, \bibfnamefont
  {D.}}, \bibinfo {author} {\bibfnamefont {J.}~\bibnamefont {Jia}}, \bibinfo
  {author} {\bibfnamefont {J.}~\bibnamefont {Yang}}, \bibinfo {author}
  {\bibfnamefont {W.}~\bibnamefont {Hong}}, \bibinfo {author} {\bibfnamefont
  {Y.}~\bibnamefont {Shu}}, \bibinfo {author} {\bibfnamefont {T.}~\bibnamefont
  {Miao}}, \bibinfo {author} {\bibfnamefont {H.}~\bibnamefont {Yan}}, \bibinfo
  {author} {\bibfnamefont {H.}~\bibnamefont {Rong}}, \bibinfo {author}
  {\bibfnamefont {P.}~\bibnamefont {Ai}}, \bibinfo {author} {\bibfnamefont
  {X.}~\bibnamefont {Zhang}}, \bibinfo {author} {\bibfnamefont
  {C.}~\bibnamefont {Yin}}, \bibinfo {author} {\bibfnamefont {J.}~\bibnamefont
  {Liu}}, \bibinfo {author} {\bibfnamefont {H.}~\bibnamefont {Chen}}, \bibinfo
  {author} {\bibfnamefont {Y.}~\bibnamefont {Yang}}, \bibinfo {author}
  {\bibfnamefont {C.}~\bibnamefont {Peng}}, \bibinfo {author} {\bibfnamefont
  {C.}~\bibnamefont {Li}}, \bibinfo {author} {\bibfnamefont {S.}~\bibnamefont
  {Zhang}}, \bibinfo {author} {\bibfnamefont {F.}~\bibnamefont {Zhang}},
  \bibinfo {author} {\bibfnamefont {F.}~\bibnamefont {Yang}}, \bibinfo {author}
  {\bibfnamefont {Z.}~\bibnamefont {Wang}}, \bibinfo {author} {\bibfnamefont
  {N.}~\bibnamefont {Zong}}, \bibinfo {author} {\bibfnamefont {L.}~\bibnamefont
  {Liu}}, \bibinfo {author} {\bibfnamefont {R.}~\bibnamefont {Li}}, \bibinfo
  {author} {\bibfnamefont {X.}~\bibnamefont {Wang}}, \bibinfo {author}
  {\bibfnamefont {Q.}~\bibnamefont {Peng}}, \bibinfo {author} {\bibfnamefont
  {H.}~\bibnamefont {Mao}}, \bibinfo {author} {\bibfnamefont {G.}~\bibnamefont
  {Liu}}, \bibinfo {author} {\bibfnamefont {S.}~\bibnamefont {Li}}, \bibinfo
  {author} {\bibfnamefont {Y.}~\bibnamefont {Chen}}, \bibinfo {author}
  {\bibfnamefont {H.}~\bibnamefont {Luo}}, \bibinfo {author} {\bibfnamefont
  {X.}~\bibnamefont {Wu}}, \bibinfo {author} {\bibfnamefont {Z.}~\bibnamefont
  {Xu}}, \bibinfo {author} {\bibfnamefont {L.}~\bibnamefont {Zhao}}, and\
  \bibinfo {author} {\bibfnamefont {X.~J.}\ \bibnamefont {Zhou}}} (\bibinfo
  {year} {2024}),\ \href {https://doi.org/10.1038/s41567-023-02348-1}
  {\bibfield  {journal} {\bibinfo  {journal} {Nat. Phys.}\ }\textbf {\bibinfo
  {volume} {20}},\ \bibinfo {pages} {571}}\BibitemShut {NoStop}%
\bibitem [{\citenamefont {Wu}\ \emph {et~al.}(2017)\citenamefont {Wu},
  \citenamefont {Wang}, \citenamefont {He}, \citenamefont {Tang}, \citenamefont
  {Liu},\ and\ \citenamefont {Cao}}]{Wu2017c}%
  \BibitemOpen
  \bibfield  {author} {\bibinfo {author} {\bibnamefont {Wu}, \bibfnamefont
  {S.-Q.}}, \bibinfo {author} {\bibfnamefont {Z.-C.}\ \bibnamefont {Wang}},
  \bibinfo {author} {\bibfnamefont {C.-Y.}\ \bibnamefont {He}}, \bibinfo
  {author} {\bibfnamefont {Z.-T.}\ \bibnamefont {Tang}}, \bibinfo {author}
  {\bibfnamefont {Y.}~\bibnamefont {Liu}}, and\ \bibinfo {author}
  {\bibfnamefont {G.-H.}\ \bibnamefont {Cao}}} (\bibinfo {year} {2017}),\ \href
  {https://doi.org/10.1103/PhysRevMaterials.1.044804} {\bibfield  {journal}
  {\bibinfo  {journal} {Phys. Rev. Mater.}\ }\textbf {\bibinfo {volume} {1}},\
  \bibinfo {pages} {044804}}\BibitemShut {NoStop}%
\bibitem [{\citenamefont {Wu}\ \emph {et~al.}(2016)\citenamefont {Wu},
  \citenamefont {Qin}, \citenamefont {Liang}, \citenamefont {Fan},\ and\
  \citenamefont {Hu}}]{wu2016topological}%
  \BibitemOpen
  \bibfield  {author} {\bibinfo {author} {\bibnamefont {Wu}, \bibfnamefont
  {X.}}, \bibinfo {author} {\bibfnamefont {S.}~\bibnamefont {Qin}}, \bibinfo
  {author} {\bibfnamefont {Y.}~\bibnamefont {Liang}}, \bibinfo {author}
  {\bibfnamefont {H.}~\bibnamefont {Fan}}, and\ \bibinfo {author}
  {\bibfnamefont {J.}~\bibnamefont {Hu}}} (\bibinfo {year} {2016}),\ \href
  {https://doi.org/10.1103/PhysRevB.93.115129} {\bibfield  {journal} {\bibinfo
  {journal} {Phys. Rev. B}\ }\textbf {\bibinfo {volume} {93}},\ \bibinfo
  {pages} {115129}}\BibitemShut {NoStop}%
\bibitem [{\citenamefont {Xia}\ \emph {et~al.}(2014)\citenamefont {Xia},
  \citenamefont {Jiao}, \citenamefont {Ye}, \citenamefont {Ge}, \citenamefont
  {Zhang}, \citenamefont {Jiang}, \citenamefont {Peng}, \citenamefont {Shen},
  \citenamefont {Fan}, \citenamefont {Cao} \emph {et~al.}}]{Xia2014}%
  \BibitemOpen
  \bibfield  {author} {\bibinfo {author} {\bibnamefont {Xia}, \bibfnamefont
  {M.}}, \bibinfo {author} {\bibfnamefont {W.}~\bibnamefont {Jiao}}, \bibinfo
  {author} {\bibfnamefont {Z.}~\bibnamefont {Ye}}, \bibinfo {author}
  {\bibfnamefont {Q.}~\bibnamefont {Ge}}, \bibinfo {author} {\bibfnamefont
  {Y.}~\bibnamefont {Zhang}}, \bibinfo {author} {\bibfnamefont
  {J.}~\bibnamefont {Jiang}}, \bibinfo {author} {\bibfnamefont
  {R.}~\bibnamefont {Peng}}, \bibinfo {author} {\bibfnamefont {X.}~\bibnamefont
  {Shen}}, \bibinfo {author} {\bibfnamefont {Q.}~\bibnamefont {Fan}}, \bibinfo
  {author} {\bibfnamefont {G.}~\bibnamefont {Cao}},  \emph {et~al.}} (\bibinfo
  {year} {2014}),\ \href {https://doi.org/10.1088/0953-8984/26/26/265701}
  {\bibfield  {journal} {\bibinfo  {journal} {J. Phys.: Condens. Matter}\
  }\textbf {\bibinfo {volume} {26}},\ \bibinfo {pages} {265701}}\BibitemShut
  {NoStop}%
\bibitem [{\citenamefont {Xiao}\ \emph {et~al.}(2022)\citenamefont {Xiao},
  \citenamefont {Zhang}, \citenamefont {Asmara}, \citenamefont {Li},
  \citenamefont {Li}, \citenamefont {Zhang}, \citenamefont {Tseng},
  \citenamefont {Dong}, \citenamefont {Wang}, \citenamefont {Chen},
  \citenamefont {Schmitt},\ and\ \citenamefont
  {Peng}}]{xiao2022dispersionless}%
  \BibitemOpen
  \bibfield  {author} {\bibinfo {author} {\bibnamefont {Xiao}, \bibfnamefont
  {Q.}}, \bibinfo {author} {\bibfnamefont {W.}~\bibnamefont {Zhang}}, \bibinfo
  {author} {\bibfnamefont {T.~C.}\ \bibnamefont {Asmara}}, \bibinfo {author}
  {\bibfnamefont {D.}~\bibnamefont {Li}}, \bibinfo {author} {\bibfnamefont
  {Q.}~\bibnamefont {Li}}, \bibinfo {author} {\bibfnamefont {S.}~\bibnamefont
  {Zhang}}, \bibinfo {author} {\bibfnamefont {Y.}~\bibnamefont {Tseng}},
  \bibinfo {author} {\bibfnamefont {X.}~\bibnamefont {Dong}}, \bibinfo {author}
  {\bibfnamefont {Y.}~\bibnamefont {Wang}}, \bibinfo {author} {\bibfnamefont
  {C.-C.}\ \bibnamefont {Chen}}, \bibinfo {author} {\bibfnamefont
  {T.}~\bibnamefont {Schmitt}}, and\ \bibinfo {author} {\bibfnamefont
  {Y.}~\bibnamefont {Peng}}} (\bibinfo {year} {2022}),\ \href
  {https://doi.org/10.1038/s41535-022-00492-0} {\bibfield  {journal} {\bibinfo
  {journal} {npj Quantum Mater.}\ }\textbf {\bibinfo {volume} {7}},\ \bibinfo
  {pages} {80}}\BibitemShut {NoStop}%
\bibitem [{\citenamefont {Xiao}\ \emph {et~al.}(2009)\citenamefont {Xiao},
  \citenamefont {Su}, \citenamefont {Meven}, \citenamefont {Mittal},
  \citenamefont {Kumar}, \citenamefont {Chatterji}, \citenamefont {Price},
  \citenamefont {Persson}, \citenamefont {Kumar}, \citenamefont {Dhar} \emph
  {et~al.}}]{Xiao2009}%
  \BibitemOpen
  \bibfield  {author} {\bibinfo {author} {\bibnamefont {Xiao}, \bibfnamefont
  {Y.}}, \bibinfo {author} {\bibfnamefont {Y.}~\bibnamefont {Su}}, \bibinfo
  {author} {\bibfnamefont {M.}~\bibnamefont {Meven}}, \bibinfo {author}
  {\bibfnamefont {R.}~\bibnamefont {Mittal}}, \bibinfo {author} {\bibfnamefont
  {C.}~\bibnamefont {Kumar}}, \bibinfo {author} {\bibfnamefont
  {T.}~\bibnamefont {Chatterji}}, \bibinfo {author} {\bibfnamefont
  {S.}~\bibnamefont {Price}}, \bibinfo {author} {\bibfnamefont
  {J.}~\bibnamefont {Persson}}, \bibinfo {author} {\bibfnamefont
  {N.}~\bibnamefont {Kumar}}, \bibinfo {author} {\bibfnamefont
  {S.}~\bibnamefont {Dhar}},  \emph {et~al.}} (\bibinfo {year} {2009}),\ \href
  {https://doi.org/10.1103/PhysRevB.80.174424} {\bibfield  {journal} {\bibinfo
  {journal} {Phys. Rev. B}\ }\textbf {\bibinfo {volume} {80}},\ \bibinfo
  {pages} {174424}}\BibitemShut {NoStop}%
\bibitem [{\citenamefont {Xie}\ \emph {et~al.}(2018{\natexlab{a}})\citenamefont
  {Xie}, \citenamefont {Gong}, \citenamefont {Ghosh}, \citenamefont {Ghosh},
  \citenamefont {Soda}, \citenamefont {Masuda}, \citenamefont {Itoh},
  \citenamefont {Bourdarot}, \citenamefont {Regnault}, \citenamefont
  {Danilkin}, \citenamefont {Li},\ and\ \citenamefont {Luo}}]{xie2018neutron}%
  \BibitemOpen
  \bibfield  {author} {\bibinfo {author} {\bibnamefont {Xie}, \bibfnamefont
  {T.}}, \bibinfo {author} {\bibfnamefont {D.}~\bibnamefont {Gong}}, \bibinfo
  {author} {\bibfnamefont {H.}~\bibnamefont {Ghosh}}, \bibinfo {author}
  {\bibfnamefont {A.}~\bibnamefont {Ghosh}}, \bibinfo {author} {\bibfnamefont
  {M.}~\bibnamefont {Soda}}, \bibinfo {author} {\bibfnamefont {T.}~\bibnamefont
  {Masuda}}, \bibinfo {author} {\bibfnamefont {S.}~\bibnamefont {Itoh}},
  \bibinfo {author} {\bibfnamefont {F.}~\bibnamefont {Bourdarot}}, \bibinfo
  {author} {\bibfnamefont {L.-P.}\ \bibnamefont {Regnault}}, \bibinfo {author}
  {\bibfnamefont {S.}~\bibnamefont {Danilkin}}, \bibinfo {author}
  {\bibfnamefont {S.}~\bibnamefont {Li}}, and\ \bibinfo {author} {\bibfnamefont
  {H.}~\bibnamefont {Luo}}} (\bibinfo {year} {2018}{\natexlab{a}}),\ \href
  {https://doi.org/10.1103/PhysRevLett.120.137001} {\bibfield  {journal}
  {\bibinfo  {journal} {Phys. Rev. Lett.}\ }\textbf {\bibinfo {volume} {120}},\
  \bibinfo {pages} {137001}}\BibitemShut {NoStop}%
\bibitem [{\citenamefont {Xie}\ \emph {et~al.}(2020)\citenamefont {Xie},
  \citenamefont {Liu}, \citenamefont {Bourdarot}, \citenamefont {Regnault},
  \citenamefont {Li},\ and\ \citenamefont {Luo}}]{xie2020spinexcitation}%
  \BibitemOpen
  \bibfield  {author} {\bibinfo {author} {\bibnamefont {Xie}, \bibfnamefont
  {T.}}, \bibinfo {author} {\bibfnamefont {C.}~\bibnamefont {Liu}}, \bibinfo
  {author} {\bibfnamefont {F.}~\bibnamefont {Bourdarot}}, \bibinfo {author}
  {\bibfnamefont {L.-P.}\ \bibnamefont {Regnault}}, \bibinfo {author}
  {\bibfnamefont {S.}~\bibnamefont {Li}}, and\ \bibinfo {author} {\bibfnamefont
  {H.}~\bibnamefont {Luo}}} (\bibinfo {year} {2020}),\ \href
  {https://doi.org/10.1103/physrevresearch.2.022018} {\bibfield  {journal}
  {\bibinfo  {journal} {Phys. Rev. Res.}\ }\textbf {\bibinfo {volume} {2}},\
  \bibinfo {pages} {022018}}\BibitemShut {NoStop}%
\bibitem [{\citenamefont {Xie}\ \emph {et~al.}(2022)\citenamefont {Xie},
  \citenamefont {Liu}, \citenamefont {Kajimoto}, \citenamefont {Ikeuchi},
  \citenamefont {Li},\ and\ \citenamefont {Luo}}]{xie2022spina}%
  \BibitemOpen
  \bibfield  {author} {\bibinfo {author} {\bibnamefont {Xie}, \bibfnamefont
  {T.}}, \bibinfo {author} {\bibfnamefont {C.}~\bibnamefont {Liu}}, \bibinfo
  {author} {\bibfnamefont {R.}~\bibnamefont {Kajimoto}}, \bibinfo {author}
  {\bibfnamefont {K.}~\bibnamefont {Ikeuchi}}, \bibinfo {author} {\bibfnamefont
  {S.}~\bibnamefont {Li}}, and\ \bibinfo {author} {\bibfnamefont
  {H.}~\bibnamefont {Luo}}} (\bibinfo {year} {2022}),\ \href
  {https://doi.org/10.1088/1361-648x/ac9441} {\bibfield  {journal} {\bibinfo
  {journal} {J. Phys.: Condens. Matter}\ }\textbf {\bibinfo {volume} {34}},\
  \bibinfo {pages} {474001}}\BibitemShut {NoStop}%
\bibitem [{\citenamefont {Xie}\ \emph {et~al.}(2018{\natexlab{b}})\citenamefont
  {Xie}, \citenamefont {Wei}, \citenamefont {Gong}, \citenamefont {Fennell},
  \citenamefont {Stuhr}, \citenamefont {Kajimoto}, \citenamefont {Ikeuchi},
  \citenamefont {Li}, \citenamefont {Hu},\ and\ \citenamefont
  {Luo}}]{xie2018odd}%
  \BibitemOpen
  \bibfield  {author} {\bibinfo {author} {\bibnamefont {Xie}, \bibfnamefont
  {T.}}, \bibinfo {author} {\bibfnamefont {Y.}~\bibnamefont {Wei}}, \bibinfo
  {author} {\bibfnamefont {D.}~\bibnamefont {Gong}}, \bibinfo {author}
  {\bibfnamefont {T.}~\bibnamefont {Fennell}}, \bibinfo {author} {\bibfnamefont
  {U.}~\bibnamefont {Stuhr}}, \bibinfo {author} {\bibfnamefont
  {R.}~\bibnamefont {Kajimoto}}, \bibinfo {author} {\bibfnamefont
  {K.}~\bibnamefont {Ikeuchi}}, \bibinfo {author} {\bibfnamefont
  {S.}~\bibnamefont {Li}}, \bibinfo {author} {\bibfnamefont {J.}~\bibnamefont
  {Hu}}, and\ \bibinfo {author} {\bibfnamefont {H.}~\bibnamefont {Luo}}}
  (\bibinfo {year} {2018}{\natexlab{b}}),\ \href
  {https://doi.org/10.1103/PhysRevLett.120.267003} {\bibfield  {journal}
  {\bibinfo  {journal} {Phys. Rev. Lett.}\ }\textbf {\bibinfo {volume} {120}},\
  \bibinfo {pages} {267003}}\BibitemShut {NoStop}%
\bibitem [{\citenamefont {Xie}\ \emph {et~al.}(2015)\citenamefont {Xie},
  \citenamefont {Cao}, \citenamefont {Zhou}, \citenamefont {Chen},
  \citenamefont {Xiang},\ and\ \citenamefont {Gong}}]{Xie2015}%
  \BibitemOpen
  \bibfield  {author} {\bibinfo {author} {\bibnamefont {Xie}, \bibfnamefont
  {Y.}}, \bibinfo {author} {\bibfnamefont {H.-Y.}\ \bibnamefont {Cao}},
  \bibinfo {author} {\bibfnamefont {Y.}~\bibnamefont {Zhou}}, \bibinfo {author}
  {\bibfnamefont {S.}~\bibnamefont {Chen}}, \bibinfo {author} {\bibfnamefont
  {H.}~\bibnamefont {Xiang}}, and\ \bibinfo {author} {\bibfnamefont {X.-G.}\
  \bibnamefont {Gong}}} (\bibinfo {year} {2015}),\ \href
  {https://doi.org/10.1038/srep10011} {\bibfield  {journal} {\bibinfo
  {journal} {Sci. Rep.}\ }\textbf {\bibinfo {volume} {5}},\ \bibinfo {pages}
  {10011}}\BibitemShut {NoStop}%
\bibitem [{\citenamefont {Xu}\ \emph {et~al.}(2019)\citenamefont {Xu},
  \citenamefont {Chen},\ and\ \citenamefont {Cao}}]{xu2019unique}%
  \BibitemOpen
  \bibfield  {author} {\bibinfo {author} {\bibnamefont {Xu}, \bibfnamefont
  {C.}}, \bibinfo {author} {\bibfnamefont {Q.}~\bibnamefont {Chen}}, and\
  \bibinfo {author} {\bibfnamefont {C.}~\bibnamefont {Cao}}} (\bibinfo {year}
  {2019}),\ \href {https://doi.org/10.1038/s42005-019-0112-1} {\bibfield
  {journal} {\bibinfo  {journal} {Commun. Phys.}\ }\textbf {\bibinfo {volume}
  {2}},\ \bibinfo {pages} {16}}\BibitemShut {NoStop}%
\bibitem [{\citenamefont {Xu}\ \emph {et~al.}(2016{\natexlab{a}})\citenamefont
  {Xu}, \citenamefont {Lian}, \citenamefont {Tang}, \citenamefont {Qi},\ and\
  \citenamefont {Zhang}}]{xu2016topological}%
  \BibitemOpen
  \bibfield  {author} {\bibinfo {author} {\bibnamefont {Xu}, \bibfnamefont
  {G.}}, \bibinfo {author} {\bibfnamefont {B.}~\bibnamefont {Lian}}, \bibinfo
  {author} {\bibfnamefont {P.}~\bibnamefont {Tang}}, \bibinfo {author}
  {\bibfnamefont {X.-L.}\ \bibnamefont {Qi}}, and\ \bibinfo {author}
  {\bibfnamefont {S.-C.}\ \bibnamefont {Zhang}}} (\bibinfo {year}
  {2016}{\natexlab{a}}),\ \href
  {https://doi.org/10.1103/PhysRevLett.117.047001} {\bibfield  {journal}
  {\bibinfo  {journal} {Phys. Rev. Lett.}\ }\textbf {\bibinfo {volume} {117}},\
  \bibinfo {pages} {047001}}\BibitemShut {NoStop}%
\bibitem [{\citenamefont {Xu}\ \emph {et~al.}(2026)\citenamefont {Xu},
  \citenamefont {Jiang}, \citenamefont {Gai}, \citenamefont {Cao},
  \citenamefont {Chen}, \citenamefont {Man}, \citenamefont {Lin}, \citenamefont
  {Deng}, \citenamefont {He}, \citenamefont {Liu}, \citenamefont {Zhao},
  \citenamefont {Lu}, \citenamefont {Chang},\ and\ \citenamefont
  {Liu}}]{Xu2026}%
  \BibitemOpen
  \bibfield  {author} {\bibinfo {author} {\bibnamefont {Xu}, \bibfnamefont
  {H.}}, \bibinfo {author} {\bibfnamefont {J.}~\bibnamefont {Jiang}}, \bibinfo
  {author} {\bibfnamefont {X.}~\bibnamefont {Gai}}, \bibinfo {author}
  {\bibfnamefont {R.-Q.}\ \bibnamefont {Cao}}, \bibinfo {author} {\bibfnamefont
  {K.}~\bibnamefont {Chen}}, \bibinfo {author} {\bibfnamefont {X.-X.}\
  \bibnamefont {Man}}, \bibinfo {author} {\bibfnamefont {H.}~\bibnamefont
  {Lin}}, \bibinfo {author} {\bibfnamefont {P.}~\bibnamefont {Deng}}, \bibinfo
  {author} {\bibfnamefont {K.}~\bibnamefont {He}}, \bibinfo {author}
  {\bibfnamefont {K.}~\bibnamefont {Liu}}, \bibinfo {author} {\bibfnamefont
  {D.}~\bibnamefont {Zhao}}, \bibinfo {author} {\bibfnamefont {Z.-Y.}\
  \bibnamefont {Lu}}, \bibinfo {author} {\bibfnamefont {K.}~\bibnamefont
  {Chang}}, and\ \bibinfo {author} {\bibfnamefont {C.}~\bibnamefont {Liu}}}
  (\bibinfo {year} {2026}),\ \href {https://doi.org/10.1021/acsnano.6c05058}
  {\bibfield  {journal} {\bibinfo  {journal} {ACS Nano}\ }\textbf {\bibinfo
  {volume} {20}},\ \bibinfo {pages} {16426}}\BibitemShut {NoStop}%
\bibitem [{\citenamefont {Xu}\ \emph {et~al.}(2016{\natexlab{b}})\citenamefont
  {Xu}, \citenamefont {Niu}, \citenamefont {Xu}, \citenamefont {Jiang},
  \citenamefont {Yao}, \citenamefont {Chen}, \citenamefont {Song},
  \citenamefont {{Abdel-Hafiez}}, \citenamefont {Chareev}, \citenamefont
  {Vasiliev}, \citenamefont {Wang}, \citenamefont {Wo}, \citenamefont {Zhao},
  \citenamefont {Peng},\ and\ \citenamefont {Feng}}]{xu2016highly}%
  \BibitemOpen
  \bibfield  {author} {\bibinfo {author} {\bibnamefont {Xu}, \bibfnamefont
  {H.~C.}}, \bibinfo {author} {\bibfnamefont {X.~H.}\ \bibnamefont {Niu}},
  \bibinfo {author} {\bibfnamefont {D.~F.}\ \bibnamefont {Xu}}, \bibinfo
  {author} {\bibfnamefont {J.}~\bibnamefont {Jiang}}, \bibinfo {author}
  {\bibfnamefont {Q.}~\bibnamefont {Yao}}, \bibinfo {author} {\bibfnamefont
  {Q.~Y.}\ \bibnamefont {Chen}}, \bibinfo {author} {\bibfnamefont
  {Q.}~\bibnamefont {Song}}, \bibinfo {author} {\bibfnamefont {M.}~\bibnamefont
  {{Abdel-Hafiez}}}, \bibinfo {author} {\bibfnamefont {D.~A.}\ \bibnamefont
  {Chareev}}, \bibinfo {author} {\bibfnamefont {A.~N.}\ \bibnamefont
  {Vasiliev}}, \bibinfo {author} {\bibfnamefont {Q.~S.}\ \bibnamefont {Wang}},
  \bibinfo {author} {\bibfnamefont {H.~L.}\ \bibnamefont {Wo}}, \bibinfo
  {author} {\bibfnamefont {J.}~\bibnamefont {Zhao}}, \bibinfo {author}
  {\bibfnamefont {R.}~\bibnamefont {Peng}}, and\ \bibinfo {author}
  {\bibfnamefont {D.~L.}\ \bibnamefont {Feng}}} (\bibinfo {year}
  {2016}{\natexlab{b}}),\ \href
  {https://doi.org/10.1103/PhysRevLett.117.157003} {\bibfield  {journal}
  {\bibinfo  {journal} {Phys. Rev. Lett.}\ }\textbf {\bibinfo {volume} {117}},\
  \bibinfo {pages} {157003}}\BibitemShut {NoStop}%
\bibitem [{\citenamefont {Xu}\ \emph {et~al.}(2022)\citenamefont {Xu},
  \citenamefont {Wu}, \citenamefont {Zheng}, \citenamefont {Yin}, \citenamefont
  {Li}, \citenamefont {Wang},\ and\ \citenamefont {Tang}}]{xu2022research}%
  \BibitemOpen
  \bibfield  {author} {\bibinfo {author} {\bibnamefont {Xu}, \bibfnamefont
  {H.-S.}}, \bibinfo {author} {\bibfnamefont {S.}~\bibnamefont {Wu}}, \bibinfo
  {author} {\bibfnamefont {H.}~\bibnamefont {Zheng}}, \bibinfo {author}
  {\bibfnamefont {R.}~\bibnamefont {Yin}}, \bibinfo {author} {\bibfnamefont
  {Y.}~\bibnamefont {Li}}, \bibinfo {author} {\bibfnamefont {X.}~\bibnamefont
  {Wang}}, and\ \bibinfo {author} {\bibfnamefont {K.}~\bibnamefont {Tang}}}
  (\bibinfo {year} {2022}),\ \href {https://doi.org/10.1007/s41061-022-00368-8}
  {\bibfield  {journal} {\bibinfo  {journal} {Topics in Current Chemistry}\
  }\textbf {\bibinfo {volume} {380}},\ \bibinfo {pages} {11}}\BibitemShut
  {NoStop}%
\bibitem [{\citenamefont {Xu}\ \emph {et~al.}(2021)\citenamefont {Xu},
  \citenamefont {Rong}, \citenamefont {Wang}, \citenamefont {Wu}, \citenamefont
  {Hu}, \citenamefont {Cai}, \citenamefont {Gao}, \citenamefont {Yan},
  \citenamefont {Li}, \citenamefont {Yin}, \citenamefont {Chen}, \citenamefont
  {Huang}, \citenamefont {Zhu}, \citenamefont {Huang}, \citenamefont {Liu},
  \citenamefont {Xu}, \citenamefont {Zhao},\ and\ \citenamefont
  {Zhou}}]{xu2021spectroscopic}%
  \BibitemOpen
  \bibfield  {author} {\bibinfo {author} {\bibnamefont {Xu}, \bibfnamefont
  {Y.}}, \bibinfo {author} {\bibfnamefont {H.}~\bibnamefont {Rong}}, \bibinfo
  {author} {\bibfnamefont {Q.}~\bibnamefont {Wang}}, \bibinfo {author}
  {\bibfnamefont {D.}~\bibnamefont {Wu}}, \bibinfo {author} {\bibfnamefont
  {Y.}~\bibnamefont {Hu}}, \bibinfo {author} {\bibfnamefont {Y.}~\bibnamefont
  {Cai}}, \bibinfo {author} {\bibfnamefont {Q.}~\bibnamefont {Gao}}, \bibinfo
  {author} {\bibfnamefont {H.}~\bibnamefont {Yan}}, \bibinfo {author}
  {\bibfnamefont {C.}~\bibnamefont {Li}}, \bibinfo {author} {\bibfnamefont
  {C.}~\bibnamefont {Yin}}, \bibinfo {author} {\bibfnamefont {H.}~\bibnamefont
  {Chen}}, \bibinfo {author} {\bibfnamefont {J.}~\bibnamefont {Huang}},
  \bibinfo {author} {\bibfnamefont {Z.}~\bibnamefont {Zhu}}, \bibinfo {author}
  {\bibfnamefont {Y.}~\bibnamefont {Huang}}, \bibinfo {author} {\bibfnamefont
  {G.}~\bibnamefont {Liu}}, \bibinfo {author} {\bibfnamefont {Z.}~\bibnamefont
  {Xu}}, \bibinfo {author} {\bibfnamefont {L.}~\bibnamefont {Zhao}}, and\
  \bibinfo {author} {\bibfnamefont {X.~J.}\ \bibnamefont {Zhou}}} (\bibinfo
  {year} {2021}),\ \href {https://doi.org/10.1038/s41467-021-23106-y}
  {\bibfield  {journal} {\bibinfo  {journal} {Nat. Commun.}\ }\textbf {\bibinfo
  {volume} {12}},\ \bibinfo {pages} {2840}}\BibitemShut {NoStop}%
\bibitem [{\citenamefont {Yakita}\ \emph {et~al.}(2014)\citenamefont {Yakita},
  \citenamefont {Ogino}, \citenamefont {Okada}, \citenamefont {Yamamoto},
  \citenamefont {Kishio}, \citenamefont {Tohei}, \citenamefont {Ikuhara},
  \citenamefont {Gotoh}, \citenamefont {Fujihisa}, \citenamefont {Kataoka}
  \emph {et~al.}}]{Yakita2014}%
  \BibitemOpen
  \bibfield  {author} {\bibinfo {author} {\bibnamefont {Yakita}, \bibfnamefont
  {H.}}, \bibinfo {author} {\bibfnamefont {H.}~\bibnamefont {Ogino}}, \bibinfo
  {author} {\bibfnamefont {T.}~\bibnamefont {Okada}}, \bibinfo {author}
  {\bibfnamefont {A.}~\bibnamefont {Yamamoto}}, \bibinfo {author}
  {\bibfnamefont {K.}~\bibnamefont {Kishio}}, \bibinfo {author} {\bibfnamefont
  {T.}~\bibnamefont {Tohei}}, \bibinfo {author} {\bibfnamefont
  {Y.}~\bibnamefont {Ikuhara}}, \bibinfo {author} {\bibfnamefont
  {Y.}~\bibnamefont {Gotoh}}, \bibinfo {author} {\bibfnamefont
  {H.}~\bibnamefont {Fujihisa}}, \bibinfo {author} {\bibfnamefont
  {K.}~\bibnamefont {Kataoka}},  \emph {et~al.}} (\bibinfo {year} {2014}),\
  \href {https://doi.org/10.1021/ja410845b} {\bibfield  {journal} {\bibinfo
  {journal} {J. Am. Chem. Soc.}\ }\textbf {\bibinfo {volume} {136}},\ \bibinfo
  {pages} {846}}\BibitemShut {NoStop}%
\bibitem [{\citenamefont {Yamakawa}\ and\ \citenamefont
  {Kontani}(2017)}]{Yamakawa2017}%
  \BibitemOpen
  \bibfield  {author} {\bibinfo {author} {\bibnamefont {Yamakawa},
  \bibfnamefont {Y.}}, and\ \bibinfo {author} {\bibfnamefont {H.}~\bibnamefont
  {Kontani}}} (\bibinfo {year} {2017}),\ \href
  {https://doi.org/10.1103/PhysRevB.96.045130} {\bibfield  {journal} {\bibinfo
  {journal} {Phys. Rev. B}\ }\textbf {\bibinfo {volume} {96}},\ \bibinfo
  {pages} {045130}}\BibitemShut {NoStop}%
\bibitem [{\citenamefont {Yamakawa}\ \emph {et~al.}(2016)\citenamefont
  {Yamakawa}, \citenamefont {Onari},\ and\ \citenamefont
  {Kontani}}]{yamakawa2016nematicity}%
  \BibitemOpen
  \bibfield  {author} {\bibinfo {author} {\bibnamefont {Yamakawa},
  \bibfnamefont {Y.}}, \bibinfo {author} {\bibfnamefont {S.}~\bibnamefont
  {Onari}}, and\ \bibinfo {author} {\bibfnamefont {H.}~\bibnamefont {Kontani}}}
  (\bibinfo {year} {2016}),\ \href {https://doi.org/10.1103/PhysRevX.6.021032}
  {\bibfield  {journal} {\bibinfo  {journal} {Phys. Rev. X}\ }\textbf {\bibinfo
  {volume} {6}},\ \bibinfo {pages} {021032}}\BibitemShut {NoStop}%
\bibitem [{\citenamefont {Yamakawa}\ \emph {et~al.}(2013)\citenamefont
  {Yamakawa}, \citenamefont {Onari}, \citenamefont {Kontani}, \citenamefont
  {Fujiwara}, \citenamefont {Iimura},\ and\ \citenamefont
  {Hosono}}]{Yamakawa2013}%
  \BibitemOpen
  \bibfield  {author} {\bibinfo {author} {\bibnamefont {Yamakawa},
  \bibfnamefont {Y.}}, \bibinfo {author} {\bibfnamefont {S.}~\bibnamefont
  {Onari}}, \bibinfo {author} {\bibfnamefont {H.}~\bibnamefont {Kontani}},
  \bibinfo {author} {\bibfnamefont {N.}~\bibnamefont {Fujiwara}}, \bibinfo
  {author} {\bibfnamefont {S.}~\bibnamefont {Iimura}}, and\ \bibinfo {author}
  {\bibfnamefont {H.}~\bibnamefont {Hosono}}} (\bibinfo {year} {2013}),\ \href
  {https://doi.org/10.1103/PhysRevB.88.041106} {\bibfield  {journal} {\bibinfo
  {journal} {Phys. Rev. B}\ }\textbf {\bibinfo {volume} {88}},\ \bibinfo
  {pages} {041106}}\BibitemShut {NoStop}%
\bibitem [{\citenamefont {Yamamoto}\ \emph {et~al.}(2024)\citenamefont
  {Yamamoto}, \citenamefont {Tokuta}, \citenamefont {Ishii}, \citenamefont
  {Yamanaka}, \citenamefont {Shimada},\ and\ \citenamefont
  {Ainslie}}]{Yamamoto2024}%
  \BibitemOpen
  \bibfield  {author} {\bibinfo {author} {\bibnamefont {Yamamoto},
  \bibfnamefont {A.}}, \bibinfo {author} {\bibfnamefont {S.}~\bibnamefont
  {Tokuta}}, \bibinfo {author} {\bibfnamefont {A.}~\bibnamefont {Ishii}},
  \bibinfo {author} {\bibfnamefont {A.}~\bibnamefont {Yamanaka}}, \bibinfo
  {author} {\bibfnamefont {Y.}~\bibnamefont {Shimada}}, and\ \bibinfo {author}
  {\bibfnamefont {M.~D.}\ \bibnamefont {Ainslie}}} (\bibinfo {year} {2024}),\
  \href {https://doi.org/10.1038/s41427-024-00549-5} {\bibfield  {journal}
  {\bibinfo  {journal} {NPG Asia Materials}\ }\textbf {\bibinfo {volume}
  {16}},\ \bibinfo {pages} {29}}\BibitemShut {NoStop}%
\bibitem [{\citenamefont {Yamamoto}\ \emph
  {et~al.}(2025{\natexlab{a}})\citenamefont {Yamamoto}, \citenamefont
  {Yamanaka}, \citenamefont {Iida}, \citenamefont {Shimada},\ and\
  \citenamefont {Hata}}]{Yamamoto2025}%
  \BibitemOpen
  \bibfield  {author} {\bibinfo {author} {\bibnamefont {Yamamoto},
  \bibfnamefont {A.}}, \bibinfo {author} {\bibfnamefont {A.}~\bibnamefont
  {Yamanaka}}, \bibinfo {author} {\bibfnamefont {K.}~\bibnamefont {Iida}},
  \bibinfo {author} {\bibfnamefont {Y.}~\bibnamefont {Shimada}}, and\ \bibinfo
  {author} {\bibfnamefont {S.}~\bibnamefont {Hata}}} (\bibinfo {year}
  {2025}{\natexlab{a}}),\ \href {https://doi.org/10.1080/14686996.2024.2436347}
  {\bibfield  {journal} {\bibinfo  {journal} {Science and Technology of Adv.
  Mater.}\ }\textbf {\bibinfo {volume} {26}},\ \bibinfo {pages}
  {2436347}}\BibitemShut {NoStop}%
\bibitem [{\citenamefont {Yamamoto}\ \emph
  {et~al.}(2025{\natexlab{b}})\citenamefont {Yamamoto}, \citenamefont
  {Ugalino}, \citenamefont {Fujii}, \citenamefont {Ohtsubo}, \citenamefont
  {Iwasawa}, \citenamefont {Kitamura}, \citenamefont {Imazono}, \citenamefont
  {Inami}, \citenamefont {Nakatani}, \citenamefont {Inaba}, \citenamefont
  {Agui}, \citenamefont {Takeuchi}, \citenamefont {Kimura}, \citenamefont
  {Takahasi}, \citenamefont {Horiba},\ and\ \citenamefont
  {Miyawaki}}]{yamamoto2025status}%
  \BibitemOpen
  \bibfield  {author} {\bibinfo {author} {\bibnamefont {Yamamoto},
  \bibfnamefont {K.}}, \bibinfo {author} {\bibfnamefont {R.}~\bibnamefont
  {Ugalino}}, \bibinfo {author} {\bibfnamefont {K.}~\bibnamefont {Fujii}},
  \bibinfo {author} {\bibfnamefont {Y.}~\bibnamefont {Ohtsubo}}, \bibinfo
  {author} {\bibfnamefont {H.}~\bibnamefont {Iwasawa}}, \bibinfo {author}
  {\bibfnamefont {M.}~\bibnamefont {Kitamura}}, \bibinfo {author}
  {\bibfnamefont {T.}~\bibnamefont {Imazono}}, \bibinfo {author} {\bibfnamefont
  {N.}~\bibnamefont {Inami}}, \bibinfo {author} {\bibfnamefont
  {T.}~\bibnamefont {Nakatani}}, \bibinfo {author} {\bibfnamefont
  {K.}~\bibnamefont {Inaba}}, \bibinfo {author} {\bibfnamefont
  {A.}~\bibnamefont {Agui}}, \bibinfo {author} {\bibfnamefont {T.}~\bibnamefont
  {Takeuchi}}, \bibinfo {author} {\bibfnamefont {H.}~\bibnamefont {Kimura}},
  \bibinfo {author} {\bibfnamefont {M.}~\bibnamefont {Takahasi}}, \bibinfo
  {author} {\bibfnamefont {K.}~\bibnamefont {Horiba}}, and\ \bibinfo {author}
  {\bibfnamefont {J.}~\bibnamefont {Miyawaki}}} (\bibinfo {year}
  {2025}{\natexlab{b}}),\ \href
  {https://doi.org/10.1088/1742-6596/3010/1/012115} {\bibfield  {journal}
  {\bibinfo  {journal} {Journal of Physics: Conference Series}\ }\textbf
  {\bibinfo {volume} {3010}},\ \bibinfo {pages} {012115}}\BibitemShut {NoStop}%
\bibitem [{\citenamefont {Yan}\ \emph {et~al.}(2015)\citenamefont {Yan},
  \citenamefont {Nandi}, \citenamefont {Saparov}, \citenamefont
  {{\v{C}}erm{\'a}k}, \citenamefont {Xiao}, \citenamefont {Su}, \citenamefont
  {Jin}, \citenamefont {Schneidewind}, \citenamefont {Br{\"u}ckel},
  \citenamefont {McCallum} \emph {et~al.}}]{Yan2015}%
  \BibitemOpen
  \bibfield  {author} {\bibinfo {author} {\bibnamefont {Yan}, \bibfnamefont
  {J.-Q.}}, \bibinfo {author} {\bibfnamefont {S.}~\bibnamefont {Nandi}},
  \bibinfo {author} {\bibfnamefont {B.}~\bibnamefont {Saparov}}, \bibinfo
  {author} {\bibfnamefont {P.}~\bibnamefont {{\v{C}}erm{\'a}k}}, \bibinfo
  {author} {\bibfnamefont {Y.}~\bibnamefont {Xiao}}, \bibinfo {author}
  {\bibfnamefont {Y.}~\bibnamefont {Su}}, \bibinfo {author} {\bibfnamefont
  {W.}~\bibnamefont {Jin}}, \bibinfo {author} {\bibfnamefont {A.}~\bibnamefont
  {Schneidewind}}, \bibinfo {author} {\bibfnamefont {T.}~\bibnamefont
  {Br{\"u}ckel}}, \bibinfo {author} {\bibfnamefont {R.}~\bibnamefont
  {McCallum}},  \emph {et~al.}} (\bibinfo {year} {2015}),\ \href
  {https://doi.org/10.1103/PhysRevB.91.024501} {\bibfield  {journal} {\bibinfo
  {journal} {Phys. Rev. B}\ }\textbf {\bibinfo {volume} {91}},\ \bibinfo
  {pages} {024501}}\BibitemShut {NoStop}%
\bibitem [{\citenamefont {Yan}\ \emph {et~al.}(2026)\citenamefont {Yan},
  \citenamefont {Wang}, \citenamefont {Xia}, \citenamefont {Paolini},
  \citenamefont {Chan}, \citenamefont {Dihingia}, \citenamefont {Rong},
  \citenamefont {Xiao}, \citenamefont {Halanayake}, \citenamefont {Song},
  \citenamefont {Gowda}, \citenamefont {Hickey}, \citenamefont {Wu},
  \citenamefont {Yu}, \citenamefont {Hirschfeld},\ and\ \citenamefont
  {Chang}}]{Yan2026}%
  \BibitemOpen
  \bibfield  {author} {\bibinfo {author} {\bibnamefont {Yan}, \bibfnamefont
  {Z.-J.}}, \bibinfo {author} {\bibfnamefont {Z.}~\bibnamefont {Wang}},
  \bibinfo {author} {\bibfnamefont {B.}~\bibnamefont {Xia}}, \bibinfo {author}
  {\bibfnamefont {S.}~\bibnamefont {Paolini}}, \bibinfo {author} {\bibfnamefont
  {Y.-T.}\ \bibnamefont {Chan}}, \bibinfo {author} {\bibfnamefont
  {N.}~\bibnamefont {Dihingia}}, \bibinfo {author} {\bibfnamefont
  {H.}~\bibnamefont {Rong}}, \bibinfo {author} {\bibfnamefont {P.}~\bibnamefont
  {Xiao}}, \bibinfo {author} {\bibfnamefont {K.~D.}\ \bibnamefont
  {Halanayake}}, \bibinfo {author} {\bibfnamefont {J.}~\bibnamefont {Song}},
  \bibinfo {author} {\bibfnamefont {V.}~\bibnamefont {Gowda}}, \bibinfo
  {author} {\bibfnamefont {D.~R.}\ \bibnamefont {Hickey}}, \bibinfo {author}
  {\bibfnamefont {W.}~\bibnamefont {Wu}}, \bibinfo {author} {\bibfnamefont
  {J.}~\bibnamefont {Yu}}, \bibinfo {author} {\bibfnamefont {P.~J.}\
  \bibnamefont {Hirschfeld}}, and\ \bibinfo {author} {\bibfnamefont {C.-Z.}\
  \bibnamefont {Chang}}} (\bibinfo {year} {2026}),\ \href
  {https://doi.org/10.1038/s41586-026-10321-0} {\bibfield  {journal} {\bibinfo
  {journal} {Nature}\ }\textbf {\bibinfo {volume} {652}},\ \bibinfo {pages}
  {342}}\BibitemShut {NoStop}%
\bibitem [{\citenamefont {Yang}\ \emph {et~al.}(2013)\citenamefont {Yang},
  \citenamefont {Wang},\ and\ \citenamefont {Lee}}]{Yang2013}%
  \BibitemOpen
  \bibfield  {author} {\bibinfo {author} {\bibnamefont {Yang}, \bibfnamefont
  {F.}}, \bibinfo {author} {\bibfnamefont {F.}~\bibnamefont {Wang}}, and\
  \bibinfo {author} {\bibfnamefont {D.-H.}\ \bibnamefont {Lee}}} (\bibinfo
  {year} {2013}),\ \href {https://doi.org/10.1103/PhysRevB.88.100504}
  {\bibfield  {journal} {\bibinfo  {journal} {Phys. Rev. B}\ }\textbf {\bibinfo
  {volume} {88}},\ \bibinfo {pages} {100504}}\BibitemShut {NoStop}%
\bibitem [{\citenamefont {Yao}\ \emph {et~al.}(2022)\citenamefont {Yao},
  \citenamefont {Mazza}, \citenamefont {Han}, \citenamefont {Yi}, \citenamefont
  {Jain}, \citenamefont {Brahlek},\ and\ \citenamefont {Oh}}]{Yao2022}%
  \BibitemOpen
  \bibfield  {author} {\bibinfo {author} {\bibnamefont {Yao}, \bibfnamefont
  {X.}}, \bibinfo {author} {\bibfnamefont {A.~R.}\ \bibnamefont {Mazza}},
  \bibinfo {author} {\bibfnamefont {M.-G.}\ \bibnamefont {Han}}, \bibinfo
  {author} {\bibfnamefont {H.~T.}\ \bibnamefont {Yi}}, \bibinfo {author}
  {\bibfnamefont {D.}~\bibnamefont {Jain}}, \bibinfo {author} {\bibfnamefont
  {M.}~\bibnamefont {Brahlek}}, and\ \bibinfo {author} {\bibfnamefont
  {S.}~\bibnamefont {Oh}}} (\bibinfo {year} {2022}),\ \href
  {https://doi.org/10.1021/acs.nanolett.2c02510} {\bibfield  {journal}
  {\bibinfo  {journal} {Nano Lett.}\ }\textbf {\bibinfo {volume} {22}},\
  \bibinfo {pages} {7522}}\BibitemShut {NoStop}%
\bibitem [{\citenamefont {Yay}\ \emph {et~al.}(2026)\citenamefont {Yay},
  \citenamefont {Meese}, \citenamefont {Kisiel}, \citenamefont {Krogstad},
  \citenamefont {Singh}, \citenamefont {Fernandes}, \citenamefont {Islam},\
  and\ \citenamefont {Fisher}}]{yay2026discovery}%
  \BibitemOpen
  \bibfield  {author} {\bibinfo {author} {\bibnamefont {Yay}, \bibfnamefont
  {K.~A.}}, \bibinfo {author} {\bibfnamefont {W.~J.}\ \bibnamefont {Meese}},
  \bibinfo {author} {\bibfnamefont {E.}~\bibnamefont {Kisiel}}, \bibinfo
  {author} {\bibfnamefont {M.~J.}\ \bibnamefont {Krogstad}}, \bibinfo {author}
  {\bibfnamefont {A.~G.}\ \bibnamefont {Singh}}, \bibinfo {author}
  {\bibfnamefont {R.~M.}\ \bibnamefont {Fernandes}}, \bibinfo {author}
  {\bibfnamefont {Z.}~\bibnamefont {Islam}}, and\ \bibinfo {author}
  {\bibfnamefont {I.~R.}\ \bibnamefont {Fisher}}} (\bibinfo {year} {2026}),\
  \href {https://doi.org/10.1126/sciadv.aec8998} {\bibfield  {journal}
  {\bibinfo  {journal} {Sci. Adv.}\ }\textbf {\bibinfo {volume} {12}},\
  \bibinfo {pages} {eaec8998}}\BibitemShut {NoStop}%
\bibitem [{\citenamefont {Yi}\ \emph {et~al.}(2018)\citenamefont {Yi},
  \citenamefont {Frano}, \citenamefont {Lu}, \citenamefont {He}, \citenamefont
  {Wang}, \citenamefont {Frandsen}, \citenamefont {Kemper}, \citenamefont {Yu},
  \citenamefont {Si}, \citenamefont {Wang}, \citenamefont {He}, \citenamefont
  {Hardy}, \citenamefont {Schweiss}, \citenamefont {Adelmann}, \citenamefont
  {Wolf}, \citenamefont {Hashimoto}, \citenamefont {Mo}, \citenamefont
  {Hussain}, \citenamefont {Le~Tacon}, \citenamefont {B{\"o}hmer},
  \citenamefont {Lee}, \citenamefont {Shen}, \citenamefont {Meingast},\ and\
  \citenamefont {Birgeneau}}]{yi2018spectral}%
  \BibitemOpen
  \bibfield  {author} {\bibinfo {author} {\bibnamefont {Yi}, \bibfnamefont
  {M.}}, \bibinfo {author} {\bibfnamefont {A.}~\bibnamefont {Frano}}, \bibinfo
  {author} {\bibfnamefont {D.~H.}\ \bibnamefont {Lu}}, \bibinfo {author}
  {\bibfnamefont {Y.}~\bibnamefont {He}}, \bibinfo {author} {\bibfnamefont
  {M.}~\bibnamefont {Wang}}, \bibinfo {author} {\bibfnamefont {B.~A.}\
  \bibnamefont {Frandsen}}, \bibinfo {author} {\bibfnamefont {A.~F.}\
  \bibnamefont {Kemper}}, \bibinfo {author} {\bibfnamefont {R.}~\bibnamefont
  {Yu}}, \bibinfo {author} {\bibfnamefont {Q.}~\bibnamefont {Si}}, \bibinfo
  {author} {\bibfnamefont {L.}~\bibnamefont {Wang}}, \bibinfo {author}
  {\bibfnamefont {M.}~\bibnamefont {He}}, \bibinfo {author} {\bibfnamefont
  {F.}~\bibnamefont {Hardy}}, \bibinfo {author} {\bibfnamefont
  {P.}~\bibnamefont {Schweiss}}, \bibinfo {author} {\bibfnamefont
  {P.}~\bibnamefont {Adelmann}}, \bibinfo {author} {\bibfnamefont
  {T.}~\bibnamefont {Wolf}}, \bibinfo {author} {\bibfnamefont {M.}~\bibnamefont
  {Hashimoto}}, \bibinfo {author} {\bibfnamefont {S.-K.}\ \bibnamefont {Mo}},
  \bibinfo {author} {\bibfnamefont {Z.}~\bibnamefont {Hussain}}, \bibinfo
  {author} {\bibfnamefont {M.}~\bibnamefont {Le~Tacon}}, \bibinfo {author}
  {\bibfnamefont {A.~E.}\ \bibnamefont {B{\"o}hmer}}, \bibinfo {author}
  {\bibfnamefont {D.-H.}\ \bibnamefont {Lee}}, \bibinfo {author} {\bibfnamefont
  {Z.-X.}\ \bibnamefont {Shen}}, \bibinfo {author} {\bibfnamefont
  {C.}~\bibnamefont {Meingast}}, and\ \bibinfo {author} {\bibfnamefont {R.~J.}\
  \bibnamefont {Birgeneau}}} (\bibinfo {year} {2018}),\ \href
  {https://doi.org/10.1103/PhysRevLett.121.127001} {\bibfield  {journal}
  {\bibinfo  {journal} {Phys. Rev. Lett.}\ }\textbf {\bibinfo {volume} {121}},\
  \bibinfo {pages} {127001}}\BibitemShut {NoStop}%
\bibitem [{\citenamefont {Yi}\ \emph {et~al.}(2015)\citenamefont {Yi},
  \citenamefont {Liu}, \citenamefont {Zhang}, \citenamefont {Yu}, \citenamefont
  {Zhu}, \citenamefont {Lee}, \citenamefont {Moore}, \citenamefont {Schmitt},
  \citenamefont {Li}, \citenamefont {Riggs}, \citenamefont {Chu}, \citenamefont
  {Lv}, \citenamefont {Hu}, \citenamefont {Hashimoto}, \citenamefont {Mo},
  \citenamefont {Hussain}, \citenamefont {Mao}, \citenamefont {Chu},
  \citenamefont {Fisher}, \citenamefont {Si}, \citenamefont {Shen},\ and\
  \citenamefont {Lu}}]{yi2015observation}%
  \BibitemOpen
  \bibfield  {author} {\bibinfo {author} {\bibnamefont {Yi}, \bibfnamefont
  {M.}}, \bibinfo {author} {\bibfnamefont {Z.-K.}\ \bibnamefont {Liu}},
  \bibinfo {author} {\bibfnamefont {Y.}~\bibnamefont {Zhang}}, \bibinfo
  {author} {\bibfnamefont {R.}~\bibnamefont {Yu}}, \bibinfo {author}
  {\bibfnamefont {J.-X.}\ \bibnamefont {Zhu}}, \bibinfo {author} {\bibfnamefont
  {J.}~\bibnamefont {Lee}}, \bibinfo {author} {\bibfnamefont {R.}~\bibnamefont
  {Moore}}, \bibinfo {author} {\bibfnamefont {F.}~\bibnamefont {Schmitt}},
  \bibinfo {author} {\bibfnamefont {W.}~\bibnamefont {Li}}, \bibinfo {author}
  {\bibfnamefont {S.}~\bibnamefont {Riggs}}, \bibinfo {author} {\bibfnamefont
  {J.-H.}\ \bibnamefont {Chu}}, \bibinfo {author} {\bibfnamefont
  {B.}~\bibnamefont {Lv}}, \bibinfo {author} {\bibfnamefont {J.}~\bibnamefont
  {Hu}}, \bibinfo {author} {\bibfnamefont {M.}~\bibnamefont {Hashimoto}},
  \bibinfo {author} {\bibfnamefont {S.-K.}\ \bibnamefont {Mo}}, \bibinfo
  {author} {\bibfnamefont {Z.}~\bibnamefont {Hussain}}, \bibinfo {author}
  {\bibfnamefont {Z.}~\bibnamefont {Mao}}, \bibinfo {author} {\bibfnamefont
  {C.}~\bibnamefont {Chu}}, \bibinfo {author} {\bibfnamefont {I.}~\bibnamefont
  {Fisher}}, \bibinfo {author} {\bibfnamefont {Q.}~\bibnamefont {Si}}, \bibinfo
  {author} {\bibfnamefont {Z.-X.}\ \bibnamefont {Shen}}, and\ \bibinfo {author}
  {\bibfnamefont {D.}~\bibnamefont {Lu}}} (\bibinfo {year} {2015}),\ \href
  {https://doi.org/10.1038/ncomms8777} {\bibfield  {journal} {\bibinfo
  {journal} {Nat. Commun.}\ }\textbf {\bibinfo {volume} {6}},\ \bibinfo {pages}
  {7777}}\BibitemShut {NoStop}%
\bibitem [{\citenamefont {Yi}\ \emph {et~al.}(2011)\citenamefont {Yi},
  \citenamefont {Lu}, \citenamefont {Chu}, \citenamefont {Analytis},
  \citenamefont {Sorini}, \citenamefont {Kemper}, \citenamefont {Moritz},
  \citenamefont {Mo}, \citenamefont {Moore}, \citenamefont {Hashimoto} \emph
  {et~al.}}]{yi2011symmetry}%
  \BibitemOpen
  \bibfield  {author} {\bibinfo {author} {\bibnamefont {Yi}, \bibfnamefont
  {M.}}, \bibinfo {author} {\bibfnamefont {D.}~\bibnamefont {Lu}}, \bibinfo
  {author} {\bibfnamefont {J.-H.}\ \bibnamefont {Chu}}, \bibinfo {author}
  {\bibfnamefont {J.~G.}\ \bibnamefont {Analytis}}, \bibinfo {author}
  {\bibfnamefont {A.~P.}\ \bibnamefont {Sorini}}, \bibinfo {author}
  {\bibfnamefont {A.~F.}\ \bibnamefont {Kemper}}, \bibinfo {author}
  {\bibfnamefont {B.}~\bibnamefont {Moritz}}, \bibinfo {author} {\bibfnamefont
  {S.-K.}\ \bibnamefont {Mo}}, \bibinfo {author} {\bibfnamefont {R.~G.}\
  \bibnamefont {Moore}}, \bibinfo {author} {\bibfnamefont {M.}~\bibnamefont
  {Hashimoto}},  \emph {et~al.}} (\bibinfo {year} {2011}),\ \href
  {https://doi.org/10.1073/pnas.1015572108} {\bibfield  {journal} {\bibinfo
  {journal} {Proc. Natl. Acad. Sci. U.S.A.}\ }\textbf {\bibinfo {volume}
  {108}},\ \bibinfo {pages} {6878}}\BibitemShut {NoStop}%
\bibitem [{\citenamefont {Yi}\ \emph {et~al.}(2013)\citenamefont {Yi},
  \citenamefont {Lu}, \citenamefont {Yu}, \citenamefont {Riggs}, \citenamefont
  {Chu}, \citenamefont {Lv}, \citenamefont {Liu}, \citenamefont {Lu},
  \citenamefont {Cui}, \citenamefont {Hashimoto}, \citenamefont {Mo},
  \citenamefont {Hussain}, \citenamefont {Chu}, \citenamefont {Fisher},
  \citenamefont {Si},\ and\ \citenamefont {Shen}}]{yi2013observation}%
  \BibitemOpen
  \bibfield  {author} {\bibinfo {author} {\bibnamefont {Yi}, \bibfnamefont
  {M.}}, \bibinfo {author} {\bibfnamefont {D.~H.}\ \bibnamefont {Lu}}, \bibinfo
  {author} {\bibfnamefont {R.}~\bibnamefont {Yu}}, \bibinfo {author}
  {\bibfnamefont {S.~C.}\ \bibnamefont {Riggs}}, \bibinfo {author}
  {\bibfnamefont {J.-H.}\ \bibnamefont {Chu}}, \bibinfo {author} {\bibfnamefont
  {B.}~\bibnamefont {Lv}}, \bibinfo {author} {\bibfnamefont {Z.~K.}\
  \bibnamefont {Liu}}, \bibinfo {author} {\bibfnamefont {M.}~\bibnamefont
  {Lu}}, \bibinfo {author} {\bibfnamefont {Y.-T.}\ \bibnamefont {Cui}},
  \bibinfo {author} {\bibfnamefont {M.}~\bibnamefont {Hashimoto}}, \bibinfo
  {author} {\bibfnamefont {S.-K.}\ \bibnamefont {Mo}}, \bibinfo {author}
  {\bibfnamefont {Z.}~\bibnamefont {Hussain}}, \bibinfo {author} {\bibfnamefont
  {C.~W.}\ \bibnamefont {Chu}}, \bibinfo {author} {\bibfnamefont {I.~R.}\
  \bibnamefont {Fisher}}, \bibinfo {author} {\bibfnamefont {Q.}~\bibnamefont
  {Si}}, and\ \bibinfo {author} {\bibfnamefont {Z.-X.}\ \bibnamefont {Shen}}}
  (\bibinfo {year} {2013}),\ \href
  {https://doi.org/10.1103/PhysRevLett.110.067003} {\bibfield  {journal}
  {\bibinfo  {journal} {Phys. Rev. Lett.}\ }\textbf {\bibinfo {volume} {110}},\
  \bibinfo {pages} {067003}}\BibitemShut {NoStop}%
\bibitem [{\citenamefont {Yi}\ \emph {et~al.}(2019)\citenamefont {Yi},
  \citenamefont {Pfau}, \citenamefont {Zhang}, \citenamefont {He},
  \citenamefont {Wu}, \citenamefont {Chen}, \citenamefont {Ye}, \citenamefont
  {Hashimoto}, \citenamefont {Yu}, \citenamefont {Si}, \citenamefont {Lee},
  \citenamefont {Dai}, \citenamefont {Shen}, \citenamefont {Lu},\ and\
  \citenamefont {Birgeneau}}]{yi2019nematic}%
  \BibitemOpen
  \bibfield  {author} {\bibinfo {author} {\bibnamefont {Yi}, \bibfnamefont
  {M.}}, \bibinfo {author} {\bibfnamefont {H.}~\bibnamefont {Pfau}}, \bibinfo
  {author} {\bibfnamefont {Y.}~\bibnamefont {Zhang}}, \bibinfo {author}
  {\bibfnamefont {Y.}~\bibnamefont {He}}, \bibinfo {author} {\bibfnamefont
  {H.}~\bibnamefont {Wu}}, \bibinfo {author} {\bibfnamefont {T.}~\bibnamefont
  {Chen}}, \bibinfo {author} {\bibfnamefont {Z.~R.}\ \bibnamefont {Ye}},
  \bibinfo {author} {\bibfnamefont {M.}~\bibnamefont {Hashimoto}}, \bibinfo
  {author} {\bibfnamefont {R.}~\bibnamefont {Yu}}, \bibinfo {author}
  {\bibfnamefont {Q.}~\bibnamefont {Si}}, \bibinfo {author} {\bibfnamefont
  {D.-H.}\ \bibnamefont {Lee}}, \bibinfo {author} {\bibfnamefont
  {P.}~\bibnamefont {Dai}}, \bibinfo {author} {\bibfnamefont {Z.-X.}\
  \bibnamefont {Shen}}, \bibinfo {author} {\bibfnamefont {D.~H.}\ \bibnamefont
  {Lu}}, and\ \bibinfo {author} {\bibfnamefont {R.~J.}\ \bibnamefont
  {Birgeneau}}} (\bibinfo {year} {2019}),\ \href
  {https://doi.org/10.1103/PhysRevX.9.041049} {\bibfield  {journal} {\bibinfo
  {journal} {Phys. Rev. X}\ }\textbf {\bibinfo {volume} {9}},\ \bibinfo {pages}
  {041049}}\BibitemShut {NoStop}%
\bibitem [{\citenamefont {Yi}\ \emph {et~al.}(2017)\citenamefont {Yi},
  \citenamefont {Zhang}, \citenamefont {Shen},\ and\ \citenamefont
  {Lu}}]{yi2017role}%
  \BibitemOpen
  \bibfield  {author} {\bibinfo {author} {\bibnamefont {Yi}, \bibfnamefont
  {M.}}, \bibinfo {author} {\bibfnamefont {Y.}~\bibnamefont {Zhang}}, \bibinfo
  {author} {\bibfnamefont {Z.-X.}\ \bibnamefont {Shen}}, and\ \bibinfo {author}
  {\bibfnamefont {D.}~\bibnamefont {Lu}}} (\bibinfo {year} {2017}),\ \href
  {https://doi.org/10.1038/s41535-017-0059-y} {\bibfield  {journal} {\bibinfo
  {journal} {npj Quantum Mater.}\ }\textbf {\bibinfo {volume} {2}},\ \bibinfo
  {pages} {57}}\BibitemShut {NoStop}%
\bibitem [{\citenamefont {Yim}\ \emph {et~al.}(2018)\citenamefont {Yim},
  \citenamefont {Trainer}, \citenamefont {Aluru}, \citenamefont {Chi},
  \citenamefont {Hardy}, \citenamefont {Liang}, \citenamefont {Bonn},\ and\
  \citenamefont {Wahl}}]{yim2018discovery}%
  \BibitemOpen
  \bibfield  {author} {\bibinfo {author} {\bibnamefont {Yim}, \bibfnamefont
  {C.~M.}}, \bibinfo {author} {\bibfnamefont {C.}~\bibnamefont {Trainer}},
  \bibinfo {author} {\bibfnamefont {R.}~\bibnamefont {Aluru}}, \bibinfo
  {author} {\bibfnamefont {S.}~\bibnamefont {Chi}}, \bibinfo {author}
  {\bibfnamefont {W.~N.}\ \bibnamefont {Hardy}}, \bibinfo {author}
  {\bibfnamefont {R.}~\bibnamefont {Liang}}, \bibinfo {author} {\bibfnamefont
  {D.}~\bibnamefont {Bonn}}, and\ \bibinfo {author} {\bibfnamefont
  {P.}~\bibnamefont {Wahl}}} (\bibinfo {year} {2018}),\ \href
  {https://doi.org/10.1038/s41467-018-04909-y} {\bibfield  {journal} {\bibinfo
  {journal} {Nat. Commun.}\ }\textbf {\bibinfo {volume} {9}},\ \bibinfo {pages}
  {2602}}\BibitemShut {NoStop}%
\bibitem [{\citenamefont {Yin}\ \emph {et~al.}(2015)\citenamefont {Yin},
  \citenamefont {Wu}, \citenamefont {Wang}, \citenamefont {Ye}, \citenamefont
  {Gong}, \citenamefont {Hou}, \citenamefont {Shan}, \citenamefont {Li},
  \citenamefont {Liang}, \citenamefont {Wu}, \citenamefont {Li}, \citenamefont
  {Ting}, \citenamefont {Wang}, \citenamefont {Hu}, \citenamefont {Hor},
  \citenamefont {Ding},\ and\ \citenamefont {Pan}}]{yin2015observation}%
  \BibitemOpen
  \bibfield  {author} {\bibinfo {author} {\bibnamefont {Yin}, \bibfnamefont
  {J.-X.}}, \bibinfo {author} {\bibfnamefont {Z.}~\bibnamefont {Wu}}, \bibinfo
  {author} {\bibfnamefont {J.-H.}\ \bibnamefont {Wang}}, \bibinfo {author}
  {\bibfnamefont {Z.-Y.}\ \bibnamefont {Ye}}, \bibinfo {author} {\bibfnamefont
  {J.}~\bibnamefont {Gong}}, \bibinfo {author} {\bibfnamefont {X.-Y.}\
  \bibnamefont {Hou}}, \bibinfo {author} {\bibfnamefont {L.}~\bibnamefont
  {Shan}}, \bibinfo {author} {\bibfnamefont {A.}~\bibnamefont {Li}}, \bibinfo
  {author} {\bibfnamefont {X.-J.}\ \bibnamefont {Liang}}, \bibinfo {author}
  {\bibfnamefont {X.-X.}\ \bibnamefont {Wu}}, \bibinfo {author} {\bibfnamefont
  {J.}~\bibnamefont {Li}}, \bibinfo {author} {\bibfnamefont {C.-S.}\
  \bibnamefont {Ting}}, \bibinfo {author} {\bibfnamefont {Z.-Q.}\ \bibnamefont
  {Wang}}, \bibinfo {author} {\bibfnamefont {J.-P.}\ \bibnamefont {Hu}},
  \bibinfo {author} {\bibfnamefont {P.-H.}\ \bibnamefont {Hor}}, \bibinfo
  {author} {\bibfnamefont {H.}~\bibnamefont {Ding}}, and\ \bibinfo {author}
  {\bibfnamefont {S.~H.}\ \bibnamefont {Pan}}} (\bibinfo {year} {2015}),\ \href
  {https://doi.org/10.1038/nphys3371} {\bibfield  {journal} {\bibinfo
  {journal} {Nat. Phys.}\ }\textbf {\bibinfo {volume} {11}},\ \bibinfo {pages}
  {543}}\BibitemShut {NoStop}%
\bibitem [{\citenamefont {Yin}\ \emph {et~al.}(2011)\citenamefont {Yin},
  \citenamefont {Haule},\ and\ \citenamefont {Kotliar}}]{yin2011kinetic}%
  \BibitemOpen
  \bibfield  {author} {\bibinfo {author} {\bibnamefont {Yin}, \bibfnamefont
  {Z.~P.}}, \bibinfo {author} {\bibfnamefont {K.}~\bibnamefont {Haule}}, and\
  \bibinfo {author} {\bibfnamefont {G.}~\bibnamefont {Kotliar}}} (\bibinfo
  {year} {2011}),\ \href {https://doi.org/10.1038/nmat3120} {\bibfield
  {journal} {\bibinfo  {journal} {Nat. Mater.}\ }\textbf {\bibinfo {volume}
  {10}},\ \bibinfo {pages} {932}}\BibitemShut {NoStop}%
\bibitem [{\citenamefont {Yin}\ \emph {et~al.}(2014)\citenamefont {Yin},
  \citenamefont {Haule},\ and\ \citenamefont {Kotliar}}]{yin2014spin}%
  \BibitemOpen
  \bibfield  {author} {\bibinfo {author} {\bibnamefont {Yin}, \bibfnamefont
  {Z.~P.}}, \bibinfo {author} {\bibfnamefont {K.}~\bibnamefont {Haule}}, and\
  \bibinfo {author} {\bibfnamefont {G.}~\bibnamefont {Kotliar}}} (\bibinfo
  {year} {2014}),\ \href {https://doi.org/10.1038/nphys3116} {\bibfield
  {journal} {\bibinfo  {journal} {Nat. Phys.}\ }\textbf {\bibinfo {volume}
  {10}},\ \bibinfo {pages} {845}}\BibitemShut {NoStop}%
\bibitem [{\citenamefont {Ying}\ \emph {et~al.}(2012)\citenamefont {Ying},
  \citenamefont {Chen}, \citenamefont {Wang}, \citenamefont {Jin},
  \citenamefont {Zhou}, \citenamefont {Lai}, \citenamefont {Zhang},\ and\
  \citenamefont {Wang}}]{Ying2012}%
  \BibitemOpen
  \bibfield  {author} {\bibinfo {author} {\bibnamefont {Ying}, \bibfnamefont
  {T.~P.}}, \bibinfo {author} {\bibfnamefont {X.~L.}\ \bibnamefont {Chen}},
  \bibinfo {author} {\bibfnamefont {G.}~\bibnamefont {Wang}}, \bibinfo {author}
  {\bibfnamefont {S.~F.}\ \bibnamefont {Jin}}, \bibinfo {author} {\bibfnamefont
  {T.~T.}\ \bibnamefont {Zhou}}, \bibinfo {author} {\bibfnamefont {X.~F.}\
  \bibnamefont {Lai}}, \bibinfo {author} {\bibfnamefont {H.}~\bibnamefont
  {Zhang}}, and\ \bibinfo {author} {\bibfnamefont {W.~Y.}\ \bibnamefont
  {Wang}}} (\bibinfo {year} {2012}),\ \href {https://doi.org/10.1038/srep00426}
  {\bibfield  {journal} {\bibinfo  {journal} {Sci. Rep.}\ }\textbf {\bibinfo
  {volume} {2}},\ \bibinfo {pages} {426}}\BibitemShut {NoStop}%
\bibitem [{\citenamefont {Ying}\ \emph {et~al.}(2018)\citenamefont {Ying},
  \citenamefont {Wang}, \citenamefont {Wu}, \citenamefont {Zhao}, \citenamefont
  {Zhang}, \citenamefont {Song}, \citenamefont {Li}, \citenamefont {Lei},
  \citenamefont {Li}, \citenamefont {Yu}, \citenamefont {Cheng}, \citenamefont
  {An}, \citenamefont {Zhang}, \citenamefont {Jia}, \citenamefont {Yang},
  \citenamefont {Chen},\ and\ \citenamefont {Li}}]{ying2018discrete}%
  \BibitemOpen
  \bibfield  {author} {\bibinfo {author} {\bibnamefont {Ying}, \bibfnamefont
  {T.~P.}}, \bibinfo {author} {\bibfnamefont {M.~X.}\ \bibnamefont {Wang}},
  \bibinfo {author} {\bibfnamefont {X.~X.}\ \bibnamefont {Wu}}, \bibinfo
  {author} {\bibfnamefont {Z.~Y.}\ \bibnamefont {Zhao}}, \bibinfo {author}
  {\bibfnamefont {Z.~Z.}\ \bibnamefont {Zhang}}, \bibinfo {author}
  {\bibfnamefont {B.~Q.}\ \bibnamefont {Song}}, \bibinfo {author}
  {\bibfnamefont {Y.~C.}\ \bibnamefont {Li}}, \bibinfo {author} {\bibfnamefont
  {B.}~\bibnamefont {Lei}}, \bibinfo {author} {\bibfnamefont {Q.}~\bibnamefont
  {Li}}, \bibinfo {author} {\bibfnamefont {Y.}~\bibnamefont {Yu}}, \bibinfo
  {author} {\bibfnamefont {E.~J.}\ \bibnamefont {Cheng}}, \bibinfo {author}
  {\bibfnamefont {Z.~H.}\ \bibnamefont {An}}, \bibinfo {author} {\bibfnamefont
  {Y.}~\bibnamefont {Zhang}}, \bibinfo {author} {\bibfnamefont {X.~Y.}\
  \bibnamefont {Jia}}, \bibinfo {author} {\bibfnamefont {W.}~\bibnamefont
  {Yang}}, \bibinfo {author} {\bibfnamefont {X.~H.}\ \bibnamefont {Chen}}, and\
  \bibinfo {author} {\bibfnamefont {S.~Y.}\ \bibnamefont {Li}}} (\bibinfo
  {year} {2018}),\ \href {https://doi.org/10.1103/PhysRevLett.121.207003}
  {\bibfield  {journal} {\bibinfo  {journal} {Phys. Rev. Lett.}\ }\textbf
  {\bibinfo {volume} {121}},\ \bibinfo {pages} {207003}}\BibitemShut {NoStop}%
\bibitem [{\citenamefont {Yu}\ \emph {et~al.}(2017)\citenamefont {Yu},
  \citenamefont {Liu}, \citenamefont {Pan}, \citenamefont {Ruan}, \citenamefont
  {Wang}, \citenamefont {Mu}, \citenamefont {Zhao}, \citenamefont {Chen},\ and\
  \citenamefont {Ren}}]{Yu2017}%
  \BibitemOpen
  \bibfield  {author} {\bibinfo {author} {\bibnamefont {Yu}, \bibfnamefont
  {J.}}, \bibinfo {author} {\bibfnamefont {T.}~\bibnamefont {Liu}}, \bibinfo
  {author} {\bibfnamefont {B.-J.}\ \bibnamefont {Pan}}, \bibinfo {author}
  {\bibfnamefont {B.-B.}\ \bibnamefont {Ruan}}, \bibinfo {author}
  {\bibfnamefont {X.-C.}\ \bibnamefont {Wang}}, \bibinfo {author}
  {\bibfnamefont {Q.-G.}\ \bibnamefont {Mu}}, \bibinfo {author} {\bibfnamefont
  {K.}~\bibnamefont {Zhao}}, \bibinfo {author} {\bibfnamefont {G.-F.}\
  \bibnamefont {Chen}}, and\ \bibinfo {author} {\bibfnamefont {Z.-A.}\
  \bibnamefont {Ren}}} (\bibinfo {year} {2017}),\ \href
  {https://doi.org/10.1016/j.scib.2016.12.015} {\bibfield  {journal} {\bibinfo
  {journal} {Sci. Bull.}\ }\textbf {\bibinfo {volume} {62}},\ \bibinfo {pages}
  {218}}\BibitemShut {NoStop}%
\bibitem [{\citenamefont {Yu}\ \emph {et~al.}(2021{\natexlab{a}})\citenamefont
  {Yu}, \citenamefont {Liu}, \citenamefont {Ruan}, \citenamefont {Zhao},
  \citenamefont {Yang}, \citenamefont {Zhou},\ and\ \citenamefont
  {Ren}}]{Yu2021}%
  \BibitemOpen
  \bibfield  {author} {\bibinfo {author} {\bibnamefont {Yu}, \bibfnamefont
  {J.}}, \bibinfo {author} {\bibfnamefont {T.}~\bibnamefont {Liu}}, \bibinfo
  {author} {\bibfnamefont {B.}~\bibnamefont {Ruan}}, \bibinfo {author}
  {\bibfnamefont {K.}~\bibnamefont {Zhao}}, \bibinfo {author} {\bibfnamefont
  {Q.}~\bibnamefont {Yang}}, \bibinfo {author} {\bibfnamefont {M.}~\bibnamefont
  {Zhou}}, and\ \bibinfo {author} {\bibfnamefont {Z.}~\bibnamefont {Ren}}}
  (\bibinfo {year} {2021}{\natexlab{a}}),\ \href
  {https://doi.org/10.1007/s11433-020-1685-3} {\bibfield  {journal} {\bibinfo
  {journal} {Science China Physics, Mechanics \& Astronomy}\ }\textbf {\bibinfo
  {volume} {64}},\ \bibinfo {pages} {267411}}\BibitemShut {NoStop}%
\bibitem [{\citenamefont {Yu}\ \emph {et~al.}(2013)\citenamefont {Yu},
  \citenamefont {Goswami}, \citenamefont {Si}, \citenamefont {Nikolic},\ and\
  \citenamefont {Zhu}}]{yu2013superconductivity}%
  \BibitemOpen
  \bibfield  {author} {\bibinfo {author} {\bibnamefont {Yu}, \bibfnamefont
  {R.}}, \bibinfo {author} {\bibfnamefont {P.}~\bibnamefont {Goswami}},
  \bibinfo {author} {\bibfnamefont {Q.}~\bibnamefont {Si}}, \bibinfo {author}
  {\bibfnamefont {P.}~\bibnamefont {Nikolic}}, and\ \bibinfo {author}
  {\bibfnamefont {J.-X.}\ \bibnamefont {Zhu}}} (\bibinfo {year} {2013}),\ \href
  {https://doi.org/10.1038/ncomms3783} {\bibfield  {journal} {\bibinfo
  {journal} {Nat. Commun.}\ }\textbf {\bibinfo {volume} {4}},\ \bibinfo {pages}
  {2783}}\BibitemShut {NoStop}%
\bibitem [{\citenamefont {Yu}\ \emph {et~al.}(2021{\natexlab{b}})\citenamefont
  {Yu}, \citenamefont {Hu}, \citenamefont {Nica}, \citenamefont {Zhu},\ and\
  \citenamefont {Si}}]{yu2021orbital}%
  \BibitemOpen
  \bibfield  {author} {\bibinfo {author} {\bibnamefont {Yu}, \bibfnamefont
  {R.}}, \bibinfo {author} {\bibfnamefont {H.}~\bibnamefont {Hu}}, \bibinfo
  {author} {\bibfnamefont {E.~M.}\ \bibnamefont {Nica}}, \bibinfo {author}
  {\bibfnamefont {J.-X.}\ \bibnamefont {Zhu}}, and\ \bibinfo {author}
  {\bibfnamefont {Q.}~\bibnamefont {Si}}} (\bibinfo {year}
  {2021}{\natexlab{b}}),\ \href {https://doi.org/10.3389/fphy.2021.578347}
  {\bibfield  {journal} {\bibinfo  {journal} {Frontiers in Physics}\ }\textbf
  {\bibinfo {volume} {9}},\ \bibinfo {pages} {578347}}\BibitemShut {NoStop}%
\bibitem [{\citenamefont {Yu}\ and\ \citenamefont {Si}(2012)}]{yu2012u}%
  \BibitemOpen
  \bibfield  {author} {\bibinfo {author} {\bibnamefont {Yu}, \bibfnamefont
  {R.}}, and\ \bibinfo {author} {\bibfnamefont {Q.}~\bibnamefont {Si}}}
  (\bibinfo {year} {2012}),\ \href {https://doi.org/10.1103/PhysRevB.86.085104}
  {\bibfield  {journal} {\bibinfo  {journal} {Phys. Rev. B}\ }\textbf {\bibinfo
  {volume} {86}},\ \bibinfo {pages} {085104}}\BibitemShut {NoStop}%
\bibitem [{\citenamefont {Yu}\ and\ \citenamefont {Si}(2013)}]{yu2013orbital}%
  \BibitemOpen
  \bibfield  {author} {\bibinfo {author} {\bibnamefont {Yu}, \bibfnamefont
  {R.}}, and\ \bibinfo {author} {\bibfnamefont {Q.}~\bibnamefont {Si}}}
  (\bibinfo {year} {2013}),\ \href
  {https://doi.org/10.1103/PhysRevLett.110.146402} {\bibfield  {journal}
  {\bibinfo  {journal} {Phys. Rev. Lett.}\ }\textbf {\bibinfo {volume} {110}},\
  \bibinfo {pages} {146402}}\BibitemShut {NoStop}%
\bibitem [{\citenamefont {Yu}\ and\ \citenamefont
  {Si}(2015)}]{yu2015antiferroquadrupolar}%
  \BibitemOpen
  \bibfield  {author} {\bibinfo {author} {\bibnamefont {Yu}, \bibfnamefont
  {R.}}, and\ \bibinfo {author} {\bibfnamefont {Q.}~\bibnamefont {Si}}}
  (\bibinfo {year} {2015}),\ \href
  {https://doi.org/10.1103/PhysRevLett.115.116401} {\bibfield  {journal}
  {\bibinfo  {journal} {Phys. Rev. Lett.}\ }\textbf {\bibinfo {volume} {115}},\
  \bibinfo {pages} {116401}}\BibitemShut {NoStop}%
\bibitem [{\citenamefont {Yu}\ \emph {et~al.}(2011)\citenamefont {Yu},
  \citenamefont {Zhu},\ and\ \citenamefont {Si}}]{yu2011mott}%
  \BibitemOpen
  \bibfield  {author} {\bibinfo {author} {\bibnamefont {Yu}, \bibfnamefont
  {R.}}, \bibinfo {author} {\bibfnamefont {J.-X.}\ \bibnamefont {Zhu}}, and\
  \bibinfo {author} {\bibfnamefont {Q.}~\bibnamefont {Si}}} (\bibinfo {year}
  {2011}),\ \href {https://doi.org/10.1103/PhysRevLett.106.186401} {\bibfield
  {journal} {\bibinfo  {journal} {Phys. Rev. Lett.}\ }\textbf {\bibinfo
  {volume} {106}},\ \bibinfo {pages} {186401}}\BibitemShut {NoStop}%
\bibitem [{\citenamefont {Yu}\ \emph {et~al.}(2014)\citenamefont {Yu},
  \citenamefont {Zhu},\ and\ \citenamefont {Si}}]{yu2014orbital}%
  \BibitemOpen
  \bibfield  {author} {\bibinfo {author} {\bibnamefont {Yu}, \bibfnamefont
  {R.}}, \bibinfo {author} {\bibfnamefont {J.-X.}\ \bibnamefont {Zhu}}, and\
  \bibinfo {author} {\bibfnamefont {Q.}~\bibnamefont {Si}}} (\bibinfo {year}
  {2014}),\ \href {https://doi.org/10.1103/PhysRevB.89.024509} {\bibfield
  {journal} {\bibinfo  {journal} {Phys. Rev. B}\ }\textbf {\bibinfo {volume}
  {89}},\ \bibinfo {pages} {024509}}\BibitemShut {NoStop}%
\bibitem [{\citenamefont {Yuan}\ \emph {et~al.}(2003)\citenamefont {Yuan},
  \citenamefont {Grosche}, \citenamefont {Deppe}, \citenamefont {Geibel},
  \citenamefont {Sparn},\ and\ \citenamefont {Steglich}}]{Yuan2003}%
  \BibitemOpen
  \bibfield  {author} {\bibinfo {author} {\bibnamefont {Yuan}, \bibfnamefont
  {H.}}, \bibinfo {author} {\bibfnamefont {F.}~\bibnamefont {Grosche}},
  \bibinfo {author} {\bibfnamefont {M.}~\bibnamefont {Deppe}}, \bibinfo
  {author} {\bibfnamefont {C.}~\bibnamefont {Geibel}}, \bibinfo {author}
  {\bibfnamefont {G.}~\bibnamefont {Sparn}}, and\ \bibinfo {author}
  {\bibfnamefont {F.}~\bibnamefont {Steglich}}} (\bibinfo {year} {2003}),\
  \href {https://doi.org/10.1126/science.1091648} {\bibfield  {journal}
  {\bibinfo  {journal} {Science}\ }\textbf {\bibinfo {volume} {302}},\ \bibinfo
  {pages} {2104}}\BibitemShut {NoStop}%
\bibitem [{\citenamefont {Yuan}\ \emph {et~al.}(2009)\citenamefont {Yuan},
  \citenamefont {Singleton}, \citenamefont {Balakirev}, \citenamefont {Baily},
  \citenamefont {Chen}, \citenamefont {Luo},\ and\ \citenamefont
  {Wang}}]{Yuan2009}%
  \BibitemOpen
  \bibfield  {author} {\bibinfo {author} {\bibnamefont {Yuan}, \bibfnamefont
  {H.}}, \bibinfo {author} {\bibfnamefont {J.}~\bibnamefont {Singleton}},
  \bibinfo {author} {\bibfnamefont {F.~F.}\ \bibnamefont {Balakirev}}, \bibinfo
  {author} {\bibfnamefont {S.~A.}\ \bibnamefont {Baily}}, \bibinfo {author}
  {\bibfnamefont {G.}~\bibnamefont {Chen}}, \bibinfo {author} {\bibfnamefont
  {J.}~\bibnamefont {Luo}}, and\ \bibinfo {author} {\bibfnamefont
  {N.}~\bibnamefont {Wang}}} (\bibinfo {year} {2009}),\ \href
  {https://doi.org/10.1038/nature07676} {\bibfield  {journal} {\bibinfo
  {journal} {Nature}\ }\textbf {\bibinfo {volume} {457}},\ \bibinfo {pages}
  {565}}\BibitemShut {NoStop}%
\bibitem [{\citenamefont {Yuan}\ \emph {et~al.}(2021)\citenamefont {Yuan},
  \citenamefont {Fan}, \citenamefont {Wang}, \citenamefont {He}, \citenamefont
  {Zhang}, \citenamefont {Xue},\ and\ \citenamefont
  {Li}}]{yuan2021incommensurate}%
  \BibitemOpen
  \bibfield  {author} {\bibinfo {author} {\bibnamefont {Yuan}, \bibfnamefont
  {Y.}}, \bibinfo {author} {\bibfnamefont {X.}~\bibnamefont {Fan}}, \bibinfo
  {author} {\bibfnamefont {X.}~\bibnamefont {Wang}}, \bibinfo {author}
  {\bibfnamefont {K.}~\bibnamefont {He}}, \bibinfo {author} {\bibfnamefont
  {Y.}~\bibnamefont {Zhang}}, \bibinfo {author} {\bibfnamefont {Q.-K.}\
  \bibnamefont {Xue}}, and\ \bibinfo {author} {\bibfnamefont {W.}~\bibnamefont
  {Li}}} (\bibinfo {year} {2021}),\ \href
  {https://doi.org/10.1038/s41467-021-22516-2} {\bibfield  {journal} {\bibinfo
  {journal} {Nat. Commun.}\ }\textbf {\bibinfo {volume} {12}},\ \bibinfo
  {pages} {2196}}\BibitemShut {NoStop}%
\bibitem [{\citenamefont {Zapf}\ and\ \citenamefont
  {Dressel}(2016)}]{Zapf2017}%
  \BibitemOpen
  \bibfield  {author} {\bibinfo {author} {\bibnamefont {Zapf}, \bibfnamefont
  {S.}}, and\ \bibinfo {author} {\bibfnamefont {M.}~\bibnamefont {Dressel}}}
  (\bibinfo {year} {2016}),\ \href
  {https://doi.org/10.1088/0034-4885/80/1/016501} {\bibfield  {journal}
  {\bibinfo  {journal} {Reports on Progress in Physics}\ }\textbf {\bibinfo
  {volume} {80}},\ \bibinfo {pages} {016501}}\BibitemShut {NoStop}%
\bibitem [{\citenamefont {Zapf}\ \emph {et~al.}(2013)\citenamefont {Zapf},
  \citenamefont {Jeevan}, \citenamefont {Ivek}, \citenamefont {Pfister},
  \citenamefont {Klingert}, \citenamefont {Jiang}, \citenamefont {Wu},
  \citenamefont {Gegenwart}, \citenamefont {Kremer},\ and\ \citenamefont
  {Dressel}}]{Zapf2013}%
  \BibitemOpen
  \bibfield  {author} {\bibinfo {author} {\bibnamefont {Zapf}, \bibfnamefont
  {S.}}, \bibinfo {author} {\bibfnamefont {H.~S.}\ \bibnamefont {Jeevan}},
  \bibinfo {author} {\bibfnamefont {T.}~\bibnamefont {Ivek}}, \bibinfo {author}
  {\bibfnamefont {F.}~\bibnamefont {Pfister}}, \bibinfo {author} {\bibfnamefont
  {F.}~\bibnamefont {Klingert}}, \bibinfo {author} {\bibfnamefont
  {S.}~\bibnamefont {Jiang}}, \bibinfo {author} {\bibfnamefont
  {D.}~\bibnamefont {Wu}}, \bibinfo {author} {\bibfnamefont {P.}~\bibnamefont
  {Gegenwart}}, \bibinfo {author} {\bibfnamefont {R.~K.}\ \bibnamefont
  {Kremer}}, and\ \bibinfo {author} {\bibfnamefont {M.}~\bibnamefont
  {Dressel}}} (\bibinfo {year} {2013}),\ \href
  {https://doi.org/10.1103/PhysRevLett.110.237002} {\bibfield  {journal}
  {\bibinfo  {journal} {Phys. Rev. Lett.}\ }\textbf {\bibinfo {volume} {110}},\
  \bibinfo {pages} {237002}}\BibitemShut {NoStop}%
\bibitem [{\citenamefont {Zapf}\ \emph {et~al.}(2011)\citenamefont {Zapf},
  \citenamefont {Wu}, \citenamefont {Bogani}, \citenamefont {Jeevan},
  \citenamefont {Gegenwart},\ and\ \citenamefont {Dressel}}]{Zapf2011}%
  \BibitemOpen
  \bibfield  {author} {\bibinfo {author} {\bibnamefont {Zapf}, \bibfnamefont
  {S.}}, \bibinfo {author} {\bibfnamefont {D.}~\bibnamefont {Wu}}, \bibinfo
  {author} {\bibfnamefont {L.}~\bibnamefont {Bogani}}, \bibinfo {author}
  {\bibfnamefont {H.~S.}\ \bibnamefont {Jeevan}}, \bibinfo {author}
  {\bibfnamefont {P.}~\bibnamefont {Gegenwart}}, and\ \bibinfo {author}
  {\bibfnamefont {M.}~\bibnamefont {Dressel}}} (\bibinfo {year} {2011}),\ \href
  {https://doi.org/10.1103/PhysRevB.84.140503} {\bibfield  {journal} {\bibinfo
  {journal} {Phys. Rev. B}\ }\textbf {\bibinfo {volume} {84}},\ \bibinfo
  {pages} {140503}}\BibitemShut {NoStop}%
\bibitem [{\citenamefont {Zhang}\ \emph
  {et~al.}(2014{\natexlab{a}})\citenamefont {Zhang}, \citenamefont {Harriger},
  \citenamefont {Yin}, \citenamefont {Lv}, \citenamefont {Wang}, \citenamefont
  {Tan}, \citenamefont {Song}, \citenamefont {Abernathy}, \citenamefont {Tian},
  \citenamefont {Egami}, \citenamefont {Haule}, \citenamefont {Kotliar},\ and\
  \citenamefont {Dai}}]{zhang2014effect}%
  \BibitemOpen
  \bibfield  {author} {\bibinfo {author} {\bibnamefont {Zhang}, \bibfnamefont
  {C.}}, \bibinfo {author} {\bibfnamefont {L.~W.}\ \bibnamefont {Harriger}},
  \bibinfo {author} {\bibfnamefont {Z.}~\bibnamefont {Yin}}, \bibinfo {author}
  {\bibfnamefont {W.}~\bibnamefont {Lv}}, \bibinfo {author} {\bibfnamefont
  {M.}~\bibnamefont {Wang}}, \bibinfo {author} {\bibfnamefont {G.}~\bibnamefont
  {Tan}}, \bibinfo {author} {\bibfnamefont {Y.}~\bibnamefont {Song}}, \bibinfo
  {author} {\bibfnamefont {D.~L.}\ \bibnamefont {Abernathy}}, \bibinfo {author}
  {\bibfnamefont {W.}~\bibnamefont {Tian}}, \bibinfo {author} {\bibfnamefont
  {T.}~\bibnamefont {Egami}}, \bibinfo {author} {\bibfnamefont
  {K.}~\bibnamefont {Haule}}, \bibinfo {author} {\bibfnamefont
  {G.}~\bibnamefont {Kotliar}}, and\ \bibinfo {author} {\bibfnamefont
  {P.}~\bibnamefont {Dai}}} (\bibinfo {year} {2014}{\natexlab{a}}),\ \href
  {https://doi.org/10.1103/PhysRevLett.112.217202} {\bibfield  {journal}
  {\bibinfo  {journal} {Phys. Rev. Lett.}\ }\textbf {\bibinfo {volume} {112}},\
  \bibinfo {pages} {217202}}\BibitemShut {NoStop}%
\bibitem [{\citenamefont {Zhang}\ \emph
  {et~al.}(2017{\natexlab{a}})\citenamefont {Zhang}, \citenamefont {Liu},
  \citenamefont {Chen}, \citenamefont {Xie}, \citenamefont {He}, \citenamefont
  {Tang}, \citenamefont {He}, \citenamefont {Li}, \citenamefont {Jia},
  \citenamefont {Rebec} \emph {et~al.}}]{Zhang2017b}%
  \BibitemOpen
  \bibfield  {author} {\bibinfo {author} {\bibnamefont {Zhang}, \bibfnamefont
  {C.}}, \bibinfo {author} {\bibfnamefont {Z.}~\bibnamefont {Liu}}, \bibinfo
  {author} {\bibfnamefont {Z.}~\bibnamefont {Chen}}, \bibinfo {author}
  {\bibfnamefont {Y.}~\bibnamefont {Xie}}, \bibinfo {author} {\bibfnamefont
  {R.}~\bibnamefont {He}}, \bibinfo {author} {\bibfnamefont {S.}~\bibnamefont
  {Tang}}, \bibinfo {author} {\bibfnamefont {J.}~\bibnamefont {He}}, \bibinfo
  {author} {\bibfnamefont {W.}~\bibnamefont {Li}}, \bibinfo {author}
  {\bibfnamefont {T.}~\bibnamefont {Jia}}, \bibinfo {author} {\bibfnamefont
  {S.~N.}\ \bibnamefont {Rebec}},  \emph {et~al.}} (\bibinfo {year}
  {2017}{\natexlab{a}}),\ \href {https://doi.org/10.1038/ncomms14468}
  {\bibfield  {journal} {\bibinfo  {journal} {Nat. Commun.}\ }\textbf {\bibinfo
  {volume} {8}},\ \bibinfo {pages} {14468}}\BibitemShut {NoStop}%
\bibitem [{\citenamefont {Zhang}\ \emph
  {et~al.}(2016{\natexlab{a}})\citenamefont {Zhang}, \citenamefont {Lv},
  \citenamefont {Tan}, \citenamefont {Song}, \citenamefont {Carr},
  \citenamefont {Chi}, \citenamefont {Matsuda}, \citenamefont {Christianson},
  \citenamefont {{Fernandez-Baca}}, \citenamefont {Harriger},\ and\
  \citenamefont {Dai}}]{zhang2016electron}%
  \BibitemOpen
  \bibfield  {author} {\bibinfo {author} {\bibnamefont {Zhang}, \bibfnamefont
  {C.}}, \bibinfo {author} {\bibfnamefont {W.}~\bibnamefont {Lv}}, \bibinfo
  {author} {\bibfnamefont {G.}~\bibnamefont {Tan}}, \bibinfo {author}
  {\bibfnamefont {Y.}~\bibnamefont {Song}}, \bibinfo {author} {\bibfnamefont
  {S.~V.}\ \bibnamefont {Carr}}, \bibinfo {author} {\bibfnamefont
  {S.}~\bibnamefont {Chi}}, \bibinfo {author} {\bibfnamefont {M.}~\bibnamefont
  {Matsuda}}, \bibinfo {author} {\bibfnamefont {A.~D.}\ \bibnamefont
  {Christianson}}, \bibinfo {author} {\bibfnamefont {J.~A.}\ \bibnamefont
  {{Fernandez-Baca}}}, \bibinfo {author} {\bibfnamefont {L.~W.}\ \bibnamefont
  {Harriger}}, and\ \bibinfo {author} {\bibfnamefont {P.}~\bibnamefont {Dai}}}
  (\bibinfo {year} {2016}{\natexlab{a}}),\ \href
  {https://doi.org/10.1103/PhysRevB.93.174522} {\bibfield  {journal} {\bibinfo
  {journal} {Phys. Rev. B}\ }\textbf {\bibinfo {volume} {93}},\ \bibinfo
  {pages} {174522}}\BibitemShut {NoStop}%
\bibitem [{\citenamefont {Zhang}\ \emph
  {et~al.}(2015{\natexlab{a}})\citenamefont {Zhang}, \citenamefont {Park},
  \citenamefont {Lu}, \citenamefont {Yu}, \citenamefont {Li}, \citenamefont
  {Zhang}, \citenamefont {Zhao}, \citenamefont {Lynn}, \citenamefont {Si},\
  and\ \citenamefont {Dai}}]{zhang2015neutron}%
  \BibitemOpen
  \bibfield  {author} {\bibinfo {author} {\bibnamefont {Zhang}, \bibfnamefont
  {C.}}, \bibinfo {author} {\bibfnamefont {J.~T.}\ \bibnamefont {Park}},
  \bibinfo {author} {\bibfnamefont {X.}~\bibnamefont {Lu}}, \bibinfo {author}
  {\bibfnamefont {R.}~\bibnamefont {Yu}}, \bibinfo {author} {\bibfnamefont
  {Y.}~\bibnamefont {Li}}, \bibinfo {author} {\bibfnamefont {W.}~\bibnamefont
  {Zhang}}, \bibinfo {author} {\bibfnamefont {Y.}~\bibnamefont {Zhao}},
  \bibinfo {author} {\bibfnamefont {J.~W.}\ \bibnamefont {Lynn}}, \bibinfo
  {author} {\bibfnamefont {Q.}~\bibnamefont {Si}}, and\ \bibinfo {author}
  {\bibfnamefont {P.}~\bibnamefont {Dai}}} (\bibinfo {year}
  {2015}{\natexlab{a}}),\ \href {https://doi.org/10.1103/PhysRevB.91.104520}
  {\bibfield  {journal} {\bibinfo  {journal} {Phys. Rev. B}\ }\textbf {\bibinfo
  {volume} {91}},\ \bibinfo {pages} {104520}}\BibitemShut {NoStop}%
\bibitem [{\citenamefont {Zhang}\ \emph
  {et~al.}(2014{\natexlab{b}})\citenamefont {Zhang}, \citenamefont {Song},
  \citenamefont {Regnault}, \citenamefont {Su}, \citenamefont {Enderle},
  \citenamefont {Kulda}, \citenamefont {Tan}, \citenamefont {Sims},
  \citenamefont {Egami}, \citenamefont {Si},\ and\ \citenamefont
  {Dai}}]{zhang2014anisotropic}%
  \BibitemOpen
  \bibfield  {author} {\bibinfo {author} {\bibnamefont {Zhang}, \bibfnamefont
  {C.}}, \bibinfo {author} {\bibfnamefont {Y.}~\bibnamefont {Song}}, \bibinfo
  {author} {\bibfnamefont {L.-P.}\ \bibnamefont {Regnault}}, \bibinfo {author}
  {\bibfnamefont {Y.}~\bibnamefont {Su}}, \bibinfo {author} {\bibfnamefont
  {M.}~\bibnamefont {Enderle}}, \bibinfo {author} {\bibfnamefont
  {J.}~\bibnamefont {Kulda}}, \bibinfo {author} {\bibfnamefont
  {G.}~\bibnamefont {Tan}}, \bibinfo {author} {\bibfnamefont {Z.~C.}\
  \bibnamefont {Sims}}, \bibinfo {author} {\bibfnamefont {T.}~\bibnamefont
  {Egami}}, \bibinfo {author} {\bibfnamefont {Q.}~\bibnamefont {Si}}, and\
  \bibinfo {author} {\bibfnamefont {P.}~\bibnamefont {Dai}}} (\bibinfo {year}
  {2014}{\natexlab{b}}),\ \href {https://doi.org/10.1103/PhysRevB.90.140502}
  {\bibfield  {journal} {\bibinfo  {journal} {Phys. Rev. B}\ }\textbf {\bibinfo
  {volume} {90}},\ \bibinfo {pages} {140502}}\BibitemShut {NoStop}%
\bibitem [{\citenamefont {Zhang}\ \emph {et~al.}(2013)\citenamefont {Zhang},
  \citenamefont {Yu}, \citenamefont {Su}, \citenamefont {Song}, \citenamefont
  {Wang}, \citenamefont {Tan}, \citenamefont {Egami}, \citenamefont
  {Fernandez-Baca}, \citenamefont {Faulhaber}, \citenamefont {Si},\ and\
  \citenamefont {Dai}}]{zhang2013measurement}%
  \BibitemOpen
  \bibfield  {author} {\bibinfo {author} {\bibnamefont {Zhang}, \bibfnamefont
  {C.}}, \bibinfo {author} {\bibfnamefont {R.}~\bibnamefont {Yu}}, \bibinfo
  {author} {\bibfnamefont {Y.}~\bibnamefont {Su}}, \bibinfo {author}
  {\bibfnamefont {Y.}~\bibnamefont {Song}}, \bibinfo {author} {\bibfnamefont
  {M.}~\bibnamefont {Wang}}, \bibinfo {author} {\bibfnamefont {G.}~\bibnamefont
  {Tan}}, \bibinfo {author} {\bibfnamefont {T.}~\bibnamefont {Egami}}, \bibinfo
  {author} {\bibfnamefont {J.~A.}\ \bibnamefont {Fernandez-Baca}}, \bibinfo
  {author} {\bibfnamefont {E.}~\bibnamefont {Faulhaber}}, \bibinfo {author}
  {\bibfnamefont {Q.}~\bibnamefont {Si}}, and\ \bibinfo {author} {\bibfnamefont
  {P.}~\bibnamefont {Dai}}} (\bibinfo {year} {2013}),\ \href
  {https://doi.org/10.1103/PhysRevLett.111.207002} {\bibfield  {journal}
  {\bibinfo  {journal} {Phys. Rev. Lett.}\ }\textbf {\bibinfo {volume} {111}},\
  \bibinfo {pages} {207002}}\BibitemShut {NoStop}%
\bibitem [{\citenamefont {Zhang}\ \emph {et~al.}(2020)\citenamefont {Zhang},
  \citenamefont {Ge}, \citenamefont {Weinert},\ and\ \citenamefont
  {Li}}]{zhang2020sign}%
  \BibitemOpen
  \bibfield  {author} {\bibinfo {author} {\bibnamefont {Zhang}, \bibfnamefont
  {H.}}, \bibinfo {author} {\bibfnamefont {Z.}~\bibnamefont {Ge}}, \bibinfo
  {author} {\bibfnamefont {M.}~\bibnamefont {Weinert}}, and\ \bibinfo {author}
  {\bibfnamefont {L.}~\bibnamefont {Li}}} (\bibinfo {year} {2020}),\ \href
  {https://doi.org/10.1038/s42005-020-0351-1} {\bibfield  {journal} {\bibinfo
  {journal} {Commun. Phys.}\ }\textbf {\bibinfo {volume} {3}},\ \bibinfo
  {pages} {75}}\BibitemShut {NoStop}%
\bibitem [{\citenamefont {Zhang}\ \emph
  {et~al.}(2017{\natexlab{b}})\citenamefont {Zhang}, \citenamefont {Zhang},
  \citenamefont {Lu}, \citenamefont {Liu}, \citenamefont {Zhou}, \citenamefont
  {Ma}, \citenamefont {Wang}, \citenamefont {Jiang}, \citenamefont {Xue},\ and\
  \citenamefont {Bao}}]{Zhang2017a}%
  \BibitemOpen
  \bibfield  {author} {\bibinfo {author} {\bibnamefont {Zhang}, \bibfnamefont
  {H.}}, \bibinfo {author} {\bibfnamefont {D.}~\bibnamefont {Zhang}}, \bibinfo
  {author} {\bibfnamefont {X.}~\bibnamefont {Lu}}, \bibinfo {author}
  {\bibfnamefont {C.}~\bibnamefont {Liu}}, \bibinfo {author} {\bibfnamefont
  {G.}~\bibnamefont {Zhou}}, \bibinfo {author} {\bibfnamefont {X.}~\bibnamefont
  {Ma}}, \bibinfo {author} {\bibfnamefont {L.}~\bibnamefont {Wang}}, \bibinfo
  {author} {\bibfnamefont {P.}~\bibnamefont {Jiang}}, \bibinfo {author}
  {\bibfnamefont {Q.-K.}\ \bibnamefont {Xue}}, and\ \bibinfo {author}
  {\bibfnamefont {X.}~\bibnamefont {Bao}}} (\bibinfo {year}
  {2017}{\natexlab{b}}),\ \href {https://doi.org/10.1038/s41467-017-00281-5}
  {\bibfield  {journal} {\bibinfo  {journal} {Nat. Commun.}\ }\textbf {\bibinfo
  {volume} {8}},\ \bibinfo {pages} {214}}\BibitemShut {NoStop}%
\bibitem [{\citenamefont {Zhang}\ \emph {et~al.}(2012)\citenamefont {Zhang},
  \citenamefont {Ying}, \citenamefont {Yan}, \citenamefont {Wang},
  \citenamefont {Wang}, \citenamefont {Xiang}, \citenamefont {Ye},
  \citenamefont {Cheng}, \citenamefont {Luo}, \citenamefont {Hu} \emph
  {et~al.}}]{Zhang2012}%
  \BibitemOpen
  \bibfield  {author} {\bibinfo {author} {\bibnamefont {Zhang}, \bibfnamefont
  {M.}}, \bibinfo {author} {\bibfnamefont {J.}~\bibnamefont {Ying}}, \bibinfo
  {author} {\bibfnamefont {Y.}~\bibnamefont {Yan}}, \bibinfo {author}
  {\bibfnamefont {A.}~\bibnamefont {Wang}}, \bibinfo {author} {\bibfnamefont
  {X.}~\bibnamefont {Wang}}, \bibinfo {author} {\bibfnamefont {Z.}~\bibnamefont
  {Xiang}}, \bibinfo {author} {\bibfnamefont {G.}~\bibnamefont {Ye}}, \bibinfo
  {author} {\bibfnamefont {P.}~\bibnamefont {Cheng}}, \bibinfo {author}
  {\bibfnamefont {X.}~\bibnamefont {Luo}}, \bibinfo {author} {\bibfnamefont
  {J.}~\bibnamefont {Hu}},  \emph {et~al.}} (\bibinfo {year} {2012}),\ \href
  {https://doi.org/10.1103/PhysRevB.85.092503} {\bibfield  {journal} {\bibinfo
  {journal} {Phys. Rev. B}\ }\textbf {\bibinfo {volume} {85}},\ \bibinfo
  {pages} {092503}}\BibitemShut {NoStop}%
\bibitem [{\citenamefont {Zhang}\ \emph
  {et~al.}(2019{\natexlab{a}})\citenamefont {Zhang}, \citenamefont {Wang},
  \citenamefont {Wu}, \citenamefont {Yaji}, \citenamefont {Ishida},
  \citenamefont {Kohama}, \citenamefont {Dai}, \citenamefont {Sun},
  \citenamefont {Bareille}, \citenamefont {Kuroda}, \citenamefont {Kondo},
  \citenamefont {Okazaki}, \citenamefont {Kindo}, \citenamefont {Wang},
  \citenamefont {Jin}, \citenamefont {Hu}, \citenamefont {Thomale},
  \citenamefont {Sumida}, \citenamefont {Wu}, \citenamefont {Miyamoto},
  \citenamefont {Okuda}, \citenamefont {Ding}, \citenamefont {Gu},
  \citenamefont {Tamegai}, \citenamefont {Kawakami}, \citenamefont {Sato},\
  and\ \citenamefont {Shin}}]{zhang2019multiple}%
  \BibitemOpen
  \bibfield  {author} {\bibinfo {author} {\bibnamefont {Zhang}, \bibfnamefont
  {P.}}, \bibinfo {author} {\bibfnamefont {Z.}~\bibnamefont {Wang}}, \bibinfo
  {author} {\bibfnamefont {X.}~\bibnamefont {Wu}}, \bibinfo {author}
  {\bibfnamefont {K.}~\bibnamefont {Yaji}}, \bibinfo {author} {\bibfnamefont
  {Y.}~\bibnamefont {Ishida}}, \bibinfo {author} {\bibfnamefont
  {Y.}~\bibnamefont {Kohama}}, \bibinfo {author} {\bibfnamefont
  {G.}~\bibnamefont {Dai}}, \bibinfo {author} {\bibfnamefont {Y.}~\bibnamefont
  {Sun}}, \bibinfo {author} {\bibfnamefont {C.}~\bibnamefont {Bareille}},
  \bibinfo {author} {\bibfnamefont {K.}~\bibnamefont {Kuroda}}, \bibinfo
  {author} {\bibfnamefont {T.}~\bibnamefont {Kondo}}, \bibinfo {author}
  {\bibfnamefont {K.}~\bibnamefont {Okazaki}}, \bibinfo {author} {\bibfnamefont
  {K.}~\bibnamefont {Kindo}}, \bibinfo {author} {\bibfnamefont
  {X.}~\bibnamefont {Wang}}, \bibinfo {author} {\bibfnamefont {C.}~\bibnamefont
  {Jin}}, \bibinfo {author} {\bibfnamefont {J.}~\bibnamefont {Hu}}, \bibinfo
  {author} {\bibfnamefont {R.}~\bibnamefont {Thomale}}, \bibinfo {author}
  {\bibfnamefont {K.}~\bibnamefont {Sumida}}, \bibinfo {author} {\bibfnamefont
  {S.}~\bibnamefont {Wu}}, \bibinfo {author} {\bibfnamefont {K.}~\bibnamefont
  {Miyamoto}}, \bibinfo {author} {\bibfnamefont {T.}~\bibnamefont {Okuda}},
  \bibinfo {author} {\bibfnamefont {H.}~\bibnamefont {Ding}}, \bibinfo {author}
  {\bibfnamefont {G.~D.}\ \bibnamefont {Gu}}, \bibinfo {author} {\bibfnamefont
  {T.}~\bibnamefont {Tamegai}}, \bibinfo {author} {\bibfnamefont
  {T.}~\bibnamefont {Kawakami}}, \bibinfo {author} {\bibfnamefont
  {M.}~\bibnamefont {Sato}}, and\ \bibinfo {author} {\bibfnamefont
  {S.}~\bibnamefont {Shin}}} (\bibinfo {year} {2019}{\natexlab{a}}),\ \href
  {https://doi.org/10.1038/s41567-018-0280-z} {\bibfield  {journal} {\bibinfo
  {journal} {Nat. Phys.}\ }\textbf {\bibinfo {volume} {15}},\ \bibinfo {pages}
  {41}}\BibitemShut {NoStop}%
\bibitem [{\citenamefont {Zhang}\ \emph
  {et~al.}(2018{\natexlab{a}})\citenamefont {Zhang}, \citenamefont {Yaji},
  \citenamefont {Hashimoto}, \citenamefont {Ota}, \citenamefont {Kondo},
  \citenamefont {Okazaki}, \citenamefont {Wang}, \citenamefont {Wen},
  \citenamefont {Gu}, \citenamefont {Ding},\ and\ \citenamefont
  {Shin}}]{zhang2018observation}%
  \BibitemOpen
  \bibfield  {author} {\bibinfo {author} {\bibnamefont {Zhang}, \bibfnamefont
  {P.}}, \bibinfo {author} {\bibfnamefont {K.}~\bibnamefont {Yaji}}, \bibinfo
  {author} {\bibfnamefont {T.}~\bibnamefont {Hashimoto}}, \bibinfo {author}
  {\bibfnamefont {Y.}~\bibnamefont {Ota}}, \bibinfo {author} {\bibfnamefont
  {T.}~\bibnamefont {Kondo}}, \bibinfo {author} {\bibfnamefont
  {K.}~\bibnamefont {Okazaki}}, \bibinfo {author} {\bibfnamefont
  {Z.}~\bibnamefont {Wang}}, \bibinfo {author} {\bibfnamefont {J.}~\bibnamefont
  {Wen}}, \bibinfo {author} {\bibfnamefont {G.~D.}\ \bibnamefont {Gu}},
  \bibinfo {author} {\bibfnamefont {H.}~\bibnamefont {Ding}}, and\ \bibinfo
  {author} {\bibfnamefont {S.}~\bibnamefont {Shin}}} (\bibinfo {year}
  {2018}{\natexlab{a}}),\ \href {https://doi.org/10.1126/science.aan4596}
  {\bibfield  {journal} {\bibinfo  {journal} {Science}\ }\textbf {\bibinfo
  {volume} {360}},\ \bibinfo {pages} {182}}\BibitemShut {NoStop}%
\bibitem [{\citenamefont {Zhang}\ \emph
  {et~al.}(2018{\natexlab{b}})\citenamefont {Zhang}, \citenamefont {Wang},
  \citenamefont {Maier}, \citenamefont {Wang}, \citenamefont {Stone},
  \citenamefont {Chi}, \citenamefont {Winn},\ and\ \citenamefont
  {Dai}}]{zhang2018neutron}%
  \BibitemOpen
  \bibfield  {author} {\bibinfo {author} {\bibnamefont {Zhang}, \bibfnamefont
  {R.}}, \bibinfo {author} {\bibfnamefont {W.}~\bibnamefont {Wang}}, \bibinfo
  {author} {\bibfnamefont {T.~A.}\ \bibnamefont {Maier}}, \bibinfo {author}
  {\bibfnamefont {M.}~\bibnamefont {Wang}}, \bibinfo {author} {\bibfnamefont
  {M.~B.}\ \bibnamefont {Stone}}, \bibinfo {author} {\bibfnamefont
  {S.}~\bibnamefont {Chi}}, \bibinfo {author} {\bibfnamefont {B.}~\bibnamefont
  {Winn}}, and\ \bibinfo {author} {\bibfnamefont {P.}~\bibnamefont {Dai}}}
  (\bibinfo {year} {2018}{\natexlab{b}}),\ \href
  {https://doi.org/10.1103/PhysRevB.98.060502} {\bibfield  {journal} {\bibinfo
  {journal} {Phys. Rev. B}\ }\textbf {\bibinfo {volume} {98}},\ \bibinfo
  {pages} {060502}}\BibitemShut {NoStop}%
\bibitem [{\citenamefont {Zhang}\ \emph
  {et~al.}(2019{\natexlab{b}})\citenamefont {Zhang}, \citenamefont {Wei},
  \citenamefont {Guan}, \citenamefont {Zhu}, \citenamefont {Qin}, \citenamefont
  {Wang}, \citenamefont {Zhang}, \citenamefont {Plummer}, \citenamefont {Zhu},
  \citenamefont {Zhang},\ and\ \citenamefont {Guo}}]{zhang2019enhanced}%
  \BibitemOpen
  \bibfield  {author} {\bibinfo {author} {\bibnamefont {Zhang}, \bibfnamefont
  {S.}}, \bibinfo {author} {\bibfnamefont {T.}~\bibnamefont {Wei}}, \bibinfo
  {author} {\bibfnamefont {J.}~\bibnamefont {Guan}}, \bibinfo {author}
  {\bibfnamefont {Q.}~\bibnamefont {Zhu}}, \bibinfo {author} {\bibfnamefont
  {W.}~\bibnamefont {Qin}}, \bibinfo {author} {\bibfnamefont {W.}~\bibnamefont
  {Wang}}, \bibinfo {author} {\bibfnamefont {J.}~\bibnamefont {Zhang}},
  \bibinfo {author} {\bibfnamefont {E.~W.}\ \bibnamefont {Plummer}}, \bibinfo
  {author} {\bibfnamefont {X.}~\bibnamefont {Zhu}}, \bibinfo {author}
  {\bibfnamefont {Z.}~\bibnamefont {Zhang}}, and\ \bibinfo {author}
  {\bibfnamefont {J.}~\bibnamefont {Guo}}} (\bibinfo {year}
  {2019}{\natexlab{b}}),\ \href
  {https://doi.org/10.1103/PhysRevLett.122.066802} {\bibfield  {journal}
  {\bibinfo  {journal} {Phys. Rev. Lett.}\ }\textbf {\bibinfo {volume} {122}},\
  \bibinfo {pages} {066802}}\BibitemShut {NoStop}%
\bibitem [{\citenamefont {Zhang}\ \emph {et~al.}(2021)\citenamefont {Zhang},
  \citenamefont {Bao}, \citenamefont {Chen}, \citenamefont {Li}, \citenamefont
  {Lu}, \citenamefont {Hu}, \citenamefont {Yang}, \citenamefont {Zhao},
  \citenamefont {Yan}, \citenamefont {Dong}, \citenamefont {Wang},
  \citenamefont {Zhang},\ and\ \citenamefont {Feng}}]{zhang2021observation}%
  \BibitemOpen
  \bibfield  {author} {\bibinfo {author} {\bibnamefont {Zhang}, \bibfnamefont
  {T.}}, \bibinfo {author} {\bibfnamefont {W.}~\bibnamefont {Bao}}, \bibinfo
  {author} {\bibfnamefont {C.}~\bibnamefont {Chen}}, \bibinfo {author}
  {\bibfnamefont {D.}~\bibnamefont {Li}}, \bibinfo {author} {\bibfnamefont
  {Z.}~\bibnamefont {Lu}}, \bibinfo {author} {\bibfnamefont {Y.}~\bibnamefont
  {Hu}}, \bibinfo {author} {\bibfnamefont {W.}~\bibnamefont {Yang}}, \bibinfo
  {author} {\bibfnamefont {D.}~\bibnamefont {Zhao}}, \bibinfo {author}
  {\bibfnamefont {Y.}~\bibnamefont {Yan}}, \bibinfo {author} {\bibfnamefont
  {X.}~\bibnamefont {Dong}}, \bibinfo {author} {\bibfnamefont {Q.-H.}\
  \bibnamefont {Wang}}, \bibinfo {author} {\bibfnamefont {T.}~\bibnamefont
  {Zhang}}, and\ \bibinfo {author} {\bibfnamefont {D.}~\bibnamefont {Feng}}}
  (\bibinfo {year} {2021}),\ \href
  {https://doi.org/10.1103/PhysRevLett.126.127001} {\bibfield  {journal}
  {\bibinfo  {journal} {Phys. Rev. Lett.}\ }\textbf {\bibinfo {volume} {126}},\
  \bibinfo {pages} {127001}}\BibitemShut {NoStop}%
\bibitem [{\citenamefont {Zhang}\ \emph
  {et~al.}(2014{\natexlab{c}})\citenamefont {Zhang}, \citenamefont {Li},
  \citenamefont {Li}, \citenamefont {Zhang}, \citenamefont {Peng},
  \citenamefont {Tang}, \citenamefont {Wang}, \citenamefont {He}, \citenamefont
  {Chen}, \citenamefont {Wang} \emph {et~al.}}]{Zhang2014a}%
  \BibitemOpen
  \bibfield  {author} {\bibinfo {author} {\bibnamefont {Zhang}, \bibfnamefont
  {W.}}, \bibinfo {author} {\bibfnamefont {Z.}~\bibnamefont {Li}}, \bibinfo
  {author} {\bibfnamefont {F.}~\bibnamefont {Li}}, \bibinfo {author}
  {\bibfnamefont {H.}~\bibnamefont {Zhang}}, \bibinfo {author} {\bibfnamefont
  {J.}~\bibnamefont {Peng}}, \bibinfo {author} {\bibfnamefont {C.}~\bibnamefont
  {Tang}}, \bibinfo {author} {\bibfnamefont {Q.}~\bibnamefont {Wang}}, \bibinfo
  {author} {\bibfnamefont {K.}~\bibnamefont {He}}, \bibinfo {author}
  {\bibfnamefont {X.}~\bibnamefont {Chen}}, \bibinfo {author} {\bibfnamefont
  {L.}~\bibnamefont {Wang}},  \emph {et~al.}} (\bibinfo {year}
  {2014}{\natexlab{c}}),\ \href {https://doi.org/10.1103/PhysRevB.89.060506}
  {\bibfield  {journal} {\bibinfo  {journal} {Phys. Rev. B}\ }\textbf {\bibinfo
  {volume} {89}},\ \bibinfo {pages} {060506}}\BibitemShut {NoStop}%
\bibitem [{\citenamefont {Zhang}\ \emph
  {et~al.}(2016{\natexlab{b}})\citenamefont {Zhang}, \citenamefont {Park},
  \citenamefont {Lu}, \citenamefont {Wei}, \citenamefont {Ma}, \citenamefont
  {Hao}, \citenamefont {Dai}, \citenamefont {Meng}, \citenamefont {Yang},
  \citenamefont {Luo},\ and\ \citenamefont {Li}}]{zhang2016effect}%
  \BibitemOpen
  \bibfield  {author} {\bibinfo {author} {\bibnamefont {Zhang}, \bibfnamefont
  {W.}}, \bibinfo {author} {\bibfnamefont {J.~T.}\ \bibnamefont {Park}},
  \bibinfo {author} {\bibfnamefont {X.}~\bibnamefont {Lu}}, \bibinfo {author}
  {\bibfnamefont {Y.}~\bibnamefont {Wei}}, \bibinfo {author} {\bibfnamefont
  {X.}~\bibnamefont {Ma}}, \bibinfo {author} {\bibfnamefont {L.}~\bibnamefont
  {Hao}}, \bibinfo {author} {\bibfnamefont {P.}~\bibnamefont {Dai}}, \bibinfo
  {author} {\bibfnamefont {Z.~Y.}\ \bibnamefont {Meng}}, \bibinfo {author}
  {\bibfnamefont {Y.-f.}\ \bibnamefont {Yang}}, \bibinfo {author}
  {\bibfnamefont {H.}~\bibnamefont {Luo}}, and\ \bibinfo {author}
  {\bibfnamefont {S.}~\bibnamefont {Li}}} (\bibinfo {year}
  {2016}{\natexlab{b}}),\ \href
  {https://doi.org/10.1103/PhysRevLett.117.227003} {\bibfield  {journal}
  {\bibinfo  {journal} {Phys. Rev. Lett.}\ }\textbf {\bibinfo {volume} {117}},\
  \bibinfo {pages} {227003}}\BibitemShut {NoStop}%
\bibitem [{\citenamefont {Zhang}\ \emph
  {et~al.}(2014{\natexlab{d}})\citenamefont {Zhang}, \citenamefont {Sun},
  \citenamefont {Zhang}, \citenamefont {Li}, \citenamefont {Guo}, \citenamefont
  {Zhao}, \citenamefont {Zhang}, \citenamefont {Peng}, \citenamefont {Xing},
  \citenamefont {Wang}, \citenamefont {Fujita}, \citenamefont {Hirata},
  \citenamefont {Li}, \citenamefont {Ding}, \citenamefont {Tang}, \citenamefont
  {Wang}, \citenamefont {Wang}, \citenamefont {He}, \citenamefont {Ji},
  \citenamefont {Chen}, \citenamefont {Wang}, \citenamefont {Xia},
  \citenamefont {Li}, \citenamefont {Wang}, \citenamefont {Wang}, \citenamefont
  {Wang}, \citenamefont {Chen}, \citenamefont {Xue},\ and\ \citenamefont
  {Ma}}]{zhang2014direct}%
  \BibitemOpen
  \bibfield  {author} {\bibinfo {author} {\bibnamefont {Zhang}, \bibfnamefont
  {W.-H.}}, \bibinfo {author} {\bibfnamefont {Y.}~\bibnamefont {Sun}}, \bibinfo
  {author} {\bibfnamefont {J.-S.}\ \bibnamefont {Zhang}}, \bibinfo {author}
  {\bibfnamefont {F.-S.}\ \bibnamefont {Li}}, \bibinfo {author} {\bibfnamefont
  {M.-H.}\ \bibnamefont {Guo}}, \bibinfo {author} {\bibfnamefont {Y.-F.}\
  \bibnamefont {Zhao}}, \bibinfo {author} {\bibfnamefont {H.-M.}\ \bibnamefont
  {Zhang}}, \bibinfo {author} {\bibfnamefont {J.-P.}\ \bibnamefont {Peng}},
  \bibinfo {author} {\bibfnamefont {Y.}~\bibnamefont {Xing}}, \bibinfo {author}
  {\bibfnamefont {H.-C.}\ \bibnamefont {Wang}}, \bibinfo {author}
  {\bibfnamefont {T.}~\bibnamefont {Fujita}}, \bibinfo {author} {\bibfnamefont
  {A.}~\bibnamefont {Hirata}}, \bibinfo {author} {\bibfnamefont
  {Z.}~\bibnamefont {Li}}, \bibinfo {author} {\bibfnamefont {H.}~\bibnamefont
  {Ding}}, \bibinfo {author} {\bibfnamefont {C.-J.}\ \bibnamefont {Tang}},
  \bibinfo {author} {\bibfnamefont {M.}~\bibnamefont {Wang}}, \bibinfo {author}
  {\bibfnamefont {Q.-Y.}\ \bibnamefont {Wang}}, \bibinfo {author}
  {\bibfnamefont {K.}~\bibnamefont {He}}, \bibinfo {author} {\bibfnamefont
  {S.-H.}\ \bibnamefont {Ji}}, \bibinfo {author} {\bibfnamefont
  {X.}~\bibnamefont {Chen}}, \bibinfo {author} {\bibfnamefont {J.-F.}\
  \bibnamefont {Wang}}, \bibinfo {author} {\bibfnamefont {Z.-C.}\ \bibnamefont
  {Xia}}, \bibinfo {author} {\bibfnamefont {L.}~\bibnamefont {Li}}, \bibinfo
  {author} {\bibfnamefont {Y.-Y.}\ \bibnamefont {Wang}}, \bibinfo {author}
  {\bibfnamefont {J.}~\bibnamefont {Wang}}, \bibinfo {author} {\bibfnamefont
  {L.-L.}\ \bibnamefont {Wang}}, \bibinfo {author} {\bibfnamefont {M.-W.}\
  \bibnamefont {Chen}}, \bibinfo {author} {\bibfnamefont {Q.-K.}\ \bibnamefont
  {Xue}}, and\ \bibinfo {author} {\bibfnamefont {X.-C.}\ \bibnamefont {Ma}}}
  (\bibinfo {year} {2014}{\natexlab{d}}),\ \href
  {https://doi.org/10.1088/0256-307X/31/1/017401} {\bibfield  {journal}
  {\bibinfo  {journal} {Chin. Phys. Lett.}\ }\textbf {\bibinfo {volume} {31}},\
  \bibinfo {pages} {017401}}\BibitemShut {NoStop}%
\bibitem [{\citenamefont {Zhang}\ \emph
  {et~al.}(2017{\natexlab{c}})\citenamefont {Zhang}, \citenamefont {Oguro},
  \citenamefont {Yao}, \citenamefont {Dong}, \citenamefont {Xu}, \citenamefont
  {Wang}, \citenamefont {Awaji}, \citenamefont {Watanabe},\ and\ \citenamefont
  {Ma}}]{Zhang2017c}%
  \BibitemOpen
  \bibfield  {author} {\bibinfo {author} {\bibnamefont {Zhang}, \bibfnamefont
  {X.}}, \bibinfo {author} {\bibfnamefont {H.}~\bibnamefont {Oguro}}, \bibinfo
  {author} {\bibfnamefont {C.}~\bibnamefont {Yao}}, \bibinfo {author}
  {\bibfnamefont {C.}~\bibnamefont {Dong}}, \bibinfo {author} {\bibfnamefont
  {Z.}~\bibnamefont {Xu}}, \bibinfo {author} {\bibfnamefont {D.}~\bibnamefont
  {Wang}}, \bibinfo {author} {\bibfnamefont {S.}~\bibnamefont {Awaji}},
  \bibinfo {author} {\bibfnamefont {K.}~\bibnamefont {Watanabe}}, and\ \bibinfo
  {author} {\bibfnamefont {Y.}~\bibnamefont {Ma}}} (\bibinfo {year}
  {2017}{\natexlab{c}}),\ \href {https://doi.org/10.1109/TASC.2017.2650408}
  {\bibfield  {journal} {\bibinfo  {journal} {IEEE Transactions on Applied
  Superconductivity}\ }\textbf {\bibinfo {volume} {27}},\ \bibinfo {pages}
  {1}}\BibitemShut {NoStop}%
\bibitem [{\citenamefont {Zhang}\ \emph
  {et~al.}(2016{\natexlab{c}})\citenamefont {Zhang}, \citenamefont {Lee},
  \citenamefont {Moore}, \citenamefont {Li}, \citenamefont {Yi}, \citenamefont
  {Hashimoto}, \citenamefont {Lu}, \citenamefont {Devereaux}, \citenamefont
  {Lee},\ and\ \citenamefont {Shen}}]{zhang2016superconducting}%
  \BibitemOpen
  \bibfield  {author} {\bibinfo {author} {\bibnamefont {Zhang}, \bibfnamefont
  {Y.}}, \bibinfo {author} {\bibfnamefont {J.~J.}\ \bibnamefont {Lee}},
  \bibinfo {author} {\bibfnamefont {R.~G.}\ \bibnamefont {Moore}}, \bibinfo
  {author} {\bibfnamefont {W.}~\bibnamefont {Li}}, \bibinfo {author}
  {\bibfnamefont {M.}~\bibnamefont {Yi}}, \bibinfo {author} {\bibfnamefont
  {M.}~\bibnamefont {Hashimoto}}, \bibinfo {author} {\bibfnamefont {D.~H.}\
  \bibnamefont {Lu}}, \bibinfo {author} {\bibfnamefont {T.~P.}\ \bibnamefont
  {Devereaux}}, \bibinfo {author} {\bibfnamefont {D.-H.}\ \bibnamefont {Lee}},
  and\ \bibinfo {author} {\bibfnamefont {Z.-X.}\ \bibnamefont {Shen}}}
  (\bibinfo {year} {2016}{\natexlab{c}}),\ \href
  {https://doi.org/10.1103/PhysRevLett.117.117001} {\bibfield  {journal}
  {\bibinfo  {journal} {Phys. Rev. Lett.}\ }\textbf {\bibinfo {volume} {117}},\
  \bibinfo {pages} {117001}}\BibitemShut {NoStop}%
\bibitem [{\citenamefont {Zhang}\ \emph {et~al.}(2010)\citenamefont {Zhang},
  \citenamefont {Yang}, \citenamefont {Chen}, \citenamefont {Zhou},
  \citenamefont {Wang}, \citenamefont {Chen}, \citenamefont {Arita},
  \citenamefont {Shimada}, \citenamefont {Namatame}, \citenamefont {Taniguchi}
  \emph {et~al.}}]{Zhang2010}%
  \BibitemOpen
  \bibfield  {author} {\bibinfo {author} {\bibnamefont {Zhang}, \bibfnamefont
  {Y.}}, \bibinfo {author} {\bibfnamefont {L.}~\bibnamefont {Yang}}, \bibinfo
  {author} {\bibfnamefont {F.}~\bibnamefont {Chen}}, \bibinfo {author}
  {\bibfnamefont {B.}~\bibnamefont {Zhou}}, \bibinfo {author} {\bibfnamefont
  {X.}~\bibnamefont {Wang}}, \bibinfo {author} {\bibfnamefont {X.}~\bibnamefont
  {Chen}}, \bibinfo {author} {\bibfnamefont {M.}~\bibnamefont {Arita}},
  \bibinfo {author} {\bibfnamefont {K.}~\bibnamefont {Shimada}}, \bibinfo
  {author} {\bibfnamefont {H.}~\bibnamefont {Namatame}}, \bibinfo {author}
  {\bibfnamefont {M.}~\bibnamefont {Taniguchi}},  \emph {et~al.}} (\bibinfo
  {year} {2010}),\ \href {https://doi.org/10.1103/PhysRevLett.105.117003}
  {\bibfield  {journal} {\bibinfo  {journal} {Phys. Rev. Lett.}\ }\textbf
  {\bibinfo {volume} {105}},\ \bibinfo {pages} {117003}}\BibitemShut {NoStop}%
\bibitem [{\citenamefont {Zhang}\ \emph {et~al.}(2011)\citenamefont {Zhang},
  \citenamefont {Yang}, \citenamefont {Xu}, \citenamefont {Ye}, \citenamefont
  {Chen}, \citenamefont {He}, \citenamefont {Xu}, \citenamefont {Jiang},
  \citenamefont {Xie}, \citenamefont {Ying} \emph {et~al.}}]{Zhang2011}%
  \BibitemOpen
  \bibfield  {author} {\bibinfo {author} {\bibnamefont {Zhang}, \bibfnamefont
  {Y.}}, \bibinfo {author} {\bibfnamefont {L.}~\bibnamefont {Yang}}, \bibinfo
  {author} {\bibfnamefont {M.}~\bibnamefont {Xu}}, \bibinfo {author}
  {\bibfnamefont {Z.}~\bibnamefont {Ye}}, \bibinfo {author} {\bibfnamefont
  {F.}~\bibnamefont {Chen}}, \bibinfo {author} {\bibfnamefont {C.}~\bibnamefont
  {He}}, \bibinfo {author} {\bibfnamefont {H.}~\bibnamefont {Xu}}, \bibinfo
  {author} {\bibfnamefont {J.}~\bibnamefont {Jiang}}, \bibinfo {author}
  {\bibfnamefont {B.}~\bibnamefont {Xie}}, \bibinfo {author} {\bibfnamefont
  {J.}~\bibnamefont {Ying}},  \emph {et~al.}} (\bibinfo {year} {2011}),\ \href
  {https://doi.org/10.1038/nmat2981} {\bibfield  {journal} {\bibinfo  {journal}
  {Nat. Mater.}\ }\textbf {\bibinfo {volume} {10}},\ \bibinfo {pages}
  {273}}\BibitemShut {NoStop}%
\bibitem [{\citenamefont {Zhang}\ \emph
  {et~al.}(2015{\natexlab{b}})\citenamefont {Zhang}, \citenamefont {Wang},
  \citenamefont {Song}, \citenamefont {Liu}, \citenamefont {Peng},
  \citenamefont {Moler}, \citenamefont {Feng},\ and\ \citenamefont
  {Wang}}]{Zhang2015}%
  \BibitemOpen
  \bibfield  {author} {\bibinfo {author} {\bibnamefont {Zhang}, \bibfnamefont
  {Z.}}, \bibinfo {author} {\bibfnamefont {Y.-H.}\ \bibnamefont {Wang}},
  \bibinfo {author} {\bibfnamefont {Q.}~\bibnamefont {Song}}, \bibinfo {author}
  {\bibfnamefont {C.}~\bibnamefont {Liu}}, \bibinfo {author} {\bibfnamefont
  {R.}~\bibnamefont {Peng}}, \bibinfo {author} {\bibfnamefont {K.}~\bibnamefont
  {Moler}}, \bibinfo {author} {\bibfnamefont {D.}~\bibnamefont {Feng}}, and\
  \bibinfo {author} {\bibfnamefont {Y.}~\bibnamefont {Wang}}} (\bibinfo {year}
  {2015}{\natexlab{b}}),\ \href {https://doi.org/10.1007/s11434-015-0842-8}
  {\bibfield  {journal} {\bibinfo  {journal} {Sci. Bull.}\ }\textbf {\bibinfo
  {volume} {60}},\ \bibinfo {pages} {1301}}\BibitemShut {NoStop}%
\bibitem [{\citenamefont {Zhao}\ \emph
  {et~al.}(2023{\natexlab{a}})\citenamefont {Zhao}, \citenamefont {Blackwell},
  \citenamefont {Thinel}, \citenamefont {Handa}, \citenamefont {Ishida},
  \citenamefont {Zhu}, \citenamefont {Iyo}, \citenamefont {Eisaki},
  \citenamefont {Pasupathy},\ and\ \citenamefont {Fujita}}]{zhao2023smectic}%
  \BibitemOpen
  \bibfield  {author} {\bibinfo {author} {\bibnamefont {Zhao}, \bibfnamefont
  {H.}}, \bibinfo {author} {\bibfnamefont {R.}~\bibnamefont {Blackwell}},
  \bibinfo {author} {\bibfnamefont {M.}~\bibnamefont {Thinel}}, \bibinfo
  {author} {\bibfnamefont {T.}~\bibnamefont {Handa}}, \bibinfo {author}
  {\bibfnamefont {S.}~\bibnamefont {Ishida}}, \bibinfo {author} {\bibfnamefont
  {X.}~\bibnamefont {Zhu}}, \bibinfo {author} {\bibfnamefont {A.}~\bibnamefont
  {Iyo}}, \bibinfo {author} {\bibfnamefont {H.}~\bibnamefont {Eisaki}},
  \bibinfo {author} {\bibfnamefont {A.~N.}\ \bibnamefont {Pasupathy}}, and\
  \bibinfo {author} {\bibfnamefont {K.}~\bibnamefont {Fujita}}} (\bibinfo
  {year} {2023}{\natexlab{a}}),\ \href
  {https://doi.org/10.1038/s41586-023-06103-7} {\bibfield  {journal} {\bibinfo
  {journal} {Nature}\ }\textbf {\bibinfo {volume} {618}},\ \bibinfo {pages}
  {940}}\BibitemShut {NoStop}%
\bibitem [{\citenamefont {Zhao}\ \emph {et~al.}(2008)\citenamefont {Zhao},
  \citenamefont {Huang}, \citenamefont {de~la Cruz}, \citenamefont {Li},
  \citenamefont {Lynn}, \citenamefont {Chen}, \citenamefont {Green},
  \citenamefont {Chen}, \citenamefont {Li}, \citenamefont {Li}, \citenamefont
  {Luo}, \citenamefont {Wang},\ and\ \citenamefont
  {Dai}}]{WOS:000261127100016}%
  \BibitemOpen
  \bibfield  {author} {\bibinfo {author} {\bibnamefont {Zhao}, \bibfnamefont
  {J.}}, \bibinfo {author} {\bibfnamefont {Q.}~\bibnamefont {Huang}}, \bibinfo
  {author} {\bibfnamefont {C.}~\bibnamefont {de~la Cruz}}, \bibinfo {author}
  {\bibfnamefont {S.}~\bibnamefont {Li}}, \bibinfo {author} {\bibfnamefont
  {J.~W.}\ \bibnamefont {Lynn}}, \bibinfo {author} {\bibfnamefont
  {Y.}~\bibnamefont {Chen}}, \bibinfo {author} {\bibfnamefont {M.~A.}\
  \bibnamefont {Green}}, \bibinfo {author} {\bibfnamefont {G.~F.}\ \bibnamefont
  {Chen}}, \bibinfo {author} {\bibfnamefont {G.}~\bibnamefont {Li}}, \bibinfo
  {author} {\bibfnamefont {Z.}~\bibnamefont {Li}}, \bibinfo {author}
  {\bibfnamefont {J.~L.}\ \bibnamefont {Luo}}, \bibinfo {author} {\bibfnamefont
  {N.~L.}\ \bibnamefont {Wang}}, and\ \bibinfo {author} {\bibfnamefont
  {P.}~\bibnamefont {Dai}}} (\bibinfo {year} {2008}),\ \href
  {https://doi.org/10.1038/nmat2315} {\bibfield  {journal} {\bibinfo  {journal}
  {Nat. Mater.}\ }\textbf {\bibinfo {volume} {7}},\ \bibinfo {pages}
  {953}}\BibitemShut {NoStop}%
\bibitem [{\citenamefont {Zhao}\ \emph {et~al.}(2016)\citenamefont {Zhao},
  \citenamefont {Liang}, \citenamefont {Yuan}, \citenamefont {Hu},
  \citenamefont {Liu}, \citenamefont {Huang}, \citenamefont {He}, \citenamefont
  {Shen}, \citenamefont {Xu}, \citenamefont {Liu}, \citenamefont {Yu},
  \citenamefont {Liu}, \citenamefont {Zhou}, \citenamefont {Huang},
  \citenamefont {Dong}, \citenamefont {Zhou}, \citenamefont {Liu},
  \citenamefont {Lu}, \citenamefont {Zhao}, \citenamefont {Chen}, \citenamefont
  {Xu},\ and\ \citenamefont {Zhou}}]{zhao2016common}%
  \BibitemOpen
  \bibfield  {author} {\bibinfo {author} {\bibnamefont {Zhao}, \bibfnamefont
  {L.}}, \bibinfo {author} {\bibfnamefont {A.}~\bibnamefont {Liang}}, \bibinfo
  {author} {\bibfnamefont {D.}~\bibnamefont {Yuan}}, \bibinfo {author}
  {\bibfnamefont {Y.}~\bibnamefont {Hu}}, \bibinfo {author} {\bibfnamefont
  {D.}~\bibnamefont {Liu}}, \bibinfo {author} {\bibfnamefont {J.}~\bibnamefont
  {Huang}}, \bibinfo {author} {\bibfnamefont {S.}~\bibnamefont {He}}, \bibinfo
  {author} {\bibfnamefont {B.}~\bibnamefont {Shen}}, \bibinfo {author}
  {\bibfnamefont {Y.}~\bibnamefont {Xu}}, \bibinfo {author} {\bibfnamefont
  {X.}~\bibnamefont {Liu}}, \bibinfo {author} {\bibfnamefont {L.}~\bibnamefont
  {Yu}}, \bibinfo {author} {\bibfnamefont {G.}~\bibnamefont {Liu}}, \bibinfo
  {author} {\bibfnamefont {H.}~\bibnamefont {Zhou}}, \bibinfo {author}
  {\bibfnamefont {Y.}~\bibnamefont {Huang}}, \bibinfo {author} {\bibfnamefont
  {X.}~\bibnamefont {Dong}}, \bibinfo {author} {\bibfnamefont {F.}~\bibnamefont
  {Zhou}}, \bibinfo {author} {\bibfnamefont {K.}~\bibnamefont {Liu}}, \bibinfo
  {author} {\bibfnamefont {Z.}~\bibnamefont {Lu}}, \bibinfo {author}
  {\bibfnamefont {Z.}~\bibnamefont {Zhao}}, \bibinfo {author} {\bibfnamefont
  {C.}~\bibnamefont {Chen}}, \bibinfo {author} {\bibfnamefont {Z.}~\bibnamefont
  {Xu}}, and\ \bibinfo {author} {\bibfnamefont {X.~J.}\ \bibnamefont {Zhou}}}
  (\bibinfo {year} {2016}),\ \href {https://doi.org/10.1038/ncomms10608}
  {\bibfield  {journal} {\bibinfo  {journal} {Nat. Commun.}\ }\textbf {\bibinfo
  {volume} {7}},\ \bibinfo {pages} {10608}}\BibitemShut {NoStop}%
\bibitem [{\citenamefont {Zhao}\ \emph
  {et~al.}(2023{\natexlab{b}})\citenamefont {Zhao}, \citenamefont {Hu},
  \citenamefont {Fu}, \citenamefont {Zhou}, \citenamefont {Gu}, \citenamefont
  {Tan}, \citenamefont {Lu},\ and\ \citenamefont {Dai}}]{zhao2023uniaxial}%
  \BibitemOpen
  \bibfield  {author} {\bibinfo {author} {\bibnamefont {Zhao}, \bibfnamefont
  {Z.}}, \bibinfo {author} {\bibfnamefont {D.}~\bibnamefont {Hu}}, \bibinfo
  {author} {\bibfnamefont {X.}~\bibnamefont {Fu}}, \bibinfo {author}
  {\bibfnamefont {K.}~\bibnamefont {Zhou}}, \bibinfo {author} {\bibfnamefont
  {Y.}~\bibnamefont {Gu}}, \bibinfo {author} {\bibfnamefont {G.}~\bibnamefont
  {Tan}}, \bibinfo {author} {\bibfnamefont {X.}~\bibnamefont {Lu}}, and\
  \bibinfo {author} {\bibfnamefont {P.}~\bibnamefont {Dai}}} (\bibinfo {year}
  {2023}{\natexlab{b}}),\ \href {https://doi.org/10.48550/arXiv.2305.04424}
  {\enquote {\bibinfo {title} {{Uniaxial-Strain Tuning of the Intertwined
  Orders in BaFe$_2$(As$_{1-x}$P$_{x}$)$_2$}},}\ }\Eprint
  {https://arxiv.org/abs/2305.04424} {arXiv:2305.04424 [cond-mat.supr-con]}
  \BibitemShut {NoStop}%
\bibitem [{\citenamefont {Zhicheng}\ \emph {et~al.}(2021)\citenamefont
  {Zhicheng}, \citenamefont {Siqi}, \citenamefont {Liangwen},\ and\
  \citenamefont {Guanghan}}]{Wang2021}%
  \BibitemOpen
  \bibfield  {author} {\bibinfo {author} {\bibnamefont {Zhicheng},
  \bibfnamefont {W.}}, \bibinfo {author} {\bibfnamefont {W.}~\bibnamefont
  {Siqi}}, \bibinfo {author} {\bibfnamefont {J.}~\bibnamefont {Liangwen}}, and\
  \bibinfo {author} {\bibfnamefont {C.}~\bibnamefont {Guanghan}}} (\bibinfo
  {year} {2021}),\ \href {https://doi.org/10.1007/s12274-021-3716-1} {\bibfield
   {journal} {\bibinfo  {journal} {Nano Res.}\ }\textbf {\bibinfo {volume}
  {14}},\ \bibinfo {pages} {3629}}\BibitemShut {NoStop}%
\bibitem [{\citenamefont {Zhou}\ \emph {et~al.}(2016)\citenamefont {Zhou},
  \citenamefont {Zhang}, \citenamefont {Liu}, \citenamefont {Tang},
  \citenamefont {Wang}, \citenamefont {Li}, \citenamefont {Song}, \citenamefont
  {Ji}, \citenamefont {He}, \citenamefont {Wang} \emph {et~al.}}]{Zhou2016}%
  \BibitemOpen
  \bibfield  {author} {\bibinfo {author} {\bibnamefont {Zhou}, \bibfnamefont
  {G.}}, \bibinfo {author} {\bibfnamefont {D.}~\bibnamefont {Zhang}}, \bibinfo
  {author} {\bibfnamefont {C.}~\bibnamefont {Liu}}, \bibinfo {author}
  {\bibfnamefont {C.}~\bibnamefont {Tang}}, \bibinfo {author} {\bibfnamefont
  {X.}~\bibnamefont {Wang}}, \bibinfo {author} {\bibfnamefont {Z.}~\bibnamefont
  {Li}}, \bibinfo {author} {\bibfnamefont {C.}~\bibnamefont {Song}}, \bibinfo
  {author} {\bibfnamefont {S.}~\bibnamefont {Ji}}, \bibinfo {author}
  {\bibfnamefont {K.}~\bibnamefont {He}}, \bibinfo {author} {\bibfnamefont
  {L.}~\bibnamefont {Wang}},  \emph {et~al.}} (\bibinfo {year} {2016}),\ \href
  {https://doi.org/10.1063/1.4950964} {\bibfield  {journal} {\bibinfo
  {journal} {Appl. Phys. Lett.}\ }\textbf {\bibinfo {volume} {108}},\ \bibinfo
  {pages} {202603}}\BibitemShut {NoStop}%
\bibitem [{\citenamefont {Zhou}\ \emph {et~al.}(2013)\citenamefont {Zhou},
  \citenamefont {Huang}, \citenamefont {Monney}, \citenamefont {Dai},
  \citenamefont {Strocov}, \citenamefont {Wang}, \citenamefont {Chen},
  \citenamefont {Zhang}, \citenamefont {Dai}, \citenamefont {Patthey},
  \citenamefont {van~den Brink}, \citenamefont {Ding},\ and\ \citenamefont
  {Schmitt}}]{zhou2013persistent}%
  \BibitemOpen
  \bibfield  {author} {\bibinfo {author} {\bibnamefont {Zhou}, \bibfnamefont
  {K.-J.}}, \bibinfo {author} {\bibfnamefont {Y.-B.}\ \bibnamefont {Huang}},
  \bibinfo {author} {\bibfnamefont {C.}~\bibnamefont {Monney}}, \bibinfo
  {author} {\bibfnamefont {X.}~\bibnamefont {Dai}}, \bibinfo {author}
  {\bibfnamefont {V.~N.}\ \bibnamefont {Strocov}}, \bibinfo {author}
  {\bibfnamefont {N.-L.}\ \bibnamefont {Wang}}, \bibinfo {author}
  {\bibfnamefont {Z.-G.}\ \bibnamefont {Chen}}, \bibinfo {author}
  {\bibfnamefont {C.}~\bibnamefont {Zhang}}, \bibinfo {author} {\bibfnamefont
  {P.}~\bibnamefont {Dai}}, \bibinfo {author} {\bibfnamefont {L.}~\bibnamefont
  {Patthey}}, \bibinfo {author} {\bibfnamefont {J.}~\bibnamefont {van~den
  Brink}}, \bibinfo {author} {\bibfnamefont {H.}~\bibnamefont {Ding}}, and\
  \bibinfo {author} {\bibfnamefont {T.}~\bibnamefont {Schmitt}}} (\bibinfo
  {year} {2013}),\ \href {https://doi.org/10.1038/ncomms2428} {\bibfield
  {journal} {\bibinfo  {journal} {Nat. Commun.}\ }\textbf {\bibinfo {volume}
  {4}},\ \bibinfo {pages} {1470}}\BibitemShut {NoStop}%
\bibitem [{\citenamefont {Zhou}\ \emph {et~al.}(2022)\citenamefont {Zhou},
  \citenamefont {Walters}, \citenamefont {Garcia-Fernandez}, \citenamefont
  {Rice}, \citenamefont {Hand}, \citenamefont {Nag}, \citenamefont {Li},
  \citenamefont {Agrestini}, \citenamefont {Garland}, \citenamefont {Wang},
  \citenamefont {Alcock}, \citenamefont {Nistea}, \citenamefont {Nutter},
  \citenamefont {Rubies}, \citenamefont {Knap}, \citenamefont {Gaughran},
  \citenamefont {Yuan}, \citenamefont {Chang}, \citenamefont {Emmins},\ and\
  \citenamefont {Howell}}]{zhou2022i21}%
  \BibitemOpen
  \bibfield  {author} {\bibinfo {author} {\bibnamefont {Zhou}, \bibfnamefont
  {K.-J.}}, \bibinfo {author} {\bibfnamefont {A.}~\bibnamefont {Walters}},
  \bibinfo {author} {\bibfnamefont {M.}~\bibnamefont {Garcia-Fernandez}},
  \bibinfo {author} {\bibfnamefont {T.}~\bibnamefont {Rice}}, \bibinfo {author}
  {\bibfnamefont {M.}~\bibnamefont {Hand}}, \bibinfo {author} {\bibfnamefont
  {A.}~\bibnamefont {Nag}}, \bibinfo {author} {\bibfnamefont {J.}~\bibnamefont
  {Li}}, \bibinfo {author} {\bibfnamefont {S.}~\bibnamefont {Agrestini}},
  \bibinfo {author} {\bibfnamefont {P.}~\bibnamefont {Garland}}, \bibinfo
  {author} {\bibfnamefont {H.}~\bibnamefont {Wang}}, \bibinfo {author}
  {\bibfnamefont {S.}~\bibnamefont {Alcock}}, \bibinfo {author} {\bibfnamefont
  {I.}~\bibnamefont {Nistea}}, \bibinfo {author} {\bibfnamefont
  {B.}~\bibnamefont {Nutter}}, \bibinfo {author} {\bibfnamefont
  {N.}~\bibnamefont {Rubies}}, \bibinfo {author} {\bibfnamefont
  {G.}~\bibnamefont {Knap}}, \bibinfo {author} {\bibfnamefont {M.}~\bibnamefont
  {Gaughran}}, \bibinfo {author} {\bibfnamefont {F.}~\bibnamefont {Yuan}},
  \bibinfo {author} {\bibfnamefont {P.}~\bibnamefont {Chang}}, \bibinfo
  {author} {\bibfnamefont {J.}~\bibnamefont {Emmins}}, and\ \bibinfo {author}
  {\bibfnamefont {G.}~\bibnamefont {Howell}}} (\bibinfo {year} {2022}),\ \href
  {https://doi.org/10.1107/S1600577522000601} {\bibfield  {journal} {\bibinfo
  {journal} {Journal of Synchrotron Radiation}\ }\textbf {\bibinfo {volume}
  {29}},\ \bibinfo {pages} {563}}\BibitemShut {NoStop}%
\bibitem [{\citenamefont {Zhou}\ \emph {et~al.}(2021)\citenamefont {Zhou},
  \citenamefont {Sun}, \citenamefont {Xi}, \citenamefont {Wang}, \citenamefont
  {Zhang}, \citenamefont {Zhang}, \citenamefont {Zhang}, \citenamefont {Xu},
  \citenamefont {Pan}, \citenamefont {Feng}, \citenamefont {Meng},
  \citenamefont {Yi}, \citenamefont {Pi}, \citenamefont {Tamegai},
  \citenamefont {Xing},\ and\ \citenamefont {Shi}}]{zhou2021disorder}%
  \BibitemOpen
  \bibfield  {author} {\bibinfo {author} {\bibnamefont {Zhou}, \bibfnamefont
  {N.}}, \bibinfo {author} {\bibfnamefont {Y.}~\bibnamefont {Sun}}, \bibinfo
  {author} {\bibfnamefont {C.~Y.}\ \bibnamefont {Xi}}, \bibinfo {author}
  {\bibfnamefont {Z.~S.}\ \bibnamefont {Wang}}, \bibinfo {author}
  {\bibfnamefont {J.~L.}\ \bibnamefont {Zhang}}, \bibinfo {author}
  {\bibfnamefont {Y.}~\bibnamefont {Zhang}}, \bibinfo {author} {\bibfnamefont
  {Y.~F.}\ \bibnamefont {Zhang}}, \bibinfo {author} {\bibfnamefont {C.~Q.}\
  \bibnamefont {Xu}}, \bibinfo {author} {\bibfnamefont {Y.~Q.}\ \bibnamefont
  {Pan}}, \bibinfo {author} {\bibfnamefont {J.~J.}\ \bibnamefont {Feng}},
  \bibinfo {author} {\bibfnamefont {Y.}~\bibnamefont {Meng}}, \bibinfo {author}
  {\bibfnamefont {X.~L.}\ \bibnamefont {Yi}}, \bibinfo {author} {\bibfnamefont
  {L.}~\bibnamefont {Pi}}, \bibinfo {author} {\bibfnamefont {T.}~\bibnamefont
  {Tamegai}}, \bibinfo {author} {\bibfnamefont {X.}~\bibnamefont {Xing}}, and\
  \bibinfo {author} {\bibfnamefont {Z.}~\bibnamefont {Shi}}} (\bibinfo {year}
  {2021}),\ \href {https://doi.org/10.1103/PhysRevB.104.L140504} {\bibfield
  {journal} {\bibinfo  {journal} {Phys. Rev. B}\ }\textbf {\bibinfo {volume}
  {104}},\ \bibinfo {pages} {L140504}}\BibitemShut {NoStop}%
\bibitem [{\citenamefont {Zhou}\ \emph {et~al.}(2017)\citenamefont {Zhou},
  \citenamefont {Eckberg}, \citenamefont {Wilfong}, \citenamefont {Liou},
  \citenamefont {Vivanco}, \citenamefont {Paglione},\ and\ \citenamefont
  {Rodriguez}}]{Zhou2017}%
  \BibitemOpen
  \bibfield  {author} {\bibinfo {author} {\bibnamefont {Zhou}, \bibfnamefont
  {X.}}, \bibinfo {author} {\bibfnamefont {C.}~\bibnamefont {Eckberg}},
  \bibinfo {author} {\bibfnamefont {B.}~\bibnamefont {Wilfong}}, \bibinfo
  {author} {\bibfnamefont {S.-C.}\ \bibnamefont {Liou}}, \bibinfo {author}
  {\bibfnamefont {H.~K.}\ \bibnamefont {Vivanco}}, \bibinfo {author}
  {\bibfnamefont {J.}~\bibnamefont {Paglione}}, and\ \bibinfo {author}
  {\bibfnamefont {E.~E.}\ \bibnamefont {Rodriguez}}} (\bibinfo {year} {2017}),\
  \href {https://doi.org/10.1039/c6sc05268a} {\bibfield  {journal} {\bibinfo
  {journal} {Chemical Science}\ }\textbf {\bibinfo {volume} {8}},\ \bibinfo
  {pages} {3781}}\BibitemShut {NoStop}%
\bibitem [{\citenamefont {Zhu}\ \emph {et~al.}(2020)\citenamefont {Zhu},
  \citenamefont {Kong}, \citenamefont {Cao}, \citenamefont {Chen},
  \citenamefont {Papaj}, \citenamefont {Du}, \citenamefont {Xing},
  \citenamefont {Liu}, \citenamefont {Wang}, \citenamefont {Shen},
  \citenamefont {Yang}, \citenamefont {Schneeloch}, \citenamefont {Zhong},
  \citenamefont {Gu}, \citenamefont {Fu}, \citenamefont {Zhang}, \citenamefont
  {Ding},\ and\ \citenamefont {Gao}}]{zhu2020nearly}%
  \BibitemOpen
  \bibfield  {author} {\bibinfo {author} {\bibnamefont {Zhu}, \bibfnamefont
  {S.}}, \bibinfo {author} {\bibfnamefont {L.}~\bibnamefont {Kong}}, \bibinfo
  {author} {\bibfnamefont {L.}~\bibnamefont {Cao}}, \bibinfo {author}
  {\bibfnamefont {H.}~\bibnamefont {Chen}}, \bibinfo {author} {\bibfnamefont
  {M.}~\bibnamefont {Papaj}}, \bibinfo {author} {\bibfnamefont
  {S.}~\bibnamefont {Du}}, \bibinfo {author} {\bibfnamefont {Y.}~\bibnamefont
  {Xing}}, \bibinfo {author} {\bibfnamefont {W.}~\bibnamefont {Liu}}, \bibinfo
  {author} {\bibfnamefont {D.}~\bibnamefont {Wang}}, \bibinfo {author}
  {\bibfnamefont {C.}~\bibnamefont {Shen}}, \bibinfo {author} {\bibfnamefont
  {F.}~\bibnamefont {Yang}}, \bibinfo {author} {\bibfnamefont {J.}~\bibnamefont
  {Schneeloch}}, \bibinfo {author} {\bibfnamefont {R.}~\bibnamefont {Zhong}},
  \bibinfo {author} {\bibfnamefont {G.}~\bibnamefont {Gu}}, \bibinfo {author}
  {\bibfnamefont {L.}~\bibnamefont {Fu}}, \bibinfo {author} {\bibfnamefont
  {Y.-Y.}\ \bibnamefont {Zhang}}, \bibinfo {author} {\bibfnamefont
  {H.}~\bibnamefont {Ding}}, and\ \bibinfo {author} {\bibfnamefont {H.-J.}\
  \bibnamefont {Gao}}} (\bibinfo {year} {2020}),\ \href
  {https://doi.org/10.1126/science.aax0274} {\bibfield  {journal} {\bibinfo
  {journal} {Science}\ }\textbf {\bibinfo {volume} {367}},\ \bibinfo {pages}
  {189}}\BibitemShut {NoStop}%
\bibitem [{\citenamefont {Zhu}\ \emph {et~al.}(2009{\natexlab{a}})\citenamefont
  {Zhu}, \citenamefont {Han}, \citenamefont {Cheng}, \citenamefont {Mu},
  \citenamefont {Shen}, \citenamefont {Zeng},\ and\ \citenamefont
  {Wen}}]{Zhu2009}%
  \BibitemOpen
  \bibfield  {author} {\bibinfo {author} {\bibnamefont {Zhu}, \bibfnamefont
  {X.}}, \bibinfo {author} {\bibfnamefont {F.}~\bibnamefont {Han}}, \bibinfo
  {author} {\bibfnamefont {P.}~\bibnamefont {Cheng}}, \bibinfo {author}
  {\bibfnamefont {G.}~\bibnamefont {Mu}}, \bibinfo {author} {\bibfnamefont
  {B.}~\bibnamefont {Shen}}, \bibinfo {author} {\bibfnamefont {B.}~\bibnamefont
  {Zeng}}, and\ \bibinfo {author} {\bibfnamefont {H.-H.}\ \bibnamefont {Wen}}}
  (\bibinfo {year} {2009}{\natexlab{a}}),\ \href
  {https://doi.org/10.1016/j.physc.2009.03.029} {\bibfield  {journal} {\bibinfo
   {journal} {Physica C: Superconductivity}\ }\textbf {\bibinfo {volume}
  {469}},\ \bibinfo {pages} {381}}\BibitemShut {NoStop}%
\bibitem [{\citenamefont {Zhu}\ \emph {et~al.}(2009{\natexlab{b}})\citenamefont
  {Zhu}, \citenamefont {Han}, \citenamefont {Mu}, \citenamefont {Cheng},
  \citenamefont {Shen}, \citenamefont {Zeng},\ and\ \citenamefont
  {Wen}}]{Zhu2009b}%
  \BibitemOpen
  \bibfield  {author} {\bibinfo {author} {\bibnamefont {Zhu}, \bibfnamefont
  {X.}}, \bibinfo {author} {\bibfnamefont {F.}~\bibnamefont {Han}}, \bibinfo
  {author} {\bibfnamefont {G.}~\bibnamefont {Mu}}, \bibinfo {author}
  {\bibfnamefont {P.}~\bibnamefont {Cheng}}, \bibinfo {author} {\bibfnamefont
  {B.}~\bibnamefont {Shen}}, \bibinfo {author} {\bibfnamefont {B.}~\bibnamefont
  {Zeng}}, and\ \bibinfo {author} {\bibfnamefont {H.-H.}\ \bibnamefont {Wen}}}
  (\bibinfo {year} {2009}{\natexlab{b}}),\ \href
  {https://doi.org/10.1103/PhysRevB.79.220512} {\bibfield  {journal} {\bibinfo
  {journal} {Phys. Rev. B}\ }\textbf {\bibinfo {volume} {79}},\ \bibinfo
  {pages} {220512}}\BibitemShut {NoStop}%
\bibitem [{\citenamefont {Zhu}\ \emph {et~al.}(2009{\natexlab{c}})\citenamefont
  {Zhu}, \citenamefont {Han}, \citenamefont {Mu}, \citenamefont {Zeng},
  \citenamefont {Cheng}, \citenamefont {Shen},\ and\ \citenamefont
  {Wen}}]{Zhu2009a}%
  \BibitemOpen
  \bibfield  {author} {\bibinfo {author} {\bibnamefont {Zhu}, \bibfnamefont
  {X.}}, \bibinfo {author} {\bibfnamefont {F.}~\bibnamefont {Han}}, \bibinfo
  {author} {\bibfnamefont {G.}~\bibnamefont {Mu}}, \bibinfo {author}
  {\bibfnamefont {B.}~\bibnamefont {Zeng}}, \bibinfo {author} {\bibfnamefont
  {P.}~\bibnamefont {Cheng}}, \bibinfo {author} {\bibfnamefont
  {B.}~\bibnamefont {Shen}}, and\ \bibinfo {author} {\bibfnamefont {H.-H.}\
  \bibnamefont {Wen}}} (\bibinfo {year} {2009}{\natexlab{c}}),\ \href
  {https://doi.org/10.1103/PhysRevB.79.024516} {\bibfield  {journal} {\bibinfo
  {journal} {Phys. Rev. B}\ }\textbf {\bibinfo {volume} {79}},\ \bibinfo
  {pages} {024516}}\BibitemShut {NoStop}%
\bibitem [{\citenamefont {Zou}\ \emph {et~al.}(2021)\citenamefont {Zou},
  \citenamefont {Fu}, \citenamefont {Wu}, \citenamefont {Li}, \citenamefont
  {Parker}, \citenamefont {Sefat},\ and\ \citenamefont
  {Gai}}]{zou2021competitive}%
  \BibitemOpen
  \bibfield  {author} {\bibinfo {author} {\bibnamefont {Zou}, \bibfnamefont
  {Q.}}, \bibinfo {author} {\bibfnamefont {M.}~\bibnamefont {Fu}}, \bibinfo
  {author} {\bibfnamefont {Z.}~\bibnamefont {Wu}}, \bibinfo {author}
  {\bibfnamefont {L.}~\bibnamefont {Li}}, \bibinfo {author} {\bibfnamefont
  {D.~S.}\ \bibnamefont {Parker}}, \bibinfo {author} {\bibfnamefont {A.~S.}\
  \bibnamefont {Sefat}}, and\ \bibinfo {author} {\bibfnamefont
  {Z.}~\bibnamefont {Gai}}} (\bibinfo {year} {2021}),\ \href
  {https://doi.org/10.1038/s41535-021-00385-8} {\bibfield  {journal} {\bibinfo
  {journal} {npj Quantum Mater.}\ }\textbf {\bibinfo {volume} {6}},\ \bibinfo
  {pages} {89}}\BibitemShut {NoStop}%
\bibitem [{\citenamefont {Zou}\ \emph {et~al.}(2014)\citenamefont {Zou},
  \citenamefont {Feng}, \citenamefont {Logg}, \citenamefont {Chen},
  \citenamefont {Lampronti},\ and\ \citenamefont {Grosche}}]{zou2014fermi}%
  \BibitemOpen
  \bibfield  {author} {\bibinfo {author} {\bibnamefont {Zou}, \bibfnamefont
  {Y.}}, \bibinfo {author} {\bibfnamefont {Z.}~\bibnamefont {Feng}}, \bibinfo
  {author} {\bibfnamefont {P.~W.}\ \bibnamefont {Logg}}, \bibinfo {author}
  {\bibfnamefont {J.}~\bibnamefont {Chen}}, \bibinfo {author} {\bibfnamefont
  {G.}~\bibnamefont {Lampronti}}, and\ \bibinfo {author} {\bibfnamefont
  {F.~M.}\ \bibnamefont {Grosche}}} (\bibinfo {year} {2014}),\ \href
  {https://doi.org/10.1002/pssr.201409418} {\bibfield  {journal} {\bibinfo
  {journal} {Phys. Status Solidi RRL}\ }\textbf {\bibinfo {volume} {8}},\
  \bibinfo {pages} {928–930}}\BibitemShut {NoStop}%
\end{thebibliography}

%

\end{document}